\documentclass[review]{elsarticle}

\biboptions{sort&compress}

\usepackage[colorlinks,citecolor=blue,linktoc=all,linkcolor=cyan]{hyperref}
\usepackage{graphicx}

\usepackage[T1]{fontenc}
\usepackage{dsfont}               
\usepackage{mathrsfs}             
\usepackage{slashed}              
\usepackage{amsmath}
\usepackage{amssymb}
\usepackage{amsbsy}
\usepackage{amsfonts,bm,braket}
\usepackage{bbold}
\usepackage{subcaption,wrapfig}

\numberwithin{equation}{section}
\numberwithin{table}{section}
\numberwithin{figure}{section}

\journal{Progress in Particle and Nuclear Physics}

\usepackage{titlesec}
\usepackage{sectsty}
\titleformat{\section}{\normalfont\Large\bfseries}{\thesection}{1em}{}
\titleformat{\subsection}{\normalfont\large\bfseries}{\thesubsection}{1em}{}
\titleformat{\subsubsection}{\normalfont\normalsize\bfseries}{\thesubsubsection}{1em}{}
\def\be{\begin{eqnarray}}
	\def\ee{\end{eqnarray}}
\newcommand{\Tr}{\rm Tr}
\def\lsim{\raise0.3ex\hbox{$<$\kern-0.75em\raise-1.1ex\hbox{$\sim$}}}
\def\gsim{\raise0.3ex\hbox{$>$\kern-0.75em\raise-1.1ex\hbox{$\sim$}}}
\def\({\left(}
\def\){\right)}
\def\hmu{\hat\mu}
\def\bea {\begin{eqnarray}}
\def\eea {\end{eqnarray}}
\def\sumintb{\sum\!\!\!\!\!\!\!\!\!\int\limits}
\def\sumintf{\sum\!\!\!\!\!\!\!\!\!\!\int\limits}
\def\sumintbb{\sum\!\!\!\!\!\!\!\!\!\int\limits}

\def\Za{\frac{\zeta'(-1)}{\zeta(-1)}}
\def\Zc{\frac{\zeta'(-3)}{\zeta(-3)}}

\def \del{\partial}
\def\mn {\mu\nu}
\def\sp{\shortparallel}
\def\om {\omega}
\def\ti {\tilde}
\def\mn {\mu\nu}
\def\mr {\mu\rho}

\def\nn{\nonumber\\}

\def\L{\ln\frac{\hat\Lambda}{2}}
\def\Lg{\ln\frac{\hat\Lambda_g}{2}}
\def\[{\left[}
\def\]{\right]}  
\allowdisplaybreaks

\begin{document}
	\setlength{\abovedisplayskip}{2.pt}
	\setlength{\belowdisplayskip}{2.pt}
	
\begin{frontmatter}
		
\title{Hard Thermal Loop - theory and applications}
\author[nhad]{Najmul Haque\corref{mycorrespondingauthor}}
\cortext[mycorrespondingauthor]{Corresponding author}
\ead{nhaque@niser.ac.in}
\address[nhad]{School of Physical Sciences, National Institute of Science Education and Research,
	An OCC of Homi Bhabha National Institute, Jatni-752050, India}
		
\author[mgmad1,mgmad2]{Munshi G.  Mustafa}

\address[mgmad1]{Visiting Professor, Department of Physics,  IIT Bombay, Pawai,  Mumbai 400076, India}
\address[mgmad2]{Ex-Senior Professor, Theory Division, Saha Institute of Nuclear Physics, Homi Bhabha National Institute, 1/AF, Bidhannagar, Kolkata 700064, India}


\begin{abstract}
In this review, we present the key aspects of modern thermal perturbation theory based on the hard thermal loop (HTL) approximation, including its theoretical foundations and applications within quantum electrodynamics (QED) and quantum chromodynamics (QCD) plasmas. To maintain conciseness, we focus on scenarios in thermal equilibrium, examining a variety of physical quantities and settings. Specifically, we explore both bulk thermodynamic properties and real-time observables in high-temperature domains relevant to heavy-ion physics.
\end{abstract}
		
\begin{keyword}
Finite temperature field theory\sep Matsubara sum\sep Hard Thermal Loop Resummation \sep Quantum Chromodynamics (QCD) \sep QCD Thermodynamics  \sep Relativistic Heavy-Ion collisions \sep Quark-Gluon Plasma (QGP)
\end{keyword}
		
	\end{frontmatter}
	
\thispagestyle{empty}
\setcounter{tocdepth}{2}
\tableofcontents

	\section{Introduction}\label{intro}
	\vspace{-0.2cm}
In this review, we aim to explore the properties of matter under extreme conditions, such as high temperature and/or density, where both relativistic and quantum mechanical effects become significant. Quantum field theory (QFT) provides the natural framework for studying matter in such conditions. The conventional field theory is formalised in vacuum at zero temperature, providing a framework to describe a wide range of phenomena observed in experiments --- a tool for tackling complex many-body problems. The theoretical predictions under this framework have been successful in describing experimental data. Moreover, with appropriate modifications, it also holds relevance in atomic, nuclear, and condensed matter physics.

In the conditions described above, thermal effects are important and the particles propagate in a medium of other particles excited in these conditions instead of the vacuum.  It is quite obvious to ponder when and what extent effects arising due to thermal background are relevant, and which new phenomena could arise due to it.  Therefore, we are led into the framework of thermal field theory~\cite{Matsubara:1955ws,Schwinger:1960qe,Keldysh:1964ud,Ghiglieri:2020dpq}. It is a framework to deal with complicated many body problems among its components  at finite temperature and chemical potential. 

Thermal field theory is in itself a very broad subject. Conditions described above can arise in the early universe and cosmology, heavy-ion collisions in terrestrial particle accelerators and compact astrophysical objects like neutron star cores or neutron star mergers. Thermal field theory has also been used in symmetry restoration in theories with spontaneously broken symmetry, in thermal neutrino production and oscillations, leptogenesis, thermal axion production,   ${\cal N}=4$ supersymmetry Yang-Mills theory, string theory, blackhole physics and anti de-sitter space/conformal field theory (Ads/CFT) correspondence. It has extensively been used in condensed matter physics. For detail references of the above phenomena, we refer to another review by one of us~\cite{Mustafa:2022got}.

We know that the perturbation theory, or more broadly a class of weak-coupling methods, based on expanding the functional integrals that define different physical quantities in powers (and logarithms) of a coupling constant. In bare perturbation theory (BPT), both static and dynamic quantities can be computed by expanding in coupling constant around the free theory. This approach works in hard scale ($p \sim 2\pi T\sim T$) regime that uses free propagators and vertices, and the contribution appears in even power of coupling ($g^{2n}$, $n$ is the number of loop). However, it becomes apparent that a strict expansion in the coupling constant only converges for temperatures several orders of magnitude higher than those relevant for any physical theory. The source of the poor convergence arises from contributions originating from soft momenta. Also the simplistic application of BPT often leads to infrared and/or collinear singularities, with the magnitude and sign sometimes becoming gauge-dependent. It is worth noting that the infrared problems are associated with bosonic excitations but not with fermionic excitations, as the fermionic expansion parameter stays finite for $p <T$. This in turn signals sensitivity to soft region of the phase space, where BPT breaks down. These issues are typically associated to contributions from soft collective excitations, characterised by momentum scales such as $gT$, $g^2T$ or with nearly collinear (small-angle scatterings). These observations collectively suggest that the BPT requires improvement to accurately compute static and dynamic quantities, where one needs a way of reorganizing the perturbative series to treats the soft sector with greater care. 

An important issue in perturbative calculations is the necessity of resumming diagrams of all loop orders to reach a result valid to a specific power of the coupling constant for a given theory. Furthermore, additional resummations are often necessary to remedy poorly converging results. This can be done through the effective field theories like dimensional reduction (DR) approach~\cite{Gross:1980br,Appelquist:1981vg,Nadkarni:1982kb,Nadkarni:1988fh,Nadkarni:1988pb} and hard thermal loop (HTL) resummation~\cite{Braaten:1989mz,Braaten:1991gm,Braaten:1990az,Taylor:1990ia,Frenkel:1991ts,Barton:1989fk} techniques. In HTL resummations, the scalar fields and the electric component of gauge fields are screened at the Debye scale, where momentum, $p\sim gT$. Conversely, the magnetic component of gauge field is screened only non-perturbatively~\cite{Linde:1978px,Linde:1980ts} at the scale where momentum, $p\sim g^2T$, known as the ultra-soft scale or non-perturbative magnetic scale, which must be determined non-perturbatively. In the present review, we will discuss both DR and HTL approach but mainly concentrate on HTL approximation including its theoretical 
foundation for theories, viz., scalar, QED and QCD. There are various ways of reorganising the finite temperature/chemical potential perturbative series. For scalar field theories, one can use $``$screened perturbation theory (SPT)~\cite{Karsch:1997gj,Chiku:1998kd,Andersen:2000yj,Andersen:2001ez,Andersen:2008bz}'' which was inspired in part by variational perturbation theory~\cite{Blaizot:1999ip,Blaizot:1999ap,Blaizot:2000fc}. For gauge theories, however, it is not possible to use a scalar gluon/photon mass.  As a result, a gauge-invariant generalisation of SPT called $``$hard-thermal-loop perturbation theory (HTLpt)~\cite{Andersen:1999fw,Andersen:1999sf,Andersen:1999va,Andersen:2002ey,Andersen:2003zk,Andersen:2009tw,Andersen:2009tc,Andersen:2010ct,Andersen:2010wu,Andersen:2011sf,Andersen:2011ug,Haque:2012my,Mogliacci:2013mca,Haque:2013qta,Haque:2013sja,Haque:2014rua,Andersen:2015eoa}'' was developed. The HTLpt framework allows for systematic analytic reorganisation of perturbative series based on the HTL effective Lagrangian. Moreover, it proves highly effective in computing both static and dynamical quantities.
 
 This review is structured as follows: Section~\ref{ftft} outlines thermal field theory in both imaginary and real-time formalisms. We then discuss methods for performing frequency summations in imaginary time using the contour integral and Saclay methods, along with exploring the relationship between functional integration and the partition function. Additionally, we examine the general structure of fermion and gauge boson two-point functions, scale separation, and the limitations of BPT at finite temperature. In Section~\ref{eft}, we delve into effective field theories (DR and HTL approaches) at finite temperature. Section~\ref{scalar_htl} focuses on scalar HTL resummation of $N$-point functions, leading to the derivation of the HTL-improved scalar Lagrangian. Continuing this approach, Section~\ref{qed_htl}  presents QED $N$-point functions via HTL resummation, yielding the HTL-improved QED Lagrangian. Section~\ref{qcd_htl} extends this analysis to calculate QCD $N$-point functions using HTL resummation, resulting in the HTL-improved QCD Lagrangian. Building on the HTL-improved Lagrangian for scalar, QED, and QCD theories, Section~\ref{ipt_chap} develops an improved perturbation theory. In Section~\ref{em_chap}, we employ HTLpt to investigate electromagnetic particle production (real and virtual photons), followed by an exploration of mesonic spectral and correlation functions in Section~\ref{meson}. Section~\ref{chapter:thermodynamics} addresses QCD thermodynamics, while Sections~\ref{damp} and~\ref{el} discuss the damping rates and energy loss of heavy leptons and partons in both QED and QCD plasmas. Finally, Section~\ref{sec:sum} offers a summary and outlook of the present review.
%
%
	
	\section{Finite Temperature Field Theory}
	\label{ftft}
	\vspace{-0.2cm}
Thermal Field Theory combines Quantum Field Theory and Statistical Mechanics to address complex many-body problems involving interactions among components at finite temperature and chemical potential. It serves as a framework to describe a large ensemble of multiple interacting particles (including gauge interactions) in a thermal environment. It has also been used to describe the occurance of new processes that were not present in vacuum field theory, phase transition involving symmetry restoration in theories with spontaneously broken symmetries, understanding the early universe's evolution, studying black hole physics (Hawking radiation), exploring the AdS/CFT correspondence, and analyzing phase transitions in condensed matter systems. Notably, thermal field theory has found extensive use in probing the physical properties of Quark-Gluon Plasma (QGP) formed during high-energy relativistic heavy-ion collisions. The QGP is a thermalised state of matter in which quasi-free quarks and gluons are deconfined from hadrons, thus allowing colour degrees of freedom to manifest over a significantly larger volume than the mere hadronic volume. In order to understand the properties of a QGP and to make unambiguous predictions about the signature of QGP formation, a thorough understanding of QGP is indispensable. For this purpose, one needs to use Quantum Chromodynamics (QCD) at finite temperature and chemical potential. 

There are {\sf two} well defined Formalism in Thermal Field Theory, namely,
(i) Imaginary time (Matsubara) formalism~\cite{Matsubara:1955ws}, and (ii) Real time (Schwinger-Keldysh) formalism~\cite{Schwinger:1960qe,Moller:1960cva,Keldysh:1964ud}.
\subsection{Imaginary Time Formalism}
\label{ift}
\subsubsection{Connection to imaginary time and Matsubara formalism}
\label{connection}
For a given Schr\"odinger operator, ${\cal A}$, the Heisenberg operator, ${\cal A}_H(t)$ can be written as
\begin{equation}
{\cal A}_H(t) = e^{i{\cal H}t} \, {\cal A} \,  e^{-i{\cal H}t} \, \, .
\label{eq15}
\end{equation}
The thermal correlation function of two operators can also be written~\cite{Das:1997gg,Mustafa:2022got}  as
\begin{eqnarray}
\langle {\cal A}_H(t) {\cal B}_H(t') \rangle_\beta  
&=& {\cal Z}^{-1}(\beta) {\rm{Tr}}\left [ e^{-\beta {\cal H}} {\cal A}_H(t) 
{\cal B}_H(t') \right ] 
= \langle {\cal B}_H(t') {\cal A}_H(t+i\beta) \rangle_\beta
\, \, . \label{eq16}
\end{eqnarray}
To obtain Eq.~\eqref{eq16}, we have used Eq.~\eqref{eq15} and inserted the unit operator $1 =  e^{-\beta {\cal H}} e^{\beta {\cal H}}$ and used the cyclic properties of the trace. The Eq.~\eqref{eq16} is known as Kubo-Martin-Schwinger (KMS) relation. The relation Eq.~\eqref{eq16} holds irrespective of Grassmann parities of the operators, {\it viz.}, for bosonic as well as fermionic operator. This KMS relation will lead to periodicity for boson and anti-periodicity for fermions because of commutation and  anti-commutation relations, respectively. It is noteworthy that the evolution operator $e^{-\beta {\cal H}}$ acquires the form of a time evaluation operator ($e^{-i{\cal H}t}$) for imaginary time ($\beta=it$) through analytic continuation. While this may seem coincidental, there might exist a deeper connection that remains undiscovered.

		\begin{wrapfigure}{r}{0.35\textwidth}
			\begin{center}
								\vspace{-0.7cm}
				\includegraphics[width=5cm,height=4cm]{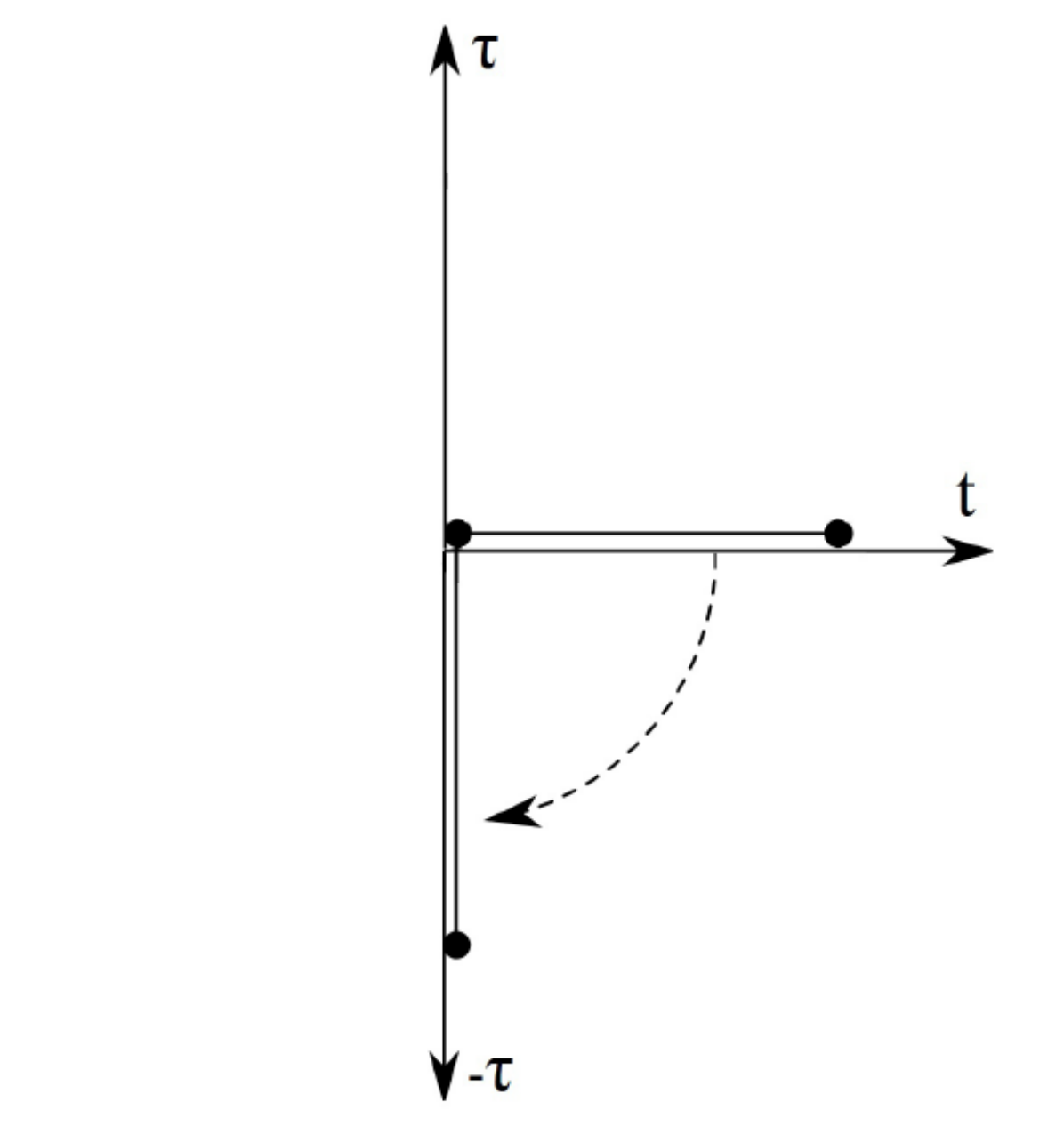}
				\vspace{-0.5cm}
				\caption{The Wick rotation in the  imaginary time axis: $t=-i\tau$.}
				\label{wick_rot}
			\end{center}
		\end{wrapfigure}
This  implies that the temperature and the imaginary time is related by $\beta=it$ as shown in Fig.~\ref{wick_rot}. This is known as {\it Wick rotation}. As $\beta=1/T, \beta$ becomes finite at finite temperature. By employing the time evolution operator, one can derive the  $S-$Matrix, Feynman rules, and Feynman diagrams.
The Matsubara (imaginary time) formalism provides a means of evaluating the partition function and other quantities through a diagrammatic method as done zero-temperature Field Theory.
\subsubsection{Periodicity (Anti-periodicity) of the Green's function}
\label{period}
The thermal Green's function~\cite{Das:1997gg,Mustafa:2022got} for $\tau > \tau'$:
\begin{eqnarray}
G_\beta(\bm{\vec x},\bm{\vec x}';\tau,\tau')&=& {\cal Z}^{-1}(\beta) \ {\rm{Tr}} 
\left ( e^{-\beta {\cal H}} {\cal T}\left [ \Phi_H(\bm{\vec x},\tau) \Phi_H(\bm{\vec x}',\tau') \right]\right ) 
= \pm \ G_\beta(\bm{\vec x},\bm{\vec x'};\tau ,\tau'+\beta),
\label{eq26}
\end{eqnarray}
where we have used same procedure as \eqref{eq16}  and  the time evolution of the state:
 $\Phi_H(\bm{\vec x'},\tau'+\beta)= e^{\beta {\cal H}} \Phi_H(\bm{\vec x'},\tau') e^{-\beta {\cal H}} $ to obtain~\eqref{eq26}.
Since the Green's function changes sign for Dirac field after one period of $\beta$ becomes negative, 
this indicates that the Dirac fields must be antiperiodic in imaginary time whereas bosonic fields becomes positive and are periodic as they do not change sign: $\Phi(\bm{\vec x}, \tau)= \pm \Phi(\bm{\vec x}, \tau+\beta)$.    
Additionally, we didn't touch upon spatial direction, it remains unaffected: $-\infty \le {x}\le \infty \Rightarrow$ open.
Now, in $T=0$ (Minkowski space-time), both space and time remain open: $-\infty \le {x}\le \infty$ and $-\infty\le \tau \le \infty$. Topology: Structure of space-time at $R^4=R^3\times R^1$ (both space and time open) and both space and time are in equal footing. Conversely, in $T\ne 0 $ (Euclidean space; imaginary time):  space remains open $-\infty \le {x}\le \infty \Rightarrow R^3$.; time remains closed: $0\le \tau \le \beta$ $\Rightarrow R^1 \rightarrow S^1$ (circle). Topology: Structure of space-time at $T\ne0$  $R^4=R^3\times R^1 \Rightarrow R^3\times S^1$. This changes the temporal components leaving the spatial components unchanged, leading to the  decoupling of space and time and causing the theory to lose Lorentz invariance.
Moreover, the introduction of a chemical potential can be achieved by modifying the temporal component of the gauge field, using a substitution $\partial_0-i\mu$  in the Lagrangian. This substitution affects only the temporal component, leaving the spatial components unchanged, while also decoupling space and time, thus rendering the theory that is no more Lorentz invariant. Besides explicitly breaking Lorentz invariance, the presence of a chemical potential may also break other internal symmetries.
Now, at $T $ and  $ \mu \ne 0$  field theory is equivalent to quantising a quantum system in finite box $\Rightarrow$ one dimensional box in $\tau$ direction ($0\le \tau \le \beta$) but space remains open, i.e., $ R^3\times S^1$ in which Lorentz invariance is broken. 
This finiteness of time leads to discrete Matsubara frequency $\omega_n$~\cite{Das:1997gg,Mustafa:2022got} as the characteristics of the imaginary time and is represented by 
\be
k_0=ik_4=i\omega_n= \left \{ \begin{array}{ll}
                  \frac{2n \pi i }{\beta} & \mbox{for boson ,} \nonumber \\
                  \frac{(2n+1)\pi i}{\beta} & \mbox{for fermion.}\nonumber
		   \end{array} 
		\right.  \label{eq37} 
\ee
\subsubsection{Frequency sum in contour integral method}
\label{freq_sum}
To compute the partition function and matrix element corresponding to a given Feynman diagram in a theory,it is necessary to perform frequency sums. There are two types of frequency sums: bosonic and fermionic.

\noindent{\bf A) Bosonic frequency Sum:}

\begin{figure}[h]
\hspace{.5in}
\includegraphics[width=15cm,height=4.5cm]{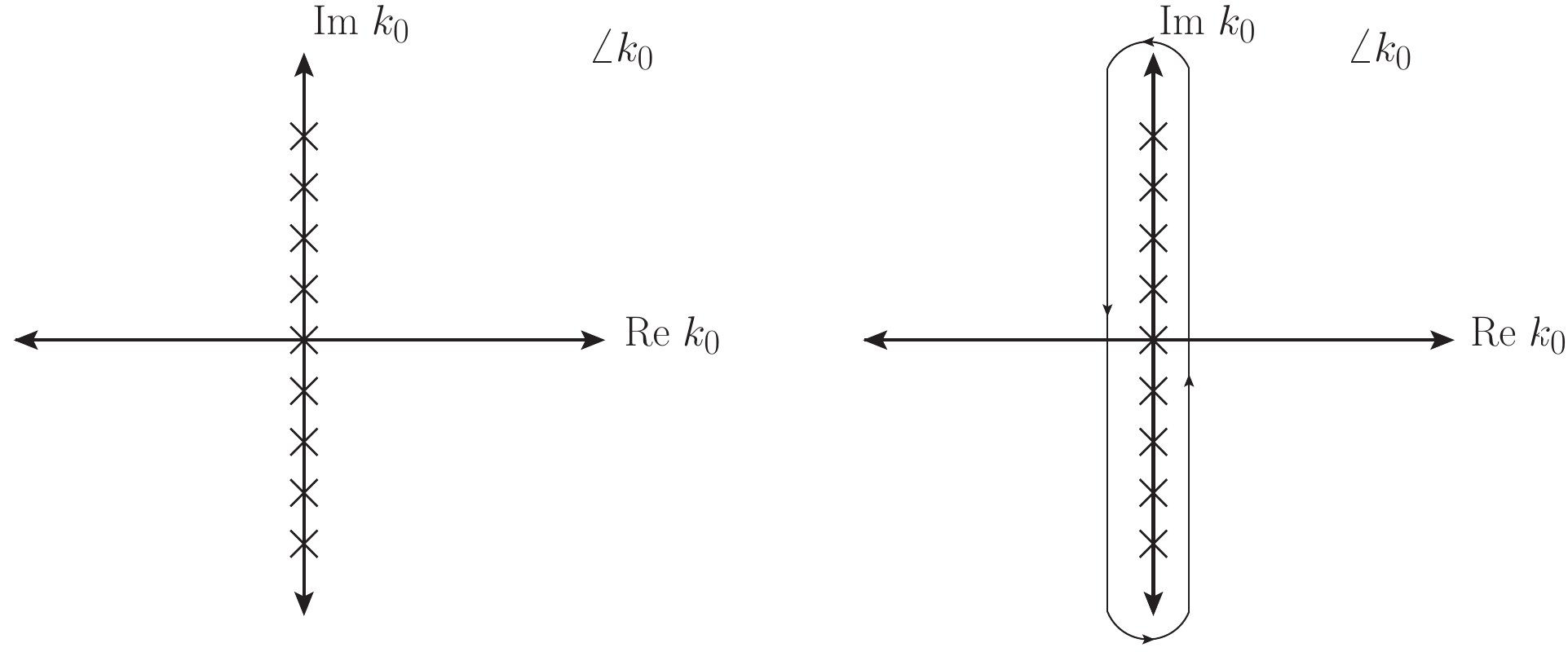}
\caption{Poles of $\coth(\beta k_0/2)$  at $k_0=2\pi i n T; n=0, \pm1, \pm 2 \cdots $, in complex $k_0$ plane.}
\label{coth_poles_fig}
\end{figure}

In general,  the form of the bosonic frequency sum can be written as
\be
\frac{1}{\beta} \sum_{n=-\infty}^{n=+\infty} f(k_0=i\omega_n=2\pi i n T) 
=\frac{1}{\beta} \sum_{n=-\infty}^{n=+\infty} f(k_0=i\omega_n=2\pi i n T) \,{\rm{Res}}\left[ \frac{\beta}{2} \coth\left ( \frac{\beta k_0}{2}\right )\right], \label{sum_intb0}
\ee
where $k_0$ is the fourth (temporal) component of momentum in Minkowski space-time and the function $f(k_0)$ is a meromorphic function
We know the hyperbolic cotangent has poles at $\coth(n\pi i)$ (see Fig.~\ref{coth_poles_fig}) at 
\be
\coth\left ( \frac{\beta k_0}{2}\right ) = \coth(n\pi i) \, \, \Rightarrow \, \, k_0= \frac{2\pi i n }{\beta} = i\omega_n
\ee
with residues $2/\beta$, and ${\rm{Res}}\left[\frac{\beta}{2} \coth\left ( \frac{\beta k_0}{2}\right )\right]$ will lead to unity. Therefore,  one can insert  hyperbolic cotangent with suitable argument.
Then one can write the RHS of~\eqref{sum_intb0} without any loss of generality as 
\begin{eqnarray}
\frac{1}{\beta} \sum_{n=-\infty}^{n=+\infty} f(k_0) \,\textrm{Res}\left[  \frac{\beta}{2} \coth\left ( \frac{\beta k_0}{2}\right )\right] 
&=&\frac{1}{\beta} \sum_{n=-\infty}^{n=+\infty} \frac{\beta}{2} \ {\rm{Res}}\left[ f(k_0) \,  \coth\left ( \frac{\beta k_0}{2}\right );
 \, \, \Rightarrow \, {\rm {poles:}} \, \, k_0= i\omega_n  = \frac{2\pi i n }{\beta} \right ].
\end{eqnarray}
Employing the residue theorem in reverse, the sum over residues is possible to expresse as an integral over a contour 
$C$  in $\angle k_0 $ enclosing the poles of the meromorphic function $f(k_0)$ but excluding the poles of  the hyperbolic cotangent ($k_0=i\omega_n=2\pi i T$) as~\cite{Kapusta:2006pm,Mustafa:2022got}
\begin{eqnarray}
\frac{1}{\beta} \sum_{n=-\infty}^{n=+\infty} f(k_0=i\omega_n) &=&
 \frac{1}{\beta} \sum_{n=-\infty}^{n=+\infty} {\rm{Res}} \left[f(k_0) \, \frac{\beta}{2} \coth\left ( \frac{\beta k_0}{2}\right )\right]  
 = \frac{1}{2\pi i} \oint_{C_1 \cup C_2} dk_0 \, f(k_0) \frac{1}{2} \, \coth\left ( \frac{\beta k_0}{2}\right ) \nn
 \!&=& \!\!
\frac{1}{2\pi i} \int_{-i\infty}^{+i\infty} \,  dk_0 \, \frac{1}{2}\Big [f(k_0)+f(-k_0)\Big] 
 +  \frac{1}{2\pi i} \int_{\epsilon-i\infty}^{\epsilon+i\infty} \,  dk_0 \Big [f(k_0)+f(-k_0)\Big] 
\frac{1}{{\exp(\beta k_0) - 1}} .
 \label{sum_intb2}
\end{eqnarray}

\begin{figure}[h]
\begin{center}
\includegraphics[width=6cm,height=5cm]{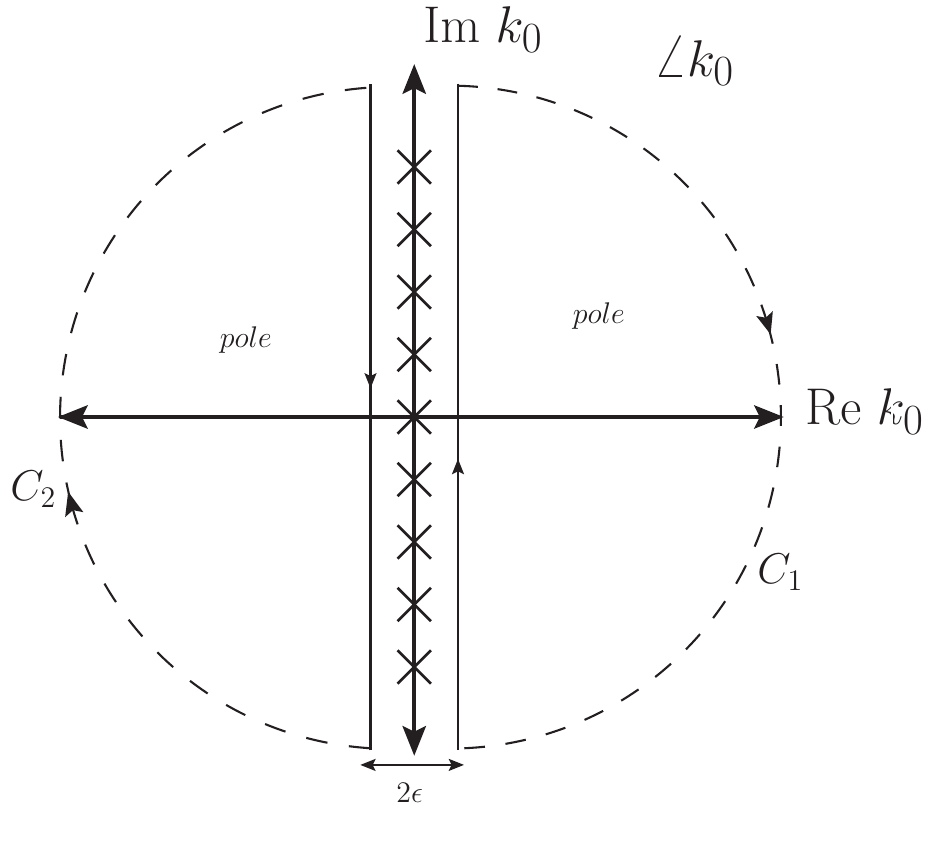}
\vspace*{-0.5cm}
\caption{Contours  $C_1$ and $C_2$ that include the poles of the meromorphic function  $f(k_0)$ 
in complex $k_0$ plane. The contours are also shifted by an amount  $\pm \epsilon$ from the  ${\rm{Im}} k_0$ line to exclude the poles of $\coth(\beta k_0/2)$ at $k_0=2\pi i n T$. }
\label{coth_poles_fig1}
\end{center}
\end{figure}

Now, some important points to note in Eq.~\eqref{sum_intb2}:
\begin{enumerate}
\item [$\bullet$] $\left[\exp(\beta k_0) - 1\right ]^{-1}$ vis-a-vis $\coth(\beta k_0/2)$ has series of poles 
at $k_0=i\omega_n=2\pi i T$ and is bounded and  analytic everywhere except at poles.
 
\item [$\bullet$] The function $f(k_0=i\omega_n)$ is a meromorphic and it has simple poles without any essential singularities or branch cuts.

\item [$\bullet$] The simple poles of $f(k_0=i\omega_n)$ should not coincide with the series of poles of $\left[\exp(\beta k_0) - 1\right ]^{-1}$
$\Rightarrow \, f(k_0=i\omega_n)$ should not have singularity along the imaginary $k_0$ axis.

\item[$\bullet$] The contour $C$ can be divided into two half circles in complex $k_0$ plane  $C_1$ and $C_2$ ~\cite{Kapusta:2006pm} 
without enclosing the poles of the  $\left[\exp(\beta k_0) - 1\right ]^{-1}$ vis-a-vis $\coth(\beta k_0/2)$ but
the contours $C_1$ and $C_2$ should enclose poles of $f(k_0)$ as shown in Fig.~\ref{coth_poles_fig1}, provided 
the meromorphic function $f(k_0)$ should decrease fast to achieve convergence.

\item [$\bullet$] If all these properties are satisfied, then $ T \sum_{n=-\infty}^{n=+\infty} f(k_0=i\omega_n)$ can be replaced by
contour integration and this is equivalent to  switching (analytically continuing) from Euclidean time (discrete frequency in Euclidean space) to real time (continuous frequency in Minkowski space-time).
\item[$\bullet$] In the second line of \eqref{sum_intb2},  the first term on the right-hand side represents the vacuum contribution, while the second term corresponds to the matter contribution.
\end{enumerate}
 {\bf B) Fermionic frequency sum for zero chemical potential ($\mu=0$):}
 
 Following the same way as the  bosonic case, the fermionic frequency sum with discrete frequency $\omega_n=(2n+1)  \pi  T$ can be obtained~\cite{Kapusta:2006pm,Mustafa:2022got}
\bea
\frac{1}{\beta} \sum_{n=-\infty}^{n=+\infty} f(k_0=i\om_n) &=& \frac{1}{2\pi i} \oint_{C_1 \cup C_2} dk_0 \, f(k_0) \frac{1}{2} \, 
\tanh\left ( \frac{\beta k_0}{2}\right )\nn
&&\hspace{-1.5cm}=\, \frac{1}{2\pi i} \int_{-i\infty}^{+i\infty} \,  d(k_0) \, \frac{1}{2}\Big [f(k_0)+f(-k_0)\Big] 
+  \frac{1}{2\pi i} \int_{\epsilon-i\infty}^{\epsilon+i\infty} \,  d(k_0) \Big [f(k_0)+f(-k_0)\Big] 
\frac{1}{{\exp(\beta k_0) + 1}} . \label{sum_intf0}
\eea

\vspace{0.2cm}
\noindent{\bf C) Fermionic frequency sum in presence of a chemical potential $\mu$:}\\
\noindent In the similar manner, one can obtain the fermionic frequency sum in presence of chemical potential $\mu$ as~\cite{Kapusta:2006pm}
\begin{eqnarray}
&&T \sum_{n=-\infty}^{n=+\infty} f(k_0=i\omega_n+\mu) =
 \frac{1}{2\pi i} \oint_{C} dk_0 \, f(k_0) \frac{1}{2} \, \tanh\left ( \frac{\beta k_0-\mu}{2}\right )  
=\frac{1}{2\pi i} \oint_{C} \,  d(k_0) \, f(k_0) +\frac{1}{2\pi i} \int_{-i\infty}^{+i\infty} \,  d(k_0) \, f(k_0) \nn
&&\hspace{3cm}\,-  \frac{1}{2\pi i} \int_{-i\infty+\mu+\epsilon}^{i\infty+\mu+\epsilon} \,  d(k_0)f(k_0) \frac{1}{{e^{\beta (k_0-\mu)} + 1}} 
 -  \frac{1}{2\pi i} \int_{-i\infty+\mu-\epsilon}^{i\infty+\mu-\epsilon} \,  d(k_0)  f(k_0)
\frac{1}{{e^{\beta(\mu- k_0)} + 1}} . \label{sum_intf7}
\end{eqnarray}
The first term corresponds to $T$-independent and yields the $T=0$ finite-density contributions whereas the second term corresponds to vacuum contribution. The last two terms correspond to fermion and antifermion contributions and vanish at $T=0$.

\subsubsection{Frequency sum in Saclay method}
\label{saclay}
Here, we present a convenient approach called Saclay method~\cite{Pisarski:1987wc} to evaluate the frequency sum involving two or more  propagators in a loop diagram. The scalar part of the bosonic and fermionic  propagator in momentum space are
\begin{eqnarray}
\Delta_{\left (B\atop F\right )} (K) &=& \frac{1}{K^2-m^2}=\frac{1} {k_0^2-E_k^2}
= \frac{1}{2E_k} \left (\frac{1}{k_0-E_k}-\frac{1}{k_0+E_k} \right ) 
= \sum_{s=\pm 1}  \frac{s}{2E} \,\, \frac{1}{k_0-sE_k} \, \label{sac0}
\end{eqnarray}
where $E_k=\sqrt{k^2+m^2}$ is the particle energy. Now, the bosonic propagator can also be written as
\begin{equation}
\Delta_B(K)=-\int_{0}^{\beta} d\tau e^{k_0\tau} \Delta_B(\tau,E_k) \, ,\label{sac1}
\end{equation}
where 
\begin{eqnarray}
\Delta_B(\tau,E_k) \!\!\!\! &=& \!\!\!\! -T\!\!\!\!  \sum_{k_0=2n\pi i T}\!\!\!\! \!\! e^{-k_0\tau} \Delta_B(K) \nn
&=& \frac{1}{2E_k} \Big [ \{ 1+n_B(E_k)\} e^{-E_k\tau} + n_B(E_k) e^{E_k\tau} \Big ] 
 = \sum_{s=\pm1} \frac{s}{2E_k} \Big [1+n_B(sE_k)\Big ] e^{-sE_k\tau}  . \ \ \ \  \label{sac2} 
\end{eqnarray}
The above equation is derived from~\eqref{sum_intb2} using contour integration. The Bose-Einstein distribution $n_B(E_k)$ is 
 given as
\begin{equation}
n_B(E_k)= \frac{1}{e^{\beta E_k}-1} \, . \label{sac3}
\end{equation}
The fermionic propagator is represented as
$
S(K) =(\slashed{K} +m)\Delta_F (K)\, . \label{sac4}
$
The quantity $\Delta_F(K)$ can be written in a mixed representation as
\begin{equation}
\Delta_F(K) = - \int_{0}^{\beta} d\tau e^{k_0\tau} \Delta_F(\tau,E_k) \, , \label{sac3a}
\end{equation}
where 
\begin{eqnarray}
\Delta_F(\tau, E_k) \!\!\!\!  &=& \!\!\!\!  -T\!\!\!\!\!\! \sum_{k_o=(2n+1)\pi i T}\!\! \!\!\!\!\!\! e^{-k_0\tau} \Delta_F(K) 
= \frac{1}{2E_k}\left[\left(1-n^+_F(E_k)\right)e^{-E_k\tau}-n_F^-(E_k)e^{E_k\tau}\right]
\!\! = \!\!\!\!  \sum_{s=\pm 1} \frac{s}{2E_k} \Big [1-n_F^+(sE_k)\Big ] e^{-sE_k\tau} ,\hspace{1cm} \label{sac4a}
\end{eqnarray}
where the Fermi-Dirac distribution is given as
$
n_F^\pm(E_k)= 1/\left(e^{\beta (E_k \mp \mu)}+1 \right)\, , \label{sac5}
$
where $(+)ve$ sign in the superscript  represents fermion whereas $(-)ve$ sign represents antifermion. Below we present the results of a few frequency sums in presence of chemical potential using Saclay method as those will be required to compute  2-point, 3-point and 4-point functions in theory like scalar, QED and QCD:
%
\begin{flalign}
&{\sf A) } \qquad T\sum_{k_0=(2n+1)\pi i T} \Delta_F(K) \Delta_F(Q=P-K) =-  \sum_{s_1,s_2} \frac{s_1s_2}{4E_kE_q}  \frac{1-n_F^+(s_1E_k)-n_F^-(s_2E_q)} {p_0-s_1E_k-s_2E_q} \, .& \label{sac6}
\end{flalign}
\vspace{-0.2cm}
\begin{flalign}
&{\sf B) }\qquad T\sum_{k_0=2n\pi i T} \Delta_B(K) \Delta_B(Q=P-K) =-  \sum_{s_1,s_2} \frac{s_1s_2}{4E_kE_q}  \frac{1+n_B(s_1E_k)+n_B(s_2E_q)} {p_0-s_1E_k-s_2E_q} \, .  & \label{sac6a}
\end{flalign}
The Eq.~\eqref{sac6a} is obtained  by replacing $n_F^\pm=-n_B $ in \eqref{sac6}.
\vspace{0.0cm}
%
 \begin{flalign}
&{\sf C)} \qquad T\sum_{k_0=(2n+1)\pi i T}k_0 \Delta_F(K) \Delta_F(Q=P-K) = - \sum_{s_1,s_2} \frac{s_2}{4E_q}  \frac{1-n_F^+(s_1E_k)-n_F^-(s_2E_q)} {p_0-s_1E_k-s_2E_q} \, .& \label{sac9}
 \end{flalign}
 \vspace{-0.2cm}
 %
  \begin{flalign}
&{\sf D)} \qquad T\sum_{k_0=2n\pi i T}k_0 \Delta_B(K) \Delta_B(Q=P-K)  = - \sum_{s_1,s_2} \frac{s_2}{4E_q}  \frac{1+n_B(s_1E_k)+n_B(s_2E_q)} {p_0-s_1E_k-s_2E_q} \, . &\label{sac9a}
 \end{flalign}
  \vspace{-0.2cm}
\begin{flalign}
&{\sf E)}\qquad T\sum_{k_0=2\pi i T} \Delta_B(K) \Delta_F(Q=P-K) 
=-  \sum_{s_1,s_2} \frac{s_1s_2}{4E_kE_q}  \frac{1+n_B(s_1E_k)-n_F^-(s_2E_q)} {p_0-s_1E_k-s_2E_q} \, . &\label{sac10}
\end{flalign}
\vspace{-.2cm}
\begin{flalign}
&{\sf F)} \qquad T\sum_{k_0=2n\pi i T}k_0 \Delta_B(K) \Delta_F(Q=P-K) = - \sum_{s_1,s_2} \frac{s_2}{4E_q}  \frac{1+n_B(s_1E_k)-n_F^-(s_2E_q)} {p_0-s_1E_k-s_2E_q} \, .& \label{sac10a}
 \end{flalign}
 \vspace{-0.2cm}
  \begin{flalign}
&{\sf G)} \qquad T \sum_{k_0=2n\pi i T} k^2_0 \Delta_B(K) \Delta_F(Q) =   
  - \sum_{s_1,s_2 } \frac{s_1s_2 }{4E_kE_q}  E_k^2 \,\,
  \frac{1+n_B\left(s_1E_k\right)-n_F^-\left(s_2E_q\right)}{p_0-s_1E_k-s_2E_q}  - \sum_{s_2} \frac{s_2}{2E_q}  \left(1-n_F^-(s_2E_q)\right) .&
  \label{sac17}
 \end{flalign}
 \vspace{-0.2cm}
\begin{flalign}
&{\sf H)} \qquad T\sum_{k_0=2\pi in\beta}\Delta_B(K)\Delta_F(P_1-K)\Delta_F(P_2-K) =  \sum_{s,s_1,s_2}  \frac{ss_1s_2}{8EE_1E_2} 
\,\,  \frac{1}{(p_{10}-p_{20}) - s_1 E_1  + s_2E_2}   \nn 
&\hspace{7.8cm}\times  \left [ \frac{1+n_B\left ( sE \right ) -n_F^-\left(s_1E_1\right )} {p_{10}-sE-s_1E_1} 
-  \frac{1+n_B\left(sE \right ) -n_F^-\left(s_2E_2\right )}{p_{20}-sE-s_2E_2}\right ] \, .& \label{sac21}
\end{flalign}
In Eq.~\eqref{sac21}, $K=(k_0,\bm{\vec k})$ is loop momentum and  $P_1=(p_{10},\bm{\vec p_1})$ and  $P_2=(p_{20},\bm{\vec p_2})$ are external momenta. $E=\sqrt{p_1^2+m^2}$ and $E_1=\sqrt{p_2^2+m_1^2}.$
\vspace{0.0cm}
\begin{flalign}
&{\sf I) }\qquad T \sum_{k_0=2\pi in\beta}k_0 \, \Delta_B(K)\Delta_B(P_1-K)\Delta_B(P_2-K) =  \sum_{s,s_1,s_2}  \frac{s_1s_2}{8E_1E_2} 
\,\,  \frac{1}{(p_{10}-p_{20}) - s_1 E_1  + s_2E_2}   \nn 
&\hspace{7.8cm}\times  \left [ \frac{1+n_B\left ( sE \right ) +n_B\left(s_1E_1\right)} {p_{10}-sE-s_1E_1} 
-  \frac{1+n_B\left(sE \right ) +n_B\left(s_2E_2\right)}{p_{20}-sE-s_2E_2}\right ] \, . &\label{sac23a}
\end{flalign}
\begin{flalign}
	&{\sf J) }\qquad T \sum_{k_0=2\pi in\beta}k_0 \, \Delta_B(K)\Delta_F(P_1-K)\Delta_F(P_2-K) =  \sum_{s,s_1,s_2}  \frac{s_1s_2}{8E_1E_2} 
	\,\,  \frac{1}{(p_{10}-p_{20}) - s_1 E_1  + s_2E_2}   \nn 
	&\hspace{7.8cm}\times  \left [ \frac{1+n_B\left ( sE \right ) -n_F^-\left(s_1E_1\right )} {p_{10}-sE-s_1E_1} 
	-  \frac{1+n_B\left(sE \right ) -n_F^-\left(s_2E_2\right )}{p_{20}-sE-s_2E_2}\right ] \, . &\label{sac23}
\end{flalign}
\vspace{-0.2cm}
\begin{flalign}
&{\sf K)}\qquad T \sum_{k_0=2\pi in\beta}k^2_0 \, \Delta_B(K)\Delta_F(P_1-K)\Delta_F(P_2-K) = - \sum_{s,s_1,s_2} 
\,\,  \frac{1}{(p_{10}-p_{20}) - s_1 E_1  + s_2E_2} \left [ \frac{s_1s_2}{4E_1E_2} \left(n_F^+(s_1E_1) -n_F^-(s_2E_2)\right) \right. \nn 
&\hspace{5.5cm}-\left. E_k^2 \ \frac{ss_1s_2}{8EE_1E_2}  \left ( \frac{1+n_B\left ( sE \right ) -n_F^-\left(s_1E_1\right )} {p_{10}-sE-s_1E_1} 
-  \frac{1+n_B\left(sE \right ) -n_F^-\left(s_2E_2\right )}{p_{20}-sE-s_2E_2}\right )\right ] \, . &\label{sac25}
\end{flalign}
\vspace{-0.4cm}
\begin{flalign}
&{\sf L) }\qquad  T \sum_{k_0=2\pi i n T}\Delta_B(K)\Delta_B(P_1-K)\Delta_B(P_2-K) =  \sum_{s,s_1,s_2}  \frac{ss_1s_2}{8EE_1E_2} 
\,\,  \frac{1}{(p_{10}-p_{20}) - s_1 E_1  + s_2E_2}   \nn 
&\hspace{7.8cm}\times  \left [ \frac{1+n_B\left ( sE \right ) +n_B\left(s_1E_1\right )} {p_{10}-sE-s_1E_1} 
-  \frac{1+n_B\left(sE \right ) + n_B\left(s_2E_2\right )}{p_{20}-sE-s_2E_2}\right ] \, . & \label{sac26}
\end{flalign}
\vspace{-0.4cm}
\begin{flalign}
&{\sf M)}\quad T \sum_{k_0=2\pi i n T}\Delta_F(K-P_1)\Delta_F(P_2-K)\Delta_F(P_1-K+Q_1) =  \sum_{s,s_1,s_2}  \frac{ss_1s_2}{8EE_1E_2} 
\,\,  \frac{1}{(p_{20}-p_{10}-q_{10}) - s_1 E_1  + s_2E_2}   \nn 
&\hspace{6cm}\times  \left[ \frac{1-n_F^+\left( sE \right) -n_F^-\left(s_1E_1\right)} {p_{20}-p_{10}-sE-s_1E_1} 
-  \frac{1-n_F^+\left(sE \right)  - n_F^-\left(s_2E_2\right )}{q_{10}-sE-s_2E_2}\right ] \, . & \label{sac27}
\end{flalign}
In Eq.~\eqref{sac27} and in the following three equations, $K$ bosonic loop momentum, $Q_1$ is incoming bosonic external momentum, $P_1$ is the incoming external fermionic momenta and $P_2$ is the outgoing external fermionic momenta.
\vspace{0.0cm}
\begin{flalign}
& {\sf N)} \quad T  \sum_{k_0=2\pi i n T} k_0 \Delta_F(K-P_1)\Delta_F(P_2-K)\Delta_F(P_1-K+Q_1) =  -\sum_{s,s_1,s_2}  \frac{s_1s_2}{8E_1E_2} 
\,\,  \frac{1}{(p_{20}-p_{10}-q_{10}) - s_1 E_1  + s_2E_2}   \nn 
&\hspace{7.8cm}\times  \left [ \frac{1-n_F^+\left ( sE \right ) -n_F^-\left(s_1E_1\right)} {p_{20}-p_{10}-sE-s_1E_1} 
-  \frac{1-n_F^+\left(sE \right ) - n_F^-\left(s_2E_2\right )}{q_{10}-sE-s_2E_2}\right ] \, . & \label{sac28}
\end{flalign}
\vspace{-0.2cm}
\begin{flalign}
&{\sf O)} \quad	T \sum_{k_0=2\pi inT}  \Delta_B(K) \Delta_F(P_1-K)\Delta_F(P_2-K)\Delta_F(P_1-K+Q_1)= \sum_{s,s_1,s_2,s_3}  \frac{ss_1s_2s_3}{8EE_1E_2E_3} \  \frac{1}{p_{10}-p_{20} - s_1 E_1  + s_2E_2}   \nn 
	&\hspace{3cm}\times\Bigg[  \frac{1}{p_{20}-p_{10}-q_{10}-s_2E_2+s_3E_3}\left( \frac{1+n_B\left ( sE \right ) -n_F^-\left(s_2E_2\right)} {p_{20}-sE-s_2E_2}  -  \frac{1+n_B\left(sE \right) - n_F^-\left(s_3E_3\right)}{p_{10}+q_{10}-sE-s_3E_3}\right) \nn
	&\hspace{4.cm}+  \ \frac{1}{q_{10}+s_1E_1- s_3E_3}\left( \frac{1+n_B\left ( sE \right ) -n_F^-\left(s_1E_1\right)} {p_{10}-sE-s_1E_1}  -  \frac{1+n_B\left(sE \right) - n_F^-\left(s_3E_3\right)}{p_{10}+q_{10}-sE-s_3E_3}\right)  \Bigg]. & \label{sac29}
\end{flalign}
\vspace{-.2cm}
\begin{flalign}
&{\sf P) }\quad	T \sum_{k_0=2\pi inT} k_0 \Delta_B(K) \Delta_F(P_1-K)\Delta_F(P_2-K)\Delta_F(P_1-K+Q_1)= -\sum_{s,s_1,s_2,s_3}  \frac{s_1s_2s_3}{8E_1E_2E_3} \  \frac{1}{p_{10}-p_{20} - s_1 E_1  + s_2E_2}   \nn 
	&\hspace{3cm}\times\Bigg[  \frac{1}{p_{20}-p_{10}-q_{10}-s_2E_2+s_3E_3}\left( \frac{1+n_B\left ( sE \right ) -n_F^-\left(s_2E_2\right)} {p_{20}-sE-s_2E_2}  -  \frac{1+n_B\left(sE \right) - n_F^-\left(s_3E_3\right)}{p_{10}+q_{10}-sE-s_3E_3}\right) \nn
	&\hspace{4cm}+  \ \frac{1}{q_{10}+s_1E_1- s_3E_3}\left( \frac{1+n_B\left ( sE \right ) -n_F^-\left(s_1E_1\right)} {p_{10}-sE-s_1E_1}  -  \frac{1+n_B\left(sE \right) - n_F^-\left(s_3E_3\right)}{p_{10}+q_{10}-sE-s_3E_3}\right)  \Bigg]. & \label{sac30}
\end{flalign}
\vspace{-.2cm}
\begin{flalign}
	&{\sf Q) }\quad	T \sum_{k_0=2\pi inT} \Delta_F(K) \Delta_F(K-S)\Delta_F(K-S-R)\Delta_F(P+K)=\sum_{s,s_1,s_2,s_3} \frac{s s_1 s_2 s_3 }{ 16 E E_1 E_2 E_3}\frac{1}{ p_0 + s_0-E_1 s_1-E_3 s_3}\nn
&\hspace{3cm}\times	\left[\frac{1}{r_0+E_1 s_1- s_2E_2 }\left( \frac{1-n_F^-(sE)-n_F^+(s_2 E_2 )}{s_0+r_0-sE -s_2E_2 }-\frac{1-n_F^-(sE)- n_F^+(s_1E_1)}{s_0-sE-s_1 E_1} \right)\right.\nn
&\hspace{3.5cm}\left.+\frac{1}{p_0+s_0+r_0-s_2 E_2 -s_3 E_3 }\left( \frac{n_F^-(sE)-n_F^-(s_3E_3)}{p_0+sE-s_3 E_3}-\frac{1-n_F^-(sE )-n_F^+(s_2E_2)}{s_0+r_0-sE -s_2E_2 } \right)\right].\label{sum_ffff}
\end{flalign}
In Eq.~\eqref{sum_ffff}, $P,S,R$ are the external outgoing bosonic momenta. Additionally, $E=k,E_1=|\bm{\vec k}-\bm{\vec s}|,E_2=|\bm{\vec k}-\bm{\vec s}-\bm{\vec r}|,E_3=|\bm{\vec p}+\bm{\vec k}| $.
\subsubsection{Functional integration and the partition function relation}
We will address a statistical thermodynamics problem wherein the system returns to its initial state after undergoing a time evolution from $t=0$ to $t$. This transition can be expressed in a functional form as $\langle \phi_a|e^{-i{\cal H}t}|\phi_b\rangle$, under the assumption that the Hamiltonian is time-independent, which simplifies the transition amplitude from one state to another.


The transition amplitude can be expressed in Minkowski space-time as  
\be
\left \langle \phi_a \left| e^{-i{\cal H} t}\right | \phi_a \right\rangle = \int {\cal D}\pi \, \int {\cal D}\phi\exp\left[ i \int_0^t\, dt \int \, d^3 {x} \Big(\pi(X)\, \partial_t \phi(X)-{\cal H}_d(\pi(X ),\phi(X)) \Big) \right ], \label{fi1}
\ee
where $\cal D$ is the functional or path integral runs over all possible paths of momentum $\pi(X)$ and field $\phi(x)$. These fields are not restricted by boundary conditions while going from initial time $t=0$ to final time $t_f=t$. 

The Hamiltonian of the system is ${\cal H}=\int  {\cal H}_d \, d^3 {x}$ along with the Hamiltonian 
density\footnote{If there is a conserved charge density ${\cal N}(\pi,\phi)$ one should also include it as ${\cal H}_d -\mu{\cal N}(\pi,\phi)$, where $\mu$ is the associated  chemical potential.} is given as
\begin{equation}
 {\cal H}_d = \pi(X) \partial_t \phi(X) -{\cal L} (\phi(X) {\dot \phi(X)}), \label{fi2}
\end{equation}
where ${\cal L}$ is Lagrangian density in Minkowski space-time. Using (\ref{fi2}) in (\ref{fi1}) one can write the transition amplitude in
${\cal L}$ and/or action ${\cal S}$ as
\begin{eqnarray}
  \left \langle \phi_a \left| e^{-i{\cal H} t}\right | \phi_a \right\rangle &=&  \int {\cal D}\phi\ 
  \exp\left[{i\int_0^t\, dt \int \, d^3 {x}\, {\cal L}}\right]=  \int {\cal D}\phi\ 
  e^{i \int  d^4 {x} \,  {\cal L}} = \int {\cal D}\phi\ 
  e^{i {\cal S}[\phi]}, \label{fi3}
\end{eqnarray}
where action in Minkowski space-time is written as
\begin{equation}
 {\cal S}[\phi] = \int  d^4 {x} \,  {\cal L} = \int_0^t\, dt \int \, d^3 {x}\, {\cal L} . \label{fi4} 
\end{equation}
Now, the partition function reads as
$
{\cal Z} ={\rm{Tr}}\rho \ = {\rm{Tr}} \left (  e^{-\beta {\cal H}} \right ) 
= \sum_n \ \  \left \langle  n  \left |  e^{-\beta {\cal H}} 
\right | n  \right \rangle  , \label{fi3a} 
$
where the summation over $n$ includes all the possible energy eigenstates of the system in Hilbert space. In the continuum case, the summation becomes an integral, and the eigenstates $|\phi\rangle$ form a complete set, each with energy $E_\phi$ . Thus, the partition function becomes
\begin{equation}
{\cal Z} = \int \ d\phi \  \left \langle \phi  \left |  e^{-\beta {\cal H}} 
\right | \phi  \right \rangle   = \int d\phi e^{-\beta E_\phi}  \ . \label{fi4a} 
\end{equation}
If one compares Eq.~\eqref{fi3} and~\eqref{fi4a}, there is a striking similarity between the path integral formulation of the transition amplitude in quantum mechanics and the partition function in statistical mechanics provided 
\begin{enumerate}
 \item [$\bullet$] The time interval $[0,t]$ in the transition amplitude described in equation in Eq.~\eqref{fi3} takes the role of $\beta$ in the
 partition function with interval $[0,\beta$] along with $\tau=it$ . This process, known as Wick rotation, involves a $90^\circ$ rotation of the integration path in the complex plane
(see Fig.~\ref{wick_rot}).
 
\item [$\bullet$] The field $\phi$ adheres  periodic or anti-periodic boundary condition, $\phi(x,0)=\pm\phi(x, \beta)$, 
as previously discussed in subsec~\ref{period}.
\end{enumerate}
With this, the transition amplitude can be viewed as the partition function within the path integral approach as
\begin{eqnarray}
 {\cal Z} ={\rm{Tr}}\rho \ &=& {\rm{Tr}} \left (  e^{-\beta {\cal H}} \right )  
= \int \ d\phi \  \left \langle \phi  \left |  e^{-\beta {\cal H}}  \right | \phi  \right \rangle 
=  \int {\cal D}\phi\ 
   e^{i\int\limits_0^t \, dt \int \, d^3 {x}\, {\cal L}}  \label{fi5} \nonumber\\
&{=\atop t\rightarrow -i\tau}& \, \int\limits_{\phi(x,0)=\pm\phi(x, \beta)} {\cal D}\phi\ 
  e^{\int_0^\beta\, d(it) \int \, d^3 {x}\, {\cal L}(t \rightarrow -i\tau)} 
= \, \int\limits_{\phi(x,0)=\pm\phi(x, \beta)} {\cal D}\phi\ 
   e^{\int\limits_0^\beta\, d\tau \int \, d^3 {x}\, {\cal L}(t \rightarrow -i\tau)} . \label{fi6}
\end{eqnarray}
At this point we would like to note that the partition function can be directly computed in Euclidean time $\tau$ and discrete frequency $i\omega_n$ directly using Eq.~\eqref{fi6}. Alternatively, one can also convert the Minkowski action in Eq.~\eqref{fi5} to momentum space and then replace the four momentum integral by the frequency sum.
\subsection{Real Time Formalism}
\vspace{-0.2cm}
Imaginary time formalism is applied to study the static and equilibrium properties of a system. The time dependence appears non-trivially through the analytical continuation. This formalism cannot describe the out of equilibrium systems. On the other hand, real time formalism is appropriate framework to handle the out of equilibrium systems and also to study the dynamical situations such as phase transition or evolution of the universe. 

The real time formalism was originally proposed by Schwinger and Keldysh~\cite{Schwinger:1960qe,Keldysh:1964ud}. Here, we briefly discuss the real time method~\cite{Das:1997gg, Bellac:2011kqa,Mallik:2016anp}. For better understanding we start with quantum mechanics. At finite temperature $\beta^{-1}=T$, the one important quantity is the partition function
\bea
{\Tr} e^{-\beta H}=\int_{-\infty}^{\infty} dq \braket{q,t|e^{-\beta H}|q,t}.
\label{partition_func}
\eea
This is the analogous to the transition amplitude for a system from $q$ at time $t$ to $q'$ at time $t'$ in a vacuum theory
\bea
\braket{q',t'|q,t}=\braket{q',t'|e^{-i H(t'-t)}|q,t},
\eea
where $H$ is the Hamiltonian of the system. The operator $e^{-iH(t'-t)}$ corresponds that the system evolves from time $t$ to $t'$. Similarly, drawing the analogy with the vacuum case, one can think that the operator $e^{-\beta H}=e^{-iH(\tau-i\beta-\tau)}$ evolves the system from time $\tau$ to $\tau-i\beta$ in the complex time path $C$.

Thermal scalar field propagator is defined as 
\bea
D(X,X')&=&i\langle T\phi(X)\phi(X')\rangle
=\theta(\tau-\tau')i \langle \phi(X)\phi(X')\rangle+\theta(\tau'-\tau)i \langle \phi(X')\phi(X)\rangle,\nonumber\\
&=&\theta(\tau-\tau')D_+(X,X')+\theta(\tau'-\tau)D_-(X,X').
\eea
Here, $\tau$ and $\tau'$ are the points on the contour. $\theta(\tau-\tau')$ is the contour-odered theta function. $D_+(\bm{\vec  x}, \bm{\vec x'}; \tau,\tau')$ is defined in domain $-\beta\leq \text{Im}(\tau-\tau')\leq 0$ whereas, $D_-(\bm{\vec  x}, \bm{\vec x'}; \tau,\tau')$ is defined in domain $\beta\geq \text{Im}(\tau-\tau')\geq 0$. The thermal propagator satisfies the equation
\be
(\square^2+m^2)D(X,X')=\delta^4(X-X'),
\ee
with boundary condition
$
D_-(\bm{\vec x},\bm{\vec x'};\tau,\tau')=D_+(\bm{\vec x},\bm{\vec x'};\tau-i\beta,\tau'),
$
known as KMS relation (shown in Eq.~\eqref{eq16}). Going to the spatial Fourier transform space and solving the differential equation one can easily get
\bea
D(\bm{\vec k};\tau-\tau')=\frac{i}{2\omega}\bigg\{\big[\theta(\tau-\tau')+n_B\big]e^{-i\omega(\tau-\tau')}+\big[\theta(\tau'-\tau)+n_B\big]e^{i\omega(\tau-\tau')}\bigg\},
\eea
where $n_B(\omega)=1/(e^{\beta \omega}-1)$ is the single particle thermal distribution function with $\omega^2=\bm{\vec k}^2+m^2$. 
\begin{wrapfigure}{r}{0.4\textwidth}
	\centering
	\includegraphics[width=8cm,height=4cm]{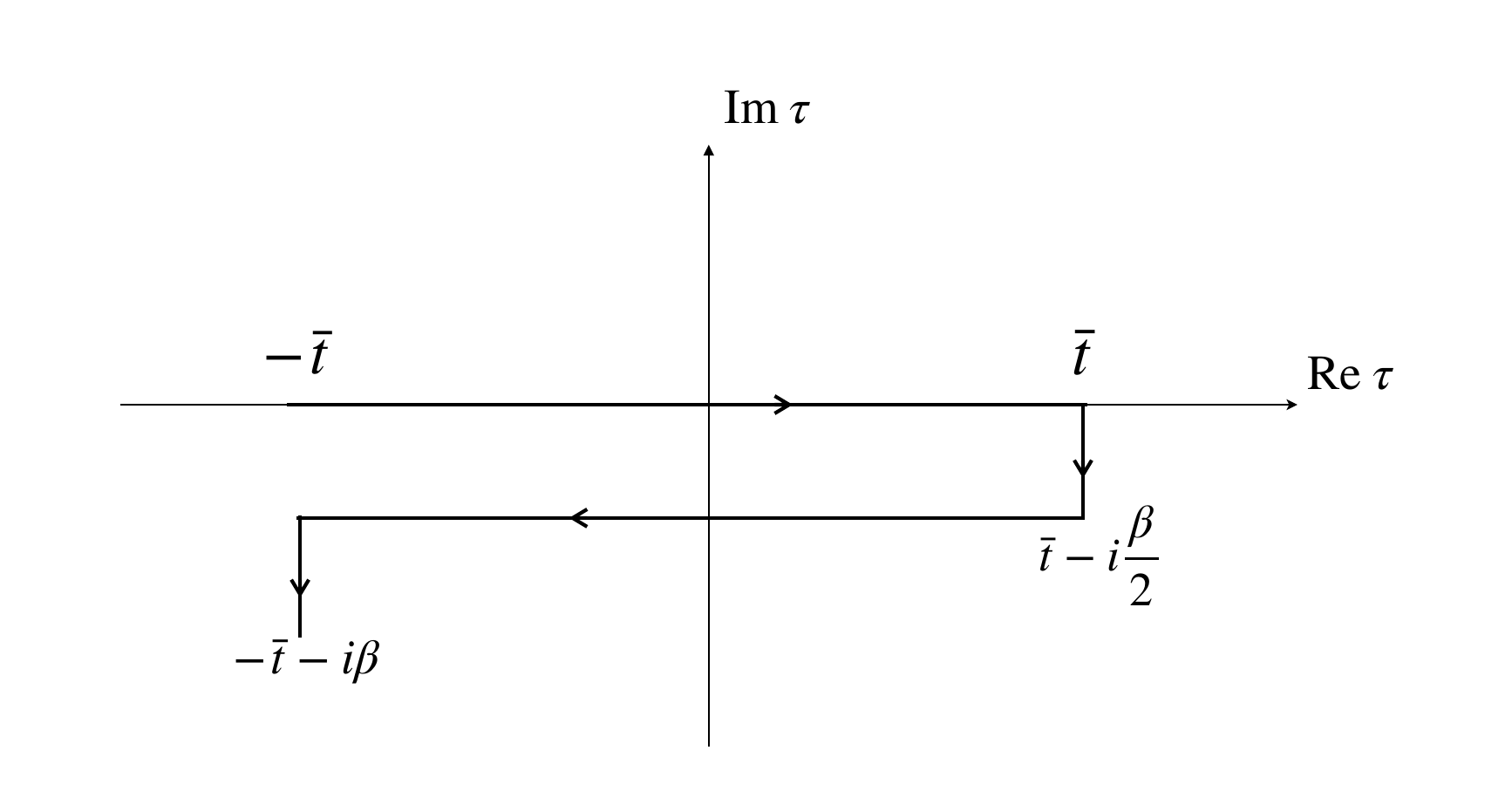}
	\vspace{-1cm}
	\caption{The contour for the real time formalism.}
	\label{contour_realtime}
\end{wrapfigure}
Now we need to specify the contour and the suitable choice is shown in Fig.~\ref{contour_realtime}. The path traverses from $-\bar t$ to $+\bar t$ in real axis, then it continues in vertical path from $\bar t$ to $\bar t-i\beta/2$. Then it runs parallel to real axis upto $-\bar t-i\beta/2$ and finally takes the vertical path up to $-\bar t-i\beta$.

Note that $D(\bm{\vec k}, \tau_h,\tau_v)\rightarrow 0$ by Riemann-Lebesgue lemma where $\tau_v$ and $\tau_h$ are points  on vertical and horizontal lines, respectively. Both points on the vertical lines are not considered as we are interested only in the propagator having points on the real axis. So, we are left with two lines parallel to real axes and the propagator can be written in form of $2\times 2$ matrix. The propagator is given in $ij$-th element as 
\bea
D(\bm{\vec k};\tau_i-\tau'_j)=\frac{i}{2\omega}\bigg\{\big[\theta(\tau_i-\tau'_j)+n_B\big]e^{-i\omega(\tau_i-\tau'_j)}+\big[\theta(\tau'_j-\tau_i)+n_B\big]e^{i\omega(\tau_i-\tau'_j)}\bigg\}.
\eea
Now, we can write down the contour ordering in terms of the usual time $(t)$ ordering. When $\tau $ and $\tau'$ are on line $1$ (real axis from $-\bar t$ to $\bar t$), $\theta(\tau_1-\tau_1')=\theta(t-t')$. If $\tau $ and $\tau'$ are on line $2$ (real axis from $\bar t-i\beta/2$ to $-\bar t-i\beta/2$), it is written as $\theta(\tau_2-\tau_2')=\theta(t'-t)$. When the points are on different lines, we note that $\theta(\tau_1-\tau_2')=0$ and $\theta(\tau_2-\tau_1')=1$.

Now, the components of propagator in the momentum space are written by taking temporal Fourier transform as
\bea
D_{ij}(\bm{\vec k},k_0)=\int_{-\infty}^{\infty} e^{ik_0 (t-t')} D(\bm{\vec k};\tau_i,\tau_j')dt.
\eea
Performing the integration, they can take the forms
\begin{subequations}
\begin{align}
D_{11}&=\Delta_F(k)+2i\pi n_B(\omega)\delta(k^2-m^2),\\
D_{12}&=2i\pi \sqrt{n_B(\omega)(1+n_B(\omega))}\delta(k^2-m^2),
\end{align}
\end{subequations}
whereas the other components are written as
$
D_{21}=D_{12},\  \mbox{and}\ 
D_{22}=-D_{11}^*.
$
Finally, we can write the momentum space scalar propagator in terms of matrix as
\bea
D(k_0,\bm{\vec k})=
\begin{bmatrix}
	\Delta_F(k)+2i\pi n_B\delta(k^2-m^2) & 2i\pi \sqrt{n_B(1+n_B)}\delta(k^2-m^2)\\
	2i\pi \sqrt{n_B(1+n_B)}\delta(k^2-m^2) & 	-\Delta_F(k)^*+2i\pi n_B\delta(k^2-m^2) 
\end{bmatrix},
\eea
where $	\Delta_F(k)$ is the free scalar propagator.
\vspace{-0.2cm}
\subsection[General Structure of Fermionic Two point functions]{General Structure of Fermionic Two point functions at finite temperature}
\label{gse}
\vspace{-0.2cm}
\subsubsection{Fermion self-energy}
\label{fgse}
\vspace{-0.2cm}
A theory possessing only massless fermions and gauge bosons, exhibits chirality invariance across all orders, alongside preserving parity invariance. At finite temperature, chiral invariance has two implications: i) absence of ${\bar \psi}\psi$ coupling in 
any finite order of perturbation theory,  ii) the general form of the fermion self-energy, as per Ref.~~\cite{Weldon:1982bn}, can be expressed as
$
\Sigma (P)=-{\cal A} P\!\!\!\!\slash, \label{gse0}
$
where $\slashed{P}=\gamma^\mu P_\mu$ is the fermion momentum $P\equiv(p_0=\omega, \bm{\vec p})$, with $p=|\bm{\vec p}|$. Here, ${\cal A}$  represents a Lorentz invariant structure function dependent on $P^2$.

The effective propagator reads as (see subsec~\ref{fgp} below)
\be
S(P) = \frac{1}{P\!\!\!\! \slash -\Sigma(P)} = \frac{P \!\!\!\! \slash}{(1+{\cal A})P^2}. \label{gse1}
\ee
The poles, $P^2=0$, lieon the light cone $\omega=p$. The term $(1+{\cal A})$ modifies the residues.

At finite temperature, the aforementioned point i) remains valid, whereas point ii) does not hold. At finite temperatures, the system does not reside in a vacuum due to the  presence of antiparticles in equal numbers to particles at such high temperatures, constituting a heat bath that introduces a specific Lorentz frame. Hence, the heat bath has a four velocity $u^\mu =(1,0,0,0)$ with $u^\mu u_\mu=1$. Consequently, the most general ansatz for the fermion self-energy takes the form~\cite{Weldon:1982bn}:
\bea
\Sigma(P) &=& -{\cal A} \slashed{P}  - {\cal B}u\!\!\!\slash \, ,\label{gse2} 
\eea
where ${\cal B}$ is another Lorentz invariant structure function in addition to ${\cal A}$.
Since $P^2=\omega^2-p^2$, one can interpret $\omega=p_0=P^\mu u_\mu$ and $p=\left (P^\mu u_\mu -P^2\right)^{1/2}$ as Lorentz invariant energy and momentum, respectively. The Lorentz invariant structure functions are obtained following the calculation in Ref.~\cite{Mustafa:2022got} as
\be
{\cal A}(\omega,p) = \frac{1}{4} \frac {{\rm{Tr}} \left [ \Sigma P\!\!\!\! \slash \, \right ] - \left( P\cdot u\right ) {\rm{Tr}} \left [\Sigma u\!\!\! \slash\right ]  }
{\left (P\cdot u\right )^2 - P^2} ,\ \qquad \text{and} \qquad 
{\cal B}(\omega,p) = \frac{1}{4} \frac  {P^2 {\rm{Tr}} \left [\Sigma u\!\!\! \slash\right ] - \left( P\cdot u\right ) {\rm{Tr}} \left [ \Sigma P\!\!\!\! \slash\,\right ]  }
{\left (P\cdot u\right )^2 - P^2} .\label{gse10}
\ee
\subsubsection{Fermion propagator}
\label{fgp}
\begin{figure}[h]
\begin{center}
\includegraphics[scale=.7]{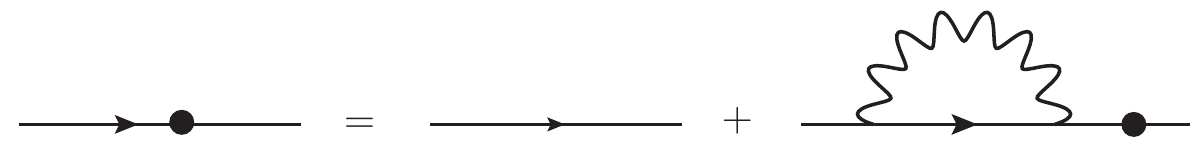}
 \caption{Pictorial representation of Dyson-Schwinger equation for an effective fermion propagator.}
 \label{dyson}
 \end{center}
 \end{figure}
 \vspace{-0.3cm}
In  Fig.~\ref{dyson},  we represent the effective propagator by $S(P)$ and the bare fermion propagator by $S_0(P)$ and fermion self-energy as $\Sigma(P)$. Following Fig.\ref{dyson}  the effective fermion propagator~\cite{Mustafa:2022got} can be written as,
\bea
S(P) &=& S_0(P) +S_0(P) \Sigma (P) S(P)  \nonumber \\
S(P) S^{-1}(P) &=& S_0(P) S^{-1}(P) +S_0(P) \Sigma (P) S(P) S^{-1}(P)  \nonumber \\
S^{-1}(P) &=& \slashed{P} -\Sigma (P) \, , \label{gse15}
\eea
which is known as fermionic Dyson-Schwinger equation.  Now, the effective fermion propagator  can be obtained from Eq.~\eqref{gse15} as
\be
S(P)=\frac{1}{P\!\!\!\! \slash -\Sigma(P)} .\label{gse14}
\ee
Using Eq.~\eqref{gse2} one can write the effective propagator as
\be
S(P)=\frac{1}{(1+{\cal A})P\!\!\!\! \slash+{\cal B}u\!\!\! \slash} =\frac{(1+{\cal A})P\!\!\!\! \slash+{\cal B}u\!\!\! \slash}{[(1+{\cal A})P\!\!\!\! \slash+{\cal B}u\!\!\! \slash]^2}
= \frac{P\!\!\!\! \slash -\Sigma(P)} {\cal D} =\frac{S^{-1}(P)}{\cal D} ,\label{gse16}
\ee
where the Lorentz invariant quantity ${\cal D}$ is given~\cite{Mustafa:2022got} as
\bea
{\cal D}{(p,u)} &=& \left [(1+{\cal A})P\!\!\!\! \slash+{\cal B}\slashed{u}\right ]^2 
= (1+{\cal A})^2 \slashed{P}^2 +2(1+{\cal A}){\cal B} P\cdot u + {\cal B}^2 .\label{gse17}
\eea
In the rest frame of heat bath, Eq.~\eqref{gse17} reads~\cite{Mustafa:2022got} as
\bea
{\cal D}{(p,\omega)} &=& {\cal D}_+{\cal D}_- , \label{gse18}
\eea
where
\be
{\cal D}_\pm (p,\omega)= (1+{\cal A})(\omega\mp p)+{\cal B} . \label{gse19}
\ee
In free case, ${\cal A}={\cal B}=0$ and Eq.~\eqref{gse19} becomes
\be
d_\pm (p,\omega)= \omega\mp p. \label{gse19a}
\ee
Combining \eqref{gse18} and \eqref{gse16}, one can write~\cite{Mustafa:2022got} the effective propagator as
\be
S(P)=\frac{S^{-1}(P)}{{\cal D}_+{\cal D}_-} .\label{gse20}
\ee
We can write~\cite{Mustafa:2022got} the self-energy in~\eqref{gse2} as
\bea
\Sigma(P) &=& -\frac{1}{2} \left[\left({\cal A}(\omega+p)+{\cal B} \right )(\gamma_0 - {\vec \gamma}\cdot \bm{\hat { p}}) 
+\left({\cal A}(\omega-p)+{\cal B} \right )(\gamma_0 + \vec{\bm \gamma}\cdot \bm{\hat { p}}) \right ].\label{gse21} 
\eea
Now we can also write~\cite{Mustafa:2022got}
\bea
\slashed{P} &=& \gamma_0\omega -p \vec{\bm \gamma}\cdot \bm{\hat { p}}  
=\frac{1}{2}\left [(\omega-p)(\gamma_0 + \vec{\bm \gamma}\cdot \bm{\hat { p}}) +(\omega+p)(\gamma_0 - \vec{\bm \gamma}\cdot \bm{\hat { p}})  \right ]. \label{gse22}
\eea
In the rest frame of the heat bath, the inverse of the effective propagator in \eqref{gse15} can now be written as
\bea
S^{-1}(P)&=&{P\!\!\!\! \slash -\Sigma(P)}
= \frac{1}{2} (\gamma_0 + \vec{\bm \gamma}\cdot \bm{\hat  p}) {\cal D}_+ + \frac{1}{2} (\gamma_0 - {\vec \gamma}\cdot \bm{\hat { p}}) {\cal D}_-
 \label{gse23}
\eea
Using \eqref{gse23} in \eqref{gse20}, one finally obtains the effective fermion propagator as
\be
S(P) =  \frac{1}{2} \frac{\gamma_0 - \vec{\bm \gamma}\cdot \bm {\hat { p}}} {{\cal D}_+(\omega,p)}+ \frac{1}{2} \frac{\gamma_0 + \vec{\bm \gamma}\cdot \bm {\hat { p}}} {{\cal D}_- (\omega,p)} ,
 \label{gse24}
 \ee
which is decomposed in helicity eigenstates. The charge invariance demands that ${\cal D}_\pm(-\omega, p)=-{\cal D}_\mp(\omega, p)$ which implies that ${\cal A}(-\omega,p) ={\cal A}(\omega,p) $ and ${\cal B}(-\omega,p) =-{\cal B}(\omega,p) $. ${\cal D}(\omega,p)$ has an imaginary part for space like 
momenta $P \, (p_0^2<p^2)$, it is also useful to define the parity properties for both real and imaginary parts of ${\cal D}(\omega,p)$  as ${\mathrm {Re}} {\cal D}_+(-\omega,p) =  -{\mathrm  {Re}} {\cal D}_-(\omega,p) $ and ${\mathrm {Im}} {\cal D}_+(-\omega,p) =  {\mathrm  {Im}} {\cal D}_-(\omega,p) $.

In free fermion case, the  propagator becomes
\be
S_0(P) =  \frac{1}{2} \frac{\gamma_0 - \vec{\bm \gamma}\cdot \bm{\hat { p}}} {d_+(\omega,p)}+ \frac{1}{2} \frac{\gamma_0 + \vec{\bm \gamma}\cdot \bm{\hat { p}}} {d_-(\omega,p) } .
 \label{gse25}
 \ee

 \subsection[General Structure of a Gauge Boson Two-point Function]{General Structure of a Gauge Boson Two-point Function at finite temperature}
 \label{gsb}
\subsubsection{Tensor decomposition}
\label{cd}
The gauge boson self-energy in vacuum follows the general structure:
\bea
\Pi^{\mu\nu}(P^2) = V^{\mu\nu}\Pi(P^2),
\eea
where the  Lorentz invariant form factor $\Pi(P^2)$ is depends solely on the four-scalar $P^2$. The vacuum projection operator is
\bea
V^{\mu\nu} = \eta^{\mu\nu}-\frac{P^\mu P^\nu}{P^2},
\eea
and it satisfies the gauge invariance through the transversality condition:
\bea
P_\mu \Pi^{\mu\nu} = 0 ,
\label{trans_cond}
\eea
where $\eta^{\mu\nu}\equiv(1,-1,-1,-1)$ and $P\equiv(\omega, \bm{\vec p})$.  Additionally, it is symmetric under the exchange of $\mu$ and $\nu$ as
$
\Pi_{\mu\nu}(P^2)=\Pi_{\nu\mu}(P^2) .
$
The presence of the heat bath or the finite temperature ($\beta=1/T$) breaks the Lorentz invariance. To construct a general covariant structure of the gauge boson self-energy at finite temperature, all four vectors and tensors must be considered, including $P^\mu$ and  $ \eta^{\mu\nu}$ from vacuum, along with the four-velocity $u^\mu$ of the heat bath. From these, four types of tensors can be formed: $P^\mu P^\nu, P^\mu u^\nu + u^\mu P^\nu, u^\mu u^\nu$ and $\eta^{\mu\nu}$~\cite{Das:1997gg,Weldon:1982aq}. These four tensors can yield two independent tensors by virtue of two constraints provided by the transversality condition in (\ref{trans_cond}). One can form two mutually orthogonal projection tensors from these two independent tensors, ensuring the construction of  Lorentz-invariant structure of the gauge boson two-point functions at finite temperature. 
To construct two such tensors, we define the Lorentz scalars, vectors and tensors that characterise the heat bath:
\be
u^\mu =(1,0,0,0), \qquad\text{and} \qquad P^\mu u_\mu=P\cdot u=\omega .\label{scal1}
\ee
Similar to vaccum, we can define $\tilde{\eta}^{\mu\nu}$ transverse to $u^{\mu}$ as
 \begin{subequations}
 \begin{align}
\tilde{\eta}^{\mu\nu} &= \eta^{\mu\nu} - u^\mu u^\nu \, \label{td1} , \\
 u_{\mu}\tilde{\eta}^{\mu\nu} &= u_{\mu}\eta^{\mu\nu} - u_{\mu}u^\mu u^\nu  = u^\nu - u^\nu = 0 \, . \label{td2}
\end{align}
\end{subequations}
 So $u^\mu$ and $\tilde{\eta}^{\mu\nu}$ are transverse. Now, any four vector can be decomposed into parallel and orthogonal component with respect to $u^\mu$:
 \begin{subequations}
 \begin{align}
 P^\mu_\sp &=(P\cdot u)u^\mu = \omega u^\mu, \qquad \mbox{and} \qquad
 P^\mu_\perp =\tilde{P}^\mu = P^\mu -  P^\mu_\sp = P^\mu-\omega u^\mu \, . \label{td2b}\\
\tilde{P}^2 &= \left(P^\mu - \om u^\mu\right)\left(P_{\mu} - \om u_{\mu}\right)
 = P^2 - \omega^2 - \om^2 + \om^2
 = P^2 - \omega^2
 = -p^2 \, . \label{td2c}
\end{align}
\end{subequations}
We can also define any four vector parallel and perpendicular to $P^\mu$
\be
u^\mu_\sp = \frac{(P\cdot u)P^\mu}{P^2} = \frac{\om P^\mu}{P^2} \, ,  \,\,\,\, {\mbox{and}}\,\,\,\,
\bar{u}^\mu  \equiv u^\mu_\perp = u^\mu - u^\mu_\sp = u^\mu - \frac{\om P^\mu}{P^2} \, . \label{td4}
\ee
So, projection of the momentum $P^\mu$ on four vector $\bar{u}^\mu$ becomes $P^\mu \bar{u}_{\mu} = 0$. Finally, one can constract two mutualy orthogonal projection tensors~\cite{Mustafa:2022got} as
\begin{subequations}
 \begin{align}
A^{\mn} &= \ti{\eta}^{\mn} - \frac{\ti{P}^\mu \ti{P}^\nu}{\ti{P}^2} \nn
&= \frac{1}{P^2 - \om^2}\left[(P^2 - \om^2)(\eta^{\mn} - u^\mu u^\nu) - P^\mu P^\nu - \om^2 u^\mu u^\nu + \om(P^\mu u^\nu + u^\mu P^\nu)\right] \, , \label{A_exp}\\
B^{\mn} &= \frac{P^2}{\ti{P}^2} \bar{u}^\mu \bar{u}^\nu  = \frac{\bar{u}^\mu\bar{u}^\nu}{\bar{u}^2} \nn
&= \frac{1}{P^2\left(P^2 - \om^2\right)}\left[P^4 u^\mu u^\nu + \om^2 P^\mu P^\nu - \om P^2(P^\mu u^\nu + u^\mu P^\nu)\right]\, . \label{B_exp}
\end{align}
It is possible to show from Eqs.~\eqref{A_exp} and ~\eqref{B_exp} that
\end{subequations}
\be
A^{\mn} + B^{\mn} = V^{\mn} = \eta^{\mn} - \frac{P^\mu P^\nu}{P^2} \, . \label{td8}
\ee
One can see that the two independent second rank symmetric tensors at finite temperature have been constructed from $\eta^{\mn},P^\mu P^\nu,u^\mu u^\nu, {\mbox{and}} \, P^\mu u^\nu + u^\mu P^\nu$ which 
are orthogonal to $P^\mu$.

\subsubsection{General structure of self-energy of a gauge boson }
\label{gbs}
The self-energy of a vector particle in a medium (finite temperature/density) can be expressed as
\be
\Pi_{\mn}= \Pi_T(\om,p)A_{\mn} + \Pi_L(\om,p)B_{\mn} , \label{gen_exp}
\ee
which satisfies current conservation $P^\mu\Pi_{\mn}=0$.
Substituting $A_{00}$ and $B_{00}$ from  \eqref{A_exp} and \eqref{B_exp}, respectively, in $\Pi_{00}$ in \eqref{gen_exp}, one can obtain~\cite{Das:1997gg,Mustafa:2022got} 
\bea
\Pi_L(\om,p) &=& \left(-\frac{P^2}{p^2}\right)\Pi_{00}(\om,p) \, . \label{pi_L} 
\eea
Now, operating $\eta^{\mn}$ on self-energy in \eqref{gen_exp}, then using \eqref{A_exp} and \eqref{B_exp} one can obtain~\cite{Das:1997gg,Mustafa:2022got}
\be
\Pi_T(\om,p) = \frac{1}{D-2}\left[{\Pi}^{\mu}_{\mu}(\om,p) - \Pi_L(\om,p)\right] =\frac{1}{2}\left[{\Pi}^{\mu}_{\mu}(\om,p) - \Pi_L(\om,p)\right].\label{pi_T} 
\ee
where $D=4$, the space-time dimension.
\subsubsection{General structure of a massless vector gauge boson propagator in covariant gauge}
\label{gbp}
  \vspace{-0.cm}
 \begin{figure}[h]
 \begin{center}
  \includegraphics[scale=1]{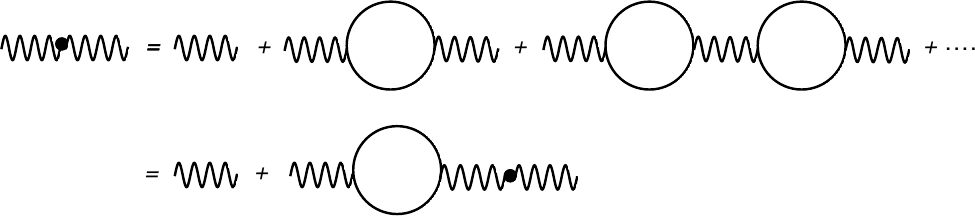}
  \vspace{-0.2cm}
  \caption{Effective gauge boson propagator.}
  \label{dyson_eqn_pic}
  \end{center}
 \end{figure}
   \vspace{-0.cm}
We represent the full propagator by $D_{\mn}$, the bare propagator by $D^0_{\mn}$ and each blob is self-energy $\Pi_{\mn}$ in  Fig.~\ref{dyson_eqn_pic}  of a gauge boson. 
We can write the full propagator in Fig. \ref{dyson_eqn_pic}  as
\bea
D^{\mr} &=& {D^{0}}^{\mr} + {D^{0}}^{\mu\alpha} \Pi_{\alpha\beta}{D}^{\beta\rho} \nonumber\\
D^{\mu\rho}D^{-1}_{\rho \nu} &=& {D^{0}}^{\mu\rho}D^{-1}_{\rho \nu} + {D^{0}}^{\mu\alpha}\Pi_{\alpha\beta}D^{\beta\rho}D^{-1}_{\rho\nu}\nonumber\\
D^{-1}_{\nu\gamma} &=& (D^{0}_{\nu\gamma})^{-1} - \Pi_{\nu\gamma}\, , \label{dys}
\eea
which is known as Dyson-Schwinger equation for a gauge boson. For massless gauge boson the propagator is given as
\be
D^0_{\mn} = -\frac{\eta_{\mn}}{P^2} + {(1-\xi)} \frac{P_\mu P_\nu}{P^4} \, , \label{gsp1}
\ee
where $\xi=1$ in Feynman gauge and $\xi=0$ in Landau gauge.
Following the details calculation in Refs.~\cite{Das:1997gg,Mustafa:2022got} one obtains the effective propagator of  an interacting gauge boson in the presence of thermal medium as
\be
{D_{\mn} = -\frac{\xi}{P^4}P_\mu P_\nu - \frac{1}{P^2+\Pi_T}A_{\mn} - \frac{1}{P^2+\Pi_L}B_{\mn}} \, . \label{gsp14}
\ee
We note here that the general structure of two-point functions for both fermion and gauge boson are applicable to both QED and QCD.
\subsection{Scale Separation at Finite Temperature}
\label{scale_sepa}
\vspace{-0.2cm}
At temperatures significantly higher than any intrinsic mass scale of a given theory, and where the coupling $g$ is less than unity, a hierarchy of energy scales emerges in the system. Within this hierarchy, three distinct energy scales arise: hard, soft (electric), and ultra-soft (magnetic). Below, we outline some key features of these scales:
\vspace{-0.2cm}
\subsubsection{ Hard scale:}
\vspace{-0.2cm}
\label{hs} 
\begin{list}{$\bullet$}{\leftmargin=0.6cm\itemsep=-1pt}
\item Scale of thermal fluctuations: momenta $\sim T$; length $\sim 1/T$.
\item Inverse mass of non-static field modes ($p_0\ne 0$).
\item Purely perturbative contributions appear in  even power of coupling as $g^{2n}$, where $n$ is the number of loop.
\end{list}
\vspace{-0.3cm}
\subsubsection {Soft (electric) scale:} 
\label{ss}
\vspace{-0.2cm}
\begin{list}{$\bullet$}{\leftmargin=0.6cm\itemsep=-1pt}
\item Scale of static chromoelectric fluctuations: momenta $\sim gT$; length $\sim 1/gT$.
\item Generates inverse Debye screening (electric screening) mass of longitudinal gauge field ($A_0$).
\item Requires resummation of an infinite subset of diagrams.
\item Contribution appears in even and odd powers of $g$ and $\log (g) $  ({viz.}, $g, \ g^2, \ g^3, g^4\log(g), \cdots$).
\end{list}
\vspace{-0.3cm}
\subsubsection{Ultra-soft (magnetic) scale:}  
\label{uss}
\vspace{-0.3cm}
\begin{list}{$\bullet$}{\leftmargin=0.6cm\itemsep=-1pt}
\item Static chromomagnetic fluctuations: momenta $\sim g^2T$; length $\sim 1/g^2T$.
\item Generates inverse magnetic mass.
\item Generates a single non-perturbative contribution to pressure starting at 4-loop order (Linde Problem).
\end{list}
\vspace{-0.4cm}
\subsection{Bare Perturbation Theory}
\label{bpt}
\vspace{-0.2cm}
At high temperatures and/or densities, matter is simple as the interaction of the theory is weaken and simplicity emerges in such extreme situations. Both static and dynamic quantities can be determined by expanding the coupling constant around the free theory. This suggests that bare perturbation theory (BPT) is effective in the hard scale regime, utilizing bare propagators and vertices, with contributions appearing in even powers of the coupling ($g^{2n}$). However, BPT is incomplete and poses serious issues, including gauge-dependent results and infrared (IR) singularities in certain quantities, as elaborated below:
\begin{list}{$\bullet$}{\leftmargin=0.25cm\itemsep=3pt}
\item Historically, it was observed that the gluon damping rate shows gauge dependence. Several such results are listed below:
\be
\gamma_{\rm g}=a \frac{g^2T}{8\pi}; \, \, \, a= \left \{ \begin{array}{ll}
                   1& \mbox{for Coloumb gauge~\cite{lopez,Kajantie:1982xx,Heinz:1986kh} and Temporal axial gauge~\cite{Kajantie:1982xx,Heinz:1986kh},} \nonumber \\                   
                   -5 & \mbox{for Feynman gauge~\cite{lopez},}  \nonumber\\
                   -11& \mbox{for Feynman gauge ~\cite{Hansson:1987um},} \nonumber\\
                    -\frac{45}{4}& \mbox{for Landau gauge ~\cite{Hansson:1987um},} \nonumber\\
                     -\frac{27}{4}& \mbox{for Covariant gauge ~\cite{Kobes:1987bi}.} \nonumber\\
		   \end{array} 
		\right.  
\ee
\item IR divergences due to the absence of electric screening scale:    
\begin{figure}[htp]
\vspace*{-0.1in}
\begin{center}
\hspace*{-2in}\includegraphics[width=6cm,height=2.7cm]{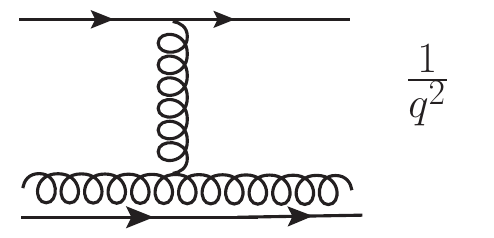}
\end{center}
\vspace*{-2.4cm}
\hspace{4.0in} { 
$ \ \ \left \{ \begin{array}{ll} q\rightarrow 0, & \mbox {\rm{quardratically}}
\\
& \mbox {\rm{ divergent}} 
\end{array}
\right.$}
\end{figure}

\item IR divergences due to the absence of magnetic screening scale (Linde problem)~\cite{Linde:1978px,Linde:1980ts}:

Let us consider a generic $(l+1)$-loop contribution to the self-energy which is shown in Fig.~\ref{linde_dia}. To evaluate the leading infrared divergent contribution we can assume that all 
propagators in Fig.~\ref{linde_dia}  are static, i.e. consider only contribution from zero Matsubara modes. In this case neglecting all tensorial structures the contribution of this 
diagram is
 \begin{figure}[h]
\begin{center}
\includegraphics[width=8cm,height=2.6cm]{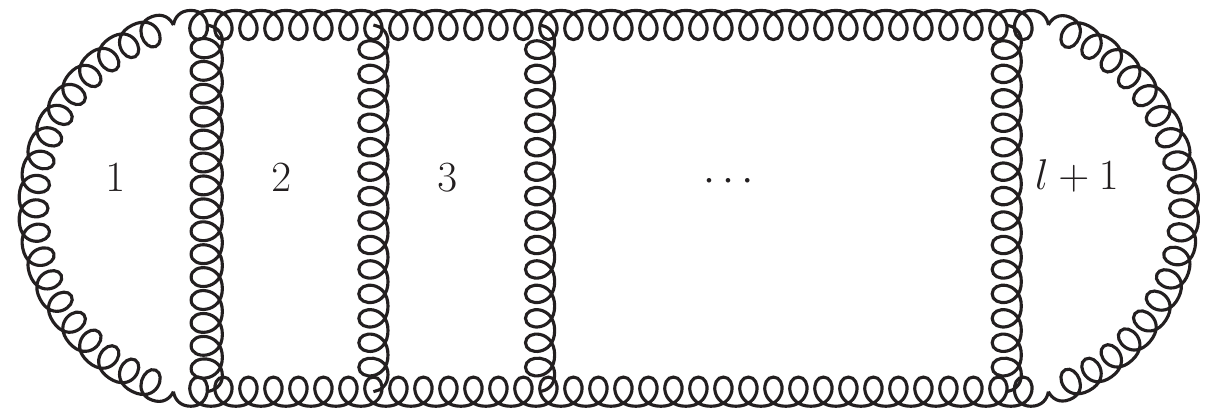}
\caption{Divergent $(l+1)$-loop contribution to the 2-point function.}
\label{linde_dia}
\end{center}
\end{figure}
\be
\underbrace{g^{2l}}_{\mbox{vertex}} \underbrace{\left (\int d^3p \right )^l}_{\mbox{loop integral}} \underbrace{p^{2l}}_{\mbox{vertex}}  
\underbrace{ \left (p^2+m^2\right )^{-3l}}_{\mbox{propagator}} \, ,\label{bpt1}
\ee
In the above expression we have introduced a magnetic mass since one might expect that such a mass scale is generated by higher loop contribution. Using simple power counting arguments one can see that for $ l = 2$ the corresponding contribution becomes
$
g^4T^2 \ln \frac{T}{m} \, , \label{bpt2}
$
which is IR finite due to electric screening mass scale $\sim gT$.

For $l=3$, it becomes
$g^6T^4 \ln \frac{T}{m} \, , \label{bpt3}$
and for $l>3$, it becomes
$
g^6T^4 \left(\frac{g^2T}{m} \right )^{l-3} \, . \label{bpt4}$  Additionally, for longitudinal gluons, the electric screening mass $\sim \, gT$ and thus for $l>3$ it becomes
$g^{l+3}T^4  \, . \label{bpt5}$

This $g^{l+3}$-behaviour is different from expected $g^{2l}$-behaviour and this causes a breakdown of the BPT. Now, if one assumes a 
magnetic screening mass $\sim\, g^2T$ for transverse gluons then it becomes $g^6T^4  \,  \label{bpt5a}$ which is independent of $l$. This indicates that for $l>3$ all diagrams of order $g^6$ are independent of the order of perturbation theory and one needs to resum infinite number of diagrams in $g^6$. The causes a complete break down of the perturbation theory in general.
\end{list}

All these observations suggest that the BPT requires improvement to compute the static and dynamic quantities. This can be done through the effective field theories like DR approach~\cite{Gross:1980br,Appelquist:1981vg,Nadkarni:1982kb,Nadkarni:1988fh,Nadkarni:1988pb}
 and HTL resummation~\cite{Braaten:1989mz,Braaten:1991gm,Braaten:1990az,Taylor:1990ia,Frenkel:1991ts,Barton:1989fk}  techniques.

	\section{Effective Field Theory at Finite Temperature}
	\label{eft}
	\vspace{-0.2cm}
\subsection{Dimensional Reduction}
\label{dr}
\vspace{-0.2cm}
As discussed in subsec.~\ref{scale_sepa} there exists a hierarchy of three momentum scales that play an important role in static properties of matter at high $T$. These scales are hard, soft (electric) and ultra-soft (magnetic). A method has been developed that systematically unravel the contributions from these various momentum scales. 
The method is based on the construction of effective field theories that reproduce static observables at successively longer distance scales $R\gg 1/T$. This effective-field-theory approach is based on an old idea called  dimensional reduction (DR)~\cite{Gross:1980br,Appelquist:1981vg,Nadkarni:1982kb,Nadkarni:1988fh,Nadkarni:1988pb}. According to this idea, the static properties of a $(3+1)$-dimensional field theory at high temperature can be expressed in terms of an effective field theory in three space dimensions.

Because of the periodicity (anti-periodicity) conditions on the fields as discussed in subsec.~\ref{period}, they can be decomposed into Fourier modes in imaginary time $\tau$, with Matsubara frequencies $\omega_n=2\pi n T$ for bosons and $\omega_n=(2n+1)\pi T$  for fermions. As can be seen the fermions do not have a  zero Matsubara mode and are already screened  at scale $\pi T$ and the same goes for the non-zero bosonic modes. The relevant degrees of freedom left are the Matsubara zero modes of the bosonic fields. Other way, the contribution to a correlator from the exchange of a Fourier mode with frequency $\omega_n$ falls off at large $R$ like $\exp(-|\omega_n|R)$. 
Thus, the only modes whose contributions do not exponentially diminish at distances greater than $1/T$ are the $n = 0$ modes of the bosons. This suggests the strategy of integrating out the fermionic modes and the nonzero modes of the bosons to obtain an effective theory for the bosonic zero modes. This process is known as the dimensional reduction, resulting in a $3$-dimensional Euclidean field theory with only the bosonic fields, which reproduces the static correlators of the original theory at distances $R\gg1/T$.

Constructing the DR effective theory by integrating out degrees of freedom beyond leading order in the coupling constant is cumbersome. Once the appropriate $3$-dimensional fields and their symmetries have been identified, a more effective strategy is to use methods of effective field theory. One formulates the most general Lagrangian ${\cal L}_{\mbox{eff}}$ for the $3$-dimensional fields respecting the symmetries. Although this effective Lagrangian contains infinitely many parameters, they are not arbitrary.
By computing static correlators in the full theory and comparing them with the corresponding correlators in the effective theory at distances $R \gg 1/T$, one can determine the parameters of ${\cal L}_{\mbox{eff}}$ in terms of $T$ and the parameters of the original theory. Notably, this matching procedure does not necessarily require explicit determination of the relation between the fields in the effective theory and the fundamental fields.  It is important to highlight that DR effective theories are powerful tools for calculating equilibrium thermodynamic quantities at high temperature and chemical potential within the imaginary time formalism, but they are not effective at low temperatures.
\subsubsection{Scalar theory}
\label{dr_scal}
For a massless scalar field with $\Phi^4$ interaction the Euclidean Lagrangian density is given as
\be
{\cal L}= \frac{1}{2} \left (\partial_\tau \Phi \right )^2 + \frac{1}{2} \left (\bm{\nabla} \Phi\right )^2 +\frac{1}{4!} g^2 \Phi^4 +\delta{\cal L}. \label{dr1}
\ee
After DR, the effective $3$-dimensional Lagrangian~\cite{Braaten:1995cm}  that describes a scalar field $\phi(\bm{\vec x})$  as
\be
{\cal L}_{\mbox{eff}} = \frac{1}{2} \left (\bm{\nabla} \phi(\bm {\vec x})\right )^2  +\frac{1}{2} m^2 (\Lambda) \phi(\bm {\vec x})^2+\frac{1}{4!} \lambda(\Lambda) \phi^4(\bm{\vec x})
+\delta{\cal L}, \label{dr2}
\ee
where $\delta{\cal L}$ includes all other local terms that are consistent with the symmetries. The parameters $m^2(\Lambda)$ and $\lambda(\Lambda)$ and many other parameters in~\eqref{dr2} depend on the ultraviolet cut-off $\Lambda$, the temperature $T$ and the coupling constant $g^2$. This effective theory for massless $\Phi^4$ theory has been used~\cite{Braaten:1995cm} to compute the free energy to order $g^5$ and the screening mass to order $g^4$. The accuracy of these calculations has also been  improved  to order $g^6\log g$  for the free energy and to order $g^5 \log g$ for the screening mass by using renormalization group equations for the parameters of the effective theory.

\subsubsection{Gauge theory}
\label{dr_gauge}
\vspace{-0.2cm}
 As discussed in in subsec.~\ref{bpt}  that the application of the  BPT at finite temperature is obstructed by infrared divergences  of thermal field theory. In scalar field theory, these IR problems can be cured by appropriate resummations which take into account the presence of the electric screening scale $gT$. In  gauge theories like QED and QCD, however, there exists also another sort of IR problems associated with static magnetic fields. Static magnetic fields are not screened at leading order of perturbation theory: the IR limit of the $1$-loop transverse self-energy vanishes in all gauges. Up to order $g^5$, the free energy can be calculated using a resummation of perturbation theory that takes into account the screening of the chromoelectric force at distances of order $1/(gT)$. There is also a qualitatively new effect that arises at order $g^6$. This is because the loop expansion for the free energy breaks down at this order in $g^6$ due to the well known Linde problem~\cite{Linde:1978px,Linde:1980ts} as discussed in subsec~\ref{bpt}. Since the chromomagnetic force is not screened at the scale $gT$, and this causes a breakdown in the resummed perturbation expansion at order $g^6$. 
This was  solved by constructing a sequence of two effective field theories that are equivalent to thermal QCD over 
successively longer length scales.

$\bullet$ The first effective theory, known as elctrostatic-QCD (EQCD), describes  the static properties of thermal QCD at distances of order $1/(gT)$ or larger. The effective Lagrangian~\cite{Braaten:1995jr,Kajantie:2002wa} is given as
\bea
{\cal L}_{\mbox{\tiny EQCD}} &=&\frac{1}{4} G_{ij}^a G_{ij}^a +\frac{1}{2} (D_iA_0)^a (D_iA_0)^a +\frac{1}{2}m_E^2A_0^aA_0^a +\frac{1}{8} (A_0^aA_0^a)^2 
+ \delta{\cal L}_{\mbox{\tiny EQCD}} \, , \label{dr3}
\eea
where $G_{ij}^a =\partial_i A_j^a - \partial_j A_i^a +g_E f^{abc} A_i^bA_j^c$ is the magnetostatic field strength with coupling $g_E$. The  term $\delta{\cal L}_{\mbox{\tiny EQCD}}$ contains all other gauge invariant operators of dimension $3$ and higher which can be built from $A_0$ and $A_i$.  Static gauge invariant correlation functions in full QCD can be reproduced in EQCD by tuning the gauge coupling constant $g_E$, the mass parameter $m^2_E$, the coupling constant $\lambda_E$, and the parameters in $\delta{\cal L}_{\mbox{\tiny EQCD}}$ as functions of $g$, $T$ as $g_E\equiv {\sqrt T}g$, $m_E\sim gT$ and 
$\lambda_E\sim g^2$.

$\bullet$ The second effective theory, called magnetostatic QCD (MQCD),  describes the static properties of QCD at distances of order $1/(g^2T)$ or larger and the effective lagrangian~\cite{Braaten:1995jr} is given as
\be
{\cal L}_{\mbox{\tiny MQCD}} = \frac{1}{4} G_{ij}^a G_{ij}^a+ + \delta{\cal L}_{\mbox{\tiny MQCD}} \, , \label{dr4}
\ee
where $G_{ij}^a$ is a magnetostatic field strength with coupling $g_M$ which differs from $g_E$ by perturbative correction. The  term $\delta{\cal L}_{\mbox{\tiny MQCD}}$ contains all other gauge invariant operators of dimension $5$ and higher which can be built out of $A_i^a$.

 Using this approach, the free energy~\cite{Braaten:1995jr} can be divided into contributions from the momentum scales $T$, $gT$, and $g^2T$, with well-defined weak coupling expansions that begin at order $g^0$, $g^3$, and $g^6$, respectively. The contributions from the $T$ and $gT$ scales can be computed using perturbative methods. For the contribution from the scale $g^2T$, the coefficients in the weak-coupling expansion can be determined using lattice simulations of pure-gauge QCD in $3$ Euclidean dimensions. 
 \subsection{Hard Thermal Loop (HTL) Resummation}
 \label{htlr}
 \subsubsection{Hard thermal loops}
 \label{htlr1}
 \vspace{-0.2cm}
 We have discussed  in subsec.~\ref{bpt}  that the application of the  BPT at finite temperature is obstructed by gauge dependent results and IR divergences  of thermal field theory. This happens because the calculations done in BPT were incomplete in order of the coupling constant as certain classes of diagrams were not taken into account. These diagrams  are of higher order in the loop expansion which contribute to the same order in the coupling constant as the one loop diagram~\cite{Braaten:1989mz}. 
\begin{wrapfigure}[8]{r}{0.48\textwidth}
\begin{center}
	\vspace{-1cm}
\includegraphics[scale=.33]{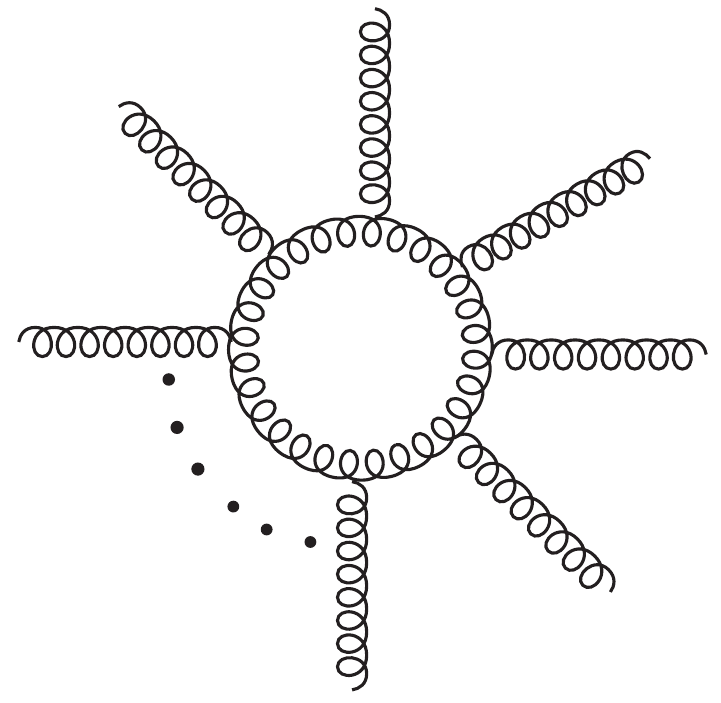}
\hspace{0.2cm}
\includegraphics[scale=.33]{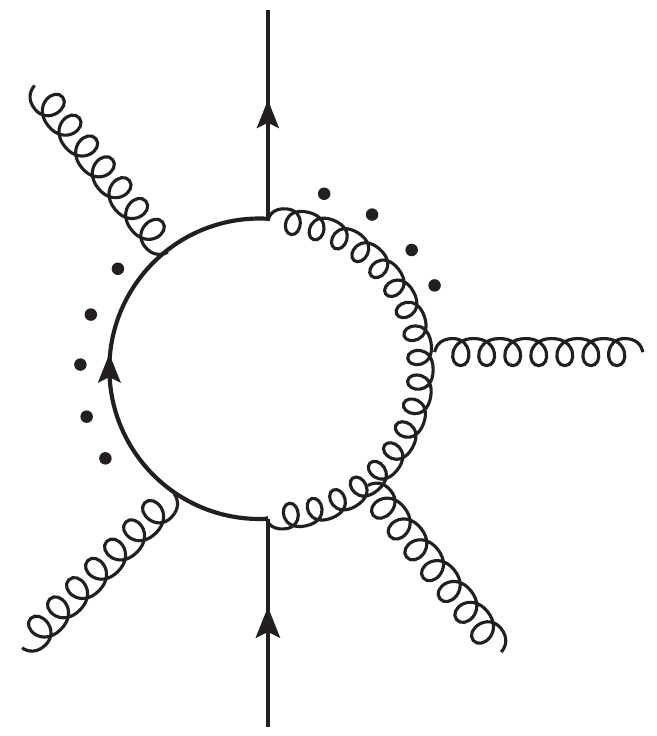}
\vspace{-0.4cm}
\caption{$N$-point graph with $N$ external gluons (left) and  $(N-2)$ external gluons and a quark pair(right).}
\label{ngq}
\end{center}
\end{wrapfigure}
These relevant diagrams can be identified by distinguishing the hard scale ($\sim T$) for loop momenta and the soft (electric) scale ($\sim gT$) for external momenta. 
In general hard thermal loops can only arise from diagrams~\footnote{There can be another diagram having quark loop with $N$ external gluons.} with either $N$ external gluon legs as shown in the left panel of Fig~\ref{ngq} or $(N-2)$ external gluon legs and an external quark pair for any $N \ge2$ as shown in right panel of Fig.~\ref{ngq}. As for example we consider $N$-point function with $N$ external gluons in left panel of Fig.~\ref{ngq}.  
In this case neglecting all tensorial structures the contribution of this  can be written as
\bea
\sim\! \underbrace{ \oint}_{\mbox{loop integral}} \! \underbrace{{\mbox{Tr}} \Big[K^{\mu 1}\cdots K^{\mu N} \Big]}_{\mbox{vertices}}
\underbrace {\Delta \left (P_1-K\right ) \cdots \Delta (P_{N-1}-K)}_{\mbox{propagators}} \, . \label{htl1}
\eea
Similarly, the contribution from the right panel of Fig.~\ref{ngq} can be written as
\bea
&\sim& \underbrace{ \oint}_{\mbox{loop integral}}  \underbrace{{\mbox{Tr}} \Big[K^{\mu 1}\cdots K^{\mu (N-1)} \Big]}_{\mbox{vertices}}
\underbrace {\Delta \left (P_1-K\right ) \cdots \Delta (P_{M-1}-K){\tilde \Delta }(P_{M}-K) \cdots {\tilde \Delta }(P_{N-1}-K)}_{\mbox{propagators}} \, . \label{htl4}
\eea
where $K$ is the loop momentum, $P$ is the external momentum, $\Delta$ is bosonic propagator and $\tilde \Delta$ is fermion propagator. In Eqs.~\eqref{htl1} and~\eqref{htl4} $P_i$, $i=1\cdots \cdots , N-1$ are linear combinations of the external momenta. In Eq.~\eqref{htl1} the power of $K$ in the numerator come from three gluon vertices and in Eq.~ \eqref{htl4} the $(N-2)$ powers of $K$ come from either the three vertices of only gluons or quark propagator and $(N-2)$ indices $\mu_i$ come from the external gluons and one is contracted with a gamma matrix.

Using Saclay method defined in subsec~\ref{saclay} the $\tau$-integration can be entangled and complicated to evaluate. This is because beyond $N=2$ this method is not so convenient for one-loop calculations. However, this is still an important calculational technique in higher loop order calculation in gauge theories. It should be easier to use the operator based technique of Espinosa and collaborators obtained in Refs.~\cite{Espinosa:2003af,Espinosa:2005gq} where one can scalarise the Feynman integrals. Looking at the results for the thermal sums  one finds that for large $N$, the number of different terms is also very large~\footnote{For examples with low $N$, it can be seen that many terms from the general form will cancel out.}. However, according to the cutting rule argument presented by Braaten and Pisarski~\cite{Braaten:1989mz}, one keeps terms that produce hard thermal loops from integrals in Eqs.~\eqref{htl1} and~\eqref{htl4}.
   
After performing frequency sums over Matsubara frequencies, the contribution of \eqref{htl1} can be written as
\bea
&\sim& \int  d^3k \frac{K^{\mu 1} \cdots K^{\mu N}}{E_kE_{p_1} \cdots E_{p_{N-1}-k}} \Big( n_B(E_k)-n_B(E_{p_1-k })\Big )\Big[\Big (p^0_1-E_k +E_{{p_1}-k}\Big ) 
\cdots \Big (p^0_{N-1}-E_k +E_{{p_{N-1}}-k}\Big )  \Big]^{-1}  \, , \label{htl2a}
\eea
where $n_B$ is the Bose-Einstein distribution.
Similarly, the contribution of \eqref{htl4} can also be written as
\bea
&\sim& \int  d^3k \frac{K^{\mu 1} \cdots K^{\mu {N-1}}}{E_kE_{p_1} \cdots E_{p_{N-1}-k}} \Big( n_B(E_k)-n_F(E_{p_1-k })\Big )\Big[\Big (p^0_1-E_k +E_{{p_1}-k}\Big ) 
\cdots \Big (p^0_{N-1}-E_k +E_{{p_{N-1}}-k}\Big )  \Big]^{-1}   ,\ \ \label{htl5}
\eea 
where $n_F$ is the Fermi-Dirac distribution.

Now we make the power counting below:
\begin{enumerate}
\item[$\bullet$] Here one considers that the loop momentum to be hard or more precisely that the largest contribution 
to the loop integral comes from the region that the loop momentum is hard, i.e., the magnitude satisfies  $k\sim T$ and $E_k\sim E_{p-k}\sim k \sim T$ for massless particles (see Fig.~\ref{scal_htl}). This is known as hard thermal loops (HTL) approximation\cite{Braaten:1989mz}.
\item[$\bullet$] Difference of distribution functions in HTL approximation:

(a)
\begin{equation}
n_B(E_k)-n_B(E_{p-k}) \approx n_B(E_k)-n_B(E_k) +\bm{\vec p}\cdot \bm{\hat k} \frac{dn_B(E_k)}{dk} = \frac{p\cos\theta}{T}n_B(E_k) \Big (1+n_B(E_k)\Big )\approx \frac{P}{T},
\end{equation}
 where $\theta$ is the angle between $P$ and $K$.

(b) 
\begin{equation}
n_B(E_k)-n_F(E_{p-k }) \approx  n_B(E_k)-n_F(E_{k }) \approx {\textrm{constant}} \, , \label{htl5a}
\end{equation}
where we have used the Taylor expansion of the distribution functions as
\begin{subequations}
\begin{align}
n_B\left(\frac{k}{T}\right) = \frac{T}{k}-\frac{1}{2}  +{\cal O}\left(\frac{k}{T}\right) \, , \qquad\text{and}\qquad
n_F\left(\frac{k}{T}\right) &= \frac{1}{2}-\frac{k}{4T}  +{\cal O}\left(\left[\frac{k}{T}\right]^2\right ) \, , \label{befd_exp}
\end{align}
\end{subequations}

\item[$\bullet$] The Landau damping factor $\Big[\Big (p^0_1-E_k +E_{{p_1}-k}\Big ) \cdots \Big (p^0_{N-1}-E_k +E_{{p_{N-1}}-k}\Big )  \Big]^{-1}  \sim \frac{1}{P^{N-1}}$  because
$p_0\mp k\pm |\bm{\vec p}-\bm{\vec k}| = \omega\mp k\pm (k- \bm{\vec p}\cdot \bm{\hat k}) \approx \omega \mp \bm{\vec p}\cdot \bm{\hat k} =P\cdot (\pm {\widehat K}) =  P\cos \theta \approx  P$ where we have used $\pm \widehat K=(1,\pm \bm{\hat k})$, a light like vector.
\item[$\bullet$] Similarly $p_0\pm k\pm |\bm{\vec p}-\bm{\vec k}|= \om \pm k\pm (k- \bm{\vec p}\cdot \bm{\hat k}) \approx \pm 2k$.
    \item[$\bullet$] The integral $ \int  d^3k \sim T^3$.
\item[$\bullet$] Each vertex $\sim g$.
\end{enumerate}

Using these power counting, the amplitude in Eq.~\eqref{htl2a}  becomes
\bea
&\sim& g^N\,\, T^3\,\,  \frac{T^N}{T^N}\,\, \frac{P}{T}\,\, \frac{1}{P^{N-1}}  
\sim \frac{g^2T^2}{P^2} \, \,  \frac{g^{N-2}}{P^{N-4}} \sim \frac{g^2T^2}{P^2} \, \, \times \,\,   \bm{\mbox{\Big(Tree level amplitude with $N$ gluons \Big)}} \, . \label{htl3}
\eea

Using power counting as before, one can write the final contribution from Eq.~\eqref{htl5} as
\bea
\sim g^N\,\, T^3\,  \frac{T^{N-1}}{T^N}\,  \frac{1}{P^{N-1}}  \sim \frac{g^2T^2}{P^2} \, \,  \frac{g^{N-2}}{P^{N-3}}
 \sim \frac{g^2T^2}{P^2} \,  \times \,   \bm{\mbox{\Big(Tree level amplitude with $(N-2)$ gluons and a quark pair\Big)}} \, . \nonumber
\eea
If the external momentum is hard ($\sim T$) then both the amplitudes are, respectively, suppressed by $g^2$ of the  $``${\bf Tree Level Amplitude}''.
But when the external momentum is soft, $P\sim gT$, then the  both amplitudes, from Fig.~\ref{ngq} become 
equivalent to $``${\bf Tree Level Amplitude}''. Based on the above analyses, it is evident that higher-order diagrams in the loop expansion contribute to same order in coupling as the one-loop diagrams, by distinguishing the hard ($\sim T$) contribution arising from loop momenta and the soft contribution ($\sim gT$) arising from external momenta. The thermal corrections arise from all orders of perturbation theory. This indicates the necessity to consider these diagrams if the physical quantity is sensitive to the soft (electric) scale. The effective theory, developed around hard thermal loops, {\it resums such diagrams}.
\subsubsection{Main features of HTL approximation:}
\label{htlr2}
\vspace{-0.2cm}
\begin{enumerate}[\leftmargin=2cm]
\item[$\bullet$] At high temperature certain classes of diagrams can be  taken care by distinguishing hard  internal (loop) momenta ($\sim T$) and soft external momenta ($\sim gT$) which were missing in BPT. 
\item[$\bullet$]It is essential to resum those HTL diagrams in geometrical series to calculate the effective propagators and vertices in one-loop. They are connected by Ward-Takahashi identity in QED and by Slanov-Taylor identity in QCD. The process of this resummation will be discussed in details in sections~\ref{scalar_htl}, \ref{qed_htl} and ~\ref{qcd_htl}.
\item[$\bullet$] At the same time, medium effects such as electric screening mass, thermal mass, collective behaviour of quasiparticles and Landau damping, are taken into account due to resummations. 
\item[$\bullet$] The effective $N$-point functions  can be utilised in perturbation theory, resulting in an effective perturbation theory know as HTL perturbation theory (HTLpt), which will be discussed in details in subsec.~\ref{htlpt}. This effective perturbation theory yields gauge independent results and is also complete  up to a certain order of the coupling (depending on the quantity under consideration). For example it is $g^5$ order in QCD thermodynamics.
\item[$\bullet$] In scalar field theory, the IR problems are resolved through appropriate resummations, accounting for the presence of the electric screening (Debye) mass of the order $gT$. In gauge theories such as QED and QCD, the IR singularities are improved due to the electric scale. However, another type of IR issues arises in connection with static magnetic fields. At the leading order of HTLpt,  static magnetic fields are not screened because the 1-loop transverse gluon self-energy vanishes in the IR limit in all gauges. Up to order $g^5$, the quantities can be calculated using HTLpt, which considers the screening of the chromoelectric scale but breaks down at $g^6$ order due to the Linde problem, as discussed in subsec.~\ref{bpt}.
\end{enumerate}
Here we now note the chronology of the development of HTL.  The photon HTL was first calculated by Silin~\cite{silin} in plasma physics, and then by Fradkin~\cite{Fradkin} using field theoretic methods. The electron HTL was first discussed by Klimov~\cite{Klimov:1981ka} who found a long wavelength mode called plasmino, and then by Weldon~\cite{Weldon:1982bn,Weldon:1989ys}. The gluon HTL was computed by Kalashnikov and Klimov~\cite{Kalashnikov:1979kq,Klimov:1982bv} and then by  Weldon~\cite{Weldon:1982aq}. Nevertheless, the importance of HTL was fully realised and rigorously advocated by Braaten and Pisarski~\cite{Braaten:1989mz},  and Frenkel and Taylor~\cite{Taylor:1990ia,Frenkel:1991ts} and Barton~\cite{Barton:1989fk}. The HTL for massive fermion was computed by Petitgirard~\cite{Petitgirard:1991mf}
and by Baym, Blaizot and Svetitsky~\cite{Baym:1992eu}. The gauge independence of HTL was studied by Kobes, Kunstatter and Rebhan~\cite{Kobes:1990xf}. The magnetic mass in QED and scalar QED is studied by Blaizot, lancu and Parwani ~\cite{Blaizot:1995kg}. Recent developments in HTL will be discussed in this review.
\subsubsection{Approximately self-consistent HTL resummation}
\label{htlr4}
\vspace{-0.2cm}
An additional approach~\cite{Blaizot:1999ip,Blaizot:1999ap,Blaizot:2000fc} based on self-consistent approximations using the skeleton representation of the thermodynamic functional~\cite{Luttinger:1960ua,deDominicis:1964zz}, suitable only for computing various thermodynamic quantities. It primarily reorganises the perturbation theory by considering the so-called $2$-loop $\Phi$-derivable~\cite{Baym:1962sx} approximation, for which it is found out that the first derivatives of the thermodynamic potential, the entropy and the quark densities, take a rather simple, effectively one-loop form~\cite{Vanderheyden:1998ph}, but in terms of fully dressed two point functions (self-energies and propagators). Since the exact self-energy  is not known,  it can be regarded as a variational function. The $\Phi$-derivable prescription~\cite{Baym:1962sx} is to truncate the perturbative expansion for the thermodynamic functional and to determine self-energy self-consistently as a stationary point of thermodynamic potential. This gives an integral equation for self-energy which is difficult to solve numerically, except in cases where it is momentum independent. But in gauge theories, an exact solution would anyhow be unsatisfactory because $\Phi$-derivable approximations in general do not respect gauge invariance. Therefore, it is proposed~\cite{Blaizot:1999ip,Blaizot:1999ap,Blaizot:2000fc} to use a gauge independent but only approximately self-consistent two point functions as obtained from HTL resummation. Using these HTL  two-point functions the expression for equilibrium thermodynamic quantities obtained from the 2-loop $\Phi$-derivable approximation give a gauge-independent and utra-violate (UV) finite results. Furthermore, the thermodynamic quantities acquire nonperturbative effects in the coupling, contain the correct leading order (LO) and the next-to-leading order (NLO) effects of interactions in accordance with HTLpt, originating from kinematical regimes where HTL serves as a justifiable approximation.
	\section{Scalar Theory}
	\label{scalar_htl}
	\vspace{-0.2cm}
Before leaping into to gauge theory, it is instructive to start with the simplest interacting quantum field theory, namely scalar $\Phi^4$ theory.
\vspace{-0.3cm}
 \subsection{Lagrangian}
\label{scalar_lag}
In this subsection, we discuss about the simplest interacting scalar field theory, namely, a single massless scalar field with a $\Phi^4$ interaction. The Lagrangian for such scalar field is expressed as
\be
{\mathcal L}\;=\;\frac{1}{2}(\partial_{\mu}\Phi)^2+\frac{1}{24}{g^2}\Phi^4 +\Delta{\mathcal L} \; ,
\label{sl}
\ee
where $g$ denotes the coupling constant, and $\Delta{\mathcal L} $ represents the renormalised counterterms.
The Lagrangian density in \eqref{sl} can be divided into a free part and an interacting part as
\be
{\mathcal L}_{0}\!\!&=&\!\!\frac{1}{2}(\partial_{\mu}\Phi)^2\; \hspace*{0.5in} \Rightarrow 
 \hspace*{1.3in} \,{\includegraphics[scale=0.7]{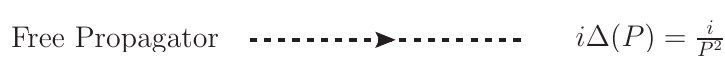}} \ee
   \vspace*{-.7cm}
 \be
  \vspace*{0cm}
{\mathcal L}_{\rm int}\!\!&=&\!\!\frac{1}{24}g^2\Phi^4\;   \hspace*{0.4in} \Rightarrow 
\hspace{2.5in} {} \,\, \vcenter{\hbox{\includegraphics[scale=0.4]{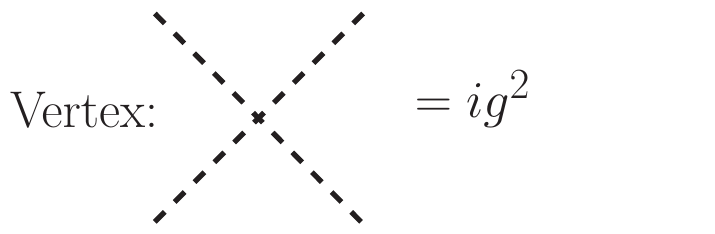}}}\,\,\,\, \, , 
\ee
where $P\equiv (p_0,\bm{\vec p})$ is the Euclidean four-momentum. The Euclidean energy $p_0$ takes discrete values: $p_0=2n\pi iT$ for bosons and $p_0=(2n+1)\pi iT$ for fermions, where $n$ is an integer. The radiative corrections are then calculated in a loop expansion, which is equivalent to a power series in $g^2$. It becomes apparent that the perturbative expansion fails at finite temperature, and the weak-coupling expansion transforms into an expansion in $g$ rather than $g^{2n}$, where $n$ is the number of loops.
\subsection{Self-Energy}
\label{scalar_se}
We will first calculate the self-energy by evaluating the relevant diagrams. 
\subsubsection{One-loop self-energy}
\label{sc_1loop}
\begin{wrapfigure}[6]{r}{0.3\textwidth}
	\begin{center}
		\vspace{-1.6cm}
		\includegraphics[height=3.cm]{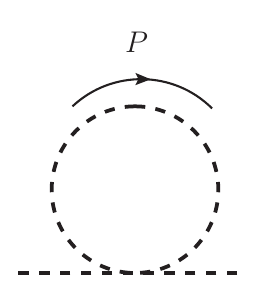}
	\end{center}
	\vspace{-0.3cm}
	\caption[One loop scalar self-energy]{One-loop scalar self-energy graph.}
	\label{thmass}
\end{wrapfigure}

	\vspace{0.2cm}
The Feynman diagram that contributes to the 1-loop self-energy  is the tadpole diagram as shown in Fig.~\ref{thmass}.
The one-loop diagram is independent of the external momentum and the resulting integral expression is
\be
\Pi^{(1)}\!\!&=&\!\!\frac{1}{2}g^2\sumintb_P\frac{1}{P^2} \; 
 =\frac{1}{24}g^2T^2 \; 
\equiv (m^{\textrm s}_{\textrm{th}})^2 \, , \label{fm}
\ee
where the superscript indicates the number of loops. 

Equation~(\ref{fm}) represents the leading order thermal mass of scalar field with $\Phi^4$ interaction. The sum-integral over $P$ represents a summation over Matsubara frequencies and integration of spatial momenta in $d=3-2\epsilon$ dimensions.  The sum-integral in Eq.~\eqref{fm} has ultraviolet power divergences that are set to zero in dimensional regularization. We are then left with the finite result~(\ref{fm}), indicating that thermal fluctuations induce a mass for the scalar field of the order $gT$. The thermal mass $m^{\textrm s}_{\textrm{th}}$ is analogous to the Debye mass, well-known in non-relativistic QED plasma.
\subsubsection{Two-loop self-energy}
\label{sc_2loop}
\begin{wrapfigure}{r}{0.5\textwidth}
	\begin{center}
		\includegraphics[scale=0.8]{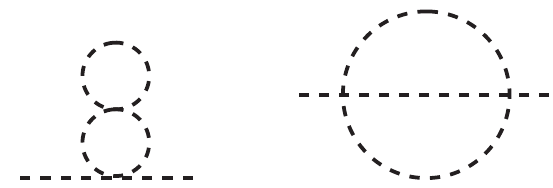}
		\vspace{-0.3cm}
		\caption[2-loop scalar self-energy]{Double-bubble and sunset diagrams in 2-loop scalar self-energy.}
		\label{2lself}
	\end{center}
\end{wrapfigure}

There are two 2-loop diagrams as shown in  Fig.~\ref{2lself}  The first one is known as double-bubble diagram whereas the second one is known as sunset diagram. We first consider the double-bubble diagram in Fig.~\ref{2lself}. 

		\vspace{0.1cm}
This diagram is also independent of the external momentum and gives the following sum-integral
\be
\Pi^{(2)}\;=\;-\frac{1}{4}g^4\sumintbb_{PQ}\frac{1}{P^2}\frac{1}{Q^4} \;.
\ee
This integral is IR divergent. The problem stems from the middle loop with two propagators. In order to isolate the source of the divergence, we look at the contribution from the zeroth Matsubara mode to the $Q$ integration
\be
-\frac{1}{4}g^4\sumintb_P\frac{1}{P^2}T\int_{\bm{ q}}\frac{1}{q^4}  = -\frac{1}{4}g^4\sumintb_P\frac{1}{P^2}T\int_{0}^\infty \frac{dq}{q^2} \, .
\label{g3}
\eea 
The integral over $q$ in \eqref{g3} behaves like $1/q^2$ and quadratic IR  divergent as $q \rightarrow 0$. This infrared divergence indicates that naive perturbation theory breaks down at finite temperature. However, one can cure this IR divergence by $``$resumming'' the leading order correction into the propagator for the $Q$ propagator (see subsec~\ref{sc_prop_htl}) as
\be
\Delta^\star(Q)\;=\;\frac{1}{Q^2-(m^{\textrm s}_{\textrm{th}})^2}\;,
\label{impprop}
\ee
where $m_{\rm{th}}^{\rm s}$ can be obtained from \eqref{fm} as $m^{\textrm s}_{\textrm{th}}=gT/\sqrt{24} \ll T$. The mass term in the propagator~(\ref{impprop}) provides an IR cutoff of order $gT$. The contribution from~(\ref{g3}) would then be
\be
-\frac{1}{4}g^4\sumintb_P\frac{1}{P^2}T\int_{\bf q}\frac{1}{(q^2+(m^{\textrm s}_{\textrm{th}})^2)^2}
\;=\;-\frac{1}{4}g^4\left(\frac{T^2}{12}\right)\left(\frac{T}{8\pi m^{\textrm s}_{\textrm{th}}}\right)
+{\mathcal O}\left(g^4m^{\textrm s}_{\textrm{th}}T\right)\;.
\ee
Since $m\sim gT$,  the leading term is ${\mathcal O}(g^3)$ and the subleading term is  ${\mathcal O}(g^5)$ . We now note that the leading contribution from the double-bubble contributes is in the  order $g^3T^2$ to the self-energy but not at order $g^4T^2$. This is different than the BPT in which all terms are even powers of the form $g^{2n}$. 

We now note that the $``$sunset diagram'' is IR finite and contributes at order $g^4$ and, hence, is subleading.
\subsubsection{Three-loop self-energy}
\label{sc_3loop}
\begin{figure}[h]
	\begin{center}
		\includegraphics[height=3.5cm,width=6cm]{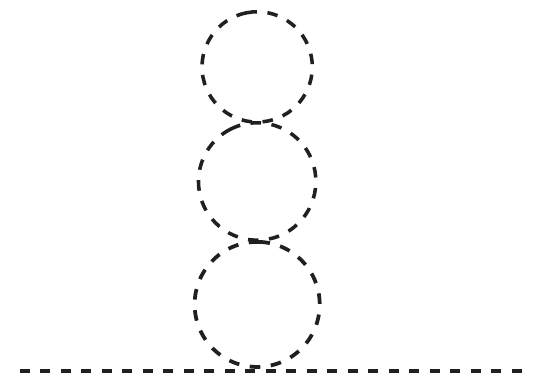}
		\includegraphics[height=3.5cm,width=5cm]{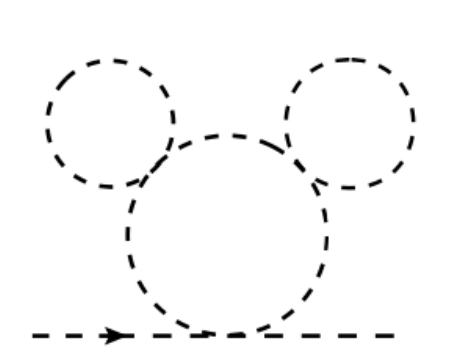}
	\end{center}
	\vspace{-0.4cm}
	\caption[Three-loop diagram]{\textit{Left panel}: Subleading order three-loop self-energy diagram; \textit{Right panel}: Leading order three-bubble  diagram in scalar self-energy.}
	\label{sc_3l}
\end{figure}

		\vspace{-0.2cm}
On the other hand, the three-loop diagram as shown in left panel  of Fig.~\ref{sc_3l}  is known as snowman graph and contributes at the  order $g^4T^2$ and hence subleading. 
Next we consider three-bubble diagram as shown in the right panel of Fig.~\ref{sc_3l} and  the leading term is ${\mathcal O}(g^3)$ but not of the ${\mathcal O}(g^6)$ as one might have expected from BPT.  \emph{Similarly, one can show that the diagrams with any number of bubbles like} Fig.~\ref{sc_3l} \emph{are all of ${\mathcal O}(g^3)$}. 

\subsection{Resummation of Bubble Diagrams in HTL Approximation}
\label{sc_bub_resum}
\begin{figure}[htb]
	\begin{center}
			\vspace{-0.3cm}
		\includegraphics[height=3.5cm,width=16cm]{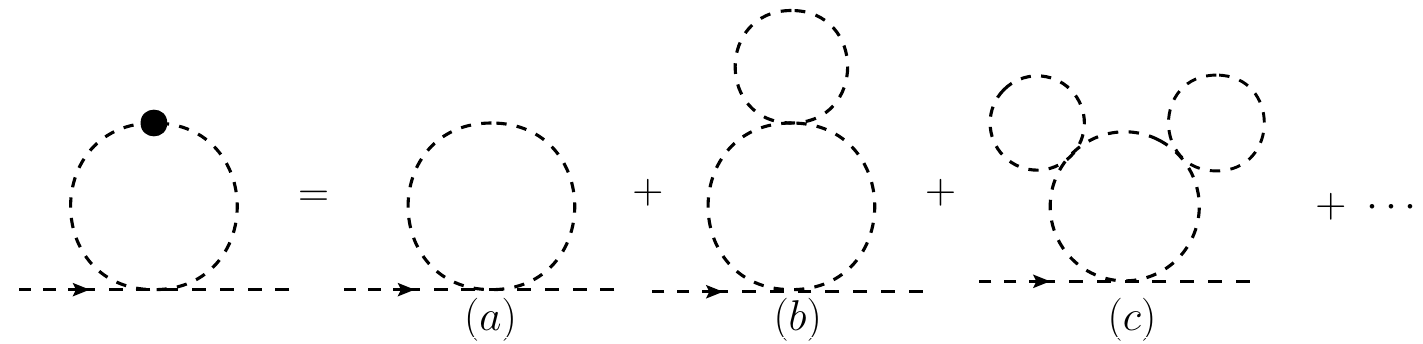}
	\end{center}
	\vspace{-0.5cm}
	\caption[Resummation of bubble diagram]{Resummation of bubble diagrams. The diagram ($a$) is of ${\mathcal O}(g^2)$ whereas 
	those in ($b$), ($c$) and higher order bubble graphs are of ${\mathcal O}(g^3)$.}
	\label{resum_bub}
\end{figure}

As seen from previous subsecetions that the BPT, clearly, breaks down since the  ${\mathcal O}(g^3)$ correction to the self-energy receives contributions from all loop orders.  This is because going to higher loops orders one identifies a class of diagrams that all contribute at ${\mathcal O}(g^3)$. They are the so-called $``$bubblegraphs'' which are distinct from, e.g. the sunset graph or the snowman graph. Therefore, one only needs to resum a subset of all possible bubble  graphs as shown in Fig.~\ref{resum_bub} in order to obtain a consistent expansion in $g$.

Summing the entire series one obtain the effective 1-loop scalar self-energy as
\be
\Pi^{(1)\star}&=&\!\!\frac{g^2}{24}T^2\left[1-\frac{g\sqrt{6}}{4\pi}+{\mathcal O}\left(g^2\right)\right]\; . \label{eff_se}
\ee
The ${\mathcal O}(g^2)$ corresponds to the 1-loop self-energy diagram in Fig.~\ref{resum_bub}(a) and the ${\mathcal O}(g^3)$ corresponds to the summation of the bubble diagrams in Fig.~\ref{resum_bub}(b), Fig.~~\ref{resum_bub}(c) and higher order bubble diagrams.

It turns out that one can also obtain the same result by simply putting the massive thermal effective propagator as given in~\eqref{impprop} in 1-loop self-energy as given in the left hand side of Fig.~\ref{resum_bub}. Then one  obtains
\be
\nonumber
\Pi^{(1)\star} \!\!&=&\!\!\frac{1}{2}g^2\sumintb_P\frac{1}{P^2-(m^{\textrm s}_{\textrm{th}})^2} 
=\frac{1}{2}g^2\Big[-T\int_{\bf p}\frac{1}{p^2+(m^{\textrm s}_{\textrm{th}})^2}+
\sumintb_P^{\prime}\frac{1}{P^2}
+{\mathcal O}\left((m^{\textrm s}_{\textrm{th}})^2\right)\Big] \\
\!\!&=&\!\!\frac{g^2}{24}T^2\left[1-\frac{g\sqrt{6}}{4\pi}+{\mathcal O}\left(g^2\right)\right]\; , \label{same}
\ee
where the prime on the sum-integral indicates that we have excluded the $n=0$ mode from the sum over the Matsubara frequencies. The order $g^3$ corresponds to the summation of the bubble diagrams in Fig.~\ref{resum_bub}, which is obtained by expanding the effective propagator in \eqref{impprop} around $m^{\textrm s}_{\textrm{th}}=0$. Thus by taking the thermal mass into account, one is resumming an infinite set of bubble diagrams from all orders in perturbation theory.

\subsection{Resummed Scalar Propagator in HTL Approximation}
\label{sc_prop_htl}
\begin{figure}[htb]
	\begin{center}
		\includegraphics[height=2cm,width=16cm]{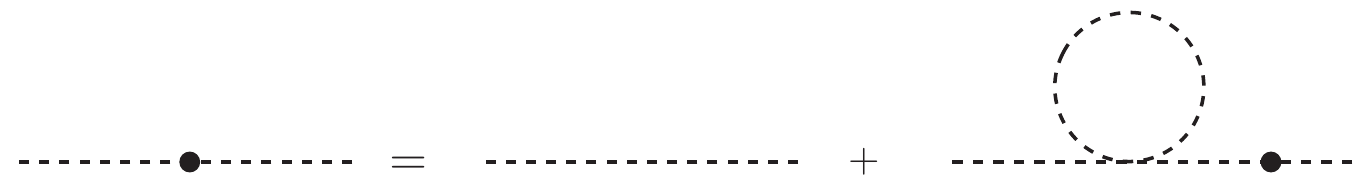}
	\end{center}
	\vspace{-0.4cm}
	\caption[Resummation of propagator]{Resummation of effective scalar propagator in 1-loop scalar self-energy.}
	\label{eff_prop}
\end{figure}

We have found in subsec~\ref{sc_bub_resum} that, at the level of the propagator (2-point function), there is an infinite set of graphs that contribute at next-to-leading order, i.e., the ${\mathcal O}(g^3)$ in the scalar self-energy. This continues as one proceeds to higher orders. The summation of these graphs can be accomplished in a more straightforward manner by using the  HTL resummed propagator as shown in Fig.~\ref{eff_prop}. 
The full propagator in 1-loop resummation is given by Dyson-Schwinger equation following Fig.~\ref{eff_prop} as
$
\Delta^\star = \Delta + \Delta \Pi^{(1)}\Delta^\star \, .
\label{resum_prop}
$
Solving this one obtains the effective scalar propagator as
\be
\Delta^\star(P)\;&=& \;\frac{1}{P^2-\Pi^{(1)}}\; .
\label{resum_prop1}
\ee
	The bare one loop scalar self-energy in \eqref{fm} is the first example of HTL approximation.  To see this, we consider the bare one-loop self-energy in Fig.~\ref{thmass} can be written from the first line of \eqref{fm} as 	\bea
	\Pi^{(1)}\!\!&=&\!\!\frac{1}{2}g^2\sumintb_P\frac{1}{P^2} \; =\; \frac{g^2}{2} T \sum_{p_0=2\pi i n T} \int \frac{d^3p}{(2\pi)^3} \frac{1}{p_0^2-\bm{\vec p}^2}\,\, , 
	\eea
\begin{wrapfigure}[15]{r}{0.4\textwidth}
	\vspace{-3mm}
	\begin{center}
		\includegraphics[scale=.5]{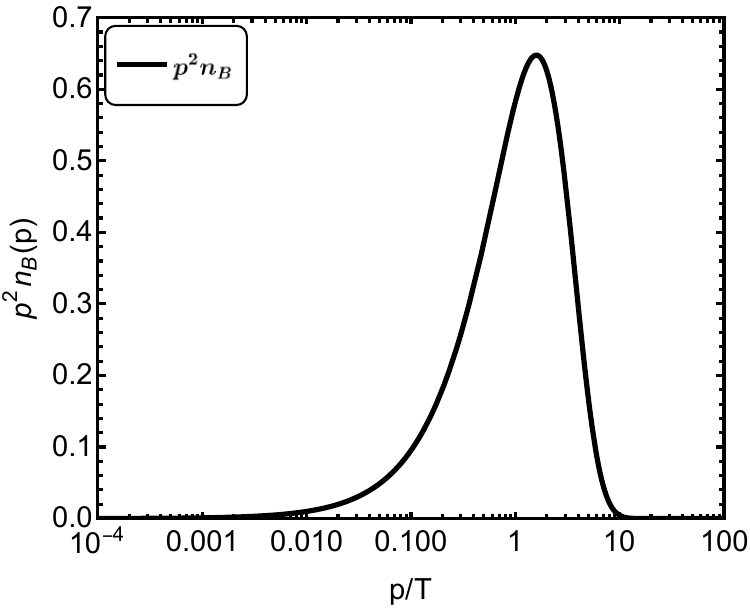}
	\end{center}
	\vspace{-0.6cm}
	\caption{HTL  contribution dominates the one-loop scalar self-energy at the hard loop momenta scale $p\sim T$.}	
	\label{scal_htl}
\end{wrapfigure}
	\vspace{0.cm}
	which after bosonic Matsubara  frequency sum transforms into an integral involving the Bose-Einstein distribution function, $n_B(p)$,  as
	\bea
	\Pi^{(1)}\!\!&\propto& \frac{g^2}{T}\int_{0}^{\infty}dp \, p^2 \, n_B(p) \, . \label{sc_1loop_htl}
	\eea
	The integral in~\eqref{sc_1loop_htl} is found to dominate at hard loop momenta, $p\sim T$ which is displayed in Fig.~\ref{scal_htl}.
Now, the effective propagator in HTL can now be written as
\be
\Delta^\star(P)\;= \;\frac{1}{P^2-(m^{\textrm s}_{\textrm{th}})^2}\;.\hfill
\label{resum_prop2}
\eea
where $m$ can be obtained from \eqref{fm} as $m_{\rm{th}}^s=gT/\sqrt{24}$. If the momenta of the propagator are of order $T$ or {\it hard}, it is evident that the thermal mass acts as a perturbation and can be neglected. However, when the momenta of the propagator are of order $gT$ or {\it soft}, the thermal mass becomes as significant as the bare inverse propagator and cannot be ignored. Therefore, by considering the thermal mass, one effectively resums an infinite set of diagrams in scalar self-energy from all orders of perturbation theory as described in~\eqref{same} in subsec~\ref{sc_bub_resum}.
\subsection{$N$-point Functions in $\Phi^4$-Theory in HTL Approximation}
\label{vertex}
\vspace{-0.2cm}
In the preceding section, we explored two-point function in scalar field theory. Upon calculating higher-order $N$-point functions in scalar theory, it becomes evident that the one-loop correction to the four-point function at high temperatures behaves~\cite{Pisarski:1990ds} as
$
\Gamma^{(4)}\;\propto\;g^4\log\left(T/p\right)\;,
$
where $p$ represents the external spatial momentum. Consequently, the loop correction to the four-point function increases logarithmically with temperature, $T$. Hence, it is always down by $g^4\log(1/g)$ as the external momentum $p$ is soft in HTL approximation, suggesting that employing a bare vertex suffices. 

More generally, one finds that the only HTL in scalar $\Phi^4$-field theory is the tadpole diagram in Fig.~\ref{thmass} and resummation is taken care of by including the thermal mass in the propagator in subsec~\ref{sc_prop_htl}. However, in gauge theories, the scenario is much more complicated than the scalar theory as we shall discuss in the next two sections~\ref{qed_htl} and \ref{qcd_htl}, respectively.
%
\subsection{HTL Improved Scalar Lagrangian}
\label{htl_sc_lag}
\vspace{-0.2cm}
In the previous subsection~\ref{scalar_se},  we have seen that when evaluating higher order vacuum diagrams, one encounters divergences in the IR regime. These divergences are encountered by using  only the propagators with zero Matsubara frequencies which do not have an IR cutoff of order $T$. This problem can be solved by using the thermal mass, $m$ in \eqref{fm} as a IR  cutoff, which is then incorporated into the theory by changing the Lagrangian in~\eqref{sl} to
$
{\mathcal L}_{\textrm{eff}} = {\mathcal L}_0 + {\mathcal L}_{\textrm{int}} +\Delta{\mathcal L} \, , \label{sc_eef}
$
where 
\begin{subequations}
\begin{align}
{\mathcal L}_0  &= \frac{1}{2}(\partial_{\mu}\Phi)^2-\frac{1}{2}{(m^{\textrm s}_{\textrm{th}})^2}\Phi^2  \, , \label{sc_free} \\
 {\mathcal L}_{\textrm{int}} &= - \frac{1}{24}{g^2}\Phi^4  +  \frac{1}{2}{(m^{\textrm s}_{\textrm{th}})^2}\Phi^2   \, . \label{sc_int}
\end{align}
\end{subequations}
We note that the physics of the problem at hand do not change when this is done, since the same term is added and subtracted in the Lagrangian.  However, in the interaction Lagrangian in~\eqref{sc_int}, the $(m^{\textrm s}_{\textrm{th}})^2$-term can also be interpreted as an interaction that couples two scalar fields. Due to this it is expected that in addition to diagrams consisting of loops and four-point vertices, new diagrams will be encountered.
	\section{Quantum Electrodynamics (QED)}
	\label{qed_htl}
	\vspace{-0.2cm}
Quantum electrodynamics (QED) is an abelian gauge theory. The symmetry group is  $U(1)$ abelian group, which is also a commutative group. In QED, the interaction between two spin $1/2$ fermionic fields is mediated by electromagnetic field photon,  which is a gauge field.
\vspace{-0.2cm}
\subsection{Lagrangian}
\label{qed_lag}
\vspace{-0.2cm}
The QED Lagrangian density describing massless fermions, electromagnetic field and
interaction between them as
\bea
{\cal L}_{\mbox{\tiny {QED}}}
&=& -\frac{1}{4} F_{\mu\nu}F^{\mu\nu} -\frac{1}{2\xi} (\partial_\mu A^\mu)^2 + {\bar \psi} 
i \gamma^\mu \partial_\mu \psi  -e\bar \psi \gamma^\mu\psi A_\mu +\delta{\mathcal L}_{\textrm{\tiny QED}},
\label{qed23}
\eea
where $\psi$ and $\bar \psi$ are Dirac fields. $F_{\mu\nu}$ is known as electromagnetic field tensor and is given as
$
F_{\mu\nu}= \partial_\mu A_\nu -\partial_\nu A_\mu, \label{qed6}
$
This is  a gauge invariant quantity and also antisymmetric, $F_{\mu\nu}=-F_{\nu\mu}$, under the exchange of the Lorentz 
indices $\mu\leftrightarrow \nu$. $A_\mu$ is the gauge field (photon) and $\xi$ is the gauge fixing parameter which takes $1$ in Feynman gauge and $0$ in Landau gauge. $\delta{\mathcal L}_{\textrm{\tiny QED}}$ is the renormalised counter terms.
\subsubsection{Feynman rules}
\label{fey_qed} 
\vspace{-0.2cm}
We note the following from Eq.~\eqref{qed23}:
\vspace{-0.2cm}
\begin{enumerate}
\item[$\bullet$] The first term corresponds to the free Lagrangian density of a gauge field (photon).
\item[$\bullet$] The second one corresponds to  the gauge fixing term.  
\item[$\bullet$]  The first and the second terms together lead  to a free photon propagator  as 
 \item[]
 \includegraphics[scale=0.6]{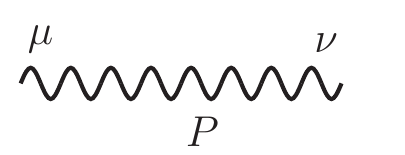}
 \vspace*{-0.65in}
\bea
\hspace*{2.4in} \Rightarrow\,\,\,\,\,\, iD_0^{\mn}(P)= \frac{i}{P^2}\left[ - \eta^{\mu\nu}  + \left(1-\xi\right )\frac{P^\mu P^\nu}{P^2} \right ]. \label{qed22}
\eea
\item[]
\item[$\bullet$] The third term is the Lagrangian density that describes the free propagation of a massless fermion (electron). The free
massless electron propagator reads as 
 \item[]
\includegraphics[scale=.7]{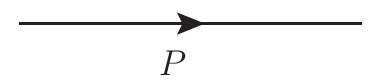}
\vspace*{-0.5in}
\bea
\hspace*{2.4in} \Rightarrow \, \, \, \, \, iS_0(P)= \frac{i}{P\!\!\! \! \slash} \, . \label{qed1}
\eea
\vspace{-0.5cm}
\item[$\bullet$] The fourth term arises from the local $U(1)$ gauge symmetry and represents to the interaction Lagrangian density of fermionic and gauge field, where photon field interact with electron field via the dimensionless coupling parameter $e$. By employing this interaction in perturbative theory, one can compute Feynman diagrams for the theory. The electron-photon vertex in QED is represented as
\item[]
\hspace*{1.2in}\includegraphics[height=2.5cm,width=4cm]{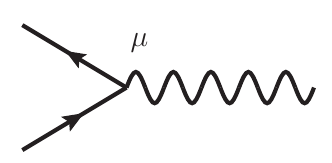}
 \vspace*{-0.8in}
\bea
\hspace*{0.3in} \Rightarrow \,\,\, - ie\gamma^\mu. \label{qed2}
\eea
 \vspace*{-.0cm}
\end{enumerate}
\subsection{One-loop Electron Self-energy and Structure Functions in HTL approximation}
\label{ssf}
\begin{wrapfigure}[7]{r}{0.35\textwidth}
 \vspace*{-0.2cm}
 	\includegraphics[width=6cm,height=2.7cm]{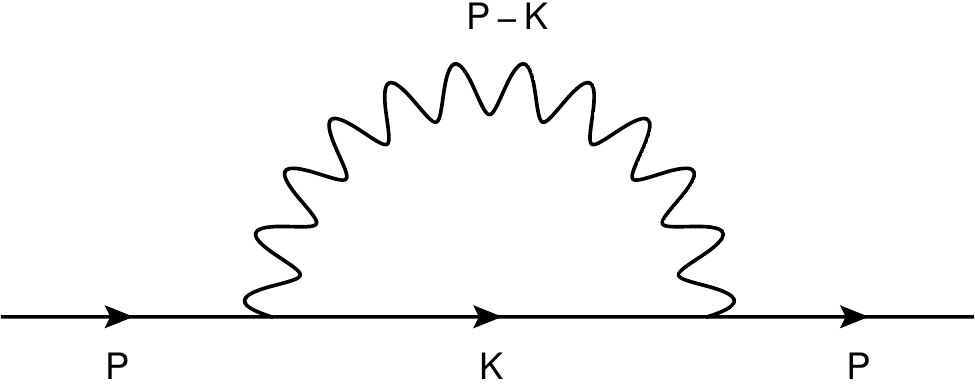}
	\vspace*{-0.2cm}
 	 \caption{One-loop electron self-energy diagram.}
 	\label{ele_se}
\end{wrapfigure}

Using Feynman rules defined in subsec~\ref{fey_qed}  the QED one-loop electron self-energy in Fig.~\ref{ele_se} can be written as 
\bea
\Sigma^{\textrm e}(P) &=& - T \sumintf_{\{K\}} (-ie\gamma_\mu) \frac{i \slashed{K}}{K^2}(-ie\gamma_\nu) \frac{-i\eta^{\mn}}{(P-K)^2 }
= -2 e^2 T \sumintf_{\{K\}}  \frac{\slashed{K}}{K^2 \  Q^2 }\, ,  \label{se0}
\eea
where $Q=(P-K)$ and $\sum\!\!\!\!\!\!\!\!\!\int\limits_{\{K\}}$ is a fermionic sum-integral and $e$ is the dimensionless coupling constant in QED. The superscript $\textrm e$ denotes electron's self-energy. The detail calculations have done in a review in Ref.~\cite{Mustafa:2022got} by one of us. We will just quote the final
expression for relevant quantities.

Using Eq.~\eqref{se0}  in Eq.~\eqref{gse10},  the structure functions~\cite{Mustafa:2022got} in the rest frame of the heat bath within HTL approximation can be written  as
\begin{subequations}
 \begin{align}
 {\cal A}^{\textrm e}(\omega,p) &=-\frac{(m^{\textrm e}_{\textrm{th}})^2}{p^2}   \int \frac{ d\Omega}{4\pi}  
\frac{{\bm{{\vec p}\cdot {\hat k}}}}{\omega-{\bm{\vec p \cdot \hat k}} } \nn
&= -\frac{(m^{\textrm e}_{\textrm{th}})^2}{p^2}   \left\langle \frac{{\bm{{\vec p}\cdot {\hat k}}}}{P\cdot \hat K} \right \rangle_{\bf \hat k } 
= \,  \frac{(m^{\textrm e}_{\textrm{th}})^2}{p^2} \left [ 1- \frac{\omega}{2p} \ln \left (\frac{\omega+p}{\omega-p}\right )\right ] \, , \label{se1} \\
\mathcal{B}^{\textrm e}(\omega,p)
& \, =\frac{(m^{\textrm e}_{\textrm{th}})^2}{p^2} \left \langle \frac{(P\cdot u)(\bm{\vec p\cdot\hat{k}})-p^2}{P\cdot\hat{K}} \right \rangle_{\bf \hat k}\nn
& = \frac{(m^{\textrm e}_{\textrm{th}})^2}{p} \left [ -\frac{\omega}{p} + \left ( \frac{\omega^2}{p^2} -1\right ) \frac{1}{2} \ln \left (\frac{\omega+p}{\omega-p}\right ) \right] \, .
\label{se2} 
\end{align}
\end{subequations}
 where ${\hat K}=(1, \bm{\hat  k})$ is a light like vector, the angular braces represent the average over the directions specified by the light like vector ${\hat K}$ and the thermal mass of the electron is obtained as
\bea
(m^{\textrm e}_{\textrm{th}})^2 =\frac{e^2}{4\pi^2}\int kdk\left[n_F^+(k)+n_F^-(k)+2n_B(k)\right]= \frac{e^2 T^2}{8}\left[1+\frac{\mu^2}{\pi^2T^2}\right] \, .   \label{se3} 
\eea
The electron self-energy in the rest frame of the heat bath can be written~\cite{Mustafa:2022got} within HTL approximation from Eq.~\eqref{se0} as
\bea
\Sigma^{\textrm e}(P) &=& {(m^{\textrm e}_{\textrm{th}})^2} \left \langle \frac{\hat{\slashed{K}}}{P\cdot \hat{K}} \right \rangle_{\bf \hat k} \,  
=\frac{(m^{\textrm e}_{\textrm{th}})^2}{2p}  \ln \left (\frac{\omega+p}{\omega-p}\right )  \gamma_0  +
 \frac{(m^{\textrm e}_{\textrm{th}})^2}{p} \left [ 1- \frac{\omega}{2p} \ln \left (\frac{\omega+p}{\omega-p}\right ) \right ] \left(\vec{\bm \gamma} \cdot \bm{\hat p} \right )\, .  \label{se6c}
\eea
This result can also be obtained using Eq.~\eqref{se1} and~\eqref{se2} in~\eqref{gse2} in the rest frame of the heat bath.
\subsection{Effective Electron Propagator and Collective Modes in HTL Approximation}
\label{disp_quasi}
One can write the effective electron propagator from Eq.~\eqref{gse24} in HTL approximation in presence of a thermal medium as
\be
S^{\textrm e}(P) =  \frac{1}{2} \frac{\gamma_0 - \vec{\bm \gamma}\cdot \hat{\bm p}} {{\cal D}^{\textrm e}_+(\omega,p)}+ \frac{1}{2} \frac{\gamma_0 + \vec{\bm\gamma}\cdot \bm {\hat { p}}} {{\cal D}^{\textrm e}_- (\omega,p)} .
\label{dfg0}
\ee
 We can obtain ${\cal D}^{\textrm e}_\pm(\omega,p)$ by combining Eqs.~\eqref{gse19},~\eqref{se1} and~\eqref{se1} as 
\bea
{\cal D}^{\textrm e}_\pm(\omega,p)=\omega \mp p -  \frac{(m^{\textrm e}_{\textrm{th}})^2}{p}  \left [ \frac{1}{2}
\left ( 1\mp \frac{\omega}{p}\right ) \ln \left (\frac{\omega+p}{\omega-p}\right )\pm 1\right ]\, . 
\label{dfq}
\eea
\begin{wrapfigure}[11]{r}{0.35\textwidth}
	\vspace{-0.1cm}
	\begin{center}
		\vspace{-0.6cm}
		\includegraphics[width=6cm,height=5.7cm]{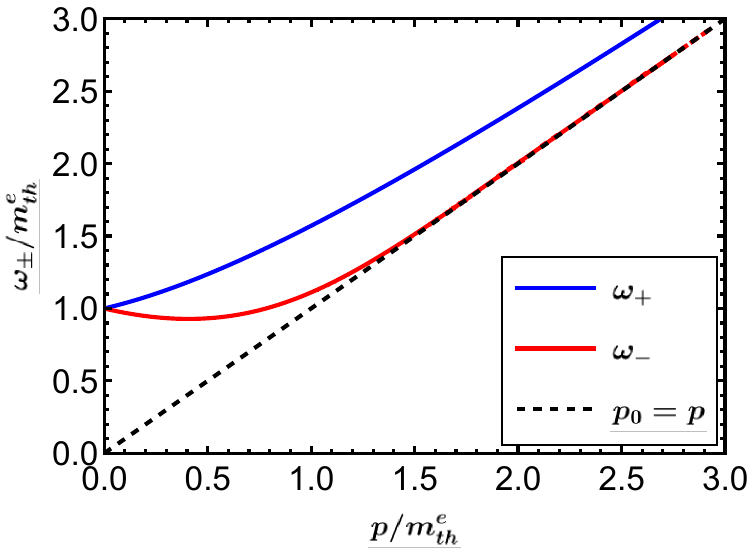}
		\vspace{-0.4cm}
		\caption{Plot of quasiparticles dispersion in HTL approximation.}
		\label{disp_plot}
	\end{center}
\end{wrapfigure}
While the effective propagator in~\eqref{dfg0} exhibits chiral symmetry, the poles of ${\cal D}^{\textrm e}_\pm(\omega,p)$ do not occur at light cone, $\omega=\pm p$. This indicates that the poles of the effective fermion propagator are located away from the light cone in the timelike domain. This discrepancy arises due to the additional term ${\cal B}\gamma_0$ appearing in self-energy in~\eqref{gse2} due to the breaking of Lorentz invariance at finite temperature~\cite{Weldon:1982bn}. 

The zeros of ${\cal D}^{\textrm e}_\pm(\omega,p)$ define dispersion property of a quark in the thermal bath. ${\cal D}^{\textrm e}_+(\omega,p)=0$ exhibits two ploes at $\omega=\omega_+(p)$ and $\omega=-\omega_-(p)$, while ${\cal D}^{\textrm e}_-(\omega,p)=0$ has two ploes at $\omega=\omega_-(p)$ and $\omega=-\omega_+(p)$. Fig.~\ref{disp_plot} only illustrates the positive energy solutions. A mode with energy $\omega_+$  signifies the in-medium propagation of a particle excitation, which is a Dirac spinors and eigenstate of $(\gamma_0 - {\vec \gamma}\cdot \bm{\hat { p}})$ with chirality to helicity ratio $+1$. On the other hand, there exists a novel long-wavelength mode known as {\em plasmino} with energy $\omega_-$  and eigenstate of  $(\gamma_0 + {\vec \gamma}\cdot \bm{\hat { p}})$ with chirality to helicity ratio $-1$. The $\omega_-$ branch has a minimum at low momentum and then approaches free dispersion curve at large momentum.  In addition, ${\cal D}^{\textrm e}_\pm(\omega,p)$ contains a discontinuous part corresponding to {\em Landau Damping} (LD) due to the presence of Logarithmic term in Eq.~\ref{dfq}. 

Below we present the approximate analytic solutions of $\omega_\pm(p)$ for small and large values of momentum $p$. For small values of momentum ($p \ll  m^{\textrm e}_{\textrm{th}}$), the dispersion relations are
 \begin{subequations}
 \begin{align}
\omega_+(p) &\, \approx \, m^{\textrm e}_{\textrm{th}} +\frac{1}{3}p + \frac{1}{3}\frac{p^2}{m^{\textrm e}_{\textrm{th}}}-\frac{16}{135}\frac{p^3}{(m^{\textrm e}_{\textrm{th}})^2} \, ,  \label{dfq1}\\ 
\omega_-(p)
&\, \approx \, m^{\textrm e}_{\textrm{th}} -\frac{1}{3}p + \frac{1}{3}\frac{p^2}{m^{\textrm e}_{\textrm{th}}}+\frac{16}{135}\frac{p^3}{(m^{\textrm e}_{\textrm{th}})^2} \, , \label{dfq2} 
\end{align}
\end{subequations}
whereas for large values of momentum ($ m^{\textrm e}_{\textrm{th}} \ll p \ll T$), one obtains
 \begin{subequations}
 \begin{align}
\omega_+(p) &\, \approx \, p +\frac{(m^{\textrm e}_{\textrm{th}})^2}{p } +  \frac{(m^{\textrm e}_{\textrm{th}})^4}{2p^3}\ln \frac{(m^{\textrm e}_{\textrm{th}})^2}{2p^2}
+ \frac{(m^{\textrm e}_{\textrm{th}})^6}{4p^5}\left[ \ln^2  \frac{(m^{\textrm e}_{\textrm{th}})^2}{2p^2} +\ln  \frac{(m^{\textrm e}_{\textrm{th}})^2}{2p^2} -1\right ] \, ,\label{dfq3}\\ 
\omega_-(p)
&\, \approx \,  p +2p\exp\left (-\frac{2p^2+(m^{\textrm e}_{\textrm{th}})^2}{(m^{\textrm e}_{\textrm{th}})^2}\right )  \, . \label{dfq4} 
\end{align}
\end{subequations}

\subsection{Spectral Representation of Fermion Propagator}
\label{spec_prop}
The in-medium electron propagator \eqref{dfg0} with ${\cal D}^{\rm e}_\pm(\omega,p)$ are given in Eq.~\eqref{dfq}.
According to Eq.~\eqref{gse19a},  ${\cal D}^{\rm e}_\pm(\omega,p)=d_\pm(\omega,p)=\omega\pm p$, for free massless case. The corresponding free spectral function can be obtained following the Eq.~\eqref{bpy_04} as
\bea
\rho^f_\pm(\omega,k) &=& \lim_{\epsilon \rightarrow 0} \frac{1}{\pi}{\textrm{Im}}\left . \frac{1}{d_\pm(\omega,p)}\right |_{\omega\rightarrow \omega+i\epsilon} 
=  \lim_{\epsilon \rightarrow 0} \frac{1}{\pi}{\textrm{Im}} \frac{1}{\omega\mp p+i\epsilon} 
=\frac{\delta(\omega\mp p)}{\left | \frac{\ d(\omega\mp p)}{d\omega}\right |} = \delta(\omega\mp p) \, . \label{spec1}
\eea
As discussed in subsec~\ref{disp_quasi} that ${\cal D}^{\rm e}_\pm(\omega,p)$ has solutions at $\omega_\pm(k)$ and $-\omega_\mp(p)$ and a cut part due to space like momentum,  $\omega^2<p^2$. In-medium spectral function corresponding to the effective electron propagator in Eq.~\eqref{dfg0} will have both pole and cut contribution as
$
\rho_\pm(\omega,p) = \rho^{\textrm{pole}}_\pm (\omega,p) + \rho^{\textrm{cut}}_\pm (\omega,p) \, . \label{spec2}
$
The pole part of the spectral function can be obtained following Eq.~\eqref{bpy_04} as
\bea
 \rho^{\textrm{pole}}_\pm (\omega,p) &=& \lim_{\epsilon \rightarrow 0} \frac{1}{\pi}{\textrm{Im}}\left . \frac{1}{{\cal D}^{\rm e}_\pm(\omega,p)}\right |_{\omega\rightarrow \omega+i\epsilon} 
\!\!\!\!\! = \frac{\delta(\omega -\omega_\pm)}{\left | \frac{d{\cal D^{\rm e}_\pm}}{d\omega}\right |_{\omega=\omega_\pm}} 
 +  \frac{\delta(\omega +\omega_\mp)}{\left | \frac{d{\cal D^{\rm e}_\mp}}{d\omega}\right |_{\omega=-\omega_\mp}} \nn
 &=& \frac{(\omega^2-p^2)}{2m^2_{\textrm{th}} } \left [\delta(\omega -\omega_\pm) +\delta(\omega +\omega_\mp)\right ]\, . \label{spec3}
 \eea
For $\omega^2<p^2$, there is a discontinuity in $\ln\frac{\omega+p}{\omega-p}$ as $\ln{y}=\ln\left |y \right |-i\pi \ ,$ which leads to the spectral  function, $\rho^{\textrm{cut}}_\pm(\omega,p)$, corresponding to the discontinuity in ${\cal D}^{\rm e}_\pm(\omega,p)$ can be obtained from \eqref{bpy_02} as
\begin{eqnarray}
\rho^{\textrm{cut}}_\pm(\omega,p)&= &\frac{1}{\pi} \lim_{\epsilon \rightarrow 0} \textrm{Im}\left.\frac{1}{{\cal D}_\pm(\omega,p)}
\right |_{{\omega\rightarrow\omega+i\epsilon}\atop {\omega < p}}\nn
&=&  \frac{\frac{ m^2_{\textrm{th}}}{2p}\left(1\mp\frac{\omega}{p}\right)
\Theta(p^2-\omega^2)}
{\left[\omega\mp 
p-\frac{m^2_{\textrm{th}}}{p}\left(\pm 1+ \frac{p\mp \omega}{2p}\ln\frac{p+\omega}{p-\omega}
\right)\right]^2+\left[\frac{\pi m^2_{\textrm{th}}}{2p}\left(1\mp\frac{\omega}{ p}\right)
\right]^2}
= \beta_\pm(\omega,p)\Theta(p^2-\omega^2)  . \label{spec4}
\end{eqnarray}
\subsection[Evaluation of $\Pi_L$ and $\Pi_T$ ]{Evaluation of $\Pi_L$ and $\Pi_T$  from One-loop Photon Self-energy in HTL Approximation}
\label{pse1l}
\begin{wrapfigure}[6]{r}{0.32\textwidth}
	\vspace{-0.4cm}
	\includegraphics[scale=0.4]{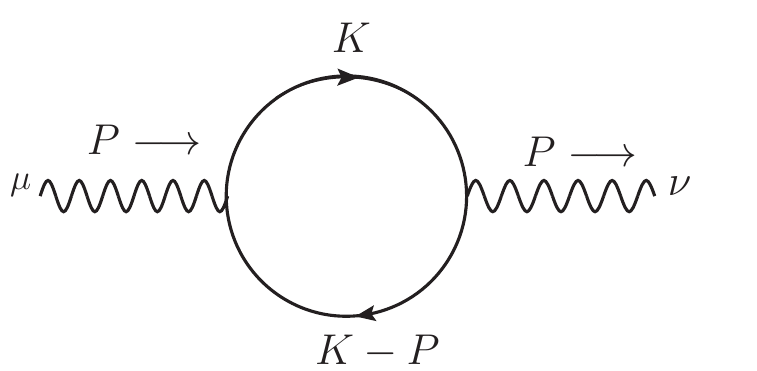}
	\vspace{-.3cm}
	\caption{One-loop photon self-energy diagram.}
	\label{photon_self}
\end{wrapfigure}
Using Feyman rules defined in subsec~\ref{fey_qed} in Fig.~\ref{photon_self} and the performing the Dirac trace the photon self -  energy can be written as
\bea
\Pi^\gamma_{\mn} &=& -  \! \int\! \frac{d^4K}{(2\pi)^4}  {\Tr} \left [\left(-ie\gamma_\mu\right)\frac{i}{ \slashed K}(-ie \gamma_\nu) \frac{i}{\slashed K -\slashed P} \right]\nn
&&\hspace{-1.5cm}=-4e^2\!\int\! \frac{d^4K}{(2\pi)^4} \frac{2 K_\mu K_\nu   - \eta_{\mn}K^2 + \eta_{\mn}K\cdot P- (K_\mu P_\nu + P_\mu K_\nu)}{K^2(P-K)^2} ,\,\label{pel2}
\eea
where superscript $\gamma$ in self-energy denotes photon.
Now in HTL approximation one can neglect external soft momentum ($\sim gT$) in the numerator, i.e., one or higher power of $P$. So, we get~\cite{silin,Fradkin} ,
\bea
\Pi^\gamma_{\mn} 
\approx -8e^2\int \frac{d^4K}{(2\pi)^4} \left[K_\mu K_\nu \Delta_F(K)\Delta_F(Q)\right] + 4 e^2 \eta_{\mn}\int \frac{d^4K}{(2\pi)^4} \Delta_F(Q) \, , \label{psl3}
\eea
where the scalar part of  the fermion propagators are $\Delta_F(K)={1}/{K^2}$  and $\Delta_F(Q)={1}/{(P-K)^2}={1}/{Q^2}$ with $Q=P-K$. We note that the details calculation within HTL approximation have been done in a review article~\cite{Mustafa:2022got} by one of us. We will just quote the final expression for various components of photon self-energy. We first write down traced part of the photon self-energy~\cite{Mustafa:2022got} without vacuum part as
\bea
(\Pi^\gamma)_\mu^\mu&=& - \frac{2e^2}{\pi^2} \int_0^\infty k \ \left[n_F^-(k)+n_F^-\right] \ dk  =- \frac{e^2T^2}{3}\left[1+\frac{3\mu^2}{\pi^2T^2}\right] = -(m^\gamma_D)^2 \, , \label{psl9}
\eea
where the Debye mass in QED is given as $(m^\gamma_D)^2 = \frac{e^2T^2}{3}\left[1+\frac{3\mu^2}{\pi^2T^2}\right]$.
Now we write down the time-time component of photon self-energy~\cite{Mustafa:2022got} as
\be
\Pi^\gamma_{00}(\om,p) =- 4e^2 \int_0^\infty  \frac{k^2\ dk}{2\pi^2} \frac{d n_F(k)}{d k } \int \frac{d\Omega}{4\pi} 
  \frac{\bm{\vec p \cdot \hat k}}{\omega - {\bm{\vec p \cdot \hat k}}}  
= (m^\gamma_D)^2  \left\langle \, \frac{\om-P\cdot {\hat K}}{P\cdot {\hat K}} \right \rangle_{\bf \hat k} \, 
= (m^\gamma_D)^2  \left [ \frac{\om}{2p}\ln \frac{\om+p}{\om-p}-1\right ] \, . \label{psl13}
\ee
Following the same procedure, one can obtain the space-space part of the photon self-energy from~\eqref{psl3} as~\cite{Mustafa:2022got}
\bea
\Pi^\gamma_{ij}(p_0,p)
&=& \frac{e^2T^2}{3} \,  \int \frac{d\Omega}{4\pi}  \, {\hat k}_i {\hat k}_j \left ( \frac{p_0}{\omega - {\bm{\vec p \cdot \hat k}}} \right ) 
= (m^\gamma_D)^2  \left \langle \, {\hat k}_i {\hat k}_j \left ( \frac{p_0}{P\cdot {\hat K}} \right ) \right \rangle_{\bf \hat k} \, ,\label{psl13i}
\eea
and also the time-space part as
\bea
\Pi^\gamma_{0i}(p_0,p) &=& (m^\gamma_D)^2  \int \frac{d\Omega}{4\pi}  \,  \left ( \frac{p_0}{P\cdot {\hat K}} \right )\, {\hat k}_i \nonumber 
= (m^\gamma_D)^2  \left\langle \,\left ( \frac{p_0}{P\cdot {\hat K}} \right )  {\hat k}_i \right \rangle_{\bf \hat k} \, . \label{psl13ii}
\eea
Using Eq.~\eqref{psl13} in Eq.~\eqref{pi_L},  one obtains the longitudinal component of the photon self-energy as
\bea
\Pi^\gamma_L (\om,p)&=& - \frac{P^2}{p^2} \Pi^\gamma_{00}(\om,p) 
=\frac{(m^\gamma_D)^2P^2}{p^2} \left [ 1- \frac{\om}{2p}\ln \frac{\om+p}{\om-p}\right ] \, . \label{psl14}
\eea
Now using \eqref{psl9} and \eqref{psl14} in \eqref{pi_T}, one obtains the transverse component of the photon self-energy  as
\bea
\Pi^\gamma_T(\om,p) &=& \frac{1}{2} \left [(\Pi^\gamma)_\mu^\mu -\Pi^\gamma_L \right ] 
= - \frac{(m^\gamma_D)^2\om^2}{2p^2} \left [1+ \frac{p^2-\om^2}{2\om p} \ln \frac{\om+p}{\om-p} \right ] \, . \label{psl15}
\eea
We now note that in the IR limit ($\om \rightarrow 0$), one gets
\begin{subequations}
 \begin{align}
 \lim_{\om \rightarrow 0} \Pi^\gamma_L(\om,p) & \, = \lim_{\om \rightarrow 0} -\frac{P^2}{p^2}\Pi^\gamma_{00}(\om,p) =-(m^\gamma_D)^2  \, , \label{psl16}  \\
 \lim_{\om \rightarrow 0} \Pi^\gamma_T(\om,p) & \, = 0 \, . \label{psl17}
 \end{align}
\end{subequations}
 The Eq.~\eqref{psl16} is the Debye electric screening mass of photon that acts as a IR regulator at the static electric scale ($\sim eT$).  Conversely, Eq.~\eqref{psl17} indicates that there is no screening for magnetic fields as the one-loop photon transverse self-energy in leading order HTL approximation vanishes in the IR limit and provides no magnetic screening mass for photon.

\subsection[Effective Photon Propagator and Collective Excitations]{Effective Photon Propagator and Collective Excitations  in HTL Approximation}
\label{drtm}
\begin{wrapfigure}[9]{r}{0.4\textwidth}
	\vspace{-0.5cm}
	\includegraphics[height=6cm,width=6.5cm]{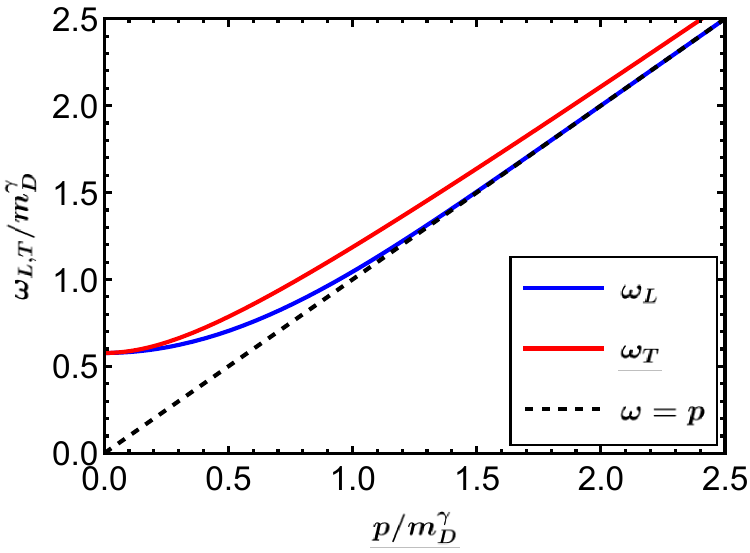}
	\vspace{-0.5cm}
	\caption{Dispersion of photon in HTL approximation in thermal medium.}
	\label{disp_rel_thermal}
\end{wrapfigure}

One write down  the HTL effective propagator for photon from Eq.~\eqref{gsp14} in presence of thermal medium as
\be
D^\gamma_{\mn} = -\left [\frac{\xi}{P^4}P_\mu P_\nu + \frac{1}{P^2+\Pi^\gamma_T}A_{\mn} + \frac{1}{P^2+\Pi^\gamma_L}B_{\mn} \right ]\, . \label{eff_photon}
\ee
Now we can find the dispersion relations using the HTL effective propagator in Eq.~\eqref{eff_photon} in the  thermal medium. The poles of the propagator give the dispersion relations:
 \begin{subequations}
 \begin{align}
P^2+\Pi^\gamma_L&=0\,  \implies
\om^2-p^2+\frac{P^2(m^\gamma_D)^2}{p^2}\left[1-\frac{\om}{2p}\ln{\frac{\om+p}{\om-p}}\right] =0\, ,
\label{Pi_L_disp}\\
P^2+\Pi^\gamma_T&=0\, \implies
\om^2-p^2-\frac{(m^\gamma_D)^2}{2}\frac{\om^2}{p^2}\left[1+\frac{p^2-\om^2}{2\om p}\ln{\frac{\om+p}{\om-p}}\right] =0 \, . 
\label{Pi_T_disp}
\end{align}
\end{subequations}
The first condition in Eq.~\eqref{Pi_L_disp} gives the longitudinal mode of propagation $\om_L$ known as plasmon and the second condition in Eq.~\eqref{Pi_T_disp} gives the transverse mode $\om_T$, which is doubly degenerate. The plasmon mode arises due to the  presence of the thermal medium. The dispersion relations for photon at one loop are plotted in Fig.~\ref{disp_rel_thermal}. Energy of photon splits into two modes due to the breaking of Lorentz symmetry in presence of thermal medium. The longitudenal one is a long wavelength mode which reduces to free dispersion very fast than the transverse one. 

It is possible to find out the approximate solutions of $\om_{T\atop L}$ for small and large values of momentum.  For small momentum, $p\ll m_D^\gamma$,  one finds
\begin{subequations}
 \begin{align}
 \om_T &\approx  \frac{m_D^\gamma}{\sqrt 3} \left [ 1+\frac{9}{5} \frac{p^2}{(m_D^\gamma)^2} - \frac{81}{35} \frac{p^4}{(m_D^\gamma)^4} + \frac{792}{125} \frac{p^6}{(m_D^\gamma)^6}\right ] \, ,\label{psl18} \\
 \om_L &\approx  \frac{m_D^\gamma}{\sqrt 3} \left [ 1+\frac{9}{10} \frac{p^2}{(m_D^\gamma)^2} - \frac{27}{280} \frac{p^4}{(m_D^\gamma)^4} + \frac{9}{2000} \frac{p^6}{(m_D^\gamma)^6}\right ] \, .\label{psl19} 
 \end{align}
\end{subequations}
For momentum, $p\gg m_D^\gamma$,  one finds
\begin{subequations}
 \begin{align}
 \om_T &\approx  p+\frac{(m_D^\gamma)^2} {2p} + \frac{(m_D^\gamma)^4} {32p^3}\left[3-2\ln \frac{8p^2}{ (m_D^\gamma)^2} \right ] \nn
\ &+\  \frac{(m_D^\gamma)^6} {128p^5}
 \left[2\ln^2  \frac{8p^2}{ (m_D^\gamma)^2} -10\ln \frac{8p^2}{ (m_D^\gamma)^2} + 7\right ]  \, ,\label{psl20} \\
 \om_L &\approx  p+2p\exp \left [  - \frac{2(p^2+(m_D^\gamma)^2)}{(m_D^\gamma)^2}\right ]\, .\label{psl21} 
 \end{align}
\end{subequations}
\subsection{Spectral Representation of Gauge Boson Propagator}
\label{srgbp}
 \vspace{-0.2cm}
 The effective propagator of a photon in presence of thermal medium is given in~\eqref{eff_photon} as discussed in subsec.~\ref{drtm} that $P^2+\Pi_{L}=0$ has solutions at $\om=\pm\om_{L}$ and $P^2+\Pi_{T}=0$ has solutions at $\om=\pm\om_{T}$.  Both also have a cut part due to space like momentum $\om^2<p^2$. In-medium spectral function corresponding to the effective photon propagator in \eqref{eff_photon} will have both pole and cut contribution as
\be
 \rho_L(\om,p) = \rho_L^{\textrm{pole}}(\om,p) + \rho_L^{\textrm{cut}}(\om,p)\, , \qquad\text{and}\qquad
  \rho_T(\om,p) = \rho_T^{\textrm{pole}}(\om,p) + \rho_T^{\textrm{cut}}(\om,p)\, . \label{srgbp5}
\ee
The pole part of the longitudinal  spectral function can be obtained using~\eqref{bpy_04} as
\bea
 \rho^{\textrm{pole}}_L (\omega,p) &=& \lim_{\epsilon \rightarrow 0} \frac{1}{\pi}{\textrm{Im}}\left . \frac{1}{P^2+\Pi_L}\right |_{\omega\rightarrow \omega+i\epsilon} \nn
& = &\frac{\om}{p^2+(m_D^\gamma)^2-\om^2}\left[ \delta(\om-\om_L)+\delta(\om+\om_L)\right ] . \label{srgbp6}
 \eea
Similarly, the pole part corresponding to the transverse spectral function can be obtained using~\eqref{bpy_04} as
\bea
 \rho^{\textrm{pole}}_T (\omega,p) &=& \lim_{\epsilon \rightarrow 0} \frac{1}{\pi}{\textrm{Im}}\left . \frac{1}{P^2+\Pi_T}\right |_{\omega\rightarrow \omega+i\epsilon} \nn
& =& \frac{\om(\om^2-p^2)}{(m_D^\gamma)^2\om^2+p^2(\om^2-p^2)}\left[ \delta(\om-\om_T)+\delta(\om+\om_T)\right ] . \label{srgbp7}
 \eea
 For $\omega^2<p^2$, there is a discontinuity in $\ln\frac{\omega+p}{\omega-p}$ as $\ln{y}=\ln\left |y \right |-i\pi \ ,$ which leads to the spectral  function, $\rho^{\textrm{cut}}_{L,T}(\omega,p)$, corresponding  to the discontinuity in $P^2+\Pi_{L,T}$ . The cut contribution to the longitudinal spectral function can be obtained using Eq.~\eqref{bpy_02} as
\begin{eqnarray}
\rho^{\textrm{cut}}_L(\omega,p) 
&&\frac{(1-x^2)x\, m_D^2\, \Theta(1-x^2)/2} {\left [p^2(1-x^2)+(1-x^2)(m_D^\gamma)^2-(1-x^2)x \, \frac{(m_D^\gamma)^2}{2}\ln\left |\frac{x+1}{x-1} \right |\right ]^2
+\left[\frac{1}{2} \pi (1-x^2)x \, (m_D^\gamma)^2 \right]^2}  \nn
&=&\beta_L(x)\Theta(1-x^2)  \, , \label{srgbp8}
\end{eqnarray}
where $x=\om/p$. Similarly, the cut part of the transverse spectral function can be obtained using~\eqref{bpy_02} as
\begin{eqnarray}
\rho^{\textrm{cut}}_T(\omega,p) 
&=& - \frac{(1-x^2)x\, (m_D^\gamma)^2\, \Theta(1-x^2)/4} {\left [p^2(1-x^2)+\frac{1}{2} (m_D^\gamma)^2 \left(x^2+\frac{(1-x^2)x}{2} \,\ln\left |\frac{x+1}{x-1} \right |\right )\right ]^2
+\left[\frac{1}{4}\pi (1-x^2)x \, (m_D^\gamma)^2 \right]^2} \nn
&=&\beta_T(x)\Theta(1-x^2)  \, . \label{srgbp9}
\end{eqnarray}

\subsection{Electron-Photon Vertex in HTL Approximation}
\subsubsection{Three-point function: electron-photon vertex in HTL approximation}
\label{3pt_qed}
\begin{figure}[tbh]
	\begin{center}
	\includegraphics[scale=.5]{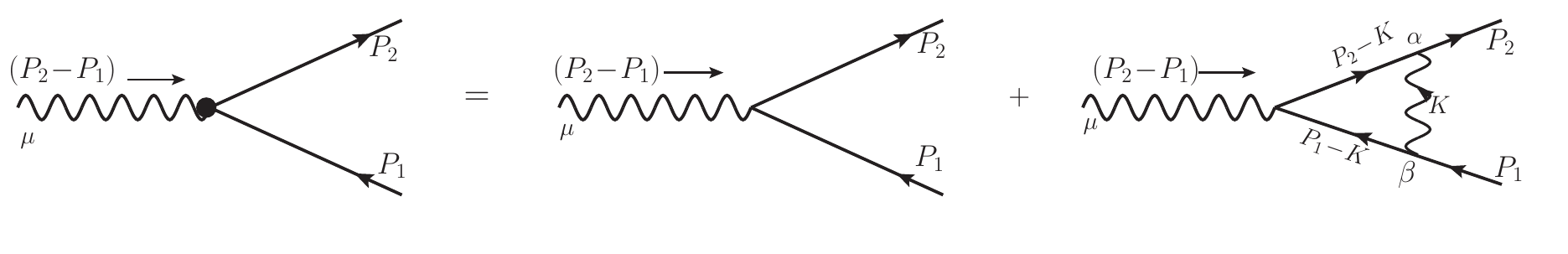}
	\vspace{-0.5cm}
	\caption{One-loop HTL effective three-point QED vertex}
        \label{htl_eegamma_vert}
       \end{center}
\end{figure}
	\vspace{-0.2cm}
The one-loop correction to the three-point electron-photon vertex is shown in Fig.~\ref{htl_eegamma_vert}. Using Feyman rules defined in subsec~\ref{fey_qed} in Feynman gauge, the one-loop expression for the three-point vertex in Minkowski space becomes
\begin{eqnarray}
\delta\Gamma_\mu (P_1,P_2;P_1-P_2)&=& -e^2  \int \frac{d^4K}{(2\pi)^4} \left[ \gamma_\alpha \left(\slashed{K} - \slashed{P}_2\right)
 \gamma_\mu \left(\slashed{K} - \slashed{P}_1\right) \gamma^\alpha \right]  \Delta_B(K){\Delta}_F(P_2-K){\Delta}_F(P_1-K)\, .
\label{vert_3pt}
\end{eqnarray}
In HTL approxomation, the external momenta in the numerator can be neglected and~\eqref{vert_3pt} is simplified as
\begin{eqnarray}
\delta\Gamma_\mu \left (P_1,P_2;P_1-P_2\right ) 
  &=& e^2 \int\frac{d^4K}{(2\pi)^4}  \left[ 4\slashed{K} {K}_\mu-2K^2\gamma_\mu \right ]
 \Delta_B(K){\Delta}_F(P_2-K){\Delta}_F(P_1-K) , \label{vert_3pt1}
\end{eqnarray}
where we have used the  relations in Minkowski space 
$\slashed{K} \gamma_\mu \slashed{K} =2\slashed{K}K_\mu -K^2 \gamma_\mu $ and $\gamma_\alpha \gamma_\mu \gamma^\alpha = -2\gamma_\mu$.
%
Now using Saclay method for frequency sum derived in \eqref{sac6} and using  the following  simplifications within approximation~\cite{Braaten:1989mz,silin,Fradkin} for massless case ($k=E_k$ and $ q = | \bm{\vec p} -  \bm{\vec k} | $)
\begin{subequations}
 \begin{align}
 q \, = \,\left | \bm{\vec p} -  \bm{\vec k} \right | &\, = \, \sqrt{p^2+k^2-2pk\cos\theta} =\sqrt{p^2+k^2-2pk c}\approx k-pc =k-{\bm{\vec p \cdot \hat k}}\, ,
 \label{psl8a}\\   
 n_F(q) & \, = n_F(k-{\bm{\vec p \cdot \hat k}}) \approx n_F(k) - {\bm{\vec p \cdot \hat k}} \frac{d n_F(k)}{d k }\,  ,
\label{psl9a}  \\
\om \pm k\pm q & \, = \om \pm k \pm k \mp \,  {\bm{\vec p \cdot \hat k}} \approx \pm 2k \, ,
\label{psl10}  \\
\omega \pm k\mp q & \, = \omega \pm k \mp k \pm \,  {\bm{\vec p \cdot \hat k}} \approx \omega \pm  {\bm{\vec p \cdot \hat k}} \, ,
\label{psl11} 
\end{align}
\end{subequations}
Using the approximations in Eq.~\eqref{psl8a},~\eqref{psl9a},~\eqref{psl10} and,~\eqref{psl11},the second term in Eq.~\eqref{vert_3pt1} becomes proportional to $\ln T$, which is subleading in $T$ and can be neglected. So, the leading order term in Eq.~\eqref{vert_3pt1} becomes
 \begin{eqnarray}
\delta\Gamma_\mu \left (P_1,P_2;P_1-P_2\right ) &=& 4 e^2 \int\frac{d^4K}{(2\pi)^4}  \slashed{K} {K}_\mu
 \Delta_B(K){\Delta}_F(P_2-K){\Delta}_F(P_1-K) . \label{vert_3pt9}
\end{eqnarray}
The spatial and temporal parts of the three-point function can be written as
\begin{eqnarray}
\delta\Gamma_i \left (P_1,P_2;P_1-P_2\right ) &=& 4 e^2T \sum_{k_0=2n\pi i T} \int\frac{d^3k}{(2\pi)^3} 
 \left [ \gamma_0k_0- \left(\boldsymbol {\vec \gamma}\cdot \bm{\vec k}\right) \right ]k_i
 \Delta_B(K){\Delta}_F(P_2-K){\Delta}_F(P_1-K) . \label{vert_3pt10}\\
\delta\Gamma_0 \left (P_1,P_2;P_1-P_2\right ) &=& 4 e^2T \sum_{k_0=2n\pi i T} \int\frac{d^3k}{(2\pi)^3} 
 \left [ \gamma_0k_0^2- \left(\boldsymbol {\vec \gamma}\cdot \bm{\vec k}\right)k_0 \right ]
 \Delta_B(K){\Delta}_F(P_2-K){\Delta}_F(P_1-K) . \label{vert_3pt11}
\end{eqnarray}
Now we need to perform following three types of frequency sum:
\begin{subequations}
\begin{align}
S_1 = T  \sum_{k_0=2n\pi i T}   \Delta_B(K){\Delta}_F(P_2-K){\Delta}_F(P_1-K) , \label{vert_3pt12}\\
S_2 = T  \sum_{k_0=2n\pi i T}  k_0 \Delta_B(K){\Delta}_F(P_2-K){\Delta}_F(P_1-K) , \label{vert_3pt13}\\
S_3 = T  \sum_{k_0=2n\pi i T}  k^2_0 \Delta_B(K){\Delta}_F(P_2-K){\Delta}_F(P_1-K). \label{vert_3pt14}
\end{align}
\end{subequations}
For massless particles $E=k$, $E_1 =q_1=|\bm{\vec p_1} -\bm{\vec k}|$ and $E_2 =q_2=|\bm{\vec p_2} -\bm{\vec k}|$.
the frequency sum can be performed using the results given in Eq.~\eqref{sac21}. Then performing the sum over $s, s_1, s_2= \pm1$, one will get  eight terms but all of them would not contribute to $T^2$ order. 
It turns out that the $T^2$ behaviour comes from $S_1$ if one performs the sum over $s$, $s_1$ and $s_2$ in \eqref{vert_3pt12} with the condition $s=-s_1=-s_2$ and Eq.~\eqref{vert_3pt12} becomes
\begin{eqnarray}
S_1=  \frac{1}{8kq_1q_2} \left \{  \frac{1}{D_1} \left[ \frac{n_B(k)+n_F^+(q_1)}{p_{10}-k+q_1} - \frac{n_B(k)+n_F^+(q_2)}{p_{20}-k+q_2} \right] 
+ \frac{1}{D_2} \left[ \frac{n_B(k)+n_F^-(q_1)}{p_{10}+k-q_1} - \frac{n_B(k)+n_F^-(q_2)}{p_{20}+k-q_2} \right] \right \} \, ,\label{vert_3pt19}
\end{eqnarray}
where $D_1=(p_{10}-p_{20} +q_1-q_2)$ and $D_2=(p_{10}-p_{20} -q_1+q_2)$. Now, again using HTL approximation in \eqref{psl8a} to \eqref{psl11}, one can simplify \eqref{vert_3pt19} as
\begin{eqnarray}
S_1 &=& - \frac{1}{8k^3} \left [ \frac{n_B(k)+n_F^+(k)}{ (P_1\cdot \hat K ) ( P_2\cdot \hat K )} + \frac{n_B(k)+n_F^-(k)}{ (P_1\cdot \hat{K}' ) ( P_2\cdot \hat{K}' )} \right ] \, , \label{vert_3pt20}
\end{eqnarray} 
where $\hat{K}_\mu\equiv(1,\hat{k})$ and $\hat{K}'_\mu\equiv(1,-\hat{k})$. The second frequency sum in Eq.~\eqref{vert_3pt13} can be performed using the result in Eq.~\eqref{sac23} and then performing the sum over $s$, $s_1$ and $s_2$ with the restriction
 $s=-s_1=-s_2$ and using  HTL approximation, one gets
\begin{eqnarray}
S_2 &=& - \frac{1}{8k^2} \left [ \frac{n_B(k)+n_F^+(k)}{ (P_1\cdot \hat K ) ( P_2\cdot \hat K )} -  \frac{n_B(k)+n_F^-(k)}{ (P_1\cdot \hat{K}' ) ( P_2\cdot \hat{K}' )} \right ] \, .  \label{vert_3pt22}
\end{eqnarray}
The third frequency sum in~\eqref{vert_3pt14} can be performed using the result in Eq.~\eqref{sac25}, then performing the sum over $s$, $s_1$ and $s_2$ with the restriction $s=-s_1=-s_2$ and using  HTL approximation, one gets
\begin{eqnarray}
S_3 &=&   \frac{1}{8k} \left [ \frac{n_B(k)+n_F^+(k)}{ (P_1\cdot \hat K ) ( P_2\cdot \hat K )} + \frac{n_B(k)+n_F^-(k)}{ (P_1\cdot \hat{K}' ) ( P_2\cdot \hat{K}' )} \right ]  = k^2 S_1 \, .  \label{vert_3pt24}
\end{eqnarray}
The spatial part of the three-point function in Eq.~\eqref{vert_3pt10} can be written using Eqs.~\eqref{vert_3pt20} and~\eqref{vert_3pt22} as
 \begin{eqnarray}
\delta\Gamma_i  &=& -  \frac{e^2}{4\pi^2} \int_0^{\infty} \!dk \, k_i [2n_B(k)+n_F^+(k)+n_F^-(k)]
  \int \frac{d\Omega}{4\pi} \frac{\gamma_0 - \boldsymbol{\vec \gamma}\cdot \bm{\hat k}}{ (P_1\cdot \hat K ) ( P_2\cdot \hat K )} \nn
&  =& - \left (m^{\textrm{e}}_{\textrm {th}}\right )^2 \int \frac{d\Omega}{4\pi} \frac{{\hat k_i} \hat{\slashed{K}} }{ (P_1\cdot \hat K ) ( P_2\cdot \hat K )}
. \label{vert_3pt25}
\end{eqnarray}
The temporal part of the three-point function in Eq.~\eqref{vert_3pt11} can be written using Eq.~\eqref{vert_3pt22} and~\eqref{vert_3pt24} as
 \begin{eqnarray}
\delta\Gamma_0 &=& -  \frac{e^2}{4\pi^2} \int_0^{\infty} dk \, k [2n_B(k)+n_F^+(k)+n_F^-(k)]
  \int \frac{d\Omega}{4\pi} \frac{ \gamma_0 - \boldsymbol{\vec \gamma}\cdot \bm{\hat k}}{ (P_1\cdot \hat K ) ( P_2\cdot \hat K )} \nn
&  =& - \left (m^{\textrm{e}}_{\textrm {th}}\right )^2 \int \frac{d\Omega}{4\pi} \frac{\hat{\slashed{K}} }{ (P_1\cdot \hat K ) ( P_2\cdot \hat K )}
. \label{vert_3pt26}
\end{eqnarray}
Using Eqs.~\eqref{vert_3pt25} and~\eqref{vert_3pt26} one can write the three-point function in compact notation as
\begin{eqnarray}
\delta\Gamma_\mu \left (P_1,P_2;P_1-P_2\right )  &=& - \left (m^{\textrm{e}}_{\textrm {th}}\right )^2 \int \frac{d\Omega}{4\pi}
 \frac{{\hat K_\mu} \hat{\slashed{K}} }{ (P_1\cdot \hat K ) ( P_2\cdot \hat K )} 
 =- \left (m^{\textrm{e}}_{\textrm {th}}\right )^2  \left \langle \frac{{\hat K_\mu} \hat{\slashed{K}} }{ (P_1\cdot \hat K ) ( P_2\cdot \hat K )}  \right \rangle_{\bm{\hat k}}  \, , \label{vert_3pt27}
\end{eqnarray}
We now check the Ward-Takahashi identity for QED as
\begin{eqnarray}
\left (P_1 -P_2\right )_\mu \delta\Gamma^\mu \left (P_1,P_2;P_1-P_2\right )  &=& \left (P_1 -P_2\right )_\mu 
 \left[- \left (m^{\textrm{e}}_{\textrm {th}}\right )^2  \left \langle \frac{{\hat K^\mu} \hat{\slashed{K}} }{ (P_1\cdot \hat K ) ( P_2\cdot \hat K )}  \right \rangle_{\bm{\hat k}} \right] 
 = \Sigma^{\textrm e} (P_1)- \Sigma^{\textrm e} (P_2) \,  , \label{vert_3pt28}
\end{eqnarray}
where the electron self-energy $\Sigma^{\textrm e} (P)$ is obtained in \eqref{se6c}. The 3-point function in QED is
\be
\Gamma_\mu(P_1,P_2;P_1-P_2)=\gamma_\mu - \delta\Gamma_\mu = \gamma_\mu
+ \left (m^{\textrm{e}}_{\textrm {th}}\right )^2  \left \langle \frac{{\hat K_\mu} \hat{\slashed{K}} }{ (P_1\cdot \hat K ) ( P_2\cdot \hat K )}  \right \rangle_{\bm{\hat k}}  \, , \label{vert_3pt29}
\ee

\subsubsection{Four-point function: two photon-two electron vertex in HTL approximation}
\label{4pt_qed}
\begin{figure}[tbh]
	\begin{center}
		\includegraphics[scale=.4]{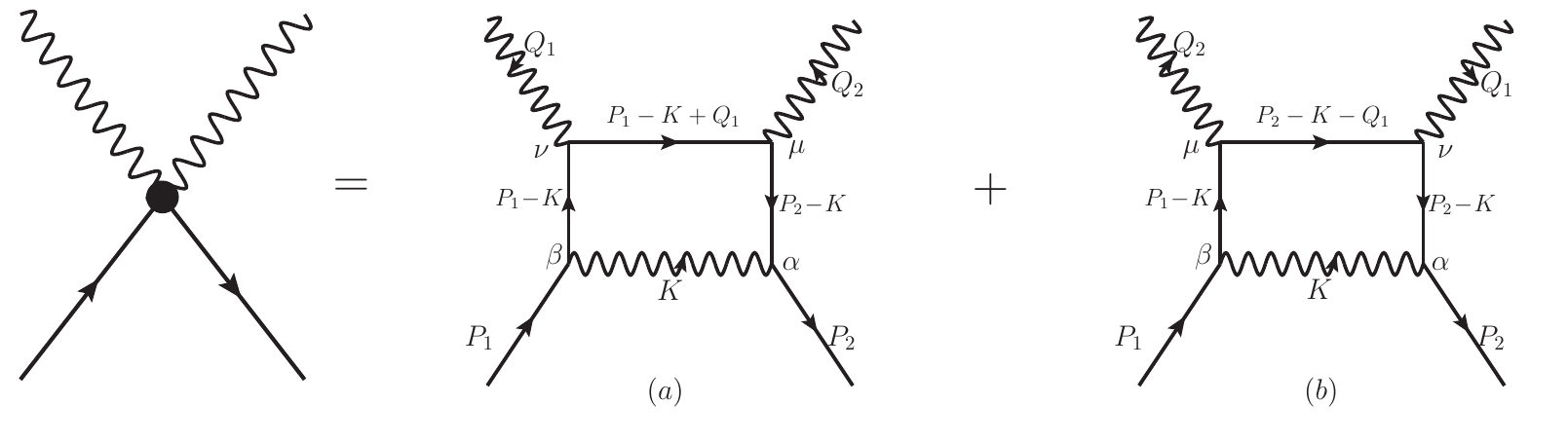}
		\vspace{-0.4cm}
		\caption{One-loop HTL correction to the four-point QED vertex}
			\label{htl_4pt_qed}
	\end{center}
\end{figure}
\vspace{-0.3cm}
Following Feynman rules in subsec~\ref{fey_qed}, the 4-point function in Fig.~\ref{htl_4pt_qed}$(a)$ can be obtained as
\begin{eqnarray}	
	\Gamma_{\mu \nu}^{(a)}\left(P_1, P_2, Q_1\right) &=&- e^2\int \frac{d^4K}{(2\pi)^4}\left[ \gamma_\alpha \left(\slashed{K} - \slashed{P}_2\right) \gamma_\mu  \left(\slashed{K} - \slashed{P}_1 - \slashed{Q}_1\right) \gamma_\nu \left(\slashed{K} - \slashed{P}_1\right) \gamma^\alpha  \right] \nn
	&&\hspace{2.0cm} \ \times\  \Delta_B(K)\Delta_F(P_2-K) \Delta_F(P_1-K+Q_1) \Delta_F(P_1-K). 
	\label{Gamma_munu_a}
\end{eqnarray}	
In HTL approximation, the Eq.~\eqref{Gamma_munu_a} can be simplified as
\begin{eqnarray}	
  \Gamma_{\mu \nu}^{(a)}\left(P_1, P_2, Q_1\right)  
 &=& - e^2T \sum_{k_0} \int\frac{d^3k}{(2\pi)^3} (2\gamma^\mu K^\nu + \gamma^\mu \gamma^\nu \slashed{K} )\Delta_F(P_2-K) 
 \Delta_F(P_1-K+Q_1) \Delta_F(K-P_1) \nn
 &+&  8e^2 T  \sum_{k_0} \int\frac{d^3k}{(2\pi)^3}   K^\mu K^\nu \slashed{K} \Delta_B(K)\Delta_F(K-P_2) \Delta_F(P_1-K+Q_1) \Delta_F(K-P_1) \nn
 &=&   \Gamma_{\mu \nu}^{(1a)}\left(P_1, P_2, Q_1\right) +   \Gamma_{\mu \nu}^{(2a)}\left(P_1, P_2, Q_1\right) \, , \label{gamma1}
\end{eqnarray}	
where $k_0=2n \pi i T$. The above Eq.~\eqref{gamma1} involves 3  and 4 propagators.
Now, using~\eqref{sac27} with
 $E=r=| \bm{\vec k} -\bm{\vec p_1}|=k-\bm{\hat{k}\cdot \vec{p}_1}$,  $E_1 =r_1=|\bm{ \vec p_2 -\vec{k}}|=k-\bm{\hat{k}\cdot \vec{p}_2}$, and
 $E_2 =r_2=| \bm{\vec p_1} -\bm{\vec k}+\bm{\vec{q_1}}|=k-\bm{\hat{k}\cdot(\vec{p}_1+\vec{q}_1)}$ for massless particles, the frequency sum involving 
 3 propagators, $ \Gamma_{\mu \nu}^{(1a)}\left(P_1, P_2, Q_1\right)$ can be performed.  Then performing
the sum over $s, s_1, s_2= \pm1$, one will get eight terms but all of which would not contribute. As we have seen for 3-point function earlier, here also
six terms under HTL approximation contribute to the non-leading order in $T$.
Now, under HTL approximation the momentum integration  of the remaining two terms 
of  $ T \sum_{k_0}\int_{\bm{k}}  k_i \Delta_F(K-P_1)  \Delta_F(P_2-K) \Delta_F(P_1-K +Q_1)$  yields the following result:
\be
\int \frac{1}{8k^3} k_i\ d^3k \frac{dn_F}{dk}\sim \int\ dk \frac{dn_F}{dk} \sim T^0, \label{gamma5}
\ee
which can be safely neglected.
So, the first term,  $\Gamma_{\mu\nu}^{(1a)}(P_1,P_2,Q_1)$,  in \eqref{gamma1} does not contribute in HTL approximation.
Now, we will calculate the second term $\Gamma_{\mu \nu}^{(2a)}\left(P_1, P_2, Q_1\right) $ of equation~\eqref{gamma1}.  
To calculate the frequency sum of $\Gamma_{\mu \nu}^{(2a)}\left(P_1, P_2, Q_1\right) $, we can define the following tensor 
\be
\mathcal{T}^{a}_{\mu\nu\beta} &=& 
T \sum_{k_0} \int\frac{d^3k}{\left(2\pi\right)^3} K_\mu K_\nu K_\beta
\Delta_B(K) {\Delta_F}(P_1-K) {\Delta_F}(P_2-K) {\Delta_F}(P_1-K+Q_1),
\label{tensor_4pt_a}
\ee
whose components are needed to be computed: 
\begin{subequations}
\begin{align}
X_0^a&=\sum_{k_0=2\pi i n T}\Delta_B(K)\Delta_F(P_1-K) \Delta_F(P_2-K) \Delta_F(P_1-K+Q_1)\, ,  \label{gamma6a} \\
X_1^a&=\sum_{k_0=2\pi i n T}k_0\Delta_B(K)\Delta_F(P_1-K) \Delta_F(P_2-K) \Delta_F(P_1-K+Q_1) \, ,  \label{gamma6b} \\
X_2^a&=\sum_{k_0=2\pi i n T}k_0^2\Delta_B(K)\Delta_F(P_1-K) \Delta_F(P_2-K) \Delta_F(P_1-K+Q_1)=k^2X_0^a\, ,  \label{gamma6c} \\
X_3^a&=\sum_{k_0=2\pi i n T}k_0^3\Delta_B(K)\Delta_F(P_1-K) \Delta_F(P_2-K) \Delta_F(P_1-K+Q_1)=k^2X_1^a \, .  \label{gamma6d} 
\end{align}
\end{subequations}
The Matsubara sums needed to evaluate the above equations are given in ~\eqref{sac29} and~\eqref{sac30}.  Using \eqref{sac29} with $E=k; \ E_1=k-\bm{\vec{p}_1\cdot\hat{k}}; \ E_2=k-\bm{\vec{p}_2\cdot\hat{k}};\ E_3=k-\bm{(\vec{p}_1+\vec{q}_1)\cdot\hat{k}}$ for massless particles,  then performing sum over $-s=s_1=s_2=s_3$ and keeping only $T^2$ order terms within HTL approximation, we get
\be
X^a_0 
 = - \frac{1}{16k^4} \left[ \frac{n_B(k)+n_F^+(k)}{P_1\!\cdot\!\hat{K}\  P_2\!\cdot\!\hat{K} \ (P_1+Q_1)\!\cdot\!\hat{K}} -  \frac{n_B(k)+n_F^-(k)}{P_1\!\cdot\!\hat{K}' \ P_2\!\cdot\!\hat{K}' \ (P_1+Q_1)\!\cdot\!\hat{K}'}\right].
 \label{X0_4pt_final}
\ee
The remaining  frequency sums in~\eqref{gamma6b} to~\eqref{gamma6d}  can be obtained as
\begin{subequations}
\begin{align}
X^a_1 &=  - \frac{1}{8k^3} \left[ \frac{n_B(k)+n_F^+(k)}{P_1\!\cdot\!\hat{K}\  P_2\!\cdot\!\hat{K}\ (P_1+Q_1)\!\cdot\!\hat{K}} + \frac{n_B(k)+n_F^-(k)}{P_1\!\cdot\!\hat{K}' \ P_2\!\cdot\!\hat{K}' \ (P_1+Q_1)\!\cdot\!\hat{K}'}\right] \, , \label{X1_4pt_final}\\
X^a_2 &= k^2X^a_0
= - \frac{1}{8k^2} \left[ \frac{n_B(k)+n_F^+(k)}{P_1\!\cdot\!\hat{K}\  P_2\!\cdot\!\hat{K} \ (P_1+Q_1)\!\cdot\!\hat{K}} -  \frac{n_B(k)+n_F^-(k)}{P_1\!\cdot\!\hat{K}' \ P_2\!\cdot\!\hat{K}' \ (P_1+Q_1)\!\cdot\!\hat{K}'}\right]\, , \label{X2_4pt_final}\\
X^a_3 &= k^2X^a_0
=- \frac{1}{8k} \left[ \frac{n_B(k)+n_F^+(k)}{P_1\!\cdot\!\hat{K}\  P_2\!\cdot\!\hat{K} \ (P_1+Q_1)\!\cdot\!\hat{K}} + \frac{n_B(k)+n_F^-(k)}{P_1\!\cdot\!\hat{K}' \ P_2\!\cdot\!\hat{K}' \ (P_1+Q_1)\cdot\hat{K}'}\right]\, .  \label{X3_4pt_final}
\end{align}
\end{subequations}
Using the results of the frequency sums from equations~\eqref{X0_4pt_final}~to~\eqref{X3_4pt_final}, then tensor in \eqref{tensor_4pt_a} becomes
\be
\mathcal{T}^{a}_{\mu\nu\beta} 
&=&-\int \frac{k\ dk}{32\pi^2} \left[2n_B(k) + n_F^+(k) +  n_F^-(k)\right] \int \frac{d\Omega}{4\pi}\  \frac{\hat{K}_\mu \hat{K}_\nu \hat{K}_\beta }{P_1\!\cdot\!\hat{K}\  P_2\!\cdot\!\hat{K}\ (P_1+Q_1)\!\cdot\!\hat{K}} \nn
&=&-\frac{T^2}{64}\left(T^2+\frac{\mu^2}{\pi^2}\right)\int \frac{d\Omega}{4\pi}\  \frac{\hat{K}_\mu \hat{K}_\nu \hat{K}_\beta }{P_1\!\cdot\!\hat{K}  \ P_2\!\cdot\!\hat{K}\ (P_1+Q_1)\!\cdot\!\hat{K}}. \label{gamma7}
\ee
So, the first diagram of the four point vertex (leading order) in Fig.~\ref{htl_4pt_qed}$(a)$ given in~\eqref{gamma1}  becomes
\be
  \Gamma_{\mu \nu}^{(a)}\left(P_1, P_2, Q_1\right)&=& 
  -(m_{\rm{th}}^e)^2\int \frac{d\Omega}{4\pi}\  
  \frac{\hat{K}_\mu \hat{K}_\nu \hat{\slashed{K}} }{P_1\!\cdot\!\hat{K}  \ P_2\!\cdot\!\hat{K} \ (P_1+Q_1)\!\cdot\!\hat{K}}. \label{gamma7a}
\ee
The second diagram of  Fig.~\ref{htl_4pt_qed}$(b)$ can be calculated in the similar manner and it becomes
\be
\Gamma_{\mu \nu}^{(b)}\left(P_1, P_2, Q_1\right)&=&-(m_{\rm{th}}^e)^2\int \frac{d\Omega}{4\pi}\  
\frac{\hat{K}_\mu \hat{K}_\nu \slashed{K} }{P_1\!\cdot\!\hat{K}\  P_2\!\cdot\!\hat{K}\ (P_2 - Q_1)\!\cdot\!\hat{K}}.
\ee
So, the QED four-point vertex within te HTL approximation from the two diagrams of Fig.~\ref{htl_4pt_qed} becomes
\be
\Gamma_{\mu \nu}\left(P_1, P_2, Q_1\right)&=&\Gamma_{\mu \nu}^{(a)}\left(P_1, P_2, Q_1\right)+\Gamma_{\mu \nu}^{(b)}\left(P_1, P_2, Q_1\right)\nn
&=& -(m_{\rm{th}}^e)^2\int \frac{d\Omega}{4\pi}\  
\frac{\hat{K}_\mu \hat{K}_\nu \hat{\slashed{K}}}{\left(P_1+Q_1\right)\!\cdot\!\hat{K}\   \left(P_2-Q_1\right)\!\cdot\!\hat{K} }\left[\frac{1}{P_1\cdot\hat{K}} + \frac{1}{P_2 \cdot\hat{K}}  \right].
\label{qed_4pt_final}
\ee
The QED four-point function in Eq.~\eqref{qed_4pt_final} satisfies the Ward identity, namely,
\be
Q_{1}^ \mu \Gamma_{\mu \nu}\left(P_1, P_2, Q_1\right)=\Gamma_\nu\left(P_1, P_2-Q_1\right)-\Gamma_\nu\left(P_1+Q_1, P_2\right).
\ee
Additionally, $
\Gamma_{\mu \mu} \simeq 0
$
Lastly, we now note that since the photons do not self-interact the three and four photon vertex do not exist. So the non-zero $N$-point HTL functions in QED are photon self-energy ($\Pi^\gamma_{\mn}$),  the electron self-energy ($\Sigma^{\rm e}$) and $N$-photon- two electron vertex.

\subsection{HTL Improved QED Lagrangian}
\label{qed_imp}
Following the results obtained in previous subsections~\ref{ssf}, \ref{pse1l}, \ref{3pt_qed} and \ref{4pt_qed}, the HTL contribution to the QED 
Lagrangian~\cite{Braaten:1991gm,Taylor:1990ia,Frenkel:1991ts} can be written as
\be
{\cal L}_{\textrm{\tiny QED}}^{\textrm {\tiny HTL}}  = {\cal L}_{\textrm{\tiny QED}} + {\cal L}_{\textrm {\tiny HTL}}  \, \label{qed_htl_lag1}
\ee
where $ {\cal L}_{\textrm{QED}}$ is usual QED lagrangian in vacuum as given in \eqref{qed23}. The HTL contribution to effective QED Lagrangian can be written
in compact form as
 \be
 {\cal L}_{\textrm {\tiny HTL}}=i \left(m^{\textrm e}_\textrm{th}\right)^2\bar\psi\gamma^\mu\left\langle\frac{\hat{K}_\mu}{\hat{K}\!\cdot\! D}\right\rangle_{\bm{\hat k}}\psi-\frac{1}{2}
 \left(m^\gamma_D\right)^2 \left ( F_{\mu\alpha}\left\langle\frac{\hat{K}^\alpha \hat{K}_\beta}{(\hat{K}\!\cdot\! D)^2}\right\rangle_{\bm{\hat k}} F^{\mu\beta}\right )\, ,
\label{qed_htl_lag2}
 \ee
where $F$ is the electromagnetic field strength,  $D$ is the covariant derivative. The two parameters $m^\gamma_D$ and $m^{\textrm e}_\textrm{th}$ are, respectively, the Debye screening mass of photon and the thermal electron mass which take into account  the 
screening effects. These  are, respectively, given in \eqref{psl9} and \eqref{se3}. The Lagrangian is non-local and gauge symmetric, which forces the presence of the covariant derivative 
in the denominator of \eqref{qed_htl_lag2} and makes it also non-linear. It is also clear from the first term in \eqref{qed_htl_lag2} there are no HTL with more than two electron lines. 
When expanded in powers of electron and photon fields,\eqref{qed_htl_lag2} generates an infinite series of non-local self-energy and vertex corrections (viz., HTL $N$-point QED 
functions obtained in subsections~\ref{ssf}, \ref{pse1l}, \ref{3pt_qed} and \ref{4pt_qed}) to the bare one. These $N$-point functions are related by Ward-Takahashi identity. If it is 
expanded in terms of electron fields, \eqref{qed_htl_lag2}  will lead to electron self-energy as obtained in \eqref{se6c} whereas when expanded in two fermions and leading order in 
gauge field, one gets back $3$-point electron-photon vertex as obtained in \eqref{vert_3pt27} and two electrons and quadratic order in gauge lead to $4$-point function (two electron-two 
photon vertex) as obtained in~\eqref{qed_4pt_final}. It is evident that the second term in \eqref{qed_htl_lag2} is bilinear in photon field, HTL Green's function with $N$ external photons ($N\ge 3$) and no  external electron lines are zero ~\cite{Bellac:2011kqa}. Very recently, the complete correction to the HTL Lagrangian up to NLO for electron-positron plasma is calculated in Ref.~\cite{Carignano:2019ofj}.

	\section{Quantum Chromodynamics (QCD)}
	\label{qcd_htl}
	\vspace{-0.3cm}
QCD is the theory of strong interaction which is a non-abelian $SU(3)$ gauge theory. It describes the quarks and gluons in the similar  way as QED does for electrons and photons. The major difference between the two theories is that QCD  contains three colour charges in fundamental representation. Consequently, a  quark can be represented by a vector with  three colour states. The interaction among colour charges is facilitated by gauge fields, notably the gluon. The special unitary group $SU(3)$ has $3^2-1=8$ generators, the number of charge mediating particles, corresponding to eight gluons in QCD. Unlike QED in photon,  the gluons in QCD exhibit self-interaction.
\vspace{-0.2cm}
\subsection{Lagrangian}
\label{qcd_lag_chap}
\vspace{-0.2cm}
The QCD Lagrangian is written as
\be
{\cal L}_{\textrm{\tiny QCD}}={\bar \psi}_{if}(i\slashed{D}-m_q) \psi_{if}
-\frac{1}{4} G_{\mn}^aG^{\mn}_a 
+{\cal L}_{\textrm{gf}}+{\cal L}_{\textrm{gh}}  + \delta{\cal L}_{ \textrm{\tiny QCD}}
\, .\label{qcd1}
\ee
We note that $\psi_{if}$ is quark spinor with colour indices $i = (\textrm {r,  g,  b})$  and $f$ indicates the flavour index and $m_q$ is the current quark mass.  $\delta{\cal L}_{ \textrm{\tiny QCD}}$ is the renormalised counter terms. The Lagrangian in Eq.~\eqref{qcd1} is quite similar to QED Lagrangian in Eq.~\eqref{qed23}, with the differences that the electromagnetic field strength tensor $F^{\mn}$, is replaced by the gluonic field strength tensor $G^a_{\mn}$ and an another set of indices corresponding to colour degrees of freedom. The gauge covariant derivative is 
 \be
 D_\mu = \partial_\mu + ig t_a A^a_\mu \, , \label{qcd_0}
 \ee
 where $g$ is the strong coupling constant, $t_a$ are the generators of $SU(3)$ gauge group and $A_\mu^a$ is gluon a field. The $t_a$ is defined as $ t_a = \frac{1}{2} \lambda_a$, where $\lambda_a$ are the  Gell-Mann matrices which were not there in QED.  $\lambda_a$ matrices change the colour of the interacting particles.  These matrices are traceless and obey the commutation relation $[\lambda_a,\, \lambda_b]=i f_{abc}\lambda^c $ with normalisation relation ${\textrm{Tr}}(\lambda_a\lambda_b)=2\delta_{ab}$, $f^{abc}$ is the structure constant of the group which is number. It is also totally antisymmetric and vanishes  if two of the indices become same.  Now, $G^a_{\mn}$ indicates the gluon field tensor which is invariant under gauge transformation and can be defined as
$
G_a^{\mn}=F_a^{\mn} - g f_{abc} A_b^\mu A_c^\nu \, , \label{qcd2}
$
where the first term is similar to the QED field tensor but with colour index and the second term represents the self interaction of gluons. In contrast to QED, the mediator gluons have colour charge which enable them to interact among themselves. Now the gluonic part of the Lagrangian can be written as
\be
{\cal L}_{ \textrm {\tiny G}} =-\frac{1}{4}G_{a\mn}{G^{\mn}_a}=-\frac{1}{4}F_{a\mn}{F^{\mn}_a}+g f_{abc}A_{a\mu}A_{b\nu}\partial^\mu A_c^\nu 
-\frac{1}{4}g^2 f_{abc} f_{alm}A^\mu_b A^\nu_c A_{l\mu} A_{m\nu} \, . \label{qcd3}
\ee
Finally, the gauge fixing and ghost terms in covariant gauge are, respectively, given as
\begin{subequations}
 \begin{align}
{\cal L}_{\textrm{gf}}&=-\frac{1}{2\xi} (\partial_\mu A_a^\mu)^2 \, , \label{qcd4} \\
{\cal L}_{\textrm{gh}}&=-{\bar C}_a\partial^2C_a - g f_{abc}{\bar C}_a \partial_\mu(A^\mu_b C_c) \, , \label{qcd5}
\end{align}
\end{subequations}
where $\xi$ is the gauge fixing parameter and $C$ is the ghost field which is a Grassmann variable. The gauge fixing  
Lagrangian in Eq.~\eqref{qcd4} is required to eliminate the unphysical degrees of freedom present in the theory. The ghost Lagrangian in covariant gauge is given in~\eqref{qcd5} which depends on the choice of the gauge fixing term. 
 
 Combining Eqs.~\eqref{qcd_0},~\eqref{qcd3},~\eqref{qcd4} and~\eqref{qcd5}, the QCD Lagrangian in Eq.~\eqref{qcd1} becomes
 \bea
{\cal L}_{\textrm{\tiny QCD}}&=& -\frac{1}{4}F_{a\mn}{F^{\mn}_a} -\frac{1}{2\xi} (\partial_\mu A_a^\mu)^2  +
{\bar \psi}_{if}(i\slashed{\partial}-m_q) \psi_{if}
-g({\bar \psi}_{if} t_a\psi_{if})\slashed{A}^a
+g f_{abc}A_{a\mu}A_{b\nu}\partial^\mu A_c^\nu  \nonumber \\
&& -\frac{1}{4}g^2 f_{abc} f_{alm}A^\mu_b A^\nu_c A_{l\mu} A_{m\nu}
-{\bar C}_a\partial^2C_a - g f_{abc}{\bar C}_a \partial_\mu(A^\mu_b C_c) + \delta{\cal L}\,
\, .\label{qcd_lag}
\eea
\subsubsection{Feynman rules}
\label{fey_qcd}
We note the following from Eq.~\eqref{qcd_lag}:
\begin{enumerate}
\item[$\bullet$] The first term describes the Lagrangian for eight non-interacting, massless spin $1$ gluon fields. 
The second one corresponds to  the gauge fixing term. These two terms together lead  to a  propagator for free gluon as 
 \item[]
 \vspace{.1cm}
\hbox{\hspace{1cm}\includegraphics[scale=.5]{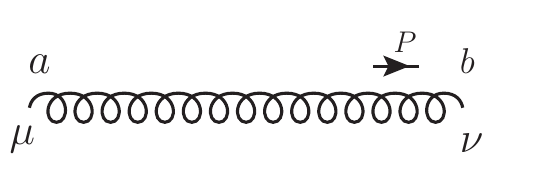}}
 \vspace*{-.8in}
\bea
\hspace*{2in} \Rightarrow\,\,\,\,\,\, i(D_0(P))^{ab}_{\mn} = \frac{i \delta^{ab}}{P^2}\left[  -\eta_{\mu\nu}  + \left(1-\xi\right )\frac{P_\mu P_\nu}{P^2} \right ] 
= i\delta^{ab} (D_0(P))_{\mn}. \label{qcd_6a}
\eea
\vspace{-.3cm}
\item[$\bullet$] The third term is the Lagrangian density that describes the free propagation of a non-interacting quark with current mass $m_q$. The free quark propagator reads as 
 \item[]
 \vspace{.0cm}
\includegraphics[scale=.5]{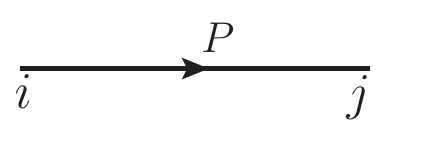}
\vspace*{-0.6in}
\bea
\hspace*{2.4in} \Rightarrow \, \, \, \, \, 
i(S_0(P))^i_j= \frac{i \delta^i_j}{P\!\!\! \! \slash -m_q} =  i\delta^i_j S_0(P) \, . \label{qcd_6b}
\eea
 \vspace{-.3cm}
\item[$\bullet$] The fourth term originates from the local gauge symmetry and corresponds to the Lagrangian density of quark and gluon fields  interaction through the dimensionless running coupling $g$. The quark-gluon $3$ vertex is represented as
 \vspace*{-0.8cm}
\includegraphics[scale=0.4]{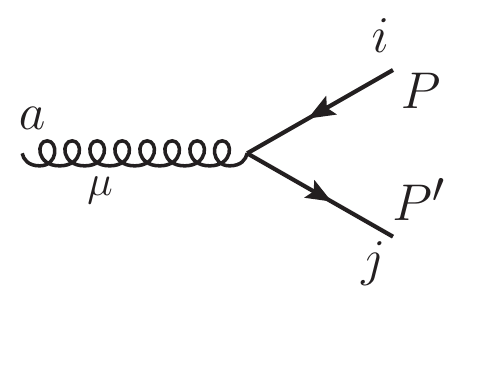}
 \vspace*{-0.6in}
\bea
\hspace*{0.6in} \Rightarrow \,\,\, - ig\gamma^\mu (t^a)^i_j. \label{qcd_6c}
\eea
\item[]
\vspace{-0.1cm}
\item[$\bullet$] The fifth term corresponds to the Lagrangian density of three gluons self-interaction  which has no analogue in QED, as photons do not self-interact. This produces three-gluon vertex in perturbation theory as
\item[]
\includegraphics[scale=0.3]{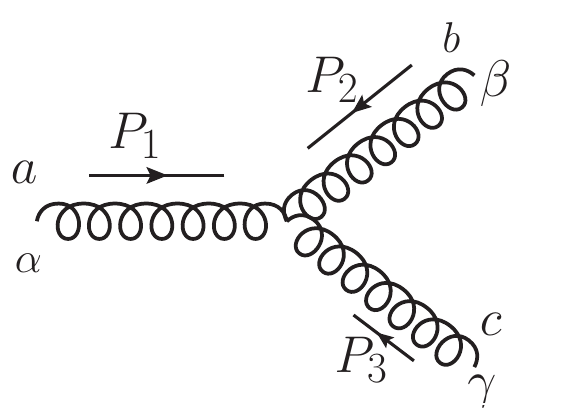}
 \vspace*{-0.68in}
\bea
\hspace*{0.7in} \Rightarrow \,\,\, - g f^{abc} \Big [ \eta^{\alpha\beta} \left(P_1-P_2\right)^\gamma +\eta^{\beta\gamma} \left(P_2-P_3\right)^\alpha 
+  \eta^{\gamma\alpha} \left(P_3-P_1\right)^\beta\Big ]. \label{qcd_6d}
\eea
\item[]
\item[$\bullet$] The sixth term  corresponds to the Lagrangian density of four gluons self-interaction which has also no analogue in QED,  as photons do not self-interact. This produces four-gluon vertex in perturbation theory as 
\item[]
\includegraphics[scale=0.3]{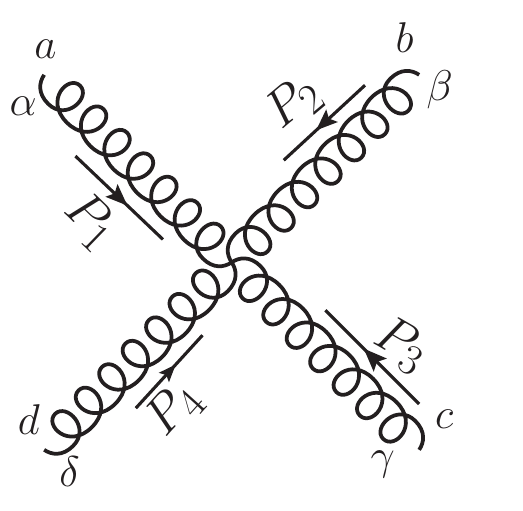}
 \vspace*{-1.0in}
\bea
\hspace*{0.7in} \Rightarrow \,\,\, 
-ig^2 \left [ \begin{array}{l}
                  \,\, \, \, f^{abe}f^{cde} \left(\eta^{\alpha\gamma} \eta^{\beta\delta}-\eta^{\alpha\delta} \eta^{\beta\gamma} \right )   \\
                  + f^{ace}f^{bde} \left(\eta^{\alpha\beta} \eta^{\gamma\delta}-\eta^{\alpha\delta} \eta^{\gamma\beta} \right )  \\   
                   + f^{ade}f^{bce} \left(\eta^{\alpha\beta} \eta^{\delta\gamma}-\eta^{\alpha\gamma} \eta^{\delta\beta} \right )  
		   \end{array} 
		\right ]. \label{qcd_6e}
		\eea
\vspace{.0cm}
\item[$\bullet$] The seventh term  corresponds to the Lagrangian density of ghost field and the propagator is given as
\item[]
\includegraphics[scale=.6]{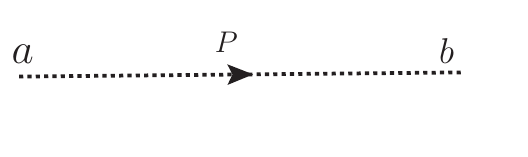}
 \vspace*{-.6in}
\bea
\hspace*{-.5in} \Rightarrow \,\,\,  \frac{i\delta^{ab}}{P^2} \, .  \label{qcd_6f}
		\eea
\item[$\bullet$] The eighth term  corresponds to the interaction Lagrangian density between ghost and gluon and the ghost-gluon interaction vertex is given as
\item[]
\vspace{0.1cm}
\includegraphics[width=4cm, height=2cm]{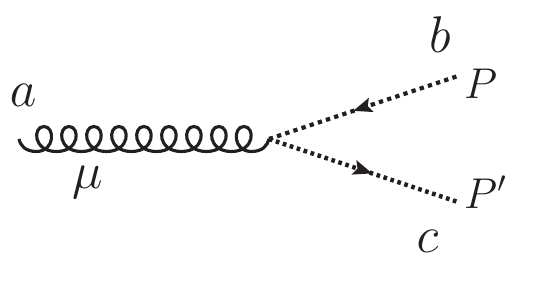}
 \vspace*{-0.8in}
\bea
\hspace*{0.6in} \Rightarrow \,\,\, -  g f^{abc}P^{\prime\mu}. \label{qcd_6g}
\eea
\vspace{0cm}
\end{enumerate}
\subsection{Gluon Two-point Functions in HTL Approximation}
\label{glu_2pt}
\vspace{-0.2cm}
We begin by noting that the previous discussion on QED in section~\ref{qed_htl} would offer a good starting point for QCD, given the similarity in their formalism. While QCD is a non-Abelian $SU(3)$ gauge theory, QED is a $U(1)$ gauge theory. Consequently, extending QED results to QCD primarily involve considering group-theoretical factors, as elaborated below.
\subsubsection{Gluon self-energy }
\label{gsf}
\begin{figure}[tbh]
\begin{center}
\includegraphics[width=17cm,height=3.0cm]{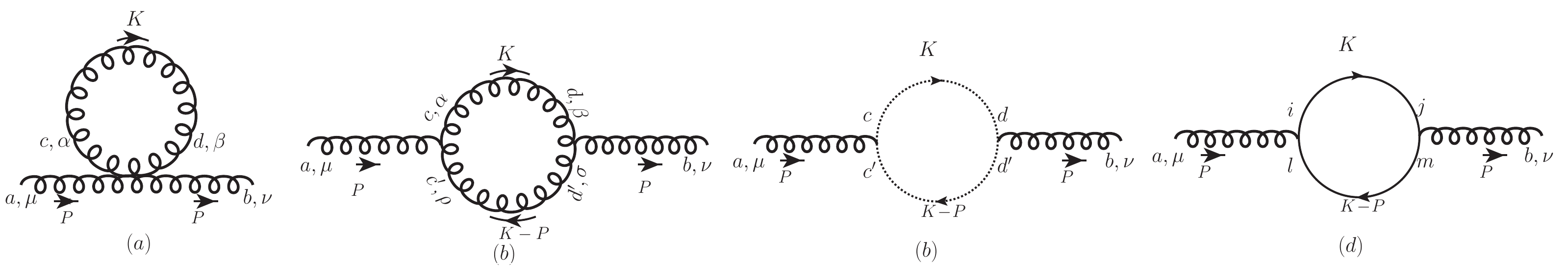}
\end{center}
\vspace{-.8cm}
\caption{One-loop gluon self-energy containing four diagrams: (a) corresponds to gluon loop arising from four gluons interaction, (b) represents gluon loop arising from three gluons interaction, (c) is the ghost loop and (d) is the quark loop.}
\label{gluon_self}
\end{figure}

\vspace{-0.cm}
We calculate the one-loop gluon self-energy~\cite{Mustafa:2022got,Kalashnikov:1979kq,Klimov:1982bv,Weldon:1982aq} and the relevant diagrams are given in Fig.~\ref{gluon_self}. 
\vspace{-.3cm}
The contribution of the first diagram (tadpole) in Fig.~\ref{gluon_self}(a) in Feynman gauge using the Feynman rules defined in subsec~\ref{fey_qcd} can be written as
\bea
\Pi_{\mn}^{({\textrm a})}(P) \delta^{ab} &=& \frac{1}{2} \int \frac{d^4K}{(2\pi)^4}  \frac{-i\eta^{\alpha\beta} \delta^{cd}}{K^2} (-ig^2)
 \Big [f^{cde}f^{bae} \left(\eta_{\alpha\nu} \eta_{\beta\mu} - \eta_{\alpha\mu} \eta_{\beta \nu} \right)  \nn
&&  + f^{cbe}f^{dae} \left(\eta_{\alpha\beta} \eta_{\nu\mu}-\eta_{\alpha\mu} \eta_{\nu\beta} \right) 
 + f^{cae}f^{dbe} \left(\eta_{\alpha\beta} \eta_{\mu\nu}-\eta_{\alpha\nu} \eta_{\mu\beta} \right) \Big ]  \nonumber \\ 
&=& -3g^2 \eta_{\mu\nu}  \int \frac{d^4K}{(2\pi)^4}   \Big [ f^{cae}f^{cbe} \Big ]  \Delta_B(K) 
= -3g^2 C_A  \delta^{ab} \eta_{\mu\nu}  \int \frac{d^4K}{(2\pi)^4} \Delta_B(K) 
 \, , \label{qcd6}
\eea
where $g$ is the strong or QCD coupling constant, $C_A\, (N_c=3)$ is the group factor and $\Delta_B(K)=1/K^2$ is the scalar part of the gluon propagator.

The contribution of the second diagram (gluon loop) in Fig.~\ref{gluon_self}(b)  can be written using the Feynman rules defined subsec~\ref{fey_qcd} as
\bea
\Pi_{\mn}^{({\textrm b})}(P)\delta^{ab} &=& \frac{1}{2} \int \frac{d^4K}{(2\pi)^4}  \frac{-i \eta^{\alpha\beta} \delta^{cd}}{K^2}\ 
 \frac{-i\eta^{\sigma\rho}\delta^{c'd'}}{(K-P)^2} 
 \times \left(-g f^{acc'} \right) \Big [\eta_{\mu\alpha} \left (P+K\right )_\rho  +\eta_{\alpha\rho} \left (-2K+P\right )_\mu  
+\eta_{\rho\mu} \left ( K-2P\right )_\alpha  \Big]  \nonumber \\
&& \times \left(-g f^{bdd'} \right) \Big [\eta_{\nu\beta} \left (-P-K\right )_\sigma  +\eta_{\beta\sigma} \left (2K-P\right )_\nu  +\eta_{\sigma\nu} \left (-K+2P\right )_\beta \Big] \nonumber \\
&=& -\frac{1}{2}g^2  \int \frac{d^4K}{(2\pi)^4} \Delta_B(K) \Delta_B(P-K)   f^{acc'} f^{bcc'} \, {\cal T}_{\mu\nu} = -\frac{C_A}{2}g^2 \delta^{ab} \int \frac{d^4K}{(2\pi)^4} \Delta_B(K) \Delta_B(P-K)  {\cal T}_{\mu\nu}\,  \label{qcd6a}
\eea
where $\Delta_B(P-k)=1/(P-K)^2$, and the tensorial structure ${\cal T}_{\mu\nu}$ is given as
\be
{\cal T}_{\mu\nu}\!\!\!&=&\!\!\!\eta^{\alpha\beta} \eta^{\sigma\rho}  \Big [\eta_{\mu\alpha} \left (P+K\right )_\rho  +\eta_{\alpha\rho} \left (-2K+P\right )_\mu  
+\eta_{\rho\mu} \left ( K-2P\right )_\alpha \Big]  
  \Big [\eta_{\nu\beta} \left (-P-K\right )_\sigma  +\eta_{\beta\sigma} \left (2K-P\right )_\nu  +\eta_{\sigma\nu} \left (-K+2P\right )_\beta  \Big] \nn
  &\approx&-2\Big [5K_\mu K_\nu + \eta_{\mu\nu} K^2 \Big ] \, , \label{qcd6b} 
\ee
where the soft external momentum ($P \sim  gT$) in the numerator, i.e., one or higher power of $P$ are neglected. Now combining ~\eqref{qcd6b} with~\eqref{qcd6a}, 
one obtains 
\be
\Pi_{\mn}^{({\textrm b})}(P)\delta^{ab}= g^2 C_A \delta^{ab} \int \frac{d^4K}{(2\pi)^4} 5K_\mu K_\nu \Delta_B(K) \Delta_B(P-K) 
+g^2 C_A \delta^{ab}\eta_{\mn}\int  \frac{d^4K}{(2\pi)^4} K^2 \Delta_B(K) \Delta_B(P-K) \, . \label{qcd7}
\ee
The contribution of the third diagram (ghost loop) in Fig.~\ref{gluon_self}(c)  can be obtained using the Feynman rules defined subsec~\ref{fey_qcd} as
\bea
\Pi_{\mn}^{({\textrm c})}(P)\delta^{ab}\!\!\!\!\! &=&\!\!\!\!\! - \int  \frac{d^4K}{(2\pi)^4} \Big (-g f^{ac'c} \Big) K_\mu \ \frac{i\delta^{cd}}{K^2}
 \Big (-g f^{bdd'} \Big) (K-P)_\nu \frac{i\delta^{d'c'}}{(K-P)^2} \nn
 &&\hspace{-1.5cm}= -g^2 \! \int\!  \frac{d^4K}{(2\pi)^4} K_\mu (K-P)_\nu f^{acc'}f^{bcc'}  \Delta_B(K) \Delta_B(P-K) 
\approx -g^2 C_A \delta^{ab} \!\int\!  \frac{d^4K}{(2\pi)^4} K_\mu K_\nu \Delta_B(K) \Delta_B(P-K)  .\, \label{qcd8} 
\eea
Combining these three diagrams, one can write
\bea
\Pi_{\mn}^{{\textrm{(a+b+c)}}}(P)
&= & g^2C_A  \int  \frac{d^4K}{(2\pi)^4} \left [4 K_\mu K_\nu -2K^2\eta_{\mn} \right]\Delta_B(K) \Delta_B(P-K) \, . \label{qcd9}
\eea
Using the result of Matsubara sum listed in sec~\ref{saclay} and following the procedure explained in Ref.~\cite{Mustafa:2022got}, we get
\bea
\Pi_{\mu}^{\mu{\textrm{(a+b+c)}}}(P)& =& - \frac{C_Ag^2T^2}{3},\qquad\Pi^{\textrm (a+b+c)}_{00}(\om,p) =\frac{C_Ag^2T^2}{3} \left \langle \, \frac{\om-P\cdot {\hat K}}{P\cdot {\hat K}}\right \rangle_{\bm{\hat k}} , \nn
\Pi^{\textrm (a+b+c)}_{ij}(p_0,p)&=& \frac{C_Ag^2T^2}{3} \left \langle \,  \frac{p_0\ {\hat k}_i {\hat k}_j }{P\cdot {\hat K}} \right \rangle_{\bm{\hat k}} \, ,\qquad \mbox{and} \qquad \Pi^{\textrm (a+b+c)}_{0i}(p_0,p) = \frac{C_Ag^2T^2}{3}  \left\langle  \frac{p_0\  {\hat k}_i }{P\cdot {\hat K}}   \right \rangle_{\bm{\hat k}} 
\eea
Now the contribution of the fourth diagram (quark loop) in Fig.~\ref{gluon_self}(d)  can be obtained using the Feynman rules defined subsec~\ref{fey_qcd} as
\bea
\Pi_{\mn}^{({\textrm d})}(P)\delta^{ab} &=&-  \! \int\! \frac{d^4K}{(2\pi)^4}  {\Tr} \left [\left(-igt^a_{ij}\gamma_\mu\right)\frac{i}{ \slashed K}(-ig \gamma_\nu t^b_{ji}) \frac{i}{\slashed K -\slashed P} \right]\nn
&=& - g^2 N_f\! \int\! \frac{d^4K}{(2\pi)^4} {\Tr}(t^at^b)  {\Tr} \left [\frac{\gamma_\mu \slashed K \gamma_\nu (\slashed K -\slashed P)}
  {K^2 (P-K)^2}\right ]  \nn
& \approx& -g^2 \delta^{ab}\frac{N_f}{2}\! \!\int \!\! \frac{d^4K}{(2\pi)^4} \!\left[8 K_\mu K_\nu -4K^2\eta_{\mn} \right]\Delta_F(K) \Delta_F(P-K)  ,\ \  \label{qcd10}
\eea
where $N_f$ is the number of quark flavour and  $\Delta_F(K)=1/K^2$ and $\Delta_F(P-K)=1/(P-K)^2$ are the scalar part of the quark propagator. 

Note that the fourth diagram (quark loop) in Fig.~\ref{gluon_self}(d) differs from the QED photon self-energy only by a factor $N_f/2$ only. So, we skip the details calculation here and used the photon self-energy calculation done in sec.~\ref{pse1l}. 

Doing so and adding the contribution of diagram $a,b,c$, the Lorentz contacted gluon self-energy from the four diagrams become  
\bea
(\Pi^{\textrm g})_\mu^\mu&=& -\frac{ g^2 T^2}{3} \left[C_A+\frac{N_f}{2}\left(1+{3\mu^2 \over \pi^2T^2}\right)\right]= -(m^{\textrm g}_D)^2 \, , \label{qcd16i}
\eea
and the QCD Debye mass is given as
\be
(m_D^{\textrm g})^2=\frac{ g^2 T^2}{3} \left[C_A+\frac{N_f}{2}\left(1+{3\mu^2\over\pi^2T^2}\right)\right] \, . \label{qcd17}
\ee
One can also obtain temporal component of self energy from~\eqref{psl13}as
\bea
\Pi^{\textrm g}_{00}(\om,p) &=&(m^{\textrm g}_D)^2 \left \langle \, \frac{\om-P\cdot {\hat K}}{P\cdot {\hat K}}\right \rangle_{\bm{\hat k}} 
=  (m^{\textrm g}_D)^2  \left [ \frac{\om}{2p}\ln \frac{\om+p}{\om-p}-1\right ] \, , \label{qcd18}
\eea
and spatial components of self-energy from \eqref{psl13i} and \eqref{psl13ii}, respectively as
\begin{subequations}
\begin{align}
\Pi^{\textrm g}_{ij}(p_0,p)= (m^{\textrm g}_D)^2  \left \langle  \frac{p_0\  {\hat k}_i {\hat k}_j }{P\cdot {\hat K}}  \right \rangle_{\bm{\hat k}} \, ,\qquad \mbox{and} \qquad \Pi^{\textrm g}_{0i}(p_0,p) = (m^{\textrm g}_D)^2  \left\langle \frac{p_0\  {\hat k}_i }{P\cdot {\hat K}}   \right \rangle_{\bm{\hat k}} \, . \label{qcd20}
\end{align}
\end{subequations}
Note that, to go from QED to QCD or photon to gluon, one should only change
\be
e^2\left(1+{3\mu^2\over\pi^2T^2}\right) \Rightarrow g^2\left[C_A+\frac{N_f}{2}\left(1+{3\mu^2\over\pi^2T^2}\right)\right] \, . \label{qcd16}
\ee
Using~\eqref{qcd18} in~\eqref{pi_L}  one obtains the longitudinal component of the gluon self-energy as
\bea
\Pi^{\textrm g}_L (\om,p)&=& - \frac{P^2}{p^2} \Pi^{\textrm g}_{00}(\om,p) 
=\frac{(m^{\textrm g}_D)^2P^2}{p^2} \left [ 1- \frac{\om}{2p}\ln \frac{\om+p}{\om-p}\right ] \, . \label{qcd21}
\eea
Now using \eqref{qcd16i} and \eqref{qcd21} in \eqref{pi_T}, one obtains the transverse component of the photon self-energy  as
\bea
\Pi^{\textrm g}_T(\om,p) &=& \frac{1}{2} \left [(\Pi^{\textrm g})_\mu^\mu -\Pi^{\textrm g}_L \right ] \, 
= - \frac{(m^{\textrm g}_D)^2\om^2}{2p^2} \left [1+ \frac{p^2-\om^2}{2\om p} \ln \frac{\om+p}{\om-p} \right ] \, . \label{qcd22}
\eea
We now note that in the IR limit ($\om \rightarrow 0$), one gets
\begin{subequations}
 \begin{align}
 \lim_{\om \rightarrow 0} \Pi^{\textrm g}_L(\om,p) & \, = \lim_{\om \rightarrow 0} -\frac{P^2}{p^2}\Pi^{\textrm g}_{00}(\om,p) =-(m^{\textrm g}_D)^2  \, , \label{qcd23}  \\
 \lim_{\om \rightarrow 0} \Pi^{\textrm g}_T(\om,p) & \, = 0 \, . \label{qcd24}
 \end{align}
\end{subequations}
 Equation~\eqref{qcd23} represents the Debye electric screening mass of the gluon, serving as an infrared (IR) regulator at the static electric scale($\sim gT$). Conversely, Eq.~\eqref{qcd24} indicates the absence of magnetic screening, as the one-loop gluon transverse self-energy in leading order HTL approximation vanishes in the IR limit, providing no magnetic screening mass for gluon.
 \vspace{-0.4cm}
\begin{figure}[h]
\begin{center}
\includegraphics[scale=0.6]{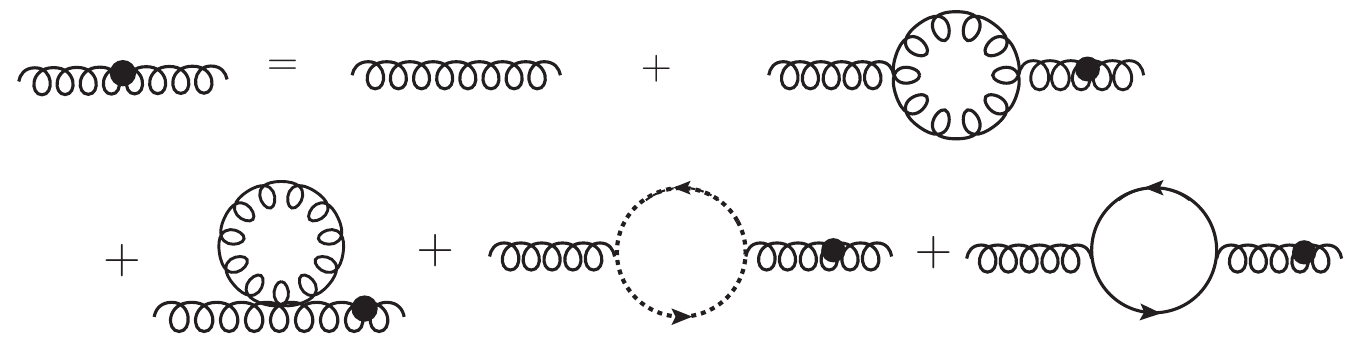}
\end{center}
\vspace{-0.4cm}
\caption{The Dyson-Schwinger equation for the HTL effective gluon propagator.}
\label{gluon_prop}
\end{figure}
\vspace{-0.3cm}
 \subsubsection{Effective gluon propagator and collective excitations}
\label{g_eff_prop}
One obtains the HTL effective gluon propagator using Eq.~\eqref{gsp14} and following Dyson-Schwinger equation in Fig.~\ref{gluon_prop} in presence of a thermal medium~\cite{Das:1997gg,Mustafa:2022got} as
\be
D^{ab}_{\mn}(P) = -\delta^{ab}\left [ \frac{\xi}{P^4}P_\mu P_\nu + \frac{1}{P^2+\Pi^{\textrm g}_T}A_{\mn} +\frac{1}{P^2+\Pi^{\textrm g}_L}B_{\mn} \right ] \, . \label{qcd25}
\ee
Now we can find the gluon dispersion relations using the HTL effective propagator in Eq.~\eqref{qcd25} in the thermal medium. The poles of the propagator give the dispersion relations:
 \begin{subequations}
 \begin{align}
P^2+\Pi^{\textrm g}_L&=0\, \,\, \implies
\om^2-p^2+\frac{P^2(m^{\textrm g}_D)^2}{p^2}\left[1-\frac{\om}{2p}\ln{\frac{\om+p}{\om-p}}\right] =0\, ,
\label{Pi_L_g_disp}\\
P^2+\Pi^{\textrm g}_T&=0\, \, \,\implies
\om^2-p^2-\frac{(m^{\textrm g}_D)^2}{2}\frac{\om^2}{p^2}\left[1+\frac{p^2-\om^2}{2\om p}\ln{\frac{\om+p}{\om-p}}\right] =0 \, .
\label{Pi_T_g_disp}
\end{align}
\end{subequations}
Therefore, we note that the collective excitations in a QCD plasma  are same as QED plasma excitations as displayed in Fig.~\ref{disp_rel_thermal}. One can  find out the approximate solutions of $\om_{T\atop L}$ for small and large values of momentum by replacing $m_D^\gamma$  by $m_D^{\rm g}$ in the respective equation in Eq.~\eqref{psl18},~\eqref{psl19},~\eqref{psl20} and  \eqref{psl21}. One can also obtain the spectral representation of  the effective gluon propagators from Eq.~\eqref{srgbp5} by replacing $m_D^\gamma$ by $m_D^{\rm g}$.  
\subsection{Quark Two-point Functions in HTL Approximation}
\label{quark_two}
\subsubsection{Quark self-energy}
\label{q_se}
\begin{wrapfigure}{R}{0.4\textwidth}
\begin{center}
 \vspace*{-.3in}
\includegraphics[scale=0.23]{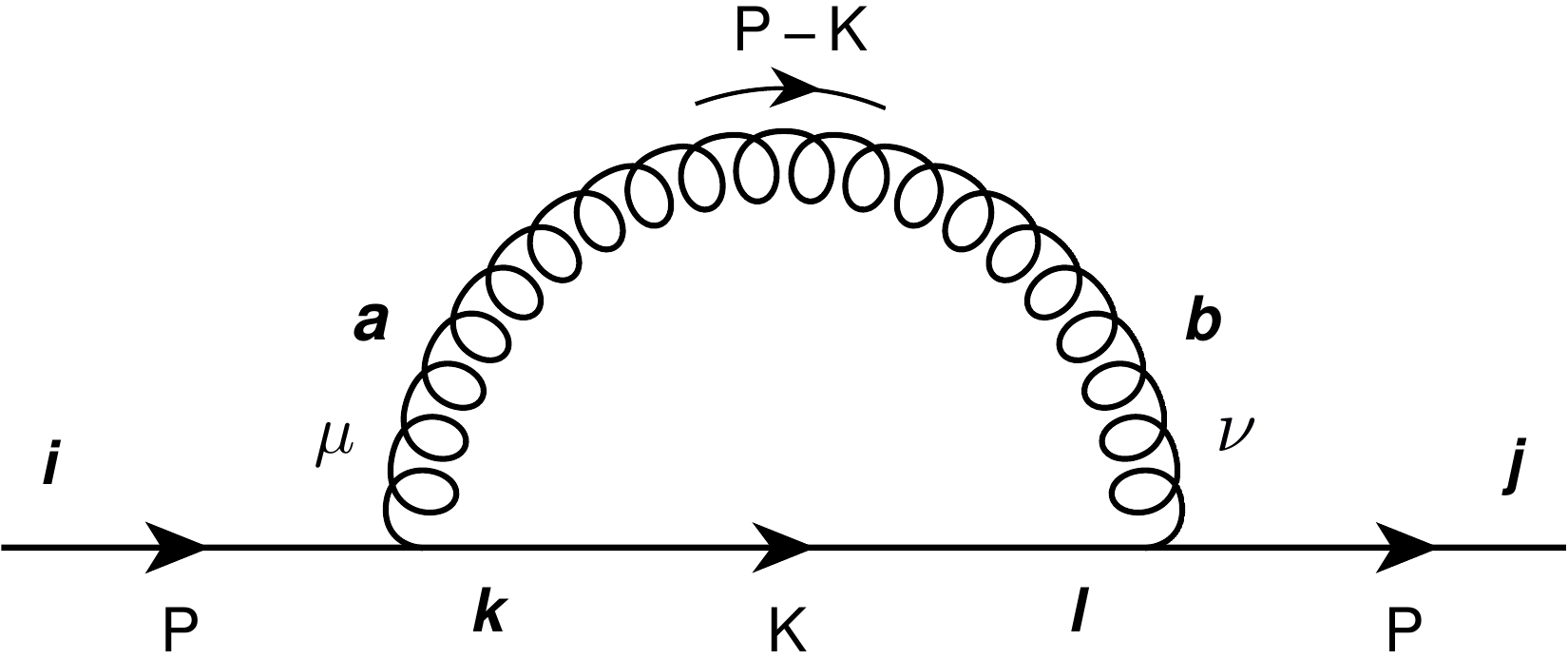}
\end{center}
\vspace{-0.5cm}
\caption{One-loop quark self-energy diagram.}
\label{quark_self}
\end{wrapfigure}
The evaluation of the quark self-energy at one-loop order is even simpler. This is because there is only one diagram (Fig~\ref{quark_self}), which is quite similar to electron self-energy  (Fig~\ref{ele_se}), where the internal photon line is replaced by a gluon line. The one-loop quark self-energy in Feynman gauge with zero current quark mass can be written using Feynman rules defined in in subsec~\ref{fey_qcd} as 
\bea
\Sigma^{\textrm q}(P)\delta^i_j &=&  - T \sumintb_{\{K\}}\,\,\, (-ig \gamma_\mu (t^a)^{i}_k) \frac{i \slashed{K} \delta^k_l}{K^2}(-ig\gamma_\nu (t^b)^{l}_j) 
\frac{(-i\delta^{ab}\eta^{\mn})}{(P-K)^2 } 
= -2g^2 C_F \delta^{i}_j  T \sumintf_{\{K\}}  \slashed{K} \Delta_F(K)\Delta_B(Q),  \label{qcd26}
\eea
where $Q=P-K$. We have also 
used the identity $(t^a t^a)^{i}_j =\frac{Nc^2-1}{2N_c}\delta^{i}_j = C_F\delta^{i}_j$ with $N_c=3$ and $ \gamma_\mu \slashed{K}\gamma^\mu=-2 \slashed{K}$. The superscript $\textrm q$ corresponds to quark. After performing the frequency sum and $k$-integration as done  in subsec.~\ref{ssf},  the quark self-energy becomes~\cite{Mustafa:2022got} 
\bea
\Sigma^{\textrm q}(P) &=& \frac{g^2C_F}{4\pi^2}\int kdk\left[n_F^+(k)+n_F^-(k)+2n_B(k)\right] \int\frac{d\Omega}{4\pi} \frac{\hat{\slashed{K}}}{P\cdot \hat{K}}\nn
& =& \frac{1}{8}C_Fg^2T^2\left(1+\frac{\mu^2}{\pi^2T^2}\right) \left \langle \frac{\hat{\slashed{K}}}{P\cdot \hat{K}}\right \rangle_{\bm{\hat k}}
= (m^{\textrm q}_{\textrm{th}})^2 \left \langle \frac{\hat{\slashed{K}}}{P\cdot \hat{K}}\right \rangle_{\bm{\hat k}}\, ,   \label{qcd27}
\eea
when compared with the electron self-energy in Eq.~\eqref{se6c}
the only difference is overall group factor. So one obtains QCD results for quark from QED results of electron by replacing
$
e^2 \Rightarrow g^2  C_F\, . \label{qcd29}
$
In QED, $e$ is the coupling constant and the group factor, $C_F=1$ for $U(1)$ gauge group whereas $g$ is coupling constant in QCD with group factor $C_F=4/3$ in $SU(3)$ gauge group. With this one can transform electron thermal  mass in \eqref{se3}  to quark thermal  mass as
\be
(m^{\textrm q}_{\textrm{th}})^2 = C_F\frac{ g^2 T^2}{8}\left(1+{\mu^2\over\pi^2T^2}\right)  =  \frac{4}{3}\frac{ g^2 T^2}{8} \left(1+{\mu^2\over\pi^2T^2}\right)   = \frac{ g^2 T^2}{6}\left(1+{\mu^2\over\pi^2T^2}\right) \ \, . \label{qcd30}
\ee
We obtain the quark self-energy~\cite{Mustafa:2022got,Weldon:1982bn,Weldon:1989ys} by replacing the electron thermal mass $m^{\textrm e}_{\textrm{th}}$ 
by the quark thermal mass $m^{\textrm q}_{\textrm{th}}$ in Eq.~\eqref{se6c} as
\bea
\Sigma^{\textrm q}(P)&=&  \frac{(m^{\textrm q}_{\textrm{th}})^2}{2p}  \ln \left (\frac{\omega+p}{\omega-p}\right )  \gamma_0  +
 \frac{(m^{\textrm q}_{\textrm{th}})^2}{p} \left [ 1- \frac{\omega}{2p} \ln \left (\frac{\omega+p}{\omega-p}\right ) \right ] \left(\vec{\bm{\gamma}} \cdot \bm{\hat p} \right)\, . 
\label{qcd31}
\eea
One can obtain~\cite{Mustafa:2022got}  the structure constants ${\mathcal A}^{\textrm q}$ and ${\mathcal B}^{\textrm q}$ for quarks associated with the general expression for fermion self-energy from~\eqref{se1} and~\eqref{se2}, respectively, as
\begin{subequations}
\begin{align}
\mathcal{A}^{\textrm q}(\omega,p) \, & = \,  \frac{(m^{\textrm q}_{\textrm{th}})^2}{p^2} \left [ 1- \frac{\omega}{2p} \ln \left (\frac{\omega+p}{\omega-p}\right )\right ]  \, ,  \label{qcd32} \\
\mathcal{B}^{\textrm q}(\omega,p)
&\, = \,  \frac{(m^{\textrm q}_{\textrm{th}})^2}{p} \left [ -\frac{\omega}{p} + \left ( \frac{\omega^2}{p^2} -1\right ) \frac{1}{2} \ln \left (\frac{\omega+p}{\omega-p}\right ) \right] \, .
\label{qcd33} 
\end{align}
\end{subequations}
\subsubsection{Effective quark propagator and collective excitations}
\label{q_eff_prop}
\begin{figure}[tbh]
\begin{center}
\includegraphics[scale=0.7]{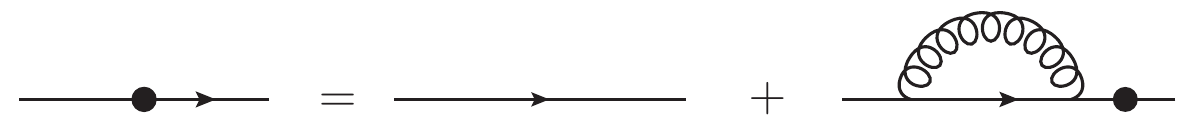}
\end{center}
\vspace{-.4cm}
\caption{The Dyson-Schwinger equation for the HTL effective quark propagator.}
\label{quark_prop}
\end{figure}
One can write the HTL effective quark propagator from~\eqref{gse24} following Dyson-Schwinger equation in Fig.~\ref{quark_prop} in presence of a thermal medium~\cite{Mustafa:2022got}  as
\be
S^{\textrm q}(P) =  \frac{1}{2} \frac{(\gamma_0 - {\vec \gamma}\cdot \bm {\hat { p}})} {{\cal D}^{\textrm q}_+(\omega,p)}+ \frac{1}{2} \frac{(\gamma_0 
+ {\vec \gamma}\cdot \bm {\hat { p}})} {{\cal D}^{\textrm q}_- (\omega,p)} 
 \label{qcd34}
 \ee
We can obtain ${\cal D}^{\textrm q}_\pm(\omega,p)$ by combining \eqref{gse19}, \eqref{qcd32} and \eqref{qcd33} as 
\bea
{\cal D}^{\textrm q}_\pm(\omega,p) &=&(\omega \mp p) -  \frac{(m^{\textrm q}_{\textrm{th}})^2}{p}  \left [ \frac{1}{2}
 \left ( 1\mp \frac{\omega}{p}\right ) \ln \left (\frac{\omega+p}{\omega-p}\right )\pm 1\right ]\, . 
\label{qcd35}
\eea
The zeros of ${\cal D}^{\textrm q}_\pm(\omega,p)$ in \eqref{qcd34} define the dispersion and collective properties of a quark in the thermal bath. These properties are similar to those in QED as discussed in subsec~\ref{disp_quasi} with the dispersion curves in Fig.~\ref{disp_plot}. Hence, we learn about the collective excitations in a QCD plasma from the knowledge gained from QED plasma excitations. One can also obtain the  approximate analytic solutions of $\omega_\pm(p)$ for small and large values of momentum $p$ from 
Eqs.~\eqref{dfq1},~\eqref{dfq2},~\eqref{dfq3} and~\eqref{dfq4}. One can also obtain the spectral representation of 
the effective quark propagator from \eqref{spec2} by replacing $m_{\rm{th}}^{\rm e}$ by $m_{\rm{th}}^{\rm q}$. We want to mention here that the ${\cal D}^{\textrm q}_\pm(\omega,p)$ obtained in Eq.~\eqref{qcd35} considering vanishing quark masses. In Ref.~\cite{Haque:2018eph}, the dispersion relation of a strange quark is studied considering finite strange quark mass. Additionally, NLO quark self-energy and corresponding dispersion laws are studied in Ref.~\cite{Sumit:2022bor} in recent time.
\subsection{Three-point Functions in HTL Approximation}
\label{qcd_three}
\subsubsection{Quark-Gluon three vertex}
\label{qg_three}
\begin{figure}[tbh]          
\begin{center}
\includegraphics[scale=0.55]{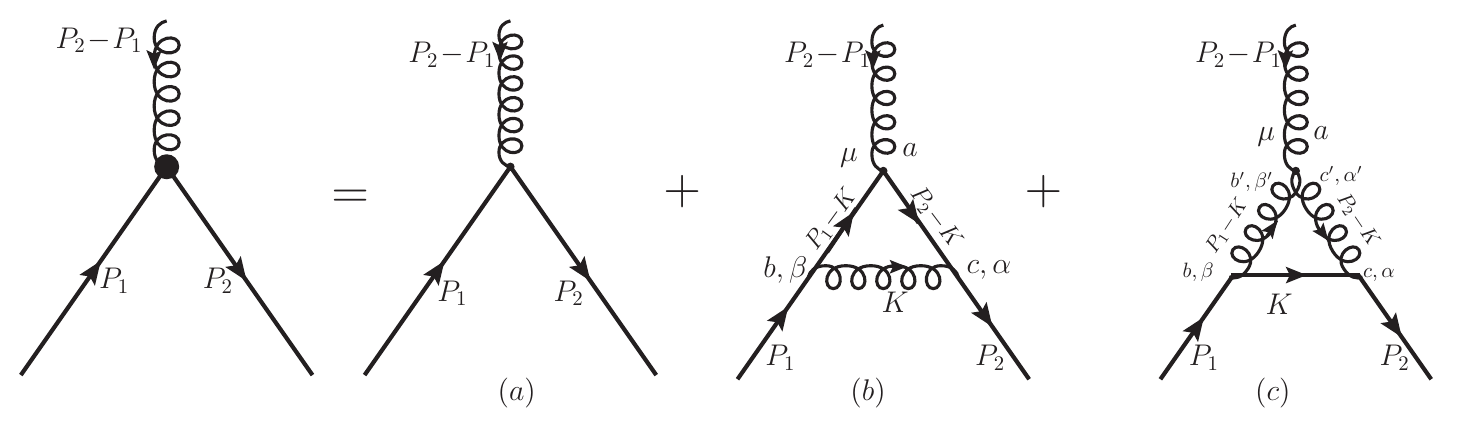}
\end{center}
\vspace*{-0.3in}
\caption{Two quarks-one gluon  HTL effective vertex.}
\label{htl_qg_vert}
\end{figure}
The one-loop approximation for the three-point quark-gluon vertex is shown in fig.~\ref{htl_qg_vert}. The diagram ($a$) of Fig~(\ref{htl_qg_vert}) is the free vertex as given in Eq.~\eqref{qcd_6c}. The diagram ($b$) of Fig~(\ref{htl_qg_vert}) is similar to the three-point QED vertex and can be written from the Feynman diagram as
\be
-ig \delta\Gamma_\mu^{(b)}t^a_{ij}&=&- i \int \frac{d^4K}{(2\pi)^4}\left[\left(-ig\gamma_\alpha\ t^c_{ii'}\right) \frac{i} {\left(\slashed{K} - \slashed{P}_2\right)} \left(-ig\gamma_\mu\ t^a_{i'j'}\right)\frac{i} {\left(\slashed{K} - \slashed{P}_1\right)}\left (-ig\gamma_\beta t^b_{jj'}\right)\frac{\left(-i\eta^{\beta\alpha} \delta^{bc}\right)}{K^2} \right  ] \nn
\delta\Gamma_\mu^{(b)}t^a_{ij}
&\approx& 4\left(C_F-C_A/2 \right)g^2t^a_{ij}\int \frac{d^4K}{(2\pi)^4}  K_\mu\slashed{K}\Delta_B(K){\Delta}_F(P_2-K){\Delta}_F(P_1-K) \, , \label{htl_3pt_qg}
\ee
where we used the relation $\left(t^b_{ii'}t^b_{jj'}t^a_{i'j'}\right)=-\frac{1}{2N_c}t^a_{ij}=(C_F-C_A/2)t^a_{ij}$ and HTL approximation for external soft momenta $P_1$ and $P_2$ in the numerator. 
Using the QED result from Eq.~\eqref{vert_3pt27}, we can write down
\be
\delta\Gamma^{(b)} _\mu\left (P_1,P_2;P_1-P_2\right ) = -\frac{(C_F-C_A/2)g^2}{8}\left(T^2+\frac{\mu^2}{\pi^2}\right) \left\langle \frac{\hat{\slashed{K}}{K}_\mu }{ (P_1\cdot \hat K ) ( P_2\cdot \hat K )} \right \rangle_{\bm{\hat k}}. \label{htl_3pt_qg1}
\ee
The diagram $(c)$ is a pure QCD diagram and can be calculated as 
\be
-ig\delta\Gamma_\mu^{(c)}t^a_{ij}&=&- i \int \frac{d^4K}{(2\pi)^4}\left(-ig\gamma_\alpha\, t^c_{i\ell}\right) \frac{-i\eta^{\alpha\alpha'}\delta^{cc'}} {\left(K - P_2\right)^2} \left[ - g f^{ab'c'} \left\{\eta^{\mu\beta'} \left(P_2-2P_1+K\right)^{\alpha'} +\eta^{\beta'\alpha'} \left(P_1+P_2-2K\right)^\mu \right.\right.\nn
&&\left.\left.+  \eta^{\alpha'\mu} \left(P_1-2P_2+K\right)^{\beta'}\right\}\right] \frac{-i\eta^{\beta\beta'}\delta^{bb'}} {\left(K - P_1\right)^2}\left (-ig\gamma_\beta t^{b}_{\ell j}\right)\frac{i}{\slashed{K}}  \nn
\delta\Gamma_\mu^{(c)}t^a_{ij} 
&\approx&-\frac{C_A}{2}g^2t^a_{ij}  \int\frac{d^4K}{(2\pi)^4} 8K_\mu\slashed{K}\Delta_B(P_1-K)\Delta_B(P_2-K)\Delta_F(K)\nn
\delta\Gamma_\mu^{(c)}&=& - \frac{C_Ag^2}{16}\left(T^2+\frac{\mu^2}{\pi^2}\right)  \int \frac{d\Omega}{4\pi} \frac{\hat{K}_\mu\hat{\slashed{K}} }{ (P_1\cdot \hat K ) ( P_2\cdot \hat K )}.
\ee
where we have used $(t^c_{i\ell} f^{abc} t^b_{\ell j})=(C_A/2) t^a_{i j}$ and neglected the soft external momenta.  So, three-point quark-gluon vertex in the HTL approximation up to one-loop is obtained as
\be
\Gamma_\mu^{qqg}=\gamma_\mu-\delta\Gamma_\mu
=\gamma_\mu + \frac{C_Fg^2}{8}\left(T^2+\frac{\mu^2}{\pi^2}\right) \left\langle  \frac{\hat{K}_\mu\hat{\slashed{K}} }{ (P_1\cdot \hat K ) ( P_2\cdot \hat K )} \right \rangle_{\hat K}
=\gamma_\mu + \left(m_{th}^q\right)^2  \left\langle  \frac{\hat{K}_\mu\hat{\slashed{K}} }{ (P_1\cdot \hat K ) ( P_2\cdot \hat K )} \right \rangle_{\bm{\hat k}}\, ,   \label{qqg_3pt}
\ee
which satisfies the Ward identity as
$
(P_1-P_2)^\mu \Gamma_\mu^{qqg}= {\slashed{P}_1} - {\slashed{P}_2} +\Sigma^{\rm q}(P_2) - \Sigma^{\rm q}(P_1) =S_{\rm q}^{-1}(P_1) - S_{\rm q}^{-1}(P_2)\, . \label{ward_3pt}
$
\subsubsection{Three-Gluon  vertex}
\label{g_three}
\begin{figure}[tbh!]          
\begin{center}
\includegraphics[scale=0.52]{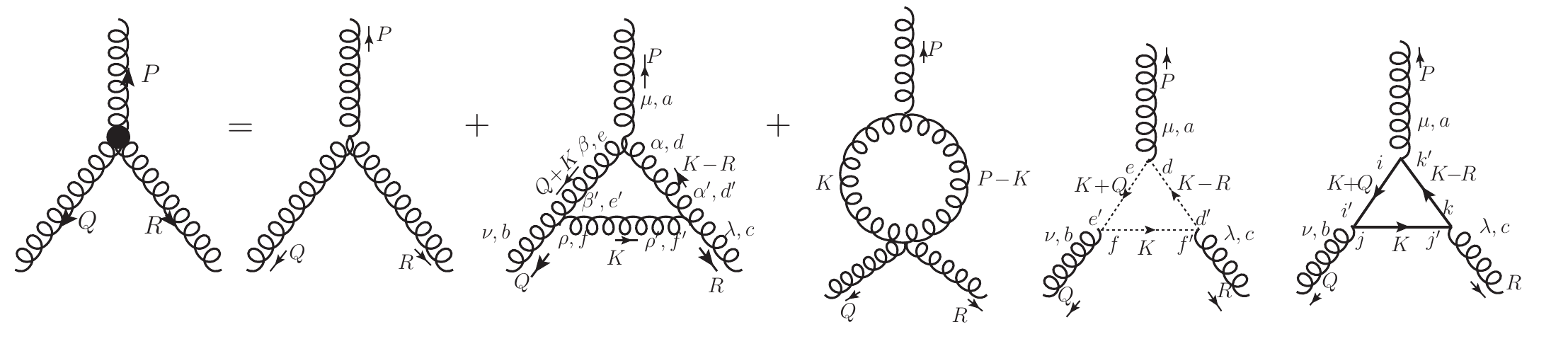}
\end{center}
\vspace*{-0.26in}
\caption{Three-gluon  HTL effective vertex.}
\label{htl_3g_vert}
\end{figure}
The first diagram of the three gluon vertex is the bare vertex and it can be written from Eq.~\eqref{qcd_6d} for three outgoing momentum $P,Q,R$ as
\be
 i\Gamma_{abc,0}^{\mu\nu\lambda}= g f^{abc} \Big [ \eta^{\alpha\beta} \left(P-Q\right)^\gamma +\eta^{\beta\gamma} \left(Q-R\right)^\alpha 
+  \eta^{\gamma\alpha} \left(R-P\right)^\beta\Big ]
\ee
The three-gluon vertex for gluons with outgoing momenta $P$, $Q$, and $R$, Lorentz indices $\mu$, $\nu$, and $\lambda$, and colour indices $a$, $b$, and $c$ is
$
i\Gamma_{abc,0}^{\mu\nu\lambda}(P,Q,R)=gf_{abc,0} \Gamma^{\mu\nu\lambda}(P, Q, R)\;,
$
where $f^{abc}$ are the structure constants. 

Now, consider the form of the HTL three gluon vertext as
\be
\Gamma^{\mu\nu\lambda}(P,Q,R)&=& \Gamma^{\mu\nu\lambda}_0(P,Q,R)- m_D^2{\cal T}^{\mu\nu\lambda}(P, Q, R)\;\nn
&=& \eta^{\mu\nu}(P-Q)^{\lambda}+ \eta^{\nu\lambda}(Q-R)^{\mu}
+ \eta^{\lambda\mu}(R-P)^{\nu} - m_D^2{\cal T}^{\mu\nu\lambda}(P, Q, R)\;.
\label{Gam3}
\ee
%
The contribution from the gluon loop in the HTL correction term ${\cal T}^{\mu\nu\lambda}$ can be expressed in Feynman gauge as
\be
{\cal T}^{\mu\nu\lambda}_{abc,gl}&=&\int \frac{d^4K}{(2\pi)^4}i\Gamma_{ade,0}^{\mu\alpha\beta}(P,R-K,Q+K) \left( \frac{-i\eta^{\alpha\alpha'} \delta^{dd'}}{(K-R)^2}\right) i\Gamma_{d'cf',0}^{\alpha'\lambda\rho'}(K-R,R,-K) \nn
&\times&\left( \frac{-i\eta^{\rho'\rho} \delta^{f'f}}{K^2}\right) i\Gamma_{f'be',0}^{\rho\nu\beta'}(K,Q,-Q-K)  \left( \frac{-i\eta^{\beta'\beta} \delta^{e'e}}{(K+Q)^2}\right)\nn
&=&ig^3\left[f^{ade}f^{dcf}f^{fbe}\right] \int \frac{d^4K}{(2\pi)^4}\Gamma^{\mu\alpha\beta}_0(P,R-K,Q+K)\Gamma^{\alpha\lambda\rho}_0(K-R,R,-K)\Gamma^{\rho\nu\beta}(K,Q,-Q-K)\nn
&&\hspace{2cm}\times\Delta_B(R-K)\Delta_B(K)\Delta_B(K+Q)\nn
&\approx&ig^3\frac{C_A}{2}f^{abc} \int \frac{d^4K}{(2\pi)^4}\Bigg[-2K^2\left(\eta^{\mn}K^\lambda+\eta^{\nu\lambda}K^\mu+\eta^{\lambda\mu}K^\nu\right)\Delta_B(K)\Delta_B(R-K)\Delta_B(K+Q)\nn
&&\hspace{2cm}-18K^\mu K^\nu K^\lambda\Delta_B(K)\Delta_B(R-K)\Delta_B(K+Q)\Big]\nn
&\approx&-9ig^3C_Af^{abc} \int \frac{d^4K}{(2\pi)^4} K^\mu K^\nu K^\lambda\Delta_B(K)\Delta_B(R-K)\Delta_B(K+Q)\nn
&=&-igf^{abc}{\cal T}^{\mu\nu\lambda}_{gl},
\ee
with 
\be
{\cal T}^{\mu\nu\lambda}_{gl}=9g^2C_A \int \frac{d^4K}{(2\pi)^4} K^\mu K^\nu K^\lambda\Delta_B(K)\Delta_B(R-K)\Delta_B(K+Q).
\ee
Now, the contribution form the third diagram (four-gluon vertex diagram) is non-leading in $T^2$ and can be neglected. Next, in ghost-loop contribution to the three-gluon vertex is
\be
{\cal T}^{\mu\nu\lambda}_{abc,gh}&=&2\int \frac{d^4K}{(2\pi)^4} \left(-gf^{ade}(K+Q)^{\mu}\right)\frac{i\delta^{dd'}}{(K-R)^2} \left(-gf^{cf'd'}(K-R)^{\lambda}\right)\frac{i\delta^{ff'}}{K^2} \left(-gf^{be'f}K^{\nu}\right)\frac{i\delta^{ee'}}{(K+Q)^2} \nn
&=&2ig^3\times\frac{1}{2}C_Af^{abc} \int \frac{d^4K}{(2\pi)^4} K^\mu K^\nu K^\lambda\Delta_B(K)\Delta_B(R-K)\Delta_B(K+Q)\nn
&=&-igf^{abc}{\cal T}^{\mu\nu\lambda}_{gh}
\ee
with 
\be
{\cal T}^{\mu\nu\lambda}_{gh}=-g^2C_A \int \frac{d^4K}{(2\pi)^4} K^\mu K^\nu K^\lambda\Delta_B(K)\Delta_B(R-K)\Delta_B(K+Q).
\ee
So, the total contribution from gluon and ghost loop becomes
\be
{\cal T}^{\mu\nu\lambda}_{g}&=&{\cal T}^{\mu\nu\lambda}_{gl}+{\cal T}^{\mu\nu\lambda}_{gh}\nn
&=&8g^2C_A \int \frac{d^4K}{(2\pi)^4} K^\mu K^\nu K^\lambda\Delta_B(K)\Delta_B(R-K)\Delta_B(K+Q).
\ee
Now, $`000$' component of ${\cal T}^{\mu\nu\lambda}_{g}$ becomes
\be
{\cal T}^{000}_g&=&8g^2C_A \int \frac{d^4K}{(2\pi)^4} k_0^3\Delta_B(K)\Delta_B(R-K)\Delta_B(K+Q)\nn
&\approx& 9g^2C_A \int \frac{d^4K}{(2\pi)^4} k_0k^2\Delta_B(K)\Delta_B(R-K)\Delta_B(K+Q)\nn
&=& 8g^2C_A \int \frac{d^3k}{(2\pi)^3}\sum_{s,s_1,s_2}  \frac{k^2s_1s_2}{8E_{|k-r|}E_{|\vec{\bm{k}}+\vec{\bm{q}}|}} 
\,\,  \frac{1}{(r_0+q_0) - s_1 E_{|\vec{\bm{k}}-\vec{\bm{r}}|}  + s_2E_{|\vec{\bm{k}}+\vec{\bm{q}}|}}   \nn
&&\hspace{1cm}\times  \left [ \frac{1+n_B\left ( sk \right ) +n_B\left(s_1E_{|\vec{\bm{k}}-\vec{\bm{r}}|}\right)} {r_{0}-sk-s_1E_{|\vec{\bm{k}}-\vec{\bm{r}}|}} 
-  \frac{1+n_B\left(sk \right ) +n_B\left(s_2E_{|\vec{\bm{k}}+\vec{\bm{q}}|}\right)}{-q_{0}-sk-s_2E_{|\vec{\bm{k}}+\vec{\bm{q}}|}}\right ] \nn
&=& 8g^2C_A \int \frac{d^3k}{(2\pi)^3}  \frac{1}{8} \Bigg\{
 \frac{1}{(r_0+q_0) + (\vec{\bm r}+\vec{\bm q})\cdot\hat{\bm k}}   
\left [ \frac{1+n_B\left ( -k \right ) +n_B(k-\vec{\bm r}\cdot\hat{\bm k})} {r_{0}+\vec{\bm r}\cdot\hat{\bm k}} 
-  \frac{1+n_B\left(-k \right ) +n_B(k+\vec{\bm q}\cdot\hat{\bm k})}{-q_{0}-\vec{\bm q}\cdot\hat{\bm k}}\right ] \nn
&&\hspace{2.5cm}+ \frac{1}{(r_0+q_0) - (\vec{\bm r}+\vec{\bm q})\cdot\hat{\bm k}}   \left[ \frac{1+n_B\left ( k \right ) +n_B(-k+\vec{\bm r}\cdot\hat{\bm k})} {r_{0}-\vec{\bm r}\cdot\hat{\bm k}} 
-  \frac{1+n_B\left(k \right) +n_B(-k-\vec{\bm q}\cdot\hat{\bm k})}{-q_{0}+\vec{\bm q}\cdot\hat{\bm k}}\right ]\Bigg\}\nn
&=& g^2C_A \int \frac{d^3k}{(2\pi)^3}  \frac{dn_B(k)}{dk} \Bigg\{
\frac{1}{(r_0+q_0) + (\vec{\bm r}+\vec{\bm q})\cdot\vec{\bm k}}   
\left [ \frac{r_0} {r_{0}+\vec{\bm r}\cdot\hat{\bm k}} 
-\frac{q_0}{q_{0} + \vec{\bm q}\cdot\hat{\bm k}}\right ] \nn
&&\hspace{2.5cm}+ \frac{1}{(r_0+q_0) - (\vec{\bm r}+\vec{\bm q})\cdot\hat{\bm k}}   
\left [\frac{r_0} {r_{0}-\vec{\bm r}\cdot\hat{\bm k}} 
-\frac{q_0}{q_{0} - \vec{\bm q}\cdot\hat{\bm k}}\right ]\Bigg\}\nn
&=& 2g^2C_A \int \frac{d^3k}{(2\pi)^3}  \frac{dn_B(k)}{dk}
\frac{1}{(r_0+q_0) - (\vec{\bm r}+\vec{\bm q})\cdot\vec{\bm k}}   
\left [ \frac{r_0} {r_{0}-\vec{\bm r}\cdot\hat{\bm k}} -\frac{q_0}{q_{0}- \vec{\bm q}\cdot\hat{\bm k}}\right ] \nn
&=&\frac{C_Ag^2T^2}{3} \int\frac{d\Omega}{4\pi} \frac{1}{P\cdot\hat{K}}\left[\frac{R\cdot u}{R\cdot\hat{K}}-\frac{Q\cdot u}{Q\cdot\hat{K}}\right],
\ee
where $u^\mu=(1,\vec{\bm0})$.  Additionally, the momentum conservation $P+Q+R=0$ is used in the last step.

Following the similar procedure, the other components of the ${\cal T}^{\mu\nu\lambda}_g$ can be calculated and one gets
\be
{\cal T}^{\mu\nu\lambda}_g=\frac{C_Ag^2T^2}{3}  \Bigg\langle \hat{K}^{\mu} \hat{K}^{\nu} \hat{K}^{\lambda} \left( {R\!\cdot\!u\over (R\!\cdot\!\hat{K})\;(P\!\cdot\!\hat{K})}
- {Q\!\cdot\!u\over\! (Q\cdot\!\hat{K})\;(P\!\cdot\!\hat{K})} \right)
\Bigg\rangle_{\hat {\bm k}}\;.
\ee

Now, the quark-loop contribution becomes
\be
{\cal T}^{\mu\nu\lambda}_{abc,q}&=&-2\int \frac{d^4K}{(2\pi)^4}\left(-ig\gamma_\mu t^a_{ik'}\right) \left( \frac{i \delta^{ii'}}{\slashed{Q}+\slashed{K}}\right) \left(-ig\gamma_\nu t^b_{i'j}\right) \left( \frac{i \delta^{jj'}}{\slashed{K}}\right) \left(-ig\gamma_\lambda t^c_{j'k}\right) \left( \frac{i \delta^{kk'}}{\slashed{K}-\slashed{R}}\right) \nn
&\approx&-2g^3N_ftr(t^at^ct^b) \int \frac{d^4K}{(2\pi)^4}tr[\gamma_\mu\slashed{K}\gamma_\nu\slashed{K}\gamma_\lambda\slashed{K}]\Delta_F(Q+K)\Delta_F{K}\Delta_F(K-R)\nn 
&\approx&-2g^3\frac{N_f}{4}\left(d^{acb}+if^{acb}\right) \int \frac{d^4K}{(2\pi)^4}16K^\mu K^\nu K^\lambda \Delta_F(K)\Delta_F(K-R)\Delta_F(K+Q)
\ee
Denoting ${\cal T}^{\mu\nu\lambda}_{abc,q}=-ig{\cal T}^{\mu\nu\lambda}_{q}f^{abc}$, one gets
\be
{\cal T}^{\mu\nu\lambda}_{q}&=& -8N_fg^2\int\frac{d^4K}{(2\pi)^4} K^\mu K^\nu K^\lambda \Delta_F(Q+K)\Delta_F{K}\Delta_F(K-R)
\ee
Now, 
\be
{\cal T}^{000}_{q} &=&-8N_fg^2\int\frac{d^4K}{(2\pi)^4} k_0^3 \Delta_F(Q+K)\Delta_F{K}\Delta_F(K-R) \nn
&=&-8N_fg^2\int\frac{d^4K}{(2\pi)^4} k_0k^2 \Delta_F(Q+K)\Delta_F{K}\Delta_F(K-R) \nn
&=& -8N_fg^2 \int \frac{d^3k}{(2\pi)^3}\sum_{s,s_1,s_2}  \frac{k^2s_1s_2}{8E_{|\vec{\bm{k}}-\vec{\bm{r}}|}E_{|\vec{\bm{k}}+\vec{\bm{q}}|}} 
\,\,  \frac{1}{(r_0+q_0) - s_1 E_{|\vec{\bm{k}}-\vec{\bm{r}}|}  + s_2E_{|\vec{\bm{k}}+\vec{\bm{q}}|}}   
\left [ \frac{1-n_F^+\left ( sk \right ) -n_F^-\left(s_1E_{|\vec{\bm{k}}-\vec{\bm{r}}|}\right)} {r_{0}-sk-s_1E_{|\vec{\bm{k}}-\vec{\bm{r}}|}} \right.\nn
&&\left.\hspace{3cm}-  \frac{1-n_F^+\left(sk \right ) -n_F^-\left(s_2E_{|\vec{\bm{k}}+\vec{\bm{q}}|}\right)}{-q_{0}-sk-s_2E_{|\vec{\bm{k}}+\vec{\bm{q}}|}}\right ] \nn
&=& -N_fg^2 \int \frac{d^3k}{(2\pi)^3}   \Bigg\{
\frac{1}{(r_0+q_0) + (\vec{\bm r}+\vec{\bm q})\cdot\hat{\bm k}}   
\left [ \frac{1-n_F^+\left ( -k \right ) -n_F^-(k-\vec{\bm r}\cdot\hat{\bm k})} {r_{0}+\vec{\bm r}\cdot\hat{\bm k}} -  \frac{1-n_F^+\left(-k \right ) -n_F^-(k+\vec{\bm q}\cdot\hat{\bm k})}{-q_{0}-\vec{\bm q}\cdot\hat{\bm k}}\right ] \nn
&&\hspace{2.5cm}+ \frac{1}{(r_0+q_0) - (\vec{\bm r}+\vec{\bm q})\cdot\hat{\bm k}}   
\left [ \frac{1-n_F^+\left ( k \right ) -n_F^-(-k+\vec{\bm r}\cdot\hat{\bm k})} {r_{0}-\vec{\bm r}\cdot\hat{\bm k}} 
-  \frac{1-n_F^+\left(k \right) -n_F^-(-k-\vec{\bm q}\cdot\hat{\bm k})}{-q_{0}+\vec{\bm q}\cdot\hat{\bm k}}\right ]\Bigg\}\nn
&=& N_fg^2 \int \frac{d^3k}{(2\pi)^3}   \Bigg\{\frac{dn_F^-(k)}{dk}
\frac{1}{(r_0+q_0) + (\vec{\bm r}+\vec{\bm q})\cdot\vec{\bm k}}   
\left [ \frac{r_0} {r_{0}+\vec{\bm r}\cdot\hat{\bm k}} 
-\frac{q_0}{q_{0} + \vec{\bm q}\cdot\hat{\bm k}}\right ] \nn
&&\hspace{2.5cm}+\frac{dn_F^+(k)}{dk} \frac{1}{(r_0+q_0) - (\vec{\bm r}+\vec{\bm q})\cdot\hat{\bm k}}   
\left [\frac{r_0} {r_{0}-\vec{\bm r}\cdot\hat{\bm k}} 
-\frac{q_0}{q_{0} - \vec{\bm q}\cdot\hat{\bm k}}\right ]\Bigg\}\nn
&=&g^2N_f \int \frac{d^3k}{(2\pi)^3}  \frac{d}{dk}[n_F^+(k)+n_F^-(k)]
\frac{1}{(r_0+q_0) - (\vec{\bm r}+\vec{\bm q})\cdot\vec{\bm k}}   
\left [ \frac{r_0} {r_{0}-\vec{\bm r}\cdot\hat{\bm k}} -\frac{q_0}{q_{0}- \vec{\bm q}\cdot\hat{\bm k}}\right ]\nn
&=&g^2N_f \left(-\frac{T^2}{6} - \frac{\mu^2}{2 \pi^2}\right) \int\frac{d\Omega}{4\pi} \frac{1}{-(P\cdot\hat{K})}\left[\frac{R\cdot u}{R\cdot\hat{K}}-\frac{Q\cdot u}{Q\cdot\hat{K}}\right]\nn
&=&\frac{g^2N_f}{6}\left(T^2+\frac{3\mu^2}{\pi^2}\right)  \Bigg\langle
\left( {R\!\cdot\!u\over (R\!\cdot\!\hat{K})\;(P\!\cdot\!\hat{K})}
- {Q\!\cdot\!u\over\! (Q\cdot\!\hat{K})\;(P\!\cdot\!\hat{K})} \right)
\Bigg\rangle_{\hat {\bm k}}\;.
\ee
Similarly, one can calculate the other components of ${\cal T}^{\mu\nu\lambda}_{q}$ and collecting all the components, one gets
\be
{\cal T}^{\mu\nu\lambda}_{q}= \frac{g^2N_f}{6}\left(T^2+\frac{3\mu^2}{\pi^2}\right)  \Bigg\langle \hat{K}^{\mu} \hat{K}^{\nu} \hat{K}^{\lambda}
\left( {R\!\cdot\!u\over (R\!\cdot\!\hat{K})\;(P\!\cdot\!\hat{K})}
- {Q\!\cdot\!u\over\! (Q\cdot\!\hat{K})\;(P\!\cdot\!\hat{K})} \right)
\Bigg\rangle_{\hat {\bm k}}\;.
\ee
Combining gluon and quark contributions, the tensor ${\cal T}^{\mu\nu\lambda}$ in the HTL correction term is defined only for $P+Q+R=0$ as
\be
-m_D^2{\cal T}^{\mu\nu\lambda}(P,Q,R)&=&{\cal T}^{\mu\nu\lambda}_{g}+{\cal T}^{\mu\nu\lambda}_{q}\nn
{\cal T}^{\mu\nu\lambda}(P,Q,R) &=& - \Bigg\langle \hat{K}^{\mu} \hat{K}^{\nu} \hat{K}^{\lambda}
\left( {R\!\cdot\!u\over (R\!\cdot\!\hat{K})\;(P\!\cdot\!\hat{K})}
- {Q\!\cdot\!u\over\! (Q\cdot\!\hat{K})\;(P\!\cdot\!\hat{K})} \right)
\Bigg\rangle_{\hat {\bm k}}\;.
\label{T3-def}
\ee
%
The tensor ${\cal T}^{\mu\nu\lambda}(P,Q,R)$ is completely symmetric in its three indices and traceless in any pair of indices: $g_{\mu\nu}{\cal T}^{\mu\nu\lambda}=0$. It is odd (even) under odd (even) permutations of the momenta $P$, $Q$, and
$R$. It satisfies the ``Ward identity''
\be
Q_{\mu}{\cal T}^{\mu\nu\lambda}(P,Q,R) \;=\;
{\cal T}^{\nu\lambda}(P+Q, R)-
{\cal T}^{\nu\lambda}(P, R+Q)\;.
\label{ward-t3}
\ee
%
The three-gluon vertex tensor therefore satisfies the Ward identity
$
P_{\mu}\Gamma^{\mu\nu\lambda}(P,Q, R) \;=\;
\Delta_{\infty}^{-1}(Q)^{\nu\lambda}-\Delta_{\infty}^{-1}(R)^{\nu\lambda}\;.
\label{ward-3}
$
%

\subsection{Four-point  Functions in HTL Approximation}
\label{four_pt}
\subsubsection{Four-Gluon  vertex}               
\label{g_four}
\begin{figure}[tbh]
\begin{center}
\includegraphics[scale=0.45]{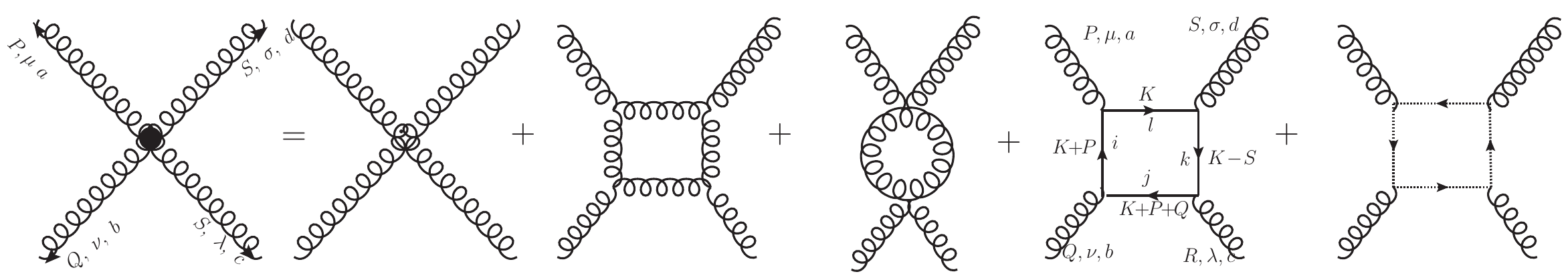}
\end{center}
\vspace*{-0.3in}
\caption{Four-gluon  HTL effective vertex.}
\label{htl_4g_vert}
\end{figure}

The quark-loop diagram can be calculated as
\be
i\mathcal{T}^{\mu\nu\lambda\sigma}_{abcd,q}&=&-i\int\frac{d^4K}{(2\pi)^4}\left(-ig\gamma^\mu t^a_{il}\right)(i(\slashed{K}+\slashed{P})(-ig\gamma^\nu t^b_{ij})(i(\slashed{K}+\slashed{P}+\slashed{Q}))(-ig\gamma^\lambda t^c_{jk})(i(\slashed{K}-\slashed{S}))(-ig\gamma^\sigma t^d_{kl})(i\slashed{K})\nn
&&\times\Delta_F(K+P)\Delta_F(K+P+Q)\Delta_F(K-S)\Delta_F(K)
\nn
&=&-ig^4N_f\mbox{Tr}[t^at^dt^ct^b] \int\frac{d^4K}{(2\pi)^4}\text{Tr}[\gamma^\mu\slashed{K}\gamma^\nu\slashed{K}\gamma^\lambda\slashed{K}\gamma^\sigma\slashed{K}] \Delta_F(K)\Delta_F(K-S)\Delta_F(K+P+Q)\Delta_F(P+K)
\nn
&=&-ig^4N_f \mbox{Tr}[t^at^dt^ct^b] \int\frac{d^4K}{(2\pi)^4}32K^\mu K^\nu K^\lambda K^\sigma\Delta_F(K)\Delta_F(K-S)\Delta_F(K-S-R)\Delta_F(P+K)\nn
&=& -32ig^4N_f  \mbox{Tr}[t^at^dt^ct^b] \mathcal{T}^{\mu\nu\lambda\sigma}_{q},
\ee
with 
\be
\mathcal{T}^{\mu\nu\lambda\sigma}_{q}=\int\frac{d^4K}{(2\pi)^4}K^\mu K^\nu K^\lambda K^\sigma\Delta_F(K)\Delta_F(K-S)\Delta_F(K-S-R)\Delta_F(P+K).
\ee
Now, the `$0000$' component of the tensor $\mathcal{T}^{\mu\nu\lambda\sigma}_{q}$ becomes
\be
\mathcal{T}^{0000}_{q}&=&\int\frac{d^4K}{(2\pi)^4} k^4\Delta_F(K)\Delta_F(K-S)\Delta_F(K-S-R)\Delta_F(P+K)\nn
&\approx& \int\frac{d^4K}{(2\pi)^4} k^4\Delta_F(K)\Delta_F(K-S)\Delta_F(K-S-R)\Delta_F(P+K).
\ee
Using the result of Matsubara sum from Eq.~\eqref{sum_ffff},  $\mathcal{T}^{0000}_{q}$ becomes
\be
&=&\sum_{s,s_1,s_2,s_3} \int\frac{d^3k}{(2\pi)^3} k^4 \frac{s s_1 s_2 s_3 }{ 16 E E_1 E_2 E_3}\frac{1}{ p_0 + s_0-E_1 s_1-E_3 s_3}\nn
&&\times	\left[\frac{1}{r_0+E_1 s_1- s_2E_2 }\left( \frac{1-n_F^-(sE)-n_F^+(s_2 E_2 )}{s_0+r_0-sE -s_2E_2 }-\frac{1-n_F^-(sE)- n_F^+(s_1E_1)}{s_0-sE-s_1 E_1} \right)\right.\nn
&&\left.+\frac{1}{p_0+s_0+r_0-s_2 E_2 -s_3 E_3 }\left( \frac{n_F^-(sE)-n_F^-(s_3E_3)}{p_0+sE-s_3 E_3}-\frac{1-n_F^-(sE )-n_F^+(s_2E_2)}{s_0+r_0-sE -s_2E_2 } \right)\right]\nn
&\approx& \int\frac{d^3k}{(2\pi)^3} k^4 \frac{(-)(+)(+)(-) }{ 16k^4}\frac{1}{ p_0 + s_0-E_1 +E_3 }\nn
&&\times	\left[\frac{1}{r_0+E_1 - E_2 }\left( \frac{1-n_F^-(-E)-n_F( E_2 )}{s_0+r_0+E_0 -E_2 }-\frac{1-n_F^-(-E)- n_F^+(E_1)}{s_0+E_0- E_1} \right)\right.\nn
&&\left.+\frac{1}{p_0+s_0+r_0- E_2 + E_3 }\left( \frac{n_F^-(-E)-n_F^-(-E_3)}{p_0-E+ E_3}-\frac{1-n_F^-(-E )-n_F^+(E_2)}{s_0+r_0+E -E_2 } \right)\right]\nn
 &&+\int\frac{d^3k}{(2\pi)^3} k^4 \frac{(+)(-)(-)(+) }{ 16k^4}\frac{1}{ p_0 + s_0+E_1 -E_3 }\nn
&&\times	\left[\frac{1}{r_0-E_1 + E_2 }\left( \frac{1-n_F^-(E)-n_F^+(- E_2 )}{s_0+r_0-E +E_2 }-\frac{1-n_F^-(E)- n_F^+(-E_1)}{s_0-E+ E_1} \right)\right.\nn
&&\left.+\frac{1}{p_0+s_0+r_0+ E_2 - E_3 }\left( \frac{n_F^-(E)-n_F^-(E_3)}{p_0+E- E_3}-\frac{1-n_F^-(E )-n_F^+(-E_2)}{s_0+r_0-E +E_2 } \right)\right].
\ee
Now, substituting
\be
E&=&k;\ E_1=|\bm{\vec{k}}-\bm{\vec{s}}|\approx k-\hat{\bm k}\cdot\bm{\vec{s}}\nn
E_2&=&|\bm{\vec{k}}-\bm{\vec{s}}-\bm{\vec{r}}|\approx k-\hat{\bm k}\cdot(\bm{\vec{s}+\vec{r}});\ \ \ 
E_3=|\bm{\vec{p}}+\bm{\vec{k}}| \approx k+\hat{\bm k}\cdot\bm{\vec{p}},
\ee
one gets
\be
\mathcal{T}^{0000}_{q}&=&\frac{1}{16} \int\frac{d^3k}{(2\pi)^3}\frac{1}{p_0+s_0+\hat{\bm k}\cdot (\bm{\vec{ p}+\vec{s}})}\left[\frac{1}{r_0+\hat{\bm k}\cdot\bm{\vec{r}}}\left(\frac{n_F^+(k)-n_F^+(k-\hat{\bm k}\cdot(\bm {\vec{s}+\vec{r}}))}{s_0+r_0+\bm{\hat{k}\cdot(\vec{s}+\vec{r})}}-\frac{n_F^+(k)-n_F^+(k-\bm{\hat{k}\cdot \vec{s}})}{s_0+\bm{\hat{k}\cdot\vec{s}}}\right)\right.\nn
&&+\left.\frac{1}{p_0+s_0+r_0+\bm{\hat{k}\cdot(\vec{p}+\vec{s}+\vec{r})}}\left( \frac{-n_F^+(k)+n_F^+(k+\bm{\hat{k}\cdot\vec{p}})}{p_0+\bm{\hat{k}\cdot\vec{ p}}}-\frac{n_F^+(k)-n_F^+(k-\bm{\hat{k}\cdot(\vec{s}+\vec{r})})}{s_0+r_0+\bm{\hat{k}\cdot(\vec{s}+\vec{r})}}\right) \right]\nn
&+&\frac{1}{16} \int\frac{d^3k}{(2\pi)^3}\frac{1}{p_0+s_0-\hat{\bm k}\cdot (\bm{\vec{ p}+\vec{s}})}\left[\frac{1}{r_0-\hat{\bm k}\cdot\bm{\vec{r}}}\left(\frac{-n_F^-(k)+n_F^-(k-\bm{{\hat k}\cdot( {\vec{q}+\vec{r}})})}{s_0+r_0-\bm{\hat{k}\cdot(\vec{s}+\vec{r})}}-\frac{n_F^-(k-\bm{\hat{k}\cdot \vec{s}})-n_F^-(k)}{s_0-\bm{\hat{k}\cdot\vec{s}}}\right)\right.\nn
&&+\left.\frac{1}{p_0+s_0+r_0-\bm{\hat{k}\cdot(\vec{p}+\vec{s}+\vec{r})}}\left( \frac{n_F^-(k)-n_F^-(k+\bm{\hat{k}\cdot\vec{p}})}{p_0-\bm{\hat{k}\cdot \vec{p}}}-\frac{-n_F^-(k)+n_F^-(k-\bm{\hat{k}\cdot(\vec{s}+\vec{r})})}{s_0+r_0-\bm{\hat{k}\cdot(\vec{s}+\vec{r})}}\right) \right]\nn
&=&\frac{1}{16}  \int\frac{d^3k}{(2\pi)^3}\frac{d}{dk}[n_F^+(k)+n_F^-(k)]\frac{1}{(P+S)\cdot\hat{K}}\nn
&&\times\left[\frac{1}{R\cdot\hat{K}}\left(\frac{-\bm{{\hat k}\cdot( {\vec{s}+\vec{r}})}}{(S+R)\cdot\hat{K}}-\frac{-\bm{\hat{k}\cdot \vec{s}}}{S\cdot\hat{K}}\right) + \frac{1}{(P+R+S)\cdot\hat{K}}\left( \frac{-\bm{\hat{k}\cdot \vec{p}}}{P\cdot\hat{K}}-\frac{-\bm{\hat{k}\cdot(\vec{s}+\vec{r})}}{(S+R)\cdot\hat{K}}\right) \right]\nn
&=&-\frac{1}{16}\left(\frac{T^2}{6} + \frac{\mu^2}{2\pi^2}\right)\left\langle \frac{1}{(P+S)\cdot\hat{K}} \left[\frac{1}{R\cdot\hat{K}}\left(-\frac{s_0+r_0}{(S+R)\cdot\hat{K}}+\frac{s_0}{S\cdot\hat{K}}\right) + \frac{1}{(P+R+S)\cdot\hat{K}}\left( -\frac{ p_0}{P\cdot\hat{K}}+\frac{s_0+r_0}{(S+R)\cdot\hat{K}}\right) \right]\right\rangle_{\hat{\bm k}}\nn
&=&-\frac{1}{96} \left(T^2+\frac{\mu^2}{2\pi^2}\right) \left\langle \left[-\frac{p_0}{P\!\cdot\!\hat{K}\; Q\!\cdot\!\hat{K}\; (Q+R)\!\cdot\!\hat{K}} - \frac{s_0}{S\!\cdot\!\hat{K}\; R\!\cdot\!\hat{K}\; (Q+R)\!\cdot\!\hat{K}\; }+\frac{s_0+r_0}{Q\!\cdot\!\hat{K}\;  R\!\cdot\!\hat{K}\; (R+S)\!\cdot\!\hat{K}}\right]\right\rangle_{\hat{\bm k}}\nn
&=&\frac{1}{96} \left(T^2+\frac{\mu^2}{2\pi^2}\right) \left\langle \left[\frac{P\cdot u}{P\!\cdot\!\hat{K}\; Q\!\cdot\!\hat{K}\; (Q+R)\!\cdot\!\hat{K}} +\frac{(P+Q)\cdot u}{Q\!\cdot\!\hat{K}\;  R\!\cdot\!\hat{K}\; (R+S)\!\cdot\!\hat{K}} + \frac{(P+Q+R)\cdot u}{R\!\cdot\!\hat{K}\; S\!\cdot\!\hat{K}\; (S+P)\!\cdot\!\hat{K} } \right]\right\rangle_{\hat{\bm k}}.
\ee
One can calculate the other components of $\mathcal{T}^{\mu\nu\lambda\sigma}_{abcd,q}$ in the similar manner. By doing so,  the contribution from the quark loop becomes
\be
\mathcal{T}^{\mu\nu\lambda\sigma}_{abcd,q}&=&\frac{32g^2N_f}{96}\left(T^2+\frac{\mu^2}{2\pi^2}\right)\mbox{Tr}[t^at^dt^ct^b]\nn
&\times&\left\langle \hat{K}^\mu\hat{K}^\nu\hat{K}^\lambda\hat{K}^\sigma\left[\frac{P\cdot u}{P\!\cdot\!\hat{K}\; Q\!\cdot\!\hat{K}\; (Q+R)\!\cdot\!\hat{K}} +\frac{(P+Q)\cdot u}{Q\!\cdot\!\hat{K}\;  R\!\cdot\!\hat{K}\; (R+S)\!\cdot\!\hat{K}} + \frac{(P+Q+R)\cdot u}{R\!\cdot\!\hat{K}\; S\!\cdot\!\hat{K}\; (S+P)\!\cdot\!\hat{K} } \right]\right\rangle_{\hat{\bm k}}.
\label{Tau_q1_final}
\ee
Apart from the quark diagram shown in Fig.~\ref{htl_4g_vert}, another quark-loop diagram will also contribute to the four-gluon vertex. In that diagram, the the direction of the flow of charge will be opposite. One can explicitly calculate the second quark-loop diagram and can show that the only the color trace term differs from \eqref{Tau_q1_final}. Adding the two contributions, the total contribution from the quark loop becomes
\be
\mathcal{T}^{\mu\nu\lambda\sigma}_{abcd,q}&=&g^2\frac{N_f}{3}\left(T^2+\frac{\mu^2}{2\pi^2}\right)\mbox{Tr}[t^at^dt^ct^b+t^at^dt^ct^b]\nn
&\times&\left\langle \hat{K}^\mu\hat{K}^\nu\hat{K}^\lambda\hat{K}^\sigma\left[\frac{P\cdot u}{P\!\cdot\!\hat{K}\; Q\!\cdot\!\hat{K}\; (Q+R)\!\cdot\!\hat{K}} +\frac{(P+Q)\cdot u}{Q\!\cdot\!\hat{K}\;  R\!\cdot\!\hat{K}\; (R+S)\!\cdot\!\hat{K}} + \frac{(P+Q+R)\cdot u}{R\!\cdot\!\hat{K}\; S\!\cdot\!\hat{K}\; (S+P)\!\cdot\!\hat{K} } \right]\right\rangle_{\hat{\bm k}}.
\label{Tau_q_final}
\ee
Earlier we have shown in all the $N$-point calculations that the combined contribution from the gluon and ghost only differs by a factor with the quark contribution and  it contibute to the Debye/thermal quark mass. One can show explicitly that this is also true for the four-point gluon vertex. So, the four-point gluon vertex within HTL approximation becomes
\be
\mathcal{T}^{\mu\nu\lambda\sigma}_{abcd}&=&2g^2\left[C_AT^2+\frac{N_f}{3}\left(T^2+\frac{\mu^2}{2\pi^2}\right)\right]\mbox{Tr}[t^at^dt^ct^b+t^at^dt^ct^b]\nn
&\times&\left\langle \hat{K}^\mu\hat{K}^\nu\hat{K}^\lambda\hat{K}^\sigma\left[\frac{P\cdot u}{P\!\cdot\!\hat{K}\; Q\!\cdot\!\hat{K}\; (Q+R)\!\cdot\!\hat{K}} +\frac{(P+Q)\cdot u}{Q\!\cdot\!\hat{K}\;  R\!\cdot\!\hat{K}\; (R+S)\!\cdot\!\hat{K}} + \frac{(P+Q+R)\cdot u}{R\!\cdot\!\hat{K}\; S\!\cdot\!\hat{K}\; (S+P)\!\cdot\!\hat{K} } \right]\right\rangle_{\hat{\bm k}}\nn
&=&2(m_D^g)^2\ \mbox{Tr}[t^at^dt^ct^b+t^at^dt^ct^b]\nn
&\times&\left\langle \hat{K}^\mu\hat{K}^\nu\hat{K}^\lambda\hat{K}^\sigma\left[\frac{P\cdot u}{P\!\cdot\!\hat{K}\; Q\!\cdot\!\hat{K}\; (Q+R)\!\cdot\!\hat{K}} +\frac{(P+Q)\cdot u}{Q\!\cdot\!\hat{K}\;  R\!\cdot\!\hat{K}\; (R+S)\!\cdot\!\hat{K}} + \frac{(P+Q+R)\cdot u}{R\!\cdot\!\hat{K}\; S\!\cdot\!\hat{K}\; (S+P)\!\cdot\!\hat{K} } \right]\right\rangle_{\hat{\bm k}}\nn
&=&2(m_D^g)^2\ \mbox{Tr}[t^at^dt^ct^b+t^at^dt^ct^b]{\cal T}^{\mu\nu\lambda\sigma}(P, Q, R, S)
\label{Tau_qg_final}
\ee

Considering all the symmetry factor, the four-gluon vertex for gluons with outgoing momenta $P$, $Q$, $R$, and $S$, Lorentz indices $\mu$, $\nu$, $\lambda$, and $\sigma$,  and colour indices $a$, $b$, $c$, and $d$ becomes
\bea
i\Gamma^{\mu\nu\lambda\sigma}_{abcd}(P, Q, R, S)& =&- ig^2\big\{ f_{abx}f_{xcd} \left(\eta^{\mu\lambda}\eta^{\nu\sigma} - \eta^{\mu\sigma}\eta^{\nu\lambda}\right)
+ 2(m_D^g)^2 {\Tr}\left[t^a\left(t^bt^ct^d+t^d t^c t^b \right)\right]{\cal T}^{\mu\nu\lambda\sigma}(P, Q, R, S)
\big\} \nn
&& + \; 2 \; \mbox{cyclic permutations}\;,
\eea
%
where the cyclic permutations are of $(Q,\nu,b)$, $(R,\lambda,c)$, and $(S,\sigma,d)$. The matrices $t^a$ are the fundamental representation of the $SU(3)$ algebra with the standard normalization ${\Tr}(t^a t^b) = {1 \over 2} \delta^{ab}$. 

The tensor ${\cal T}^{\mu\nu\lambda\sigma}$ in the HTL correction term in Eq.~\eqref{Tau_qg_final} is defined for $P+Q+R+S=0$ as
\be
{\cal T}^{\mu\nu\lambda\sigma}(P, Q, R, S) =
\Bigg\langle\! \hat{K}^{\mu} \hat{K}^{\nu} \hat{K}^{\lambda} \hat{K}^{\sigma}
\left( {P\!\cdot\!u \over P\!\cdot\!\hat{K} \; Q\!\cdot\!\hat{K} \; (Q+R)\!\cdot\!\hat{K}}
 + {(P + Q)\!\cdot\!u\over Q\!\cdot\!K\;R\!\cdot\!\hat{K}\;(R + S)\!\cdot\!\hat{K}}
+{(P + Q + R)\!\cdot\!u\over R\!\cdot\!\hat{K}\; S\!\cdot\!\hat{K}\;(S + P)\!\cdot\!K}\right)
\!\Bigg\rangle_{\bf\hat k}.
\ \
\label{T4-def}
\ee
%
This tensor is totally symmetric in its four indices and traceless in any pair of indices: $g_{\mu\nu}{\cal T}^{\mu\nu\lambda\sigma}=0$. It is even under cyclic or anti-cyclic permutations of the momenta $P$, $Q$, $R$, and $S$. It satisfies the ``Ward identity''
\be
Q_{\mu}{\cal T}^{\mu\nu\lambda\sigma}(P, Q, R, S)&=& {\cal T}^{\nu\lambda\sigma}(P + Q, R, S) - {\cal T}^{\nu\lambda\sigma}(P,R + Q, S)
\label{ward-t4}
\ee
%
and the ``Bianchi identity''
\be\nonumber
{\cal T}^{\mu\nu\lambda\sigma}(P,Q,R,S) + {\cal T}^{\mu\nu\lambda\sigma}(P,R,S,Q)+
{\cal T}^{\mu\nu\lambda\sigma}(P,S,Q,R)=0\;.
\label{Bianchi}
\ee
When its colour indices are traced in pairs, the four-gluon vertex becomes particularly simple:
\be
\delta^{ab} \delta^{cd} i \Gamma_{abcd}^{\mu\nu\lambda\sigma}(P, Q, R, S)
&=& -i g^2 N_c (N_c^2-1) \Gamma^{\mu\nu,\lambda\sigma}(P, Q, R, S) \;,
\ee
where the colour-traced four-gluon vertex tensor is
\be
\Gamma^{\mu\nu,\lambda\sigma}(P , Q, R ,S)&=&
2\eta^{\mu\nu}\eta^{\lambda\sigma} - \eta^{\mu\lambda}\eta^{\nu\sigma} - \eta^{\mu\sigma}\eta^{\nu\lambda}-(m_D^g)^2{\cal T}^{\mu\nu\lambda\sigma}(P,S,Q,R)\;.
\label{Gam4}
\ee
Please take note of the ordering of momenta within the arguments of the tensor ${\cal T}^{\mu\nu\lambda\sigma}$, which comes from the application of the Bianchi identity (\ref{Bianchi}). The tensor (\ref{Gam4}) exhibits symmetric
upon the interchange of $\mu$ and $\nu$, under the interchange of $\lambda$ and $\sigma$, and under the interchange of $(\mu,\nu)$ and $(\lambda,\sigma)$. Additionally, it maintains symmetry upon the interchange of $P$ and $Q$, $R$ and $S$, and $(P,Q)$ and $(R,S)$. Furthermore, it satisfies the Ward identity:
\be
P_{\mu}\Gamma^{\mu\nu,\lambda\sigma}(P, Q, R, S)
&=&\Gamma^{\nu\lambda\sigma}(Q, R + P, S) - \Gamma^{\nu\lambda\sigma}(Q, R, S + P)\;.
\label{ward-4}
\ee
%

\subsubsection{Two Quark-Two Gluon  HTL effective vertex.}
\label{htl_2g_2q_vert}
\begin{figure}[tbh]
\begin{center}                            
\includegraphics[scale=0.4]{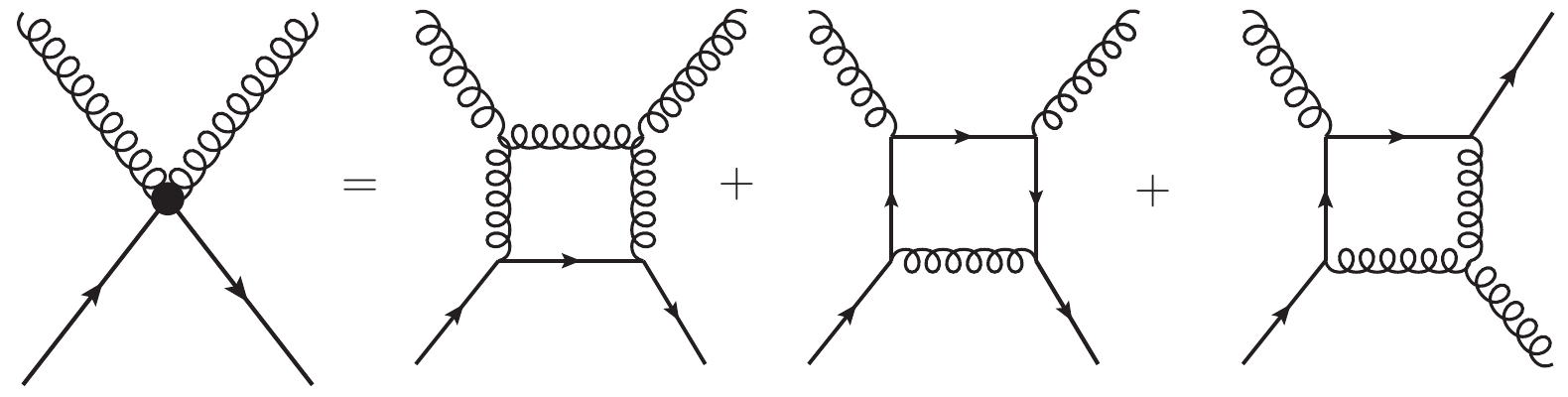}
\end{center}
\vspace{-0.45cm}
\caption{ 2quark-2gluon  HTL effective vertex.}
\label{2g_2q_vert}
\end{figure}         
 We define the quark-gluon four-point vertex with outgoing gluon momenta $P$ and $Q$, incoming quark momentum $R$, and outgoing quark momentum $S$. It reads
  \be
 \delta^{ab} \Gamma^{\mu\nu}_{abij}(P,Q,R,S) \!\!&=&\!\! 
 - g^2 m_q^2 C_F \delta_{ij} \tilde{\cal T}^{\mu\nu}(P,Q,R,S)  
 \equiv g^2 C_F \delta_{ij} \Gamma^{\mu\nu}  \, ,
 \label{4qgv}
 \ee
 There is no tree-level term. The tensor in the HTL correction term is only defined for $P+Q-R+S=0$, as
 \be\nonumber
 \tilde{{\cal T}}^{\mu\nu}(P,Q,R,S)
 +\left\langle
 \hat{K}^{\mu}\hat{K}^{\nu}\left(\frac{1}{R\!\cdot\!\hat{K}}+\frac{1}{S\!\cdot\!\hat{K}}\right)
 \frac{\hat{\slashed{K}}}{(R-P)\!\cdot\!\hat{K}\;(S+P)\!\cdot\!\hat{K}}
 \right\rangle_{\bm{\hat k}} .
 \ee
 This tensor is symmetric in $\mu$ and $\nu$ and is traceless.
 It satisfies the Ward identity:
 \be
 \!\!P_{\mu}\Gamma^{\mu\nu}(P,Q,R,S)\!=\!\Gamma^{\nu}(Q,R-P,S)-
 \Gamma^{\nu}(Q,R,S+P)\,.
 \label{qward2}
 \ee
 
\subsection{HTL Improved QCD Lagrangian}
\label{qcd_imp}
The HTL contribution to the QCD  Lagrangian~\cite{Braaten:1991gm,Taylor:1990ia,Frenkel:1991ts} can be written as
\be
{\cal L}_{\textrm{\tiny QCD}}^{\textrm {\tiny HTL}}  = {\cal L}_{\textrm{\tiny QCD}} + {\cal L}_{\textrm {\tiny HTL}}  \, \label{htl_qcd_lag1}
\ee
where $ {\cal L}_{\textrm{QCD}}$ is usual QCD lagrangian in vacuum as given in Eq.\eqref{qcd1}. Following the results obtained in previous subsections ~\ref{gsf}, \ref{q_se}, \ref{qg_three}, \ref{g_three}, \ref {g_four} and \ref{htl_2g_2q_vert}  for $N$-point functions, the HTL improved Lagrangian can be written in compact form as~\cite{Braaten:1991gm,Taylor:1990ia,Frenkel:1991ts}  as
 \be
 {\cal L}_{\tiny {\textrm {HTL}}}=i (m^{\rm q}_\textrm{th})^2\bar\psi\gamma^\mu\left\langle\frac{\hat{K}_\mu}{\hat{K}\cdot\! D}\right\rangle_{\hat{\bf k}}\psi-\frac{1}{2}
 (m^{\rm g}_D)^2 {\rm Tr}\left ( G_{\mu\alpha}\left\langle\frac{\hat{K}^\alpha \hat{K}_\beta}{(\hat{K}\cdot\! D)^2}\right\rangle_{\hat{\bf k}} G^{\mu\beta}\right )\, ,
\label{htl_qcd_lag2}      
 \ee
where $G$ is the gluon field strength. The ${\Tr}$ is over colour indices. The two parameters $m^{\rm g}_D$ and $m^{\rm q}_\textrm{th}$ are, respectively, the Debye screening mass and the thermal quark mass which take into account  the screening effects.  The Lagrangian is non-local and gauge symmetric, which forces the presence of the covariant derivative in the denominator of \eqref{htl_qcd_lag2}, which makes it also non-linear. When expanded in powers of quark and gluon fields, Eq.~\eqref{htl_qcd_lag2} generates an infinite series of non-local self-energy and vertex corrections (viz., HTL $N$-point QCD functions) obtained in subsections~\ref{gsf}, \ref{q_se}, \ref{qg_three}, \ref{g_three}, \ref {g_four} and \ref{htl_2g_2q_vert} to the bare one. If it is expanded in terms of quark fields,~\eqref{htl_qcd_lag2} will lead to quark self-energy as obtained in~\eqref{qcd27} whereas when expanded in two fermions and leading order in gauge field, one gets back $3$-point quark-gluon vertex as obtained in \eqref{qqg_3pt} and two quarks and quadratic order in gauge lead to $4$-point function (two quark-two gluon vertex) as obtained in subsec~\ref{htl_2g_2q_vert}. If expanded in quartic order in gauge field one obtains $4$-gluon vertex as given in \eqref{Gam4}. These $N$-point functions we derived so far  are related by Ward-Takahashi  identity. We also note that the ghost-gluon vertex remains same as bare on because there is HTL correction to it.

       \section{Improved Perturbation Theory}
        \label{ipt_chap}
        	\vspace{-0.2cm}
\subsection{Screened Perturbation Theory (SPT)}
 \label{spt}
 The scalar field  Lagrangian in $\Phi^4$ theory is given by \eqref{sl} and the HTL improved one is given in \eqref{sc_eef}.  The counterterm $\Delta {\cal L}$ there has to
 be adjusted through dimension regularisation to cancel the divergences order by order in $g^2$ as the conventional perturbative expansion in powers of $g^2$  generates ultraviolet divergences.  However, as we have  in  finite temperature the conventional perturbative expansion also generates infrared divergences. We seen in section~\ref{scalar_htl} that these divergences can be removed by resumming the higher order diagrams that generate a thermal mass of order $m\equiv m_{\rm{th}}^s = gT/\sqrt{24}$ for the scalar particle. This resummation changes the perturbative series from an expansion in powers of $g^2$ to an expansion in powers of $(g^2)^{1/2} = g$.  Based on this a new perturbative scheme for massless scalar field theory with a $\Phi^4$-interaction, known as screened perturbation theory (SPT), was proposed in Ref.~\cite{Karsch:1997gj} for the evaluation of thermodynamic quantities in field theories at finite temperature by taking into account exactly the phenomenon of screening or thermal mass $m$ in scalar propagator. The SPT was made more systematic in Ref.~\cite{Chiku:1998kd} by applying to a spontaneously broken scalar field theory. The Lagrangian density is written~\cite{Chiku:1998kd,Andersen:2000yj,Andersen:2001ez,Andersen:2008bz} as
\bea
{\cal L}^{\rm{SPT}} &=& -{\cal E}_0 + \frac{1}{2}\partial_\mu\Phi\partial^\mu\Phi -\frac{1}{2}\left (m^2-m_1^2\right)\Phi^2 +\frac{g^2}{24} \Phi^4 +\Delta {\cal L} +
\Delta{\cal L}^{\rm{SPT}} \, , \label{spt1}
\eea
where ${\cal E}_0$ is vacuum energy density term. We note that the mass term is added and subtracted like \eqref{sc_eef} in little different way. If ${\cal E}_0=0$ and
$m=m_1$, the one gets the free Lagrangian  \eqref{sl}. The SPT is defined by considering $m^2$ to be order of $1$ and $m_1^2$ to be order of $g^2$, and then expanding 
systematically in powers of $g^2$ and setting $m^2_1=m^2$ at the end of the calculation. This reorganises the perturbation theory in which the expansion is about the free field 
theory defined by
\be
{\cal L}^{\rm{SPT}}_{0} &=  -{\cal E}_0 + \frac{1}{2}\partial_\mu\Phi\partial^\mu\Phi -\frac{1}{2} m^2 \Phi^2 \, \label{spt2} 
\ee
along with the interaction part
\be
 {\cal L}^{\rm{SPT}}_{\rm{int}}= \frac{1}{2}m_1^2\Phi^2  +\frac{g^2}{24} \Phi^4 +\Delta {\cal L} +
\Delta{\cal L}^{\rm{SPT}} \, . \label{spt3}
\ee
In addition to $\Phi^4$ interaction, $m_1^2$ term can also be interpreted as an interaction that couples two $\Phi$ fields.
The SPT also produces utraviolet divergences, but they would be  cancelled by the counter term $\Delta{\cal L}^{\rm{SPT}} $.
If one uses dimensional regularisation and minimal subtraction, the coefficients of these operators are polynomial in $g^2$ and 
$(m^2-m_1^2)$, and the explicit  expression of $\Delta{\cal L}^{\rm{SPT}}$ is given in Refs.~\cite{Andersen:2001ez,Andersen:2008bz}.
This SPT has been used to calculate  the thermodynamic quantities~\cite{Karsch:1997gj,Andersen:2000yj,Andersen:2001ez,Andersen:2008bz}. in scalar $\Phi^4$-theory.
  
\subsection{Hard Thermal Loop Perturbation Theory (HTLpt)}
\label{htlpt}
A  formalism was developed  for computing the thermodynamic quantities rigorously, which controls the related IR problems via a reorganization of finite temperature perturbation 
theory and the HTL approximation is called Hard Thermal Loop perturbation theory 
(HTLpt)~\cite{Andersen:1999fw,Andersen:1999sf,Andersen:1999va,Andersen:2002ey,Andersen:2003zk,Andersen:2009tw,Andersen:2009tc,Andersen:2010ct,Andersen:2010wu,Andersen:2011sf,Andersen:2011ug,Haque:2012my,Mogliacci:2013mca,Haque:2013qta,Haque:2013sja,Haque:2014rua,Andersen:2015eoa}. The HTLpt framework allows for a systematic 
analytic reorganization of perturbative series based on the HTL effective Lagrangian. Further more, it is manifestly gauge
invariant and also very useful to calculate both static and dynamical quantities.  
The HTLpt approach is an extension of the simpler screened perturbation  theory which has been applied to scalar field 
theories~\cite{Karsch:1997gj,Chiku:1998kd,Andersen:2000yj,Andersen:2001ez,Andersen:2008bz}.

The HTLpt Lagrangian density can be written as
 \be
 {\cal L}_{\tiny \rm {HTLpt}}=\left.\left ({\cal L}_{\rm {QED}\atop{\rm{QCD}}}+(1-\delta) {\cal L}_{\rm HTL}\right)\right|_{g\rightarrow\sqrt{\delta}g}+
 \Delta{\cal L}_{\rm HTL}
\label{htl2} 
\ee
 where $\Delta{\cal L}_{\rm HTL}$ is additional counterterm needed to cancel the UV divergences generated in HTLpt. The HTL improved Lagrangian, ${\cal L}_{\rm HTL}$ is given in \eqref{qed_htl_lag2} for QED and in \eqref{htl_qcd_lag2} for QCD, respectively. HTLpt is defined by considering $\delta$ as a formal expansion 
 parameter. By incorporating the HTL improvement term~\eqref{htl2} into the QED or QCD Lagrangian in \eqref{qed_htl_lag2} or  in \eqref{htl_qcd_lag2}, HTLpt consistently shifts the perturbative expansion from  an ideal gas of massless particles, to a gas of massive quasiparticles which are the more apt physical degrees of freedom at high temperature and chemical potential.
 It is noteworthy to mention that the HTLpt Lagrangian \eqref{htl2} transforms into the QED or QCD Lagrangian in \eqref{qed23} or in~\eqref{qcd_lag} when $\delta=1$.

In HTLpt, the physical quantities are calculated through an expansion in powers of $\delta$, terminating at some specified order, and $\delta = 1$ is set. As previously noted, this signifies a rearrangement of the perturbation series in which the screening effects through $m_D^2$ and $m_\textrm{th}^2$ terms in Eq.~\ref{htl2} have been considered to all orders but then systematically subtracted out at higher orders in perturbation theory by the $\delta m_D^2$ and $\delta m_\textrm{th}^2$ terms in Eq.~\ref{htl2}. One usually expands to orders $\delta^0$, $\delta^1$, $\delta^2$, respectively, for computing leading order (LO), next-to-leading order (NLO), and next-to-next-leading order (NNLO) results. Note that HTLpt is gauge invariant order-by-order in the $\delta$-expansion and, consequently, the results obtained are gauge independent.

If the $\delta$-expansion  could be computed to all orders the results would not depend on $m_D$ and $m_\textrm{th}$ when one puts $\delta=1$. However, any truncation of the $\delta$-expansion results $m_D$ and $m_\textrm{th}$ dependent outcomes. Hence, a prescription is necessary to determine $m_D$ and $m_\textrm{th}$ as a function of temperature $(T)$, chemical potential $(\mu)$, and strong coupling constant $(\alpha_s)$. There are various prescriptions and some of them had been discussed in~\cite{Andersen:2011sf} at zero chemical potential. The HTL perturbation expansion produces UV divergences. In QCD perturbation theory, renormalizability restricts the UV divergences in such a way that they can be eliminated by the counterterm Lagrangian $\Delta {\cal L}_{\rm QCD}$. Usually the renormalization of HTLpt can be considered by adding a counterterm Lagrangian $\Delta{\cal L}_{\rm HTL}$  in (\ref{htl2}). However, there is no such proof yet that the HTLpt is renormalizable, so the general structure of the UV divergences remain unknown. The most optimistic scenario is that HTLpt would be renormalizable, such that the UV divergences in the physical observables can all be eliminated through proper counterterms.

 The HTLpt in QCD has been used to study the various physical quantities relevant for understanding the properties of QGP, viz., the thermodynamic properties~\cite{Andersen:1999fw,Andersen:1999sf,Andersen:1999va,Andersen:2002ey,Andersen:2003zk,Andersen:2009tw,Andersen:2009tc,Andersen:2010ct,Andersen:2010wu,Andersen:2011sf,Andersen:2011ug,Haque:2012my,Mogliacci:2013mca,Haque:2013qta,Haque:2013sja,Haque:2014rua,Andersen:2015eoa,Haque:2010rb,Haque:2011iz,Chakraborty:2001kx,Chakraborty:2002yt,Chakraborty:2003uw}, 
dilepton production rate~\cite{McLerran:1984ay,Baier:1988xv,Braaten:1990wp,Greiner:2010zg,Ghisoiu:2014mha,Ghiglieri:2014kma,Ghiglieri:2015nba,Aurenche:1998nw,Aurenche:1999ec,Karsch:2000gi,Thoma:1997dk,Aurenche:2002pc,Aurenche:2002wq,Carrington:2007gt}, photon production rate~\cite{Kapusta:1991qp,Baier:1991em,Aurenche:1998nw,Aurenche:1999ec,Arnold:2001ba,Arnold:2001ms,Peitzmann:2001mz},
single quark and quark-antiquark potentials~\cite{Mustafa:2004hf,Mustafa:2005je,Chakraborty:2006md,Chakraborty:2007ug,Laine:2006ns,Dumitru:2007hy,Dumitru:2009ni,Thakur:2013nia}, fermion damping rate~\cite{Pisarski:1993rf,Braaten:1992gd,Peigne:1993ky,Thoma:1995ju,Thoma:2000dc,Thoma:1990fm,Thoma:1993vs}, photon damping rate~\cite{Thoma:1994fd,Abada:2011cc}, gluon damping rate~\cite{Braaten:1989kk,Braaten:1990it,Pisarski:1990ds} and  parton energy-loss~\cite{Thoma:1990fm,Braaten:1991jj,Braaten:1991we,Thoma:1991jum,Arnold:2002ja,Jeon:2003gi,Mustafa:2004dr,Mustafa:2003vh,Chakraborty:2006db,Qin:2007rn,Ghiglieri:2015ala,Guo:2024mgh} and 
 plasma instabilities~\cite{Mrowczynski:2000ed,Romatschke:2003ms,Romatschke:2004jh,Rebhan:2004ur,Schenke:2006fz,Rebhan:2008uj,Attems:2012js}. Additionally, the transport theory based on the HTL perturbation theory has been used to developed semi-classical transport theory~\cite{Litim:1999ns,Litim:2001db} that addresess the Debye mass and damping rate~\cite{Litim:1999id} and the high density QCD phase with color superconductivity~\cite{Litim:2001je}.

Another area where the methods can be applied  is cosmology. There are relations between thermal photon production and right-handed neutrino production. We refer to Refs.~\cite {Drewes:2017zyw,Biondini:2017rpb} for recent reviews which put the calculations in the physical context of this extension of the Standard Model (SM).  The photon corresponds to the right-handed neutrino, as they are both singlets under the gauge groups of the plasma. The quarks correspond to the Higgs doublet and left-handed leptons, which couple to the right-handed neutrino via a Yukawa coupling. Finally, gluons correspond to the electroweak gauge bosons, which interact with the active leptons and scalars. We then wish to highlight some significant applications:  the leading-order collinear production rate was determined in the symmetric phase in Refs.~\cite {Besak:2012qm,Anisimov:2010gy}, with the latter paper also obtaining the fermionic sum rule, which also enters the production rate for ultra-relativistic right-handed neutrinos. The collinear rate in the broken phase was derived in Ref.~\cite{Ghiglieri:2016xye} -- which also derived the vector boson HTLs of the SM in that phase at zero chemical potential; chemical potentials were considered in both phases in Refs.~\cite{Ghiglieri:2018wbs,Ghiglieri:2017gjz}. It has also been used to calculate active neutrino soft scattering at NLO in the broken phase. The HTLpt has also been applied to compute thermal axion production~\cite{Graf:2010tv} and lepton asymmetry during leptogenesis~\cite{Kiessig:2011ga,Kiessig:2011fw}.
  
Very recently, HTL is extended to calculate the free energy of $\mathcal{N} = 4$ supersymmetric Yang-Mills (SYM) in four spacetime dimensions in Refs.~\cite{Du:2020odw,Du:2021jai,Andersen:2021bgw}. Also for other applications of $\mathcal{N} = 4$ SYM theory, we refer to recent review in Ref.~\cite{Ghiglieri:2020dpq}. Long-time ago there was also effort  to calculate the chemical equilibration time of light and also in particular for semi-light strange quarks in the quark-gluon plasma by means of HTLpt in order to obtain a more accurate number than from standard semi-phenomenological kinetic theory. As far as our knowledge is concerned there were not much success to receive such a gauge-independent and infrared-safe HTL improved rate.
 
In the following sections we will apply HTLpt  to study various properties of mostly hot and dense  QCD plasma (QGP) and a few example for QED plasma.

	\section{Electromagnetic Particle Production Rate: both virtual and real photons}\label{em_chap}
	\vspace{-0.2cm}
Electromagnetically interacting particles --- real photons and virtual photons (dileptons) --- serve as excellent probes of the thermodynamic state of evolving strongly interacting matter produced in ultra-relativistic nucleus-nucleus collisions. This is because electromagnetic interactions are strong enough to lead to a detectable signal and yet are weak enough to allow the emitted photons and leptons escape from the finite nuclear system without further interactions. It's important to note that the nature of emission of real and virtual photon crucially depends on the size of the hot thermal system from which they are emitted. If the system is sufficiently large, photons will undergo rescattering and thermalisation, and resulting the momentum space distribution following the Planck distribution. The corresponding emission rate will then be that of black body radiation which depends only on the temperature and the area of the emitting body but not on its microscopic properties. Since the typical size of systems produced in heavy ion collisions is much less than the mean free path of the photons, they are likely to escape the hot zone without interacting further and the emission rate in this case depends on the dynamics of the thermal constituents through the imaginary part of the photon self energy. Hence, the spectra of photons and dileptons can provide information about the properties of the constituents from which they originated~\cite{Jackson:2019yao,Ghiglieri:2016tvj}.

For most purposes, the emission rates of photons and dileptons can be obtained within a classical framework. Employing the kinetic theory approach, photon emission rates~\cite{Ruuskanen:1992hh,Gutbrod1993ParticlePI} can be calculated in terms of the equilibration rate of photons in a thermal bath. This rate is related to the exclusive kinetic rates of emission and absorption associated with squared matrix element of the various physical processes in the system. However, 
in Ref.~\cite{Feinberg:1976ua}, Feinberg has shown that emission rates can be related to the electromagnetic current correlation function in a thermalised system in a quantum picture and, more importantly, in a nonperturbative manner. Generally, the production rate of a particle weakly interacting with the constituents of the thermal bath(the constituents may interact strongly among themselves, the explicit form of their coupling strength is not important) can always be expressed in terms of the discontinuities or imaginary parts of the self-energies of that particle~\cite{Gutbrod1993ParticlePI,Weldon:1990iw,McLerran:1984ay,Gale:1990pn,Kobes:1985kc,Kobes:1986za}. In this section, we briefly outline the connection between the emission rates of virtual (in subsec~\ref{dilep_chap}) and real (in subsec~\ref{photon_chap}) photons  and the spectral function of the photon, which is connected with the discontinuities in self- energies in a thermal system~\cite{Weldon:1990iw,McLerran:1984ay,Gale:1990pn}. Subsequently, utilising this connection, we will compute virtual and real photon rates from hot and dense QCD matter (QGP) produced in ultra-relativistic.
\subsection{Dilepton Production Rate}
\label{dilep_chap}
Thermal dileptons ($q{\bar q}\rightarrow \gamma^*\rightarrow l^+l^-$, where  $\bar {q}(q)$ are (anti)quark, $\gamma^*$ denotes virtual photon, and $l^+l^-$ are lepton pair) emitted from the fireball in ultrarelativistic heavy-ion collisions might serve as a promising signature~\cite{Ruuskanen:1991au,McLerran:1984ay,Cleymans:1986na,Cleymans:1993jm,Cleymans:1992gb,Alam:1996fd} for the QGP formation in such collisions. Unlike hadronic signals, dileptons and photons carry direct information about the early phase of the fireball, since they do not interact with the surrounding medium after their production. Therefore, they can be used as a direct probe for the QGP. However, a significant background arises from hadronic decays. Thus, it is desirable to identify  specific features in the dilepton spectrum that could indicate the presence of deconfined matter. Indeed perturbative calculations~\cite{Braaten:1990wp,Karsch:2000gi} have revealed distinct structures (van Hove peaks~\cite{hove} and gaps) in the production rate of low-mass dileptons caused by non-trivial in-medium quark dispersion relations. In the following subsec~\ref{dil_th} we briefly discuss  the dilepton production rate from a thermal medium.
\subsubsection{ Dilepton emission rate in presence of a thermal medium}
\label{dil_th}
\vspace{-0.2cm}
The dilepton multiplicity per unit space-time volume is given~\cite{Weldon:1990iw} as
\bea
\frac{dR}{d^4x}&=& 2\pi e^2 e^{-\beta 
p_0}L_{\mu\nu}\rho^{\mu\nu}\frac{d^3 \bm{\vec q}_1}{(2\pi)^3E_1}\frac{d^3\bm{\vec q}_2}{(2\pi)^3E_2}, \label{d1}
\eea
where $e$ is the electromagnetic coupling, $\bm{\vec q}_i$ and $E_i$ with $i=1,2$ are three momentum and energy  of the lepton pairs.
The photonic tensor or the electromagnetic spectral function in thermal medium can be written as
\bea
\rho^{\mu\nu}(p_0,\bm{\vec p}) &=& -\frac{1}{\pi}\frac{e^{\beta p_0}}{e^{\beta 
p_0}-1}\textrm{Im}\left[D^{\mu\nu}(p_0,\bm{\vec p})\right]\equiv 
-\frac{1}{\pi}\frac{e^{\beta p_0}}{e^{\beta 
p_0}-1}~\frac{1}{P^4}\textrm{Im}\left[\Pi^{\mu\nu}(p_0,\bm{\vec  p})\right], \label{d2}
\eea
where $`\textrm{Im}$' stands for imaginary part,  $\Pi^{\mu\nu}$ is the two point current-current correlation function or the self-energy of photon and 
$D^{\mu\nu}$ represents the photon propagator. 
Here we have used the  relation~\cite{Weldon:1990iw} 
 \bea
D^{\mu\nu}(p_0,\bm{\vec p}) = \frac{1}{P^4}\Pi^{\mu\nu}(p_0,\bm{\vec p}) \, ,
\label{d2i}
 \eea
where $P \equiv (p_0,\bm{\vec p})$ is the four momenta of the photon. Also  the leptonic tensor in terms of Dirac spinors is given by
\bea
L_{\mu\nu} &=& \frac{1}{4} 
\sum\limits_{\mathrm{spins}}\mathrm{Tr}\left[\bar{u}(Q_2)\gamma_\mu 
v(Q_1)\bar{v}(Q_1)\gamma_\nu u(Q_2)\right] 
= Q_{1\mu}Q_{2\nu}+Q_{1\nu}Q_{2\mu}-(Q_1\cdot Q_2+m_l^2)g_{\mu\nu},
\label{d3}
\eea
where $Q_i\equiv (q_0, \bm{\vec q}_i)$ is the four momentum of the $i$-th lepton and $m_l$ is the mass of the lepton.
Now inserting $\int d^4P\, \delta^4(Q_1+Q_2-P)=1$, one can write the dilepton 
multiplicity from \eqref{d1} as
\bea
\frac{dR}{d^4x}\
&=& 2\pi e^2 e^{-\beta p_0}\int d^4P \,  \delta^4(Q_1+Q_2-P)
L_{\mu\nu}\rho^{\mu\nu}\frac{d^3 \bm{\vec q}_1}{(2\pi)^3E_1}\frac{d^3\bm{\vec q}_2}{(2\pi)^3E_2}. \label{d4} 
\eea
Using the identity
\bea
\int\frac{d^3 \bm{\vec q}_1}{E_1}\frac{d^3 \bm{\vec q}_2}{E_2}
\delta^4(Q_1+Q_2-P)\, L_{\mu\nu} = 
\frac{2\pi}{3} 
\left(1+\frac{2m_l^2}{P^2}\right)\sqrt{1-\frac{4m_l^2}{P^2}}
\left(P_\mu P_\nu-P^2g_{\mu\nu}\right) 
=\frac{2\pi}{3}F_1(m_l,P^2)\left(P_\mu P_\nu-P^2g_{\mu\nu}\right),\ \label{d5}
\eea
the dilepton production rate in~\eqref{d4} comes out to be
\bea
\frac{dR}{d^4xd^4P}  =
\frac{\alpha} {12\pi^4}\frac{n_B(p_0)}{P^2} F_1(m_l,P^2) \  \textrm{Im} \left [\Pi^{\mu}_{\mu}(p_0, \bm{\vec p})\right] 
= \frac{\alpha} {12\pi^4}\frac{n_B(p_0)}{P^2} F _1(m_l,P^2) \  \frac{1}{2i} {\textrm{Disc}} \left [\Pi^{\mu}_{\mu}(p_0, \bm{\vec p})\right], \label{d6}
\eea 
where $n_B(p_0)=(e^{p_0/T}-1)^{-1}$ and $e^2=4\pi\alpha$, $\alpha$ is the electromagnetic coupling constant. We have also used the transversality 
condition $P_\mu\Pi^{\mu\nu}=0$. The invariant mass of the lepton pair is defined as 
$
M^2\equiv P^2(=p_0^2-|\bm{\vec p}|^2=\omega^2-|\bm{\vec p}|^2). \label{d6a}
$
We also note that for massless lepton ($m_l=0$) $F_1(m_l,P^2)=1$. The quantity $\textrm{Im} \left [\Pi^{\mu}_{\mu}(p_0, \bm{\vec p})\right] $ contains 
information about the constituents of the thermal bath and thus is of great relevance.

The \eqref{d6} is the familiar result most widely used for the dilepton emission rate from a thermal medium. 
It must be emphasized that this relation is valid only to ${\cal O}(e^2)$ since it does not account for the possible reinteractions of the virtual photon 
on its way out of the thermal bath. The possibility of emission of more than one photon has also been neglected here. 
However, the expression is true to all orders in strong interaction. 

\subsubsection{ Hard dilepton production rate (Born Rate ${\cal O}(\alpha^2)$)}
\label{born}
The production of hard dileptons with total energy $\om\sim T$ can be  computed by standard perturbative techniques~\cite{Cleymans:1986na}. The leading-order contribution (Born rate)  to the dilepton rate comes from the one-loop photon diagram ($\Pi_{\mn}$ ) with a quark loop as given in Fig.~\ref{photon_self}. The contribution of one loop photon self-energy diagram in Fig.~\ref{photon_self} can be written as
\be
\Pi_{\mn}(P) = -  N_c \sum_f e_f^2  \int \frac{d^4K}{(2\pi)^4} {\Tr }\left [\left(-ie\gamma_\mu\right ) iS_0(K) \left(-ie\gamma_\nu \right ) iS_0(Q) \right ] 
= -\frac{5}{3} e^2 T \sum_{\{k_0 \}} \int  \frac{d^3k}{2\pi)^3} 
{\Tr }\left [\gamma_\mu S_0(K) \gamma_\nu S_0(Q) \right ] \, ,\label{born1}
\ee
where $Q=P-K$ and $S_0$ is free quark propagator given in ~\eqref{gse25}. The self-energy is summed up for two massless quark flavours $u$ and $d$.  Now, using ~\eqref{gse25} in~\eqref{born1} and then performing the Dirac traces, one gets
\bea
\Pi^{\mu}_{\mu}(\om,p) &=& \frac{10}{3} e^2 T \sum_{\{k_0\}} \int  \frac{d^3k}{(2\pi)^3} \left \{ \left (1-\bm{\hat k} \cdot \bm{\hat q}\right) 
\left [ \frac{1}{d_+(\om_1, k)d_+(\om_2,q)} + \frac{1}{d_-(\om_1,k)d_-(\om_2,q)}\right]\right.\nn
&&\hspace{2cm}+\, \left.\left (1+\bm{\hat k} \cdot \bm{\hat q}\right) 
\left [ \frac{1}{d_+(\om_1,k)d_-(\om_2,q)} + \frac{1}{d_-(\om_1,k)d_+(\om_2,q)}\right]\right \} \, . \label{born3}
\eea
where $d_\pm$ are given in \eqref{gse19a}.
Using Braaten-Pisarski-Yuan (BPY)  prescription~\cite{Braaten:1990wp}  derived in \eqref{bpy11}, one can find the imaginary part of \eqref{born3} as
\bea
&&\textrm{Im} \Pi^{\mu}_{\mu}(\om,p) = \frac{10}{3} \pi e^2 \left (1-e^{\beta\om}\right ) \int  \frac{d^3k}{(2\pi)^3} 
\int_{-\infty}^{+\infty} d\omega_1 \int_{-\infty}^{+\infty} d\omega_2\, 
n_F(\omega_1) n_F(\omega_2) \left(\omega-\omega_1-\omega_2 \right )   \bigg\{  \left (1-\bm{\hat k} \cdot \bm{\hat q}\right) \nn
&&\times \Big[ \rho^f_+(\om_1,k)\rho^f_+(\om_2,q)+ \rho^f_-(\om_1,k)\rho^f_-(\om_2,q)\Big] 
+  \left (1+\bm{\hat k} \cdot \bm{\hat q}\right)\Big[ \rho^f_+(\om_1,k)\rho^f_-(\om_2,q)+ \rho^f_-(\om_1,k)\rho^f_+(\om_2,q)\Big] \bigg \} \, ,\label{born5}
\eea
where $\beta=1/T$,  $\om $ is the energy of the external boson and 
where the free quark spectral functions $ \rho^f_\pm(x,y)$ are obtained in  \eqref{spec1}.
 One can perform $\om_1$ and $\om_2$ integration in Eq.~\eqref{born5} using delta functions of $ \rho^f_\pm(x,y)$. Then we note that the energy conservation allows only this $\delta(\om-k-q)$ and other three are prohibited. We know that  $x=\cos\theta$, $\theta$ is $\angle \bm {\vec p, \vec q}$. Using $\bm{\hat q\cdot \hat k}=(px-k)/q$ one can write Eq.~\eqref{born5} as
\bea
\textrm{Im} \Pi^{\mu}_{\mu}(\om,p) &=& 
 -\frac{5}{6\pi}  e^2 \left (e^{\beta\om}- 1\right ) \int_0^\infty k^2dk \int_{-1}^{1} dx  \left (1- \frac{px-k}{q} \right)n_F(k) n_F(q) 
\delta(\om-k-q) \, . \label{born9}
\eea
We note here that the $ \delta(\om-k-q)$ in Eq.~\eqref{born9} corresponds to a process of quark and antiquark annihilation to a virtual photon which decays to a lepton pair ($q\bar q\rightarrow \gamma^*\rightarrow l^+l^-$). Now, we make a variable change from $x\rightarrow q$ with limits  $x=-1 \,\, \Rightarrow \,\,\, q_{\text{max}}=p+k $
\be
x&=+1 \,\,\,\, \Rightarrow  \,\,\,\, q_{\text{min} }= \left( p^2+k^2-2pk\right)^{\frac{1}{2}} =
\left \{ \begin{array}{ll}
                  p-k & \mbox{ if} \,\,\,\,  k<p \nn
                  k-p &\mbox{ if} \,\,\,\,  p <k   \nn
		   \end{array}   
		   \right. 
\ee
and from the zeros of the argument of $\delta(\om-k-q)$ one gets $ (\om-p)/2<k<(\om+p)/2$. Now,  first performing $q$ integration using $\delta$-function and then $k$ integration, one can simplify the expression of $\textrm{Im} \Pi^{\mu}_{\mu}(\om,p)$ and gets the dilepton production rate~\cite{Cleymans:1986na} from Eq.~\eqref{d6} for massless lepton as 
\bea
\frac{dR}{d^4xd^4P} = \frac{dN}{d^4P} &=&
\frac{\alpha} {12\pi^4}\frac{n_B(\om)}{M^2}   \frac{10\alpha}{3}  T \frac{M^2}{p}  \ln\left[\frac{\cosh\frac{\om+p}{4T}}{\cosh\frac{\om-p}{4T}}\right] 
= \frac{5\alpha^2}{18\pi^4} n_B(\om) \frac{T}{p} \ln\left[\frac{\cosh\frac{\om+p}{4T}}{\cosh\frac{\om-p}{4T}}\right] \, . \label{born12}
\eea 
Now, we would like to compute the dilepton production rate for photon momentum ${\bm{\vec p}}=0$. This leads to ${\bm{\vec q}}=-\bm{\vec k}$, ${\bm{\hat q}}=-\bm{\hat k}$ and
$ q=|-\bm{\vec k}|=k$ and
the dileption rate~\cite{Cleymans:1986na} for zero photon momentum becomes
\bea
\left.\frac{dR}{d^4xd^4P} \right |_{\bm{\vec p}=0}&=&
\frac{\alpha} {12\pi^4}\frac{n_B(\om)}{\om^2}   \frac{5\alpha}{3}  \left (e^{\beta\om}-1\right ) \om^2 n^2_F\left(\frac{\om}{2}\right)\,  
= \frac{5\alpha^2}{36\pi^4}  n^2_F\left(\frac{\om}{2}\right)\, . \label{born14}
\eea 
Now the Born-term for $\bm {\vec p}=0$ and $\om\ll T$ is simply obtained from~\eqref{born14} and given~\cite{Greiner:2010zg} by 
\begin{equation}
	\left. \frac{dR}{d^4xd^4P}\right |_{\bm{\vec p}=0}=\frac{5\alpha^2}{144\pi^4}=1.90\times 10^{-8} \ .
	\label{soft18}
\end{equation}
We note  that the production rate for soft dileptons could differ by orders of magnitude~\cite{Kapusta:1984rz} from the naive prediction from quark-antiquark annihilation. Here one needs to apply the HTL resummation techniques~\cite{Braaten:1989mz,Braaten:1989kk,Pisarski:1988vd} to obtain a systematic calculation of the production rate. 
\subsubsection{Soft dilepton production rate in 1-loop HTL approximation}
\label{bpy_soft}
\vspace{-0.cm}
As discussed above when the photon energy $\om\sim gT$ is soft, the one-loop calculation is not complete. There are hard-thermal-loop corrections that are as large as the results obtain in \eqref{born12} and \eqref{born14}. These are resummed into a diagrammatic expansion using the HTL approximation methods developed in Ref.~\cite{Braaten:1989mz,Braaten:1989kk,Pisarski:1988vd} and also discussed in subsec~\ref{htlr}. 
\begin{wrapfigure}{r}{.5\textwidth}
	\begin{center}
		\vspace{-0.6cm}
		\includegraphics[scale=0.4]{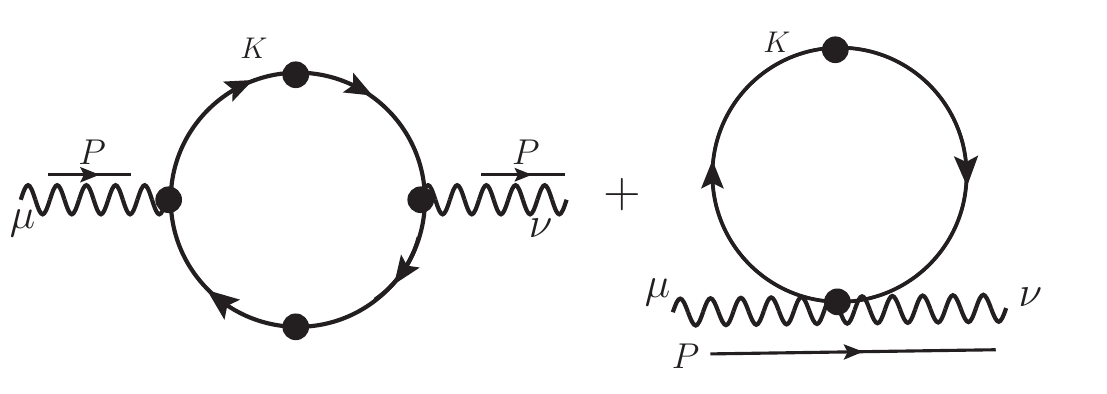}
	\end{center}
	\vspace{-0.8cm}
	\caption{The diagram in (a) corresponds to one-loop photon self-energy and (b) corresponds to photon tadpole diagram for soft dilepton production in HTL approximation.}
	\label{soft_dilep}
\end{wrapfigure}
For lines carrying soft momentum, bare propagators are replaced by effective propagators, which resum the self-energy corrections from hard thermal loops. If all external lines entering a vertex are soft, an effective vertex, which includes a vertex correction from a hard thermal loop, is required as shown in Fig.~\ref{soft_dilep}. The soft dilepton rate has been calculated in HTL approximation following Fig.~\ref{soft_dilep} in Ref.~\cite{Braaten:1990wp}.

\begin{wrapfigure}[14]{r}{0.45\textwidth}
   	\begin{center}
   		\includegraphics[width=7cm, height=6cm]{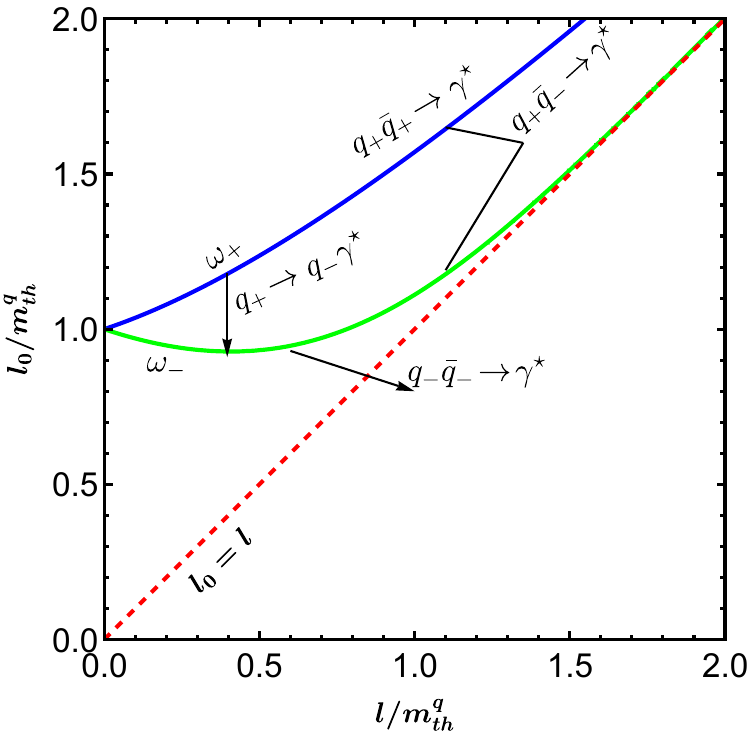}
   	\end{center}
   	\vspace{-0.6cm}
   	\caption{Displays quasi-quark dispersion relation in HTL approximation along with various physical processes arising from it which are discussed
   		in \eqref{soft11a} to \eqref{soft11f}.}
   	\label{disp_quark_htl}
   \end{wrapfigure}

The photon self-energy in 1-loop HTL approximation in Fig.~\ref{soft_dilep} can be written as
\bea
\Pi_{\mn}(P) &=& \Pi_{\mn}^{\textrm{a}}(P) +  \Pi_{\mn}^{\textrm{b}}(P)\, , \label{soft1}
\eea
The contribution of the second term  $\Pi_{\mn}^{\textrm{b}}(P) $ in Eq.~\eqref{soft1}vanishes  as $\Gamma^{\mu\mu}=0$ but it is required to verify the Ward identity $P^\mu \Pi_{\mn}=0$. Then one can write the Eq.~\eqref{soft1} as 
\bea
\Pi_{\mn}(P) &=&  -\frac{5}{3} e^2 T\sum_{\{k_0\}}\int \frac{d^3k}{(2\pi)^3}
 {\Tr} \Big[S^{\textrm{q}}(K) \Gamma_\mu(K-P,-K;P)S^{\textrm{q}}(Q) \Gamma_\nu(K,P-K;-P)\Big] \, , \label{soft1a}
\eea
where it is summed up for two massless quark flavours ($u$ and $d$).
Now, the effective quark propagator is given in Eq.~\eqref{qcd34}  and the effective quark-photon vertex is given in Eq.~\eqref{vert_3pt29} where $m_{\rm{th}}^{\rm e}$ is to be replaced by $m_{\rm{th}}^{\rm q}$.
%
As discussed in subsec~\ref{q_eff_prop} the poles of the effective quark propagator,  ${\cal D}^{\textrm q}_\pm(l_0,l)$ defines dispersion property of a quark in the thermal bath.
${\cal D}^{\textrm q}_+(l_0,l)=0$ has two poles at $l_0=\omega_+(l)$ and $l_0=-\omega_-(l)$ whereas 
 ${\cal D}^{\textrm q}_-(l_0,l)=0$ has two poles at $l_0=\omega_-(l)$ and $l_0=-\omega_+(l)$.  
  Only the positive energy solutions, $l_0>0$ are displayed in Fig.~\ref{disp_quark_htl}.   A mode with energy $\omega_+$  represents the in-medium propagation  of a normal quark excitation with a thermal mass. We denote this quasiparticle as $q_+$.  This is a Dirac spinors and eigenstate of $(\gamma_0 - \vec{\bm \gamma}\cdot \bm{\hat { l}})$ with chirality to helicity ratio $+1$.
   On the other hand  there is a new long wavelength mode known as {\em plasmino} with energy $\omega_-$  with no analog at zero temperature. We denote this quasiparticle by $q_-$ and  it is a eigenstate of $(\gamma_0 + \vec{\bm \gamma}\cdot \bm{\hat { l}})$ with chirality to helicity ratio $-1$.  Note that both $q_+$ and $q_-$ modes have same mass: $\om_+(0)=\om_-(0) = m^{\textrm q}_{\textrm{th}}$. The $\om_+$ branch increases monotonically where as the $\omega_-$ branch has a minimum at low momentum. Both modes approach free dispersion curve (the light cone) from above  at large momentum.  In addition, due to space like momentum,  $l_0^2<l^2$. ${\cal D}^{\textrm q}_\pm(l_0,l)$ contains a discontinuous
part corresponding to {\em Landau Damping} (LD) due to the presence of Logarithmic term in the quark propagator in Eq.~\ref{qcd34}. In-medium spectral function corresponding to the effective fermion propagator in \eqref{qcd34} will have both pole and cut contribution. It has  been obtained in Eq.~\eqref{spec2} in the subsec~\ref{spec_prop} in which $m_{\rm{th}}^{\rm e}$ is to be replaced by $m_{\rm{th}}^{\rm q}$.

The effective 3-point function $\Gamma_\mu$ is the sum of bare vertex $\gamma_\mu$ and one-loop correction from HTL $\delta\Gamma_\mu$ and given in \eqref{vert_3pt29}. For photon momentum $\bm{\vec p}=0$, it reduces to
\begin{subequations}
\begin{align}
\Gamma_0(K-P,K;P) &= \left[1-\frac{(m^{\textrm q}_{\textrm{th}})^2}{p_0k} \delta Q_0\right]\gamma_0 + \frac{(m^{\textrm q}_{\textrm{th}})^2}{p_0k} 
\delta Q_1 \bm{\hat k} \cdot \boldsymbol \vec{\bm \gamma} , \label{soft3c} \\
\Gamma_i(K-P,K;P) &=  \left[1-\frac{(m^{\textrm q}_{\textrm{th}})^2}{p_0k} \left(\delta Q_0-\frac{1}{3} \delta Q_2\right)\right]\gamma_i +  \frac{(m^{\textrm q}_{\textrm{th}})^2}{p_0k} \delta Q_1 {\hat k}_i  \gamma_0 
-\frac{(m^{\textrm q}_{\textrm{th}})^2}{p_0k} {\hat k}_i \bm{\hat k} \cdot \boldsymbol \vec{\bm \gamma} , \label{soft3d}  
\end{align}
\end{subequations}
where
$\delta Q_n= Q_n\left(\frac{k_0}{k}\right)-Q_n\left(\frac{k_0-p_0}{k}\right)$\, ,
$Q_0\left(x\right) = \frac{1}{2} \ln \frac{x+1}{x-1} \,$ , 
$Q_1\left(x\right)= \left[x Q_0\left(x\right)-1 \right]\,$ ,  and 
$Q_2(x) = \frac{1}{2}\left[3xQ_1(x)-Q_0(x)\right].$ 
Now, using Eq.~\eqref{soft1a}, \eqref{soft3c} and \eqref{soft3d} in Eq.~\eqref{soft1a}, one can perform traces over Dirac indices. Then using \eqref{bpy11}, one obtains the imaginary parts of the photon self-energy at $\bm{\vec p}=0$. Then using the imaginary part
we can write the dilepton production rate from~\eqref{d6}  for massless leptons~\cite{Braaten:1990wp} at $\bm{\vec p}=0$ as
\bea
\left. \frac{dN}{d^4Xd^4P} \right |_ {\bm{\vec p}=0}&=& \frac{10\alpha^2}{9\pi^4} \frac{1}{\om^2} \int^\infty_0 k^2 dk  \int^\infty_{-\infty} d\om_1 
 \int^\infty_{-\infty}  d\om_2 n_F(\om_1) n_F(\om_2) \delta\left(\om-\om_1-\om_2\right) \nn
& &\hspace{-1.0cm}\times \Bigg\{4\left(1-\frac{\om_1^2-\om_2^2}{2k\om}\right)^2\rho_+(\om_1,k)\rho_-(\om_2,k) + \left(1+\frac{\om_1^2+\om_2^2-2k^2-2(m^{\textrm{q}}_{\textrm{th}})^2}{2k\om}\right)^2\rho_+(\om_1,k)\rho_+(\om_2,k) \nn
&+&  \left(1-\frac{\om_1^2+\om_2^2-2k^2-2(m^{\textrm{q}}_{\textrm{th}})^2}{2k\om}\right)^2 \rho_-(\om_1,k)\rho_-(\om_2,k)+\Theta(k^2-\om_1^2) \frac{(m^{\textrm{q}}_{\textrm{th}})^2}{4k\om^2} \left(1-\frac{\om_1^2}{k^2} \right) \nn
 &&\hspace{-1.5cm}\times \left[\left(1+\frac{\om_1}{k}\right)\rho_+(\om_2,k) +\left(1-\frac{\om_1}{k}\right)\rho_-(\om_2,k) \right] \Bigg\}
  =\left. \frac{dR}{d^4Xd^4P} \right |_ {\bm{\vec p}=0}^{\textrm{PP}} + \left. \frac{dR}{d^4xd^4P} \right |_ {\bm{\vec p}=0}^{\textrm{PC}}+ \left. \frac{dR}{d^4xd^4P} \right |_ {\bm{\vec p}=0}^{\textrm{CC}} \! . \label{soft10}
\eea
We note that the quark spectral functions $\rho_\pm$ have both pole and cut contribution. So the terms with two powers of $\rho_\pm$ will lead to three types of contributions: pole-pole (PP),
pole-cut (PC) and cut-cut (CC) as written in last line of \eqref{soft10}.
Now we note that the pole-pole part of the dilepton rate will have eight terms but only 4 terms will be allowed by the energy conservation and they correspond to various process
of dilepton producrion which are indicated in Fig.~\ref{disp_quark_htl}:
\begin{subequations}
\begin{align}
\delta\left(\om-\om_+-\om_-\right) &\Rightarrow  \om=\om_++\om_- \Rightarrow q_+{\bar q_-} \rightarrow\gamma^*\rightarrow l^+l^-: \, \,  {\textrm{annihilation of $q_+$ and 
$\bar q_-$ modes}} \label{soft11a}\\
\delta(\om-2\om_+) &\Rightarrow \om-2\om_+ =0 \Rightarrow q_+{\bar q_+}\rightarrow \gamma* \rightarrow  l^+l^- : \, \, {\textrm{annihilation of $q_+$ and 
$\bar q_+$ modes}}. \label{soft11c}\\
\delta(\om+\om_--\om_+)&\Rightarrow \om=\om_+-\om_- \Rightarrow q_+\rightarrow q_- \gamma^*\rightarrow q_-l^+l^-: \, \, {\textrm{$q_+$ mode  makes a transition 
to $q_-$ with a $\gamma^*$}}. \label{soft11d}\\
\delta(\om-2\om_-) &\Rightarrow \om-2\om_- =0 \Rightarrow q_-{\bar q_-}\rightarrow \gamma* \rightarrow  l^+l^- : \, \, {\textrm{annihilation of $q_-$ and 
$\bar q_-$ modes}}. \label{soft11f}
\end{align}
\end{subequations}
 Now the pole-pole contribution in~\eqref{soft10} becomes
\bea
 \left. \frac{dR}{d^4xd^4P} \right |_ {\bm{\vec p}=0}^{\textrm{PP}} &=&  \frac{10\alpha^2}{9\pi^4} \frac{1}{\om^2}  
 \left[ k_1^2  n_F(\om_+(k_1)) n_F(\om_-(k_1)) \left(1-\frac{\om_+^2(k_1)-\om_-^2(k_1)}{2k_1\om}\right)^2 \frac{(\om_+^2(k_1)-(k_1)^2)(\om_-^2(k_1)-(k_1)^2)} {(m^{\textrm{q}}_{\textrm{th}})^4} \right.\nn
&&\hspace{-1cm}\left. \times \left| -\frac{d}{dk}\left[\om_+(k)+\om_-(k)\right]\right |_{k=k_1^i}^{-1} +  \sum^2_{i=1} (k_2^i)^2\left[1-n_F(\om_-(k_2^i))\right]n_F(\om_+(k_2^i))
 \frac{(\om^2_+(k_2^i)-(k_2^i)^2)(\om_-^2(k_2^i)-k_s^2)}{2(m^{\textrm{q}}_{\textrm{th}})^4} \right. \nn
&&\left. \times \left(1+\frac{\om_-^2(k_2^i)+\om_+^2(k_2^i)-2(k_2^i)^2-2(m^{\textrm{q}}_{\textrm{th}})^2}{2k_2^i\om}\right)^2 
 \left| -\frac{d}{dk}\left[\om_+(k)-\om_-(k)\right]\right |_{k=k_2^i}^{-1}\right. \nn
&&\left. + k_3^2 n_F^2(\om_+(k_3)) \left(1+\frac{\om_+^2(k_3)-k_3^2-(m^{\textrm{q}}_{\textrm{th}})^2}{k_3\om}\right)^2  \frac{(\om^2_+(k_3)-k_3^2)^2}{4(m^{\textrm{q}}_{\textrm{th}})^4}
\left| -2\frac{d}{dk}\left[\om_+(k)\right]\right |_{k=k_3}^{-1}\right.  \nn
&&\hspace{-1cm}\left. + \sum^2_{i=1} (k_4^i)^2 n_F^2(\om_-(k_4^i)) \left(1-\frac{\om_-^2(k_4^i)-(k_4^i)^2-(m^{\textrm{q}}_{\textrm{th}})^2}{k_4^i\om}\right)^2  \frac{(\om^2_-(k_4^i)-(k_4^i)^2)^2}{4(m^{\textrm{q}}_{\textrm{th}})^4}
\left| -2\frac{d}{dk}\left[\om_-(k)\right]\right |_{k=k_4^i}^{-1}\right] \, . \label{soft13}
\eea
We note that
 $k_1$ is the solutions of $\om-\om_+(k)-\om_-(k)=0$, 
$k_2^i$ is the solutions of $\om-\om_+(k)+\om_-(k)=0$,
$k_3$ is the solutions of $\om-2\om_+(k)=0$,
$k_4^i$ is the solutions of $\om-2\om_-(k)=0$ and
$d\left[\om_\pm(k)\right]/dk$ is obtained from dispersion relation ${\cal D}_\pm^{\textrm{q}}(k_0,k)=0$. 
Now the contribution of the pole-cut to the dilepton rate can be written as
\bea
 \left. \frac{dR}{d^4xd^4P} \right |_ {\bm{\vec p}=0}^{\textrm{PC}} &=&  \frac{10\alpha^2}{9\pi^4} \frac{1}{\om^2}  
 \int^\infty_0 k^2 dk  \Bigg\{\frac{(\om_+^2-k^2)} {2(m^{\textrm{q}}_{\textrm{th}})^2} n_F(\om_+)n_F(\om-\om_+)    \Theta\left(k^2-(\om-\om_+)^2\right) \nn
 &&\hspace{-1.5cm} \times \left[4  \left(1-\frac{\om_+^2-(\om-\om_+)^2}{2k\om}\right)^2 \beta_-(\om-\om_+, k) +2  \left(1+\frac{(\om-\om_+)^2+\om_+^2-2k^2-2(m^{\textrm{q}}_{\textrm{th}})^2}{2k\om}\right)^2 \beta_+(\om-\om_+, k) \right.  \nn
 &&+\left.\frac{(m^{\textrm{q}}_{\textrm{th}})^2}{4k\om^2} \left(1-\frac{(\om-\om_+)^2}{k^2}\right ) \left(1+\frac{\om-\om_+}{k}\right )\right]
 + \frac{(\om_-^2-k^2)} {2(m^{\textrm{q}}_{\textrm{th}})^2} n_F(\om_-)n_F(\om-\om_-)  \Theta\left(k^2-(\om-\om_-)^2\right) \nn
  &&\hspace{-1.5cm} \times \left[4  \left(1-\frac{(\om-\om_-)^2-\om_-^2}{2k\om}\right)^2 \beta_+(\om-\om_-, k) +2  \left(1-\frac{(\om-\om_-)^2+\om_-^2-2k^2-2(m^{\textrm{q}}_{\textrm{th}})^2}{2k\om}\right)^2 \beta_-(\om-\om_-, k) \right. \nn
  &&\hspace{6.6cm} + \left.\frac{(m^{\textrm{q}}_{\textrm{th}})^2}{4k\om^2} \left(1-\frac{(\om-\om_-)^2}{k^2}\right ) \left(1-\frac{\om-\om_-}{k}\right)\right] \Bigg\} \, ,\label{soft15}
 \eea
where we kept only those terms allowed by $\Theta$-functions.

Now the contribution of the cut-cut to the dilepton rate can be written as
\bea
 \left. \frac{dR}{d^4xd^4P} \right |_ {\bm{\vec p}=0}^{\textrm{CC}} &=&  \frac{10\alpha^2}{9\pi^4} \frac{1}{\om^2}  
 \int^\infty_0 k^2 dk \int^k_{-k}  d\om_1  n_F(\om_1) n_F(\om-\om_1) \Theta\left(k^2-(\om-\om_1)^2\right) 
 \Bigg[ 4  \left(1-\frac{\om_1^2-(\om-\om_1)^2}{2k\om}\right)^2 \nn
 &&\hspace{-0.5cm}\left. \times \beta_+(\om_1, k)  \beta_-(\om-\om_1, k) +
 \left(1+\frac{\om_1^2+(\om-\om_1)^2-2k^2-2(m^{\textrm{q}}_{\textrm{th}})^2}{2k\om}\right)^2  \beta_+(\om_1, k)  \beta_+(\om-\om_1, k)\right.\nn
 &&\left. +\left(1-\frac{(\om-\om_1)^2+\om_1^2-2k^2-2(m^{\textrm{q}}_{\textrm{th}})^2}{2k\om}\right)^2  \beta_-(\om_1, k)  \beta_-(\om-\om_1, k) 
 +\frac{(m^{\textrm{q}}_{\textrm{th}})^2}{4k\om^2} \left(1-\frac{\om_1^2}{k^2}\right )  \right.\nn
 &&\times \left\{ \left(1+\frac{\om_1}{k}\right ) \beta_+(\om-\om_1, k) +\left(1-\frac{\om_1}{k}\right ) \beta_-(\om-\om_1, k)  \right\}\Bigg] \, .\label{soft16}
 \eea
 \begin{figure}[htb]
\begin{center}
\includegraphics[width=8.5cm,height=5.5cm]{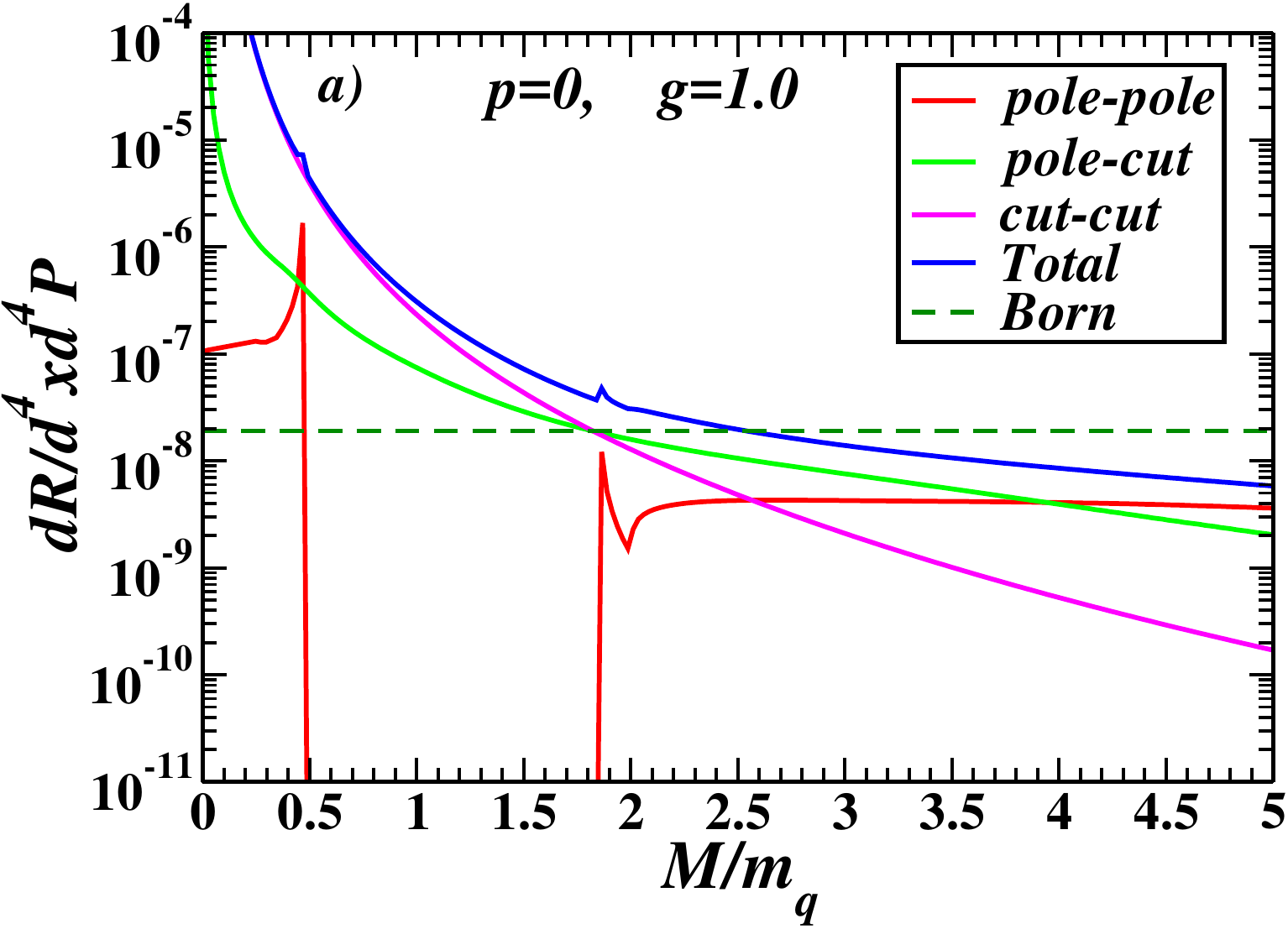}
\includegraphics[width=8.5cm,height=5.5cm]{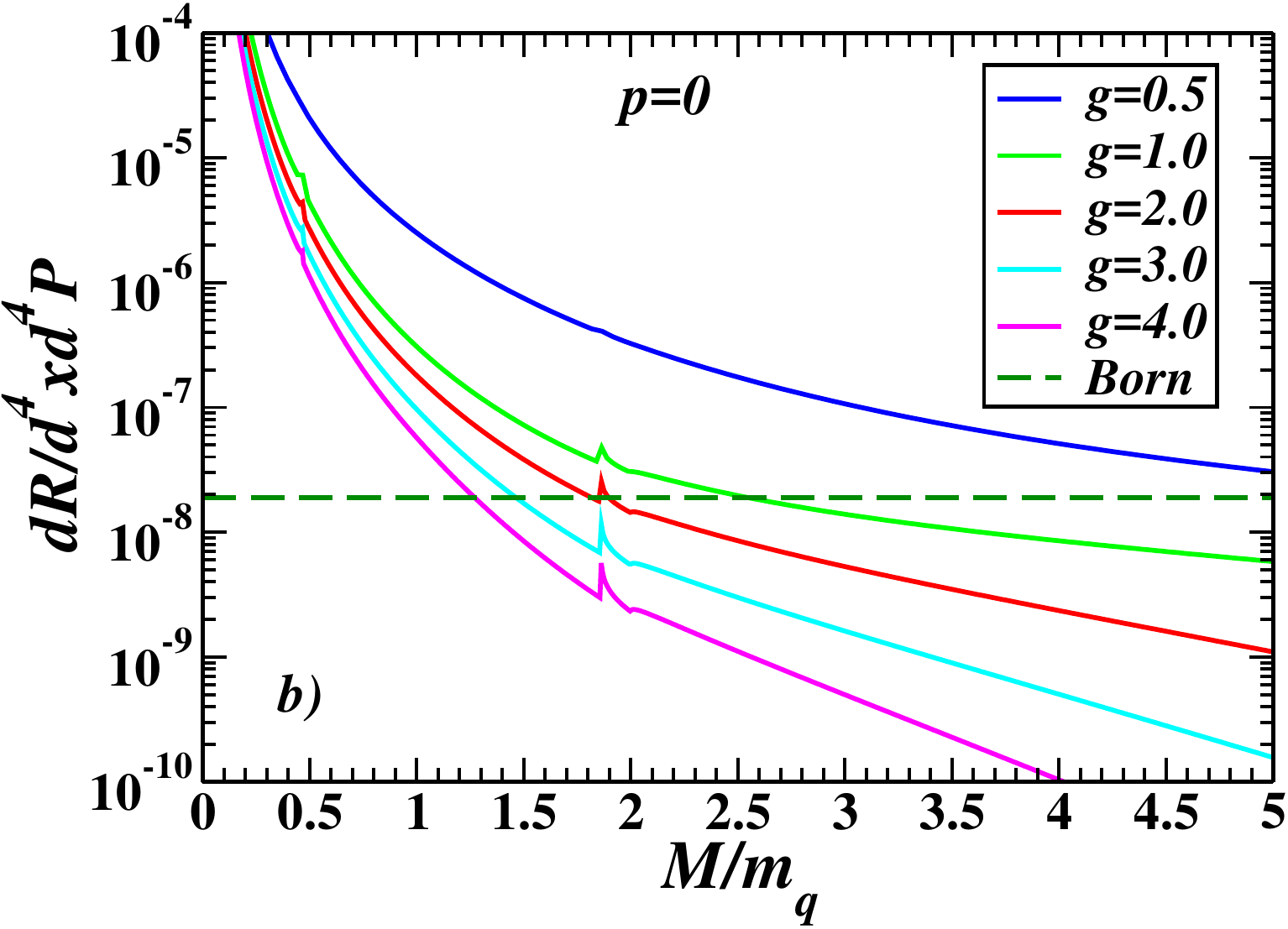}
\end{center}
\vspace{-0.5cm}
\caption{{\textit{Left panel}: (a)} The soft dilepton production rate from 1-loop HTL approximation at photon momentum $\bm{\vec p}=0$ as function of $M/m_{\rm q}=M/m^{\textrm{q}}_{\textrm{th}}$ where $M$ is the invariant mass of the photon, $M=\sqrt{{\om^2 - \bm{\vec p}^2}}=\om$ and the thermal mass of quark, $m_{q}=m^{\textrm{q}}_{\textrm{th}}$.  {\textit{Right panel}: (b)} Total 1-loop rate for various $g$ values. The Born rate  with $\om\ll T$ is a constant as given in Eq.~\eqref{soft18}. These figures are adopted from
Ref.~\cite{Greiner:2010zg}.}
\label{soft_dilep_rate}
\end{figure}
  The contribution of various individual processes from pole-pole part  to the soft dilepton production rate in 1-loop HTL approximation~\cite{Braaten:1990wp} are displayed in the  Fig.~\ref{soft_dilep_rate}$(a)$ with red curve. The transition process in Eq.~\eqref{soft11d}\, $q_+\rightarrow q_-\gamma*$, begins at the energy $M=\omega=0$ and ends up with a van-Hove peak (van Hove singularities)  where all of the transitions from $q_+$ branch are directed towards the minimum of the $q_-$ branch. A van Hove peak~\cite{hove} appears where the density of states diverges due to the vanishing group velocity, $\frac{d\om}{dk}$ at the maximum $M=\om=\left[\om_+(k)-\om_-(k)\right]$. The $q_+\rightarrow q_-\gamma*$ process terminates at van Hove peak after which there is a gap because no other processes are active in this invariant mass regime. Then, the annihilation of the two plasmino modes in \eqref{soft11f}, $q_-{\bar q}_-\rightarrow \gamma^*$, opens up with again a van-Hove peak at $\omega=2 \, \times $ the minimum energy of the plasmino mode ($\om_-$) as 
 $\frac{d\om}{dk}$ vanishes. The contribution of this process decreases exponentially.  At $\omega=2m^{\textrm{q}}_{\textrm{th}}$, the annihilation processes involving usual quark modes in \eqref{soft11c}, $q_+{\bar q}_+\rightarrow \gamma^*$, and that of a quark and a plasmino mode in \eqref{soft11a}, $q_+{\bar q}_-\rightarrow \gamma^*$, begin.  However, the former one ($q_+{\bar q}_+\rightarrow \gamma^*$) grows with the energy and would converge to the usual Born rate (leading order perturbative rate)~\cite{Cleymans:1986na} at high mass whereas the later one ($q_+{\bar q}_-\rightarrow \gamma^*$) initially grows at a very fast rate, but then decreases slowly and finally drops very quickly. The behaviour of the latter process can easily be understood from the dispersion properties of quark and plasmino mode. We also note that such structures were also found in 
nonperturbative dilepton rates~\cite{Mustafa:2022got,Mustafa:1999dt,Mustafa:2002pb,Mustafa:1999cp,Peshier:1999dt,Bandyopadhyay:2015wua}
 and also in mesonic spectral function~\cite{Karsch:2000gi}.

In addition to the pole-pole contributions, there are also contributions from pole-cut (green curve) and cut-cut (purple curve) due to the cuts in the effective quark propagator. 
Because the spectral density along the cut is a smooth function, these contributions do not produce any dramatic structure. At low energies, the pole-cut term grows like $1/\om^2$ and the cut-cut term grows like $1/\om^4$, and they completely overwhelm the structure due to the pole-pole terms.The total rate from pole-pole, pole-cut and cut-cut contributions are shown in blue curve in Fig.~\ref{soft_dilep_rate}$(a)$.  For comparison, the naive prediction in Eq.~\eqref{born14} is shown as a green dashed line in Fig.~\ref{soft_dilep_rate}$(a)$. 

Now one can easily understand that the physical origin of the pole-cut and cut-cut contributions is Landau damping, where a soft quark interacts with hard quarks and gluons. The physical processes which give the pole-cut contributions~\cite{Braaten:1990wp} involve the absorption of a soft quasiparticle $q_\pm$  by a hard particle, 
with a soft virtual photon radiated from the $q_\pm$ or from a hard quark or antiquark. Such processes  can be represented as  $q_\pm G\rightarrow Q\gamma^*$ and $q_\pm {\bar Q}\rightarrow  G\gamma^*$,  where $Q$  is a hard quark and $G$ is a hard gluon. On the other hand,  the cut-cut terms arise from processes~\cite{Braaten:1990wp} in 
which two hard particles scatter by exchange of a soft quark, while radiating a soft virtual photon; for example, $QG \rightarrow GQ\gamma^*$ or $Q{\bar Q}\rightarrow GG\gamma^*$. 
In Fig.~\ref{soft_dilep_rate}$(b)$ we represent the total 1-loop HTL rate for various values of the strong coupling $g$ where the energy gaps are smoothened due to the pole- cut and cut-cut contributions. Also the structures due to the van Hove singularities become also less prominent in the total contributions. For comparison the naive prediction in Eq.~\eqref{born14} is shown as a green dashed line in Fig.~\ref{soft_dilep_rate}$(b)$.  For $\om\sim m^{\textrm{q}}_{\textrm{th}}$, the total rate exceeds the naive prediction by orders of magnitude.

However, these corrections are not sufficient, and two-loop diagrams within the HTL perturbation scheme~\cite{Aurenche:1998nw,Aurenche:1999ec} contribute to the same order and are even larger than the one-loop results~\cite{Braaten:1990wp}. In the following subsec~\ref{2loop_soft}, we discuss two-loop HTLpt dilepton rate.
\subsubsection{ Dilepton production rate in 2-loop HTL approximation}
\begin{figure}[h]
\begin{center}
\includegraphics[scale=0.38]{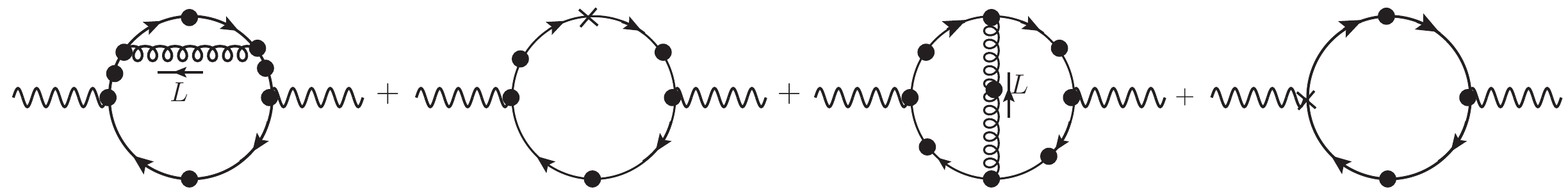}
\end{center}
\vspace{-0.5cm}
\caption{Two-loop contributions to the dilepton rate. A black dot denotes HTL propagator or vertex. Crosses are HTL counter terms.}
\label{two_loop_dilep}
\end{figure}

\vspace{-0.2cm}
We summarize in this subsection the complete results of the calculation of the dilepton rate up to two loops in the HTL perturbative expansion as given in Fig.~\ref{two_loop_dilep}. This rate comprises the one-loop result of Ref.~\cite{Braaten:1990wp} discussed in previous subsec~\ref{bpy_soft}, the two-loop bremsstrahlung contribution derived in Ref.~\cite{Aurenche:1998nw} and finally the contribution of the Compton and annihilation processes  calculated in Ref.~\cite{Aurenche:1999ec}. 
The total one- and two-loop rate at $\bm{\vec p}=0$ and $\om\ll T$ in the leading logarithm, {\em i.e.}, $\ln(1/g)$ approximation reads~\cite{Aurenche:1998nw,Aurenche:1999ec} 
\be
\left. \frac{dR}{d^4xd^4P}\right |_{\bm{\vec p}=0}&=&\frac{5\alpha^2}{9\pi^6}\frac{(m^{\textrm{q}}_{\textrm{th}})^2}{\om^2} \left [\frac{3(m^{\textrm{q}}_{\textrm{th}})^2}{\om^2}\left\{
\frac{\pi^2}{12}\ln \left(\frac{T^2}{(m^{\textrm{q}}_{\textrm{th}})^2}\right)+ 
\ln\left(\frac{T^2}{(m^{\textrm{g}}_{\textrm{th}})^2}\right)\right\}\right.\nn
&&\left.+\frac{\pi^2}{4}\ln \left (\frac{\om T}{\om^2+(m^{\textrm{q}}_{\textrm{th}})^2} \right )  +2\ln \left (\frac{\om T}{\om^2+(m^{\textrm{g}}_{\textrm{th}})^2} \right ) 
\right ]  , \ \ \label{soft17}
\ee
\label{2loop_soft}
\begin{figure}
	\vspace{-0.9cm}
	\begin{center}
		\includegraphics[width=8cm,height=5cm]{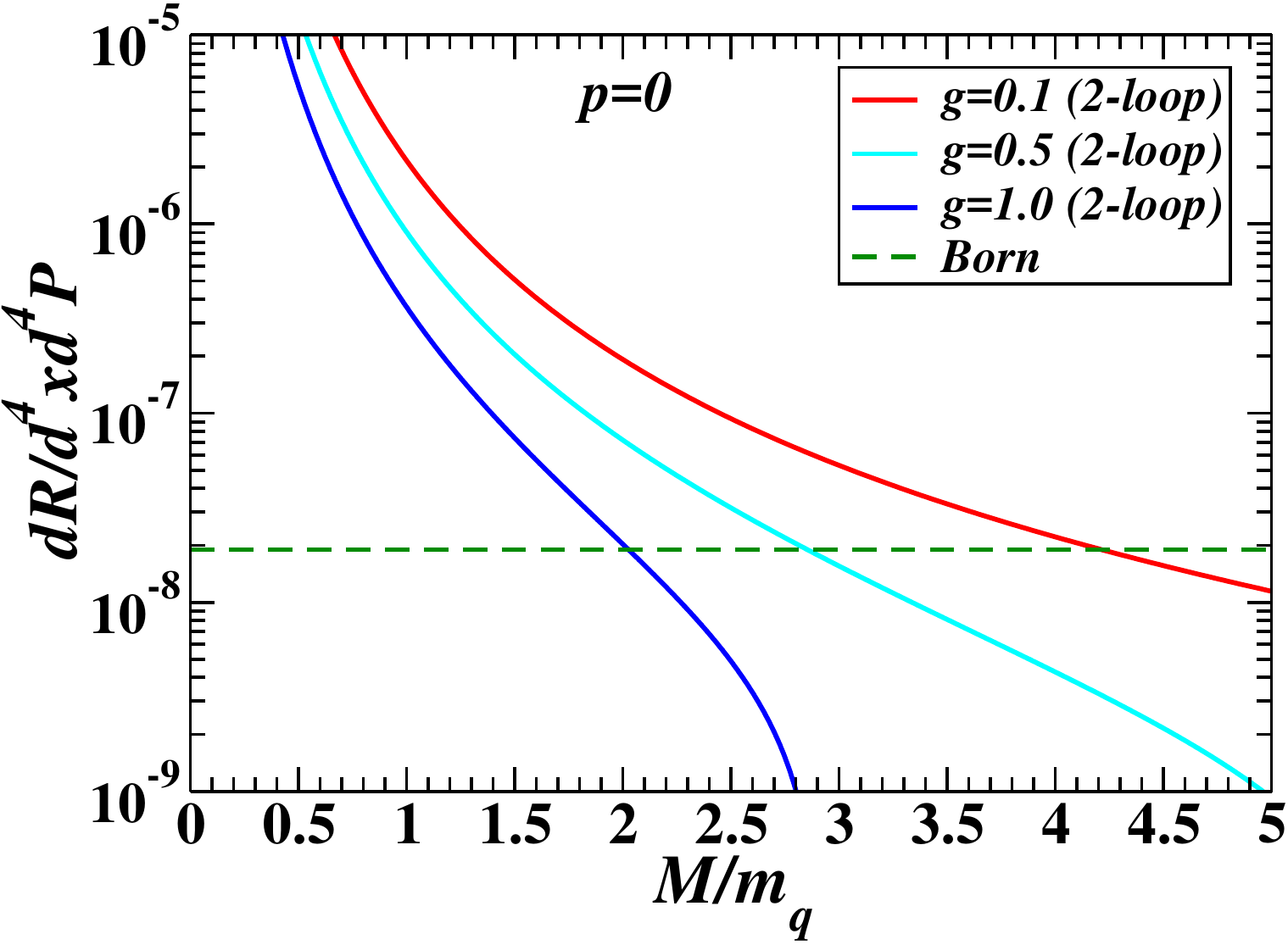}
	\end{center}
	\vspace*{-0.3in}
	\caption{The figure depicts the complete 2-loop dilepton rate for small invariant masses at zero momentum alongside the Born-rate  (shown as a dashed line), plotted against the scaled invariant photon mass $M/m^{\textrm{q}}_{\textrm{th}}$ where the quark thermal mass, $m_{\rm q}=m^{\rm q}_{\rm{th}}$. This figure is taken from Ref.~\cite{Greiner:2010zg}.}
	\vspace*{0.3in}
	\label{fig_2loop}
\end{figure}
where the Debye mass for gluon with two quark flavours is given in Eq.~\eqref{qcd17} as $(m^{\textrm g}_{\textrm{th}})^2=(m^{\textrm{g}}_{{D}})^2/3=8(m^{\textrm{q}}_{\textrm{th}})^2/3$ with  $m^{\textrm{q}}_{\textrm{th}}=gT/\sqrt 6$ is  given in Eq.~\eqref{qcd30}. 
 Now, the first term inside the square brackets comes from the cut-cut contribution
in the one-loop HTL calculation whereas the second term is the bremsstrahlung coming from the two-loop diagrams. The third and fourth terms combine the pole-cut contribution in the 1-loop as well as the $L^2>0$ pieces at 2-loop and associated counter term. 
\vspace{0.0cm}
We also note that the expression in~\eqref{soft17} is of the same order in $g$ as the Born-term for soft $\om\sim gT$.  
In Fig.~\ref{fig_2loop}, we compare the Born-rate with the complete two-loop rate for a range $g$ values. It is clear from Fig.~\ref{fig_2loop} that the $2$-loop rate predominates in the perturbative regime ($g\leq 1$) over the Born-term for low mass domain, $\om=M/m^{\textrm{q}}_{\textrm{th}}\leq 2$. However, the van Hove singularities present in one-loop do not appear as they are washed out due to the leading logarithm approximation within the two-loop HTLpt.
\subsubsection{ $\alpha_s$ correction to the Born rate in 1-loop and 2-loop HTL approximation}
\label{2loop_alpha_s}
For further information, it is also to be noted that the semi hard ($\om=M\sim T$ and $p\gg T$) dilepton rate, i.e., the $\alpha_s$ correction to the Born rate has also been calculated within the HTLpt method~\cite{Thoma:1997dk,Aurenche:2002pc,Aurenche:2002wq,Carrington:2007gt} in 1-loop and 2-loop.  In Ref.~\cite{Thoma:1997dk} the thermal dilepton rate has been calculated from a 1-loop graph, with one hard bare quark propagator, and one HTL quark propagator which is taken to be asymptotically hard. 
\begin{figure}[h]
\centering
\begin{subfigure}[h]{0.46\textwidth}
	\centering
	\includegraphics[width=\textwidth]{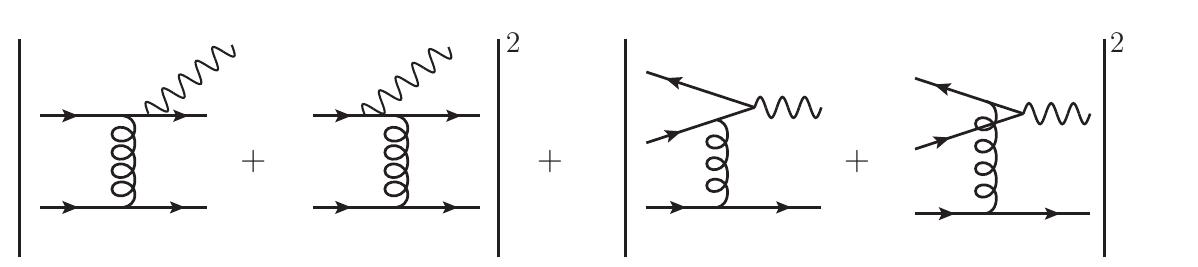}
	\caption{The bremsstrahlung and off-shell annihilation processes from 2-loop diagram}
	\label{off_brems}
\end{subfigure}
\hspace{0.4cm}
\begin{subfigure}[h]{0.46\textwidth}
	\centering
	\includegraphics[width=\textwidth]{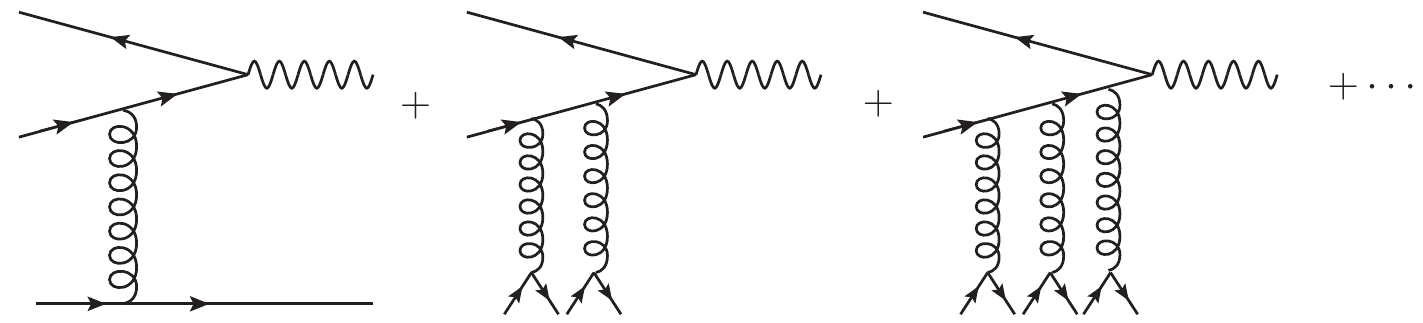}
	\caption{Some multiple rescattering diagrams}
	\label{mult_lpm}
\end{subfigure}
\caption{Various scattering diagrams are shown.}
\end{figure}
\begin{figure}[htb]
\begin{center}
\includegraphics[scale=.4]{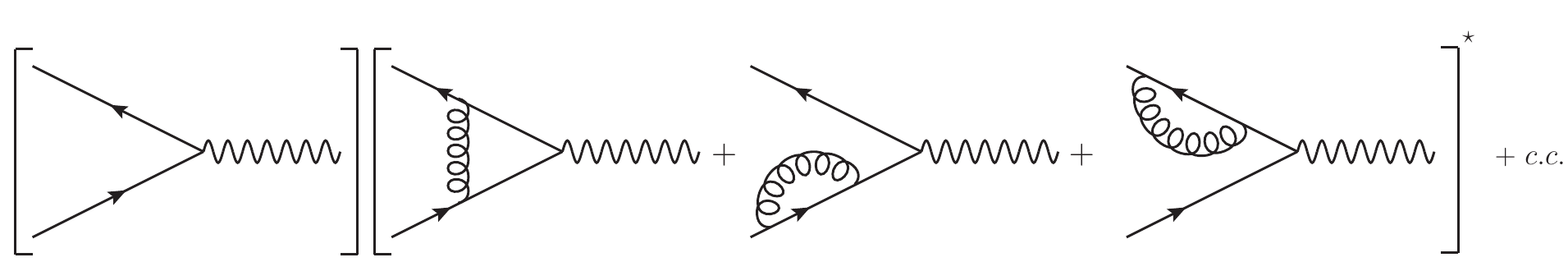}
\end{center}
\vspace*{-0.2in}
\caption{The interference between the Born level amplitude and 1-loop corrections to them.}
\label{inter_born}
\end{figure}
This corresponds to a decay into one quasiquark with thermal mass $m^{\textrm{q}}_{\textrm{th}}$ and one massless quark, or the annihilation of a massive and massless quark pair. Because of using one hard bare propagator gives the wrong threshold in the momentum-integral which are obtained using a theta function. This was corrected
\footnote{where the contribution from $2 \rightarrow 2$ processes to the production of virtual photons of ${\cal O}(\alpha_s)$ have been calculated.} 
in Ref.~\cite{Carrington:2007gt} by including the asymptotic mass  on the hard quark propagator is given as
\be
M^2_\infty =2 \left(m^{\textrm{q}}_{\textrm{th}}\right)^2. \label{asym_mass}
\ee
Since the effective vertex corrections was not considered~\cite{Thoma:1997dk}, the dilepton rate has missed some leading order contributions. However, numerical results showed only a slight modification. 

In Refs.~\cite{Aurenche:2002pc,Aurenche:2002wq} the dilepton rate was calculated from  bremsstrahlung  and the associated off-shell annihilation  processes as shown in Fig.~\ref{off_brems}, including multiple scattering contributions in Fig.~\ref{mult_lpm} and interference terms in Fig.~\ref{inter_born}, all of which are contained in the 
Landau-Pomeranchuk-Migdal (LPM)
\footnote{The emission of photons or gluons from a high energetic charged particle is suppressed by finite formation time: a photon or gluon can not be emitted if there is not enough time for photon or gluon to be on mass shell before  the next scattering takes place. This is called LPM effect~\cite{Landau:1953ivy,Landau:1965ksp,Migdal:1956tc}.}
 resummation~\cite{Landau:1953ivy,Landau:1965ksp,Migdal:1956tc}. 
These appear at the two-loop level but because of a strong collinear singularity in the processes in Fig.~\ref{off_brems} there is collinear  enhancement, 
powers of $T^2/M_\infty^2 \sim 1/g^2$ are generated so that they contribute at the leading order in the strong coupling. Higher order diagrams with a ladder topology involve the same enhancement mechanism and therefore they also contribute at leading order~\cite{Aurenche:2002wq}. The contribution evaluated in Refs.~\cite{Aurenche:2002pc,Aurenche:2002wq} completes the ${\cal O}(\alpha_s)$  of the rate of low-mass lepton pairs
~\footnote{We note that the outcome cannot be expressed in closed analytic form, but requires a numerical solution of an inhomogeneous Schr\"odinger-type equation with a light-cone potential describing interactions.}
produced at large momentum ($\sim \pi T$) in a plasma in equilibrium. This rate should then be added to the contributions of the Born term ($q\bar{q} \rightarrow \gamma^*$) and of the $2 \rightarrow 2$ processes ($q{\bar q}\rightarrow  g\gamma^*$ and $qg \rightarrow q\gamma^*$) which appear at one loop in the perturbative expansion of the HTL effective theory, had already been calculated in Refs.~\cite{Thoma:1997dk,Altherr:1992th}. Production of the pairs in multiple scattering processes are found to be important and even dominate the rate in a wide range of masses. An interesting result is the absence of a threshold in the mass dependence of the virtual photon: the rescattering processes completely wash out the $q{\bar q}$ annihilation threshold naively expected. We further note that the dilepton production rate from a QCD plasma at a temperature above a few hundred MeV are evaluated~\cite{Laine:2013vma} in complete $\alpha_s$ order beyond HTLpt but  including their dependence on a non-zero momentum with respect to the heat bath.  
The main problem in applying perturbative results discussed above to realistic situations is the fact that $g$ is not small but rather we have 
$g\sim 1.5-2.5$. Close to the critical temperature, $T_c$, even $g$ could be as high as $6$~\cite{Peshier:1994zf, Levai:1997yx}. Hence the different momentum  scales are not distinctly separated in the real sense and, even if one still believes in perturbative results (see Figs.~\ref{soft_dilep_rate}, \ref{fig_2loop} and $\alpha_s$ corrected rates~\cite{Thoma:1997dk,Aurenche:2002pc,Aurenche:2002wq,Carrington:2007gt}) at least qualitatively, it is not clear which of the above rates applies to heavy-ion collisions. However, in all cases there are substantial corrections to the Born-rate. The perturbative rates within their uncertainties in various regime probably suggest that the Born-rate may not be sufficient for describing the low mass dilepton spectrum.  
%
\begin{wrapfigure}[12]{r}{0.45\textwidth}
	\begin{center}
		\includegraphics[width=8cm, height=5cm]{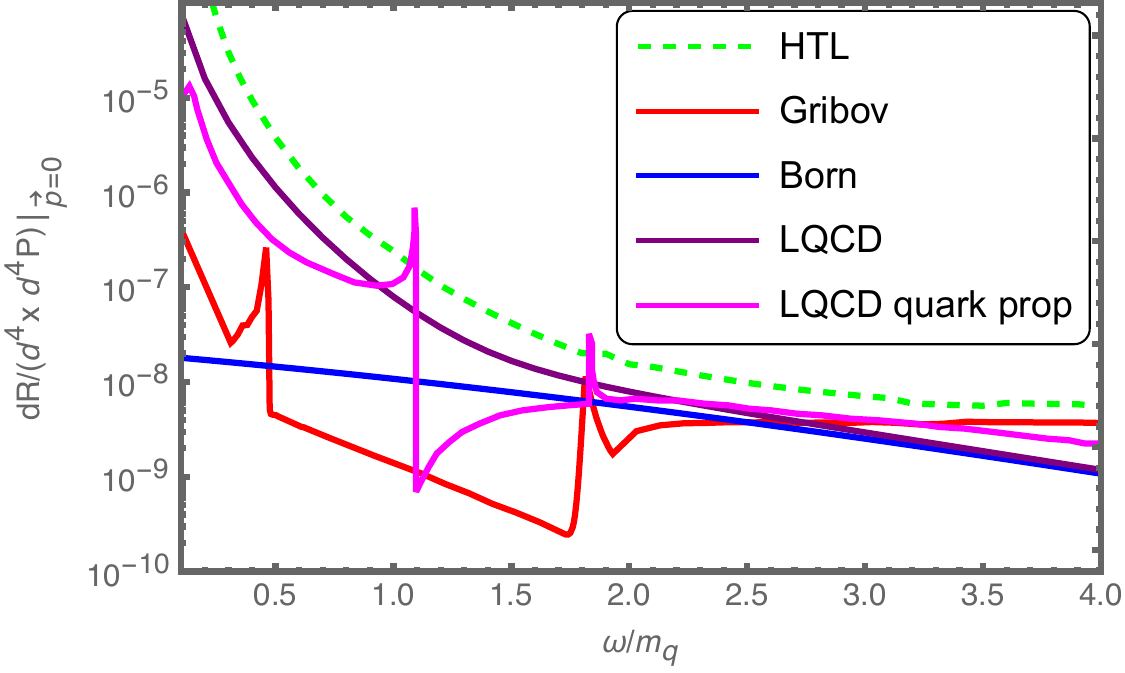}
	\end{center}
	\vspace{-0.6cm}
	\caption{ Comparison of dilepton rates from 1-loop HTL, nonperturbative Gribov action, Born, and LQCD. We note that $m_{\rm q}=m^{\rm q}_{\rm{th}}$}.
	\label{dilep_comp}
\end{wrapfigure}

In Fig.~\ref{dilep_comp} we displayed both perturbative and nonperturbative dilepton rates. The green dashed line represents total  1-loop HTL results.  The blue line corresponds to Born rate. The red line represents the dilepton rate obtained in Refs.~\cite{Mustafa:2022got,Bandyopadhyay:2015wua} using Gribov action~\cite{Gribov:1978npb,Zwanziger:1989mf,Vandersickel:2012tz}.  
We also note that the dilepton rate at $\bm{\vec p}=0$ in quenched lattice QCD (LQCD)~\cite{Kitazawa:2009uw,Kaczmarek:2012mb}  from the vector spectral function, for the first time, was obtained in Refs.~\cite{Ding:2010ga,Karsch:2001uw} represented by brown line in Fig.~\ref{dilep_comp}. It was found that the rate matches with the Born rate at large $\om$  whereas at  low $\om$ it does not show any structures.  This is because the vector spectral function in continuous time is obtained from the correlator in the finite set of discrete Euclidean time using a probabilistic maximum-entropy-method (MEM)~\cite{Asakawa:2000tr,Nakahara:1999vy}  with a somewhat ad hoc continuous ansatz for the spectral function at low  $\om$ and  a free spectral function for large $\om$. Since the MEM analyses are very sensitive to the prior assumption of the structure of the spectral function at small $\om$, the error is expected to be significant at small $\om$. We further  note that, recently, the dilepton rate~\cite{Kim:2015poa} at $\bm{\vec p}=0$, represented by purple line in Fig.~\ref{dilep_comp}, has been obtained using the improved vector spectral function constructed with the two-pole ansatz by analysing the LQCD  propagator in the quenched approximation~\cite{Kitazawa:2009uw,Kaczmarek:2012mb}.  The rate  is found to be smooth below $\omega/m^{\text q}_{\text{th}}<1$. However, interestingly unlike Refs.~\cite{Ding:2010ga,Karsch:2001uw}, it shows some structures in the mass range $1<\omega/m^{\text q}_{\text{th}} < 2 $ similar to that of Ref.~\cite{Mustafa:2022got,Bandyopadhyay:2015wua} in the mass range $0.5 <\omega/m^{\text q}_{\text{th}} < 2 $ with the inclusion of the QCD propagators within the nonperturbative Gribov-Zwanziger action~\cite{Gribov:1978npb,Zwanziger:1989mf,Vandersickel:2012tz}.  Nevertheless, the computation of the dilepton rate in LQCD involves various intricacies and uncertainties. The fundamental difficulties in performing the necessary analytic continuation in LQCD. Until LQCD overcomes the uncertainties and difficulties in the computation of the vector spectral function, one needs to depend, at this juncture, on the prediction of the effective (non)perturbative approaches for the dilepton rate at low mass in particular.

\subsection{Photon Production Rate}
\label{photon_chap}
\vspace{-0.2cm}
The theoretical calculations of the photon emission from a thermal system has a long history, leading to the advancements in quantum physics. In astrophysics, the detection of electromagnetic radiation from stars' hot surfaces and other celestial objects, including the Cosmic Microwave Background (CMB) radiation, provides crucial information like temperature, size, and chemical composition. Deviations from pure black-body spectra are particularly intriguing, offering insights into universe composition, evolution, and structure formation via CMB analysis~\cite{Wright:1992tf}. However, photon emission from the nuclear fireball formed in relativistic heavy-ion collisions differs from macroscopic stellar objects. While photons from stars become thermalised upon leaving the surface, those from a Quark-Gluon Plasma (QGP) follow a different pattern. In a QGP, photons are emitted similarly to any thermal source, originating from quarks with electric charge. Due to energy-momentum conservation, these quarks must interact with QGP's thermal particles to emit photons, making it impossible to observe in an ideal, non-interacting QGP. Nonetheless, interactions (both strong and electromagnetic) are always present within the QGP. Direct photons serve as a promising signature for QGP formation in relativistic heavy-ion collisions~\cite{Shuryak:1978ij,Kajantie:1981wg,Halzen:1981kz,Kajantie:1982nj,Hwa:1985xg,Staadt:1985uc,Neubert:1989hu}, emphasising their production from the QGP.
This is elaborated upon in the subsequent section, along with their significance as electromagnetic probes of the QGP~\cite{Alam:1996fd,Peitzmann:2001mz}. We will now discuss  the formulation real photon emission rates from a thermal medium in following subsec~\ref{pho_th}. 
\vspace{-0.2cm}
\subsubsection{ Photon emission rate in presence of a thermal medium}
\label{pho_th}
This relation which expresses the dilepton multiplicity~\cite{Weldon:1990iw} per unit space-time volume in terms of the spectral function of the photon in the medium is an important result as given in \eqref{d1} as
\bea
\frac{dR}{d^4x}&=& 2\pi e^2 e^{-\beta 
p_0}L_{\mu\nu}\rho^{\mu\nu}\frac{d^3 \bm{\vec q}_1}{(2\pi)^3E_1}\frac{d^3\bm{\vec q}_2}{(2\pi)^3E_2} \nn
&=&-2 e^2 \frac{L_{\mn}}{P^4} \frac{d^3 \bm{\vec q}_1}{(2\pi)^3E_1}\frac{d^3\bm{\vec q}_2}{(2\pi)^3E_2} \frac{1}{e^{\beta p_0}-1}
 \textrm{Im}\Big[\Pi^{\mn}(p_0,\bm{\vec  p})\Big] \, ,\label{ph1}
\eea
where $e$ is the electromagnetic coupling, $\bm{\vec q}_i$ and $E_i$ with $i=1,2$ are three momentum and energy  of the lepton pairs.  
$`\textrm{Im}$' stands for imaginary part,  $\Pi^{\mu\nu}$ is the two point current-current correlation function or the self-energy of photon  and $P \equiv (E=p_0,\bm{\vec p})$ is the four momenta of the photon.

\begin{figure}[h]
\begin{center}
\includegraphics[scale=0.37]{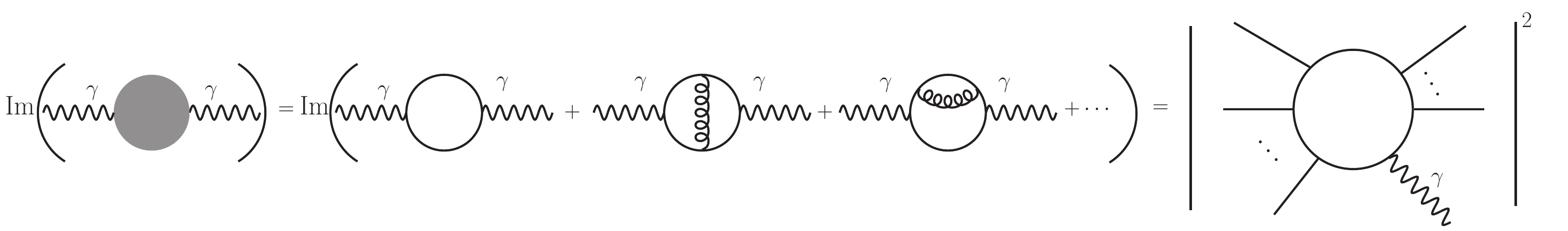}
\end{center}
\vspace{-0.8cm}
\caption{The optical theorem in thermal field theory.}
\label{cutkowski}
\end{figure}

Now, we note that the real photon multiplicity per unit volume can be obtained from Eq.~\eqref{ph1} with following modifications: The factor $\left(e^2 L_{\mn}/{P^4}\right)$ which arises from the lepton spin sum of the square modulus of the product of the electromagnetic vertex $\gamma^* \rightarrow l^+ l^-$, 
the leptonic current $L_{\mn}$ involving Dirac spinors and the square of the photon propagator $1/P^4$. For real photon this factor is to be replaced by the factor $- g_{\mn}$ that 
comes from the photon polarisation sum ${e^2 L_{\mn}}/{P^4}\Longrightarrow \sum_{\textrm{polarisation}} \epsilon_\mu\epsilon_\nu=-g_{\mn}$. The two body phase space factor for the lepton pair ($ l^+ l^-$), $d^3 \bm{\vec q}_1/\left[(2\pi)^3E_1\right]  d^3 \bm{\vec q}_2/\left[(2\pi)^3E_2\right] $, has to be replaced by one body phase space for real photon ($\gamma$), $d^3 \bm{\vec p}/\left[(2\pi)^3E\right]$. With this two replacements,  the real photon multiplicity per unit volume can be obtained from Eq.~\eqref{ph1} as
\bea
\frac{dR}{d^4x}&=& 2 \, g_{\mn} \, \frac{1}{e^{\beta E}-1}\, \textrm{Im}\Big[\Pi^{\mn}(E,\bm{\vec  p})\Big] \frac{d^3 \bm{\vec p}}{(2\pi)^3E}\, . \label{ph2}
\eea
Finally, one can obtain the differential real photon emission rate as
\be
E\frac{dR}{d^4x d^3 \bm{\vec p}} = \frac{2 }{(2\pi)^3} \, n_B(E) \, \textrm{Im}\Big[\Pi^{\mu}_{\mu}(E,\bm{\vec  p})\Big]   . \label{ph3}
\ee
We note that the real photon emission rate is correct up to order ${\cal O}(e^2)$ in electromagnetic interaction since it does not account for the possible reinteractions of the real photon on its way out of the thermal bath. But this is exact, in principle, to all orders in strong interaction. However,  for all practical purposes one is able to evaluate up to a finite order of loop expansion. It is clear from the above that in order to calculate the photon  emission rate from a thermal system one needs to evaluate the imaginary part of the photon self-energy. Now, 
we note that  the Cutkosky thermal cutting rules~\cite{Das:1997gg,Weldon:1983jn,Kobes:1985kc,Kobes:1986za,Gelis:1997zv} presents a systematic method to compute the imaginary part of a Feynman diagram. The Cutkosky rule expresses the imaginary part of the $l$-loop amplitude in terms of physical amplitude of lower order [($l-1)$-loop or lower] 
which is pictorially represented in Fig.~\ref{cutkowski}. When the imaginary part of the self energy is calculated up to and including $l$ order loops, then one obtains the photon emission rate for the reaction $m$ particles $\rightarrow n$ particles $+\gamma$  where $l$ satisfies $m+ n < l+1$. This approach becomes equivalent to the relativistic kinetic theory 
which has been used to the photon emission rates~\cite{Ruuskanen:1992hh,Gutbrod1993ParticlePI,Gale:1987ki} in terms of the equilibration rate of photons in a thermal bath. The imaginary part of the photon self-energy tensor in this case is related to the exclusive kinetic rates of emission and absorption associated with the phase space of each particles including the photon, a Bose-Einstein or Fermi-Dirac distribution for each particle in the initial state, and a Bose-enhancement or Pauli-suppression factor for each particle in the final state, the squared matrix element of the various physical processes in the system and corresponding degeneracy factor for a given physical process.
The differential production rate for a particle $X$ with four momentum $P\equiv \left ( E, \bm{\vec p}\right )$ can be written in kinetic theory as
\bea
\frac{dR}{d^4x d^3\bm{\vec p}} &=& \frac{1}{2E(2\pi)^3} \int \frac{d^3\bm{\vec p}_1}{2E_1(2\pi)^3} n_1(E_1) \cdots  \frac{d^3\bm{\vec p}_m}{2E_m(2\pi)^3}  n_m(E_m)
 \int \frac{d^3{\bm{\vec p}}^{\,\, \prime}_1}{2E^\prime_1(2\pi)^3}  \left (1\pm n^\prime_1(E^\prime_1)\right ) \cdots  \frac{d^3\bm{\vec p}^{\,\,\prime}_n}{2E^\prime_n(2\pi)^3} \nn
 && \times  \left (1\pm n^\prime_n(E^\prime_n)\right )  (2\pi)^4 \delta\left(\sum_{i=1}^{m}P_i-\sum_{f=1}^{n}P^\prime_f-P\right)  \left|{\cal M}\right |^2 \, , \label{ph4}
\eea
where ${\cal M}$ is the matrix element of the basic process of $m$ particles $\rightarrow n$ particles $+X$.  $n_i(E_i)$ and $n^\prime_f(E^\prime_f )$  are  a Bose-Einstein or Fermi-Dirac distribution for initial state particle $i$ and final state particle $f$ but excluding the produced particle $X$, respectively.  $P_i$ is the four momentum of initial particle $i$ whereas
$P^\prime_f$ is that for final state particle $f$ except the produced particle $X$. The factor $ \left (1\pm n^\prime_f(E_f^\prime)\right ) $ is either a Bose-enhancement  with $(+)$ve sign or a Pauli-suppression with $(-)$ve sign  for each particle in the final state,  excluding the produced particle $X$.
\subsubsection{Hard photon production from QCD kinetic theory}
\label{kinetic_theory}
\begin{wrapfigure}[10]{r}{0.43\textwidth}
	\vspace{-1cm}
	\begin{center}
		\includegraphics[scale=0.25]{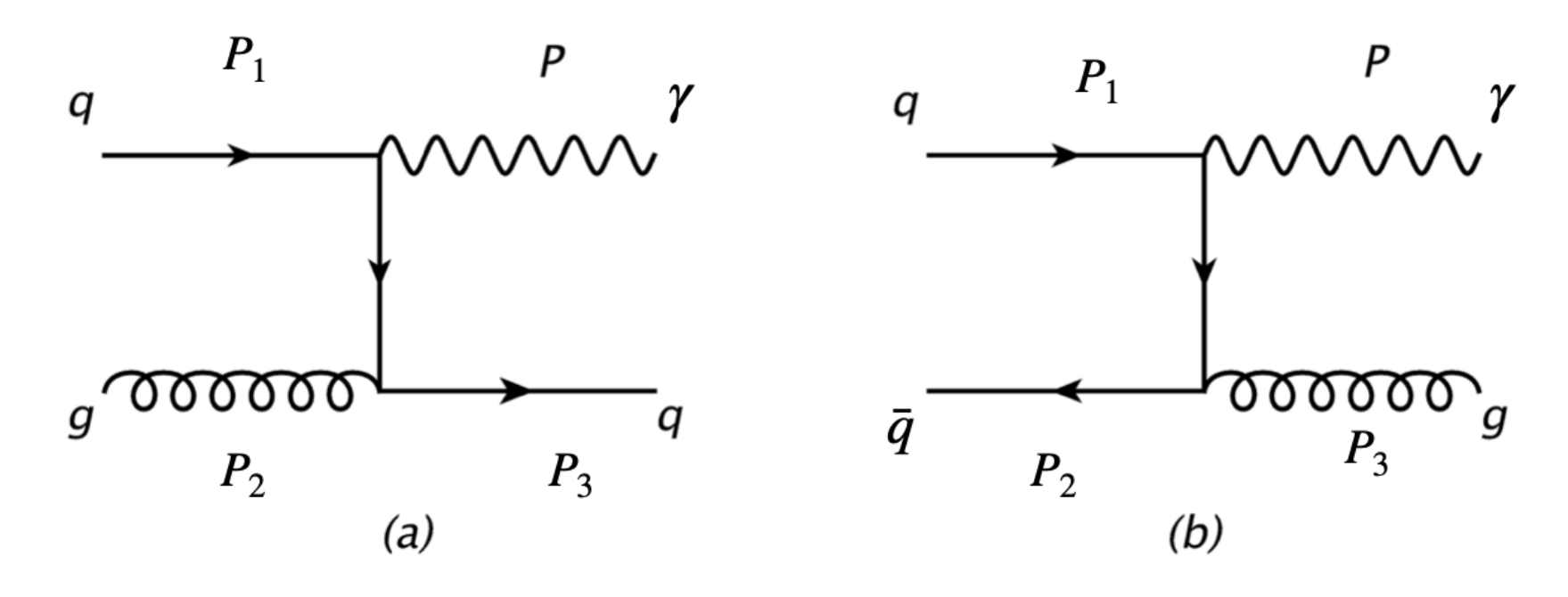}
	\end{center}
	\vspace{-0.7cm}
	\caption{Hard photon production in QCD through (a) Compton and (b) annihilation processes arising from the imaginary part of 2-loop diagrams in Fig.~\ref{cutkowski}.}
	\label{kin_hard_photon}
\end{wrapfigure}

The hard photon production was independently calculated by Kapusta et al.~\cite{Kapusta:1991qp} and Baier et al.~\cite{Baier:1991em}. In Fig.~\ref{cutkowski} there is sum of imaginary part of photon self-energies in the right side of the first line which lead to following processes~\cite{Kapusta:1991qp,Baier:1991em}:
\begin{enumerate}
\item The 1-loop diagram leads to processes $q\bar{q}\rightarrow \gamma$ is not allowed as  there is no phase space for $\gamma$ because it is on mass shell and energy conservation is not satisfied.
\item The 2-loop diagrams in Fig.~\ref{cutkowski} lead to two processes: Compton ($q(\bar q)g\rightarrow q(\bar q)\gamma)$ and annihilation ($q{\bar q}\rightarrow g\gamma)$ as shown in Fig.~\ref{kin_hard_photon}. 
\item There are also certain cuts in 2-loop diagrams which lead to $g^2$ corrections to the disallowed process $q\bar{q}\rightarrow \gamma$.
\end{enumerate}

\vspace{-0.0cm}
According to the above point 2  the photon production rate can be written from \eqref{ph4} as
\bea
E\frac{dR_i}{d^4x d^3\bm{\vec p}} \!\!&= &\!\!\frac{{\cal D}_i}{2(2\pi)^3} \int \frac{d^3\bm{\vec p}_1}{2E_1(2\pi)^3} \, \frac{d^3\bm{\vec p}_2}{2E_2(2\pi)^3} \,
\frac{d^3{\bm{\vec p}}_3}{2E_3(2\pi)^3}  n_1(E_1) \nn
&&\hspace{-2cm}\times n_2(E_2) \left (1\pm n_3(E_3)\right )
 (2\pi)^4 \delta\left(P_1+P_2-P_3-P\right)  \left|{\cal M}_i\right |^2 ,\ \ \  \label{ph5}
\eea
where ${\cal M}_i$ is the matrix element of the either processes in Fig.~\ref{kin_hard_photon} and ${\cal D}_i$ is the degeneracy factor to the corresponding process. Now, 
introducing $d^3{\bm{\vec p}_3}/[2E_3] \longrightarrow \int d^4P_3 \, \Theta\left(P^0_3\right) \delta\left(P_3^2\right)$, one can write
\be
E\frac{dR_i}{d^4x d^3\bm{\vec p}} 
=  \frac{{\cal D}_i}{2(2\pi)^8}\! \int\! \frac{d^3\bm{\vec p}_1}{2E_1}  \frac{d^3\bm{\vec p}_2}{2E_2}  \delta \left[\left(P_1+P_2-P\right)^2\right]  
\Theta\left(E_1+E_2-E\right)  \left|{\cal M}_i\right |^2  n_1(E_1)  n_2(E_2) \left[1\pm n_3\left(E_1+E_2-E\right)\right]. \label{ph6}
\ee
Now, we introduce  Mandelstam variables  $s=(P_1+P_2)^2$ and $t=(P_1-P)^2$  through the relations $\int ds \, \delta[s-(P_1+P_2)^2]$ and 
 $\int dt \, \delta[t-(P_1-P)^2]$ to make the rate invariant as
\bea
E\frac{dR_i}{d^4x d^3\bm{\vec p}} \!\!\!&=&\!\! \frac{{\cal D}_i}{2(2\pi)^8} \int ds\, \, dt\,\, \frac{d^3\bm{\vec p}_1}{2E_1}  \,\,  \frac{d^3\bm{\vec p}_2}{2E_2}  \,\,
 n_1(E_1)  n_2(E_2) \left (1\pm n_3\left(E_1+E_2-E\right) \right ) 
 \left|{\cal M}_i\left(s,t\right)\right |^2 \Theta\left(E_1+E_2-E\right) \nn 
&&\times  \delta \left[\left(P_1+P_2-P\right)^2\right] \delta\left[s-(P_1+P_2)^2\right ] \delta\left[t-(P_1-P)^2\right]  \, . \label{ph7}
\eea
We now choose the coordinate system as
$P=E\left(1,0,0,1\right)$,
$P_1=E_1 \left(1,\sin\theta_1\cos\phi_1,\sin\theta_1\sin\phi_1,\cos\theta_1\right)$ and
$P_2=E_2 \left(1,\sin\theta_2\cos\phi_2,\sin\theta_2\sin\phi_2,\cos\theta_2\right)$. 
Using  the $\delta$-functions arguments in \eqref{ph7} 
\be
	\cos\theta_2=\frac{2EE_2-s-t}{2EE_2},  \quad
		\cos\theta_1=\frac{t+2EE_1}{2EE_1},  \quad\text{and}\quad
		\cos\left(\phi_1-\phi_2\right)	=\frac{-2E^2s-2EE_2t+(s+t)(2EE_1+t)}{4E^2E_1E_2 \sin\theta_1\sin\theta_2}\, .  \label{ph10abc} 
\ee
Using Eq~\eqref{ph10abc} in the the rate in~\eqref{ph7} and then performing the $\theta$-integrations using $\delta$-functions, one gets
\bea
E\frac{dR_i}{d^4x d^3\bm{\vec p}} \!\!\!&=&\!\!\! \frac{{\cal D}_i}{2(2\pi)^8} \int\! ds \, dt \left|{\cal M}_i\left(s, t\right)\right |^2 \int_{0}^{\infty} 
 \frac{E_1 dE_1}{2}   \,  \left[\frac{1}{2EE_1}\right] \,\,
 \int_{0}^{\infty} \frac{E_2 dE_2}{2}  \,  \left[\frac{1}{2EE_2}\right]    n_1(E_1)  n_2(E_2)  \left[1\pm n_3(E_1+E_2-E)\right ]  \nn
 &&\hspace{8cm}\times  \int_{0}^{2\pi} d\phi_1    \int_{0}^{2\pi} d\phi_2 \,\, \delta\left(\Phi \right) 
 \Theta\left(E_1+E_2-E\right), \label{ph12}
\eea
where the solutions of the $\delta$-functions are $\cos\theta^0_1$ and $\cos\theta^0_2$ which are obtained in~\eqref{ph10abc}. 
First the $\phi_1$ integration is done using $\delta$-function\footnote{There are two values of $\phi_1$ that satisfies 
$\cos(\phi_1-\phi2)$ in \eqref{ph10abc}, which will give a factor of 2.} in then $\phi_2$ and one gets,
\bea
E\frac{dR_i}{d^4x d^3\bm{\vec p}} = \frac{{\cal D}_i}{(2\pi)^7}\frac{1}{16E} \int ds\, \, dt\, \left|{\cal M}_i\left(s,t\right)\right |^2\, \int_{0}^{\infty} dE_1 \, d{\cal{E}} 
\,\, e^{-{\cal E}/T} \,\,\left (1\pm n_3({\cal E}-E)\right )  \,\, \Big [AE_1^2+BE_1+ C\Big ]^{-\frac{1}{2}}\, , \label{ph17}
\eea
where
$A=-s^2$, 
$B=2s(Es+2Et-{\cal E}t)$ and 
$C = st(s+t) -(Es+{\cal E}t)^2$.
We have also considered ${\cal E}=E_1+E_2$ and approximated $n_1(E_1)n_2(E_2)=e^{-(E_1+E_2)/T}=e^{-{\cal E}/T}$ which are good approximations
 for hard photons, ${\cal E}>E>T$. 
Now, one  can find out integration limits of $E_1$  and  ${\cal E}$  from $3$ delta functions in \eqref{ph7} vis-a-vis   \eqref{ph10abc}:
$  {\cal E}_{\textrm{min}}= ({s}/{4E})+E $,   ${\cal E}_{\textrm{max}}= \infty $ and
$\sqrt{AE_1^2+BE_1+ C} = s \sqrt{\left(E_1^{\textrm{max}}-E_1\right) \left(E_1-E_1^{\textrm{min}}\right)}$ with
$  E_1^{\textrm{min}}\ge  -{t}/{4E} \, \Longrightarrow \, \,  E_1^{\textrm{min}}\le E_1 \le E_1^{\textrm{max}}$. 
Using these limits we first perform $E_1$-integration and the ${\cal E}$-integration and one gets
\bea
E\frac{dR_i}{d^4x d^3\bm{\vec p}} \!\!\!&=&\!\! \frac{{\cal D}_i}{(2\pi)^6}\frac{T}{32E}\,\, e^{-{E}/T} \, \, \int \frac{ds}{s}\,\,\ln \left(1\mp e^{-\frac{s}{4ET}}\right)^\mp \, \,\int dt\, 
\left|{\cal M}_i(s,t) \right |^2\,  \, . \label{ph25}
\eea
where ($-$)ve sign corresponds to if the third particle is a boson  whereas ($+$)ve sign is  for a fermion in the final state whose origin is either a Bose-enhancement  or a Pauli-suppression. 
The differential cross-section for a given process is related to matrix element squared as
${d\sigma_i}/{dt} = {\left|{\cal M}_i \right |^2}/{(16\pi s^2)}$.
For Compton process ($qg\rightarrow q\gamma$) in Fig.~\ref{kin_hard_photon}$(a)$, the differential cross-section is given as~\cite{Kapusta:1991qp} 
\be
\frac{{d\sigma}^{qg\rightarrow q\gamma}}{dt} =-\frac{\pi\alpha\alpha_s}{3s^2}\, \frac{u^2+s^2}{su} = \frac{\pi\alpha\alpha_s}{3s^2} 
\left(\frac{s+t}{s}+\frac{s}{s+t}\right)\, = \frac{\pi\alpha\alpha_s}{3s^2} \left[1+\frac{t}{s} +\frac{t}{s+t}\right], \label{ph30}
\ee
whereas for the annihilation process ($q{\bar q}\rightarrow g\gamma$) in Fig.~\ref{kin_hard_photon}$(b)$, it is given as~\cite{Kapusta:1991qp} 
\be
\frac{{d\sigma}^{q{\bar q}\rightarrow g\gamma}}{dt} &=&\frac{8\pi\alpha\alpha_s}{9s^2}\, \frac{u^2+t^2}{ut} = -\frac{8\pi\alpha\alpha_s}{9s^2}  \left[1+\frac{s}{t} +\frac{t}{s+t}\right] 
=  -\frac{8\pi\alpha\alpha_s}{9s^2}  \left[2+\frac{s}{t} -\frac{s}{s+t}\right]\, . \label{ph31}
\ee 
We now note that the integral over $t$ will represent total cross-section for given physical process. But for massless exchanged particles in Fig.~\ref{kin_hard_photon}, 
the total cross-section will be infinite because the differential cross-sections with \eqref{ph30} and \eqref{ph31} will diverge at $t$ and/or $u=0$. 
So it requires many-body effects to regulate this divergence. In the next subsection it will be done how this divergence is removed through HTL approximation. For the time being, we drop the phase space that making the divergence  
when $t$ is soft, $t<k_c^2$, where $k_c$ is an infrared cut-off with the restriction $T^2\gg k_c^2>0$.  Now, the $t$- and $s$-integration limits  become $-s+k_c^2\le t\le-k_c^2$ and $2k_c^2\le s <\infty$. This regulates the $u$ and $t$ divergences symmetrically and obeyed the identity $s+u+t=0$  for massless particles.  
Using the limits of $s$ and $t$, the Eq.~\eqref{ph25} becomes
\bea
E\frac{dR_i}{d^4x d^3\bm{\vec p}} \!\!\!&=&\!\! \frac{{\cal D}_i}{(2\pi)^5}\frac{T}{4E}\,\, e^{-{E}/T} \, \, \int_{2k_c^2}^\infty ds \,\,\ln \left(1\mp e^{-\frac{s}{4ET}}\right)^\mp \, \,
\int_{-s+k_c^2}^{-k_c^2}  dt  
\, \,  \, s \,\, \frac{d\sigma_i(s,t)}{dt} \, . \label{ph33}
\eea
where it is evident that there is no phase space if $s \le 2k_c^2$. We can now calculate ${\cal D}_i$ for specific process as
\be
{\cal D}_{qg\rightarrow q\gamma} =\underbrace{(2\times 2)}_{\mbox{spin}} \times \underbrace{2}_{q\, \, \&\,\, 
\bar q} \times \underbrace{(3\times 8)}_{\mbox{colour}}  \times 
\underbrace{\left[\left(\frac{2}{3}\right)^2 +\left(-\frac{1}{3}\right)^2 \right]}_{\mbox{summed over}{\,\, u\,\,\& \,\, d \,\,}} 
=\frac{320}{3}\, ,
{\cal D}_{q{\bar q}\rightarrow g\gamma} = \underbrace{(2\times 2)}_{\mbox{spin}}  \times \underbrace{(3\times 
3)}_{\mbox{colour}}  \times  \frac{5}{9}=20 \, .\label{ph34b}
\ee
After perforing $s$ and $t$ integrations the Compton differential rate for hard photon production in \eqref{ph33} 
becomes
\bea
E\frac{dR^{qg\rightarrow q\gamma}}{d^4x d^3\bm{\vec p}} 
&=& \!\!\!\frac{5 \alpha\alpha_s}{54\pi^2}\, \,T^2\,\, e^{-{E}/T} \, \, 
\left[ \ln\left (\frac{4ET}{k_c^2}\right )  +   \ln 2 +\frac{1}{2} -\gamma_{\tiny{E}} +\frac{\zeta^\prime(2)}{\zeta(2)} 
\right ] \, ,\label{ph39}
\eea
whereas the annihilation differential rate for hard photon production becomes
\bea
E\frac{dR^{q\bar q \rightarrow g\gamma}}{d^4x d^3\bm{\vec p}} 
&=& \!\!\!\frac{5 \alpha\alpha_s}{27 \pi^2} \, \,T^2\,\, e^{-{E}/T} \, \, 
\left[ \ln\left (\frac{4ET}{k_c^2}\right ) - 1 -\gamma_{\tiny{E}} +\frac{\zeta^\prime(2)}{\zeta(2)} \right ] \, .\label{ph43}
\eea
The total rate for hard photon production corresponding to physical processes ($2\rightarrow 2$) in 
Fig.~~\ref{kin_hard_photon} is obtained 
by adding \eqref{ph39} and \eqref{ph43}
\bea
E\frac{dR}{d^4x d^3\bm{\vec p}} 
&=& \!\!\!\frac{5 \alpha\alpha_s}{18 \pi^2} \, \,T^2\,\, e^{-{E}/T} \, \, 
\left[ \ln\left (\frac{4ET}{k_c^2}\right ) +\frac{1}{3}\ln 2 - \frac{1}{2} -\gamma_{\tiny{E}} + \frac{\zeta^\prime(2)}{\zeta(2)} \right ] \, .\label{ph44}
\eea
which was obtained independently by Kapusta et al~\cite{Kapusta:1991qp} and Baier et al.~\cite{Baier:1991em}.

If one uses Maxwell-Boltzmann (MB)  approximation  instead of  Bose-enhancement or Pauli-suppression in the final state in \eqref{ph17} and following the same procedure 
as before one gets the total hard photon rate ~\cite{Kapusta:1991qp,Baier:1991em} as
\bea
\left.E\frac{dR}{d^4x d^3\bm{\vec p}}\right|_\text{MB} = \frac{20\alpha\alpha_s}{9\pi^4} \,T^2\,\, e^{- E/T} 
\left[ \ln\left (\frac{4ET}{k_c^2}\right )-\frac{1}{4}-\gamma_{\tiny E} \right]\, . \label{ph48}
\eea
In all these above rates, there is a logarithmic infrared sensitivity, i.e. the rate diverges logarithmically if the soft cut-off $k_c\rightarrow 0$. 
These rates agree with those reported earlier in  the the literature~\cite{Hwa:1985xg,Neubert:1989hu} at least for the dominant logarithmic term. The constant terms are different if one uses different infrared regulator such as quark mass instead of soft cut-off on the four momentum transfer of the exchanged quarks in $t$ and $u$ channels. We note that the analyses in Refs.~\cite{Hwa:1985xg,Neubert:1989hu}  no screening mechanism has been used as an infrared regulator. Instead the bare mass of the exchanged quarks, $m^{\text q}$  in Fig.~\ref{kin_hard_photon}  has been used as a regulator but the rate diverges logarithmically if the mass of the exchanged quark is zero. Therefore, Kajantie and Ruuskanen argued~\cite{Kajantie:1982nj} that the bare quark mass should be replaced by an effective thermal quark mass
$m^{\text q}_{\text{th}}$. This means that even the production of energetic photons is sensitive to many-body effects of the QGP, since the exchange of soft quarks plays an important role in the production mechanism. A systematic treatment of 
many-body effects is provided by the HTL resummation technique developed by Braaten and Pisarski~\cite{Braaten:1989mz} which was independently used in Ref.~\cite{Kapusta:1991qp} and~\cite{Baier:1991em} to regulate the soft photon production from QGP. In the following subsection~\ref{htl_ir_1loop}, we will discuss the infrared regulated soft photon production from 1-loop HTL self-energy of photon.
\subsubsection{Infrared regulated soft photon from 1-loop HTL approximation}
\label{htl_ir_1loop}
In the previous subsec~\ref{kinetic_theory} for kinetic theory calculation,  a cut of $k^2_c$ on the four momentum transfer in $t$ and $u$ was considered to regulate infrared divergence which had forced to leave out a small amount phase space in photon rate obtained in~\eqref{ph44} and \eqref{ph48} corresponding processes in Fig.~\ref{kin_hard_photon}.

\begin{wrapfigure}[9]{r}{0.45\textwidth}
	\begin{center}
		\vspace{-1.cm}
		\includegraphics[scale=0.2]{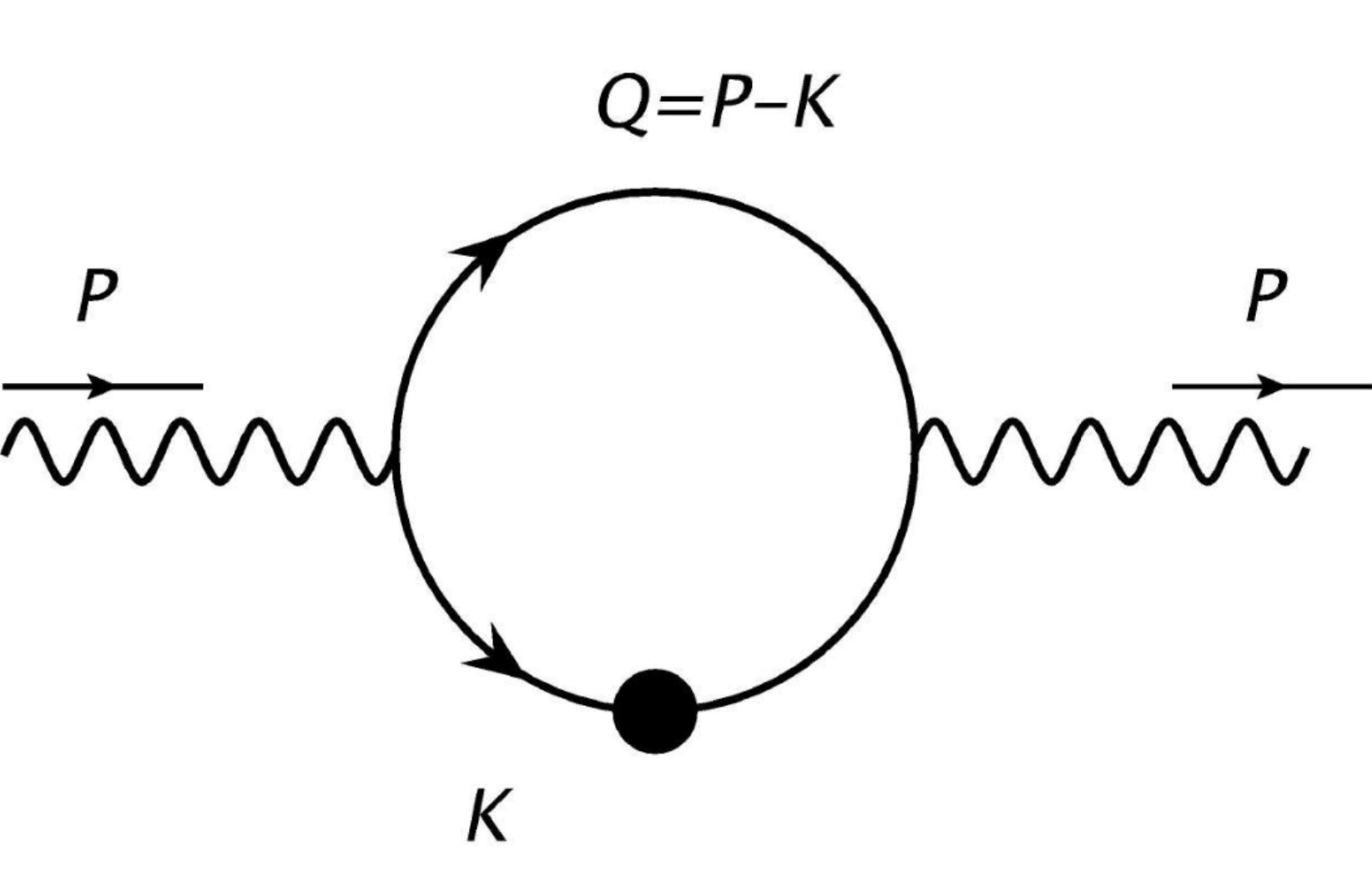}
	\end{center}
	\vspace{-0.7cm}
	\caption{1-loop photon self-energy in HTL approximation. The internal line with black blob is the soft quark propagator in HTL approximation whereas that without blob is hard quark propagator.}
	\label{ir_soft_photon}
\end{wrapfigure} 
 We need to calculate the photon rate for this small phase space~\cite{Kapusta:1991qp,Baier:1991em} for soft exchange of quarks which can be obtained from  Fig.~\ref{ir_soft_photon}. 
In Fig.~\ref{ir_soft_photon},  the dressing of one of the quark propagators is only required to get infrared regulated soft photon rate in lowest order in $\alpha_s$~\cite{Kapusta:1991qp,Baier:1991em}. It is not necessary to dress both the propagators or the vertices, as those will only introduce higher-order corrections in $\alpha_s$. Cutting through the blob (soft propagator) gives rise to processes in Fig.~\ref{kin_hard_photon}.

The contribution of 1-loop HTL photon self-energy  diagram in Fig.~\ref{ir_soft_photon} can be written as
\bea
\Pi^{\mu}_{\mu}(P) 
= -\frac{5}{3} e^2 T \sum_{\{k_0 \}} \int  \frac{d^3k}{2\pi)^3} 
{\Tr }\left [\gamma_\mu S^q(K) \gamma^\mu S^{\text q}_0(Q) \right ] ,\label{ph49}
\eea
where it is summed up for two massless quark flavours $u$ and $d$, and $S^{\text q}_0$ is free quark propagator  given in  \eqref{gse25} and the HTL quark propagator
is $S^{\text q}$ given in \eqref{qcd34}. 
Using them and performing  traces over Dirac indices, then BPY prescription~\cite{Braaten:1990wp}  derived in \eqref{bpy11}, one can write
the imaginary part of \eqref{ph49}  as
\bea
&&\textrm{Im} \Pi^{\mu}_{\mu}(E, p) = -\frac{10}{3} \pi e^2 \left (e^{\beta E}-1\right ) \int  \frac{d^3k}{(2\pi)^3} 
\int_{-\infty}^{+\infty} d\omega_1 \int_{-\infty}^{+\infty} d\omega_2\, 
n_F(\omega_1) n_F(\omega_2) \delta \left(E-\omega_1-\omega_2 \right )
 \times \bigg\{  \left (1-\bm{\hat k} \cdot \bm{\hat q}\right)\nn
&&\times\Big[ \rho_+(\om_1,k)\rho^f_+(\om_2,q)+ \rho_-(\om_1,k)\rho^f_-(\om_2,q)\Big] 
 +  \left (1+\bm{\hat k} \cdot \bm{\hat q}\right)\Big[ \rho_+(\om_1,k)\rho^f_-(\om_2,q)+ \rho_-(\om_1,k)\rho^f_+(\om_2,q)\Big] \bigg \} \, .\label{ph51}
\eea
In  subsec~\ref{spec_prop}  the free spectral functions $\rho_\pm^f$ are obtained in  \eqref{spec1} as $\rho^f_\pm(\om_2,q)=\delta(\om_2\mp q)$ whereas the HTL spectral functions $\rho_\pm(\om_1,k)$ are also obtained in \eqref{spec2} which have pole parts involving $\delta$-functions given in \eqref{spec3} and branch 
cuts  given in \eqref{spec4}. As discussed in the previous  subsec~\ref{kinetic_theory}, an infrared cut-off $k_c$ was introduced to regulate the soft phase space. Now, for Fig.~\ref{kin_hard_photon} and  Fig.~\ref{ir_soft_photon} indicates that the exchanged quark should be dressed and must satisfy~\cite{Kapusta:1991qp} 
\be
 -k_c^2\le \om_1^2-k^2 \le 0 \,\, \,\,\,\,\,  \Longrightarrow   \,\,\,\,\, 0\le k^2-\om_1^2  \le k_c^2 \,\, \,\, \,\,\,\,\,
 {\mbox{and}}\,\, \,\,\,\,\,\,\,  E\gg T \, \label{ph53a} 
\ee
We note that in the weak coupling limit $gT\ll k_c\ll T$, the arbitrary separation scale $k_c$ cancels once the hard and 
the soft contributions are added, as it should be the case for a consistent leading-order calculation~\cite{Braaten:1991dd}.
Now, one can perform the $\om_2$-integration using the on-shell $\delta$-functions in the free spectral functions 
$\rho_\pm^f$, and then neglect the other energy conserving $\delta$-function $\delta(E-\om_1+q)$ as it will not be satisfied and $x={\angle \bm{\vec k},\, \bm{\vec p}}$.  To perform the $x$-integration we change a variable $x\rightarrow q$ as done earlier for obtaining the Born rate. Then performing the $q$-integration using the restrictions in \eqref{ph53a}, one gets
\bea
\textrm{Im} \Pi^{\mu}_{\mu}(E) =- \frac{5}{12 \pi} e^2  \, \left (e^{\beta E}-1\right ) e^{-E/T}\, \int_0^\infty  dk  \int_{-\infty}^{+\infty} d\omega_1 
\Big[  (k-\om_1) \rho_+(\om_1,k)  +  (k+\om_1)  \rho_-(\om_1,k) \, \Big ] \, \Theta\left (k_c^2-k^2+\om_1^2\right)  ,\ \  \label{ph56}
\eea
We note that the pole contributions involving $\delta$-functions in the quark spectral functions $\rho^{\text{pole}}_\pm$  will not contribute in this order but that of cut part $\rho_\pm^{\text{cut}}$, will do.  Therefore, the physical process leading to the photon production from Fig.~\ref{ir_soft_photon} is related to Landau damping, i.e. the interaction of a soft quark with thermal gluons. So, the diagram of  Fig.~\ref{ir_soft_photon} contains infinitely many quark-gluon loops as the HTL quark self-energy is resummed in the effective quark propagator in Fig.~\ref{quark_prop}.

We note that the $\rho_\pm^{\text{cut}}$ has the complex momentum dependence, the integrations in \eqref{ph56}
 cannot be done analytically~\cite{Kapusta:1991qp,Baier:1991em}. However, using generalised Kramers-Kronig 
 relations, the so-called Leontovich relations~\cite{Leontovich:1961ne}, it can be shown, that only the high energy limit of the HTL quark propagator is needed~\cite{Thoma:2000ne} and it can be performed analytically  if one knows the quark response function. The result obtained~\cite{Thoma:2000ne} as
\bea
\textrm{Im} \Pi^{\mu}_{\mu}(E)
&=& \frac{10\pi}{9} \alpha\alpha_s  \, \left (e^{\beta E}-1\right ) e^{-E/T} \,\, T^2 \, \, 
\ln\left(\frac{k_c^2}{2(m_{\rm{th}}^{\rm q})^2 }\right)\, . \label{ph75}
\eea
This agrees with Refs.~\cite{Kapusta:1991qp,Baier:1991em} but the factor $1/2$ under the logarithm was obtained  numerically by using the full cut part of the spectral function. The advantage of 
the use of the Leontovich relation  is that it could be done analytically~\cite{Thoma:2000ne} if one knows the quark response function. Now, combining~\eqref{ph3}, \eqref{ph75} and~\eqref{asym_mass}, one gets the soft photon production as
\bea
E\frac{dR}{d^4x d^3\bm{\vec p}} 
&=& \!\!\!\frac{5 \alpha\alpha_s}{18 \pi^ 2} \, \,T^2\,\, e^{-{E}/T} \, \, \ln\left (\frac{k_c^2}{M_\infty^2}\right ) \, .\label{ph76}
\eea
Now, combining  the hard part in \eqref{ph44} and  the soft part in \eqref{ph76} one gets the energetic  photon production as
\bea
E\frac{dR^{\text{KB}}}{d^4x d^3\bm{\vec p}} 
&=& \!\!\!\frac{5 \alpha\alpha_s}{18 \pi^2} \, \,T^2\,\, e^{-{E}/T} \, \, 
\left[ \ln\left (\frac{4ET}{M_\infty^2}\right ) +\frac{1}{3}\ln 2 - \frac{1}{2} -\gamma_{\tiny{E}} + \frac{\zeta^\prime(2)}{\zeta(2)} \right ] \, ,\label{ph77}
\eea
where the separation scale $k_c$ that  acts as an infrared  cut-off of the hard part  cancels out as claimed in Ref.~\cite{Braaten:1991dd}. This happens because the prefactor in hard and soft rate are same. So, the HTL approximation~\cite{Braaten:1989mz} works well here to regulate the soft part.  We also note that apart from other factors the 1-loop HTL rate is proportional to $T^2/E$ which is due to phase space. KB, in the superscript of the rate above, denotes the production rate of energetic photon by  Kapusta et al.~\cite{Kapusta:1991qp} and Baier et al.~\cite{Baier:1991em}. 

It is worth noting   that for finite quark chemical potential $\mu$, one cannot use the MB approximation for the initial particle distributions in the hard contribution as done in \eqref{ph17}. Otherwise, no cancellation of the separation scale will take  when  the hard and the soft part are added. So, the photon production rate at finite $\mu$ can be determined only numerically. For $\left | \mu/T \right |<1$, the factor $T^2$ in  \eqref{ph77} has to be replaced~\cite{Weldon:1982aq} to a good accuracy simply by $T^2+\mu^2/\pi^2$. The 1-loop HTL photon production rate has also been calculated for undersaturated (chemically non-equilibrated) QGP~\cite{Shuryak:1992bt,Kampfer:1994rr,Strickland:1994rf,Traxler:1995kx,Srivastava:1996qd,Pal:1998jr,Baier:1997xc}.

Finally, we note that the result in \eqref{ph77}, for realistic values of the strong coupling constants, is incomplete and  the contributions beyond the leading logarithm (other factors inside the square brackets can be taken inside the logarithm) become important and can even dominate the leading order ${\cal O}(\alpha_s)$ coming from 1-loop HTL.  Such contributions come from higher order diagrams, i.e., 2-loop diagrams in the HTLpt describing e.g. bremsstrahlung and inelastic pair annihilation,  which have strong infrared and collinear sensitivity and contribute in ${\cal O}(\alpha_s)$. They are not included in the soft part of the polarisation tensor using the full quark propagator~\eqref{qcd34}, since they come from the exchange of a hard quark. In the following subsec~\ref{htl_ir_2loop}, we will discuss the photon production from 2-loop HTLpt.

\subsubsection{Photon production from 2-loop HTL approximation}
\label{htl_ir_2loop}
\begin{figure}[htb]
\begin{center}
\includegraphics[scale=0.65]{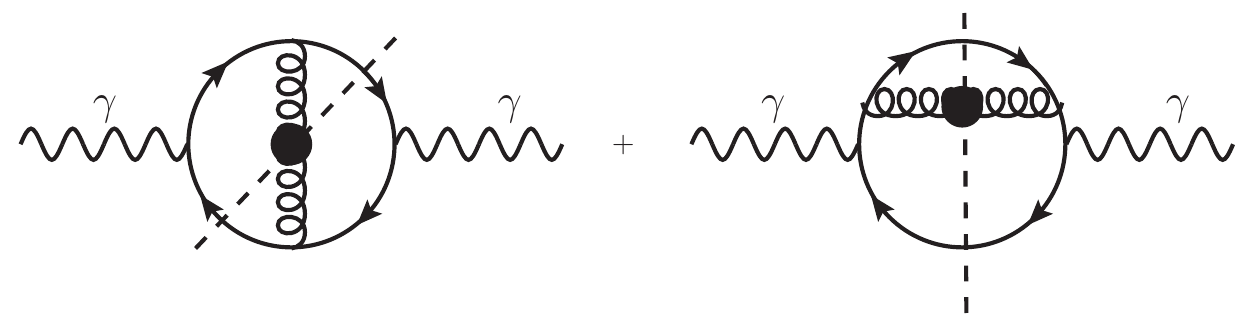}
\end{center}
\vspace{-0.4in}
\caption{Two-loop photon self-energy in HTL approximation where the exchanged gluon is soft but internal quark lines are hard. The dashed lines through the black blobs represent thermal cutting rules to obtain imaginary part.}
\label{ir_soft_2loop}
\end{figure}

As discussed in the previous two subsections~\ref{kinetic_theory} and \ref{htl_ir_1loop}, the  ${\cal O}(\alpha_s)$ HTL rate in \eqref{ph77}  were obtained by adding the 1-loop HTL contribution for soft quarks in Fig.~\ref{ir_soft_photon} and the 2-loop diagrams  in the right side of Fig.~\ref{cutkowski}, where the intermediate quark is hard.  We further note that in the 2-loop diagrams in Fig.~\ref{cutkowski}, one assumed  the exchanged gluon is also hard, because it is a thermal particle with average energy is $T$. However, if this gluon is soft, there will be a Bose enhancement factor $n_B(l_0\sim gT) \approx T/k_0 \sim 1/g$ and  this contribution should be important. Therefore, according to the HTL approximation,  one needs to replace the exchanged hard gluon in 2-loop diagrams in Fig.~\ref{cutkowski} by HTL resummed gluon as shown in  Fig.~\ref{ir_soft_2loop}. This was first realised by Aurenche  et al.~\cite{Aurenche:1998nw} which was an important realisation.

\begin{figure}[htb]
\begin{center}
\includegraphics[scale=0.55]{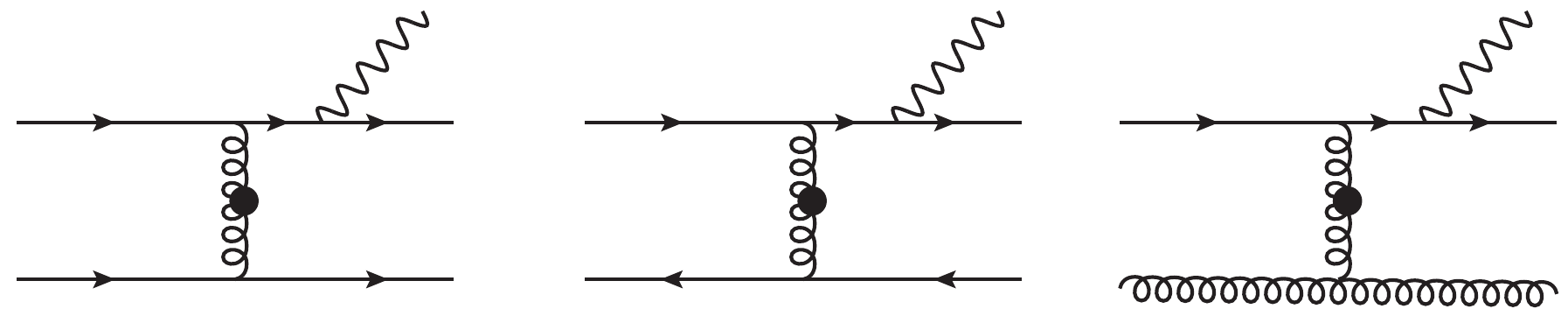}
\end{center}
\vspace{-0.2in}
\caption{Bremsstrahlung processes appearing from cutting through the black blob of the dressed gluon propagator in Fig.~\ref{ir_soft_2loop}.}
\label{brems}
\end{figure}
\begin{wrapfigure}[8]{r}{0.5\textwidth}
\begin{center}
\includegraphics[scale=0.5]{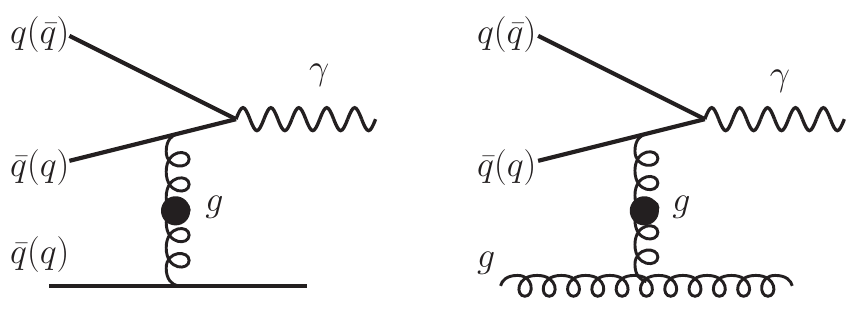}
\end{center}
\vspace{-0.3in}
\caption{Annihilation with scattering (inelastic pair annihilation)  processes appearing from cutting through the black blob of the dressed gluon propagator in Fig.~\ref{ir_soft_2loop}.}
\label{aws}
\end{wrapfigure}

The Bremsstrahlung processes in Fig.~\ref{brems} come from the thermal cutting~\cite{Das:1997gg,Kobes:1985kc,Kobes:1986za,Gelis:1997zv} through the black blob of the HTL gluon propagator in Fig.~\ref{ir_soft_2loop}, i.e., from the imaginary part of the gluon self-energy of the HTL gluon propagator corresponding to Landau damping of the time-like gluon. Since the HTL gluon propagator in Fig.~\ref{gluon_prop} contains hard quark and gluon loops, which lead to physical processes contained in the imaginary part of Fig.~\ref{ir_soft_2loop} are bremsstrahlung  in Fig.~\ref{brems}  and annihilation with scattering (inelastic pair annihilation) in Fig.\ref{aws}. Usually one expects  that the imaginary part of 2-loop diagrams should be of the order $e^2 g^4$ due to additional 3-point QCD vertices and also multiplied by a factor $T^2/M^2_\infty$. However,  due to a strong collinear singularity in the processes in Fig.~\ref{brems} and in Fig.~\ref{aws}  there is collinear enhancement, powers of $T^2/M_\infty^2 \sim 1/g^2$ are generated so that they contribute at the leading order ${\cal O}(e^2g^2)$ in the strong coupling.
\vspace{-0.1in}
\vspace{-0.3cm}
\subsubsection*{A)  Two-loop rate by Aurenche, Gelis, Kobes and Zaraket (AGKZ)}
\vspace{-0.2cm}
The 2-loop rate calculation of Fig.~\ref{ir_soft_2loop} by Aurenche, Gelis, Kobes and Zaraket (AGKZ)~\cite{Aurenche:1998nw} is very involved and tedious. However, 
 we now present the 2-loop final results of the production rates of photons ($E\gg T$) for the two processes originating from  Fig.~\ref{ir_soft_2loop}.
 The rate for the bremsstrahlung (brem) process for two massless quark flavours~\cite{Aurenche:1998nw} is
\bea
\left. E\frac{dR}{d^4x d^3\bm{\vec p}} \right |^{\text{AGKZ}}_{\text{brem}}
&=& \!\!\!\frac{40 \alpha\alpha_s}{9 \pi^5} \, \,T^2\,\, \,\, e^{-{E}/T} \, \,  \left(J_T-J_L\right)\, \ln 2 \, , \label{ph78}
\eea
whereas that for the annihilation with scattering (aws) process is obtained as
\bea
\left. E\frac{dR}{d^4x d^3\bm{\vec p}} \right |^{\text{AGKZ}}_{\text{aws}}
&=& \!\!\!\frac{40 \alpha\alpha_s}{27 \pi^5} \, \,E\,T\,\, \,\, e^{-{E}/T} \, \,  \left(J_T-J_L\right)\, , \label{ph79}
\eea
where the correct\footnote{We note that there was a numerical error in the computation of $J_T$ and $J_L$ in  Ref.~~\cite{Aurenche:1998nw} that resulted in a factor of 4 enhancement of their values: $J_T\approx 4.45$ and $J_L\approx -4.26$.} value of the constants were independently obtained in Refs.~\cite{Dutta:2001ii,Steffen:2001pv,Srivastava:1999ekv} as  $J_T\approx 1.108$ and $J_L\approx -1.064$.  The total two-loop rate is the sum of \eqref{ph78} and~\eqref{ph79}. It can seen that 2-loop HTL rates in \eqref{ph78} and \eqref{ph79} is same order ${\cal{O}}\left(\alpha\alpha_s\right)$ as 1-loop results in \eqref{ph77} due to the collinear enhancement by a power of $T^2/M_\infty^2 \sim 1/g^2$ as discussed earlier. On the other hand because of phase space the  annihilation with scattering  process is proportional to $T$ where as the bremsstrahlung process is $T^2/E$ which is same as that of annihilation and Compton process in 1-loop HTL. We also note that these HTL 2-loop photon production rates have also been generalised to chemically non-equilibrated QGP~\cite{Mustafa:2000sg,Mustafa:2001pf,Dutta:2001ii}.

However,  the quantitative analysis of 2-loop rates~\cite{Aurenche:1998nw}  given in \eqref{ph78} and \eqref{ph79} were incomplete because a suppression of the emission rate due to the LPM effect~\cite{Landau:1953ivy,Landau:1965ksp,Migdal:1956tc} arise from multiple scattering during the photon emission process, which limits the coherence length of the emitted radiation, was ignored. Nevertheless, this was  pointed out by the same authors later in Refs.~\cite{Aurenche:1999ec,Aurenche:1999tq} which had discussed the physics involved but did not attempt to make a complete calculation as claimed in Refs.~\cite{Arnold:2001ba,Arnold:2001ms}. That is, one internal propagator in the diagrams become nearly on-shell and receives ${\cal O}(1)$ corrections from the quasiparticle width. This reflects sensitivity to additional scatterings which occur during the photon emission process. However, merely including the appropriate width on the intermediate propagator is not sufficient; a consistent treatment is required for a more elaborate and detailed analysis. This  improvement were done over the rate of Aurenche et al.~\cite{Aurenche:1998nw,Aurenche:1999ec,Aurenche:1999tq}  by Arnold et al.~\cite{Arnold:2001ba,Arnold:2001ms} which we will discuss below.
\subsubsection*{B)  Two-loop rate by Arnold, Moore and Yaffe (AMY)}
\vspace{-0.2cm}
As discussed above, Arnold, Moore and Yaffe (AMY)~\cite{Arnold:2001ba} provided a consistent method to  take in account the ${\cal O}(1)$ corrections by deriving an integral equation, similar in form to a linearised kinetic equation, whose solution determines the rate of bremsstrahlung and inelastic pair annihilation processes, to leading order in $g$. 
In a companion paper~\cite{Arnold:2001ms}  the accurate solution of that integral equation was obtained for specific theories of interest (viz., QED and QCD).

A full calculation of  leading order ${\cal O}(\alpha\alpha_s)$ photon production rate, including 1-loop and 2-loop with fully considering the LPM effect in QCD, was obtained~\cite{Arnold:2001ms}  as
\bea
\left. E\frac{dR}{d^4x d^3\bm{\vec p}} \right |^{\text{AMY}}_{\text{QCD}} = \frac{5 \alpha\alpha_s}{9 \pi^2} \, \,T^2 \, n_F(E) \, \left[  \ln\left(\frac{T}{M_\infty}\right) +
\frac{1}{2} \ln\left(\frac{2E}{T}\right) +C_{2\leftrightarrow 2} \left(\frac{E}{T} \right) + C_{\text{brem}} \left(\frac{E}{T} \right)+C_{\text{aws}} \left(\frac{E}{T} \right)\right ], \label{ph80}
\eea
where 
\be
\lim_{{E}/{T}\rightarrow\infty}C_{2\leftrightarrow 2} \left(\frac{E}{T} \right)=\frac{2}{3}\ln 2 - \frac{1}{4} -\frac{\gamma_{\tiny{E}}}{2} + \frac{\zeta^\prime(2)}{2\zeta(2)} \, ,\label{ph81}
\ee
where $M_\infty$ is given in Eq.~\eqref{asym_mass}. Note that the first three terms within the square brackets (along with pre-factors) in Eq.~\eqref{ph80} are the contribution from the 1-loop HTL result and it agrees with Eq.~\eqref{ph77} in the limit  $E\gg T$. However, for arbitrary photon energy $E\gg gT$, the $C_{2\leftrightarrow 2} \left({E}/{T} \right)$, $C_{\text{brem}} \left({E}/{T} \right)$ and $C_{\text{aws}} \left({E}/{T} \right)$ involve multidimensional integrals which cannot be evaluated analytically but computed numerically. The fitted numerical results are summarised~\cite{Arnold:2001ms}  as
\begin{subequations}
\begin{align}
C_{2\leftrightarrow 2} (y) &\approx 0.041\, y^{-1}- 0.3615 +1.01 e^{-1.35\, y} \, , \label{82a} \\
C_{\text{brem}}(y)+C_{\text{aws}} (y)&\approx \sqrt{1+\frac{N_f}{6}} \left[\frac{0.548\, \ln\left(12.28+1/y \right) }{(y)^{3/2}}
+ \frac{0.133 \, y }{\sqrt{1+{y}/{16.27}}} \right] \, , \label{ph82b}
\end{align}
\end{subequations}
where $N_f$ is number of quark flavours. These fits are valid in the range $0.2 \le y\le 50$.  

The QED photon production rate based on Ref.~\cite{Arnold:2001ms} is also obtained~\cite{Mustafa:2008nk,Yaresko:2010xe} as 
\bea
\left. E\frac{dR}{d^4x d^3\bm{\vec p}} \right |^{\text{AMY}}_{\text{QED}} \!\! \!\!\!\!\!\!\!&=&  \!\!\!\! \frac{\alpha^2}{2\pi^2} \, \,T^2 \, n_F(E) \, \left[ \ln\left(\frac{T}{M_\infty}\right) +
\frac{1}{2} \ln\left(\frac{2E}{T}\right) +C_{2\leftrightarrow 2} \left(\frac{E}{T} \right)+C_{\text{brem}} \left(\frac{E}{T} \right)+C_{\text{aws}} \left(\frac{E}{T} \right)\right ]
\, , \label{ph83}
\eea
where $M^2_\infty= 2(m^{\textrm e}_{\text{th}})^2$ with $(m^{\textrm e}_{\text{th}})^2=e^2T^2/8$ is obtained in \eqref{se3}. 

We note that $C_{2\leftrightarrow 2}\left({E}/{T} \right) $ is same in QED and QCD because the interference effect cancels out. 
On the other hand the 2-loop coefficients are fitted in the range $0.2 \le y\le 30$ as
\be
C_{\text{brem}} (y)+C_{\text{aws}}(y)\approx \left[\frac{2a \, \ln\left(b+1/y \right) }{(y)^{c}}
+ \frac{2 \,d \, y }{\sqrt{1+{y}/{f}}} \right] \, , \label{ph84}
\ee
where $ a = 0.374958$, $b = 0.431304$, $c = -0.05465$, $d = 0.157472$ and $f = 1.73085$.

\begin{wrapfigure}[14]{r}{0.5\textwidth}
\begin{center}
\includegraphics[scale=0.5]{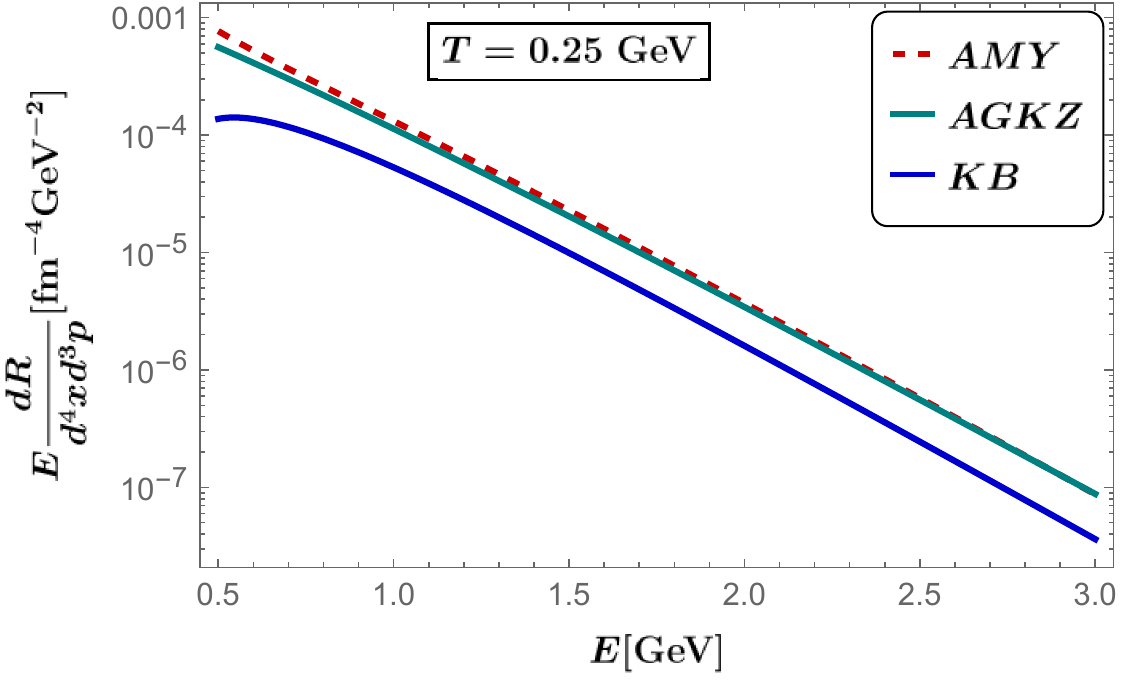}
\end{center}
\vspace*{-0.35in}
\caption{The differential photon production rate as function of photon energy for 1-loop HTLpt (KB rate) and 2-loop HTLpt (AGKZ and AMY rates) at $T=0.25$ GeV.}
\label{plot_photon_rate}
\end{wrapfigure}
In Fig.~\ref{plot_photon_rate} we have displayed the various photon differential rates: 1-loop HTLpt  by KB  given in \eqref{ph77}, 2-loop HTLpt by AGKZ in \eqref{ph78} and  \eqref{ph79} 
and also by AMY in \eqref{ph80} for $T=0.25$GeV with 1-loop running strong coupling~\cite{ParticleDataGroup:2012pjm}
\be
\alpha_s(\Lambda) = \frac{12\pi}{(33-2N_f)\ln\left({\Lambda^2}/{\Lambda^2_{\overline{\rm{MS}}}} \right) } \, . \label{ph85}
\ee 
For one-loop running we fix the scale $\Lambda_{\bar{\rm{MS}}}$ by requiring that $\alpha_s(1.5 {\rm{GeV}}) = 0.326$ which is obtained from lattice 
measurements~\cite{Bazavov:2012ka}. For one-loop running, this procedure gives $\Lambda_{\overline{\rm{MS}}}= 176$ MeV and we consider the momentum 
scale $\Lambda=2\pi T$.  One can see that two-loop contributions dominate over one-loop result in all energies even though order of coupling $({\cal O}(\alpha\alpha_s))$
is same both cases. This is because the leading  $({\cal O}(\alpha\alpha_s))$
is complete in two-loop calculations.
\subsubsection{Discussion on photon production from higher than 2-loop}
\label{higher_loop}

\begin{wrapfigure}[6]{r}{0.4\textwidth}
	\vspace*{-0.5in}
	\begin{center}
		\includegraphics[scale=0.18]{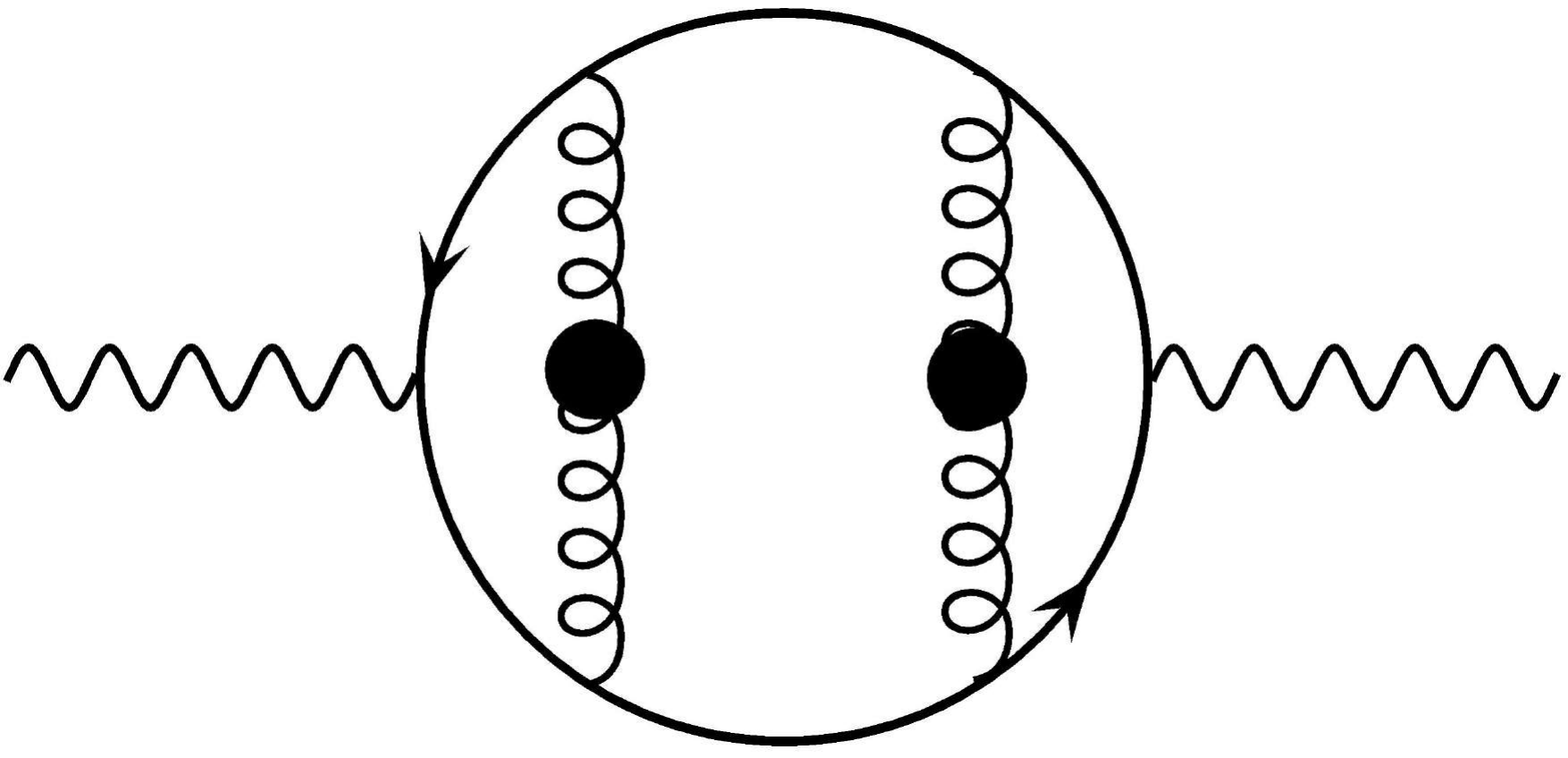}
	\end{center}
	\vspace*{-0.45in}
	\caption{3-loop HTL photon self-energy.}
	\label{3loop_phot}
\end{wrapfigure}

We have observed in previous subsec~\ref{htl_ir_2loop} that 2-loop HTL order contributes  in 1-loop HTL order, i.e., in same order in coupling. The natural question arises what would happen if one goes beyond two loop.  This question was also studied by AGKZ~\cite{Aurenche:1999tq} have also studied by considering at 3-loop HTL diagrams and one such is displayed in Fig.~\ref{3loop_phot}.  
Through power counting it is shown that the 3-loop diagram is proportional to the 2-loop diagram times a factor $g^2T/m_s$, where $m_s$ is the infrared cut-off for the additional exchanged gluon. For longitudinal gluons the $m_s$ is provided by the static electric screening of order $gT$. On the hand, for a transverse gluon this cut-off  is provided by the non-perturbative magnetic screening mass of the order $m_s=g^2T$. Hence, the 3-loop contribution becomes the same order as the 2-loop. This
means that the infinite number of higher-order diagrams contribute to the same order, $\alpha\alpha_s$, as the 2-loop HTL diagram.This argument is basically in the same line that has been utilised by Linde~\cite{Linde:1978px,Linde:1980ts} to indicate that the BPT in QCD breaks down as discussed in the subsec.~\ref{bpt}. We also note that the static magnetic fields are not screened at leading order of HTLpt since 1-loop HTL  transverse gluon self-energy  in all gauges vanishes in the infrared limit $\om\rightarrow 0$, which has been shown in \eqref{qcd24}. 

However, the power counting argument can be used carefully because according to the Kinoshita-Lee-Nauenberg (KLN) theorem~\cite{Kinoshita:1962ur,lee1964degenerate}, there may be cancellations of infrared singularities between different cuts of the diagram.  We know that the sum over the different cuts, indeed, generates a kinematical cut-off. However, this cut-off becomes smaller than the non-perturbative static magnetic screening mass if the virtuality of the photon is small. In particular, for real photons the rate is always sensitive to the static magnetic screening mass. We now note that there is no magnetic screening mass in leading order HTLpt but if one considers the nonperturbative magnetic screening mass for transverse gluons then beyond 2-loop order  infinitely, many higher-order diagrams contribute to the same order,  $\alpha\alpha_s$, as the 2-loop HTL diagram. The real photon rate can be calculated using HTLpt that takes into account the screening of the chromoelectric scale but breaks down at 3-loop order due to Linde problem. On the other hand, however, in weak coupling limit~\cite{Peitzmann:2001mz} the dileptons with an invariant mass greater than $g^2T$,  the KLN  cut-off  becomes relevant and their rate can be calculated perturbatively.

	\section{Mesonic Correlation Function}\label{meson}
	\vspace{-0.2cm}
Although lattice calculations have achieved satisfactory precision in determining hadron properties in vacuum conditions, little is known about basic hadronic parameters in a thermal medium through such first principle calculations. For instance, masses and widths at finite temperature remain largely unexplored. Typically, lattice calculations of these quantities at zero temperature involve computing correlation functions in Euclidean time, a method that extends to spatial correlation functions at finite temperature. While spatial correlation functions indicate abrupt changes in medium hadron properties above $T_c$~\cite{Detar:1987hib}, they offer only indirect evidence for modifications in, for instance, hadron masses and widths.
A more suitable approach involves a detailed analysis of temporal correlation functions at finite temperature~\cite{Hashimoto:1992np,Boyd:1994np}, restricted to the Euclidean time interval $[0,1/T]$, with interesting information on hadronic states encoded in the spectral functions for these correlators~\cite{Shuryak:1993kg}. 
 
Current lattice calculations demonstrate significant changes in the behaviour of temporal correlation functions in the high temperature QCD plasma~\cite{Karsch:2000gi,Ding:2010ga,Karsch:2001uw,Hashimoto:1992np,Boyd:1994np,Shuryak:1993kg,Boyd:1995cw,QCD-TARO:2001jaq}. However, notably near the transition temperature $T_c$, the correlation functions clearly deviate from those of freely propagating quarks.Consequently, it is crucial to understand in how far the temporal correlation functions carry information about the existence or non-existence of bound states or resonances in the plasma phase. Various calculations within the framework of low energy effective models also suggest strong modifications of hadron properties~\cite{Brown:1991kk,Friman:1997tc,Rapp:1997fs,Thoma:1999nm,Islam:2014sea} and consequently also 
of the spectral functions~\cite{Chiku:1998kd}. However, it is difficult in such model calculations to deal with the quark substructure of hadrons, which will become important at high temperature where one expects to find indications for the propagation of almost free, massless quarks.
Eventually it is the hope, that spectral methods~\cite{Jarrell:1996rrw}, which have proven successful in characterising hadron spectral functions at zero temperature~\cite{Nakahara:1999vy}, can also be extended at finite temperature. Particularly in the high temperature limit, well above the QCD phase transition temperature, it then might be appropriate to compare lattice calculations for temporal hadron correlators also with perturbative calculations. To some extent, non-perturbative information can be incorporated into such an analytic calculation by employing the HTL
resummation scheme~\cite{Braaten:1989mz}, which is successful in describing
many properties of hot and dense QCD matter~\cite{Thoma:1997bi}.

In this section, we will analyse~\cite{Karsch:2000gi} the structure of temporal correlation functions and their spectral functions within the context of HTLpt at zero spatial momentum. Indeed, the spectral functions, which one will extract from an analysis of  temporal correlators, are closely related to quark-antiquark annihilation processes in the quark-gluon plasma. For the vector channel this is linked to the dilepton production at high temperature, which has been studied in the HTL-approximation~\cite{Braaten:1990wp} and discussed in details in subsec~\ref{bpy_soft}.
The temperature dependence of the pseudoscalar correlator is related to the chiral condensate. Thermal effects on this as well as the pseudoscalar masses and dispersion relations influence the appearance or suppression of a disoriented chiral condensate which might lead to observable effects in relativistic heavy ion collisions~\cite{Schaffner-Bielich:1998mra}.

\subsection{Generalities}                                                                    
\subsubsection{Thermal meson correlation and spectral function}
\label{meson_corr}
In this subsection  we will present the framework for the calculation of thermal meson correlation functions and spectral function at finite temperature. The correlation function is  constructed  from meson currents 
$J_M (\tau,\bm{\vec x}) =\bar{q}(\tau, \bm{\vec x})\Gamma_M q(\tau, \bm{\vec x})$, 
where $\Gamma_M$ is an appropriate combination of $\gamma$-matrices
that fixes the quantum numbers of a meson channel; {\it i.e.,} $\Gamma_M =
1$, $\gamma_5$, $\gamma_\mu$, $\gamma_\mu \gamma_5$ for scalar,
pseudoscalar, vector and pseudo-vector channels, respectively. 
The two point function for a fixed momentum $\bm{\vec p}$  is defined~\cite{Hashimoto:1992np,Boyd:1994np,Boyd:1995cw} as
\be
G_M(\tau,\bm{\vec p}) &=& \int \frac{d^3\bm{\vec x}}{(2\pi)^3}\langle J_M (\tau, \bm{\vec x}) J_M^{\dagger} (0, \vec{0}) \rangle 
{ e}^{-i \bm{\vec p}\cdot \bm{\vec x}} \, , \label{tempcor0}
\ee
where the Euclidean time is restricted as $\tau \in [0,\beta=1/T]$.  Now, the thermal two-point functions in coordinate space, $G_M(\tau,\bm{\vec x})$,
can be written as~\cite{Karsch:2000gi,Ding:2010ga,Karsch:2001uw,Hashimoto:1992np,Boyd:1994np,Boyd:1995cw} as 
\begin{eqnarray}
G_M(\tau,\bm{\vec x}) &=& 
\langle J_M (\tau, \bm{\vec x}) J_M^{\dagger} (0, \vec{0}) \rangle 
=T \sum_{n=-\infty}^{\infty} \int 
{{ d}^3\bm{\vec p} \over (2 \pi)^3} \;{ e}^{-i(\omega_n \tau+ \bm{\vec p}\cdot \bm{\vec x})}\;
G_M(\omega_n,\bm{\vec p})\, ,
\label{tempcor1}
\end{eqnarray}
 and the Fourier transformed correlation function $G_M(\omega_n,\bm{\vec p})$ is given at the discrete Matsubara modes, 
$\omega_n = 2n \pi T$. The imaginary part of the momentum space correlator
gives the spectral function $\sigma_M(\omega,\bm{\vec p})$, following \eqref{bpy_03} and \eqref{bpy_04} as
\be
G_M(\omega_n,\bm{\vec p}) = -\int_{-\infty}^{\infty} { d}
\omega\; {\sigma_M(\omega,\bm{\vec p}) \over i\omega_n - \omega +i\epsilon} 
\quad \Rightarrow \quad
\sigma_M(\omega,\bm{\vec p}) = {1\over \pi} {\rm Im}\;  G_M(\omega,\bm{\vec p})\, 
\label{tempcor2ab}
\ee
We note that for vector correlator the Lorentz indices are suppressed and can be denoted by $M=(00, ii, v )$ corresponding to (temporal, spatial, vector). One can  write the vector spectral function as  $\sigma_v= \sigma_{00}+\sigma_{ii}$, where $\sigma_{ii}$ is the sum over the three space-space components and $\sigma_{00}$ is the time-time component of $\sigma_{\mn}$.

Using~\eqref{tempcor1} and~\ref{tempcor2ab}, one obtains the spectral representation of the thermal correlation functions in coordinate space at fixed momentum can be obtained as    
\begin{equation}
G_M(\tau,\bm{\vec p}) =  \int_{0}^{\infty} { d} \omega\; 
\sigma_M (\omega,\bm{\vec p})\; 
{{\cosh}\left(\omega (\tau - \beta/2)\right) \over {\sinh} (\omega \beta/2)}\, .
\label{tempcor3}
\end{equation}
We note that the Euclidean correlation function is usually restricted to vanishing three momentum, $\bm{\vec p}=0$, in the analysis of lattice gauge theory and one can write $G_M(\tau T)=G_M (\tau, {\vec {\bm 0}})$.

\subsubsection{ Density fluctuation and its response}
\label{density_fluc}
\vspace{-0.2cm}
Let $ {\cal O}_{\alpha }$ be a Heisenberg operator where $\alpha$  may be associated with a degree of freedom in the system.
In a static and uniform external field $ {\cal F}_{\alpha } $, the (induced)
expectation value of the operator ${\cal O}_\alpha \left( 0,
\bm{\vec x} \right) $ is written~\cite{Hatsuda:1994pi,Kunihiro:1991qu} as
\begin{equation}
\phi _{\alpha }\equiv \left\langle {\cal O} _{\alpha }\left
( 0,\bm{\vec x}\right) \right\rangle _{\cal F} = \frac{{\rm Tr}\left
[ {\cal O} _{\alpha }\left( 0,\bm{\vec x}\right) e^{-\beta \left
( {\cal H}+{\cal H}_{ex}\right) }\right] }{{\rm Tr}\left[ e^{-\beta
\left( {\cal H}+{\cal H}_{ex}\right) }
\right] }=\frac{1}{V}\int d^{3}x\, \left\langle {\cal O} _{\alpha }
\left( 0,\bm{\vec x}\right) \right\rangle \: , \label{dfr1}
\end{equation}
where translational invariance is assumed, $V$ is the volume of the
system and
${\cal H}_{ex} $ is given by
\begin{equation}
{\cal H}_{ex}=-\sum _{\alpha }\int d^{3}x\, {\cal O} _{\alpha }\left( 0,
\bm{\vec x}\right) {\cal F}_{\alpha }\: .\label{dfr2}
\end{equation}
The (static) susceptibility $ \chi _{\alpha \sigma } $ is defined as
the rate with which the expectation value changes
in response to that external field,
\begin{eqnarray}
\chi _{\alpha \sigma }(T) & = & \left. \frac{\partial \phi _{\alpha }}
{\partial {\cal F}_{\sigma }}\right| _{{\cal F}=0}
  = \beta \int d^{3}x\, \left\langle {\cal O} _{\alpha }\left
( 0,\bm{\vec x}\right) {\cal O} _{\sigma }
( 0,\vec{\bm 0})
 \right\rangle \: , \label{dfr3}
\end{eqnarray}
where
$\langle {\cal O}_\alpha (0,{\vec {\mathbf x}})
{\cal O}_\sigma(0,{\vec {\bm 0}})\rangle $
is the two point correlation function with operators evaluated
at equal times. There is no broken symmetry as
$
\left.\left\langle {\cal O} _{\alpha }
\left ( 0,\bm{\vec x}\right ) \right\rangle
\right|_{{\mathcal F}\rightarrow 0}
 =\left. \left\langle {\cal O} _{\sigma } 
( 0,\vec{\bm 0}) \right\rangle 
\right |_{{\mathcal F}\rightarrow 0}=0  \ . \label{dfr4}
$
\subsection{Free Meson Spectral and Correlation Function}
\label{meson_spec_free}
\begin{figure}[h]
\begin{center}
	\includegraphics[scale=0.45]{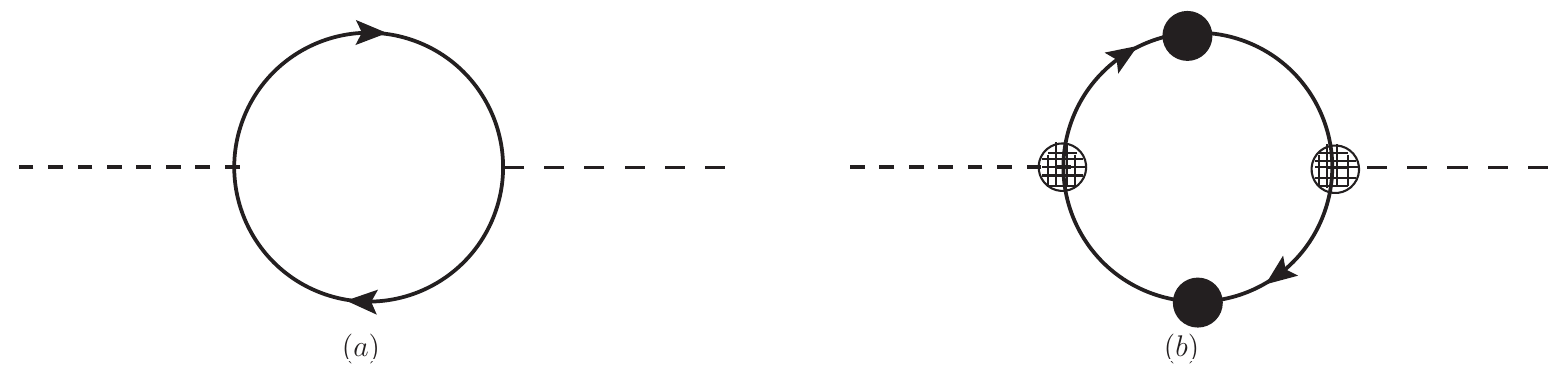}
\end{center}
\vspace{-0.45cm}
\caption{The meson self-energy diagrams with free quark propagator vertices (a)   and  in the
HTL approximation (b). 
}
\label{fig:bubble}
\end{figure}

\vspace{-0.2cm}
The starting point for a calculation of the meson spectral functions and the meson correlation functions is the momentum space representation of the latter~\cite{Florkowski:1993bq}. In leading-order perturbation theory, one evaluates the self-energy diagram depicted in Fig.~\ref{fig:bubble}(a), 
where the internal quark lines represent a bare quark propagator $S_0(k_0, \bm{\vec k})$. This propagator can be expressed in terms of its spectral function $\rho^f (\omega, \bm{\vec k},m)$ and is conveniently written in Saclay representation as:
\begin{equation}
\hspace*{-1cm}
S_0(k_0, \vec{k}) = -(\gamma_0 \; k_0 - \vec{\boldsymbol \gamma}\cdot \bm{\vec k} + m )\;
\int_0^{1/T} \;{ d}\tau {\rm e}^{k_0\tau} \int_{-\infty}^{\infty}
{\;{d}\omega} \, 
\rho^{f}(\omega, \bm{\vec k},m)\; [1-n_F(\omega)]\; { e}^{-\omega \tau},  
\label{fspec1}
\end{equation}
where $k_0=(2n+1)i\pi T$, and the free spectral function 
\begin{equation}
\rho^f (\omega, \bm{\vec k}, m) = {1 \over 2\omega}\; 
\Big[ \delta (\omega - \omega_k) + \delta (\omega + \omega_k) \Big]\quad ,
\label{fspec2}
\end{equation} 
with $\omega_k = \sqrt{\bm{\vec k}^2 + m^2}$.
The thermal meson spectral functions are then obtained from \eqref{tempcor2ab}, that is from the imaginary  part of the correlation functions in momentum space,
\begin{equation}
G_M(\omega, \bm{\vec p}) = 2 N_c 
T \sum_{k_0} \int {{d}^3k \over (2 \pi)^3} \;{\rm Tr}\biggl[ \Gamma_M \,
S_0(k_0,\bm{\vec k})) \, \Gamma_M^{\dagger} 
\, S_0^{\dagger} (\omega - k_0, \bm{\vec p}-\bm{\vec k}) \biggr]
\label{fspec3}
\end{equation}
In the case of free fermions this is easily evaluated. In the limit of vanishing external momentum one finds for the spectral functions,  
\begin{eqnarray}
\sigma_{ M}^{f} (\omega, \bm{\vec p}=0) &=& 
{N_c \over 4 \pi^2} \;
\Theta (\omega- 2m) \;  \omega^2  \;\tanh (\omega/4T)
\; \sqrt{1-\biggl({2m\over \omega}\biggr)^2 }
\,  \biggl(a_M + \biggl({2m \over \omega}\biggr)^2 b_M\biggr) 
\quad,\quad 
\label{fspec4}
\end{eqnarray} 
where different quantum number channels are characterised by the pair of parameters $(a_M, b_M)$. For the scalar (s), pseudoscalar (ps), vector (v) and pseudo-vector (pv) channels they are given by (-1,1), (1,0), (2,1) and (-2,3), 
respectively\footnote{In the vector and pseudo-vector cases
we denote by $\sigma_M$ the trace over the Lorentz indices of
$\sigma_M^{\mu \nu}$.}.  
In the massless limit the spectral functions are chirally symmetric,
$|\sigma_{\text{ps}}| = |\sigma_{\text s}|$ and $|\sigma_{\text pv}| = |\sigma_{\text v}|$.
In this case the remaining integral in \eqref{tempcor3} can  
be done analytically and one obtains for example in the pseudoscalar case~\cite{Florkowski:1993bq},
\begin{equation}
\tilde{G}_{\text {ps}} (\tilde{\tau},\bm{\vec p}=0) = 2\pi N_c\; ( 1-2\tilde{\tau})
{1+\cos^2(2\pi\tilde{\tau} ) \over \sin^3 (2\pi\tilde{\tau} )} 
+ 4 N_c \; {\cos(2\pi\tilde{\tau} ) \over \sin^2 (2\pi\tilde{\tau} )}\quad ,
\label{spec5}
\end{equation}
where $\tilde \tau =\tau T$ and $\tilde{G}_M (\tilde{\tau},\bm{\vec p})={G}_M (\tilde{\tau},\bm{\vec p})/T^3$ are the dimensionless quantities.
\vspace{-0.2cm}
\subsection{Meson Spectral Function in 1-loop HTL approximation}
\label{meson_spec_htl}
\vspace{-0.2cm}
We now aim to move beyond the free quark approximation and explore in-medium quark propagators. In the weak coupling limit ($g\ll 1$), it is consistent to employ a HTL resummed quark propagator (and quark-meson vertex) when the quark momentum is soft, i.e., of order $gT$. A consistent way in the weak coupling limit ($g\ll 1$)
is the use of a HTL resummed quark propagator (and quark-meson vertex) if the quark momentum is soft, i.e. of order $gT$. These corrections to the spectral function are of the same order as the free spectral function for small energy $\omega$. Using the HTL resummation technique important medium effects of the medium such as the effective
quark masses and the Landau damping are taken into account. 

As discussed in subsec~\ref{fgse}  that the HTL resummed quark propagator is chiral symmetric in spite of the appearance of an effective quark mass~\cite{Weldon:1982bn,Klimov:1982bv}. 
In the following we will ignore the parity of the meson states and will generically talk about scalar and vector channels
only. However, we will  show results for pseudoscalar and vector spectral functions and correlators which in our notation are
strictly positive.  To leading order one has to evaluate the HTL self-energy diagram shown in Fig.~\ref{fig:bubble}(b).
The HTL resummed quark propagator is obtained in~\eqref{qcd34}. We will discuss about the vertex for appropriate cases below.
\subsubsection{Pseudoscalar meson spectral function}
\label{ps_htl}
\vspace{-0.2cm}
In the case of the scalar and pseudoscalar spectral function the 
vertices are given by the bare vertices $\Gamma_s = 1$ and 
$\Gamma_{\text{ps}}=\gamma_5$, since contributions from a HTL resummation to
the vertices are suppressed in this case,  i.e.,  lead to higher order 
corrections, as discussed for the scalar case in the Yukawa theory~\cite{Thoma:1994yw}
and scalar QED~\cite{Kraemmer:1994az}. In the case of QCD the absence of HTL 
corrections for scalar as well as pseudoscalar vertices has been shown 
in the context of kinetic theory~\cite{Blaizot:1993be}.

In \eqref{fspec3} we use HTL resummed propagator $S$ from \eqref{qcd34}, pseudoscalar vertex $\Gamma_{\text{ps}}=\gamma_5$, performing trace over Dirac
matrices and then taking the imaginary part using BPY prescription in \eqref{bpy12}, one can obtain from \eqref{tempcor2ab} as~\cite{Karsch:2000gi} 
\begin{eqnarray}
\hspace{-1.0cm}\sigma_{\rm ps} (\omega,\bm{\vec p})
\hspace{-0.3cm} &=& \hspace{-0.3cm}
2N_c ({e}^{\omega/T} -1) \int {{ d}^3k \over (2\pi )^3} \; 
\int_{-\infty}^{\infty} { d}x \,  { d}x'
\; n_F(x) n_F(x') \delta(\omega -x-x')\nn
&\times& \hspace{-0.3cm} \left\{ (1-\bm{\hat q}\cdot \bm{\hat k})
[\rho_+(x,k) \rho_+(x',q) + \rho_-(x,k) \rho_-(x',q)]
+(1+\bm{\hat q}\cdot \bm{\hat k})
[\rho_+(x,k) \rho_-(x',q) + \rho_-(x,k) \rho_+(x',q)]\right\},
\label{spec_ps}
\end{eqnarray}
where $\bm{\vec q}=\bm{\vec p}-\bm{\vec k}$ and  the HTL resummed 
spectral functions for massless quarks is given  in \eqref{spec2}  in the subsec~\ref{spec_prop},  in which 
$m_{\rm{th}}^{\rm e}$ is to be replaced by $m_{\rm{th}}^{\rm q}$ and  have both pole and cut contributions.
Since the thermal meson correlation function has product of two  spectral functions, they will receive pole-pole 
(PP) and and cut-cut (CC) contributions as
functions
$
\sigma_{\rm ps}=\sigma^{\rm PP}_{\rm ps}+\sigma^{\rm PC}_{\rm ps}+\sigma^{\rm CC}_{\rm ps}\, . 
\label{spec_decomp}
$
One can obtain from \eqref{spec_ps} the PP,  PC and CC contributions following subsec~\ref{bpy_soft} and 
quote below explicit expressions~\cite{Karsch:2000gi} of those to the pseudoscalar spectral function. 

The pole-pole contribution ($\bm{\vec p}=0$) is given by
\begin{eqnarray}
	\hspace{-1.0cm}\sigma^{\rm PP}_{\rm ps} (\omega)\hspace{-0.3cm} &=&\hspace{-0.3cm} 
	{N_c\over 2\pi^2} \left(m_{{\rm th}}^{\text q}\right)^{-4}  (e^{\omega/T} -1)
	\biggl[ n_F^2(\omega_+(k_1)) \bigl( \omega_+^2(k_1) -k_1^2 \bigr)^2 {k_1^2\over 2| \omega_+'(k_1)|} 
	+2 \sum_{i=1}^2\bigg\{
	n_F(\omega_+(k_2^i)) \bigl[1-n_F(\omega_-(k_2^i))\bigr]\nn
	&&\hspace{-.7cm}\times \bigl(\omega_-^2(k_2^i) -(k_2^i)^2 
	\bigr) \bigl(\omega_+^2(k_2^i) -(k_2^i)^2 \bigr) \,  {(k_2^i)^2\over | \omega_+'(k_2^i)-\omega_-'(k_2^i)|}\bigg\}
	+\hspace{-0.0cm} \sum_{i=1}^2
	n_F^2(\omega_-(k_3^i)) \bigl(\omega_-^2(k_3^i) -(k_3^i)^2 \bigr)^2 
	{(k_3^i)^2\over 2| \omega_-'(k_3^i)|}\; \biggr] \, . \label{spec_pp_ps} 
\end{eqnarray}
Here $\omega_\pm (k)$ denote the quark dispersion relations for the ordinary 
quark (+) and the plasmino (-) branch as discussed in Fig.~\ref{disp_quark_htl}, 
$k_1$ is the solution of $\omega - 2 \omega_+(k_1) = 0$, $k_2^i$ and $k_3^i$
are the solutions of  $\omega - \omega_+(k_2^i)+\omega_-(k_2^i) =0$ and  
$\omega - 2  \omega_-(k_3^i) = 0$, respectively. Note that for small momenta the last two
equations can each have two solutions. Furthermore,  
$\omega_\pm'(k)\equiv ({d} \omega_\pm (x) / {d} x)|_{x=k}$.

For the pole-cut contribution we find
\begin{eqnarray}
&&\hspace{-0.7cm}\sigma^{\rm PC}_{\rm ps} (\omega) =
{2N_c\over  (\pi m_{{\rm th}}^{\text q})^2} ({e}^{\omega \over T} -1)\! \int_0^\infty\!\! dk  k^2 
\hspace{-0.cm}  \biggl[
\Theta (k^2-(\omega-\omega_+)^2) \, n_F(\omega-\omega_+)  n_F(\omega_+) \beta_+(\omega-\omega_+,k)
(\omega_+^2-k^2) +\Theta (k^2-(\omega-\omega_-)^2) \nn
&&\hspace{-0.5cm}\times
 n_F(\omega-\omega_-) n_F(\omega_-) \, \beta_-(\omega-\omega_-,k)\,
(\omega_-^2-k^2) 
\hspace{-0.05cm}~+
\Theta (k^2-(\omega+\omega_-)^2) \, n_F(\omega+\omega_-)\, [1-n_F(\omega_-)]\, \beta_+(\omega+\omega_-,k)\,
(\omega_-^2-k^2) \nn
&&\hspace{5cm}
+\Theta (k^2-(\omega+\omega_+)^2)\, n_F(\omega+\omega_+)\, [1-n_F(\omega_+)]\, \beta_-(\omega+\omega_+,k)\,
(\omega_+^2-k^2) \biggr] \, . \label{spec_ps_pc}
\end{eqnarray}
Finally we obtain for the cut-cut contribution
\begin{eqnarray}\hspace{-1.0cm}\sigma^{\rm CC}_{\rm ps} (\omega)\hspace{-0.3cm} &=&
\hspace{-0.3cm}
{2N_c\over \pi^2} ({e}^{\omega/T} -1) \int_0^\infty  dk\; k^2 \int_{-k}^{k}
dx \; n_F(x)\, n_F(\omega-x) \, \Theta(k^2-(\omega -x)^2)  \nn
&&\hspace{3.0cm} \times \biggl[
\beta_+(x,k) \, \beta_+(\omega-x,k) + \beta_-(x,k)\, \beta_-(\omega-x,k) \biggr]  \, . \label{spec_ps_cc}
\end{eqnarray}
\subsubsection{Vector meson spectral function}
\label{vec_htl}
The corresponding relation for the vector spectral function, which also includes HTL-vertex contributions,
is related to the dilepton production rate 
\begin{equation}
{\sigma_{\rm v}} (\omega , \bm{\vec p}=0) = \frac{18\pi^2N_c}{5\alpha^2}
\> \left ({ e}^{\omega /T}-1\right )\> \omega^2\>  \left. \frac{dR}
{d^4x \, d \omega \, d^3\bm{\vec p}}\right |_{\bm{\vec p}=0}\, . 
\label{spec_v}
\end{equation}
The dilepton rate has been calculated in \eqref{soft10} following Ref.~\cite{Braaten:1990wp} in the HTL approximation and  the 
physical processes related to the pole-pole, pole-cut and cut-cut contributions have been discussed in details in subsec~\ref{bpy_soft}. In particular, there are characteristic peaks that show up in the pole-pole contribution (van Hove singularities~\cite{hove}) are related to the dispersion relation\footnote{In Refs.\cite{Mustafa:1999cp,Peshier:1999dt,Bandyopadhyay:2015wua} it has been argued that the full in-medium quark propagator leads in general to two branches in the dispersion relation, of which one exhibits a minimum.}   
in plasmino branch in Fig.~\ref{disp_quark_htl}. They are caused by a diverging density 
of states which is inversely proportional to the derivative of the dispersion relations, $\omega'_{\pm}(k)$. Owing to the minimum in the plasmino branch these derivatives vanish. We also note that such structures are also found in mesonic spectral function~\cite{Karsch:2000gi}. Apart from values close to the van Hove singularities in $\sigma^{\rm PP}$ 
one finds that the cut contributions dominate the spectral function for small  values of $\tilde{\omega}=\om/T$, e.g.  for $\tilde{\omega}\; \lsim \; g(T)$.

\begin{figure}[htb]
\begin{center}
\includegraphics[width=8cm, height=5cm]{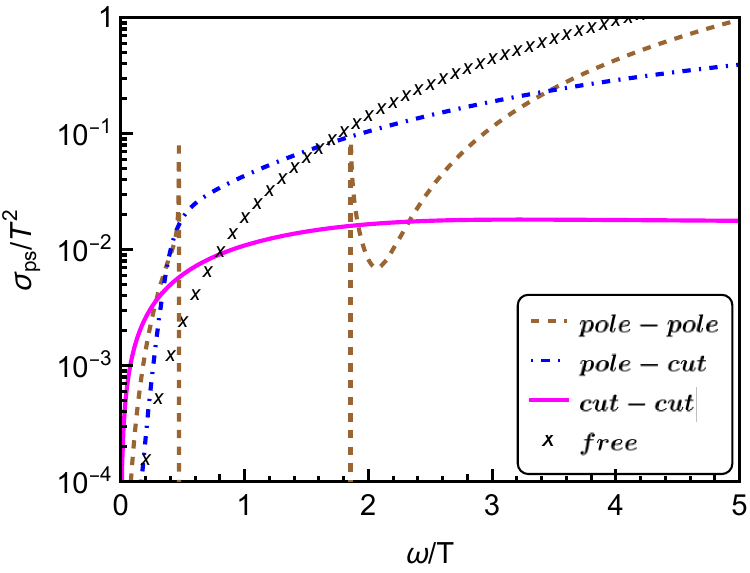}
\includegraphics[width=8cm, height=5cm]{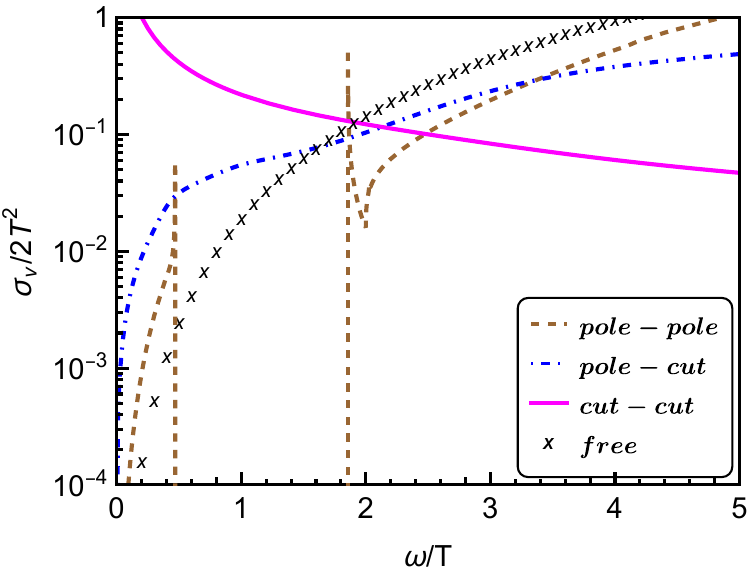}
\end{center}
\vspace{-0.55cm}
\caption{The pole-pole, pole-cut and cut-cut contributions to the 
pseudoscalar (a) and vector (b) spectral function for $\tilde{m}^{\rm q}_{\rm th}={m}^{\rm q}_{\rm th}/T=1$.
The crosses show the free meson spectral function.}
\label{fig:sigma}
\end{figure}

In Fig.~\ref{fig:sigma}, we illustrate the pole $({\sigma}^{\rm PP}/T^2)$ and cut $({\sigma}^{\rm PC}/T^2,~~{\sigma}^{\rm CC}/T^2)$ contributions to the scalar (Fig.~\ref{fig:sigma}(a)) and vector (Fig.~\ref{fig:sigma}(b)) spectral functions for 
the scenario where $\tilde{m}^{\rm q}_{\rm th}={m}^{\rm q}_{\rm th}/T=1$, extrapolating the HTL results obtained in the 
weak coupling limit to $g=\sqrt{6}$. Initially, it's unclear whether the HTL resummation method can be applied to such substantial couplings, though extrapolation to realistic values of the coupling constant has been used for various QGP observables~\cite{Thoma:1995ju}. Nevertheless, this approach enables us to qualitatively explore the influence of medium effects, and as we will demonstrate, the minor corrections we observe validate our choice of couplings. 
As depicted in Fig.~\ref{spec_v}, the pole and cut contributions affect the spectral function differently. The former predominates for large $\tilde{\omega}$. The deviations of $\sigma^{\rm HTL}$ from the free 
spectral function in this energy range, along with the threshold for $\tilde{\omega} \simeq 2~\tilde{m}^{\rm q}_{\rm th}$, is due to the presence of a finite thermal mass in the quark dispersion relation, reflecting the nearly free propagation of two quarks in the plasma. Additional interactions of these quarks with the thermal medium (Landau damping)
are represented by the cut contributions, resulting in an enhancement over the free spectral functions for small  $\tilde{\omega}$. Moreover, we observe that the pole-pole and pole-cut contributions to the spectral functions exhibit similar behaviour in the scalar and vector  channels. However, the cut-cut contribution behaves differently at low energies. While it diminishes for small $\tilde{\omega}$ in the pseudoscalar channel, it linearly diverges in the vector channel. This divergence  can be traced back to the structure of the effective HTL-vertex, which contains a collinear singularity~\cite{Baier:1993zb}. Consequently, infinitely many higher-order diagrams in the HTL expansion contribute to the same order 
in the coupling constant~\cite{Aurenche:1998nw,Aurenche:1999tq}, indicating that the low-frequency part of the vector spectral functions is inherently non-perturbative.

In this context we  note that the procedure used in Ref.~\cite{Karsch:2000gi} satisfactorily describes the soft fermionic modes, but its application to the propagation of hard quarks is not reliable. An improved version of the so-called next-to-leading approximation (NLA) scheme, which allows a better treatment of the hard fermionic modes, is proposed in Ref.~\cite{Alberico:2004we}. The meson spectral functions have also been calculated in leading order HTLpt and beyond HTLpt  for non-zero momentum~\cite{Alberico:2006wc} and chemical potential~\cite{Czerski:2008zz}. 
We further note that the vector channel spectral function and the dilepton production rate from a QCD plasma at a temperature above a few hundred MeV are evaluated~\cite{Laine:2013vma} up to next-to- leading order (NLO) beyond HTLpt but  including their dependence on a non-zero momentum with respect to the heat bath.  

\subsection{Thermal Meson Correlators in HTL Approximation}  
\subsubsection{Thermal pseudoscalar meson correlation function}
\label{corr_ps}
We will now compute the temporal correlators from the spectral functions derived in the last section using (\ref{ps_htl}). Although the temporal correlators are dominated by hard energies $\omega \sim T$, we use the spectral functions, calculated within the HTL resummation scheme, to account for medium effects from the quark propagator at least semi-empirically. 

\begin{figure}[htb]
\begin{center}
\includegraphics[width=8cm, height=5cm]{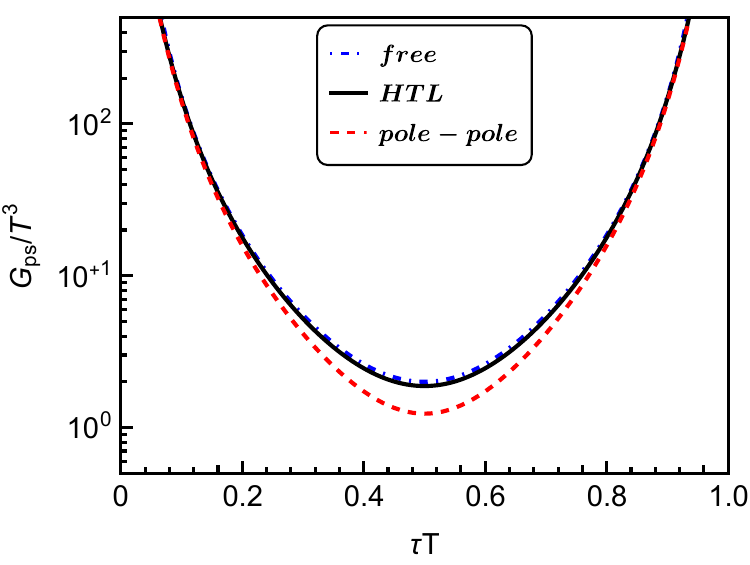} \hspace*{0.25in}
\includegraphics[width=8cm, height=5cm]{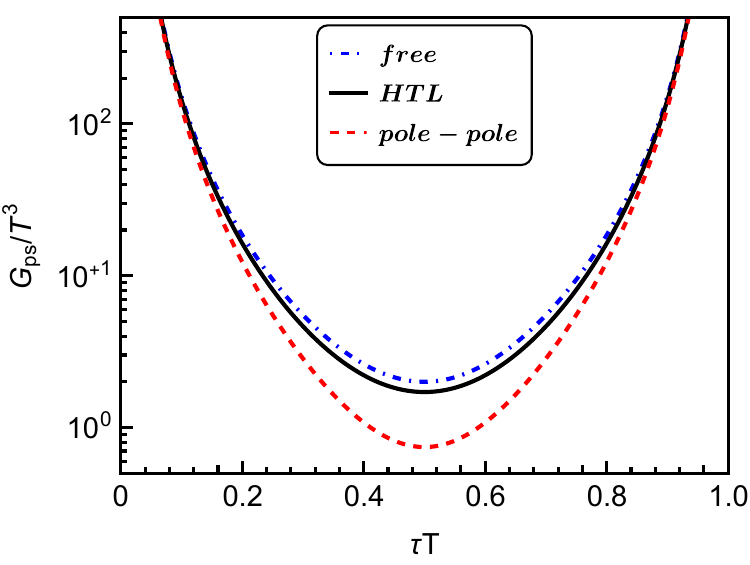}
\end{center}
\vspace{-0.5cm}
\caption{The thermal pseudo-scalar meson correlation function in the HTL approximation for $\tilde{m}^{\rm q}_{\rm th}=1$ (left) and $\tilde{m}^{\rm q}_{\rm th}=2$ (right). The curves shown the complete thermal correlator (middle line), the correlator constructed from $\sigma^{\rm PP}_{\rm s}$ only (lower line) and the free thermal correlator 
(upper line).}
\label{fig:meson}
\end{figure}

The interplay between pole and cut contributions to the HTL resummed spectral functions also impacts the behaviour of thermal meson correlation functions. The presence of a non-vanishing thermal quark mass tends to results in a more rapid decrease of the correlator in Euclidean time compared to the free massless correlator. However, the enhancement of the low-energy part, attributable to the cut contributions, counteracts this trend. This phenomenon is illustrated by the behaviour of thermal meson correlation functions in the scalar channel, as depicted in Fig.~\ref{fig:meson} for $\tilde{m}^{\rm q}_{\rm th}=1$ and $2$.   
As anticipated, the correlator constructed solely from the pole-pole contribution is steeper than the free correlator and is strongly dependent on scaled thermal mass $\tilde{m}_T$. Nevertheless, the cut contributions enhance the low energy contributions in the spectral function, thereby flattening the correlator once again. Surprisingly, for $\tilde{m}^{\rm q}_{\rm th} \simeq 1$ this seems to compensate almost completely the deviations from the free correlator introduced by the pole contributions. The difference between the free and HTL-resummed correlators is most prominent for $\tau T \simeq 1/2$, where the contribution from the low energy regime in the spectral function is most significant. Conversely, as $\tau T $ approaches to $0$ and $1$, the free and HTL resummed correlators converge, as $\lim_{\omega\rightarrow \infty} \sigma^{\rm HTL}(\omega)/\sigma^{\rm free} (\omega) = 1$.  
\subsubsection{Thermal vector meson correlation function}
\label{corr_v}
\vspace{-0.2cm}
 In subsec~\ref{meson_corr}, we have previously discussed that calculating of the vector correlators within the HTL method requires the use of effective quark-meson vertices, as illustrated in Fig.~\ref{fig:bubble}(b). This results to a linear divergence of the spectral function in the vector channel at low frequencies, which in turn renders the temporal correlator infrared divergent. In fact, although the scalar correlation functions are infrared finite, it is to be expected, that also in this case the low frequency part of the HTL resummed spectral functions will be modified significantly from contributions of higher order diagrams. It thus seems to be reasonable to consider modified correlation functions, which are less sensitive to details of the low frequency part of the spectral functions. We therefore define the subtracted correlators
$
\Delta \tilde{G}_M (\tau) \equiv \tilde{G}_M (\tau) - \tilde{G}_M (\beta/2)~.
\label{subG}
$
In the subtracted correlation functions the infrared divergences are eliminated. They are well-defined in the scalar as well as in the vector  channels. 
\begin{figure}[h]
\begin{center}
\includegraphics[width=8cm, height=5cm]{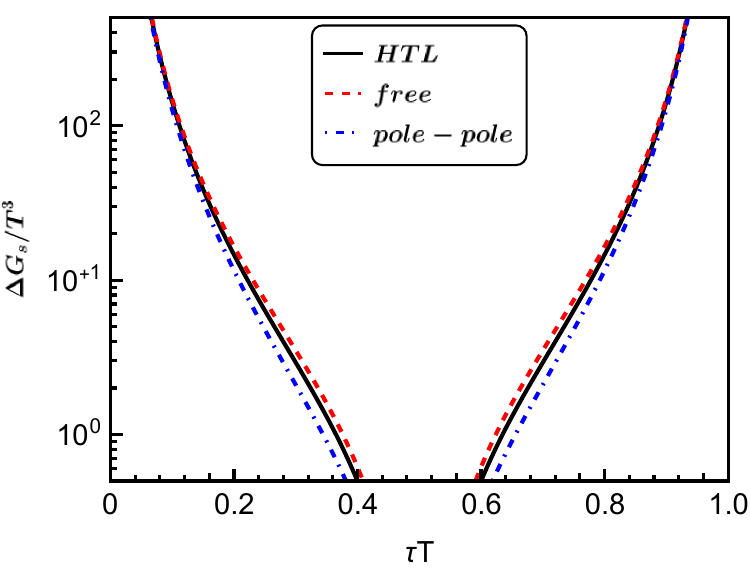} \hspace*{0.25in}
\includegraphics[width=8cm, height=5cm]{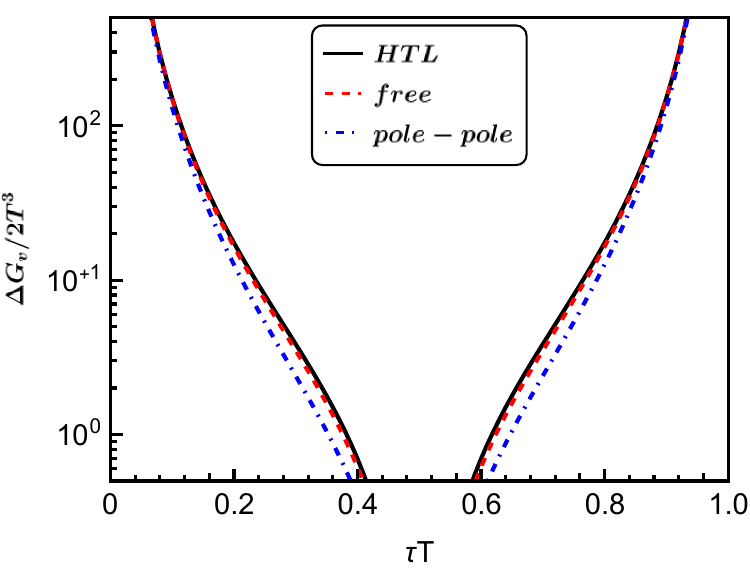}
\end{center}
\vspace{-0.6cm}
\caption{The subtracted thermal pseudo-scalar (left) and vector (right) meson 
correlation functions in HTL approximation for $\tilde{m}_T=2$.  The curves illustrate the complete thermal correlator (middle line), the correlator constructed from $\sigma^{\rm PP}_{\rm ps}$ only (lower line), and the free thermal correlator  (upper line).}
\label{fig:subtracted}
\end{figure}
In Fig.~\ref{fig:subtracted} we compare the HTL resummed subtracted correlation functions with corresponding results for the free case. This comparison reveals that following the elimination of the infrared divergent parts, the structure of the pole and cut contributions is similar in scalar and vector channels. Notably, the vector correlator appears to be even closer to the leading-order perturbative (free) correlator than the scalar correlation function. However, the primary distinction between the scalar and the vector channel using the HTL approximation lies in the differing behaviour of the cut-cut contribution to the spectral functions at low energies. Specifically, the vector spectral function diverges in the infrared limit, resulting in a singular expression for the vector correlation function. This feature is in agreement with the observation that the dilepton production rate, closely related to the vector spectral function, cannot be computed within the HTL improved perturbation scheme for small invariant masses of the order of $g^2T$.
\vspace{-0.2cm}
\subsection{Quark Number Susceptibility and Temporal Euclidean Correlation Function}
\subsubsection{Definition}
\label{qns_temp_def}
\vspace{-0.2cm}
The Quark Number Susceptibility (QNS) is a measure of the response of the quark number density with infinitesimal change in the quark chemical potential, $\mu+\delta\mu $. Under such a situation the external field, ${\cal F}_\alpha$, in ({\ref{dfr2}}) can be identified as the quark chemical potential and the operator ${\cal O}_\alpha$ as
the temporal component ($J_0$) of the vector current,  $J_\sigma(t,{\bm {\vec x}})= \overline{\psi} \Gamma_{ \sigma}\psi$, where $\Gamma^\sigma$ is in general a three point function. Then the QNS for a given quark flavour follows from~(\ref{dfr3}) as
\begin{eqnarray}
\chi_q(T) &=& \left.\frac{\partial \rho}{\partial \mu}\right |_{\mu=0}
= \left.\frac{\partial^2 {P}}{\partial \mu^2}\right |_{\mu=0}
= \int d^4x \ \left \langle J_0(0,\bm{\vec x})J_0(0,{\vec {\mathbf 0}})
\right \rangle \ =- {\lim_{\bm{\vec p}\rightarrow 0}} 
{\mbox {Re}}\ G_{00}^R(0,{\bm{\vec p}}),
\label{qns1}
\end{eqnarray}
where $G_{00}^R$ is the retarded correlation function. To obtain (\ref{qns1}) in concise form, we have used the fluctuation-dissipation theorem given as
\begin{equation}
G_{00}(\omega,{\bm{\vec p}})=-\frac{2}{1-e^{-\omega/T}} {\mbox{Im}}
G_{00}^R(\omega,{\bm{\vec p}}), \label{qns2}
\end{equation}
and the Kramers-Kronig dispersion relation
\begin{equation}
{\mbox{Re}}G_{00}^R(\omega,{\bm{\vec p}})=\int_{-\infty}^{\infty} 
\frac{d\omega^\prime}{2\pi}
 \frac{{\mbox{Im}}G_{00}^R(\omega,{\bm{\vec p}})}{\omega^\prime-\omega},
\label{qns3}
\end{equation}
where $\lim_{\bm{\vec p}\rightarrow 0}{\mbox{Im}} G_{00}^R
(\omega,{\vec{\mathbf p}})$ is proportional to $\delta(\omega)$ due to the quark number conservation~\cite{Hatsuda:1994pi,Kunihiro:1991qu}. 
Also the number density for a given quark flavour can be written  as
\begin{equation}
\rho=\frac{1}{V}
\frac{{\rm{Tr}}\left [  {\cal N} e^{-\beta \left({\cal H}-\mu {\cal N}
\right )}\right ]}
{{\rm{Tr}}\left [e^{-\beta \left({\cal H}-\mu {\cal N}\right )}\right ]} \frac{\langle {\cal N}\rangle}{V} = \frac{\partial {P}}
{\partial \mu} \ , \label{qns4}
\end{equation}
with the quark number operator, ${\cal N}=\int J_0(t, \bm{\vec x}) \ d^3x =\int {\bar \psi}(x)\Gamma_0\psi(x)d^3x$, and ${P}=\frac{T}{V} \ln {\mathcal Z}$ is the pressure and  ${\mathcal Z}$ is the partition function of a quark-antiquark gas. The quark number density vanishes if $\mu\rightarrow 0$, i.e., there is no broken CP symmetry.  Now, \eqref{dfr3} or \eqref{qns1} indicates that the thermodynamic derivatives with respect to the external source are related to the temporal component of the static correlation function associated with the number conservation of the system. This relation in (\ref{qns1}) is known as {\em the thermodynamic sum rule}.

\noindent  Owing to the quark number conservation the temporal spectral function 
$\sigma_{00}(\omega,{\vec{\mathbf 0}})$ in \eqref{tempcor2ab} becomes
\begin{equation}
\sigma_{00}(\omega,{\vec{\mathbf 0}})=\frac{1}{\pi} {\mbox{Im}} G_{00}^R
(\omega,{\vec{\mathbf 0}})= - \omega \delta(\omega) \chi_q(T) \, . \label{qns5}
\end{equation} 
Using~(\ref{qns5}) in~(\ref{tempcor3}), it is straight forward to obtain
the temporal correlation function~\cite{Haque:2011iz} as
\begin{equation}
G_{00}(\tau T)=-T\chi_q(T), \label{eq7}
\end{equation}
which is proportional to the QNS $\chi_q$ and $T$, but independent of $\tau$.

\subsubsection{Leading order quark number susceptibility in HTL approximation}
\label{lo_qns}
Now considering the fermion part of the  HTL Lagrangian in Eq.~\eqref{htl_qcd_lag1}
\be
{\cal L}_{\textrm{\tiny QCD}}^{\textrm {\tiny HTL}}  = {\cal L}_{\textrm{\tiny QCD}} + {\cal L}_{\textrm {\tiny HTL}} 
= i\bar \psi \slashed D\psi+ i(m^{\rm q}_{\rm{th}})^2 \bar \psi \left \langle \frac{\slashed K}{K \cdot D} \right \rangle_{\bm{\hat k}}  \psi 
. \label{Lhtl_q}
\ee 
We note that the covariant derivative usually contains background field or any source, $j$ depending upon the physical requirement.  To motivate the perspective we define the covariant derivative $D^\nu$ as
\be
D^\nu = \left [{\cal D}^\nu -i\delta^{\nu 0}(j+\delta j) \right] =\left[ {\tilde D}^\nu -i\delta^{\nu 0}\delta j\right]\, ,
\label{co_deriv}
\ee
where ${\tilde D}^\nu={\cal D}^\nu -i\delta^{\nu 0} j$.  We note that ${\cal D}$ contains gauge coupling 
 and $\delta j$ is an infinitismal change in external sourceto which the response of the system can be calculated, as discussed earlier. Later it can be identified with a variation of some physical quantity depending upon the requirement of the system under consideration. 
Now, expanding the HTL term in \eqref{Lhtl_q}, we can write as
\bea
{\cal L}_{\textrm{\tiny QCD}}^{\textrm {\tiny HTL}}(j+\delta j) & =& \bar \psi \left(i\slashed{\tilde D} +\Sigma\right )\psi
+ \delta j \bar \psi \Gamma_0 \psi +   \delta j^2  \bar \psi \frac{\Gamma_{00}}{2} \psi +{\cal O}(\delta j^3) 
, \label{expan}
\eea
where the various N-point functions in coordinate space are generated as
\bea
\Sigma=(m^{\rm q}_{\rm{th}})^2 \left \langle \frac{\slashed K}{iK \cdot \tilde D} \right \rangle, \,\,\,
\Gamma_0= \delta^{\nu0}\gamma_\nu - (m^{\rm q}_{\rm{th}})^2
\left \langle \frac{\slashed KK_\nu \delta^{\nu 0}}{(iK \cdot\tilde D)^2} \right \rangle , \,\,\,
\Gamma_{00}=2(m^{\rm q}_{\rm{th}})^2
\left \langle \frac{\slashed KK_\nu K_\alpha\delta^{\nu 0}\delta^{\alpha 0}}{(iK \cdot\tilde D)^3} \right \rangle \, ,
\eea
where these functions can easily be transformed into momentum space.

Now, the external source, $j$ is considered as the quark chemical potential, $\mu$. 
One can write the quark-antiquark partition function  from~\eqref{fi6} in presence of a variation of quark chemical potential $\mu+\delta\mu$ as
\begin{eqnarray}
{Z}[\beta;\mu+\delta \mu]=\int {\cal D}[\bar \psi] 
{\cal D}[{\psi}] \ e^{\int_{0}^\beta d\tau \, \int d^3x{\cal L}_{\textrm{\tiny QCD}}^{\textrm {\tiny HTL}}(\psi,{\bar \psi}; \mu+\delta \mu)}\, . \label{i3}
\end{eqnarray}
The pressure can be written~\cite{Haque:2010rb} as
\be
{P}[\beta:\mu+\delta \mu]=\frac{1}{\cal V} \ln{Z}[\beta:\mu+\delta \mu]
\ , \label{i3_1}
\ee
where the four-volume, ${\cal V}=\beta V $ with  $V$ is the three-volume.
Expanding ${P}$ in Taylor series around  $\delta j$ one can write
\begin{eqnarray}
{P}[\beta:\mu+\delta \mu] &=& {P}[\beta;\mu]+ \delta \mu \ 
 {P}'[\beta; \mu+\delta \mu]\Big |_{\delta \mu \rightarrow 0}
+\frac{{\delta \mu}^2}{2}\  {P}''[\beta; \mu+\delta \mu]
\Big |_{\delta \mu \rightarrow 0}
+ \cdots \cdots \ .
\label{i3_1t}
\end{eqnarray}
The first derivative of ${P}$ w.r.t. $\mu$ is related to the conserved density in \eqref{dfr1} whereas the second derivative is related to the conserved density fluctuation in \eqref{dfr3}. 

Now ${\cal P}'$ can be obtained as
\begin{eqnarray}
\!\!\!\!\!
\!\!\!\!\!
\left. \frac{\del{ P}[\beta; \mu+\delta \mu]}{\del \mu}
\right |_{\delta \mu \rightarrow 0}
\!\!\!\!\!
&=&\frac{1}{{\cal V}{ Z}[\beta;\mu]}\ {\int {\cal D}[\bar \psi] 
{\cal D}[\psi] 
\int_{0}^\beta d\tau \, \int d^3x \ {\bar \psi(x)} \Gamma_0[\mu] \psi (x)} \,
e^{ \left ({\int_{0}^\beta d\tau \, \int d^3x\ {\cal L}_{\textrm{\tiny QCD}}^{\textrm {\tiny HTL}}
(\psi,{\bar{\psi}}; \mu)}\right)} \, .
\label{i3_2} 
\end{eqnarray}
The full HTL quark propagator in presence of uniform $\mu$ can be written as
\begin{eqnarray}
{\cal S}_{\alpha\sigma}[\mu](x,x')&=&\frac{
\int {\cal D}[\bar \psi] \, {\cal D}[\psi] \,
\psi_{\alpha}(x)\, {\bar \psi_\sigma(x')}\, \exp \left ({i\int d^4x \, {\cal L}_{\textrm{\tiny QCD}}^{\textrm {\tiny HTL}}
(\psi,{\bar{\psi}}; \mu)}\right)}
{ \int {\cal D}[\bar \psi]\, {\cal D}[\psi] \,
 \exp \left ({i\int d^4x {\cal L}_{\textrm{\tiny QCD}}^{\textrm {\tiny HTL}}
(\psi,{\bar{\psi}};\mu)}\right)} \ .
\label{i3_fp}
\end{eqnarray}
We now note that this full HTL propagator, ${\cal S}[\mu]$, is indeed difficult to calculate and we would approximate it by 1-loop HTL resummed propagator~\eqref{qcd34}, $S^{\textrm q}[\mu]$ and also other HTL 
functions below.

Now using (\ref{i3_fp})  and performing the traces over the colour, flavour,
Dirac and coordinate indices in \eqref{i3_2}  one can write~\cite{Haque:2010rb} the quark number density as
\begin{eqnarray}
\frac{\del{P}[\beta,\mu+\delta\mu]}{\del \mu}\Big |_{\delta \mu=0}
= \ N_c\, N_f\, T  \sum_{\{k_0\}} \!\!\int\! \frac{d^3k}{(2\pi)^3} 
{\Tr}\Big [S^{\textrm q}[\mu](K)\ \Gamma_0[\mu](K,-K;0) \Big ] , 
\label{i3_tr}
\end{eqnarray}
where $`${\Tr}' indicates the trace over  
Dirac matrices,  $N_f$ is the number of massless flavours,
$N_c$ is the number of colours. 

Similarly, we obtain ${ P}''$ as 
\begin{eqnarray}
\left.\frac{\del^2{P}[\beta;\mu+\delta \mu]}{\del \mu^2}
\right|_{\delta \mu \rightarrow 0} 
&=& -N_cN_f T  \sum_{\{k_0\}} \int \frac{d^3k}{(2\pi)^3}  {\Tr} \Big [S^{\textrm q}[\mu](K)\ \Gamma_0[\mu](K,-K;0) \ S^{\textrm q} (K)\ \Gamma_0[\mu](K,-K;0) 
\nn
&&  
- S^{\textrm q}[\mu](K)\  \Gamma_{00}[\mu](K,-K;0,0) 
\Big ] \, , 
\label{i3_d2}
\end{eqnarray}
where we have also used the following relation obtained 
using the inverse of $S^{\textrm q}[\mu]$ and the Ward identity as
\begin{eqnarray}
\!\!\!\!\!\!
\!\!\!\!\!\!
\frac{\del S^{\textrm q}[\mu](K)}{\del \mu}
&=&  -S^{\textrm q}[\mu](K)\ \frac{\del {S^{\textrm q}}^{-1}[\mu](K)}
{\del \mu} \ S^{\textrm q}[\mu](K)
= -S^{\textrm q}[\mu](K)\ \Gamma_0[\mu](K,-K;0)\ S^{\textrm q}[\mu](K)
\ . \label{s0_ii} 
\end{eqnarray}
The quark number density follows from Eq.~\eqref{i3_tr} as
\begin{eqnarray}
\rho(\beta,\mu) &=& 
\frac{\del{P}(\beta,\mu)}{\del \mu}\Big |_{\delta \mu=0}
= \ N_c\, N_f\, T  \sum_{\{k_0\}} \!\!\int\! \frac{d^3k}{(2\pi)^3} 
{\Tr}\Big [S^{\textrm q}(K)\ \Gamma_0(K,-K;0) \Big ] .
\label{i3_tr_f}
\end{eqnarray}
Now, the Eq.~\eqref{i3_d2} represents the quark number susceptibility (QNS) in the limit $\mu\rightarrow 0$ as
\begin{eqnarray}
\chi(\beta)\!\!&=&\!\!\!
\left.\frac{\del{\rho}}{\del \mu}
\right|_{\mu \rightarrow 0}
=\left.\frac{\del^2{P}}{\del \mu^2}
\right|_{\mu \rightarrow 0} 
= - N_cN_f T  \sum_{\{k_0\}} \int \frac{d^3k}{(2\pi)^3}  \nn   
&\times&
{\Tr} \Big [S^\star (K)  \Gamma_0(K,-K;0)S^\star(K)\Gamma_0(K,-K;0)
 - S^\star (K) \Gamma_{00}(K,-K;0,0)\Big ], 
\label{s1}
\end{eqnarray}
where right hand side (RHS) is the temporal correlation functions at the external momentum $P=(\omega_p,|{\vec p}|)=0$, which is related to the thermodynamic derivatives known as thermodynamic sum rule.The first term in the RHS  of (\ref{s1}) is a $1$-loop self-energy whereas the second term corresponds to a tadpole in HTLpt with effective $4$-point HTL functions~\cite{Chakraborty:2001kx,Chakraborty:2003uw} as shown in Fig.~\ref{soft_dilep}. We note that in the LO QNS,  the second term in the RHS can be neglected as it contributes in higher order. However, we note that it is convenient to calculate the quark number density first and then the QNS by taking the $\mu$ derivative of it. This will be equivalent to calculate the LO QNS directly from Eq.~\eqref{s1}.
\subsubsection*{A) Free case}
\vspace{-0.2cm}
It is straightforward to obtain the quark number density for free case from Eq.~\eqref{i3_tr_f}, replacing HTL 2- and 3-point functions $S^{\rm q}$ and $\Gamma_0$, respectively by free one and it becomes
\begin{eqnarray}
\rho^f(\beta,\mu) &=& \frac{\del{P^f}(\beta,\mu)}{\del \mu}
= \ N_c\, N_f\, T  \sum_{k_0=(2n+1)i\pi T +\mu}  \int\! \frac{d^3k}{(2\pi)^3} 
{\Tr}\Big [S^{\textrm q}_0 \ \gamma_0  \Big ] \, .
\label{nod_free}
\end{eqnarray}
%
%
Using free propagator $S_0^{\rm q}$ is given in \eqref{gse25} and performing the trace over Dirac matrices, one can obtain
\begin{eqnarray}
\rho^f(T,\mu)=2N_cN_fT\int\frac{d^3k}{(2\pi)^3}\sum_{k_0=(2n+1)\pi i T+\mu}
\left[\frac{1}{k_0-k}+ \frac{1}{k_0+k}\right] .\label{f4}
\end{eqnarray}
To perform the frequency sum, we use the standard technique of contour integration as discussed in subsec~\ref{freq_sum} 
\begin{equation}
\frac{1}{2\pi i}\oint_C \, F(k_0) \, \frac{\beta}{2}\mbox{tanh}\left(\frac{\beta (k_0-\mu)}{2}
\right)dk_0 =\frac{\beta}{2}\ \frac{1}{2\pi i}\times 
(-2\pi i)\sum {\mbox{Residues}}\ . 
\label{f5}
\end{equation}
It can be seen that the first term of (\ref{f4}) has a simple pole at 
$k_0=k$ whereas the second term has a pole at $k_0=-k$.
After calculating 
the residues of those two terms, the number density becomes
\begin{eqnarray}
\rho^f(T,\mu)
&=&2N_cN_f\int\frac{d^3k}{(2\pi)^3}\Big[n_F(k-\mu)-n_F(k+\mu)\Big] \, . 
\label{f8i}
\end{eqnarray}
The QNS is obtained as 
\begin{eqnarray}
\chi^f(T)&=&\left.\frac{\del}{\del\mu}\left[\rho^f(T,\mu)\right]
\right|_{\mu=0}
=4N_cN_f\beta\int \frac{d^3k}{(2\pi)^3}n_F(k)\left(1-n_F(k)\right)
=N_fT^2 \ . 
\label{f9}
\end{eqnarray} 
\subsubsection*{B) 1-loop HTL  case}
\vspace{-0.2cm}
The HTL propagator $S^{\rm q}(K)$ is given in  \eqref{qcd34} with \eqref{qcd35}. The zero momentum limit of the 3-point HTL function can be 
obtained from the differential Ward identity~\cite{Haque:2011iz}  using \eqref{gse23} as
\begin{subequations}
\begin{align}
\Gamma_0(K,-K;0)&=\frac{\del}{\del k_0}  \left ({{S^{\rm q}}^{-1}(K)}\right )
=a\gamma_0+b\, \vec{\boldsymbol{\gamma}}\cdot\bm{\hat k} \ , \label{H2} \\
a\pm b& ={\cal D}_\pm'(k_0,k), \,\,\,\,\,\, {\mbox{with}}  \,\,\,\,\,\,
{\cal D}'_\pm  = \frac{{\cal D}_\pm}{k_0\mp k} + \frac{2(m_{\textrm{th}}^{\rm q})^2}{k_0^2-k^2} \ . \label{H3p}
\end{align}
\end{subequations}
Now  performing the  trace over Dirac matrices in \eqref{i3_tr_f}, the quark number density in HTLpt~\cite{Haque:2011iz,Haque:2010rb} becomes  
\be
\rho^{\textrm{HTL}}(T,\mu)= 2N_cN_fT\sumintf_{\{ k_o\}} \frac{d^3k}{(2\pi)^3}
\left[\frac{{\cal D}_+'}{{\cal D}_+}+
\frac{{\cal D}_-'} {{\cal D}_-}\right]
=2N_cN_fT\sumintf_{\{k_0\}}\frac{d^3k}{(2\pi)^3}\left[\frac{1}{k_0-k}
+\frac{1}{k_0+k}
 + \frac{2(m_{\textrm{th}}^{\rm q})^2}{k_0^2-k^2}\left(\frac{1}{{\cal D}_+}+\frac{1}{{\cal D}_-}
\right)\right],\, \label{H4}
\ee
where $k_0=(2n+1)i\pi T+\mu$.
We note that the first two terms correspond to the free case whereas the
third term couples both HTL  and the free contributions. 
Apart from the various poles due to particles at the light cone 
and QPs due to HTL in (\ref{H4}) it also has
LD part as ${\cal D}_\pm(k_0,k)$ contain Logarithmic terms which generate 
discontinuity for $k^2_0<k^2$, as discussed earlier. Equation (\ref{H4}) 
can be decomposed in individual contribution as
\begin{equation}
\rho^{\textrm{HTL}}(T,\mu)=\rho^{\textrm{QP}}(T,\mu)+\rho^{\textrm{LD}}(T,\mu) \ . \label{HT}
\end{equation}

\noindent {\textit{1) Quasiparticles (QPs) contribution}}

The pole part of the number density can be written as
\be
\rho^{\textrm{QP}}(T,\mu)=2N_cN_fT\int\frac{d^3k}{(2\pi)^3}
\frac{1}{2\pi i}\oint_{C'}
\left[\frac{1}{k_0-k}+\frac{1}{k_0+k} 
+ \frac{2(m_{\textrm{th}}^{\rm q})^2}{k_0^2-k^2}\left(\frac{1}{{\cal D}_+}+
\frac{1}{{\cal D}_-}\right)\right]
\ \frac{\beta}{ 2}\tanh\frac{\beta(k_0-\mu)}{2}dk_0 \ . \label{H5}
\end{eqnarray}

\begin{enumerate}
\item
First two terms in (\ref{H5}) give the same contribution as the free case
in  (\ref{f4}).
\item The first part of the third term has four simple poles 
at $k_0=\omega_+,-\omega_-,k,-k$. The last two poles ($k_0=\pm k$) due to hard
momenta are associated with the soft contributions and  get easily 
separated out.
After performing the contour integration,
its contribution can be written~\cite{Haque:2010rb} as
\begin{eqnarray}
\frac{1}{2\pi i} \oint_{C_1'} \frac{2(m_{\textrm{th}}^{\rm q})^2}{k_0^2-k^2}\frac{1}{D_+}
\frac{\beta}{2}\tanh\frac{\beta(k_0-\mu)}{2} \ dk_0 = - \frac{\beta}{2}
\left [  \tanh\frac{\beta(\omega_+-\mu)}{2}
 -  \tanh\frac{\beta(\omega_-+\mu)}{2}
-\tanh\frac{\beta(k-\mu)}{2} 
+\tanh\frac{\beta(k+\mu)}{2}\right ] 
. \nonumber
\end{eqnarray}
\item
The the second part of the third term has also four simple poles at $k_0=\omega_-,-\omega_+,-k, k$. After performing the contour integration its contribution can be written~\cite{Haque:2010rb}  as
\be
\frac{1}{2\pi i} \oint_{C_2'} \frac{2(m_{\textrm{th}}^{\rm q})^2}{k_0^2-k^2}\frac{1}{D_-}
\frac{\beta}{2}\tanh\frac{\beta(k_0-\mu)}{2} \ dk_0 = -\frac{\beta}{2}
\left [  \tanh\frac{\beta(\omega_--\mu)}{2}
-  \tanh\frac{\beta(\omega_++\mu)}{2}
-\tanh\frac{\beta(k-\mu)}{2} 
+\tanh\frac{\beta(k+\mu)}{2}\right ] . \nonumber 
\ee
\end{enumerate}
Using \eqref{f8i}, and above two results in \eqref{H5} one can obtain the HTL quasiparticle  contributions to the quark number 
density~\cite{Haque:2011iz,Haque:2010rb} as
\be
\rho^{\textrm{QP}}(T,\mu) = 2N_cN_f\!\int \!\frac{d^3k} {(2\pi)^3}\Big[
n_F(\omega_+-\mu)+n_F(\omega_--\mu)-n_F(k-\mu) -n_F(\omega_++\mu)
-n_F(\omega_-+\mu)+n_F(k+\mu) 
\Big ] ,\label{H6}
\ee
which agrees with that of the two-loop approximately self-consistent $\Phi$-derivable HTL resummation of Blaizot 
{\it et al}~\cite{Blaizot:2001vr,Blaizot:2002xz}.

The QNS in LO due to HTL QP can also be obtained~\cite{ Haque:2011iz,Haque:2010rb} from \eqref{H6} as
\be
\chi^{\textrm {QP}}(T)=\left.\frac{\del}{\del\mu}\left[\rho^{\textrm{QP}}\right]
\right|_{\mu=0} \!\!\!\!\!\!\!\!
= 4N_cN_f\beta\!\int \!\frac{d^3k} {(2\pi)^3}\Big [
n_F(\omega_+)\left(1-n_F(\omega_+)\right) 
\ + \ n_F(\omega_-)\left(1-n_F(\omega_-)\right) 
-\ n_F(k)\left(1-n_F(k)\right)
\Big ] , \label{H8} 
\ee
where the $\mu$ derivative is performed only to the explicit $\mu$
dependence~\cite{Blaizot:2001vr,Blaizot:2002xz}. Obviously (\ref{H8}) agrees exactly with that of 
the 2-loop approximately self-consistent  $\Phi$-derivable 
HTL resummation of Blaizot {\it et al}~\cite{Blaizot:2001vr,Blaizot:2002xz}. At very high $T$, $\omega_\pm\rightarrow k$ and (\ref{H8}) reduces to free
case as obtained in (\ref{f9}). 

\noindent {\textit{2) Landau damping (LD) contribution}}

The LD part of the quark number density follows from (\ref{H4}) using  \eqref{spec4}  in the subsec~\ref{spec_prop}
one gets~\cite{Haque:2011iz,Haque:2010rb}
\be
\rho^{\textrm{LD}}(T,\mu)\!\! 
=  N_cN_f\!\! \int\! \frac{d^3k}{(2\pi)^3}
\!\!\int _{-k}^k\!\! {d\omega} 
\left(\frac{2(m_{\textrm{th}}^{\rm q})^2}{k^2-\omega^2}\right) \,
\beta_+(\omega,k) \,
\Big[n_F(\omega-\mu)- n_F(\omega+\mu) \Big ]  \ .
\label{L3}
\ee
Note that we have considered $\beta_-(-\om,k)=\beta_+(\om,k)$  when
 $\om\rightarrow -\omega$. Also the LD part of the QNS becomes 
\begin{eqnarray}
\chi^{\textrm{LD}}(T)
&=&\left.\frac{\partial}{\partial \mu}\rho^{\textrm{LD}}(T,\mu)\right|_{\mu=0} =  2N_cN_f\beta \int \frac{d^3k}{(2\pi)^3}\int_{-k}^k d\omega 
\left(\frac{2(m_{\textrm{th}}^{\rm q})^2}{k^2-\omega^2}\right)\, \beta_+(\om,k) \, n\big(\omega\big)\, \big(1-n(\omega)\big)
\ , \label{L4}
\end{eqnarray}
where the $\mu$ derivative is again performed only to the explicit $\mu$
dependence. It is also to be noted that the LD contribution is of the
order of $m_q^4\equiv (m_{\textrm{th}}^{\rm q})^4$. The LD contribution can not be compared with 
that of the 2-loop approximately self-consistent  $\Phi$-derivable HTL 
resummation of Blaizot {\em et al}~\cite{Blaizot:2001vr,Blaizot:2002xz} as it does not 
have any closed form for the final expression. However, the numerical values
of both the QNS agree very well. 

\subsubsection*{C) QNS in conventional perturbation theory}
In conventional perturbation theory, for massless QCD the QNS has been 
calculated~\cite{Kapusta:2006pm,Toimela:1984xy} upto order $g^4\log(1/g)$ at $\mu=0$ as 
\begin{equation}
\frac{\chi^{\textrm{pQCD}}}{\chi_f}=1- \frac{1}{2}\left (\frac{g}{\pi}\right )^2 +
\sqrt{1+\frac{N_f}{6}}\left(\frac{g}{\pi}\right )^3-
\frac{3}{4}\left(\frac{g}{\pi}\right )^4 \log\left(\frac{1}{g}\right)+
{\cal O}(g^4)\ . \label{qns_pert}
\end{equation}
We note that for all temperatures of relevance the series decreases 
with temperature and approaches the ideal gas value from the above, 
which is due to the convergence problem of the conventional perturbation 
series. Nevertheless, the perturbative LO, $g^2$, contribution is also contained in HTL 
approximation through the $N$-point HTL functions. 
\begin{figure}[h]
\begin{center}
\includegraphics[width=8.5cm, height=5.5cm]{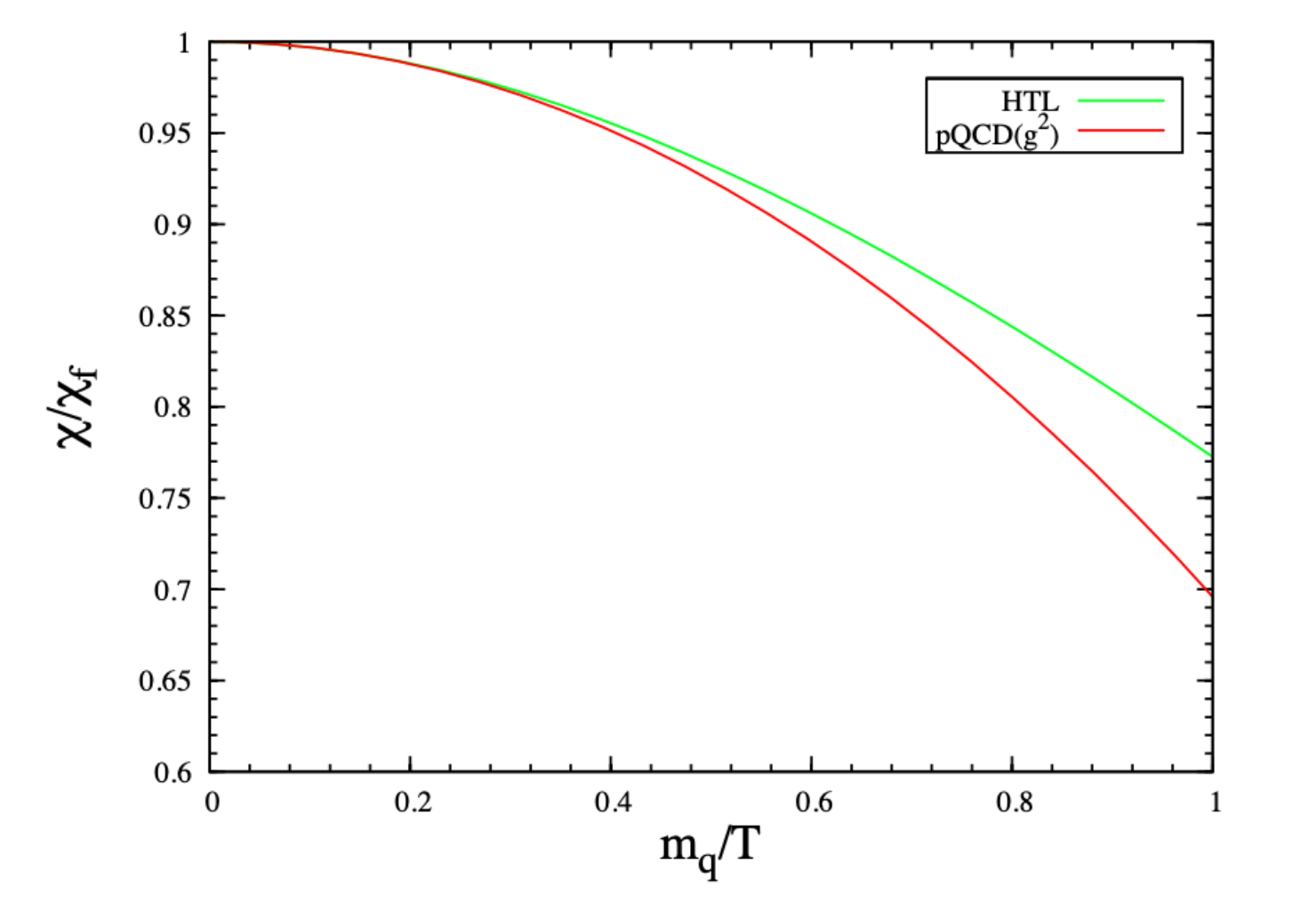} 
\includegraphics[width=8.5cm, height=5.5cm]{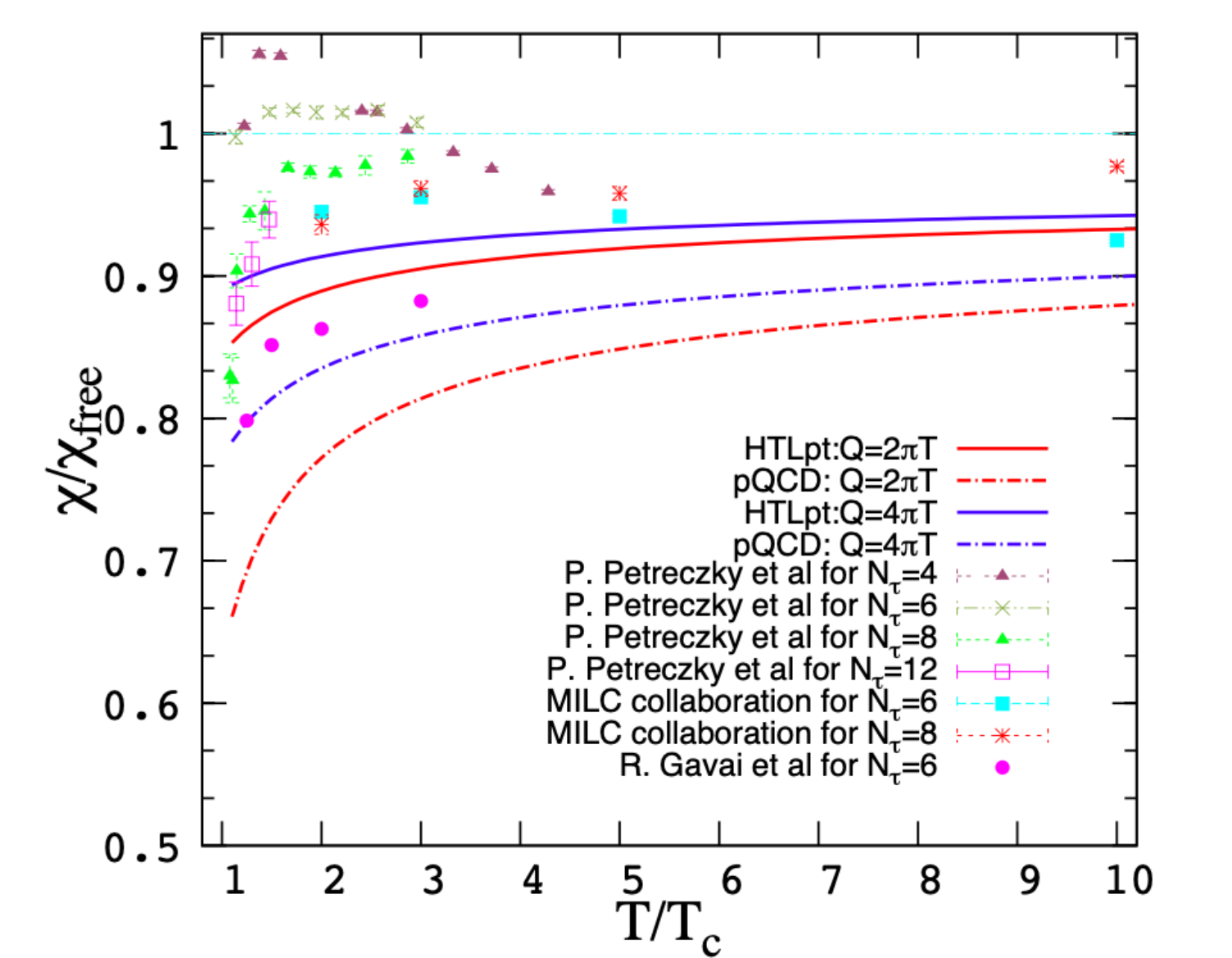}
\end{center}
\vspace{-0.5cm}
\caption{\textit{Left Panel}: The ratio of 2-flavour LO HTL to free quark QNS and also that of LO  perturbative one  as a 
function of $m_q/T\equiv m^{\rm q}_{\textrm{th}}/T$. This figure is taken from Ref.~\cite{Haque:2010rb}. \textit{Right Panel}: The 2-flavour scaled QNS with that of free one as a function of $T/T_c$.
The solid lines are for LO in HTLpt whereas the dashed lines are 
for LO (proportional to $g^2$) in pQCD~\cite{Kapusta:2006pm,Toimela:1984xy}. 
The different choices of the renormalisation scale are $\Lambda =2\pi T \ {\mbox{(red)}}, \ \
{\mbox{and}} \ \ 4\pi T \ {\mbox{(blue)}}$. The symbols represent the
various lattice data~\cite{Allton:2005gk,Petreczky:2009cr,Bazavov:2009zn,Bernard:2004je,Gavai:2001fr,Gavai:2001ie}. 
This figure is taken from Ref.~\cite{Haque:2010rb}.} 
\label{pert_g2}
\end{figure}

In the left part of Fig.~\ref{pert_g2} we display the LO HTL QNS and LO  perturbative QNS scaled with free one as a function of $m^{\rm q}_{\textrm{th}}/T$,  the effective strong coupling. In the weak coupling limit both approach unity whereas the 
HTL case has a little slower deviation from the ideal gas value than the LO of the conventional perturbative one. The latter could be termed as an improvement over the conventional perturbative results. The results are in very good agreement with that of Refs.~\cite{Blaizot:2001vr,Blaizot:2002xz}.

Now we display in the right panel of Fig.~\ref{pert_g2} the 2-flavour\footnote{We note that the QNS has a very weak flavour dependence that enters through the temperature dependence of the strong coupling as 
$\alpha_s(T)=\frac{12\pi}{(33-2N_f)\ln(Q^2/\Lambda_0^2)}$ where $Q$ is the momentum scale and $T_C=0.49\Lambda_0$. } scaled QNS in LO with that of free gas as a function of temperature that shows significant improvement over pQCD results of order $g^2$~\cite{Kapusta:2006pm,Toimela:1984xy}. Moreover, it also shows the same trend as the available lattice
results~\cite{Allton:2005gk,Petreczky:2009cr,Bazavov:2009zn,Bernard:2004je,Gavai:2001fr,Gavai:2001ie,Cheng:2009zi,Petreczky:2009at}, though there is a large variation among the various lattice results within the improved
lattice (asqtad and p4) actions~\cite{Petreczky:2009cr,Bernard:2004je} due to the higher order
discretisation of the relevant operator associated with the thermodynamic derivatives. A detailed analysis on uncertainties of the ingredients in the lattice QCD calculations is presented in Refs.~\cite{Bazavov:2009zn,Cheng:2009zi}. This calls for further investigation both on the analytic side by improving the HTL resummation schemes and on the lattice side by refining the various lattice ingredients. 

\begin{figure}[!tbh]
\begin{center}
{\includegraphics[scale=0.3]{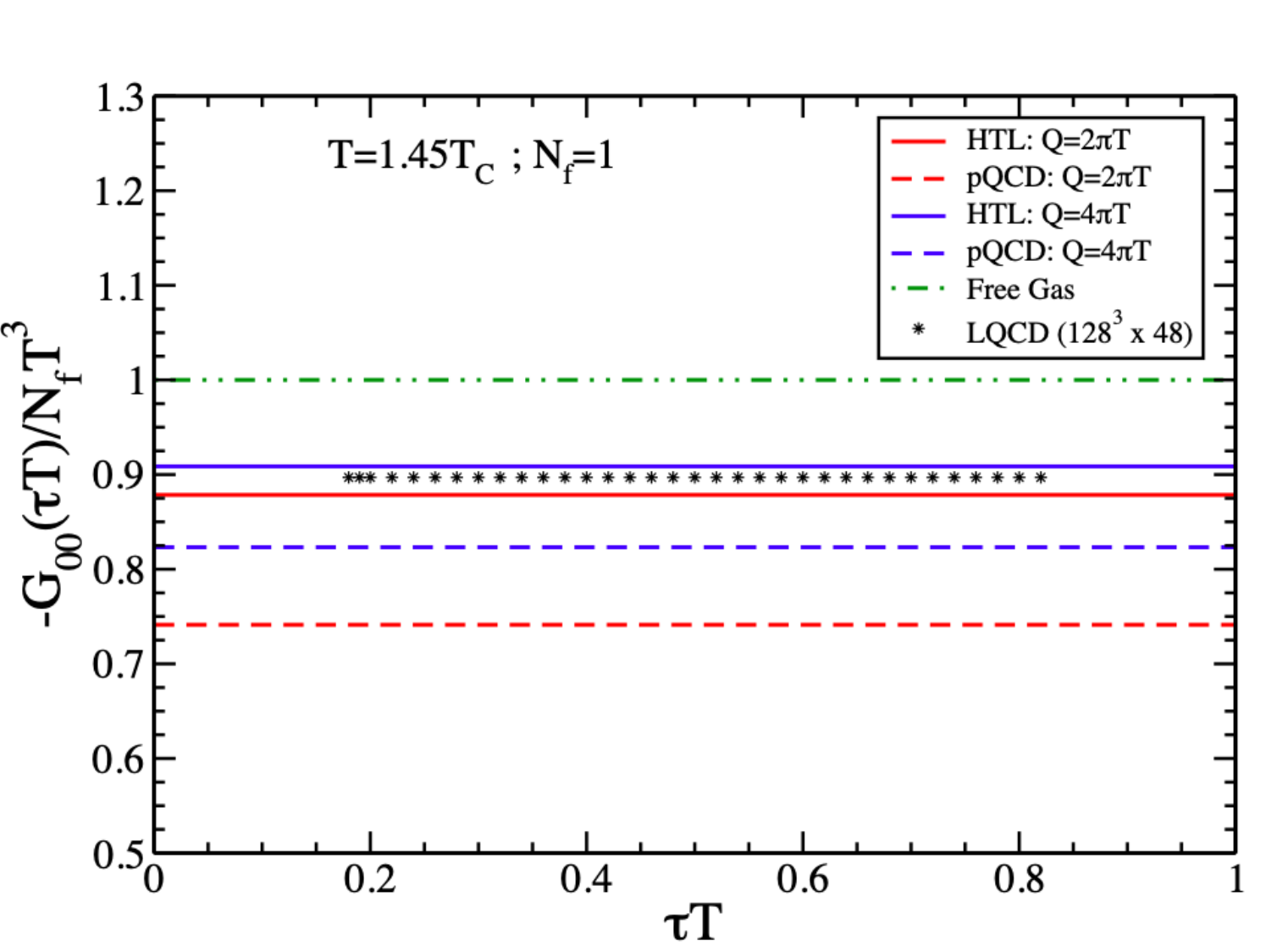}} \hspace*{0.2in}
{\includegraphics[scale=0.3]{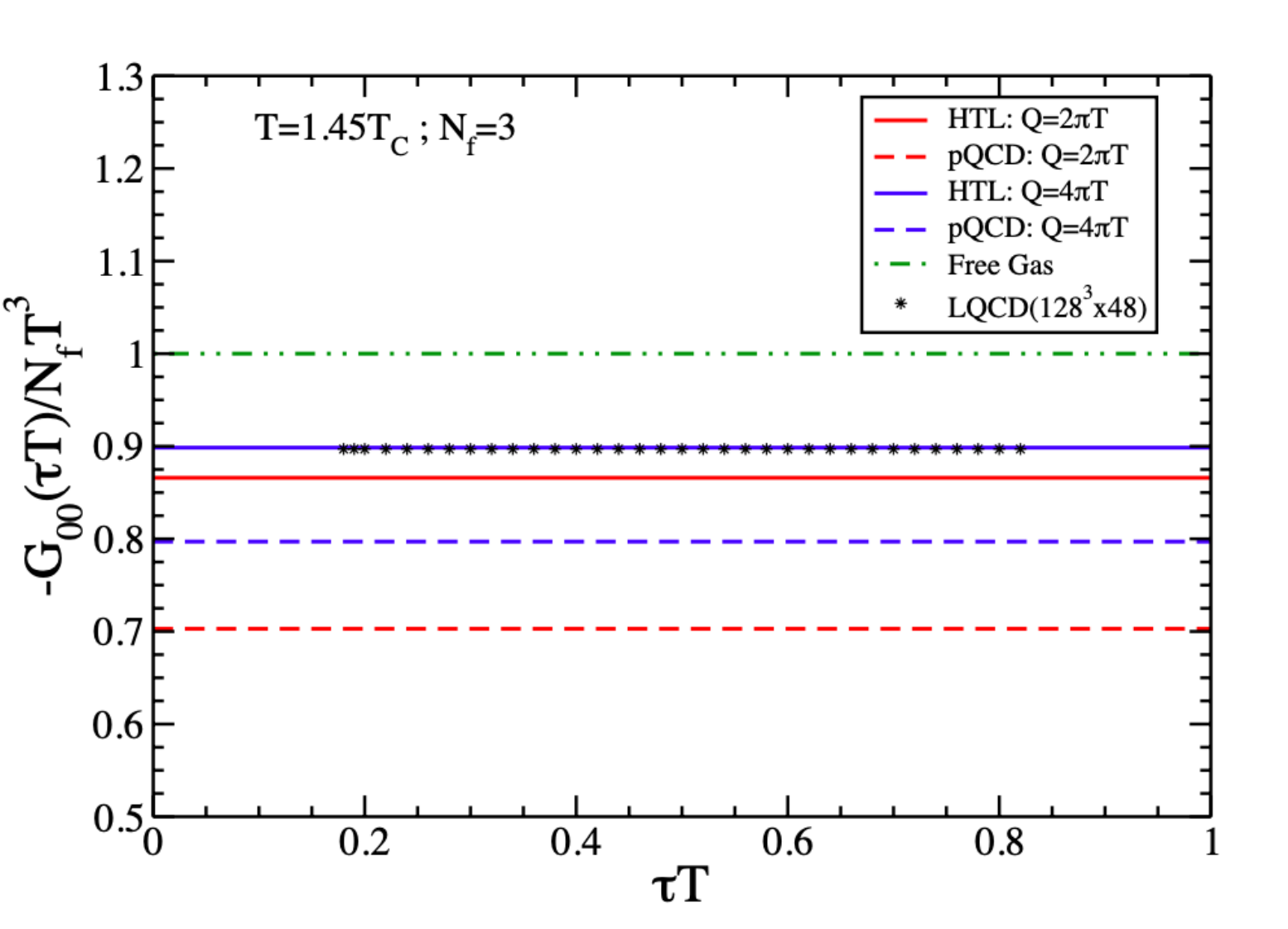}}
\end{center}
\vspace{-0.7cm}
\caption{The scaled temporal correlation function with $T^3$ for $N_f=1$ (left panel) and
$N_f=3$ (right panel) at $T=1.45T_C$ for $\Lambda=2\pi T$ (red) and $4\pi T$ (blue) as a function of scaled Euclidean time, $\tau T$. The symbols represent the recent lattice data~\cite{Ding:2010ga}. These figures are taken from Ref.~\cite{Haque:2011iz}.}
\label{corr1}
\end{figure}
Reference~\cite{Ding:2010ga} presents an enhanced lattice calculation performed within the quenched approximation of QCD, achieving a temporal correlation function determination with better than $1\%$ accuracy. Using the LO HTLpt QNS in~\eqref{HT}, we derive the temporal
correlation function in~\eqref{eq7} and compare it with the lattice data~\cite{Ding:2010ga}. Figure~\ref{corr1} illustrates the scaled temporal correlation function with $T^3$ for $N_f=1$ (left panel) and $N_f=3$ (right panel) at $T=1.45T_c$.  It is noteworthy that both HTLpt and pQCD correlation functions exhibit weak flavour dependence due to the temperature-dependent coupling, $\alpha_s$, as discussed earlier. The LO HTLpt result indicates 
an improvement over that of the pQCD one~\cite{Kapusta:2006pm,Toimela:1984xy} for 
different choices of the renormalisation scale as shown in Fig.~\ref{corr1}. Also, the HTLpt 
results show  a good agreement to that of recent lattice gauge theory calculation~\cite{Ding:2010ga} performed 
on lattices up to size $128^3\times 48$ in quenched approximation for a quark 
mass $\sim 0.1T$. Notably, unlike the dynamical spatial part of the correlation function in the vector channel, 
the temporal part does not encounter any infrared problem in the low energy part, as it is linked to the static quantity through the thermodynamic sum rule
associated with the corresponding symmetry, namely the number conservation of the system.  
Finally, it is observed that even when comparing improved lattice action (asqtad) data~\cite{Bernard:2004je}  
and recent quenched data~\cite{Ding:2010ga} for QNS, the quantitative difference 
is within $5\%$ in the temperature domain $T_C\leq T\leq 3T_C$.


The LO QNS as a response of the conserved density fluctuation in HTLpt when compared with the available lattice data with improved lattice actions~\cite{Allton:2005gk,Petreczky:2009cr,Bazavov:2009zn,Bernard:2004je,Gavai:2001fr,Gavai:2001ie},
in the literature within their wide variation shows the same trend but deviates from those in certain extent. The same HTL QNS is used to compute the temporal part of the Euclidean correlation in the vector current which agrees quite well with that of improved lattice gauge theory calculations~\cite{Ding:2010ga} recently performed within the quenched 
approximation on lattices up to size $128^3\times48$ for a quark mass  $\sim 0.1T$. It is also interesting to note that the quantitative difference between the recent quenched approximation data~\cite{Ding:2010ga}  and
the full QCD data with improved (asqtad) lattice action~\cite{Bernard:2004je} for QNS
is within $5\%$ in the temperature range $T_c\le T\le 3T_c$.  Leaving aside the difference in ingredients in various lattice calculations, one can expect that the HTLpt and lattice calculations are in close proximity for quantities associated 
with the conserved density fluctuation. We note that the QNS in LO, NLO and NNLO in HTLpt,  will be discussed 
in the next section~\ref{chapter:thermodynamics} .

	\section{QCD Thermodynamics}\label{chapter:thermodynamics}
	The determination of the equation of state (EOS) of QCD matter is extremely important in QGP phenomenology. Various effective models (see e.g. ~\cite{Peshier:1999ww,Peshier:2002ww,Pisarski:2000eq,Hatsuda:1994pi,Kunihiro:1991qu,Ratti:2005jh,Ghosh:2006qh,Ghosh:2007wy,Mukherjee:2006hq,Roessner:2006xn,Sasaki:2006ww}) exists to describe the EOS of strongly interacting matter; however, one would prefer to utilize systematic first-principles QCD methods. Currently, the most reliable approach for determining the EOS is lattice QCD~\cite{DeGrand:2006zz}. At present, lattice calculations can be carried out at arbitrary temperature, however, they are limited to relatively small chemical potentials~\cite{Gavai:2003mf,Borsanyi:2012cr}.  Alternatively, perturbative QCD (pQCD)~\cite{Bellac:2011kqa,Kapusta:2006pm,Shuryak:1977ut, Chin:1978gj,Kapusta:1979fh,Toimela:1982hv,Arnold:1994ps,Arnold:1994eb,Zhai:1995ac,Kajantie:2002wa} can be applied at high temperature and/or chemical potentials where the strong coupling ($g^2=4\pi \alpha_s$) is small in magnitude, and non-perturbative effects are expected to be small.  However, due to infrared singularities in the gauge sector, the perturbative expansion of the finite-temperature and density QCD partition function breaks down at order $g^6$ requiring 
non-perturbative input, albeit through a single numerically computable number~\cite{Kajantie:2002wa,Linde:1978px,Linde:1980ts}. Up to order $g^6\ln(1/g)$, it is feasible to calculate the necessary coefficients using analytic (resummed) perturbation theory.

Since the inception of pQCD, there has been a tremendous effort to compute the pressure order by order in the 
weak coupling expansion~\cite{Bellac:2011kqa,Kapusta:2006pm,Shuryak:1977ut, Chin:1978gj, Kapusta:1979fh,Toimela:1982hv,Arnold:1994ps,Arnold:1994eb,Zhai:1995ac,Kajantie:2002wa,Vuorinen:2003fs,Ipp:2006ij}.  The pressure has been calculated to order of $g^6\ln(1/g)$ at  zero chemical potential ($\mu=0$) and finite temperature $T$~\cite{Kajantie:2002wa} and finite chemical potential/temperature ($\mu \geq 0$ and $T \ge 0$)~\cite{Vuorinen:2003fs}. Additionally, the pressure is known to order $g^4$ for large $\mu$ and arbitrary $T$~\cite{Ipp:2006ij}. Unfortunately, it is observed that as successive perturbative orders are included, the series exhibits poor convergence, with the dependence on the renormalization scale increasing rather than diminishing. Convergence of the resulting perturbative series is only achieved at very high temperatures ($T\sim 10^5 \, T_c$). One could be tempted to say that this is due to the largeness of the QCD coupling constant at realistic  temperatures; however, in practice one finds that the relevant small quantity is, in fact, $\alpha_s/\pi$ which for phenomenologically relevant temperatures is on the order of one-tenth.  Instead, it is observed that the coefficients of $\alpha_s/\pi$ are large. This is evident when examining the weak coupling expansion of  the free energy ${\cal F} (T,\mu)$ of QGP calculated~\cite{Vuorinen:2003fs} up to order $\alpha_s^{3}\ln (\alpha_s)$ 
\subsection{LO and NLO HTL Thermodynamics}
In HTL perturbation theory, the next-to-leading order (NLO) thermodynamic potential was computed in~\cite{Andersen:2002ey, Andersen:2003zk} at finite temperature and zero chemical potential. However in view of the ongoing RHIC beam energy scan and planned FAIR experiments, there is a drive to accurately determine the thermodynamic 
functions at finite chemical potential. In this subsection we discuss the NLO pressure of quarks and gluons at finite $T$ and $\mu$.  The computation utilizes a high temperature expansion through fourth order in the ratio of the chemical potential to temperature. This allows us to reliably access the region of high temperature and small chemical potential. We compare our final result for the NLO HTLpt  pressure at finite temperature and chemical potential with state-of-the-art perturbative finite temperature QCD calculations and available lattice QCD results.
\begin{figure}[h]
	\begin{center}
		\includegraphics[width=16cm,height=5.5cm]{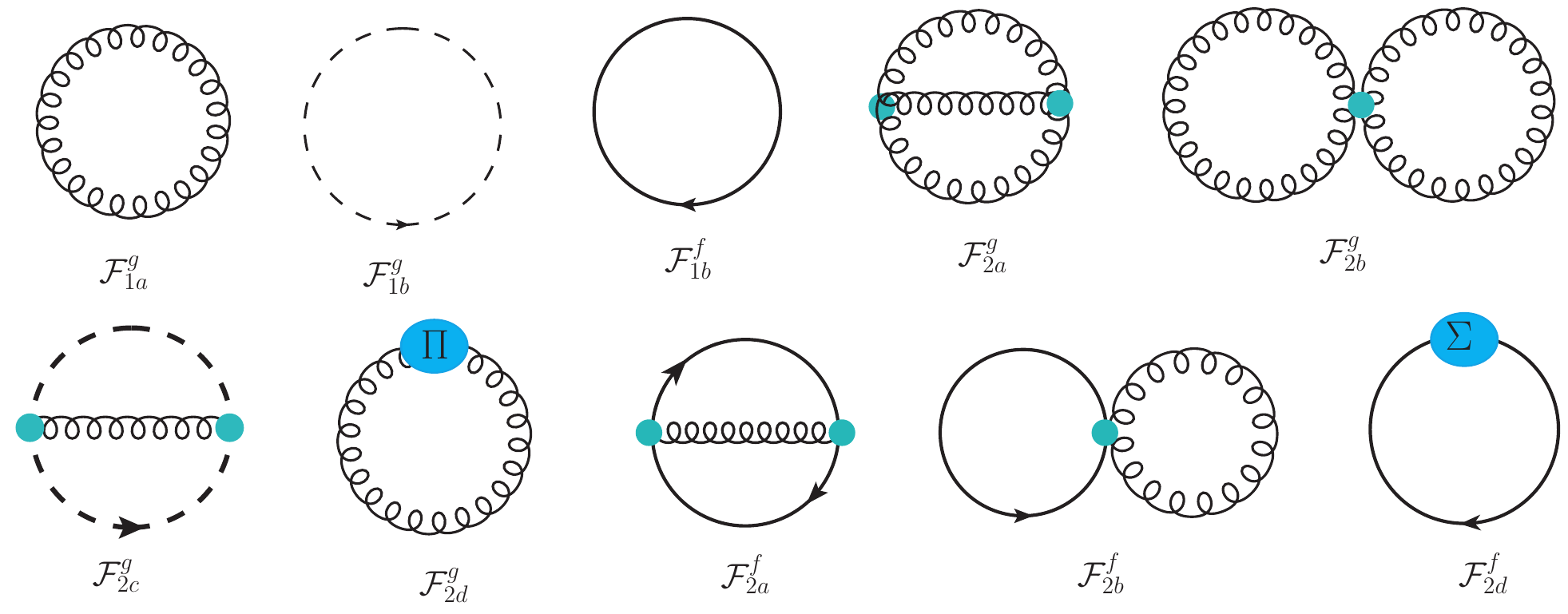}
		\vspace{-0.2cm}
		\caption{Diagrams containing fermionic lines relevant for NLO thermodynamics potential in HTLpt with finite chemical potential. Shaded circles indicate HTL $n$-point functions.}
		\label{diagramfig}
	\end{center}
\end{figure}
\vspace{-0.5cm}
\subsubsection{LO HTLpt thermodynamic potential}
The LO HTLpt thermodynamic potential, $\Omega_{\rm LO}$, for an $SU(N_c)$ gauge theory with $N_f$ massless quarks 
in the fundamental representation can be written as ~\cite{Andersen:1999sf,Andersen:2002ey, Andersen:2003zk}
\be
\Omega_{\rm LO}= d_A {\cal F}_{1a}^g +  d_F {\cal F}_{1b}^f+\Delta_0{\cal E}_0\;,
\ee
where $d_F=N_f N_c$ and $d_A=N_c^2-1$ with $N_c$ is the number of
colours. ${\cal F}_q$ and ${\cal F}_g$ are the one loop contributions to quark and gluon free energies, respectively. 
The LO counter-term is the same as in the case of zero chemical potential~\cite{Andersen:1999sf}
\be
\Delta_0{\cal E}_0 = \frac{d_A}{128\pi^2\epsilon}m_D^4\;.
\label{count0}
\ee
Here we have considered the QCD Debye mass $m_D^{\rm g}\equiv m_D$ and quark thermal mass $m_{\rm{th}}^{\rm q}\equiv m_q$.

The hard contribution to the quark free energy in one loop quark diagram in Fig~\ref{diagramfig}  becomes~\cite{Mustafa:2022got,Haque:2012my, Haque:2013qta}
\be
{\cal F}_{1b}^{f(h)}\!\!\! &=&\!\!\! -\frac{7\pi^2}{180}T^4\left(1+\frac{120}{7}\hat\mu^2 
+\frac{240}{7}\hat\mu^4\right) + \left(\frac{\Lambda}
{4\pi T}\right)^{2\epsilon} \frac{m_q^2 T^2}{6}\biggr[\left(
1+12\hat\mu^2\right)
\nn 
\!\!\!&+&\!\!\!\! \left.
\epsilon\left(2-2\ln2+2\frac{\zeta'(-1)}{\zeta(-1)}+ 24(\gamma_E+2\ln2)
\hat\mu^2 - 28\zeta(3)\hat\mu^4 + {\cal O}\left(\hat\mu^6\right)
\right)\right] +
\frac{m_q^4}{12\pi^2}(\pi^2 - 6) \, .\hspace{5mm} 
\label{Quark1loop}
\ee
Using the last diagram in the second line in Fig~\ref{diagramfig}, one obtains the hard contribution to the HTL quark counter-term~\cite{Haque:2012my, Haque:2013qta} becomes
\be
{\cal F}_{2d}^{f(h)} = -\frac{m_q^2 T^2}{6}\left(1+12\hat\mu^2\right)
-   \frac{m_q^4}{6\pi^2}(\pi^2 - 6)  \, .
\label{count}
\ee
We note that the first term in ~(\ref{count}) cancels the order-$\epsilon^0$ term in the coefficient of $m_q^2$ in (\ref{Quark1loop}). There are no soft contributions either from the leading-order quark term or from the HTL quark counter-term. Now, using the expressions of ${\cal F}_{1b}^f$ with finite quark chemical potential in (\ref{Quark1loop}) and 
${\cal F}_{1a}^g$ from Ref.~\cite{Andersen:2002ey, Andersen:2003zk} and adding the counterterm in Eq.~\eqref{count0}, we obtain the thermodynamic potential at leading order in the $\delta$-expansion~\cite{Mustafa:2022got,Haque:2012my, Haque:2013qta} as,
\be
\Omega_{\rm LO}& =& - d_A\frac{\pi^2T^4}{45}\Bigg\{1+\frac{7}{4}\frac{d_F}{d_A}\left(
1+\frac{120}{7}\hat\mu^2+\frac{240}{7}\hat\mu^4\right)
- \frac{15}{2}\left[1+\epsilon \left(2+2\frac{\zeta'(-1)}{\zeta(-1)}+2\ln{\frac{\hat\Lambda}{2}}\right)\right]\hat m_D^2 
\nn
&-&
\frac{30d_F}{d_A}\left[1+12\hat\mu^2\ +\epsilon\Big(2-2\ln2+2\frac{\zeta'(-1)}
{\zeta(-1)}+2\ln{\frac{\hat\Lambda}{2}} + 24(\gamma_E+2\ln2)\hat\mu^2 \right.
-
28\zeta(3)\hat\mu^4 + {\cal O}\left(\hat\mu^6\right)\Big)\bigg]\hat m_q^2\nn
&+&30\left(\frac{\Lambda}{2 m_D}\right)^{2\epsilon}\left[1+\frac{8}{3}\epsilon\right]
\ \hat m_D^3 + \frac{45}{8}\left(2\ln{\frac{\hat\Lambda}{2}}-7+2\gamma_E+\frac{2\pi^2}{3}\right)
\hat m_D^4 - 60\frac{d_F}{d_A}(\pi^2-6)\hat m_q^4\Bigg\}
\;,
\label{Omega-LO}
\ee
where $\hat m_D$, $\hat m_q$, $\hat\Lambda$, and $\hat \mu$ are dimensionless variables:
\be
\hat m_D &=& \frac{m_D}{2 \pi T}  \;,\ 
\hat m_q = \frac{m_q}{2 \pi T}  \;,\ 
\hat \Lambda =\frac{\Lambda}{2 \pi T}  \;,\ 
\hat \mu = \frac{\mu}{2 \pi T}  \;. 
\ee. 
In Eq.~\eqref{Omega-LO}, we have kept terms of ${\cal O}(\epsilon)$ since they will be needed for the
two-loop renormalization.

\subsubsection{Next-to-leading order thermodynamic potential}
At NLO one must consider the diagrams shown in Fig.~(\ref{diagramfig}).
The resulting NLO HTLpt thermodynamic potential can be written in the following general form~\cite{Andersen:2003zk} 
\be
\Omega_{\rm NLO}&=&\Omega_{\rm LO} + d_A \left[{\cal F}_{2a}^g+{\cal F}_{2b}^g+{\cal F}_{2c}^g + {\cal F}_{2d}^g \right] 
+d_A s_F \left[{\cal F}_{2a}^f+{\cal F}_{2b}^f \right] \nonumber \\
&&+ d_F {\cal F}_{2d}^f + \Delta_1{\cal E}_0 + \Delta_1 m_D^2\frac{\partial}{\partial m_D^2}
\Omega_{\rm LO}+\Delta_1 m_q^2\frac{\partial}{\partial m_q^2}\Omega_{\rm LO}\;,
\label{OmegaNLO}
\ee
where $s_F=N_f/2$. At NLO the terms that depend on the chemical potential  are ${\cal F}_q$, ${\cal F}_{3qg}$, ${\cal F}_{4qg}$, ${\cal F}_{qct}$,  $\Delta_1 m_q^2$, and $\Delta_1 m_D^2$ as displayed
in Fig.~(\ref{diagramfig}). The other terms, e.g. ${\cal F}_g$, ${\cal F}_{3g}$, ${\cal F}_{4g}$, 
${\cal F}_{gh}$ and ${\cal F}_{gct}$ coming from gluon and ghost loops remain the same as the $\mu=0$ case~\cite{Andersen:2002ey,	Andersen:2003zk}. 
We also add that the vacuum energy counter-term, $\Delta_1{\cal E}_0$, remains the 
same as the $\mu=0$ case whereas the  mass counter-terms, $\Delta_1 m_D^2$ and $\Delta_1m_q^2$, have to be computed for $\mu\ne0$. These counter-terms are  of order $\delta$.  This completes a general description of 
contributions one needs to compute in order to determine NLO HTLpt thermodynamic potential 
at finite chemical potential.  

Adding the contributions from the two-loop diagrams, the HTL gluon and quark counterterms, the contribution from vacuum and mass renormalizations, and the leading-order thermodynamic potential in~(\ref{Omega-LO}) we
obtain the complete expression for the QCD thermodynamic potential at next-to-leading order in HTLpt in Refs.~\cite{Haque:2012my, Haque:2013qta}:
\be
\Omega_{\rm NLO}\!\!\!&=&\!\!\!
- d_A \frac{\pi^2 T^4}{45} \Bigg\{ 
1 + \frac{7}{4} \frac{d_F}{d_A}\left(1+\frac{120}{7}
\hat\mu^2+\frac{240}{7}\hat\mu^4\right) - 15 \hat m_D^3 
-\frac{45}{4}\left(\log\frac{\hat\Lambda}{2}-\frac{7}{2}+\gamma_E+\frac{\pi^2}{3}\right)\hat m_D^4
+ 60 \frac{d_F}{d_A}\left(\pi^2-6\right)\hat m_q^4	
\nn
&&\hspace{-1.1cm}+\, \frac{\alpha_s}{\pi} \Bigg[ -\frac{5}{4}\left\{\!c_A + \frac{5s_F}{2}\!\left(\!1+\frac{72}{5}
\hat\mu^2+\frac{144}{5}\hat\mu^4\right)\!\right\}
+ 15 \left(c_A+s_F(1+12\hat\mu^2)\right)\!\hat m_D - 
\frac{55}{4}\Bigg\{ c_A\!\left(\log\frac{\hat\Lambda}{2}- \frac{36}{11}\log\hat m_D - 2.001\!\right)\nn
&&\hspace{-1.1cm}- \frac{4}{11} s_F \left[\ln\frac{\hat\Lambda}{2}-2.337 +
 (24-18\zeta(3))\left(\ln\frac{\hat\Lambda}{2} -15.662\right)\hat\mu^2
+ 120\left(\zeta(5)-\zeta(3)\right)
\left(\log\frac{\hat\Lambda}{2} -1.0811\right)\!\hat\mu^4 + 
{\cal O}\left(\hat\mu^6\right)\right] \!\!\Bigg\} \hat m_D^2  \nn
&&\hspace{-1.1cm}-\,45 \, s_F \left\{\log\frac{\hat\Lambda}{2}
+ 2.198  -44.953\hat\mu^2-\left(288 \ln{\frac{\hat\Lambda}{2}}    
+19.836\right)\hat\mu^4 + {\cal O}\left(\hat\mu^6\right)\right\} \hat m_q^2
+ \frac{165}{2}\left\{ c_A\left(\log\frac{\hat\Lambda}{2}+\frac{5}{22}+\gamma_E\right)\right.
\nn
 &&\hspace{-1.1cm}-\,
\left. \frac{4}{11} s_F \left(\log\frac{\hat\Lambda}{2}-\frac{1}{2}+\gamma_E+2\ln2 -7\zeta(3)\hat\mu^2+
31\zeta(5)\hat\mu^4 + {\cal O}\left(\hat\mu^6\right) \right)\right\}\hat m_D^3
\nn 
&&\hspace{-1.1cm}+\,
15 s_F \left(2\frac{\zeta'(-1)}{\zeta(-1)}
+2\ln \hat m_D\right)\Big[(24-18\zeta(3))\hat\mu^2 
+ 120(\zeta(5)-\zeta(3))\hat\mu^4 + 
{\cal O}\left(\hat\mu^6\right)\Big] \hat m_D^3
+ 180\,s_F\hat m_D \hat m_q^2 \Bigg]
\Bigg\} .
\label{Omega-NLO}
\ee
\subsubsection{Pressure}
\label{nlopress}
\vspace{-0.2cm}
In the preceding section we computed both LO and NLO thermodynamic potential in the presence of quark chemical potential
and temperature. All other thermodynamic quantities can be calculated using standard thermodynamic relations. The pressure
is defined as 
\be
P = -{\Omega}(T,\mu,m_q,m_D) \, ,
\label{pressure-potential}
\ee
Both LO  and NLO pressure can be obtained using ~\eqref{Omega-LO} and \eqref{Omega-NLO}, respectively.
At leading order, the weak coupling expressions for the mass parameters are 
\be 
m_D^2 &=& \frac{g^2T^2}{3} \left [ c_A +s_F\left(1+12\hat\mu^2\right)  \right ]\, ;  \hspace*{0.2in}
m_q^2 = \frac{g^2T^2}{4} \frac{c_F}{2} \left (1 + 4\hat\mu^2\right ) \, . \label{mass_lo}
\ee
The NLO pressure is  accurate up to ${\cal O}(g^3)$ and nominally accurate to ${\cal O}(g^5)$ since it was 
obtained from an expansion of two-loop thermodynamic potential in a power series in $m_D/T$ and $m_q/T$, treating both $m_D$ and $m_q$ having leading terms proportional to $g$. Using the result above, the mass parameters $m_D$ and $m_q$ for NLO can be determined by solving the two variational equations:
\be
\left.\frac{\partial \Omega_{\rm NLO}}{\partial \hat{m}_D}\right|_{m_q=\text{constant}} = 0 \,|,
\,\,\,\,\,\,\, {\mbox{and}} \,\,\,\,\,\,\,\,\,
\left.\frac{\partial \Omega_{\rm NLO}}{\partial \hat{m}_q}\right|_{m_D=\text{constant}} = 0 \, .
\ee
This leads to the following two gap equations which will be solved numerically
\begin{eqnarray}
&& \hspace{-7mm} 
45\hat m_D^2\left[1+\left(\ln\frac{\hat\Lambda}{2}-\frac{7}{2}+\gamma_E+
\frac{\pi^2}{3}\right)\hat m_D\right]
=
\frac{\alpha_s}{\pi}\Bigg\{15(c_A+s_F(1+12\hat\mu^2))
-\frac{55}{2}\left[c_A\left(\ln\frac{\hat\Lambda}{2}-\frac{36}{11}\ln{\hat m_D}-3.637\right)
\right.
\nonumber\\
&-&\left.\frac{4}{11}s_F\left\{\ln\frac{\hat\Lambda }{2}-2.333+(24-18\zeta(3))
\left(\ln\frac{\hat\Lambda }{2}-15.662\right)\hat\mu^2
+120(\zeta(5)-\zeta(3))\left(\ln\frac{\hat\Lambda }{2}-1.0811\right)\hat\mu^4
\right\}\right]\hat m_D \nn
&+&\frac{495}{2}\left[c_A\left(\ln\frac{\hat\Lambda }{2}
+\frac{5}{22}+\gamma_E\right)
\frac{4}{11}s_F\left\{\ln\frac{\hat\Lambda }{2}-\frac{1}{2}
+\gamma_E + 2\ln2-7\zeta(3)\hat\mu^2+31\zeta(5)\hat\mu^4
\right.\right.
\nonumber\\
&-&\left.\left.
\left(\frac{\zeta'(-1)}{\zeta(-1)}+\ln \hat m_D+
\frac{1}{3}\right)\!\left((24-18\zeta(3))\hat\mu ^2+120(\zeta(5)-\zeta(3))\hat\mu^4\right)\right\}\right] \hat m_D^2
+180 s_F\hat m_q^2\Bigg\}, 
\label{gap_md}
\end{eqnarray}
and
\begin{eqnarray}
\hat m_q^2&=&\frac{d_A}{8d_F\left(\pi ^2-6\right)}\frac{\alpha_s s_F}{\pi }\left[3\left(\ln\frac{\hat\Lambda }{2}
+2.198-44.953\ \hat\mu^2 -\left(288\ln\frac{\hat\Lambda }{2}+19.836\right)\hat\mu^4\right)-12\hat m_D\right]. 
\label{gap_mq}
\end{eqnarray}
\begin{figure}[h]
\includegraphics[width=9.5cm, height=7cm]{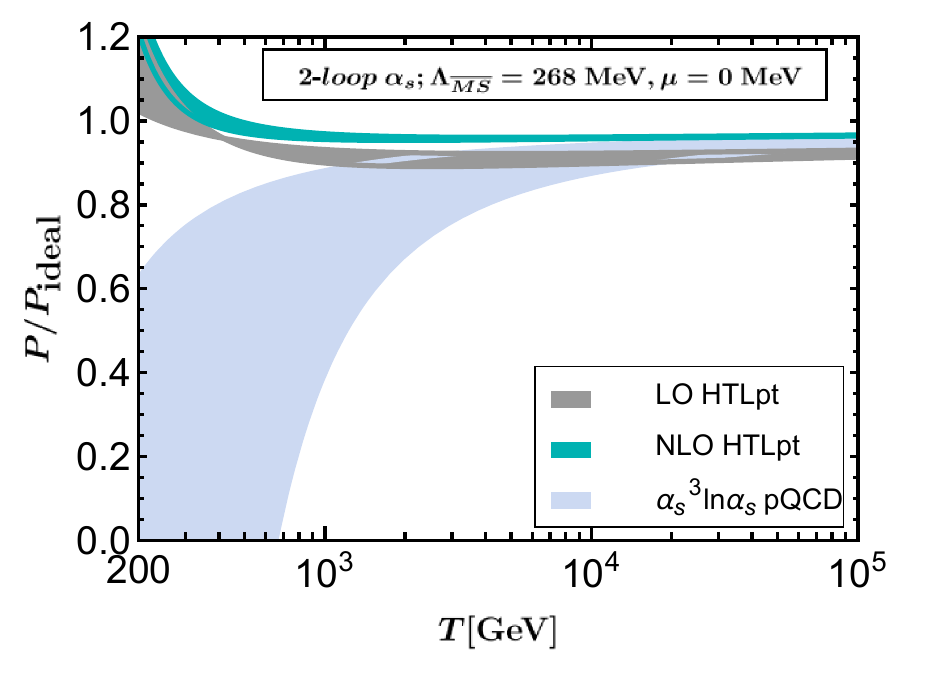}
\includegraphics[width=9.5cm, height=7cm]{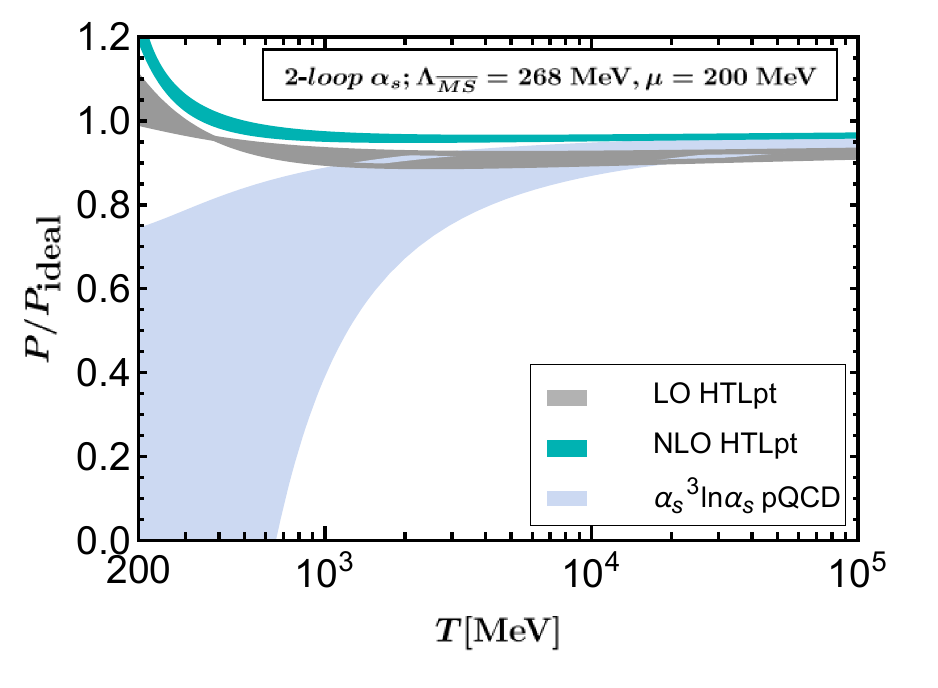}
\vspace{-1.2cm}
\caption{The NLO HTLpt pressure scaled with ideal gas pressure plotted along with four-loop	pQCD pressure~\cite{Vuorinen:2002ue,Vuorinen:2003fs} for two different values of chemical potential with $N_f=3$  and 2-loop running coupling constant $\alpha_s$. The bands are obtained by varying the renormalization scale by a factor of  2 around its central value $\Lambda=2\pi \sqrt{T^2+\mu^2/\pi^2}$ \cite{Vuorinen:2002ue,Vuorinen:2003fs,Rebhan:2003wn, Cassing:2007nb,Gardim:2009mt,Kurkela:2009gj}. 
We use $\Lambda_{\overline{\rm MS}}=290$ MeV based on recent lattice calculations~\cite{Bazavov:2012ka} of the three-loop running of $\alpha_s$.}
	\label{press_htl_nlo}
\end{figure}
The NLO HTLpt result differs from the pQCD result through order $\alpha_s^3\ln\alpha_s$ at low temperatures.
A NNLO HTLpt calculation at finite $\mu$ would agree better with pQCD $\alpha_s^3\ln\alpha_s$ as found in $\mu=0$ case~\cite{Andersen:2011sf}. The  HTLpt result 
clearly indicates a modest improvement over pQCD in respect of convergence and sensitivity of the renormalization 
scale. 
 Both LO and NLO HTLpt results are less sensitive to  the choice of the renormalization scale than the weak coupling results with the inclusion of successive orders of approximation. Comparison  with available lattice QCD data~\cite{Borsanyi:2012cr} suggests that LO and NLO  HTLpt and pQCD cannot accurately account for the lattice QCD results
below approximately $3\,T_c$; however, the results are in very good qualitative agreement with the lattice QCD results without any fine tuning. We note that  the NLO HTL pressure  is only strictly accurate to order ${\cal }(g^3)$ there is over-counting occurring at higher orders in g, namely at ${\cal }(g^4)$ and ${\cal }(g^5)$. However, one needs to go to next-to-next-to-leading order (NNLO) HTLpt (3-loop)  calculation to improve the results.
\subsubsection{Quark Number Susceptibility}
\label{qns}
\vspace{-0.2cm}
We are now in a position to obtain the second and fourth-order HTLpt QNS following second and fourth order 
$\mu$ derivative of  pressure in~\eqref{pressure-potential}. We note that the pure gluonic loops at any order do not contribute to QNS, however, gluons contribute through  the dynamical fermions through fermionic loops. This makes QNS proportional to only quark degrees of freedom.  Below we present \mbox{(semi-)}analytic expressions for both LO and NLO QNS.

To obtain the second and fourth-order quark number susceptibilities in HTLpt, one requires expressions for $m_D$,  $\frac{\partial^2}{\partial\mu^2}m_D$,  $m_q$, and $\frac{\partial^2}{\partial\mu^2}m_q$ at $\mu=0$ from Eqs.~(\ref{gap_md}) and (\ref{gap_mq}).\footnote{Note that odd derivatives with respect to $\mu$ vanish at $\mu=0$. Fourth-order derivatives at $\mu=0$	are nonzero, however, they appear as multiplicative factors of the gap equations and are therefore not required.} These derivative equations can be found in Ref.~\cite{Haque:2013qta}.
\subsubsection{LO HTLpt second-order QNS}
\vspace{-0.2cm}
An analytic expression for the LO HTLpt second-order QNS can be obtained~\cite{Haque:2013qta}  using \eqref{pressure-potential}
\begin{eqnarray}
\chi_2^{\rm LO}(T)&=&\left.\frac{\partial^2 }{\partial\mu^2}{\cal P}_{\rm LO}(T,\Lambda,\mu)\right|_{\mu=0}=\frac{1}{(2\pi T)^2}
\left.\frac{\partial^2 }{\partial\hat\mu^2}{\cal P}_{\rm LO}(T,\Lambda,\hat\mu)\right|_{\hat\mu=0}
=\frac{d_F T^2}{3}\Bigg[1-\frac{3c_F}{4}\left(\frac{g}{\pi}\right)^2 \nn
&+&\frac{c_F}{4}\sqrt{3(c_A+s_F)}\left(\frac{g}{\pi }\right)^3                 -
\frac{c_F^2}{64}\left(\pi^2-6\right)\left(\frac{g}{\pi}\right)^4 
\!\! +\!\! \frac{c_F}{16}(c_A+s_F)\left(\log\frac{\hat\Lambda
}{2}-\frac{7}{2}+\gamma_E + \frac{\pi ^2}{3}
\right)\left(\frac{g}{\pi }\right)^4\Bigg] \ , \label{chi2_lo}
\end{eqnarray}
where the LO Debye and quark masses listed in ~(\ref{mass_lo}) and their $\mu$ derivatives have been used. 


\subsubsection{LO HTLpt fourth-order QNS}

An analytic expression for the LO HTLpt fourth-order QNS can also be obtained~\cite{Haque:2013qta}  using LO pressure from
\eqref{pressure-potential}
\begin{eqnarray}
\chi_4^{\rm LO}(T)&=&\left.\frac{\partial^4 }{\partial\mu^4}{\cal P}_{\rm LO}(T,\Lambda,\mu)\right|_{\mu=0}=\frac{1}{(2\pi T)^4}
\left.\frac{\partial^4 }{\partial\hat\mu^4}{\cal P}_{\rm LO}(T,\Lambda,\hat\mu)\right|_{\hat\mu=0}
=\frac{2d_F}{\pi^2}\Bigg[1-\frac{3}{4}c_F
\left(\frac{g}{\pi }\right)^2 \nn
&+&\frac{3}{8}c_F s_F\sqrt{\frac{3}{c_A+s_F}}\left(\frac{g}{\pi
}\right)^3
-\frac{c_F^2\left(\pi^2-6\right)}{64}\left(\frac{g}{\pi}\right)^4
 +\ \frac{3}{16}c_F
s_F\left(\log\frac{\hat\Lambda}{2}-\frac{7}{2}+\gamma_E
+\frac{\pi^2}{3}\right)\left(\frac{g}{\pi}\right)^4\Bigg] \ , \label{chi4_lo}
\end{eqnarray}
where, once again, the LO Debye and quark masses listed in Eqs.~(\ref{mass_lo}) and their $\mu$ derivatives have been used. We note that both $\chi_2^{\rm LO}$ in (\ref{chi2_lo}) and $\chi_4^{\rm LO}$ in (\ref{chi4_lo}) are the same as those recently obtained by Andersen et al.~\cite{Andersen:2012wr}; however, the closed-form expressions obtained
here have not been explicitly listed therein.
\subsubsection{NLO HTLpt second and fourth-order QNS}
\vspace{-0.2cm}
A semi-analytic expression for the NLO  HTLpt second-order QNS can be obtained~\cite{Haque:2013qta} from 
NLO pressure in \eqref{pressure-potential}
\begin{eqnarray}
\chi_2^{\rm NLO}(T)&=&\left.\frac{\partial^2 }{\partial\mu^2}{\cal P}_{\rm NLO}(T,\Lambda,\mu)\right|_{\mu=0}=\frac{1}{(2\pi T)^2}
\left.\frac{\partial^2 }{\partial\hat\mu^2}{\cal P}_{\rm NLO}(T,\Lambda,\hat\mu)\right|_{\hat\mu=0}
= \frac{d_AT^2}{2}\Bigg[\frac{2}{3}\frac{d_F}{d_A} \nn
&+& \frac{\alpha_s}{\pi}s_F\Bigg\{-1
+4\ \hat m_D(0)
+\frac{2}{3}\left(\ln{\frac{\hat\Lambda}{2}-15.662}\right)
\times(4-3\zeta(3))\,\hat m_D^2(0)+ 44.953\ \hat m_q^2(0)
\nonumber\\     &+&\left[\frac{14}{3}\zeta(3)+\left(\frac{\zeta'(-1)}{\zeta(-1)}
+ \ln\hat m_D(0)\right)( 16-12\zeta(3))\right]\hat m_D^3(0)\Bigg\}\Bigg].\hspace{1cm}
\label{chi2_nlo}
\end{eqnarray}
Additionally, a semi-analytic expression for the NLO HTLpt fourth-order QNS can also be obtained~\cite{Haque:2013qta} from NLO pressure as
\begin{eqnarray}
	\chi_4^{\rm NLO}(T)
	&=&\left.\frac{\partial^4 }{\partial\mu^4}{\cal P}_{\rm NLO}(T,\Lambda,\mu)\right|_{\mu=0}
	=\frac{1}{(2\pi T)^4}\left.\frac{\partial^4 }
	{\partial\hat\mu^4}{\cal P}_{\rm NLO}(T,\Lambda,\hat\mu)\right|_{\hat\mu=0}
	=\!\!\!\!\frac{d_A}{4\pi^2}\Bigg[ 8\frac{ d_F}{d_A}
	+\frac{\alpha_s}{\pi }s_F 
	\Bigg\{-12 +6  \hat m_D''(0)
	\nonumber\\
	&+&\!\!\!\! 3 \hat m_D^2(0) 
	\left[ \left(\frac{\zeta'(-1)}{\zeta(-1)}+\ln\hat m_D(0)+\frac{1}{3}\right)
	(24-18 \zeta(3)) +7\zeta(3)\right] \hat m_D''(0)
	+\hat m_D(0)\hat m_D''(0)  \left(\ln\frac{\hat\Lambda} {2}-15.662\right) \nn
	&\times& (8-6 \zeta(3)) 
	- 4 \hat m_D^3(0) \left[31 \zeta(5)-120 \left(
	\frac{\zeta'(-1)}{\zeta(-1)}+\ln\hat m_D(0)\right) (\zeta(5)-\zeta(3))
	\right]
	\nonumber\\
	&+&\!\!\!\! 80\hat m_D^2(0)  \left(\ln\frac{\hat\Lambda} {2}-1.0811\right) (\zeta(5)-\zeta(3))
	+ 134.859\ \hat m_q(0)\hat m_q''(0)\Bigg\}\Bigg], \hspace{2cm}
	\label{chi4_nlo}
\end{eqnarray}
\begin{figure}[tbh]
		\includegraphics[width=0.51\textwidth,height=0.4\textwidth]{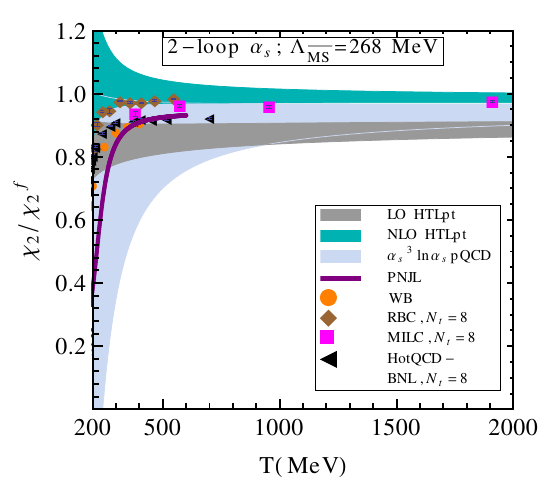}
	\hspace{-0.7cm}
		\includegraphics[width=0.51\textwidth,height=0.4\textwidth]{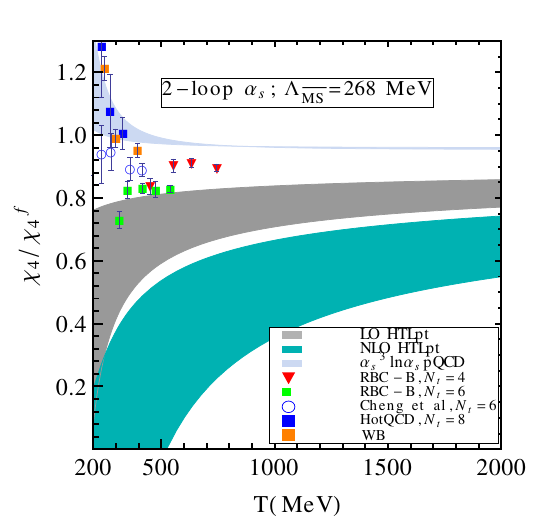}
\vspace{-0.2cm}
	\caption{{\em Left panel} : $\chi_2$  scaled by the free field value for LO (grey band) and NLO (sea green band) in 2-loop HTLpt, 4-loop pQCD (sky blue band)~\cite{Vuorinen:2003fs}, LQCD (various symbols)~\cite{Borsanyi:2012cr,Petreczky:2012rq,Petreczky:2009cr, Bernard:2004je,Borsanyi:2012rr}, and PNJL model (thick purple line)~\cite{Bhattacharyya:2010jd,Bhattacharyya:2010ef} are plotted as a function of the temperature. {\em Right panel} : $\chi_4$ scaled by the free field value for LO and NLO HTLpt as given, respectively, in (\ref{chi4_lo}) and (\ref{chi4_nlo}), 4-loop pQCD~\cite{Vuorinen:2003fs}, and LQCD~\cite{Petreczky:2009cr,Cheng:2008zh} are plotted as a function of the temperature. We used $\Lambda_{\overline{\rm MS}}=268$ MeV  and 2-loop $\alpha_s$ for HTLpt and pQCD in both the plots. Additionally, in both the plots, the bands in HTLpt and pQCD are obtained by varying the ${\overline{\rm MS}}$ renormalization scale ($\Lambda$) around its central value by a factor of two. Lattice QCD results~\cite{Petreczky:2009cr,Cheng:2008zh, Borsanyi:2012rr} are represented by symbols.}
	\label{fig_chi2}
\end{figure}
We note that no $\mu$ derivatives of the mass parameters appear in (\ref{chi2_nlo}) and, as a result, 
$\chi_2^{\rm NLO}(T)$ reduces to such a simple and compact form. This is because the second derivatives of the mass parameters with respect to $\mu$ always appear as multiplicative factors of the gap equations
and hence these contributions vanish~\cite{Haque:2013qta}.  Numerically solving for the variational masses 
one can directly compute $\chi_2^{\rm NLO}(T)$ from Eq.~\ref{chi2_nlo}. Alternatively, 
we have also computed $\chi_2^{\rm NLO}(T)$ by performing numerical differentiation of the pressure
which leads to the same result within numerical errors.

In Eq.~\eqref{chi4_nlo},  the double derivatives of the mass parameters with respect to $\mu$ survive, but the fourth derivatives of the mass parameters disappear as discussed earlier. One can now directly compute the fourth-order susceptibility by using numerical solutions of the gap equations in given in Ref.~\cite{Haque:2013qta}. Alternatively, we have also computed  $\chi_4^{\rm NLO}(T)$ by performing numerical differentiation of the pressure which leads to the same result within numerical errors.

In left panel of Fig.~(\ref{fig_chi2}), we have depicted the second-order quark number susceptibility (QNS), scaled by the corresponding free gas limit, for $N_f = 3$ as a function of temperature. As discussed earlier, the bands shown for the HTLpt and pQCD~\cite{Vuorinen:2003fs} results illustrate the sensitivity of $\chi_2$ to the choice of the renormalization scale $\Lambda$. However, $\chi_2$ in both HTLpt and pQCD depends only weakly on the chosen order of the running 
of the strong coupling and in turn only depends weakly on $\Lambda_{\overline{\rm MS}}$ , as discussed in Ref.~\cite{Haque:2013qta}. The LO HTLpt prediction for $\chi_2$ appears to reasonably agree with the available Wuppertal-Budapest LQCD data, obtained using the tree-level improved Symanzik action and a stout smeared staggered 
fermionic action with light quark masses $\sim 0.035 \, m_s$, with $m_s$ being the strange quark mass near its 
physical value. Nonetheless, there exists considerable variation among different lattice  computations~\cite{Borsanyi:2012cr,Petreczky:2009cr,Bernard:2004je} considering improved lattice actions and a 
range of quark masses.  Lowering the quark mass nearer to its physical
value~\cite{Borsanyi:2012cr} seems to have a very small effect in the temperature range, as seen from the LQCD data. The  RBC-Bielefeld collaboration~\cite{Petreczky:2012rq} data for $\chi_2$ shown in Fig.~(\ref{fig_chi2}) used  a $p4$ action whereas the MILC collaboration~\cite{Bernard:2004je} used an asqtad action. 
In both cases, the light quark mass ranges from (0.1-0.2)$\,m_s$. Results for $\chi_2$ obtained using a
nonperturbative PNJL model~\cite{Bhattacharyya:2010jd,Bhattacharyya:2010ef}, which includes an six-quark interaction, are only available very close to the phase transition temperature. In Fig.~(\ref{fig_chi2}), we observed that NLO HTLpt~(\ref{chi2_nlo}) demonstrates a modest improvement over the pQCD calculation shown, which is accurate to ${\cal O}(\alpha_s^3\ln \alpha_s)$. However, the NLO $\chi_2$ is higher than the LO one at higher temperature and it goes beyond the free gas value at lower temperatures. It should be mentioned that, although the 2-loop calculation improves upon
the LO results by rectifying over-counting which causes incorrect coefficients in the weak coupling limit, it does so by pushing the problem to higher order in $g$. The reason can be understood in the following way: in HTLpt the loop and coupling expansion are not symmetrical, therefore at a given loop order there are contributions from higher orders in coupling. Since the NLO HTL pressure and thus QNS is only strictly accurate to order ${\cal O}(g^3)$ there is over-counting occurring at higher orders in $g$, namely at ${\cal O}(g^4)$ and ${\cal O}(g^5)$. A NNLO HTLpt calculation would fix the problem through ${\cal O}(g^5)$, thereby guaranteeing that, when expanded in a strict power series in $g$, the HTLpt result would reproduce the perturbative result order-by-order through ${\cal O}(g^5)$.  

In the right panel of Fig.~(\ref{fig_chi2}), we present the fourth-order QNS ($\chi_4$) scaled by the corresponding free gas value for HTLpt as given in (\ref{chi4_lo}) and (\ref{chi4_nlo}), pQCD,  and LQCD. Both the HTLpt and pQCD results exhibit a very weak dependence on the choice of order of the running of  $\alpha_s$ and thus $\Lambda_{\overline{\rm MS}}$. Nevertheless, the HTLpt results are found to be far below the pQCD result~\cite{Vuorinen:2003fs}, which is accurate to ${\cal O}(\alpha_s^3\ln(\alpha_s))$, and the LQCD results~\cite{Petreczky:2009cr,Cheng:2008zh}. Additionally, the correction to $\chi_4$ when going from LO to NLO is quite large. This is due to the fact that the fourth order susceptibility is highly sensitive to the erroneous  ${\cal O}(g^4)$ and ${\cal O}(g^5)$ terms which appear at NLO. It is expected that carrying the HTLpt calculation to NNLO would improve this situation; however, only explicit calculation can prove this.  It's worth noting that while the pQCD result is very close to the Stefan-Boltzmann limit, the dimensional-reduction resummation method yields a fourth-order QNS  approximately $20\%$ below the Stefan-Boltzmann  limit~\cite{Andersen:2012wr}, placing it slightly above the LO HTLpt result shown in Fig.~(\ref{fig_chi2})
\vspace{-0.2cm}
\subsection{NNLO HTL Thermodynamics}
\begin{figure}[h]
\begin{center}
	\vspace{-0cm}\includegraphics[width=18cm,height=13cm]{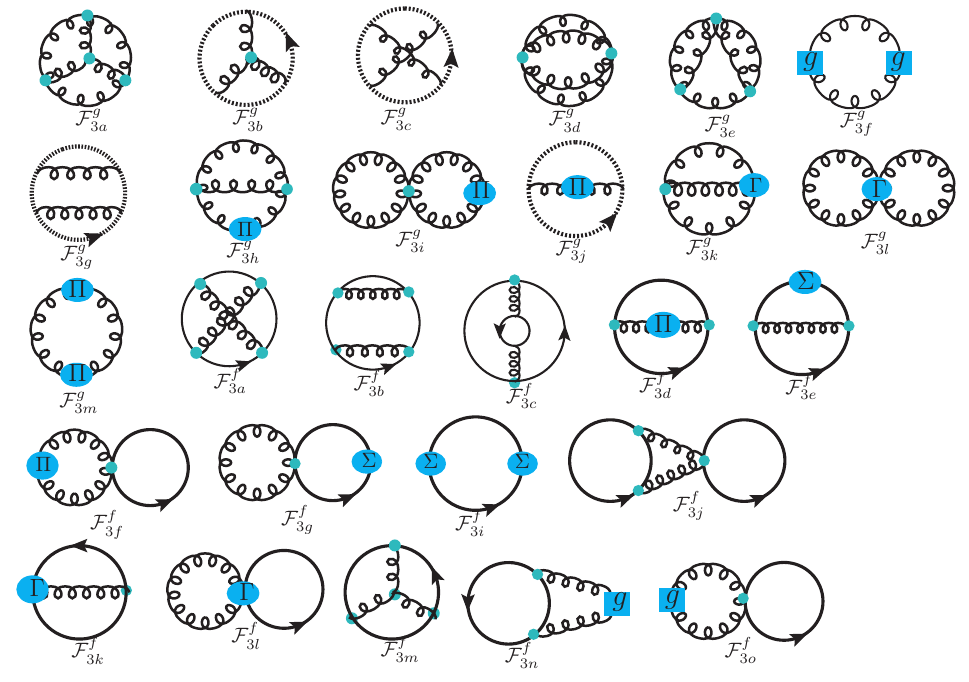}
	\caption{Three loop HTL Feynman diagrams that will contribute to the thermodynamic potential.}
	\label{feyn_diag3}
\end{center}
\end{figure}

\begin{figure}[h]
\begin{center}
	\includegraphics[width=12cm]{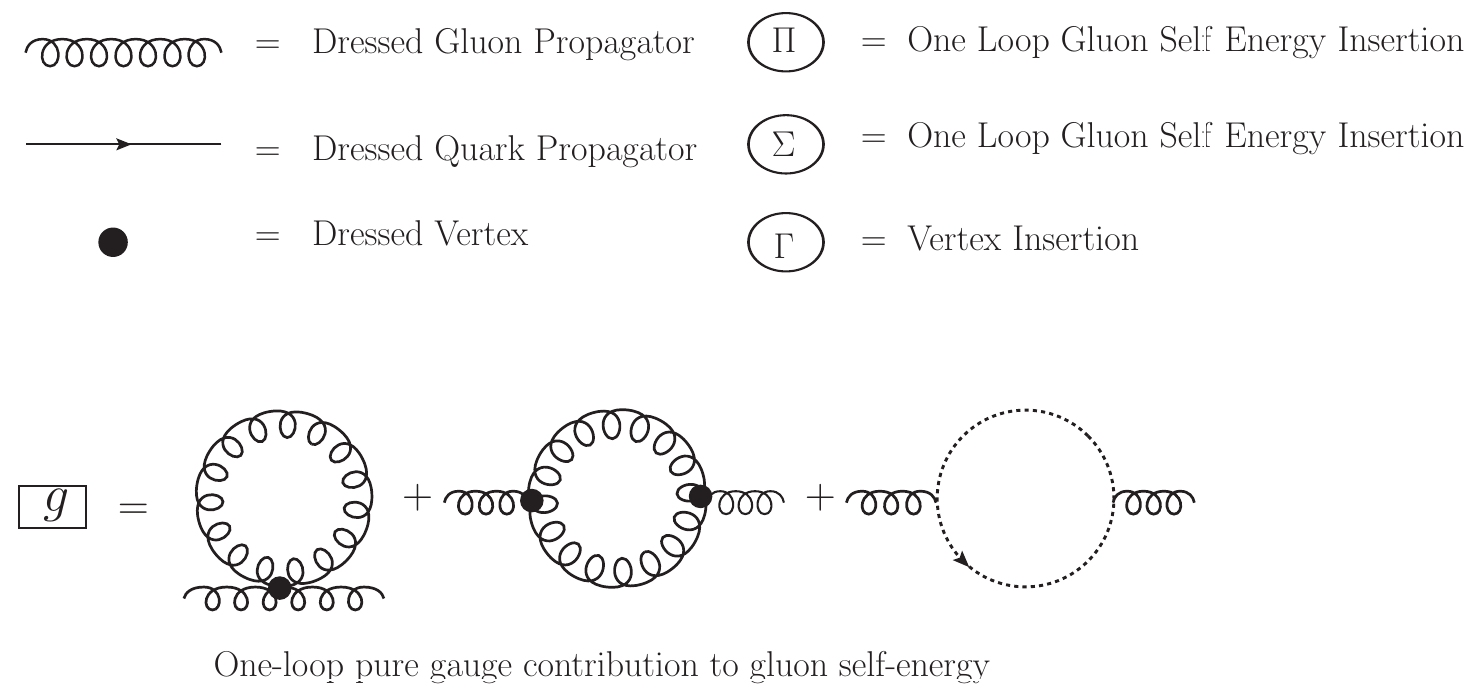}
	\caption{The shorthand notations used in Fig.~(\ref{feyn_diag3}).}
	\label{shorthand}
	\end{center}
\end{figure}

\vspace{-0.2cm}
The diagrams needed for the computation of the HTLpt thermodynamic potential through NNLO are listed in 
Figs.~(\ref{diagramfig}) and (\ref{feyn_diag3}). The shorthand notations used in Fig.~(\ref{feyn_diag3})
have been explained in Fig.~(\ref{shorthand}).

In Ref.~\cite{Andersen:2011sf} the authors computed the NNLO thermodynamic potential at zero chemical potential.  Here we extend the NNLO calculation to finite chemical potential~\cite{Haque:2013sja,Haque:2014rua}. For this purpose, one needs to only consider diagrams which contain at least one quark propagator; however, for completeness we also list the purely gluonic contributions below.  In the results we will express thermodynamic quantities in terms of dimensionless variables:  $\hat{m}_D = m_D/(2\pi T)$, $\hat{m}_q = m_q/(2\pi T)$, $\hat{\mu} = \mu/(2\pi T)$,  $\hat{\Lambda} = \Lambda/(2\pi T)$ and $\hat{\Lambda}_g = \Lambda_g/(2\pi T)$ where $\Lambda$ and  $\Lambda_g$ are renormalization scales  for gluon and fermion respectively as discussed in  Refs~\cite{Andersen:2011sf,Haque:2013sja,Haque:2014rua}.

The complete NNLO HTLpt thermodynamic potential can be expressed in terms of each diagrams of Figs.~(\ref{diagramfig})
and (\ref{feyn_diag3}) as
\bea
\Omega_{\rm NNLO}\!\!&=&\!\!d_A\left[{\cal F}_{1a}^g+{\cal F}_{1b}^g+{\cal F}_{2d}^g+{\cal F}_{3m}^g\right]+d_F\[{\cal F}_{1b}^f
+{\cal F}_{2d}^f+{\cal F}_{3i}^f\]
\hspace{-.1cm}+d_Ac_A\Big[{\cal F}_{2a}^g+{\cal F}_{2b}^g+{\cal F}_{2c}^g+{\cal F}_{3h}^g+{\cal F}_{3i}^g+{\cal F}_{3j}^g
+{\cal F}_{3k}^g+{\cal F}_{3l}^g\Big]   \nonumber\\
&&\hspace{-.8cm}+\ d_As_F\Big[{\cal F}_{2a}^f+{\cal F}_{2b}^f+{\cal F}_{3d}^f+{\cal F}_{3e}^f+{\cal F}_{3f}^f+{\cal F}_{3g}^f+
{\cal F}_{3k}^f+{\cal F}_{3l}^f\Big]   
\hspace{-.1cm}  + d_Ac_A^2\Big[{\cal F}_{3a}^g+{\cal F}_{3b}^g+{\cal F}_{3c}^g+{\cal F}_{3d}^g
+{\cal F}_{3e}^g+{\cal F}_{3f}^g+{\cal F}_{3g}^g\Big] \nn
&&\hspace{-.8cm}+\ d_As_{2F}
\Big[{\cal F}_{3a}^f+{\cal F}_{3b}^f\Big]  
\hspace{-.1cm}+d_Ac_As_{F}
\Big[-\frac{1}{2}{\cal F}_{3a}^f+{\cal F}_{3m}^f+{\cal F}_{3n}^f+{\cal F}_{3o}^f\Big]
+d_As_F^2\Big[{\cal F}_{3c}^f+{\cal F}_{3j}^f\Big]  
\hspace{-.1cm} + \Delta_0{\cal E}_0+\Delta_1{\cal E}_0+ \Delta_2{\cal E}_0\nn
&& \hspace{-.8cm} +\ \Delta_1m_D^2\frac{\partial}
{\partial m_D^2}\Omega_{\rm LO}+\Delta_1m_q^2\frac{\partial}{\partial m_q^2}\Omega_{\rm LO}
\hspace{-.1cm}+\Delta_2m_D^2\frac{\partial}
{\partial m_D^2}\Omega_{\rm LO} +\Delta_2m_q^2\frac{\partial}{\partial m_q^2}\Omega_{\rm LO} + 
\Delta_1m_D^2\frac{\partial}{\partial m_D^2}
\Omega_{\rm NLO} \hspace{-.1cm} + \Delta_1m_q^2\frac{\partial}{\partial m_q^2}\Omega_{\rm NLO}\nn
&&\hspace{-.8cm} +\ \frac{1}{2}\left[\frac{\partial^2}{(\partial m_D^2)^2}\Omega_{\rm LO}
\right]\left(\Delta_1 m_D^2\right)^2 
\hspace{-.1cm}+\frac{1}{2}\left[\frac{\partial^2}{(\partial m_q^2)^2}
\Omega_{\rm LO}\right]\left(\Delta_1 m_q^2\right)^2
+d_A\left[\frac{c_A{\cal F}_{2a+2b+2c}^g+s_F{\cal F}_{2a+2b}^f}{\alpha_s}\right]\Delta_1\alpha_s,
\label{omega_initial}
\eea
where the necessary counterterms at any order in $\delta$ can be calculated using the following counter terms
\be
\Delta{\cal E}_0&=&\frac{d_A}{128\pi^2\epsilon}(1-\delta)^2m_D^4\ ,
\qquad \Delta m_D^2=\frac{11c_A-4s_F}{12\pi\epsilon}\alpha_s\delta(1-\delta)m_D^2\ ,
\nn
\Delta m_q^2&=&\frac{3}{8\pi\epsilon}\frac{d_A}{c_A}\alpha_s \delta(1-\delta)m_q^2\ ,
\qquad
\delta\Delta\alpha_s=-\frac{11c_A-4s_F}{12\pi\epsilon}\alpha_s^2\delta^2\ .
\label{cnlo}
\ee
The expressions for the one- and two-loop diagrams above can be found in Refs.~\cite{Andersen:2002ey,Andersen:2003zk}. The expressions for the three-loop bosonic diagrams ${\cal F}_{3a}^g$--${\cal F}_{3m}^g$ are presented in section 3 of Ref.~\cite{Andersen:2010ct}. The three-loop diagrams specific to QCD, i.e. the  non-Abelian diagrams involving quarks, are given in Refs.~\cite{Haque:2013sja,Haque:2014rua}.

In Refs.~\cite{Andersen:2002ey,Andersen:2003zk} the NLO HTLpt thermodynamic potential was reduced to scalar sum-integrals. Evaluating these scalar sum-integrals exactly appears intractable, however, an approximate calculation can be achieved by expanding them in powers of $m_D/T$ and $m_q/T$ following the method developed in Ref.~\cite{Andersen:2001ez}. We will adopt the same strategy in this section, including all terms through order $g^5$ under the assumption that $m_D$ and $m_q$ are ${\cal O}(g)$ at leading order. At each loop order, the contributions can be divided into those arising from hard and soft momenta, proportional to the scales $T$ and $gT$ respectively. In one-loop diagrams, contributions are either hard $(h)$ or
soft $(s)$, while at the two-loop level, there are hard-hard $(hh)$, hard-soft $(hs)$, and soft-soft $(ss)$ contributions.
At three loops, there are hard-hard-hard $(hhh)$, hard-hard-soft $(hhs)$, hard-soft-soft $(hss)$, and soft-soft-soft $(sss)$
contributions. Detailed calculations are given in Refs.~\cite{Haque:2013sja,Haque:2014rua}.
We expand each term of Eq.~\eqref{omega_initial} for small $m_D/T$ and $m_q/T$, calculating necessary sum-integrals and three dimensional integrals that arise. For detailed calculations, we refer Refs.~\cite{Haque:2013sja,Haque:2014rua}. In this section, we summarise the results from~\cite{Haque:2013sja,Haque:2014rua} to obtain final thermodynamic potential.

We consider first  the case that all quarks have the same chemical potential $\mu_f = \mu = \mu_B/N_f$ where 
$f$ is a flavour index with $f \in \{ \mu_u, \mu_d, \mu_s, \cdots, \mu_{N_f} \}$. Here we are considering $N_f=3$, 
so $\mu_f = \mu = \mu_B/3$.  Then we will present the general result with separate chemical potentials for each flavour.
\subsubsection{NNLO result for equal chemical potentials}
When all quarks have the same chemical potential $\mu_i = \mu = \mu_B/3$ we can straightforwardly combine
the results for the various sum-integrals in three-loop  and  the  HTLpt counterterm, we obtain our final result for 
the NNLO HTLpt thermodynamic potential~\cite{Haque:2013sja,Haque:2014rua}  in the case that all quarks have the same chemical potential 
\begin{eqnarray}
\frac{\Omega_{\rm NNLO}}{\Omega_0}
&=& \frac{7}{4}\frac{d_F}{d_A}\(1+\frac{120}{7}\hmu^2+\frac{240}{7}\hmu^4\)
-\frac{s_F\alpha_s}{\pi}\bigg[\frac{5}{8}\left(1+12\hat\mu^2\right)\left(5+12\hat\mu^2\right)
-\frac{15}{2}\left(1+12\hat\mu^2\right)\hat m_D +90\hat m_q^2 \hat m_D
\nn
&&\hspace{-1.5cm}-\, \frac{15}{2}\bigg(2\ln{\frac{\hat\Lambda}{2}-1 -\aleph(z)}\Big)\hat m_D^3
\bigg]
+15 s_{2F}\left(\frac{\alpha_s}{\pi}\right)^2\bigg[{35\over 64} -{1\over2}\(1-12\hmu^2\)\frac{\zeta'(-1)}
{\zeta(-1)}+{59\over8} \hat\mu^2+ {83\over4} \hat\mu^4 - 36i\hat\mu\aleph(2,z)\nn
&&\hspace{-1.5cm}+\,6(1+8\hat\mu^2)\aleph(1,z)+3i\hat\mu(1+4\hat\mu^2)\aleph(0,z)
- \frac{3}{2}\hat m_D\left(1+12\hat\mu^2\right)\bigg] 
+\left(\frac{s_F\alpha_s}{\pi}\right)^2\left[\frac{5}{4\hat m_D}\left(1+12\hat\mu^2\right)^2 \right.\nn
&&\left.\hspace{-1.5cm}+\,30\left(1+12\hat\mu^2
\right)\frac{\hat m_q^2}{\hat m_D}
+   \frac{25}{12}\Bigg\{ \left(1 +\frac{72}{5}\hat\mu^2+\frac{144}{5}\hat\mu^4\right)\ln\frac{\hat\Lambda}{2}\right.
+\frac{1}{20}\(1+168\hmu^2+2064\hmu^4\)
+\frac{3}{5}\(1+12\hmu^2\)^2\gamma_E\nn
&&\hspace{-1.5cm} - \, \frac{8}{5}(1+12\hat\mu^2)\frac{\zeta'(-1)}{\zeta(-1)}
- \frac{34}{25}\frac{\zeta'(-3)}{\zeta(-3)}
-\frac{72}{5}\Big[8\aleph(3,z)
+3\aleph(3,2z)+ 12 i \hat\mu\,(\aleph(2,z)+\aleph(2,2z))  
- 12\hat\mu^2\aleph(1,2z)\nn
&&\hspace{-1.5cm} -\, i \hat\mu(1+12\hat\mu^2)\,\aleph(0,z)   - 2(1+8\hat\mu^2)\aleph(1,z)\Big]\Bigg\} -\left.\frac{15}{2}\(1+12\hat\mu^2\)\(2\L-1-\aleph(z)\)\hat m_D\right]
+ \left(\frac{c_A\alpha_s}{3\pi}\right)\left(\frac{s_F\alpha_s}{\pi}\right)\nn
&&\hspace{-1.5cm}\times\Bigg[\frac{15}{2\hat m_D}\(1+12\hmu^2\)
-\frac{235}{16}\Bigg\{\bigg(1+\frac{792}{47}\hat\mu^2+\frac{1584}{47}\hat\mu^4\bigg)\ln\frac{\hat\Lambda}{2}
-\frac{144}{47}\(1+12\hmu^2\)\ln\hat m_D+\frac{319}{940}\left(1+\frac{2040}{319}\hat\mu^2+\frac{38640}{319}\hat\mu^4\right)\nn
&&\hspace{-1.5cm}-\,\frac{24 \gamma_E }{47}\(1+12\mu^2\) -
\frac{44}{47}\(1+\frac{156}{11}\hmu^2\)\frac{\zeta'(-1)}{\zeta(-1)}
-\frac{268}{235}\frac{\zeta'(-3)}{\zeta(-3)}
-\frac{72}{47}\Big[4i\hat\mu\aleph(0,z)
+\left(5-92\hat\mu^2\right)\aleph(1,z)+144i\hmu\aleph(2,z)\nn
&&\hspace{-1.5cm}
+\,52\aleph(3,z)\Big]\Bigg\}+\frac{90\hat m_q^2}{\hat m_D}
+\frac{315}{4}\Bigg\{\!\!\(1+\frac{132}{7}\hmu^2\!\)\!\L+\frac{11}{7}\(1+12\hmu^2\)\gamma_E+\frac{9}{14}\!\(1+\frac{132}{9}\hmu^2\)\! +\frac{2\aleph(z)}{7}\Bigg\}\hat m_D 
\Bigg]
+ \frac{\Omega_{\rm NNLO}^{\rm YM}}{\Omega_0}  .\hspace{1.2cm}
\label{finalomega1}
\end{eqnarray}
where $\Omega_0=-d_A\pi^2T^4/45$ and 
$\Omega_{\rm NNLO}^{\rm YM}$ is the NNLO pure glue thermodynamic potential~\cite{Andersen:2010ct}
	\be
	\frac{\Omega_{\rm NNLO}^{\rm YM}}{\Omega_0} \!\!\!&=&\!\!\! 1-\frac{15}{4}\hat m_D^3+\frac{c_A\alpha_s}{3\pi}\Bigg[-\frac{15}{4}
	+\frac{45}{2}\hat m_D-\frac{135}{2}\hat m_D^2-\frac{495}{4}\(\Lg+\frac{5}{22}+\gamma_E\)  \hat m_D^3 \Bigg]
	+\(\frac{c_A\alpha_s}{3\pi}\)^2\Bigg[\frac{45}{4\hat m_D}\nn
	&&\hspace{-1.5cm} +\,\frac{1485}{4}\!\(\!\Lg-\frac{79}{44}+\gamma_E+\ln2-\frac{\pi^2}{11}\)\!\hat m_D  -\frac{165}{8}\(\!\Lg-\frac{72}{11}\ln\hat m_D-\frac{84}{55}-\frac{6\gamma_E}{11}
	-\frac{74}{11}\Za+\frac{19}{11}\Zc\)\!\Bigg] .\hspace{1.2cm}
	\label{nnloym}
	\ee
It is worth mentioning here that the NNLO pure glue thermodynamic potential in Eq.~\eqref{nnloym} appears to be independent of the chemical potential since there is no explicit dependence on it.  However, the chemical potential is implicitly present within the Debye mass, $m_D$. It arises in pure glue diagrams due to the internal quark loop in effective gluon propagators and effective vertices. It's important to note that the complete thermodynamic potential (\ref{finalomega1}) reduces to thermodynamic potential outlined in Ref.~\cite{Andersen:2011sf} as the chemical potential approaches zero. Furthermore, the aforementioned thermodynamic potential yields the correct ${\cal O}(g^5)$ 
perturbative result when expanded in a strict power series in $g$~\cite{Vuorinen:2002ue,Vuorinen:2003fs}.
\subsubsection{NNLO result -- General case}
\vspace{-0.2cm}
It is relatively straightforward to generalize the previously obtained result (\ref{finalomega1}) to the case that each quark
has a separate chemical potential $\mu_f$.  The final result~\cite{Haque:2013sja,Haque:2014rua}  is
\begin{eqnarray}
\frac{\Omega_{\rm NNLO}}{\Omega_0}
&=& \frac{7}{4}\frac{d_F}{d_A}\frac{1}{N_f}\sum\limits_f\(1+\frac{120}{7}\hmu_f^2+\frac{240}{7}\hmu_f^4\)
-\frac{s_F\alpha_s}{\pi}\frac{1}{N_f}\!\sum\limits_f\bigg[\frac{5}{8}\left(5+72\hat\mu_f^2+144\hat\mu_f^4\right)
-\frac{15}{2}\left(1+12\hat\mu_f^2\right)\hat m_D \nn
&-&\frac{15}{2}\bigg(2\ln{\frac{\hat\Lambda}{2}-1
	-\aleph(z_f)}\Big)\hat m_D^3
+90\hat m_q^2 \hat m_D\bigg] + \frac{s_{2F}}{N_f}\left(\frac{\alpha_s}{\pi}\right)^2\sum\limits_f\bigg[\frac{15}{64}\bigg\{35-32\(1-12\hmu_f^2\)\frac{\zeta'(-1)} {\zeta(-1)}+472 \hat\mu_f^2\nn
&+&1328  \hat\mu_f^4
+ 64\Big(-36i\hat\mu_f\aleph(2,z_f)+6(1+8\hat\mu_f^2)\aleph(1,z_f)+3i\hat\mu_f(1+4\hat\mu_f^2)\aleph(0,z_f)\Big)\bigg\}- \frac{45}{2}\hat m_D\left(1+12\hat\mu_f^2\right)\bigg] \nn
&+& \left(\frac{s_F\alpha_s}{\pi}\right)^2
\frac{1}{N_f}\sum\limits_{f}\frac{5}{16}\Bigg[96\left(1+12\hat\mu_f^2\right)\frac{\hat m_q^2}{\hat m_D}
+\frac{4}{3}\(1+12\hmu_f^2\)\(5+12\hat\mu_f^2\)
\ln\frac{\hat{\Lambda}}{2}
+\frac{1}{3}+4\gamma_E+8(7+12\gamma_E)\hat\mu_f^2\nn
&+&112\mu_f^4
- \frac{64}{15}\frac{\zeta^{\prime}(-3)}{\zeta(-3)}-
\frac{32}{3}(1+12\hat\mu_f^2)\frac{\zeta^{\prime}(-1)}{\zeta(-1)}
- 96\Big\{8\aleph(3,z_f)+12i\hat\mu_f\aleph(2,z_f)-2(1+2\hat\mu_f^2)\aleph(1,z_f)\nn
&-& i\hat\mu_f\aleph(0,z_f)\Big\}\Bigg]  + \left(\frac{s_F\alpha_s}{\pi}\right)^2
\frac{1}{N_f^2}\sum\limits_{f,g}\Bigg[\frac{5}{4\hat m_D}\left(1+12\hat\mu_f^2\right)\left(1+12\hat\mu_g^2\right)
+90\Bigg\{ 2\left(1 +\gamma_E\right)\hat\mu_f^2\hat\mu_g^2
-\Big\{\aleph(3,z_f+z_g)\nn
&+&\aleph(3,z_f+z_g^*)
+s_F 4i\hat\mu_f\left[\aleph(2,z_f+z_g)+\aleph(2,z_f+z_g^*)\right]
- 4\hat\mu_g^2\aleph(1,z_f)-(\hat\mu_f+\hat\mu_g)^2\aleph(1,z_f+z_g)- (\hat\mu_f-\hat\mu_g)^2\nn
&\times& \aleph(1,z_f+z_g^*)
- 4i\hat\mu_f\hat\mu_g^2\aleph(0,z_f)\Big\}\Bigg\}-\frac{15}{2}\(1+12\hat\mu_f^2\)\(2\L-1-\aleph(z_g)\)  \hat 
m_D\Bigg]
+ \left(\frac{c_A\alpha_s}{3\pi}\right)\left(\frac{s_F\alpha_s}{\pi N_f}\right)\nn
&\times& \sum\limits_f\Bigg[
-\frac{235}{16}\Bigg\{\bigg(1+\frac{792}{47}\hat\mu_f^2+\frac{1584}{47}\hat\mu_f^4\bigg)\ln\frac{\hat\Lambda}{2}
-\frac{144}{47}\(1+12\hmu_f^2\)\ln\hat m_D+\frac{319}{940}\left(1+\frac{2040}{319}\hat\mu_f^2+\frac{38640}{319}\hat\mu_f^4\right)   
\nn
&-&\frac{24 \gamma_E }{47}\(1+12\hat\mu_f^2\)-\frac{44}{47}\(1+\frac{156}{11}\hmu_f^2\)\frac{\zeta'(-1)}{\zeta(-1)}
-\frac{268}{235}\frac{\zeta'(-3)}{\zeta(-3)}
-\frac{72}{47}\Big[4i\hat\mu_f\aleph(0,z_f)+\left(5-92\hat\mu_f^2\right)\aleph(1,z_f)\nn
&+& 144i\hmu_f\aleph(2,z_f)
+52\aleph(3,z_f)\Big]\Bigg\}
+ \frac{15}{2\hat m_D}\(1+12\hmu_f^2\)+90\frac{\hat m_q^2}{\hat m_D}
+\frac{315}{4}\Bigg\{\(1+\frac{132}{7}\hmu_f^2\)\L
\nonumber\\
&+&\frac{11}{7}\(1+12\hmu_f^2\)\gamma_E+\frac{9}{14}\(1+\frac{132}{9}\hmu_f^2\)
+\frac{2}{7}\aleph(z_f)\Bigg\}\hat m_D 
\Bigg]
+ \frac{\Omega_{\rm NNLO}^{\rm YM}}{\Omega_0} \, ,
\label{finalomega}
\end{eqnarray}
where the sums over $f$ and $g$ include all quark flavours, $z_f = 1/2 - i \hat{\mu}_f$, and $\Omega_{\rm NNLO}^{\rm YM}$ is the pure-glue contribution as before.
\subsubsection{Mass prescription}
\label{pres}
\vspace{-0.2cm}
As outlined in Ref.~\cite{Andersen:2011sf,Haque:2014rua}, the two-loop perturbative electric gluon mass, first calculated by Braaten and Nieto in~\cite{Braaten:1995cm,Braaten:1995jr}, proves most appropriate for three-loop HTLpt calculations. We employ the 
Braaten-Nieto (BN) mass prescription for $m_D$ throughout the rest of this section. Initially, the two-loop perturbative mass was calculated in Refs.~\cite{Braaten:1995cm,Braaten:1995jr} for zero chemical potential, however, Vuorinen has extended it to finite chemical potential. The resulting expression for $m_D^2$ is~\cite{Vuorinen:2002ue,Vuorinen:2003fs}
\begin{eqnarray}
\hat m_D^2\!\!\!&=&\!\!\!\frac{\alpha_s}{3\pi} \Biggl\{c_A
+\frac{c_A^2\alpha_s}{12\pi}\(5+22\gamma_E+22\Lg\) + \frac{1}{N_f} \sum\limits_{f}
\Biggl[ s_F\(1+12\hmu_f^2\) +\frac{c_As_F\alpha_s}{12\pi}\(\(9+132\hmu_f^2\)+22\(1+12\hmu_f^2\)\gamma_E\right.\nn
&&\left.+\,2\(7+132\hmu_f^2\)\L+4\aleph(z_f)\) +\frac{s_F^2\alpha_s}{3\pi}\(1+12\hmu_f^2\)\(1-2\L+\aleph(z_f)\)
-\frac{3}{2}\frac{s_{2F}\alpha_s}{\pi}\(1+12\hmu_f^2\) \Biggr] \Biggr\} \, .
\end{eqnarray}
The effect of the in-medium quark mass parameter $m_q$ in thermodynamic functions is small and following
Ref.~\cite{Andersen:2011sf} we take $m_q=0$ which is the three loop variational solution.
The maximal effect on the susceptibilities comparing the perturbative quark mass, 
$\hat{m}_q^2 =c_F \alpha_s(T^2+\mu^2/\pi^2)/8\pi$, with the variational solution, $m_q=0$, is approximately 0.2\% at $T=200$ MeV. At higher temperatures, the effect is much smaller, e.g. 0.02\% at $T=1$ GeV.
\vspace{-0.2cm}
\subsubsection{Thermodynamic functions} 
\label{thermof}
\vspace{-0.2cm}
In this section, we present our final results for the NNLO HTLpt pressure, energy density, entropy density, trace anomaly, and speed of sound. We illustrate our NNLO result utilising the one loop  running coupling shown in Eq.~\eqref{ph85}.  Regarding the renormalisation scale, we use separate scales, $\Lambda_g$ and $\Lambda$, for purely-gluonic and fermionic graphs, respectively.  We set the central values of these renormalization scales to be $\Lambda_g = 2\pi T$ and $\Lambda=2\pi \sqrt{T^2+\mu^2/\pi^2}$.  In all plots, the thick lines denote the result obtained using these central values and the light-blue band signifies the variation in the result under when both scales are varied by a factor of two, e.g. $\pi T \leq \Lambda _g \leq 4 \pi T$. For all numerical results below, we employ $c_A = N_c=3$ and $N_f=3$.

\noindent \textbf{Pressure:} 
The QGP pressure can be obtained directly from the thermodynamic potential in Eq.~\ref{finalomega1} as
\be
{\cal P}(T,\Lambda,\mu)=-\Omega_{\rm NNLO}(T,\Lambda,\mu) \, ,
\ee
where $\Lambda$ above is understood to include both scales $\Lambda_g$ and $\Lambda$. We note that in the ideal gas limit, the pressure becomes
\be
{\cal P}_{\rm ideal}(T,\mu)=\frac{d_A\pi^2T^4}{45}\left[1+\frac{7}{4}\frac{d_F}{d_A}\left(1+\frac{120}{7}\hmu^2
+\frac{240}{7}\hmu^4\right)\right] .
\ee
%
\begin{figure}[h!]
	\begin{center}
		\hspace{-2mm}\includegraphics[height=6cm, width=8cm]{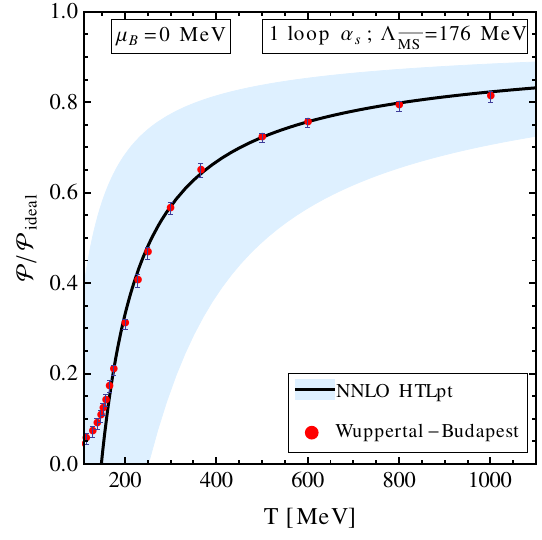}
		\includegraphics[height=6cm, width=8cm]{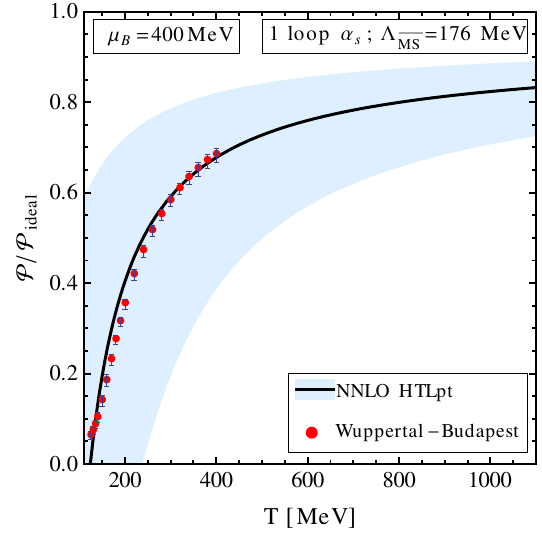}
		\vspace{-0.4cm}
	\caption[Comparison of the $N_f=2+1$, $\mu_B=0$ (left) and $\mu_B=400$ MeV (right) NNLO HTLpt 
	pressure using one-loop running coupling constant with lattice data.]{Comparison of the $N_f=2+1$, $\mu_B=0$ (left) and $\mu_B=400$ MeV (right) NNLO HTLpt pressure with lattice data from Borsanyi et al. \cite{Borsanyi:2010cj,Borsanyi:2012uq}. For the HTLpt results a one-loop running coupling constant was used.}
	\label{pres_1l}
	\end{center}
\end{figure}
In Figs.~(\ref{pres_1l}) we compare the scaled NNLO HTLpt pressure for $\mu_B=0$ (left) and $\mu_B=400$ MeV
(right) with lattice data from Refs.~\cite{Bazavov:2009zn,Borsanyi:2010cj,Borsanyi:2012uq}. The
deviations below $T\sim 200$ MeV are due to the fact that our calculation does not include hadronic degrees of freedom
which dominate at low temperatures (see e.g. fits in~\cite{Huovinen:2009yb}) or nonperturbative effects~\cite{KorthalsAltes:1999xb,Pisarski:2000eq,KorthalsAltes:2000gs,Zwanziger:2004np,Vuorinen:2006nz,deForcrand:2008aw,Fukushima:2013xsa}. Further, in order to gauge the sensitivity of the results to the order of the running coupling, in Fig.~(\ref{pres_1l}) we show the results obtained using a one-loop running coupling. We also note that  comparing the 1-loop running coupling result  results with 3-loop running coupling~\cite{Haque:2013sja,Haque:2014rua}, the sensitivity of the results to the order of the  running coupling is small for $T \gtrsim 250$ MeV~\cite{Haque:2014rua}. As a result, we will use one-loop running coupling to plot all the themodynamics quantities and it is consistent with the counterterms necessary to renormalize the NNLO thermodynamic potential (\ref{cnlo}).  

\noindent \textbf{Energy density:}
Once the pressure is determined, calculating other thermodynamic functions such as the energy density becomes straightforward by computing derivatives of the pressure with respect to temperature and chemical potential. The energy density can be derived as follows:
\begin{eqnarray}
{\cal E}=T\frac{\partial{\cal P}}{\partial T}+\mu\frac{\partial{\cal P}}{\partial \mu}-{\cal P} \, .
\end{eqnarray}
%
We note that in the ideal gas limit, the entropy density becomes
\be
{\cal E}_{\rm ideal}(T,\mu)=\frac{d_A\pi^2T^4}{15}\left[1+\frac{7}{4}\frac{d_F}{d_A}\left(1+\frac{120}{7}\hmu^2
+\frac{240}{7}\hmu^4\right)\right].
\ee
\begin{figure}[tbh]
	\begin{center}
		\hspace{-2mm}
		\includegraphics[width=8.5cm,height=6cm]{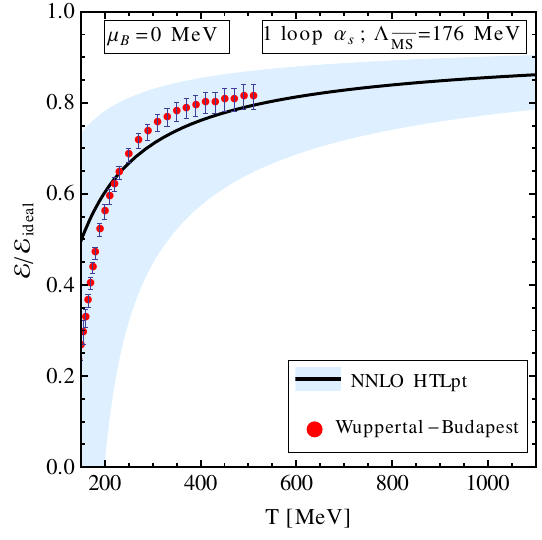}
		\includegraphics[width=8.5cm,height=6cm]{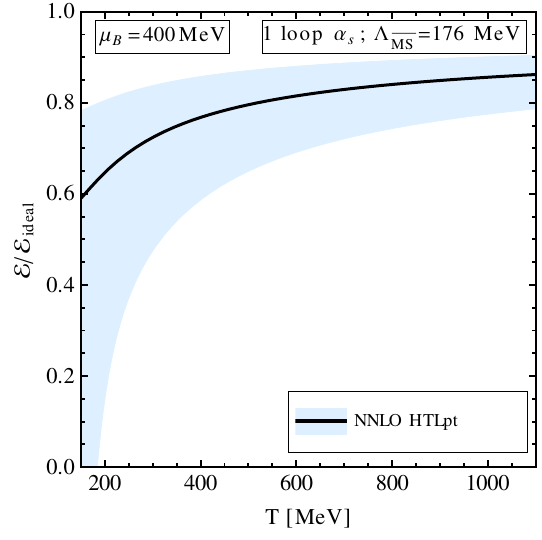}
	\caption{
		Comparison of the $N_f=2+1$, $\mu_B=0$ (left) and $\mu_B=400$ MeV (right) NNLO HTLpt energy density with lattice data. The $\mu_B=0$ lattice data are from \cite{Borsanyi:2010cj}.}
	\label{ed_1l}
\end{center}
\end{figure}
In Fig.~(\ref{ed_1l}) we plot the scaled NNLO HTLpt energy density for $\mu_B=0$ (left) and $\mu_B=400$ MeV (right) together with $\mu_B=0$ lattice data from Ref.~\cite{Borsanyi:2010cj}. As we can see from this figure, there is reasonable agreement between the NNLO HTLpt energy density and the lattice data when the central value of the scale is used.

\noindent \textbf{Entropy density:}
Similarly, we can compute the entropy density
$
{\cal S}(T,\mu) = \frac{\partial{\cal P}}{\partial T} \, .
$
We also note that in the ideal gas limit, the entropy density becomes
\be
{\cal S}_{\rm ideal}(T,\mu)=\frac{4d_A\pi^2T^3}{45}\left[1+\frac{7}{4}\frac{d_F}{d_A}\left(1+\frac{60}{7}\hmu^2\right)\right] .
\ee
\begin{figure}[tbh]
	\begin{center}
		\hspace{-2mm}\includegraphics[width=8.5cm,height=6cm]{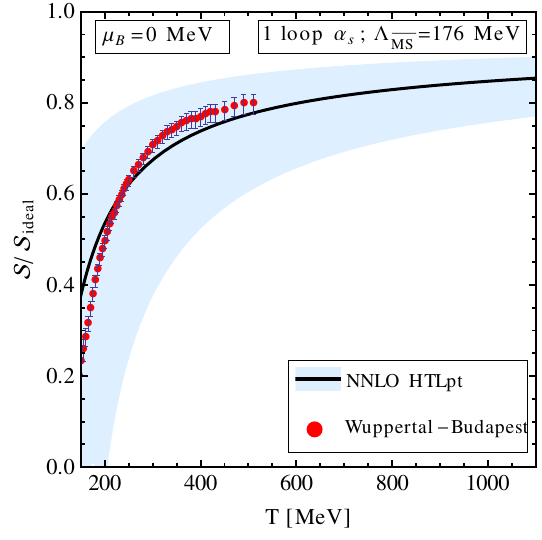}
		\includegraphics[width=8.5cm,height=6cm]{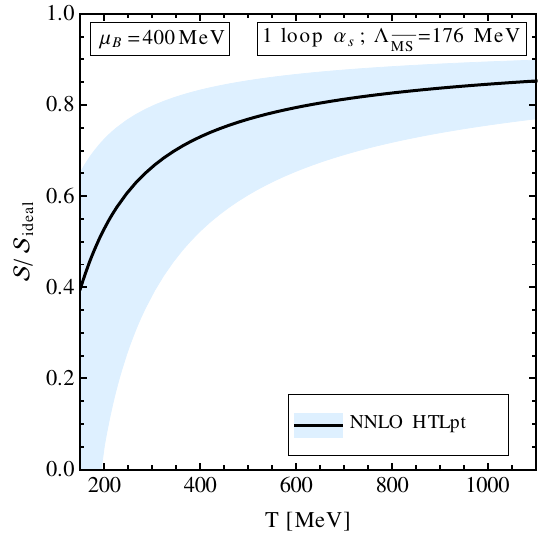}
	\caption[Comparison of the $N_f=2+1$, $\mu_B=0$ (left) and $\mu_B=400$ MeV (right) NNLO HTLpt entropy density with lattice data.]{
		Comparison of the $N_f=2+1$, $\mu_B=0$ (left) and $\mu_B=400$ MeV (right) NNLO HTLpt entropy density with lattice data.
		The $\mu_B=0$ lattice data are from \cite{Borsanyi:2010cj}.}
	\label{en_1l}
	\end{center}
\end{figure}
%
In Fig~(\ref{en_1l}) we plot the scaled NNLO HTLpt entropy density for $\mu_B=0$ (left) and $\mu_B=400$ MeV (right)
together with $\mu_B=0$ lattice data from Ref.~\cite{Borsanyi:2010cj}.  As we can see from this figure, 
there is quite good agreement between the NNLO HTLpt entropy density and the lattice data when the central value of the scale is used.

\noindent \textbf{Trace anomaly:}
Given that it is usually the trace anomaly itself that is calculated on the lattice and subsequently integrated to derive other thermodynamic properties, it is interesting to directly compare it with lattice data for the trace anomaly. The trace
anomaly ${\cal I}$ is defined as the energy density ${\cal E}$ minus three times the pressure ${\cal P}$. In the ideal gas limit, the trace anomaly approaches to 
zero as the energy density equals three times the pressure. Nonetheless, upon incorporating interactions, the trace anomaly, also known as the interaction measure, deviates from zero.
\begin{figure}[h!]
	\begin{center}
		\includegraphics[width=8.5cm,height=5.8cm]{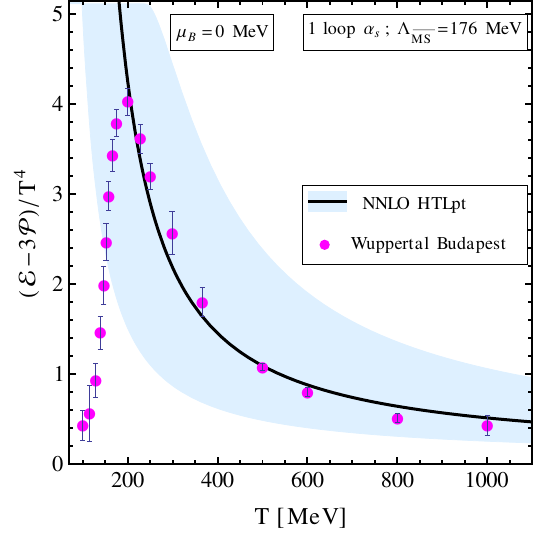}
		\includegraphics[width=8.5cm,height=5.8cm]{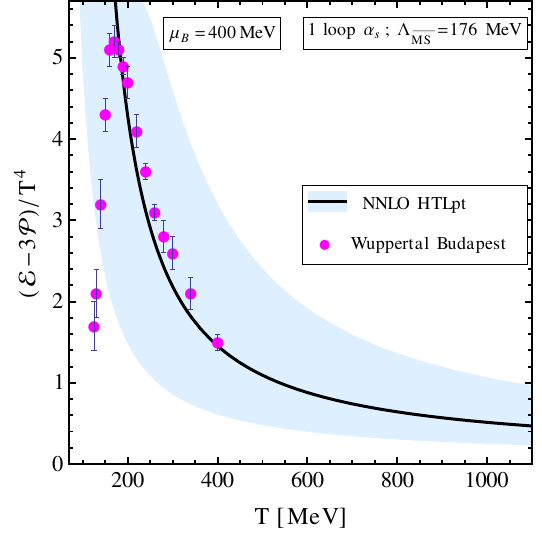}
		\vspace{-0.5cm}
	\caption{Comparison between the $N_f=2+1$, $\mu_B=0$ (left) and $\mu_B=400$ MeV (right) NNLO HTLpt 
		trace anomaly and corresponding lattice data. The lattice data for $\mu_B=0$ is from \cite{Borsanyi:2010cj}, while for $\mu_B=400$ MeV, it is from~\cite{Borsanyi:2012cr}. 
	}
	\label{ta_1l}
	\end{center}
\end{figure}

\vspace{-0.cm}
In Fig.~(\ref{ta_1l}) we plot the scaled NNLO HTLpt trace anomaly for $\mu_B=0$ (left) and $\mu_B=400$ MeV (right)
together with lattice data from Refs.~\cite{Borsanyi:2010cj} and \cite{Borsanyi:2012cr}. As we can see from this figure, there is quite good agreement between the NNLO HTLpt trace anomaly and the lattice data for $T \gtrsim 220$ MeV when the central value of the scale is used.

\noindent \textbf{Speed of sound:}
Another quantity which is phenomenologically interesting is the speed of sound.  The speed of sound is defined as
$
c_s^2=\frac{\del{\cal P}}{\del{\cal E}} \, .
$
%
\begin{figure}[h]
\begin{center}
		\hspace{-3mm}
		\includegraphics[width=8cm,height=6cm]{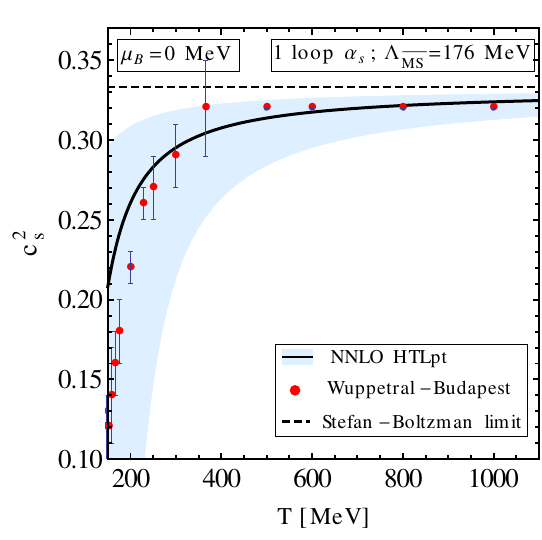}
	\hspace{-3mm}
		\includegraphics[width=8cm,height=6cm]{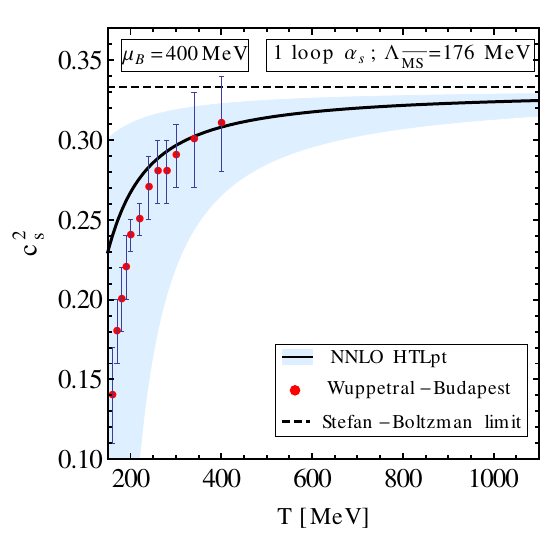}
		\vspace{-0.4cm}
	\caption{Comparison between $N_f=2+1$, $\mu_B=0$ (left) and $\mu_B=400$ MeV (right) NNLO HTLpt 	speed of sound squared with lattice data. The lattice data for $\mu_B=0$ are from~\cite{Borsanyi:2010cj}, while for $\mu_B=400$ MeV lattice data are taken from \cite{Borsanyi:2012cr}.}
	\label{cssq_1l}
	\end{center}
\end{figure}
\vspace{-0.cm}

In Fig.~(\ref{cssq_1l}) we plot the NNLO HTLpt speed of sound for $\mu_B=0$ (left) and $\mu_B=400$ MeV (right)
together with lattice data from Refs.~\cite{Borsanyi:2010cj} and \cite{Borsanyi:2012cr}.  As we can see from this figure, 
there is quite good agreement between the NNLO HTLpt speed of sound and the lattice data when the central value of the scale is used.
\vspace{-0.4cm}
\subsubsection{Quark number susceptibilities:}
\vspace{-0.2cm}
Having the complete thermodynamic potential as a function of temperature and chemical potential(s) allows us 
to calculate the quark number susceptibilities. Generally, one can introduce an idividual chemical potential 
for each quark flavour, forming $N_f$-dimensional vector $\bm{\mu}\equiv(\mu_1,\mu_2,...,\mu_{N_f})$.  
By taking derivatives of the pressure with respect to chemical potentials in this set, we obtain the quark 
number susceptibilities\,\footnote{We have specified that the derivatives should be evaluated at $\bm{\mu}=0$. 
	In general, one could define the susceptibilities at $\bm{\mu} = \bm{\mu}_0$.}
\be
\chi_{ijk\,\cdots}\left(T\right)&\equiv& \left. \frac{\partial^{i+j+k+ \, \cdots}\; {\cal P}\left(T,\bm{\mu}\right)}
{\partial\mu_u^i\, \partial\mu_d^j \, \partial\mu_s^k\, \cdots} \right|_{\bm{\mu}=0} \, .
\label{qnsdef}
\ee
Below we will use a shorthand notation for the susceptibilities by specifying derivatives by a string of quark flavours 
in superscript form, e.g. $\chi^{uu}_2 = \chi_{200}$, $\chi^{ds}_2 = \chi_{011}$, $\chi^{uudd}_4 = \chi_{220}$, etc.

When computing the derivatives with respect to the chemical potentials we treat $\Lambda$ as being a constant 
and only put the chemical potential dependence of the $\Lambda$ in after the derivatives are taken.  We have done
this in order to more closely match the procedure used to compute the susceptibilities using resummed dimensional
reduction \cite{Mogliacci:2013mca}.\footnote{One could instead put the chemical potential dependence of the $\Lambda$ in prior to taking the derivatives with respect to the chemical potentials.  If this is done, the central lines obtained
are very close to the ones obtained using the fixed-$\Lambda$ prescription, however, the scale variation typically increases in this case.}

\noindent \textbf{Baryon number susceptibilities:}
%
We begin by considering the baryon number susceptibilities.  The $n^{\rm th}$-order baryon number susceptibility is defined as
\be
\chi_B^n(T) \equiv \left.\frac{\partial^n {\cal P}}{\partial \mu_B^n}\right|_{\mu_B=0} \, .
\ee
For a three flavour system consisting of $(u,d,s)$, the baryon number susceptibilities can be related to the quark number 
susceptibilities~\cite{Petreczky:2012rq}
\be
\chi_2^B=\frac{1}{9}\[\chi_2^{uu}+\chi_2^{dd}+\chi_2^{ss}+2\chi_2^{ud}+2\chi_2^{ds}+2\chi_2^{us}\] \, ,
\label{gen_chi2}
\ee
and
\be
\chi_4^B &=& \frac{1}{81}\bigg[\chi_4^{uuuu}+\chi_4^{dddd}+\chi_4^{ssss}+4\chi_4^{uuud}+4\chi_4^{uuus}
+\ 4\chi_4^{dddu}+4\chi_4^{ddds}+4\chi_4^{sssu}+4\chi_4^{sssd}\nn
&&
\hspace{8mm}+6\chi_4^{uudd} +\ 6\chi_4^{ddss}+6\chi_4^{uuss}+12\chi_4^{uuds}+12\chi_4^{ddus}+12\chi_4^{ssud} \bigg] \, .
\hspace{5mm}
\label{gen_chi4}
\ee
%
\begin{figure}[h]
	\begin{center}
			\hspace{-10mm}
			\includegraphics[width=8.0cm,height=6cm]{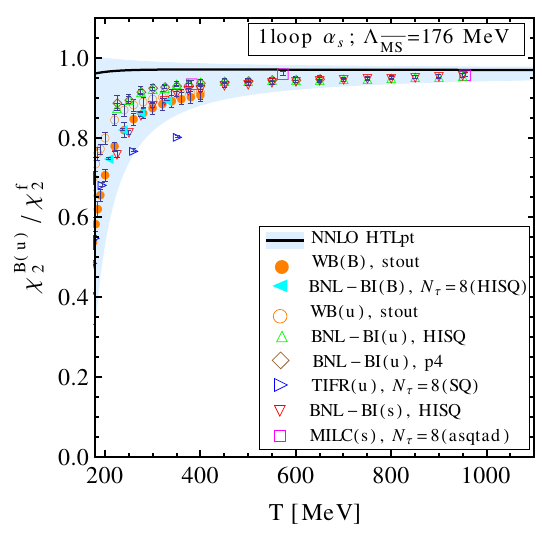}
			\includegraphics[width=8.0cm,height=6cm]{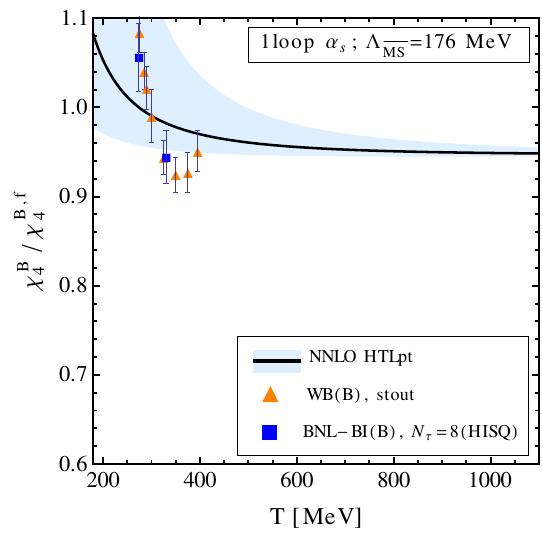}
			\vspace{-0.2cm}
		\caption{The scaled second order baryon number susceptibility compared with various lattice data (left) and the scaled fourth order baryon number susceptibility compared with various lattice data (right) using one loop running.. The lattice data labeled WB, BNL-BI(B), BNL-BI(u,s), MILC, and TIFR come from Refs.~\cite{Borsanyi:2011sw}, \cite{Bazavov:2013dta}, \cite{Bazavov:2013uja},~\cite{Bernard:2004je},  and~\cite{Datta:2014zqa}, respectively.}
		\label{qns24_1l}
	\end{center}
\end{figure}
\vspace{-0.2cm}
In Fig.~(\ref{qns24_1l}) we compare the NNLO HTLpt result for the second and forth order baryon number susceptibility with lattice data
from various groups. In the left panel of the  Fig.~(\ref{qns24_1l}), the HTLpt second-order baryon number suscepbility is compared with the lattice data for both the baryon number and quark number susceptibilites. Because, in HTLpt second-order baryon and quark number suscepbilities differ by a factor $1/3$ as shown is Eq~\eqref{chi2uu_B}. So, the scaled quark and baryon number susceptibitites are identical in HTLpt. As one can see, for this quantity, the size of the light-blue band becomes larger if one uses the three-loop running, however, the central value obtained is very close in both cases.  

When compared to the lattice data, we observe that the NNLO HTLpt prediction for the second-order baryon number susceptibity is approximately $10\%$ higher than the lattice data at $T=250$ MeV and approximately 2\% higher at $T = 800$ MeV. In this context, it is worth noting that recently the four-loop second-order baryon number susceptibility has been computed in Ref.~\cite{Mogliacci:2013mca} using the resummed dimensional reduction method. 
 Our result, taken together with the resummed dimensional reduction approach, suggest that the quark sector of the QGP can be accurately described using resummed perturbation theory for temperatures above approximately 300 MeV.
\begin{figure}[tbh!]
	\centerline{\includegraphics[width=8.5cm,height=6.5cm]{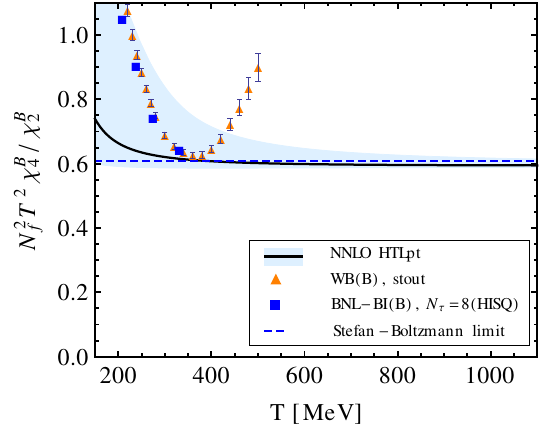}
	\includegraphics[width=8.5cm,height=6.5cm]{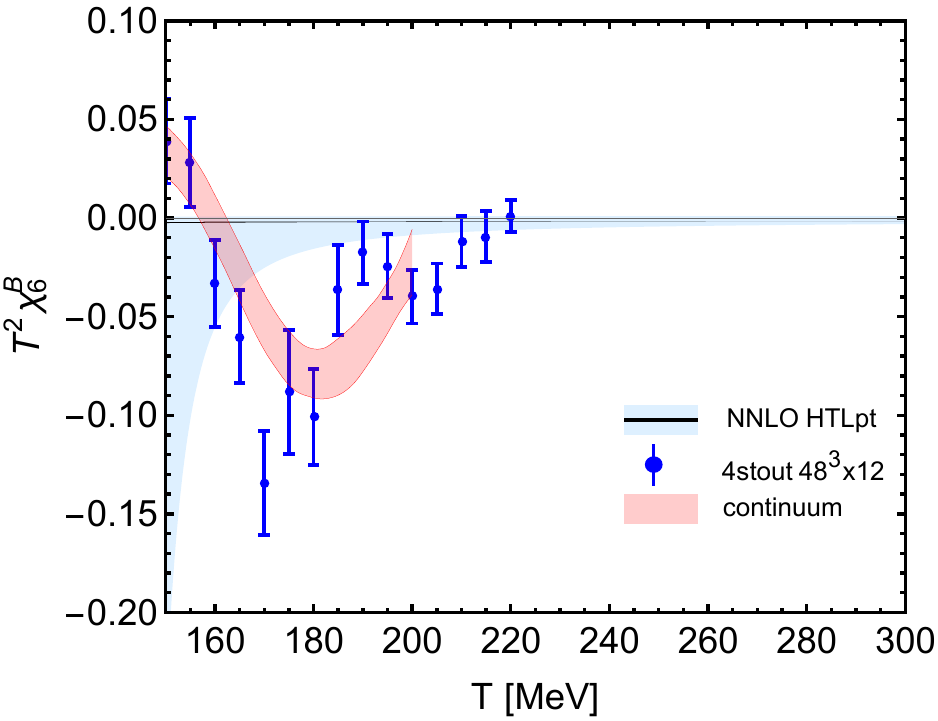}}
	\vspace{-0.4cm}
	\caption{\textit{Left panel:} 
		Comparison of the $N_f=2+1$ NNLO HTLpt ratio of the fourth to second order baryon susceptibility with lattice data. The data labelled WB and BNL-BI(B) come from Refs.~\cite{Borsanyi:2012rr,Borsanyi:2013hza} and \cite{Bazavov:2013dta}, respectively. \textit{Right Panel:}The $N_f=2+1$ NNLO HTLpt scaled sixth-order baryon susceptibility with  lattice data~\cite{Borsanyi:2018grb,Borsanyi:2024anr} as a function of temperature.
	}
	\label{qnsrat_qns6}
\end{figure}

In left panel of Fig.~(\ref{qnsrat_qns6}), we depict the scaled ratio of the fourth and second order baryon number susceptibilities as a function 
of temperature, alongside lattice data for this ratio. As evident from this figure, this ratio rapidly approaches the Stefan-Boltzmann limit if one considers the central NNLO HTLpt line. Upon comparison with the lattice data, we observe that  the NNLO HTLpt result is below the lattice data for temperatures less than approximately $300$ MeV. Without lattice data at higher temperatures, it's hard to draw a firm conclusion regarding the temperature at which HTLpt provides a good description of this quantity.
In right panel of Fig.~(\ref{qnsrat_qns6}) we show the comparison of NNLO HTLpt prediction for the sixth order baryon number susceptibility with the lattice data from Refs.~\cite{Borsanyi:2018grb,Borsanyi:2024anr}. In Ref.~\cite{Borsanyi:2018grb}, $4stout$ action is used with finite lattice $48^3\times12$; whereas continuum extrapolated data are taken from Ref.~\cite{Borsanyi:2024anr}. The HTLpt result for $\chi_6^B$ is approaching the lattice result at temperature greater than $210$ MeV. As the available lattice data for $\chi_6^B$ is available only upto $220$\ MeV, the high temperature lattice data is needed to draw a conclusion about the agreement between HTLpt and lattice data for sixth-order susceptibilities. 

\noindent \textbf{Single quark number susceptibilities:}
\begin{figure}[tbh]
\begin{center}
		\includegraphics[width=8.2cm,height=6cm]{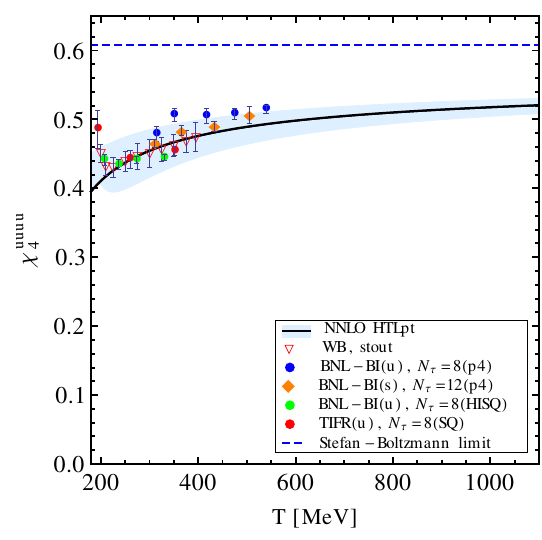}
		\includegraphics[width=8.2cm,height=6cm]{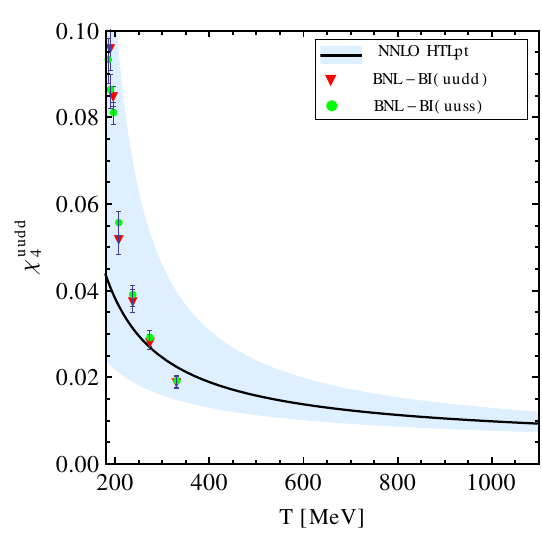}
		\vspace{-0.4cm}
	\caption{Comparing the $N_f=2+1$ NNLO HTLpt fourth order diagonal single quark number susceptibility 
		(left) and the only non-vanishing fourth order off-diagonal quark number susceptibility (right) with lattice data. In the left figure, the dashed blue line indicates the Stefan-Boltzmann limit $6/\pi^2$. The data labelled BNL-BI(uudd), BNL-BI(u,s), BNL-BI(uuss), and TIFR are sourced from Refs.~\cite{Bazavov:2013dta}, \cite{Bazavov:2013uja}, \cite{Bazavov:2012vg}, and \cite{Datta:2014zqa}, respectively.}
	\label{figsingleq}
	\end{center}
\end{figure}
We now turn our attention to the single quark number susceptibilities as defined in Eq.~\eqref{qnsdef}.  In this context, we use the general expression for the NNLO thermodynamic potential with different quark chemical potentials (Eq.~\eqref{finalomega}). The resultant susceptibilities can either be diagonal (same flavour on all derivatives) or off-diagonal (different flavour on some or all indices). In HTLpt, off-diagonal susceptibilities explicitly emerge from graphs ${\cal F}_{3c}^f$  and ${\cal F}_{3j}^f$; however, the latter vanishes when we use the variational mass prescription for the quark mass ($m_q=0$), hence we need to consider only the ${\cal F}_{3c}^f$ graph. Moreover, there are potential off-diagonal contributions coming from all HTL terms since the Debye mass receives contributions from all quark flavours.  In practice,
however, because we evaluate derivatives with respect to the various chemical potentials and then take $\mu_i \rightarrow 0$, one finds that all off-diagonal second order susceptibilities vanish in HTLpt. Consequently, for the three-flavour case, we find
$
\chi_2^{ud}=\chi_2^{ds}=\chi_2^{su}=0 \, ,
$
and, consequently, the single quark second order susceptibility is proportional to the baryon number susceptibility
\be
\chi_2^{uu}=\frac{1}{3}\chi_2^B.
\label{chi2uu_B}
\ee
For the fourth order susceptibility, there is only one non-zero off-diagonal susceptibility, namely, $\chi_4^{uudd}=\chi_4^{uuss}=\chi_4^{ddss}$, which is related to the diagonal susceptibility,
e.g. $\chi_4^{uuuu}=\chi_4^{dddd}=\chi_4^{ssss}$, as 
\be
\chi_4^{uuuu}=27\chi_4^B-6\chi_4^{uudd}.
\label{u4rel}
\ee
As a consequence, one can compute $\chi_4^{uuuu}$ directly from (\ref{finalomega}) or by computing $\chi_4^B$ using (\ref{finalomega1}) and $\chi_4^{uudd}$ using (\ref{finalomega}) and applying the above relation. In our final plots we compute $\chi_4^{uuuu}$ directly from (\ref{finalomega}), however, we have checked that we obtain the same result if we use (\ref{u4rel}) instead.

In Fig.~(\ref{figsingleq}) (left), our findings for the fourth-order single quark susceptibility $\chi_4^{uuuu}$ depicted alongside lattice data from Refs.~\cite{Bazavov:2013dta,Bazavov:2013uja,Bazavov:2012vg,Datta:2014zqa}. As illustrated in this figure, for the fourth order susceptibility, there is very good agreement with available lattice data. Furthermore, the HTLpt result exhibits small scale variation for this particular quantity. In Fig.~(\ref{figsingleq}) (right), our findings for the fourth-order off-diagonal single quark susceptibility $\chi_4^{uudd}$ are compared to lattice data. From this right panel we also see reasonably good agreement between the NNLO HTLpt result and the available lattice data.  

\begin{wrapfigure}[14]{r}{0.45\textwidth}
	\vspace{-5mm}
	\begin{center}
		{\includegraphics[width=7.5cm,height=6.5cm]{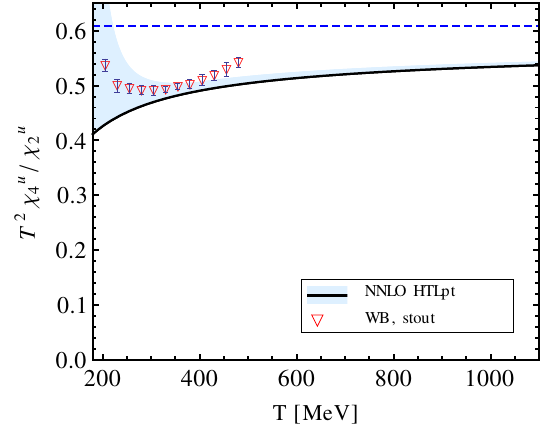}}
		\vspace{-0.4cm}
		\caption[Comparison of the $N_f=2+1$ NNLO HTLpt ratio of the fourth to second order single quark susceptibility with lattice data.]
		{Comparison of the $N_f=2+1$ NNLO HTLpt ratio of the fourth to second order single quark susceptibility with lattice data. The data labeled WB come from Refs.~\cite{Borsanyi:2012rr,
				Borsanyi:2013hza}.}
		\label{qnsrat2_1l}
	\end{center}
\end{wrapfigure}
Figure~(\ref{qnsrat2_1l}) showcases the scaled ratio of the fourth- and second-order single quark susceptibilities
Once more,  we see observed a good agreement between the NNLO HTLpt result and lattice data.  It is worth noting that for both Figures~(\ref{figsingleq}) and (\ref{qnsrat2_1l}), the lattice data are confined to relatively low temperatures. It will be intriguing to conduct comparisons with higher temperature lattice data as they become available.   

It is also possible to estimate the sixth-order quark number susceptibilities, namely, $\chi_{600}$ , $\chi_{420}$, and $\chi_{222}$ using NNLO HTLpt resummed thermodynamic potential  as shown in Eq.~\eqref{finalomega}. The estimation can be found in Ref.~\cite{Haque:2014rua}.

\noindent \textbf{The curvature of the QCD phase transition line:}
 At high temperatures and low net baryon density, numerical lattice QCD calculations find that nuclear matter becomes deconfined and chiral symmetry is restored. The resulting phase diagram of QCD encodes the temperature and chemical-potential dependence of these transitions, including the order of each phase transition. In Ref.~\cite{Haque:2020eyj}, the predictions for the second-, fourth and sixth-order curvature coefficients are obtained. 
 \begin{figure}[h!]
 	\begin{center}
 	\includegraphics[width=9cm, height=5cm]{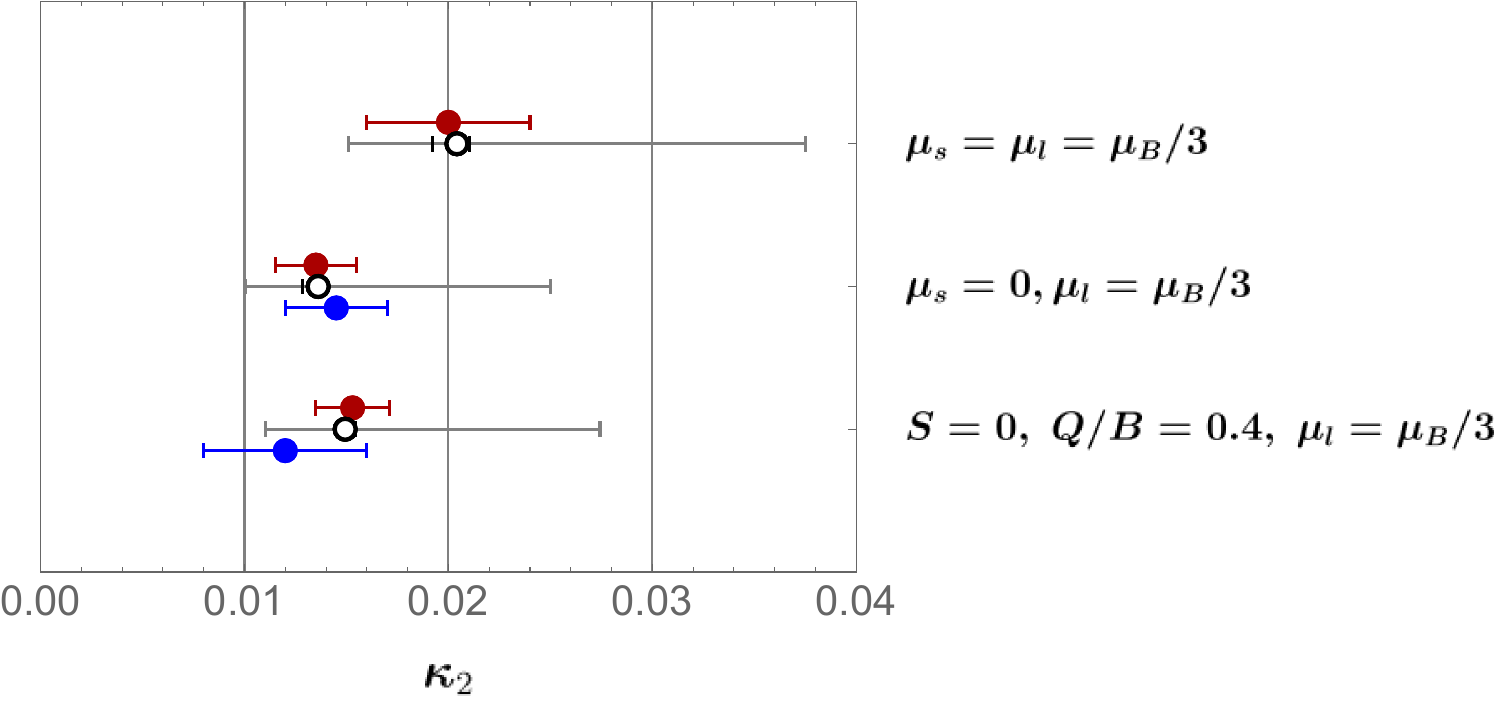}
 	 	\includegraphics[width=9cm, height=5cm]{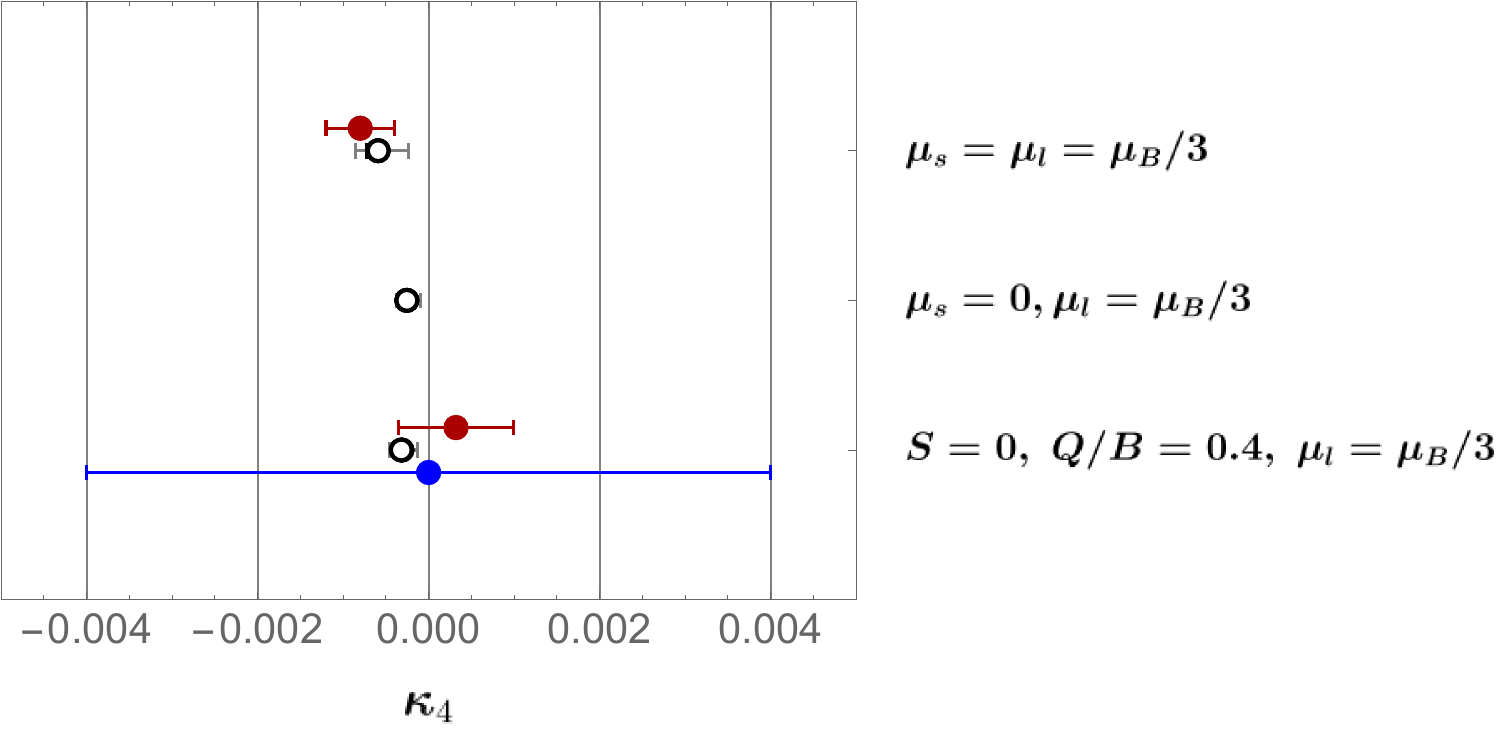}
 	 	\vspace{-0.4cm}
 	 		\caption{\textit{Left panel:} Solid circles represent lattice calculations of $\kappa_2$~\cite{Cea:2015cya,Bonati:2015bha,Bonati:2018nut,Borsanyi:2020fev,HotQCD:2018pds}, arranged from top to bottom, respectively.  Red solid circles denote findings obtained through the imaginary chemical potential method, while blue solid circles depict results obtained using Taylor expansions around $\mu_B=0$.  Open black circles signify NNLO HTLpt predictions. The black error bars associated with the HTLpt predictions result from variation of the assumed renormalization scale, whereas the gray error bars are the result of varying the thermodynamic variable used to determine the curvature. \textit{Right panel:} Solid circles are lattice calculations of $\kappa_4$ from Refs.~\cite{Borsanyi:2020fev,HotQCD:2018pds,Bonati:2014rfa}, arranged from top to bottom, respectively.  The colour scheme and error bar representation remain consistent with those in the left panel.}
 	 	\label{fig:kappa2}
 	 	\end{center}
 	\end{figure}
 	
The phase transition temperature in HTLpt is defined as the temperature at which the NNLO HTLpt resummed thermodynamic potential [Eq.~\eqref{finalomega}] goes to zero.  In Fig.~\ref{fig:kappa2}, we present three cases corresponding to (i) $\mu_s = \mu_l = \mu_B/3$, (ii) $\mu_s=0$, $\mu_l = \mu_B/3,$ and (iii) $S = 0$, $Q/B = 0.4$, $\mu_l = \mu_B/3$.  In each of the three cases, we find excellent agreement with continuum extrapolated lattice QCD results for $\kappa_2$, given current statistical uncertainties. Furthermore, we provide HTLpt predictions for $\kappa_4$ in all three cases, finding again excellent agreement with lattice extractions of this coefficient where available.

	\section{Damping Rate}\label{damp}
	\vspace{-0.2cm}
The damping rate $\gamma$ for a particle in a thermal medium has a simple physical significance. Because of continual
interactions with the medium, such a particle (or more accurately, a quasiparticle) lacks a distinct energy level and instead manifests as a resonance with width $\gamma$. Only when $\gamma$ is considerably smaller than its energy,
a quasiparticle be regarded as a true physical excitation, as only then can it propagate long enough to have
significant physical effects.

We know that the interaction rate of a particle at finite temperature indicates how often scattering of a particle with the particles at the heat bath take place. This rate is  associated with the phase space of each particles involved in scattering, a Bose-Einstein or Fermi-Dirac distribution for each particle in the initial state, and a Bose-enhancement or Pauli-suppression factor for each particle in the final state, the squared matrix element of the various physical processes in the system and corresponding degeneracy factor for a given physical process (see \eqref{ph4}). The damping rate is defined as the imaginary part of the dispersion relation of a particle which is obtained from the imaginary part of the  self-energy of that particle. Now, we note that the Cutkosky thermal cutting rules~\cite{Das:1997gg,Kobes:1985kc,Kobes:1986za,Gelis:1997zv} presents a systematic method to compute the imaginary part of a Feynman loop diagram. The Cutkosky rule expresses the imaginary part of the $l$-loop amplitude in terms of physical amplitude of lower order [($l-1$)-loop or lower]. This indicates that the damping rate and the interaction rate are closely related. This allows for a simple explanation for physical processes responsible for damping of a particle in the thermal medium. Moreover, the rates are the starting point for important physical quantities like energy loss and transport coefficients. The damping rate for light and heavy particles have been extensively studied in the literature~\cite{lopez,Kajantie:1982xx,Heinz:1986kh,Hansson:1987um,Kobes:1987bi,Pisarski:1988vd,Braaten:1989kk,Braaten:1990it,Thoma:1995ju,Lebedev:1989ev,Lebedev:1990un,Lebedev:1990kt,Thoma:1990fm,Braaten:1990ee,Burgess:1991wc,Baier:1991dy,Nakkagawa:1992ew,Rebhan:1992ca,Braaten:1992gd,Kobes:1992ys,Altherr:1992ti,Baier:1992bv,Pisarski:1993rf,Peigne:1993ky,Thoma:1993vs,Thoma:1994fd,Thoma:2000dc}. Below we define the damping rate for various kind of particles.

\subsection{General Setup}  
\label{gen_su}
\vspace{-0.2cm}
The damping or interaction rate describes the damping of a particle with the time evolution~\cite{Thoma:1995ju,Thoma:2000dc} in the form of plane wave $\exp(-i\om t )$, where $\om$ is the frequency of the particle under consideration. From the dispersion property of a particle  discussed earlier the frequency $\om$ has both real and imaginary part:
\be
\om = {\rm {Re}}\ \om +i{\rm {Im}}\ \om. \label{dm1}
\ee
Now, in this subsection we define the damping rate as  $\gamma=-{\rm {Im}}\ \om \label{dm2}$ and we write  
$
\exp(-i\om t )= \exp\left(-i {\rm {Re}}\ \om t\right) \times \exp\left(-\gamma t\right) \, . \label{dm3}
$
First we consider a scalar field and the dispersion relation can be obtained from the pole of the propagator in \eqref{resum_prop1} as
\bea
\om^2-p^2-\Pi(\om,p) =
\left ( {\rm {Re}}\ \om -i\gamma \right)^2 -p^2 - {\rm {Re}}\ \Pi \left( {\rm {Re}}\ \om -i\gamma ,k\right) -i{\rm {Im}}\ \Pi \left( {\rm {Re}}\ \om -i\gamma ,p\right)=0 . \label{dm4}
\eea
In the case of overdamping $\gamma \ll {\rm {Re}}\ \om$, one can write the real part of \eqref{dm4} as
\be
\left ( {\rm {Re}}\ \om \right)^2 -p^2 - {\rm {Re}}\ \Pi \left( {\rm {Re}}\ \om,p \right)=0 \, , \label{dm5}
\ee
which leads to the dispersion relation $\om(p) =  {\rm {Re}}\ \om$. On the other hand the imaginary part  can be written as
\be
-2i {\rm {Re}}\ \om \gamma -i{\rm {Im}}\ \Pi \left( {\rm {Re}}\ \om,p\right) =0 \,\,\,\,\,\,\, 
\Rightarrow  \,\,\,\,\,\,\,  \gamma= - \frac{1}{2\om(p)} {\rm {Im}}\ \Pi \left( \om(p),p\right)\, ,  \label{dm6}
\ee which is the the damping rate for a scalar particle.

For  gauge boson the propagator in covariant gauge is obtained in \eqref{gsp14} and it has two degenerate transverse modes and one long wavelength plasmon mode. The dispersion relations for  longitudinal plasmon mode and transverse mode   are given as
\begin{subequations}
\begin{align}
P^2+\Pi_L(\om,p)= \om^2-p^2+\Pi_L(\om,p)=0  \, , \label{dm11} \\
P^2+\Pi_T(\om,p)= \om^2-p^2+\Pi_T(\om,p)=0  \, ,\label{dm12} 
\end{align}
\end{subequations}
where $\Pi_L$ and $\Pi_T$ are longitudinal and transverse self-energy of gauge boson and defined in \eqref{pi_L} and \eqref{pi_T}. Following the same procedure as the scalar case one can obtain the damping rate of longitudinal and transverse gauge boson damping rates, respectively, as
\be
\gamma_L&=\frac{1}{2\om(p)}{\rm{Im}}\Pi_L(\om,p) \, , 
\qquad \mbox{and} \quad
\gamma_T&=\frac{1}{2\om(p)}{\rm{Im}} \Pi_T(\om,p)  \, . \label{dm13_14}
\ee
For massless fermion the dispersion can be obtained from \eqref{gse19} as
\bea
{\cal D}_\pm (p,\omega)= (1+{\cal A}(\om,p))(\omega\mp p)+{\cal B}(\om,p)=0   \,\,\,\,\,\,\, 
\Rightarrow \, \,\,\,\,\,\,\,  \om = \pm p - \frac{{\cal B}(\om,p)} {1+{\cal A}(\om,p)} \approx  \pm p - {\cal B}(\om,p) \, ,\label{dm7}
\eea
where $\cal A$ in the denominator contributes to higher order only. We note that ${\cal D}_+=0$ gives dispersion a normal quark where as ${\cal D}_-=0$ give the long wavelength dispersion of a plasmino mode. The quantities $\cal A$ and ${\cal B}$ are, respectively given in~\eqref{gse10}.
Now for no overdamping $\gamma \ll  {\rm {Re}}\ \om$, one can write the damping rate for massless quark at ${\rm {Re}}\ \om = p$  as
\be
\gamma_\pm = - {\rm {Im}}\ \om_\pm = {\rm {Im}} \ {\cal B}(p_0=\om=p,p) = -\left.\frac{1}{4p_0} {\rm {Im}} \  {\Tr}\left (P\!\!\! \! \slash \ \Sigma \right ) \right |_{\om=p} \, , \label{dm8}
\ee
where it turns out that both normal quark and plasmino have the same damping rate and one has to calculate the imaginary part of the quark self-energy on the mass shell at finite temperature.

For heavy fermion of mass $M$ and four momentum $P$, the dispersion relation can be written as
\be
P\!\!\!\! \slash +M-\Sigma(E,p)=0 \, , \label{dm9}
\ee
where $E=\sqrt{P^2+M^2}$  is energy and $\Sigma(E,p)$ is self-energy of heavy fermion. Squaring  \eqref{dm9} and then taking trace one can write
\bea
E = \frac{1}{2E}\  {\Tr}\ \left [\left (\slashed{P} +M\right )\ \Sigma(E,p) \right]   \,\,\,\,\,\,\, \,
\Rightarrow \,\,\,\,\,\,\, \, \gamma(E) = - {\rm {Im}}\ E = -\frac{1}{2E} \ {\rm {Im}} \ {\Tr}\ \left [\left (\slashed{P} +M\right )\ \Sigma(E,p) \right ]\, , \label{dm10}
\eea 
where we have neglected $\Sigma^2$ term which contributes to higher order. In the following sections, we will discuss the damping rate of various particles in QED and QCD plasma.

\subsection{Hard Damping Rate}
\label{hard_damp}
In this subsec we will discuss the damping rate for hard particles (both fermion and boson) with $M, E, p\ge T$ in BPT and HTLpt.
 \subsubsection{Hard heavy fermion damping rate}
 \label{sub:heavy}
\begin{figure}[htb]
\begin{center}
\includegraphics[height=2.5cm, width=10cm]{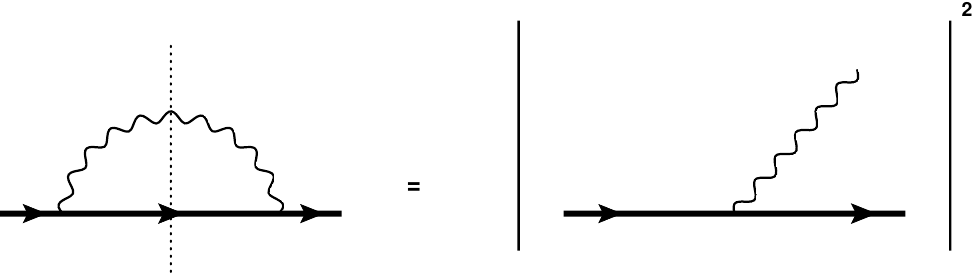}
\end{center}
\vspace{-0.7cm}
\caption{One-loop muon self-energy diagram in BPT. The bold line represent the massive lepton.}
\label{muon_1l_bpt}
\end{figure}

\begin{figure}[h]
\begin{center}
		\includegraphics[height=2.2cm, width=16cm]{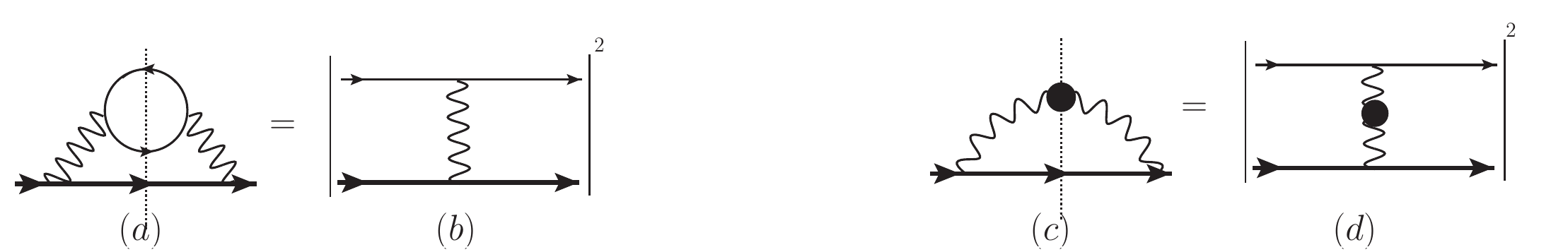}
\end{center}
\vspace{-0.5cm}
\caption{\textit{Left panel}: Two-loop muon self-energy diagram in BPT.  \text{Right panel}: Two-loop muon self-energy diagram in HTLpt. The blob corresponds to effective photon propagator. The bold line in both the panels represent the massive lepton.}
\label{muon_bpt+htl}
\end{figure}

 In this subsection we will first calculate the damping rate for heavy lepton (muon) in QED plasma using \eqref{dm10}, which requires muon self-energy. In BPT the muon  self-energy in lowest order is given in the left side of Fig.~\ref{muon_1l_bpt}. We note that this 1-loop self-energy does not have any imaginary part on-shell~\cite{Thoma:1995ju,Thoma:2000dc}. 
 The cutting rules~\cite{Das:1997gg,Kobes:1985kc,Kobes:1986za,Weldon:1983jn,Gelis:1997zv} indicates the equality of the imaginary part of self-energy to the matrix element given in right side of Fig.~\ref{muon_1l_bpt} that describe the emission or absorption of photon from a bare muon. However,  such process is forbidden due to energy-momentum conservation~\cite{Thoma:1995ju,Thoma:2000dc}. 
\vspace{-0.3cm}
 Next we consider the two-loop muon self-energy in BPT as given in the Fig.~\ref{muon_bpt+htl}a, where only bare propagator and vertices are used. The  Fig.~\ref{muon_bpt+htl}b corresponds to the scattering process $\mu e^\pm \rightarrow \mu e^\pm$ appear due to the cutting of the two-loop diagram. This processes corresponds to energy loss of muon in QED plasma containing  $e^+e^-$. We consider that the mass and momentum of heavy fermion is much larger than the mass of the electron ($m_e$):  $M,\, p\gg T\gg m_e$, which corresponds to hard scale. This can be realised for muons at a temperature 10 MeV, which might be the case in supernova explosions.
 \vspace{-0.cm}

Using naive power counting the diagrams in the Fig.~\ref{muon_bpt+htl}a and Fig.~\ref{muon_bpt+htl}b, one can find that the damping rate $\gamma_{\rm{hq}}(E)$  is order of $e^4 (\sim \alpha^2)$ and quadratically divergent due to the exchange massless photon~\cite{Thoma:1995ju,Thoma:2000dc}. This is the problem in BPT as discussed earlier.

One can improve the BPT result by using HTLpt. Here one could replace the photon propagator in one-loop self-
energy in the  Fig.~\ref{muon_1l_bpt} by an effective photon propagator as displayed in Fig.~\ref{muon_bpt+htl}c. 
We note that one does not need the effective muon propagator and vertices as the muon is hard ($M\gg T$).  The 
effective photon propagator in \eqref{gsp14} contains an infinite number of electron loops at hard momenta 
($ \gtrsim T$) and the damping mechanism is caused by the elastic scatterings $\mu e^\pm \rightarrow \mu e^\pm$ 
in QED plasma through exchange of virtual plasma modes as shown in Fig.~\ref{muon_bpt+htl}d which appears 
due to the cutting of the self-energy. The imaginary part of the muon self-energy comes from the imaginary or 
discontinuous part of the photon propagator which arises due to space like momentum $q^2>q_0^2$. Therefore, 
the damping mechanism can be thought as a virtual Landau damping of the collective photons.

The muon self-energy in HTLpt can be written from the Fig.~\ref{muon_bpt+htl}c as
\bea
\Sigma^(P) = - T \sumintb_Q \,\, (-ie\gamma^\mu) (iS_0(P') )  (-ie\gamma^\nu) (i D_{\mn}(Q) )
=-e^2  T \sumintb_Q \gamma^\mu S_0(P')  \gamma^\nu  D_{\mn}(Q)  ,  \label{dm15}
\eea
where $D^{\mn}$ is the effective photon propagator given in  \eqref{eff_photon}, $S_0(P'=P-Q)$ is the bare muon propagator given in \eqref{sac4} to \eqref{sac4a} in Saclay representation and $\gamma^\mu$ is bare three point function in QED. In Feynman gauge, using \eqref{eff_photon}, \eqref{A_exp}, \eqref{B_exp} for photon propagator,  and \eqref{sac4} for muon propagator, one can perform the trace
\bea
{\Tr} \left [\left (P\!\!\!\! \slash +M\right )\ \Sigma(P) \right ] &=& e^2T   \sumintb_Q \ \Delta_F(P') \ {\Tr}\left [\left (P\!\!\!\! \slash +M\right )\ \gamma^\mu \left (P'\!\!\!\!\! \! \slash +M\right ) \gamma^\nu \right ] \left [ A_{\mn}D_T(Q) + B_{\mn} D_L(Q) \right ] \nn
&=&4 e^2T  \sumintb_Q  \ \Delta_F(P-Q) \Big [- 2D_T(\om,q) \Big\{ p_0^2-p_0q_0+ \bm{\vec p \cdot \vec q} -\left(\bm{\vec p \cdot \hat q} \right)^2 -M^2 \Big \}\nn
&& + D_L(\om,q)\Big \{p_0^2+p^2 +p_0q_0-\bm{\vec p \cdot \vec q}+M^2 - \frac{2}{Q^2}\Big(p_0q_0- (\bm{\vec p \cdot \hat q})q\Big)^2-2(\bm{\vec p \cdot \hat q})^2  \Big\} \Big ] \, .\label{dm16}
\eea
We can perform the frequency sum using Saclay method. The heavy quark propagator is given in \eqref{sac3a} and~\eqref{sac4a}. For photon propagators one can follow the spectral representation using \eqref{bpy1}
as
\be
D_{L,T}(q_0,q) =\int_{-\infty}^{\infty} d\omega \  \frac{\rho_{L,T}(\omega,q)}{\omega-q_0-i\epsilon} \, , \label{dm17}
\ee
where $q_0=2n\pi iT$, bosonic frequency, $ \rho_{L,T}(\omega,q)$ are the spectral functions of the photon propagator and defined in \eqref{bpy_04}. They are obtained in Eq.~\eqref{srgbp5}. 
Now following  \eqref{bpy3} one can write
\be
\frac{1}{\omega-q_0-i\epsilon_1} =\big(1+n_B(\omega)\big )  \int_{0}^{\beta} d\tau \  e^{-(\omega-q_0)\tau}  \, . \label{dm18}
\ee
Combining \eqref{dm17} and \eqref{dm18}, one can write
\be
D_{L,T}(q_0,q) =  \int_{0}^{\beta} d\tau \  e^{q_0\tau}\, \int_{-\infty}^{\infty} d\omega \ \rho_{L,T}(\omega,q)  \big(1+n_B(\omega)\big )  e^{-\om\tau}\, . \label{dm19}
\ee
Using \eqref{sac3a}, \eqref{sac4a} and \eqref{dm19} in \eqref{dm16},  and performing the frequency sum $q_0$ using Saclay method that yields $\delta(\tau'-\tau)$ by which $\tau'$ integration can be done very easily. Then performing the $\tau$ integration, one gets
\bea
{\Tr} \left [\left (P\!\!\!\! \slash +M\right )\ \Sigma(P) \right ] &=& 4e^2 \int \frac{d^3q}{(2\pi )^3} \int_{-\infty}^{\infty} d\omega \,  \frac{1+n_B(\omega)}{2E'}
\left[\big(1-n_F(E')\big) \frac{e^{-\beta(E'+\om)}+1}{p_0-E'-\om} - n_F(E') \frac{e^{\beta(E'-\om)}+1}{p_0+E'-\om}\right] \nn
&&\times  \Big[\rho_L(\om,q)\Big \{p_0^2+E^2 +p_0q_0-\bm{\vec p \cdot \vec q} - \frac{2}{Q^2}\Big(p_0q_0- (\bm{\vec p \cdot \hat q})q\Big)^2-2(\bm{\vec p \cdot \hat q})^2  \Big\}  \nn
 && - 2 \rho_T(\om,q) \Big\{p_0^2-p_0q_0+ \bm{\vec p \cdot \vec q} -\left(\bm{\vec p \cdot \hat q} \right)^2 -M^2 \Big \}  \Big] \, , \label{dm20}
\eea
where we have used $e^{\beta p_0}=-1$ as $p_0=(2n+1)i\pi T$, the fermionic frequency. We also know that 
\be
\lim_{\epsilon \rightarrow 0} {\rm{Im}} \frac{1}{x+i\epsilon}=-\pi\delta(x) \, . \label{dm21}
\ee  
Using  Eqs.~\eqref{srgbp5} and~\eqref{dm21}, one can extract the imaginary part of \eqref{dm20}  by analytically continue $p_0$ in the denominator to the real value 
$p_0=\left.(E+ i \epsilon)\right|_{\epsilon\rightarrow 0}$ as
\bea
 {\Tr} \left [\left (P\!\!\!\! \slash +M\right )\ {\rm{Im}} \Sigma(P) \right ] \!\!\!&=&\!\!\! -4e^2 \pi \big(1+e^{-\beta E} \big)\int \frac{d^3q}{(2\pi )^3} \int_{-\infty}^{\infty} d\omega \, 
\Big[\big(1-n_F(E')\big) \delta(E-E'-\om) - n_F(E') \delta(E+E'-\om)\Big] \nn
&&\times  \frac{1+n_B(\omega)}{2E'} \Big[\rho^{\rm{cut}}_L(\om,q)\Big \{2E^2 +E\om-\bm{\vec p \cdot \vec q} - \frac{2}{\om^2-q^2}\Big(E\om- (\bm{\vec p \cdot \hat q})q\Big)^2-2(\bm{\vec p \cdot \hat q})^2 
 \Big\}  \nn
 && \hspace{1cm}-\, 2 \rho^{\rm{cut}}_T(\om,q) \Big\{p^2-E\om+ \bm{\vec p \cdot \vec q} -\left(\bm{\vec p \cdot \hat q} \right)^2 \Big \}  \Big] \, . \label{dm22}
\eea
We note that this is an exact expression and due to the complicated expression of the cut part of spectral densities one need to compute the integral numerically. However, for hard massive muon $p,\, M, \, E, \, E' \gg T$ and HTL approximation $\om<q<T$ one can make following simplifications:
\begin{enumerate}
\item  $e^{-E/T}$, $e^{-E'/T}$ and $n_F(E)$ will be zero.
\item $E'=\sqrt{p'^+M^2} \approx \sqrt{E^2-2\bm{\vec p \cdot \vec q}} \approx E-\bm{\vec v \cdot \vec q}$, obtained by expanding in the limit $q\ll E$, and $\bm{\vec v}= {\bm {\vec p}}/{E}$. 
\item $\delta\big (E-E'-\om\big) =\delta\big (\bm{\vec v \cdot \vec q}-\om\big)$ and $\delta\big (E-\om + E'\big)$  will never be satisfied as $E,\, E'\gg \om$.
\item ${1}/{E'} \approx {1}/{E}$. 
\item Using \eqref{befd_exp} one can write 
\be
1+n_B(\om)\approx \frac{T}{\om}+\frac{1}{2} \, . \label{dm22a}
\ee
\item The discontinuous part of photon spectral densities can be obtained with $\om <q$ from  \eqref{srgbp8} and \eqref{srgbp9}, respectively, as
\be
\rho^{\rm{cut}}_L(\om,q)\approx \frac{\left(m_D^\gamma \right)^2\om} {2q \left[q^2+\left(m_D^\gamma \right)^2 \right]^2} \, ,\quad \mbox{and}\qquad
\rho^{\rm{cut}}_T(\om,q)\approx - \frac{\left(m_D^\gamma \right)^2\om q} {4\left[q^6+ \frac{1}{16}\pi^2\om^2 \left(m_D^\gamma \right)^4 \right]} \, , \label{dm23_24}
\ee
\end{enumerate}
where $\left(m_D^\gamma \right)^2=3\left(m_{\rm{th}}^\gamma \right)^2 = e^2T^2/3$ and $(m_{\rm{th}}^\gamma)^2=e^2T^2/9$, is the thermal photon mass.

Now using points 1 to 4, then performing  the angle integration $\angle (\bm{\vec p ,\, \vec q})$ using $\delta$-function in \eqref{dm22}, one can combine the result with \eqref{dm10} to write the heavy muon damping rate as
\bea
\gamma_{\rm{hq}}(E)= &\approx&  \frac{e^2}{2\pi v} \int_{0}^{\infty} q\, dq  \int_{-vq}^{vq} d\omega \left (1+n_B(\om)\right )\,
\left[\rho^{\rm{cut}}_L(\om,q) -  \left(v^2-\frac{\om^2}{q^2}\right)\rho^{\rm{cut}}_T(\om,q)  \right] \, .  \label{dm25} 
\eea
From the integration region of $q$ and $\om$ we first estimate the contribution to damping rate when both $q$ and $\om $ hard ($\sim T$). For hard $q$ and $\om$, the spectral functions in~\eqref{dm23_24} has multiplicative factor $(m_{\rm {th}}^\gamma)^2$ and the only scale $T$. Since the damping rate has dimension of energy, the hard contribution to damping rate must be order of $e^2(m_{\rm {th}}^\gamma)^2/T\sim e^4T$. This estimate is identical to that which
would be obtained by considering the contribution to the damping rate from the tree-level scattering process in Fig.~\ref{muon_bpt+htl}b.

We next consider the contribution to $\gamma_{\rm {hq}}$  from the integration region where $q$ and $\om$ are soft($\sim eT$). Since $\om$ is of order $eT$, one can keep only the first term in  Bose enhancement factor given in \eqref{dm22a}. Using this, one can write soft contribution to the the damping rate from~\eqref{dm22} as
\bea
\gamma_{\rm{hq}}(E) \approx  \frac{e^2T}{2\pi v} \int_{0}^{\infty} q\, dq  \int_{-vq}^{vq} \frac{d\omega}{\om}  \,
\left[\rho^{\rm{cut}}_L(\om,q) -  \left(v^2-\frac{\om^2}{q^2}\right)\rho^{\rm{cut}}_T(\om,q)  \right] 
 = \gamma_L+\gamma_T \,  .  \label{dm25a}  
\eea
where
\begin{subequations}
\begin{align}
\gamma_L &=  \frac{e^2T}{2\pi} \int_{0}^{\infty}\frac{ \left(m_D^\gamma \right)^2 \ q \ dq }{\left[q^2+\left(m_D^\gamma \right)^2 \right]^2}=\frac{e^2T}{4\pi} \, , \label{dm26}\\ 
\gamma_T &= \frac{e^2T}{8\pi v} \int_{0}^{\infty} q^2 \, dq \  \int_{-vq}^{vq} d\om \  \left(v^2-\frac{\om^2}{q^2} \right)
\frac{\left(m_D^\gamma \right)^2} {q^6+ \frac{1}{16}\pi^2 \om^2 \left(m_D^\gamma \right)^4} \nn
&{=\atop v\ll 1} \frac{e^2 T}{\pi^3}\int dq \left[-\frac{4 q}{ \left(m_D^\gamma \right)^2}+\left(\frac{16 q^3}{ \left(m_D^\gamma \right)^2 \pi v}+\frac{\pi v}{ q}\right) \tan ^{-1}
\left(\frac{  \left(m_D^\gamma \right)^2 \pi v}{4 q^2}\right)\right]
\approx \frac{e^2 T v}{2 \pi  } \int\frac{dq}{q}.\,  \label{dm27}
\end{align}
\end{subequations}
After $\om$-integration, the final form of the~\eqref{dm27} is obtained by expanding  the result in soft momentum limit $q \ll T$.  The longitudinal part of muon damping rate in \eqref{dm26}, due to the exchange of longitudinal photons, is finite because the infrared singularity is screened by the electric Debye mass $m_D^\gamma $. On the other hand the transverse part in \eqref{dm27}, due to the exchange of transverse photons, is logarithmically infrared divergent. The 
reason for this  divergence in transverse photon part is due to the absence of the magnetic screening mass ($m_{\rm{mag}} \sim e^2T$) in QED as shown in~\eqref{psl17} following a discussion. Since the static magnetic fields are not screened, it is expected that non-perturbative effects screen static magnetic fields over distances of order $1/m_{\rm{mag}}=1/e^2T$ 
\cite{Linde:1978px,Linde:1980ts,Pisarski:1988vd,Thoma:2000dc}. In this case one can use an infrared ultra-soft cut-off for $q$-integration as $q\sim m_{\rm{mag}}\sim e^2T$. On the other hand, $q$ is restricted to 
soft momenta, then one can use  an upper limit or ultra-violet cut-off for $q$-integration as  $q\sim eT$.Then we obtain within the leading log approximation the final result as
\be
\gamma_T(v\ll 1)\approx \frac{e^2T}{2\pi}\, v\, \ln \frac{1}{e} \, . \label{dm28}
\ee
We note that  the results given in \eqref{dm26} and \eqref{dm28} are obtained in Feynman gauge which agree with the results 
in Coulomb gauge~\cite{Thoma:1995ju,Thoma:2000dc} and thus gauge independent. For real time formalism of muon damping rate one can see Ref.~\cite{Thoma:2000dc}. 

We now make following comments on the damping rate of hard heavy fermion:
\begin{enumerate}
\item In BPT the damping rate for hard heavy muon was proportional to $e^4$ but in HTL approximation it becomes $e^2$. This reduction is due to strong infrared singularity in the
damping rate.
\item In BPT the damping rate for heavy muon was quadratically infrared divergent whereas in HTL approximation it becomes logarithmically infrared divergent. The reduction from quadratic to logarithmic singularity is, however, an improvement over BPT, which allows to calculate the damping rate with logarithmic accuracy.
\item One can obtain damping rate for heavy quarks in QCD by replacing the QED coupling $e^2$ by $C_Fg^2$, where $C_F=4/3$, the Casimir invariant in fundamental representation in QCD. Also, one needs an infrared ultra-soft cut-off  for $q$-integration as $q\sim m_{\rm{mag}}\sim g^2T$ in QCD.
\item In case of heavy quark at rest, $v=0$, the damping rate has been derived by Pisarski in Ref.~\cite{Pisarski:1988vd}.
\item The \eqref{dm25} is valid for a hard light fermion (electron and quark) with momentum  $p\geq T$ and setting $v=1$. The terms  $e^{-E/T}$, $e^{-E'/T}$ and $n_F(E)$ will be zero as long as the thermal energy of a light fermion, $E=p\geq 3T$. Then one has to perform the $\om$-integration numerically. This has been obtained in Refs.~\cite{Thoma:1990fm} as $\gamma = \frac{C_Fg^2}{2\pi} \ln (1/g)$ to logarithmic accuracy. This result (up to a factor of $3$) was found earlier in Refs.~\cite{Lebedev:1989ev,Lebedev:1990un} and later verified in Refs.~\cite{Burgess:1991wc,Rebhan:1992ca}.
\item The damping rate for hard gluons follows from light quark just replacing $C_F$ by the Casimir invariant of adjoint representation $C_A$.  It can also be obtained by explicitly calculating the  gluon self-energy with one effective gluon propagator~\cite{Braaten:1990ee,Burgess:1991wc}.
\end{enumerate}

 \subsubsection{Hard photon damping rate}
 \label{sub:photon} 
 The gauge boson damping rate are defined in \eqref{dm13_14}. We note that for real photon one gets from~\eqref{pi_L}  $\Pi_L(p_0=p,p)=0$ and from \eqref{pi_T} $\Pi_T(p_0=p,p)= \Pi_\mu^\mu/2$. Using those one can write photon damping rate in~\eqref{dm13_14} as
 \be
 \gamma_{\rm{ph}}=\gamma_T=\frac{1}{4E}{\rm{Im}} \Pi_\mu^\mu(E=p,p)  \, . \label{dm30}
 \ee
 We know that the hard photon damping rate in lowest order BPT~\cite{Thoma:1995ju,Thoma:1994fd} is logarithmically infrared divergent for vanishing quark mass. We have also seen in subsec~\ref{photon_chap}  that the leading order hard photon production rate  has also logarithmically infrared singularity. However, a finite result in leading order for hard photon damping rate can be calculated using HTLpt akin to the hard photon production rate because it  is related to the photon damping rate. Using \eqref{ph3} one can relate the photon damping rate with the real photon differential production rate as
 \be
 \gamma_{\rm{ph}}(E=p) = \frac{(2\pi)^3}{8 \, n_B(E)} \, \frac{dR}{d^4x d^3 \bm{\vec p}} \, . \label{dm31}
 \ee
The photon production rate  has been discussed in details for both one-loop ($E\gg T$) and two-loop ($E\gg gT$) in subsec~\ref{photon_chap} using kinetic theory and HTLpt. One can use those photon production rates to obtain the photon damping rate for one-loop and two-loop. However, a direct calculation of 1-loop hard photon damping rate can be found in Ref.~\cite{Thoma:1994fd}.

\subsection{Soft  Damping Rate}
\label{soft_damp}
In this subsec we intend to discuss the damping rate for fermion and boson with soft momenta, $p~\sim gT$. As we have discussed in sections~\ref{qed_htl} and~\ref{qcd_htl} that these particles are collective modes known as long wavelength plasmon and plasmino mode.

\subsubsection{Massless quark damping rate}
\label{sub:light}
\vspace{-0.2cm}
 \begin{figure}[h]
\begin{center}
\includegraphics[width=9cm, height=2.cm]{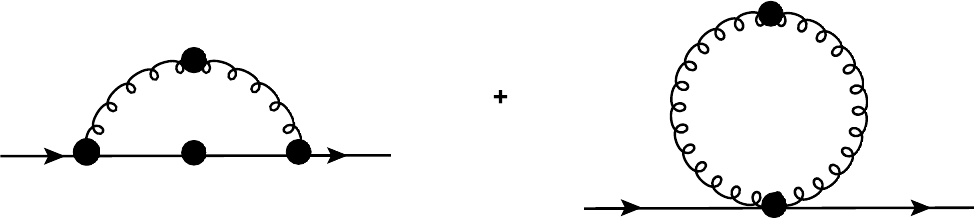}
\end{center}
\vspace{-0.7cm}
\caption{One-loop quark self-energy diagram in HTLpt, where the solid blobs represent the HTL  propagators and vertices.}
\label{dm_s_q}
\end{figure}
In Ref~\cite{Baier:1991dy} it was argued that because of  mass shell singularities there are some subtleties in the proof of gauge invariance in HTL approximation~\cite{Pisarski:1988vd,Braaten:1989mz} in covariant gauge. This put a question mark in calculating damping rate of massless quark at rest ($p=0$)  by using the  HTL approximation~\cite{Pisarski:1988vd,Braaten:1989mz} in covariant gauge as it depends on the gauge parameter. However, this issue  was resolved in Refs.~\cite{Nakkagawa:1992ew,Rebhan:1992ca}  that showed a careful treatment of the mass shell singularity requires an infrared cutoff.   It was argued  that the gauge dependent contribution to the damping rate vanishes when calculated in the presence of the cutoff,  and agrees with the formal proof of gauge invariance in Ref.~\cite{Pisarski:1988vd}.

The damping rate for soft massless quarks ($p\sim gT$) on the mass shell ($P^2=(m_{\rm{th}}^{\rm q})^2=g^2T^2/6$) can be obtained~\cite{Braaten:1992gd} at rest from \eqref{dm8} as
\be
\gamma_{\rm q}(0)=\gamma_\pm(0) = -\frac{1}{8} {\rm {Im}} \  {\Tr}\big (\gamma^0 \Sigma \left(\om_\pm=m_{\rm{th}}^q+i\epsilon,\, p=0 \right) \big) \, , \label{dm31_1}
\ee
where an extra factor of $1/2$ comes from the on mass shell condition. In the limit $p\rightarrow 0$ one gets from quark disperson relation rise $\om(0)=m_{\rm{th}}^q$. This leads to the equality of  the damping rate of normal quark mode and long wavelength plasmino mode as $\gamma_q(0)=\gamma_+(0)=\gamma_-(0)$. The soft contribution of the quark self-energy can be obtained from diagrams in Fig.~\ref{dm_s_q} where all propagators and vertices are HTL  $N$-point functions which are obtained in subsections~\ref{g_eff_prop}, \ref{q_eff_prop}, \ref{qg_three} and \ref{htl_2g_2q_vert}.  Calculation of quark self-energy in Fig.~\ref{dm_s_q} along with its imaginary part is a very tedious one. So, we refer to the final result which is gauge independent. In leading order in $g$ the soft quark damping rate was, independently, obtained by
Braaten et al~\cite{Braaten:1992gd} and by Kobes et al~\cite{Kobes:1992ys} as
$
\gamma_{\rm q}(0) = a \frac{g^2T}{12\pi} \, , \label{dm32}
$
where the coefficient $a=5.63$ and $5.71$ for two and three flavour, respectively. We note that the damping mechanism is happening due to elastic ($2\rightarrow 2$) and inelastic ($2\rightarrow 3$) processes which appear by appropriately cutting the self-energy diagram in Fig.~\ref{dm_s_q} .
 \vspace{-0.2cm}
 \subsubsection{Gluon damping rate}
 \label{sub:gluon}
  \vspace{-0.2cm}
The gauge boson damping rate are defined in~\eqref{dm13_14}.  However, the damping rate for soft gluons ($p\sim gT$) on the mass shell ($P^2=(m_{\rm{th}}^g)^2=\frac{g^2T^2}{9}(C_A+N_f/2)$) can be obtained~\cite{Heinz:1986kh,Pisarski:1988vd,Thoma:1995ju,Thoma:2000dc} at rest  from \eqref{dm13_14} as
\begin{subequations}
\begin{align} 
\gamma_T(0)&=\frac{1}{2m_{\rm{th}}^g} {\rm{Im}} \Pi_T(\om_T=m_{\rm{th}}^g+i\epsilon,p=0)  \, , \label{dm33} \\
\gamma_L(0)&=\frac{1}{2m_{\rm{th}}^g} \lim_{p\rightarrow 0} {\rm{Im}} \Pi_L(\om_L=m_{\rm{th}}^g+i\epsilon,p)  =-\frac{m_{\rm{th}}^g}{2} \lim_{p\rightarrow 0} \left[ \frac{1}{p^2}{\rm{Im}}\Pi_{00}(\om_L =m_{\rm{th}}^g+ i\epsilon,p) \right] \, , \label{dm34}
\end{align}
\end{subequations}
where in longitudinal case we have used \eqref{pi_L}. We know from the gluon dispersion relation that  $p\rightarrow 0$,  $\om_L(0)=\om_T(0)=m_{\rm{th}}^g$ and $\Pi_L(m_{\rm{th}}^g,0)=\Pi_T(m_{\rm{th}}^g,0)$ as there is no preferred direction. This leads to the equality of the damping rate of  transverse gluon and long wavelength plasmon mode as 
$\gamma_g(0)=\gamma_L(0)=\gamma_T(0)$.

\begin{figure}[htb]
\begin{center}
\includegraphics[scale=.90]{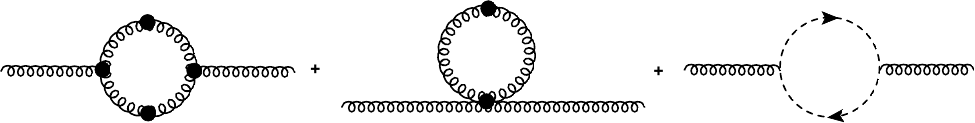}
\end{center}
\vspace{-0.6cm}
\caption{One-loop gluon self-energy diagram in HTLpt, where the solid blobs represent the HTL  propagators and vertices.}
\label{dm_s_g}
\end{figure}

Historically, way back in middle of 1980's, there was a huge controversy in computing the gluon damping rate in QCD perturbation theory at finite temperature. Much focus was made on plasmon mode.  It was found that the leading order gluon damping rate is gauge dependent both in magnitude and sign as discussed in subsec~\ref{bpt}.
This led to a doubt on the applicability of QCD perturbation theory at finite temperature.  We know both the plasmon and the transverse modes of the gluon are physical degrees of freedom, and their damping rates are determined by the position of the poles in the gluon propagator. These rates are physical quantities, and must be gauge invariant if computed correctly. This inconsistency also led Braaten and Pisarski~\cite{Braaten:1989kk,Braaten:1989mz,Pisarski:1988vd}  to develop HTL approximation and point out that those calculations~\cite{lopez,Kajantie:1982xx,Heinz:1986kh,Hansson:1987um,Kobes:1987bi}, however, are incomplete -- there are higher-loop diagrams which contribute to the same order in the coupling constant $g$ as the one-loop diagram. This implies that one needs to take into account those diagrams if the physical quantity is sensitive to the soft (electric) scale. 
The effective theory, developed around hard thermal loops, resums such diagrams. The bare propagators and vertices (N-point functions) are to be replaced by effective N-point functions as shown in Fig.~\ref{dm_s_g}. The main features of HTL resummation are discussed in details in subsec~\ref{htlr2}.

 \vspace{-0.3cm}
To obtain the soft gluon damping rate one needs soft-gluon self-energy as displayed in Fig.~\ref{dm_s_g} where the gluon propagators and vertices are HTL resumed ones. The 2-point, 3-gluon and 4-gluon functions are obtained, respectively, in subsections~\ref{g_eff_prop}, \ref{g_three} and \ref{g_four}. We need these quantities in zero momentum ($p=0$) and makes them bit simpler. We also note that there are no 
hard thermal loops for amplitudes with external ghosts, so the ghost propagator and the ghost-gluon vertex remain the same as in the bare ones. The ghost loop in Fig.~\ref{dm_s_g} is the same as the BPT. One needs to find out the imaginary part of the gluon self-energy and the calculation of which is far from trivial but can be found in Refs.~\cite{Braaten:1989kk,Braaten:1990it}. Following Refs.~\cite{Braaten:1989kk,Braaten:1990it} we quote the gluon damping rate at rest 
\be
\gamma_{\rm g}(0) =a \frac{g^2T}{8\pi} \, , \label{dm36}
\ee
where $a=1.35288+2.30000+2.98250= 6.63568$.  Imaginary parts of the self energy will involve gluon spectral densities given in~\eqref{srgbp5} which has pole part and Landau cut part.  We note here the soft gluon damping rate  will have contributions from pole-pole, pole-cut and cut-cut  like soft dilepton production rate as obtained in subsec~\ref{bpy_soft}.  This can also be visualised by cutting through the diagram in Fig.~\ref{dm_s_g}. We note that cutting through a blob means cutting a HTL self-energy which leads to Landau damping. Now, cut through two lines will lead to pole-pole contribution, cut through one blob and one line will lead to pole-cut term and cut through two blobs lead to cut-cut term . However pole-pole contribution (at $p=0$) corresponds to decay of collective modes ($g\rightarrow gg$)  which are kinematically forbidden due to energy conserving $\delta$-function.  The pole-cut term corresponds to elastic scattering of $gg\rightarrow gg$ and cut-cut term describes inelastic process $gg\rightarrow ggg$. The contribution from transverse gluon pole-cut part is $1.35288$, that from plasmon pole-cut is $2.30000$, and the cut-cut term  gives $2.98250$. Therefore, $a$ is positive number and the controversy had been settled and there was no unstability but also contributes to a deeper insight on various processes (These processes also give rise to both elastic and inelastic(Bremstrahlung) energy loss of gluons). The damping rate in \eqref{dm36} much larger than Coulomb gauge~\cite{lopez,Kajantie:1982xx,Heinz:1986kh} and temporal axial gauge~\cite{Kajantie:1982xx,Heinz:1986kh} with $a=1$. One can also calculate the damping rate of gluon at finite momentum and the calculation will be extremely non-trivial~\cite{Braaten:1989kk,Braaten:1990it}.

As discussed in subsec~\ref{sub:heavy} that the hard damping rates are logarithmically infra divergent but soft damping rates are infrared finite even though there is no magnetic screening of static magnetic field. The reason is that the damping rates at zero momentum are caused by scatterings, the exchange of transverse gluons do not cause any divergence because the magnetic processes are suppressed for zero external momentum~\cite{Thoma:1995ju,Thoma:2000dc}.
	
 \vspace{-0.2cm}
      \section{Energy Loss}\label{el}
          \vspace{-0.2cm}
The study of the energy loss of charged particles in a given medium has been rigorously examined for decades. The energy loss of a charged particle with a mass greater than that of an electron in a non-relativistic plasma is described by the famous Bethe-Bloch formula, which is based on the scattering of a charged particle with electrons in the plasma.. In case of relativistic plasma, viz., QCD plasma (QGP) and  QED plasma (electron-positron plasma) requires quantum field theoretic description. In QED plasma muon energy loss was computed in Ref.~\cite{Braaten:1991jj}. The parton energy loss is of interest for QGP as  high energetic partons (jets), produced from hard collisions at the initial stage of the relativistic Nucleus-Nucleus collisions, propagate through the thermalised QCD matter and interact with the medium particles quarks and gluon and loose energy, thus jet will be quenched. This is treated as an important signal for QGP formation~\cite{Gyulassy:1990ye} in relativistic heavy-ion collisions. The parton energy loss in QGP  was first computed by   Bjorken~\cite{Bjorken:1982tu} following  parton elastic scatterings with medium particles (quarks and gluons) in QCD matter as
\be
-\frac{dE}{dx}= \left(\frac{2}{3} \right)^{\pm 1} \, 4\pi\alpha_s^2 T^2\, \left (1+\frac{N_f}{6} \right ) \, \ln \left( \frac{q_{\rm{max}}} {q_{\rm{min}}} \right) \, , \label{el1}
\ee
where $`+$' sign is for quark and $`-$' sign is for gluon. $q_{\rm{max}}=\sqrt{4ET}$ and $q_{\rm{min}}$ is an infrared cut-off which are maximum and minimum momentum transfer in a elastic collision.  However, there is some uncertainty in choice of separation scale $q_{\rm{max}}$ and $q_{\rm{min}}$. 

We also note that the light quark formula was adopted for heavy quark~\cite{Braaten:1991we} as
\be
-\frac{dE}{dx}= \frac{ 8\pi\alpha_s^2 T^2}{3}\, \left (1+\frac{N_f}{6} \right ) \left[\frac{1}{v} -\frac{1-v^2}{2v^2} \ln\frac{1+v}{1-v}\right] \ln \left( \frac{q_{\rm{max}}} {q_{\rm{min}}} \right) \, , \label{el1a}
\ee
where for $v=p/E$ is the heavy quark velocity in the medium. The~\eqref{el1a} reduces to~\eqref{el1} when $v\rightarrow 1$ for light quarks.

Also one can estimate Bjorken formula for heavy lepton~\cite{Braaten:1991jj} from \eqref{el1a}, by substituting $g$ by $e$ and dividing by colour factor $2/3$ with $N_f=1$ for QED and neglecting the first term inside the parenthesis which comes from gluon\footnote{as we will see that for QED the process, $\mu \gamma \rightarrow \mu\gamma$, vanishes to leading order in $T/M$ due to kinematical reason},  as 
\be
-\frac{dE}{dx}= \frac{ e^4 T^2}{24\pi}\, \left[\frac{1}{v} -\frac{1-v^2}{2v^2} \ln\frac{1+v}{1-v}\right] \ln \left( \frac{q_{\rm{max}}} {q_{\rm{min}}} \right) \, . \label{el1b}
\ee
We also note that in addition to the elastic collisional energy loss there is also radiative energy loss due to inelastic collisions (Bremstrahlung). This is indeed very involved calculation which we also discuss later.
 \vspace{-0.2cm}
 \subsection{General Setup}
 \label{el_gen}
 \vspace{-0.2cm}
The infrared contribution to the energy loss~\cite{Ichimaru73} for large distance collision can be described as follows. 
\vspace{-0.3cm}
\subsubsection{Energy loss definition  from linear response theory}
\label{el_lrt}
An appropriate description~\cite{Ichimaru73} of various plasma properties can be obtained if one knows how a plasma responds to an external disturbance. Typically, to establish this response relationship in a plasma, one examines the plasma's response to an external electric field, which induces a current density. If the system remains stable against such a weak disturbance, the induced current may appropriately be expressed by that part of the response  linear in the externally disturbing field. The {\it linear response} of a plasma to an external electromagnetic field has extensively been investigated~\cite{Ichimaru73} in plasma physics, where the external current is related to the total electric field by the equation
\begin{equation}
\vec{J}^a_{\rm{ind}}(\omega,\bm{\vec q})= -\frac{i\omega}{4\pi}
\left [ \epsilon (\omega,\bm{\vec q})- \mathbb{1} \right ] 
\vec{E}^a_{\rm{tot}} (\omega,\bm {\vec  q}) , \label{el1c} 
\end{equation}
where $\epsilon (\omega,\bm{\vec q})$ is the dielectric tensor, describing the linear (chromo)electromagnetic properties of the medium, and $\mathbb{1}$ is the identity operator. Since we are interested in a system of a relativistic colour charge moving through a QCD plasma, we utilise the linear response theory for the purpose by simply assigning a colour index, $a=1\cdots 8$, to the relevant quantities herein and subsequent analysis, to take into account for quantum and non-Abelian effects.

We now briefly outline the theoretical description~\cite{Ichimaru73,Thoma:1990fm,Chakraborty:2006md} based on the linear response theory for the QGP as a finite, continuous, homogeneous and isotropic dielectric medium, which can be expressed by a dielectric tensor, $\epsilon(\omega,\bm{\vec q})$, depending on the direction only
through the momentum vector, $\vec {\mathbf  q}$. One can also construct another set of tensors of rank two from the momentum vector,  $\vec {\mathbf  q}$ which are the longitudinal projection tensor, ${\cal P}^L$ and the transverse projection tensor, ${\cal P}^T$, respectively, given as 
\begin{eqnarray}
{\cal P}^L_{ij} = \frac{q_iq_j}{q^2}, \qquad\mbox{and} \qquad {\cal P}^T_{ij} = \delta_{ij}- \frac{q_iq_j}{q^2} \ \ , \label{el1d}
\end{eqnarray}
with the general properties of the projection operators:  
$({\cal P}^L)^2 ={\cal P}^L$, $({\cal P}^T)^2 ={\cal P}^T$,
${{\cal P}^L}\cdot {{\cal P}^T} =0$ and
$({\cal P}^L)^2 + ({\cal P}^T)^2 =1$.
Now the dielectric tensor can be written  as a linear combination of  these two mutually independent components as
\begin{eqnarray}
\epsilon(\omega,\bm{\vec q}) &=& {\cal P}^L\epsilon_L(\omega,q)
\ + \ {\cal P}^T\epsilon_T(\omega,q) \, \, , \nonumber \\
\epsilon_{ij} &=& {\cal P}^L_{ij}\epsilon_L (\omega,q)\ 
+ \ {\cal P}^T_{ij}\epsilon_T(\omega,q) \ , \label{el1e} 
\end{eqnarray}
where the longitudinal, $\epsilon_L$ and the transverse, $\epsilon_T$, 
dielectric functions are given by
\begin{eqnarray}
\epsilon_L(\omega,q) = \frac{\epsilon_{ij}q_iq_j}{q^2} \, \, , \qquad \mbox{and} \qquad \epsilon_T(\omega,q) = \frac{1}{2}\left [{\rm {Tr}} 
\epsilon(\omega,\bm{\vec q})
-\epsilon_L (\omega,q) \right ] \, \, . \label{el1f}
\end{eqnarray} 
Now, we shall use the dielectric tensor in (\ref{el1e}) for a set of macroscopic equations, {\it viz.} Maxwell and continuity equations, in momentum space to obtain a relation~\cite{Ichimaru73,Thoma:1990fm,Chakraborty:2006md} between the total chromoelectric field and the external current as
\begin{eqnarray}
\left [\epsilon_L {\cal P}^L +\left ( \epsilon_T -\frac{q^2}{\omega^2}
\right ){\cal P}^T \right ] \vec{E}^a_{\rm {tot}}(\omega,\bm{\vec q})= 
\frac{4\pi} {i\omega} \vec{J}^a_{\rm {ext}}(\omega,\bm{\vec q}) 
\, \, . \label{el1g}
\end{eqnarray}
Combining ~\eqref{el1c}, ~\eqref{el1e} and~\eqref{el1g}, the induced current density can be related to the external current density as
\begin{equation}
\vec {J}^a_{\rm{ind}} (\omega,\bm{\vec q}) = \left [ \left ( 
\frac{1}{\epsilon_L}-1 \right ){\cal P}^L +\frac{1-\epsilon_T}
{\epsilon_T-q^2/\omega^2} {\cal P}^T \right ] 
\vec{J}^a_{\rm{ext}}(\omega,\bm{\vec q}) \, \, .\label{el11a}
\end{equation}
The incoming colour charged particle will induce a colour electric field in the medium which acts through Lorentz force and the incoming particle will decelerate. Since we are interested in studying the behaviour in the plasma reacting to a colour charged particle, $Q^a$ moving with a constant velocity $\bm{\vec v}$, 
we
specify the external current~\cite{Thoma:1990fm,Chakraborty:2006md} as
\begin{eqnarray}
\vec{J}^a_{\rm{ext}}&=& Q^a \bm{\vec v} \delta (\bm{\vec x}-
\vec{\mathbf v}t ) \stackrel{\rm{FT}}{=} 2\pi Q^a \bm{\vec v} 
\delta(\omega - \bm{\vec q}\cdot \bm{\vec v}) \, \, \, ,
 \label{el2} 
\end{eqnarray} 
where ${\rm{FT}}$ stands for Fourier transformation and $v=|\bm{\vec v}|$. The finite motion of
the charged particle deforms the screening charge cloud and also suffers a retarding force causing energy-loss. The soft contribution to the collisional differential energy-loss by the induced chromoelectric field 
is defined as~\cite{Ichimaru73,Thoma:1990fm}
\begin{equation}
-\frac{dE}{dx} = Q^a \frac{\bm{\vec v}}{v}\cdot {\rm {Re}}\vec{E}^a_{\rm{ind}}
(\bm{\vec x}=\bm{\vec v}t,t) \, \, ,  \label{el3}
\end{equation}
The induced electric field can be obtained from the inverse of \eqref{el1g}
which is given as
\begin{equation}
-\vec{E}^a_{\rm{ind}} = \left [ \left ( \frac{1}{\epsilon_L}-1\right ){\cal P}^L 
+\left(\frac{1}{\epsilon_T-q^2/\omega^2}-\frac{1}{1-q^2/\omega^2}  \right)  
{\cal P}^T \right ]\frac{4\pi}{i\omega} \vec{J}^a_{\rm{ext}} 
\, \,  ,  \label{el4}
\end{equation}
where ${\cal P}^L $ and ${\cal P}^T$ are, respectably, longitudinal and transverse projection operators and given in Ref.~\cite{Thoma:1990fm,Chakraborty:2006md}. Combining~\eqref{el3} and~\eqref{el4}, the contribution to the soft differential energy-loss in QGP is obtained~\cite{Thoma:1990fm,Chakraborty:2006md} as
\begin{equation}
\frac{dE}{dx} = - \frac{C\alpha_s}{2\pi^2v} \int d^3q\ \frac{\omega}{q^2}
\left [ {\rm{Im}}\epsilon^{-1}_L + \left (v^2q^2-\omega^2\right ) {\rm{Im}}\left (\omega^2\epsilon_T -q^2 \right )^{-1}\right ]_{\omega=\bm{\vec q}\cdot \bm{\vec v}} \, \,
, \label{el5}
\end{equation}
where where $C$ is the Casimir invariant for partons: for quark  $C=C_F=4/3$  and for gluon $C=C_A=3$.

The dielectric constants are
given~\cite{Thoma:1990fm,Chakraborty:2006md} as
\be
\epsilon_L(\omega,q) = 1+\frac{\Pi^{\rm g}_L(\omega,q)}{Q^2} \, \, ,
\qquad \mbox{and} \qquad
\epsilon_T(\omega,q) = 1+\frac{\Pi^{\rm g}_T(\omega,q)}{\omega^2} \, \, , \label{el67}
\ee
where $\Pi_L^{\rm g}$ and $\Pi_T^{\rm g}$ are, respectively,  given in~\eqref{qcd21} and~\eqref{qcd22}. Following this formalism the authors of  Refs.~\cite{Mrowczynski:1991da,Koike:1992xs} combined the infrared contribution along with the short distance scattering considered by Bjorken, they proposed that the Debye mass $m_D^{\rm g}$ as  the separation scale that implies infrared cut-off,  $q_{\rm{min}}=m_D^{\rm g}$. However, still there is some uncertainty in the choice of the separation scale. We will see later that such uncertainty will be removed by using HTLpt~\cite{Thoma:1990fm} and $q_{\rm{max}}$ will be taken care by kinematics at least for heavy fermions~\cite{Braaten:1991jj,Braaten:1991we}. 
\subsubsection{Energy loss definition from quantum field theory}
\label{el_ft}
We also define energy loss of a particle another way. We know that the scattering  of a particle in a medium causes energy loss. On the other hand, the scattering rate is related to the interaction rate vis-a-vis damping rate as we discussed it in sec~\ref{damp}. So energy loss is related to the interaction rate.
The interaction rate  for $2\rightarrow 2$ process can be obtained from \eqref{ph4} as
\bea
\Gamma(E) \!\!\!\!\! &=&\!\!\!\! \frac{1}{2E} \int \!\frac{d^3\bm{\vec p^{\,\prime}}}{2E'(2\pi)^3}  \int\! \frac{d^3\bm{\vec k}}{2E_k(2\pi)^3}  n_i(E_k)
 \int \frac{d^3{\bm{\vec k'}}}{2E_{k'}(2\pi)^3}  \left(1\pm n_i(E_{k'}) \right ) 
 (2\pi)^4 \delta^4\left(P +K-K'-P' \right) \frac{1}{2} \sum_{\rm{spins}}  \left|{\cal M}\right |^2 , \label{el8}\ \ \ \ \ \ 
\eea
where $P(E, \bm{\vec p})$ is four momentum of a fast moving incoming particle (jet)  and  $K=(E_k, \bm{\vec k})$ is four momentum of a incoming medium particle
depending upon QED and QCD plasma, whereas $P'=(E', \bm{\vec p^{\,\prime}})$  and $K'=(E_{k'}, \bm{\vec k'})$ are, respectively, outgoing decelerated
jet and medium particle.  ${\cal M}$ is the matrix element for sum of all possible $2\rightarrow 2$ processes.  The square of ${\cal M}$ represents the average over initial
and summation over final spin states. The distribution function $n_i$ is for medium particle depending upon fermion or boson.

Now the mean time  between two collisions of fast moving particle~\cite{Thoma:1995ju,Braaten:1991jj} is $t=1/\Gamma$, so the distance traveled by the fast moving charge with a velocity $v$ is $\Delta x= v/\Gamma$. The energy loss per collision is $E-E'=\om$, where $\om$ is the energy transfer of the exchanged gauge boson  with four momentum $Q=(\om, \bm{\vec q})$. The average energy lost by the fast moving colour charge per collision is given~\cite{Thoma:1995ju,Braaten:1991jj}  as
\be
\Delta E = \frac{1}{\Gamma}\int_0^\infty dE' (E-E') \frac{d\Gamma}{dE'}(E,E') \, , \label{el9}
\ee
where $d\Gamma/dE'$ is the differential interaction rate with respect to the final state decelerated  particle energy $E'$. The average energy loss per distance traveled is the ratio of $\Delta E$ and $\Delta x$ as
\be
-\frac{dE}{dx} = \frac{1}{v}\int_0^\infty dE' (E-E') \frac{d\Gamma}{dE'}(E,E') \, , \label{el10}
\ee
where the energy loss $dE/dx$ can be computed from \eqref{el8} by inserting $(E-E')/v =\om/v$.
\subsection{Collisional Energy Loss}
\label{coll-el}
\subsubsection{Energy loss of muon in QED plasma}
\label{muon_el}
The energy loss for muon with mass $M$ can be written from~\eqref{el10}
\be
-\frac{dE}{dx} = \frac{1}{v}\int_M^\infty dE' (E-E') \frac{d\Gamma}{dE'}(E,E') \, , \label{el11}
\ee
The energy loss  can also be written in muon self-energy. According to Refs.~\cite{Weldon:1983jn,Braaten:1991jj} one can also write the interaction rate averaged over two spin states of the muon as 
\be
\Gamma(E)=-\frac{1}{2E} \frac{e^{E/T}}{e^{E/T}+1} \sum_s\overline{u}(P,s) \, {\rm{Im}}\Sigma(E+i\epsilon,\bm{\vec p})\, u(P,s) \, , \label{el12}
\ee
where $u(P,s)$ is the spinor for a muon with four momentum $P=(E,\bm{\vec p})$ and spin $s$. Expressing the over $s$ by Dirac trace, the interaction rate becomes quite similar to damping rate as
\be
\Gamma(E)=-\frac{1}{2E}\big(1-n_F(E)) {\Tr} \big [ (P\!\!\!\! \slash +M) {\rm{Im}}\Sigma(E+i\epsilon,\bm{\vec p})\big] \, , \label{el13}
\ee
where $M$ is the mass of the heavy fermion. Now energy loss of muon can be obtained from Eq.~\eqref{el11} by inserting Eq.~\eqref{el13}. Alternatively, by using directly $\Gamma(E,E')$ from \eqref{el8} in \eqref{el11} for processes $\mu X \rightarrow \mu X'$, where $X$ and $X'$  are initial and final medium particles. They can be more than one or more electrons, positron or photons.
 
We know that the Bjorken formula for collisional energy loss for heavy quark in Eq.~\eqref{el1a} has uncertainty  due to ambiguities in the choice of $q_{\rm{min}}$ and $q_{\rm{max}}$. This ambiguity is removed ~\cite{Braaten:1991dd} in which the effect of screening is taken into account by considering a separation scale $q_c$ between hard ($q\sim T$) momenta transfer and soft ($q\sim eT$) momenta transfer, akin to the photon production rate discussed in section~\ref{em_chap}.  This should be chosen so that $eT\ll q_c \ll T$ in the $e\rightarrow 0$. The hard contribution ($q>q_c$) is calculated from tree-level processes appearing from the imaginary part of two-loop muon self-energy as displayed in Fig.~\ref{muon_bpt+htl}a. On the other hand, the soft contribution ($q_c<q$) from one-loop diagram in Fig.~\ref{muon_bpt+htl}c where the bare photon propagator is replaced by the effective photon propagator in HTL approximation. We  further note that in the weak coupling limit $eT\ll q_c\ll T$, the arbitrary separation scale $q_c$ cancels once the hard and the soft contributions are added, as it should be the case for a consistent leading-order calculation~\cite{Braaten:1991dd}.

First we would discuss the hard ($M, E, p> T$) contributions coming from processes $e^\pm \mu \rightarrow e^\pm \mu$ and the Compton processes $\gamma\mu\rightarrow \gamma \mu$ but Compton process would not contribute~\cite{Braaten:1991jj}  to leading order in $T/M$. The two processes. $e^\pm \mu \rightarrow e^\pm \mu$ have the same contribution and so one needs to multiply a factor of $2$ with the either process. The detail calculations have been done in Ref.~\cite{Braaten:1991jj} and we just quote the final compact result as
\be
- \left.\frac{dE}{dx}\right |_{\rm{hard}} = \frac{e^4T^2}{24\pi} \left[\frac{1}{v} -\frac{1-v^2}{2v^2} \ln\frac{1+v}{1-v}\right] 
\ \left [\ln  \frac{T} {q_c} +\ln  \frac{E} {M} +A_{\rm{hard}}(v) \right] , \label{el14}
\ee
where the function $A_{\rm{hard}}(v)$ can be obtained as
\be
A_{\rm{hard}}&=&\frac{3}{2}-\gamma_E +2 \ln2+\frac{\zeta '(2)}{\zeta (2)}\nn
&& \hspace{-1cm}-\,\left[\frac{ 1}{v}-\frac{1-v^2}{2v^2}\log \frac{1+v}{1-v}\right]^{-1}\left[ -\frac{1-v^2}{4v^2}\left({\rm Sp}\left(\frac{1+v}{2}\right)-{\rm Sp}\left(\frac{1-v}{2}\right)+\frac{1}{2}\ln\frac{1+v}{1-v}\ln\frac{1-v^2}{4}\right)-\frac{2}{3}v\right]  .
\ee
Note that $A_{\rm{hard}}(v)$ decreases monotonically as a function of the velocity from $1.239$ at $v=0$ to $1.072$ at $v=1$. The hard contribution to the muon energy loss depends on the separation scale $q_c$ and it is logarithmically divergent in the limit $q_c\rightarrow 0$.

Now, we discuss the soft contribution to the muon energy loss. In this case, one obtains the interaction rate from~\eqref{el13} using  the Fig~\ref{muon_bpt+htl}c 
and following the calculation in subsec~\ref{hard_damp} as
\bea
\Gamma(E)&\approx&  \frac{e^2}{2\pi v} \int_{0}^{\infty} q\, dq  \int_{-vq}^{vq} d\omega\left (1+n_B(\om)\right )
\,\left[\rho^{\rm{cut}}_L(\om,q) -  \left(v^2-\frac{\om^2}{q^2}\right)\rho^{\rm{cut}}_T(\om,q)  \right] \,  .  \label{el15} 
\eea
We first note that the hard contribution ($q,\,\om \sim T$) to the interaction rate is same as the damping rate, order $e^4T$, which has quadratic infrared singularity as discussed after Eq.~\eqref{dm25} in subsec~\ref{hard_damp}. Also the soft contribution to the interaction rate is same to the damping rate, order $e^2T$, which has logarithmic infrared singularity as discussed in subsec~\ref{hard_damp}. Thus,  the resummation of hard thermal loops may not be sufficient to calculate the interaction rate vis-a-vis the damping rate to leading order in $e$.

Using Eq.~\eqref{el15} and~\eqref{dm22a} in Eq.~\eqref{el11} one can write the soft contribution to the energy loss as
\be
- \left.\frac{dE}{dx}\right |_{\rm{soft}} = \frac{e^2}{2\pi v^2} \int_{0}^{q_c} q\, dq  \int_{-vq}^{vq}  d\omega \,\left(\frac{T}{\om}+\frac{1}{2}\right) \, \om  \,
\left[\rho^{\rm{cut}}_L(\om,q) -  \left(v^2-\frac{\om^2}{q^2}\right)\rho^{\rm{cut}}_T(\om,q)  \right] \, , \label{el16}
\ee
where we have  imposed an upper limit  $q_c$ on the momentum transfer $q$ since  $|\om|<vq$ and $\om$ is soft ($\sim eT$). We note that with the first term $T/\om$, inside the parenthesis vanishes because the omega integration becomes odd function of $\om$. The leading order contribution comes from the second term $1/2$ inside the parenthesis as
\be
- \left.\frac{dE}{dx}\right |_{\rm{soft}} = \frac{e^2}{4\pi v^2} \int_{0}^{q_c} q\, dq  \int_{-vq}^{vq}  d\omega \, \om  \,
\left[\rho^{\rm{cut}}_L(\om,q) -  \left(v^2-\frac{\om^2}{q^2}\right)\rho^{\rm{cut}}_T(\om,q)  \right] \, . \label{el17}
\ee
We get the imaginary part of the longitudinal and transverse dilelectric functions from Eq.~\eqref{el67} by using Eq.~\eqref{bpy_04} in the limit  $0< \om<q$ as
\begin{subequations}
\begin{align}
{\rm{Im}} \epsilon_L^{-1} (\om, q) &= Q^2\, {\rm{Im}} \left(\frac{1}{Q^2+\Pi_L(\om,q)} \right ) = \pi \, Q^2 \, \rho_L^{\rm{cut}}(\om, q) \approx 
-\pi\, q^2 \, \rho_L^{\rm{cut}}(\om, q) \, , \label{el17a} \\
{\rm{Im}} \big (\om^2\epsilon_T-q^2\big)^{-1} &=  {\rm{Im}} \left(\frac{1}{Q^2+\Pi_T(\om,q)} \right ) =  \pi \, \rho_T^{\rm{cut}}(\om, q) \, . \label{el17b}
\end{align}
\end{subequations} 
where the spectral functions given in~\eqref{srgbp8} and~\eqref{srgbp9} can be obtained in the limit $0 <\om<q$ as
\begin{subequations}
\begin{align}
\rho_L^{\rm{cut}}(\om, q) &=  \frac{ (m_D^{\gamma})^2}{2} \om \, q \left \{\left [q^2+  (m_D^{\gamma})^2  - \frac{(m_D^{\gamma})^2}{2} \frac{\om}{q} 
\ln \frac{q+\om}{q-\om}\right]^2 + \left[\pi\frac{ (m_D^{\gamma})^2}{2} \frac {\om}{q}\right]^2 \right\}^{-1} \, ,\label{el17d} \\
\rho_T^{\rm{cut}}(\om, q) &= \frac{ (m_D^{\gamma})^2}{4} \frac{(\om^2-q^2) \om}{q^3}
\left[\left\{q^2-\om^2+ \frac{ (m_D^{\gamma})^2\om^2}{2q^2} \left(1+\frac{q^2-\om^2}{2\om q} \ln \frac{q+\om}{q-\om}\right)\right\}^2
+ \left\{\frac{\pi (m_D^{\gamma})^2 \om(q^2-\om^2)}{4q^3}\right\}^2\right]^{-1} .\hspace{0cm} \label{el17e}
\end{align}
\end{subequations}
Using Eq.~\eqref{el17a} and~\eqref{el17b} in~\eqref{el17}, one gets 
\be
- \left.\frac{dE}{dx}\right |_{\rm{soft}} =- \frac{e^2}{4\pi^2 v^2} \int_{0}^{q_c} \frac{ dq}{q}   \int_{-vq}^{vq}  d\omega \, \om  \,
\left[{\rm{Im}} \epsilon_L^{-1} (\om, q)  +  \left(v^2 q^2-\om^2\right) {\rm{Im}} \big (\om^2\epsilon_T-q^2\big)^{-1} \right] \, . \label{el17c} 
\ee
We compare  this result with that obtained~\cite{Thoma:1990fm} in Eq.~\eqref{el5} for the QCD plasma (QGP) by changing $C_Fg^2 \rightarrow e^2$ for QED plasma and changing the angle integration $\angle(\bm{\vec q}\, , \bm{\vec v})$  to $\om$ integration, one gets same result as Eq.~\eqref{el17c}.
This indicates~\cite{Thoma:1990fm} that muon $dE/dx$ by computing  the electric field induced by a classical source consisting of a point charge moving at fixed velocity is equal to the quantum field theory description. The agreement holds only after using the small $\om$ approximation in~\eqref{el17},  which is equivalent to a high-temperature approximation. 

Now using Eq.~\eqref{el17d} and~\eqref{el17e} in Eq.~\eqref{el17} and then changing the integration variables~\cite{Braaten:1991jj}, $q$ and $\om$ to $q$ and $x=\om/q$, respectively. The integral over $q$ can then be performed analytically giving rise a logarithmic dependence on $q_c$. In the logarithmic term, the integral over $x$ can also be evaluated analytically but other $x$-integrals can be performed numerically~\cite{Braaten:1991jj}. The final result  is obtained in Ref.~\cite{Braaten:1991jj} as
\be
- \left.\frac{dE}{dx}\right |_{\rm{soft}} = \frac{e^4T^2}{24\pi} \left[\frac{1}{v} -\frac{1-v^2}{2v^2} \ln\frac{1+v}{1-v}\right] 
\ \left [\ln  \frac{q_c}{eT} +A_{\rm{soft}}(v) \right] , \label{el18}
\ee
where the function $A_{\rm{soft}}(v)$ is 
\be
A_{\rm{soft}}(v)\!\!\!&=&\!\!\!\ln3-{1\over v^2}\left[\frac{1}{v}-\frac{1-v^2}{2 v^2}\log \frac{1+v}{1-v}\right]^{-1} \left[\int_0^v x^2 \left\{\ln\frac{3 \pi  x}{2}+\frac{1}{2} \ln \left[1+Q^2_{\ell}(x)\right]+Q_\ell(x) \left(\frac{\pi }{2}-\tan ^{-1}Q_\ell(x)\right)\right\}\! dx\right.\nn
&&\left.+ \frac{1}{2}\int_0^v\frac{x^2}{1-x^2} \left(v^2-x^2\right) \left\{\ln \frac{3 \pi  x}{4} + \frac{1}{2} \ln \left(1+Q^2(x)\right)+Q_t(x) \left(\frac{\pi }{2}-\tan ^{-1}Q_t(x)\right)\right\}dx\right] .
\ee
Note that the function $A_{\rm{soft}}(v)$  begins at $0.049$ at $v =0$, increases to a maximum of $0.292$ at $v=0. 96$, and then decreases to $0.256$ at $v=1$. Alternatively, one can also compute~\cite{Braaten:1991jj}  the soft contribution from the scattering diagram with exchanged soft photon in Fig~\ref{muon_bpt+htl}d.
\begin{figure}[h]
\begin{center}
\includegraphics[height=5.8cm, width=9cm]{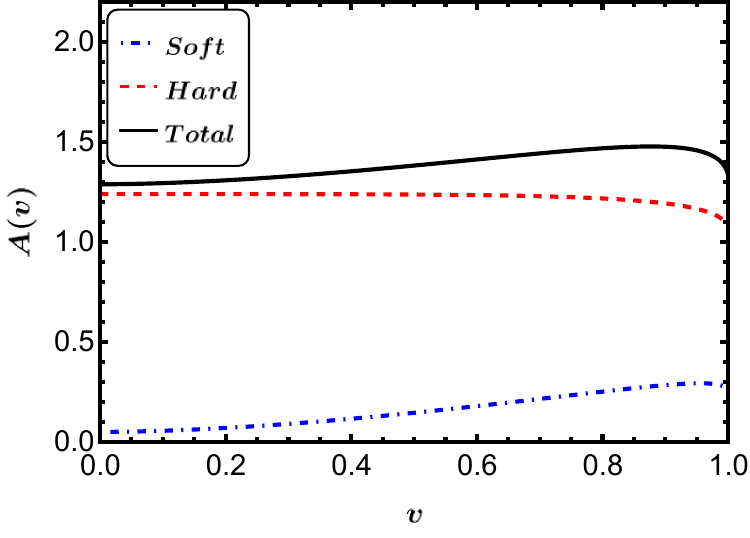}
\includegraphics[height=6cm, width=9cm]{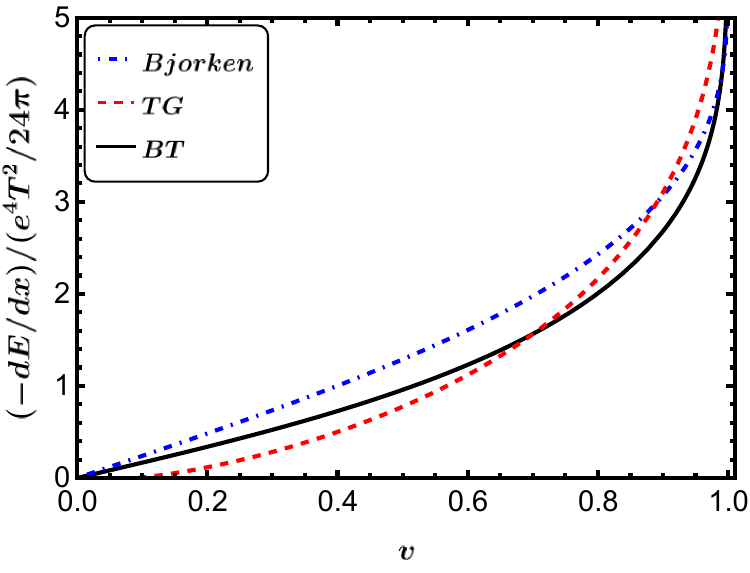}
\end{center}
\vspace*{-0.3in}
\caption{\textit{Left panel}: The function $A(v)$ defined in \eqref{el19} as a function of the muon velocity $v$ (solid curve). It is the sum of $A_{\rm{hard}}(v)$ (dashed curve) and $A_{\rm{soft}}(v)$  (dotted curve). \textit{Right Panel:} Energy loss of the muon as a function of its velocity $v$. The complete leading-order result (solid line) of BT~~\cite{Braaten:1991jj} in \eqref{el19} is compared to previous calculations by Thoma and Gyulassy (TG) (dashed curve)~\cite{Thoma:1990fm}  and Bjorken (dotted curve)~\cite{Bjorken:1982tu} in~\eqref{el1b}.}
\label{Av_de_mu}
\end{figure}

Now, adding hard contribution in Eq.~\eqref{el14} and soft contribution in Eq.~\eqref{el18}, one obtains the complete result for the energy loss of muon obtained by Braaten and Thoma (BT)~\cite{Braaten:1991jj} as
\be
- \frac{dE}{dx} = \frac{e^4T^2}{24\pi} \left[\frac{1}{v} -\frac{1-v^2}{2v^2} \ln\frac{1+v}{1-v}\right] 
\ \left [\ln  \frac{E}{M} + \ln\frac{1}{e} + A(v) \right] , \label{el19}
\ee
where $A(v)=A_{\rm{hard}}(v)+A_{\rm{soft}}(v)$. It is clearly seen from Eq.~\eqref{el19} that the separation scale, $q_c$, cancels out in final expression, leaving a logarithm of $1/e$. This indicates that $dE/dx$ has contributions from transfer ranging from the hard scale ($\sim T$) and the soft scale ($\sim eT$). The dependence of the function $A(v)$ on the velocity $v$ is shown in the left panel Fig.~\ref{Av_de_mu}, together with hard and soft contributions. $A(v)$ increases from $A(v)=1.288$ at $v=0$ to a maximum of $1.478$ at $v=0.88$ and then decreases to $A (v) = 1.328$ at $v = 1$.


In the right panel of Fig.~\ref{Av_de_mu} the scaled energy loss of muon is displayed as a function of its velocity and compared with the result of Bjorken~\cite{Bjorken:1982tu}  given in \eqref{el1b} and the soft contribution from the calculation of Thoma and Gyulassy (TG)~\cite{Thoma:1990fm}  has been obtained from Eq.~\eqref{el18} by replacing  $q_c=4Tp/(E-p+4T)$ under the logarithm. For Bjorken formula in Eq.~\eqref{el1b} we have used $q_{\max}=\sqrt{4ET}$ and $q_{\min}=m_D^\gamma=\sqrt{3} \ m_{\rm{th}}^\gamma$, the Debye screening mass in QED. Both of the previous calculations differ significantly from the complete leading-order result. Bjorken's approximation significantly overestimates $dE/dx$ over most of the range of velocity. The calculation of Thoma and Gyulassy underestimates $dE/dx$ for $v< 0.7$ and overestimates it for $v>0.7$.

We note that  at $v\rightarrow 0$, the factor in the first set of square braces in Eq.~\eqref{el19} approaches $2v/3$, implying a small positive energy loss proportional to $v$. This should not be the case on physical ground, because heavy particle with kinetic energy less than $T$ should in principle gain  energy~\cite{Braaten:1991jj,Chakraborty:2006db}. The $dE/dx$ should flip a sign at certain value of $v$ which is of the order of thermal velocity $\sqrt{T/M}$. On the other hand the energy loss given in \eqref{el19} breaks down in the $v\rightarrow 1$ limit.

Now one can also obtain the energy loss in two extreme limits~\cite{Braaten:1991jj}: $v\rightarrow 0$ and $v\rightarrow 1$, respectively,  as
\begin{subequations}
\begin{align}
- \left. \frac{dE}{dx} \right |^{v\rightarrow 0} &=-\frac{e^4T^3}{12\pi M v } \left [ \ln \frac{4\sqrt{3}}{e} + \frac{1}{2}-\gamma +\frac{\zeta^\prime(2)}{\zeta(2)}\right ] \, , \label{el19a}\\
- \left. \frac{dE}{dx} \right |^{v\rightarrow 1} &=\frac{e^4T^2}{48 \pi} \left [ \ln \frac{E}{e^2T} + 2.031\right ] .\, \label{el19b}
\end{align}
\end{subequations}
We note that the energy loss in \eqref{el19a} in $v\rightarrow 0$ is negative meaning energy gain for subthermal velocity. The factor $1/v$ indicates a divergence which appears due to kinematical reason. One can approximate the crossover velocity where energy loss flips a sign by equating $v\rightarrow 0$ limit of Eq.~\eqref{el19} with magnitude of Eq.~\eqref{el19a}. This gives rise to $v=\sqrt{3T/M}=0.55$ for $M/T=10$. On the other hand the crossover energy between the regions of validity of Eq.~\eqref{el19} and~\eqref{el19b} can be estimated by equating the $v\rightarrow 1$ limit of Eq.~\eqref{el19}  with Eq.~\eqref{el19b}. This gives $E \simeq 0.54M /T$. If $M/T=10$, the crossover velocity is $0.98$.

In summary, Ref.~\cite{Braaten:1991jj} has obtained the energy loss for a heavy lepton with mass $M \gg T$ in three regimes of velocity or energy. For subthermal velocities $v \ll \sqrt{3T/M}$, it is negative and given in Eq.~\eqref{el19a}. In the intermediate region where $v \gg \sqrt{3T/M}$, and $E < 0.54 M^2 /T$,  the energy loss is given by~\eqref{el19}. In the ultrarelativistic region $E > 0.54 M^2 /T$, it is given in Eq.~\eqref{el19b}. We also note that for collisional QED plasma the energy loss of a heavy fermion has recently been calculated in Ref.~\cite{Guo:2024mgh}.

\subsubsection{Energy loss of heavy quark in QCD plasma}
\label{heavy_el}
In Ref.~\cite{Svetitsky:1987gq,GolamMustafa:1997id,Mustafa:2004dr,Mustafa:2003vh} the diffusion of charm quarks was studied in QGP and the charm quark's drag coefficient was found to be related to  energy loss as $A(p) =(dE/dx)/p$, from  tree level calculation of elastic collisions of charm quark with medium particles light quarks and gluons. The kinematics of the scattering automatically provided the upper limit of  momentum transfer $q_{\max}$ but for infrared cut-off an {\textit{ ad hoc}}  lower limit was chosen as $q_{\min}=m_{\rm {th}}^{\rm g}$. This violets gauge invariance and causes an ambiguity  similar to Bjorken's calculation.

In the previous subsec~\ref{muon_el} the collisional energy loss of muon in leading order has been discussed within HTL approximation. In this subsec we would like to discuss the collisional energy loss~\cite{Braaten:1991we}  of a relativistic heavy quark traversing through a QGP to leading order in $g$ and 
$T/M$, 
where $M$ is mass of heavy quark. The calculation will be almost parallel to that  heavy lepton.  We assume that mass ($M$) and momentum ($p$) of heavy quarks are larger than the temperature ($M,\, p\gg T$) of the QGP. The result can be obtained in two kinematical regions: $E\ll M^2/T$ and $E\gg M^2/T$. We note that the leading order in strong coupling $g$, 
the energy loss of heavy quark originates from elastic scattering processes: $Qq(\bar q)\rightarrow Qq(\bar q)$ and $Qg\rightarrow Qg$.  Some of the contributions to $dE/dx$ can directly be obtained from the corresponding QED calculation in subsec~\ref{muon_el}  by simple substitution.

We note that in Ref.~\cite{Braaten:1991we} the maximum momentum transfer $q$ from elastic scattering processes $Qq$ and $Qg$ with energy of thermal particles ($q$ and $g$) $k$ is obtained kinematically as $q_{\max}=2k(1+k/E)/(1-v +2k/E)$. In the region $E\ll M^2/T$, $q_{\max}$ can be approximated as $2k/(1-v)$ whereas, for $E\gg M^2/T$, $q_{\max}=E$, the energy of the heavy quark. We further note that for $E\sim M^2/T$, it becomes very complicated to obtain $dE/dx$ due to the necessity of using the general expression  of $q_{\max}$.

We first consider the region $E\ll M^2/T$ and  assume that the heavy quark kinetic energy is much larger than $T$.  Some of the contributions to $dE/dx$ can be obtained from the corresponding QED calculation in subsec~\ref{muon_el}  by simple substitution. It has both hard and soft contributions. The soft contribution can be found from QED plasma
calculation~\cite{Braaten:1991jj} by replacing $e$ by $g$, multiplied by colour factor $4/3$  and thermal photon mass $m_{\rm{th}}^\gamma = eT/3$ by thermal gluon mass $m_{\rm{th}}^{\rm g} =( gT/\sqrt{3}) (1+N_f/6)^{1/2}$. The soft contribution as obtained in Ref.~\cite{Braaten:1991we} following Eq.(41) of Ref.~\cite{Braaten:1991jj}  is 
quoted below as
\bea
- \left.\frac{dE}{dx}\right |_{\rm{soft}}^{Qq+Qg}& = & \frac{g^4T^2}{6\pi}\left (1+\frac{N_f}{6}\right )\Bigg\{ \Bigg[\frac{1}{v} -\frac{1-v^2}{2v^2} \ln\frac{1+v}{1-v}\Bigg] 
\ln  \frac{q_c}{m_{\rm{th}}^{\rm g}} -\frac{1}{v^2} \int_0^v dx\, x^2 \Bigg[\ln\frac{3\pi x}{2}+\frac{1}{2}\ln \big[1+Q^2_l(x)\big] \nn
&+& Q_l(x)\left(\frac{\pi}{2}-\tan^{-1} Q_l(x)\right)\Bigg]-\frac{1}{2v^2}\int_0^v dx\, x^2 \, \frac{v^2-x^2}{1-x^2}
\Bigg[\ln\frac{3\pi x}{4}+\frac{1}{2}\ln \big[1+Q^2_t(x)\big] \nn
&& +Q_t(x)\left(\frac{\pi}{2}-\tan^{-1} Q_t(x)\right)\Bigg] \Bigg \}\, , \label{el20}
\eea
where
\be
Q_l(x)=\frac{1}{\pi} \left[-\ln\frac{1+x}{1-x}+\frac{2}{x} \right] \, ; \,\,\,\,\,\, {\mbox{and}}  \,\,\,\,\,\,
Q_t(x)=\frac{1}{\pi} \left[\ln\frac{1+x}{1-x}+\frac{2x}{1-x^2}\right] \, . \label{el20b}
\ee
Similarly, the hard contribution to the energy loss  from  processes $Qq(\bar q)\rightarrow Qq(\bar q)$, as given in Fig.~\ref{muon_bpt+htl}b where muon should 
be  replaced by heavy quark for QGP~\cite{Braaten:1991we}, can be obtained from QED plasma results
by replacing $e$ by $g$, multiplied by colour factor $2/3$ and summing over $N_f$ of the initial thermal quark in Eq.(22) of Ref.~\cite{Braaten:1991jj} as
\bea
- \left.\frac{dE}{dx}\right |_{\rm{hard}}^{Qq(\bar q)} &=&\frac{g^4T^2}{6\pi} \frac{N_f}{6} \left \{ \left[\frac{1}{v} -\frac{1-v^2}{2v^2} \ln\frac{1+v}{1-v}\right] 
\left[\ln \frac{4ET}{q_cM} + \frac{3}{2}-\gamma_E +\frac{\xi^\prime(2)}{\xi(2)} \right ]\right. \nn 
&& \left. - \frac{1-v^2}{4v^2}\left[{\rm{Sp}}\left(\frac{1+v}{2}\right)-{\rm{Sp}}\left(\frac{1-v}{2}\right) +\frac{1}{2} \ln\frac{1+v}{1-v} \ln\frac{1-v^2}{4}\right] 
-\frac{2}{3}v\right \} \, , \label{el21} 
\eea
where the Euler's constant $\gamma_E=0.57722$,  $\xi(z)$ is  Riemann zeta-function with ${\xi^\prime(2)}/{\xi(2)} =-0.56996$ and the Spencer function is given as
\be
{\rm{Sp}}(x) = -\int_0^x \frac{1}{t}\, \ln(1-t) \, dt \, . \label{el22}
\ee
We note that the separation scale $q_c$ in $\ln (4ET/q_cM)$ term in~\eqref{el22} cancels with $\ln (q_c/m_{\rm{th}}^{\rm g})$ term proportional to $N_f$  in \eqref{el21}.
\begin{figure}[htb]
\begin{center}
\includegraphics[scale=1.0]{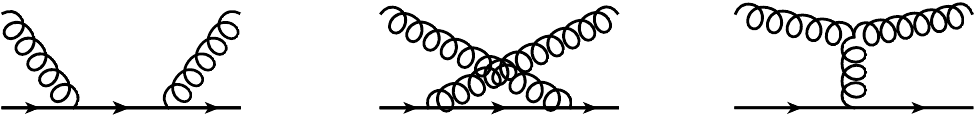}
\end{center}
\vspace{-0.3cm}
\caption{The hard scattering diagram of $Qg\rightarrow Qg$ processes.}
\label{Qg_hard}
\end{figure}
Now, one needs to compute hard scattering for $Qg\rightarrow Qg$ processes as in shown in Fig.~\ref{Qg_hard} because in QED, $\mu\gamma\rightarrow \mu\gamma$ 
process does not contribute~\cite{Braaten:1991jj} to leading order in $T/M$.  The energy loss for heavy quark for hard processes of $Qg$ scattering from $s$- and $u$-channels can be 
obtained~\cite{Braaten:1991we} as
\be
- \left.\frac{dE}{dx}\right |_{\rm{hard}}^{Qg(s+u)} =\frac{g^4T^2}{12\pi}\, \left[\frac{1}{v} -\frac{1-v^2}{2v^2} \ln\frac{1+v}{1-v}\right]  \, . \label{el23}
\ee
As seen that the scattering of $s$- and $u$-channel does not produce any infrared singularity and no separation scale $q_c$ appears in \eqref{el23}.
However, there is an infrared singularity from $t$-channel diagram, so the cut-off $q>q_c$ is required. Using $q_c<T$, one obtains~\cite{Braaten:1991we}
the result as
\bea
- \left.\frac{dE}{dx}\right |_{\rm{hard}}^{Qg(t)} &=&\frac{g^4T^2}{6\pi}  \left \{ \left[\frac{1}{v} -\frac{1-v^2}{2v^2} \ln\frac{1+v}{1-v}\right] 
\left[\ln \frac{2ET}{q_cM} + 1-\gamma_E +\frac{\xi^\prime(2)}{\xi(2)} \right ]\right. \nn 
&& \left. - \frac{1-v^2}{4v^2}\left[{\rm{Sp}}\left(\frac{1+v}{2}\right)-{\rm{Sp}}\left(\frac{1-v}{2}\right) +\frac{1}{2} \ln\frac{1+v}{1-v} \ln\frac{1-v^2}{4}\right] 
-\frac{2}{3}v\right \} \, . \label{el24} 
\eea
The final result for energy loss of a relativistic heavy quark with energy $E\ll M^2/T$, is obtained in Ref.~\cite{Braaten:1991we} by 
adding the soft contribution from \eqref{el20} and hard contributions from Eqs.~\eqref{el21}, \eqref{el23} and~\eqref{el24} as
\bea
- \frac{dE}{dx} &=&\frac{8\pi \alpha_s^2T^2}{3} \left (1+\frac{N_f}{6}\right )  \left[\frac{1}{v} -\frac{1-v^2}{2v^2} \ln\frac{1+v}{1-v}\right] 
\ln \left[ 2^{N_f/(6+N_f)} B(v) \frac{ET}{m_{\rm{th}}^{\rm g}\, M} \right ]\, , \label{el25} 
\eea
where
\be
 B(v)=\frac{1}{6}\exp \left[A_{\rm{soft}}(v)+A_{\rm{hard}}(v)- \frac{3}{8}+\frac{3}{6+N_f}\right]
\ee
is a smooth function of the velocity that increases monotonically from $B(0)=0.604$ to a maximum of $B(0.88)=0.731$ 
and then decreases to $B(1)=0.629$.

Next we consider the kinematical region $E\gg M^2/T$. Here also several QED results~\cite{Braaten:1991jj} can be generalised to QCD plasma. The soft contribution to the heavy quark energy loss from $Qq$ and $Qg$ scattering has been obtained in~\cite{Braaten:1991we} as
\bea
- \left.\frac{dE}{dx}\right |_{\rm{soft}}^{Qq+Qg}&=&\frac{g^4T^2}{6\pi} \left (1+\frac{N_f}{6}\right )
\left[\ln \frac{q_c}{m_{\rm{th}}^{\rm g}} - 0.843 \right ]\, . \label{el26} 
\eea
The hard contribution from $Qq$ scattering is obtained in Ref.~\cite{Braaten:1991we}  as
\bea
- \left.\frac{dE}{dx}\right |_{\rm{hard}}^{Qq}&=&\frac{g^4T^2}{12\pi} \frac{N_f}{6}
\left[\ln \frac{2ET}{q_c^2} + \frac{8}{3}-\gamma_E +\frac{\zeta^\prime(2)}{\zeta(2)} \right ]\, . \label{el27} 
\eea
The contributions from $s$- and $u$-channel of $Qg$ scattering vanishes~\cite{Braaten:1991we}  while that from $t$-channel is obtained as
\bea
- \left.\frac{dE}{dx}\right |_{\rm{hard}}^{Qg}&=&\frac{g^4T^2}{12\pi} 
\left[\ln \frac{ET}{q_c^2} + \frac{8}{3}-\gamma_E +\frac{\zeta^\prime(2)}{\zeta(2)} \right ]\, . \label{el28} 
\eea
Summing up the contributions from \eqref{el26} to \eqref{el28} of soft and hard scatterings, the energy loss of heavy quark with energy $E\gg M^2/T$ is 
obtained~\cite{Braaten:1991we}  as
\bea
- \frac{dE}{dx} &=&\frac{8\pi \alpha_s^2T^2}{3} \left (1+\frac{N_f}{6}\right ) 
\ln \left[ 2^{N_f/2(6+N_f)} 0.920  \frac{\sqrt{ET}}{m_{\rm{th}}^{\rm g}} \right ]\, , \label{el29} 
\eea
When $E\sim M^2/T$, the formula for the energy loss must crossover from the $v \rightarrow 1$ limit of Eq.~\eqref{el25} to Eq.~\eqref{el29}. It should be a good approximation to simply use Eq.~\eqref{el25} for some crossover energy $E_{\rm{cross}}$ and then switch over to Eq.~\eqref{el29}. One can restrict $E_{\rm{cross}}$  by demanding $dE/dx$ continuous at $E= E_{\rm{cross}} =1.80 M^2/T$ for $N_f=2$ quark favours.  

We also note that earlier the energy loss of a high energy heavy quark was earlier obtained by TG in Ref.~\cite{Thoma:1990fm}. This was obtained~\cite{Thoma:1990fm} from Eq.~\eqref{el5} as following:  first, using the cut part of  the gluon spectral functions from Eqs.~\eqref{el17d} and~\eqref{el17e} in replacing QED Debye screening mass $m_D^\gamma$ by QCD Debye screening mass $m_D^{\rm g}$ and using them in Eqs.~\eqref{el17a} and~\eqref{el17b} which are then to be used in Eq.~\eqref{el5}. Now one has to perform two integrations;  one is over the exchanged gluon momentum $q$ and the other one  is over angle $\angle (\bm{\vec v \cdot \vec q})$. It is possible to perform the integration over the momentum $q$ analytically, but the integral over the angle has to be done numerically and one ends up with the final, gauge independent result for the energy loss of a high energy heavy quark. However, one can obtain analytical result  by setting $m_D^{\rm g}=0$ in the denominator of the cut part of the spectral functions in Eqs.~\eqref{el17d} and~\eqref{el17e}. Under such approximation  the heavy quark energy loss was obtained analytically by TG~\cite{Thoma:1990fm} as 
\bea
- \frac{dE}{dx} &=&\frac{4\pi C_F \alpha_s^2T^2}{3} \ln\left( \frac{q_{\max}}{q_{\min}}\right)\, \frac{1}{v^2}
 \left[v +\frac{v^2-1}{2} \ln\frac{1+v}{1-v}\right]\, 
 , \label{el29a} 
 \eea
where $q_{\min}=m_D^{\rm g} =\sqrt{3} m_{\rm{th}}^{\rm g}$ and $q_{\max}=4pT/(E-p+4T)$. This analytic expression differs by $10\%$ from the actual result.
\begin{figure}[bbt]
\begin{center}
\includegraphics[scale=0.65]{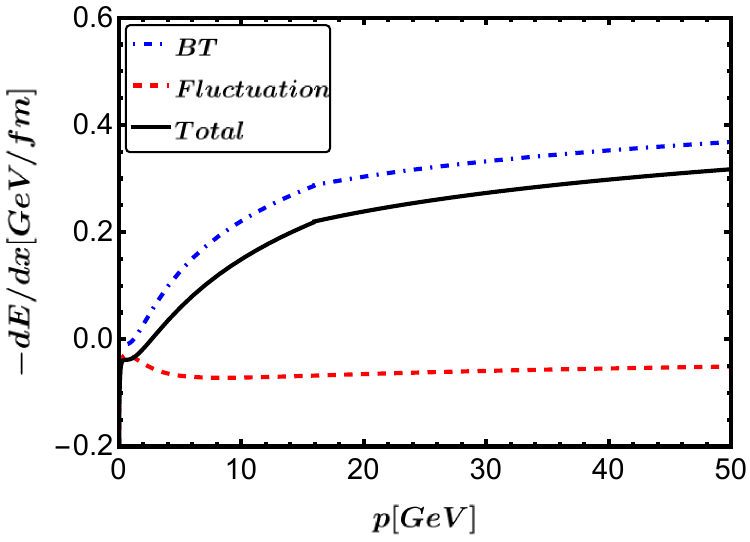}
\includegraphics[scale=0.65]{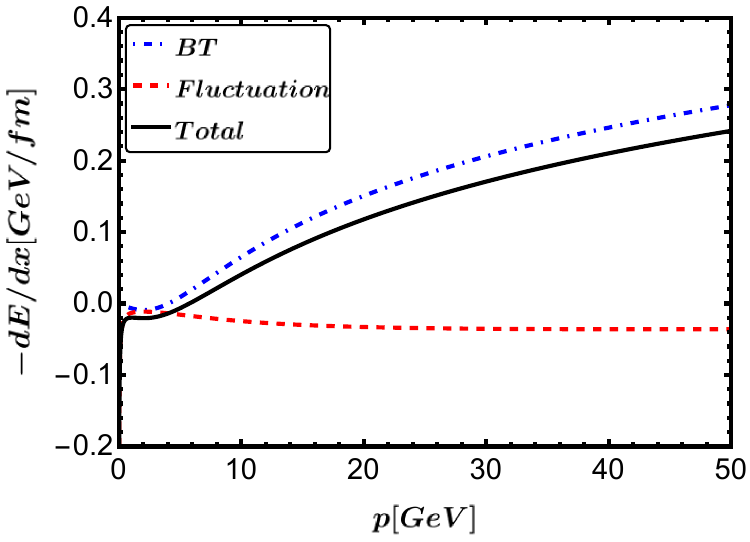}
\end{center}
\vspace{-0.5cm}
\caption{Represents the  differential energy loss for charm(left panel) and bottom (right panel) quarks as a function of their momentum. The complete result by BT~\cite{Braaten:1991we} to leading order in $g$, (solid curve) is compared to previous calculations by TG~\cite{Thoma:1990fm} (red dashed curve) and Bjorken~\cite{Bjorken:1982tu} (blue dashed-dotted curve).}
\label{dedx_c_b}
\end{figure}
The complete leading order result by BT given in~\eqref{el25} for $E\ll M^2/T$ and \eqref{el29} for $E\gg M^2/T$ is displayed by solid line in Fig.~\ref{dedx_c_b} as a function of momentum of the heavy quarks for  $T=0.25$GeV, $\alpha_s=0.2$ and $N_f=2$. For charm quark with mass $M_c=1.5$GeV in the left panel of  Fig.~\ref{dedx_c_b}, 
the crossover energy $E_{\rm{cross}}^c=16$GeV, such that the most energy range relevant for jets in heavy-ion collisions is denoted by \eqref{el29}. As discussed earlier that the discontinuity in slope at the crossover point could be avoided by a more general calculation with $E\sim M^2/T$. As can be seen from Fig.~\ref{dedx_c_b} for the charm quark the complete leading-order calculation is in reasonable agreement with that of TG~\cite{Thoma:1990fm} given in \eqref{el29a} 
 at low energies, but then crosses over to a form that is closer to Bjorken's estimate given in \eqref{el1a} at high energies.
 For bottom quark with mass $M_b=5$GeV in the right  panel of  Fig.~\ref{dedx_c_b}, the crossover energy $E_{\rm{cross}}^b=180$GeV, so the entire range of energy is covered by the formula given in \eqref{el25}. Both TG and Bjorken results overestimate the complete leading order result of BT.

As discussed for muon in subsec~\ref{muon_el}, here also the energy loss given in \eqref{el25} breaks down at thermal velocity $v\sim \sqrt{T/M}$, because the energy loss must flip sign in this region. This because  quark with $v =0 $ can only gain energy in a collision. The methods used above to compute $dE/dx$ at high energies can also be used to compute it in 
the limit $v\rightarrow 0$  as
\bea
- \frac{dE}{dx} &=& - \, \frac{16\pi \alpha_s^2T^2}{3Mv} \left (1+\frac{N_f}{6}\right )  \ln \left[ 2^{N_f/(6+N_f)} 0.604 \frac{T}{m_{\rm{th}}^{\rm g}} \right ]\, . \label{el30} 
\eea
One can approximate the crossover velocity where energy loss flips a sign by equating $v\rightarrow 0$ limit of Eq.~\eqref{el25} with magnitude of Eq.~\eqref{el30}. This gives rise to $v=\sqrt{3T/M}$.  With a plasma temperature of $0.25$ MeV as used in Fig.~\ref{dedx_c_b}, which gives rise to a momentum of $1.5$ GeV for charm quarks and $5.0$ GeV for bottom quarks.
\vspace{-0.2cm}
\subsubsection{Energy gain of heavy quark}
\vspace{-0.2cm}
In subsec~\ref{el_lrt}, a semiclassical approach was adopted to compute the collisional energy loss of a heavy quarkresulting from the medium's polarization effects~\cite{Thoma:1990fm}.  It is assumed that the energy lost by the particle per unit time or length is negligible compared to the particle's energy itself , allowing for the particle's velocity change during motion to be disregarded, i.e, the particle moves in a straight-line trajectory. The particle's energy loss is determined by the work of the retarding forces exerted on it in the plasma by the chromoelectric field generated by the particle's motion. Hence, the particle's energy loss per unit time can be derived~\cite{Thoma:1990fm,Mrowczynski:1991da,Chakraborty:2006db} from Eq.~\eqref{el3} as
\begin{equation}
-\frac{dE}{dt} = Q^a \, \bm{\vec v} \cdot {\rm {Re}}\, \vec{E}^a_{\rm{ind}}
(\bm{\vec r}=\bm{\vec v}t,t) \, \, ,  \label{el31}
\end{equation}
where  $dx=v \, dt$ and the field is evaluated at the particle's location. This formula for energy loss does not consider field fluctuations in the plasma or particle recoil in collisions. To take into  these fluctuations it is necessary to replace Eq.~\eqref{el31} with~\cite{Chakraborty:2006db},
\begin{equation}
-\frac{dE}{dt} =\left\langle Q^a {\bm{\vec v}}(t)\cdot {\rm {Re}}\, \vec{\cal E}^a_{\rm{ind}, t}
(\bm{\vec r=\vec x}=\bm{\vec v}t,t) \right \rangle  \, \, ,  \label{el32}
\end{equation}
$\left\langle\cdots\right\rangle$ denotes the statistical averaging operation. It is noteworthy that two types of averaging procedures can be conducted~\cite{Chakraborty:2006db}: i) an ensemble average {\it w.r.t} the equilibrium density matrix 
ii) a time average over random fluctuations in plasma. These two operations are commuting and only after both of them are 
performed the average quantity takes up a smooth value~\cite{Kalman61}. After performing both averages following Ref.~\cite{Chakraborty:2006db}, the energy loss coming form the fluctuations can be summarized as,
\begin{eqnarray}
-\left.\frac{dE}{dt}\right|_{\rm fl} &=& 
\frac{C_F\alpha_s}{8\pi^2 Ev^3}\int_0^{q_{\max}v}d\omega\,
\coth{\frac{\beta\omega}{2}} {F\left(\omega,q={\om\over v}\right)}
+ 
\frac{C_F\alpha_s}{8\pi^2 Ev}\int_0^{q_{\max}}dq\,q\int_0^{qv}
d\omega\,\coth{\frac{\beta\omega}{2}}G\left(\omega,q \right)\,, 
\label{el33}
\end{eqnarray} 
where the imaginary part of the longitudinal dielectric function is related to  $F\left(\omega,q\right) = 8\pi \omega^2 {\rm Im}\,
\epsilon_L/|\epsilon_L|^2$  whereas the imaginary part of the transverse dielectric function is related to $G\left(\omega,q\right) = 16 {\rm Im}\,
\epsilon_T/|\epsilon_T - q^2/\omega^2|^2$ and the upper cut-off $q_{\max}$ one \cite{Adil:2006ei} gets, 
\begin{equation}
q_{\max} =\mbox{min}\,\left\{E,\frac{2k\left(E+p\right)}
{\sqrt{M^2+2k\left(E+p\right)}}\right\}\,, \label{el34}
\end{equation}
where $k \sim T$ is the typical momentum of the thermal partons of the QGP.  Using \eqref{qcd21} and \eqref{qcd22} in \eqref{el67}, one obtains 
the dielectric functions in QCD medium  as
\begin{subequations}
\begin{align}
\epsilon_L(\om,q)&=1+\frac{(m_D^{\rm g})^2}{q^2}\left [1-\frac{\om}{2q} \ln \left| \frac{\om+q}{\om-q}\right|\right ] +i\pi \frac{\om (m_D^{\rm g})^2}{2q^3} \Theta(q^2-\om^2) \, ,
 \label{el34a} \\
\epsilon_T(\om,q)&=1- \frac{(m_D^{\rm g})^2}{2q^2}\left [1+\frac{q^2-\om^2}{2\om q}  \ln \left| \frac{\om+q}{\om-q}\right|\right ]  
+ i\pi \frac{\om (m_D^{\rm g})^2(q^2-\om^2)}{4\om q^3} \Theta(q^2-\om^2) \, .\label{el34b} 
\end{align}
\end{subequations}
The equation~Eq.~\eqref{el33} denotes energy gain, as it represents the average energy (per unit time) absorbed by a propagating particle from the heat bath. Physically, this phenomenon arises from gluon absorption. Previous studies~\cite{Wang:2001cs} have also demonstrated that thermal absorption of gluons leads to a reduction in radiative energy loss. We arrive at a somewhat similar conclusion as there, albeit in a different context.
\begin{figure}[htb]
\begin{center}
\includegraphics[scale=0.7]{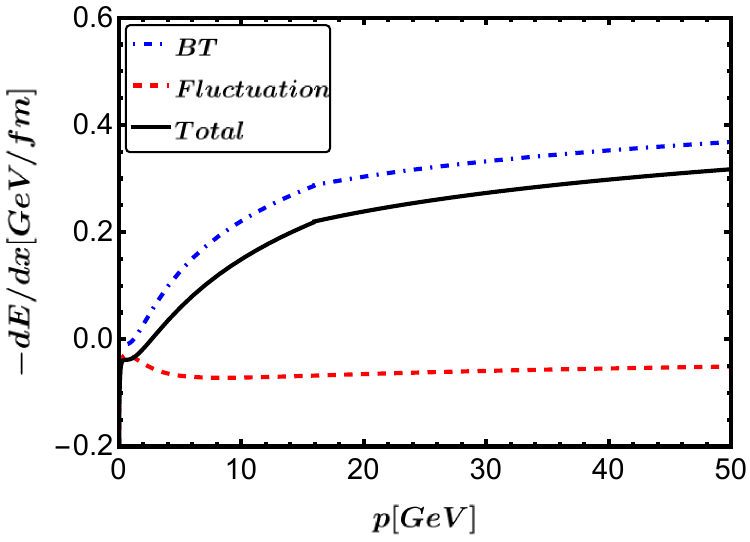}
\includegraphics[scale=0.7]{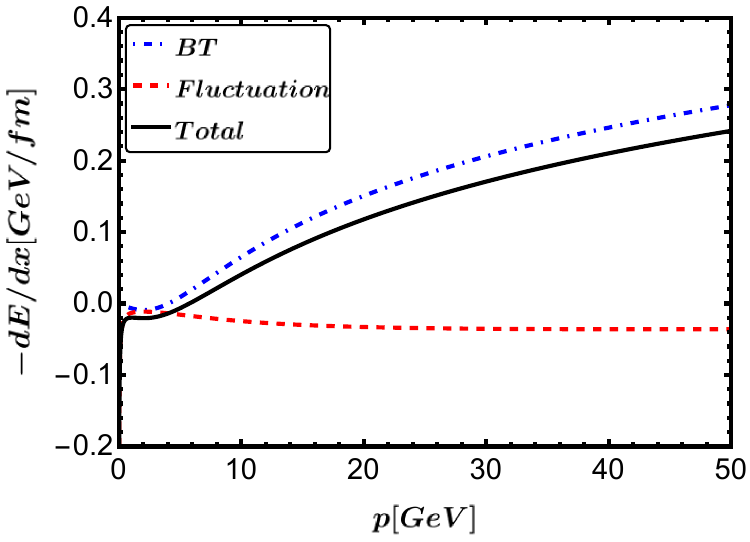}
\end{center}
\vspace{-0.5cm}
\caption{The differential energy loss for charm(left panel) and bottom (right panel) is as a function of their momentum.  
The red dashed line represents the energy gain as given~\eqref{el33} due to fluctuation whereas the total energy loss after adding energy gain is represented by solid green curve. The energy loss results~\cite{Braaten:1991we} to leading order in $g$ without fluctuation as given in~\eqref{el25} and \eqref{el29} is represented by (blue dashed curve). We have used $T=0.25$ GeV, $\alpha_s=0.2$, charm quark mass $=1.5$ GeV and bottom quark mass $=5.0$ GeV.}
\label{dedx_gain_c_b}
\end{figure}
In Fig.~\ref{dedx_gain_c_b} we  have displayed the relative collisional energy loss of a charm and bottom quark where the effect of field fluctuations is taken into account.  It is apparent that the effect of these fluctuations on heavy quark energy loss is significant at low momenta. For momenta ($4-20$) GeV the fluctuation effect reduces the collisional  loss by $(17-39)\%$ for charm and $(12-31)\%$ for bottom. At higher momenta, as it will be relevant for LHC, the relative importance of the fluctuation gain to the collisional loss decreases gradually.

We further note that the fluctuation gain~\cite{Chakraborty:2006db} diverges linearly as $v \rightarrow 0$. This can be understood by looking at the leading log part in $v \rightarrow 0$ limit
\begin{eqnarray}
-\left.\frac{dE}{dx}\right|_{\rm fl}^{\rm leading-log} &=& - \,
2\pi C_F\alpha_s^2\left(1+\frac{N_f}{6}\right)\frac{T^3}{Ev^2}
\ln{\frac{1+v}{1-v}}  \, \ln{\frac{k_{\max}}{k_{\min}}}\, 
\label{el35}
\end{eqnarray}
where $k_{\min} = m^{\rm g}_D $ is the Debye mass. The divergence here is kinematic as $\left.\frac{dE}{dt}\right|_{\rm fl}$ is finite for $v=0$ and therefore $\left.\frac{dE}{dx}\right|_{\rm fl} = \frac{1}{v} \left.\frac{dE}{dt}\right|_{\rm fl}$ diverges for $v \rightarrow 0$.
\subsubsection{Light quark energy loss}
\label{el_light}
\vspace{-0.2cm}
The energy loss for a high energy partons was also  earlier obtained by TG~\cite{Thoma:1990fm} from  Eq.~\eqref{el5} following the  same procedure as obtained for heavy quark. Using the cut part of the gluon spectral functions  from Eq.~\eqref{el17d} and \eqref{el17e} in replacing $m_D^\gamma$ by  
$m_D^{\rm g}$ and again using them in Eq.~\eqref{el17a} and~\eqref{el17b} which are  then to be used in \eqref{el5}.  Now one has to perform two integrations; one is over the exchanged gluon momentum $q$ and the other one  is over angle $\angle (\bm{\vec v \cdot \vec q})$. 
The $q$-integral  for light quark with momentum $p$ can be performed analytically which is proportional to $\ln (q_{\max}/q_{\min})$. This expression is infrared finite due to the screening factor $q_{\min}=m_D^{\rm g}=\sqrt{3}m_{\rm{th}}^{\rm g}$ in the denominator and logarithmically ultraviolet divergent for $q_{\max}=p$ for light quark, $E=p$. One has to perform the angle integration with $v=1$ for light quarks.
\begin{figure}
\begin{center}
	\vspace{-0.3cm}
\includegraphics[height=6cm, width=8.5cm]{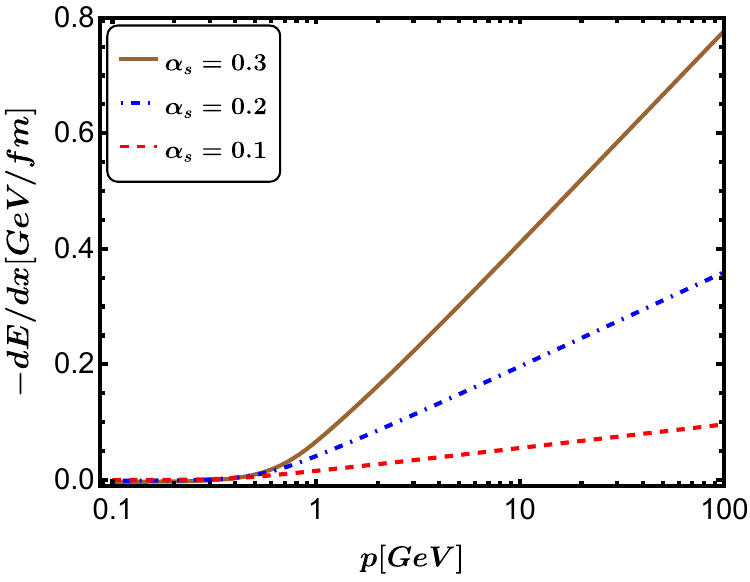}
\end{center}
\vspace{-0.5cm}
\caption{The differential energy loss for light quarks as function momenta for $T=0.25$GeV and three values of $\alpha_s=0.1,\ ,0.2, \, {\rm{and}} \, 0.3$.}
\label{dedx_light_q}
\end{figure}
 In Fig.~\ref{dedx_light_q}, the differential energy loss for a high energetic light quark is displayed as function of momentum $p$ for $T=0.25$ and three different values of $\alpha_s$ appropriate for QGP produced in ultra-relativistic heavy-ion collisions. For a given $T$ and $\alpha_s$, $dE/dx$ increases with increase of quark momentum $p$. The $dE/dx$ also increases with  $\alpha_s$ for a given momentum  $p$.
 
 However, we note that the result in Ref.~\cite{Thoma:1990fm} is incomplete since the high temperature approximation used for the dielectric functions is valid only for energies and momenta of the exchanged gluon $\om,\, q \ll T$ as shown in Eq.~\eqref{el17a} and~\eqref{el17b}. It also  suffered from an ambiguity associated with the choice of upper limit $q_{\max}$.

The energy loss of a light quark or gluon can be calculated~\cite{Thoma:1991jum} similarly as the one of a heavy quark for $ E\gg M^2/T$. The only difference is that one introduces another separation scale $T\ll {\tilde q}\ll E$ apart from the separation scale $q_c$ between hard ($q\sim T$) momenta  and soft ($q\sim gT$) momenta transfer.  
Following one loop calculation~\cite{Thoma:1991jum}, of quark self-energy with soft gluon propagator, the soft contribution ($q<q_c$) is obtained as
\bea
- \left.\frac{dE}{dx}\right |_{\rm{soft}}^{qq+qg}&=&\frac{8}{3}\, \pi \alpha_s^2T^2 \left (1+\frac{N_f}{6}\right )
\left[\ln \frac{q_c}{m_{\rm{th}}^{\rm g}} - 0.843 \right ]\, , \label{el36} 
\eea
which agrees with the heavy quark expression in \eqref{el26} in the limit $v =1$. The constant $-0.843$ comes from a numerical integration in \eqref{el5} with an upper limit 
$q_c\gg gT$.

\begin{figure}
	\begin{center}
	\includegraphics[width=8.cm,height=6.5cm]{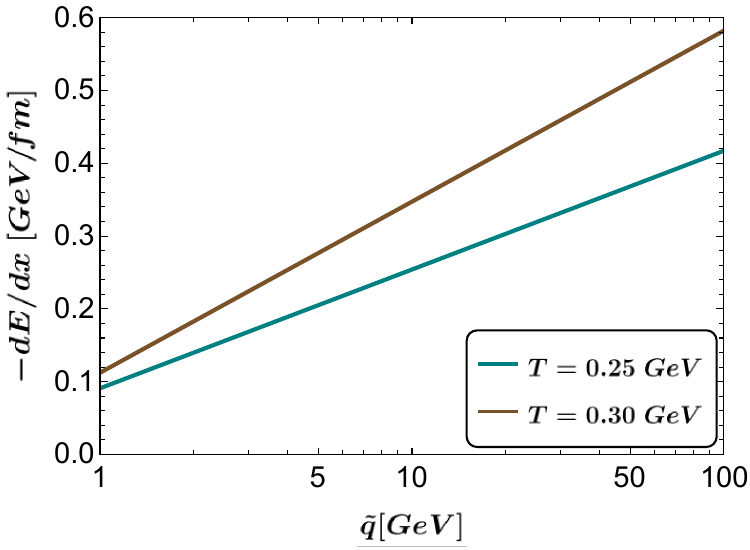}
	\caption{Displays light quark  energy loss as a function of cut-off parameter $\tilde q$ for two temperatures:  $T=0.25$ GeV (green curve) and $0.3$ GeV (brown curve) with	  $\alpha_s=0.2$ and $N_f=2$ light quark flavours.}
	\label{dedx_l_q}
	\end{center}
\end{figure}

For the hard contribution $q_c<q<{\tilde q}$, one has to consider the tree level diagrams for $qq\rightarrow qq$ and $qg\rightarrow qg$ scatterings. Following Ref.~\cite{Thoma:1991jum} one obtains hard contributions as
\begin{subequations}
\begin{align}
- \left.\frac{dE}{dx}\right |_{\rm{hard}}^{qq}&=\frac{8}{3}\, \pi \alpha_s^2T^2 \, \frac{N_f}{6}
\left[\ln \frac{\sqrt{{\tilde q}T}}{q_c} + 1.106 \right ]\, , \label{el36a} \\
- \left.\frac{dE}{dx}\right |_{\rm{hard}}^{qg}&=\frac{8}{3}\, \pi \alpha_s^2T^2 
\left[\ln \frac{\sqrt{{\tilde q}T}}{q_c} +0.760\right ]\, . \label{el36b} 
\end{align}
\end{subequations}
We note that in a single collision, the light quark looses more energy than a heavy quark~\cite{Thoma:1991jum}, so  the ultra-hard contribution ($q>\tilde q$) is not negligible. However, for $q>\tilde q$ the computation of $dE/dx$ is much more difficult because one cannot relax some approximation and more importantly  $u$- and $s$-channel diagrams become important and even singular when the exchanged gluon momentum $q$ approaches the quark jet momentum $p$. On the other hand, momentum transfer of the order of $p$ does not contribute to the energy loss of a jet, because in this case the energy is collinearly transferred to a particle of the QGP. This means that the energy remains inside the jet and is not dispersed into the QGP. So, this process does not contribute to quark jet energy loss~\cite{Thoma:1991jum}.
Now adding the soft contribution from Eq.~\eqref{el36} and hard contributions from Eqs.~\eqref{el36a} and~\eqref{el36b}, one gets the complete leading order result~\cite{Thoma:1991jum} for light quark as
\bea
- \frac{dE}{dx} =\frac{8\pi \alpha_s^2T^2}{3} \left (1+\frac{N_f}{6}\right ) 
\ln \left[ 2^{N_f/2(6+N_f)} 0.920  \frac{\sqrt{{\tilde q}T}}{m_{\rm{th}}^{\rm g}} \right ]\, , \label{el37} 
\eea
which agrees with heavy quark result given in \eqref{el29} in the limit $v\rightarrow 1$ and replacing $E$ by $\tilde q$. 
However, it is important to observe that although the separation scale $q_c$  cancels out, the scale $\tilde q$ persists. 
In Fig.~\ref{dedx_l_q}, the energy loss of a light quark is illustrated as a function of the cut-off $\tilde q$ for $\alpha_s=0.2$, $N_f=2$ and, two 
temperatures: $T=0.25$ GeV (green curve) and $0.3$ GeV (brown curve). Typical jet energies at RHIC and LHC are the order 30-40 GeV. 
Hence, $\tilde q=15$ GeV is a conservative choice. (Certainly,  $\tilde q= E/2$ is an upper limit, as in this scenario the scattered QGP particle is more energetic than the outgoing external quark). Then, one can find $\left. dE/dx\right |_{\tilde q=15 {\rm {GeV}}} = 0.23$ GeV/fm for $T=0.25$ GeV and $\left. dE/dx\right |_{\tilde q=15 {\rm {GeV}}} = 0.33$ GeV/fm  for $T=0.3$ GeV. 
Hence, it can be concluded that the collisional energy loss of a light quark in a QGP is pertinent to jet quenching~\cite{Mustafa:2003vh,Mustafa:2004dr}. Additionally, it is noteworthy that the energy loss of a high-energy gluon can be obtained by multiplying Eq.~\eqref{el37} with $9/4$.

\subsection{Radiative Energy Loss} 
\label{el_rad}
\vspace{-0.2cm}
The radiative energy loss is caused due to deceleration of a charge particle in the medium that emits a gluon known as bremsstrahlung process. The first estimate of gluon emission from a quark-quark scattering  ($qq\rightarrow qqg$) was given by Gunion-Bertsch (GB) formula~\cite{Gunion:1981qs} where the inelastic processes (bremsstrahlung) by factorising elastic process ($qq\rightarrow qq$) with a gluon emission($q\rightarrow qg$). Later it was explicitly used in Refs.~\cite{Srivastava:1996rz,Srivastava:1996qd,GolamMustafa:1997id} for chemical equlibration to derive the soft gluon emission from gluon-gluon scattering ($gg\rightarrow ggg$). Further, some important corrections to the GB formula for soft gluon were obtained in Refs.~\cite{Das:2010hs,Abir:2010kc} and also gluon emission from heavy quarks~\cite{Dokshitzer:1991fd,Dokshitzer:2001zm,Abir:2011jb}. The GB formula has also been used  for radiative energy loss of heavy quarks in QGP~\cite{Dokshitzer:2001zm,Mustafa:1997pm,Abir:2012pu}. The energy loss per collisions is not caused by the exchanged energy of a elastic scattering but by the energy of the emitted/radiated gluon. The GB formula does not take into account the medium effects, which are discussed below:
 
Here the  $2\rightarrow3$ process in BPT suffers from  infrared divergence due to the exchange massless gluon and quark.  This is  regulated through the introduction of soft  exchanged gluon propagator as done in Fig.~\ref{brems} and Fig.~\ref{aws} coming from two-loop diagrams for photon production. This diagrams are expected to be order of $e^2g^4$ due to additional 3-point QCD vertex and also multiplied by a factor $T^2/M_\infty^2$. Since this diagrams have strong collinear singularity, there will be collinear enhancement $T^2/M_\infty^2\sim 1/g^2\sim 1/\alpha_s$ is obtained so that they contribute in leading order ${\cal O}(e^2g^2)$  in the strong coupling. The order of $\alpha_s$ is reduced by one order in $\alpha_s$.  The radiative energy loss for partons in QCD becomes proportional to $\alpha_s^2$ instead of expected $\alpha_s^3$ from $2\rightarrow 3$ processes. As a consequence the radiative and collisional energy loss become of same order in $\alpha_s$ as ${\cal O}(\alpha_s^2)$.
  
 Beside the medium effect that regulates collinear singularity,  there is also  another medium effect that plays an important role in radiative energy loss which is known as LPM effect~\cite{Landau:1953ivy,Landau:1965ksp,Migdal:1956tc} as discussed in section~\ref{em_chap}. Now, we have seen the photon emission  even at leading order,  is sensitive to processes involving multiple scatterings occurring during the emission process. Diagrammatically, it manifests as the presence of interference terms involving multiple collisions, which parametrically are equally as important as those of Fig.~\ref{brems} and Fig.~\ref{aws}.  This is also true for gluon emission.
 
There are several radiative energy loss  approaches have been developed based on how the medium is modelled: 
The  approach developed by Armesto, Salgado and Wiedemann (ASW)  and their
collaborators~\cite{Armesto:2003jh, Wiedemann:2000ez,Wiedemann:2000za,Wiedemann:2000tf} 
models the medium as a series of Debye-screened, heavy, coloured scattering centres. In this approach, referred to as the 
ASW scheme, the hard parton radiates a virtual gluon which is then progressively brought on shell by a large number of soft scatterings off these heavy centres. Another approach developed by Gyulassy, Levai, Vitev (GLV) and their 
collaborators~\cite{Gyulassy:1999zd,Gyulassy:2000fs,Gyulassy:2000er,Gyulassy:2001nm,Djordjevic:2003zk,Djordjevic:2004nq},  known as GLV scheme, considers the same medium as ASW, however both the hard parton and the 
radiated gluon undergo a few but hard interactions with the centres leading to the emission of the gluon. Yet another approach by Arnold, Moore and Yaffe  (AMY) and collaborators~\cite{Arnold:2001ba,Arnold:2001ms,Arnold:2002ja,Jeon:2003gi} developed a scheme, known as AMY scheme. The medium is assumed to be composed of quark gluon quasi-particles with dispersion relations and interactions given by the HTL effective theory. Consequently, all quasi-particles in the medium have thermal masses $\sim gT$, with their scattering predominantly governed by soft scattering. The hard jet is assumed to have a virtuality scale comparable to the Debye mass or thermal mass but with energy $E\gg T$. Following this, one identifies and resums the collinear enhanced contributions emanating from the scattering and induced radiation off 
the hard parton. The entire calculation is carried out at the scale of the temperature $T$ which is assumed to be large ($T\rightarrow \infty$) so that the effective coupling $g (T )$ is small. After the calculation of the single gluon emission kernel, the AMY scheme uses rate equations to incorporate 
multiple emissions, while the GLV and the ASW use a Poisson emission Ansatz.

The AMY formalism is different from the other approaches as it naturally incorporates feedback from the medium which is missing from all other approaches. In Refs.~\cite{Arnold:2002ja} the photon emission rate has been extended to gluon radiation from partons. A consistent method was used  in deriving an integral equation, 
similar in form to a linearised kinetic equation, whose solution determines the rate of bremsstrahlung. The rates for a quark to decay into a quark and gluon,  a gluon to decay into a quark antiquark pair etc., are now used to set up a Fokker-Planck equation which describes the change in the distribution of hard partons with time~\cite{Jeon:2003gi}. There are two sets of equations which describe the change of the distribution of the sum of quarks with antiquarks, and gluons.  Then, the radiative energy loss of hard partons~\cite{Jeon:2003gi} and the transport coefficient~\cite{Majumder:2010qh} $\hat q$ were obtained in ${\cal O}(\alpha_s^2$). 
\section{Summary and outlook}\label{sec:sum}
	 \vspace{-0.2cm}
In this review we have  introduced the most important aspects of modern thermal perturbation theory based on HTL approximation including its theoretical foundations and discussed some applications within QED and QCD plasma. For compactness, we have restricted our discussion to only thermal equilibrium. In particular, we discussed both bulk thermodynamic and real-time observables, and cover the realms of high temperatures and or density, relevant for heavy-ion physics. We have started the review with an introduction that discusses the need for thermal field theory, the various scales in thermal system,  the problems of bare perturbation theory based on hard scale and need for development of reorganising the perturbation theory that carefully takes into account the soft scale of a given theory.

In section~\ref{ftft}  we briefly discussed basics of field theory with thermal background both in imaginary time and real time formalism. We presented how the frequency sum in imaginary time are performed using contour integration and Saclay method. We also discussed the relation between functional integration and the partition function. How the general structure of two point functions for a given theory get modified in thermal background have been presented in brief.  Then we discuss the subtleties at finite temperature and the scale separations. We also outlined the shortcomings with some examples of bare perturbation theory  at finite temperature. 

In section~\ref{eft} the effective field theories at finite temperature  have been discussed in details how one can avoid the problems in bare perturbation theory. We started with the dimensional reduction technique for both scalar and gauge theories. 
The dimensional reduction  has been developed that can systematically unravel the contributions from the various momentum scales (hard $\sim T$, soft $\sim gT$ and  ultra-soft $\sim g^2T$). The method is based on the construction of effective field theories that reproduce static observables at successively longer distance scales $R \gg 1/T$. According to this idea, the static properties of a ($3 + 1$)-dimensional field theory at high temperature can be expressed in terms of an effective field theory in three space dimensions. We note that the DR effective theories are very powerful tools for calculations of equilibrium thermodynamic quantities at high temperature and chemical potential 
within imaginary time formalism but do not work at low temperature. Then we discussed the hard thermal loop approximation that states the calculations done in bare perturbation theory
were incomplete in order of the coupling constant as certain classes of diagrams were not taken into account. These diagrams are of higher order in the loop expansion which contribute
to the same order in the coupling constant as the one loop diagram. Then we identified the hard thermal loops in QCD and laid down the HTL approximations. It has been shown that the 
amplitude of higher order loops can be written as $(g^2T^2/P^2)\times {\rm {Tree \,\, Level\,\, Amplitude}}$, where $P$ is the external momentum. 
If the external momentum is hard ($P\sim T$) then  the higher order amplitude is suppressed by $g^2$ of the $``$Tree Level Amplitude''. But when the external momentum is 
soft ($P\sim gT$), then the amplitude becomes equivalent to $``$Tree Level Amplitude''. So, it is evident that diagrams of higher order in loop expansion contribute to same order in 
coupling as the one-loop by distinguishing the hard scale ($\sim T$) arising from loop momenta and soft scale ($\sim gT$) from external momenta. There are thermal corrections from all 
orders of perturbation theory. This implies that one needs to take into those diagrams if the physical quantity is sensitive to the soft (electric) scale. The effective theory, developed around 
hard thermal loops, resums such diagrams. Then we discussed main features of the HTL resummation based on which  an effective field theory can be developed.
We also discussed another approach  HTL resummation which is approximately self-consistent.  It is based on self-consistent approximations using the skeleton 
representation of the thermodynamic functional, suitable  only for computing various thermodynamic quantities. It reorganises the perturbation theory by considering the so-called 
$2$-loop $\Phi$-derivable approximation, for which it turns out that the first derivatives of 
the thermodynamic potential, the entropy and the quark densities, take a rather simple, effectively one-loop form, but in terms of fully dressed two point functions. But in gauge theories,
the exact two point functions are too complicated to be solved exactly (even numerically). Therefore, it was proposed in the literature to use a gauge independent but only approximately
 self-consistent two point functions as obtained from HTL resummation. 

In section~\ref{scalar_htl} the HTL resummation of bubble diagrams of self-energy has been obtained in scalar theory with $\Phi^4$ interaction which shows that it is 
an expansion of the power of $g$ instead of the power of $g^2$. Then we discussed the effective $N$-point functions such as propagator and 4-point function within HTL approximation.  
The effective propagator receives a mass correction as $\big(P^2-(m_{\rm{th}}^s)^2\big)^{-1}$, where the  $m_{\rm{th}}^s = gT/\sqrt{24}$. If the momenta of the propagator
 is hard ($\sim T$), clearly the thermal mass is a perturbation and can be dropped. However, if the momenta of the propagator is  soft ($\sim gT$), the thermal mass 
is as large as the bare  propagator and cannot be omitted. Thus by taking the thermal mass into account, one is resumming an infinite set of diagrams in scalar self-energy 
from all orders of perturbation theory. On the other hand the one loop correction to the scalar $4$-point function $\Gamma^{(4)}$ at high temperature is proportional to $g^4\log (T/p)$. 
It is therefore always down by $g^4 \log(1/g)$ as the external momentum $p$ is soft in HTL approximation, and one can conclude that it is sufficient to use a bare vertex. 
Based on the $N$-point functions we write down the HTL improved scalar Lagrangian with a $\Phi^4$ interaction.
 
In section~\ref{qed_htl}, we address the HTL resummation of N-point functions in QED. Specifically, we derive HTL-resummed 2-point functions (electron and photon self-energies and their propagators), 3-point functions (electron-photon vertex), and 4-point functions (two electron-two photon vertex), all of which are interconnected via Ward identities. Both electrons and photons acquire thermal mass, exhibiting collective behaviors in the presence of a heat bath. Additionally, alongside the quasi-electron, a long-wavelength plasmino mode emerges, attributed to broken Lorentz invariance due to the heat bath. Landau damping occurs for electrons due to the complex nature of their self-energy, and similarly for photons, which possess complex self-energies. We analytically solve the dispersion relations for both electron and photon in two extreme momentum limits: low and high. Furthermore, we obtain the spectral representations of their respective propagators, valuable for  application purpose. The infrared limit of the longitudinal photon self-energy yields the Debye electric screening mass, serving as an infrared regulator at the static electric scale($\sim eT$). However, no screening for magnetic fields is observed, as the 1-loop transverse photon self-energy in the leading-order HTL approximation vanishes in the infrared limit, providing no magnetic screening mass for photons. Finally, leveraging insights from N-point functions, we formulate the HTL improved QED Lagrangian for quasiparticles.
 
 In section~\ref{qcd_htl} we presented the HTL resummation of QCD $N$-point functions.  We begin by noting that the discussion on QED in section~\ref{qed_htl} provided a good starting point for QCD, since these two theories follow the similar formalism except self-interaction of gluons. We know that QCD is a non-Abelian $SU(3)$ gauge theory whereas 
 QED is a $U(1)$ gauge theory. Thus the generalisation of  QED results to some of the QCD results would mostly involve group-theoretical factors. By replacing $e^2\Rightarrow g^2(C_A+N_f/2)$ in photon self-energy one obtains gluon-self energy and thus propagators. The quark self-energy  and quark-gluon three vertex 
 can be obtained from QED by replacing $e^2\Rightarrow g^2C_F$. We explicitly obtain 3-gluon vertex, 4-gluon vertex and two quark-two gluon vertex. Since the two-point functions have same structure as QED, we learn about the collective excitations in a QCD plasma from the acquired knowledge of QED plasma excitations. The QCD plasma provides the  electric screening mass to regulate infrared divergence but no magnetic screening akin to QED. In general gauge theories in HTL resummation do not provide screening of magnetic field.  Based on the $N$-point functions we write down the HTL improved QCD Lagrangian for quasiparticles.
 
In section~\ref{ipt_chap}, we discussed the improved perturbation theory for scalar, QED and QCD. Based on the HTL resummation a new perturbative scheme for massless scalar field theory with a $\Phi^4$-interaction, known as $``$screened perturbation theory'' inspired in part by variational perturbation theory, is discussed  for the evaluation of  thermodynamic quantities in field theories at finite temperature by taking into account exactly the phenomenon of screening or thermal mass  in scalar propagator. For gauge theories, however, it is not possible to use a  gluon/photon mass.  As a result, a gauge-invariant generalisation of screened perturbation theory called $``$hard-thermal-loop perturbation theory (HTLpt)'' was developed to study the thermodynamic quantities. The HTLpt framework allows for systematic analytic reorganisation of perturbative series based on the HTL effective Lagrangian. It is  very useful to calculate both static and dynamical quantities.
 
 We used HTLpt to obtain electromagnetic particle production (virtual and real photons) in section~\ref{em_chap}. In subsec~\ref{dilep_chap} we discuss the virtual photon vis-a-vis dilepton production. We calculate the hard dilepton production from 1-loop BPT that leads to quark-antiquark annihilation ($q\bar q \rightarrow \gamma^*\rightarrow l^+l^-$), known as Born rate. We note that the production rate for soft dilepton could differ by orders of magnitude 
 from the naive prediction from quark-antiquark annihilation. Using  HTL resurnmation techniques  we discussed  a systematic calculation of the soft production rate in  one-loop HTLpt where HTL effective propagators and vertices have been used. We note that the quark spectral function has both pole and cut contribution. 
 So the terms with two powers of it will lead to three types of contributions: pole-pole, pole-cut and cut-cut. The pole-pole part shows dramatic structure (van-Hove peaks)  due to van-Hove singularity
 caused by vanishing group velocity $\frac{d\om}{dk}=0$. This is understood from the nature of the dispersion relation for plasmino mode. In addition to the pole-pole contributions, there are also contributions from pole-cut  and cut-cut  due to the 
 cuts in the effective quark propagator. Because the spectral density along the cut is a smooth function, these contributions do not produce any dramatic structure. At low energies $\om$, the pole-cut term grows like $1/\om^2$ and the cut-cut term grows like $1/\om^4$, and they completely overwhelm the structure due to the pole-pole terms. However, these corrections are not sufficient, and two-loop diagrams within the HTL perturbation scheme contribute to the same order and are even larger than the one-loop results. The 2-loop rate comprises  the one-loop HTLpt result, the bremsstrahlung contribution from two-loop HTLpt  and  finally the contribution of the Compton and annihilation processes. The 2-loop rate dominates in the perturbative regime ($g \leq1$) over the Born-term for low mass domain,  $\om/m^{\rm q}_{\rm{th}} =M/m^{\rm q}_{\rm{th}} \leq 2$. However, the van Hove singularities contained in one-loop do not appear as they are washed out due to the leading logarithm approximation within the two-loop HTLpt. We also note that the 2-loop contribution is of the same order in the Born-term for soft $\om \sim gT$ . Also the $\alpha_s$ correction to the Born rate has been obtained in 1-loop and 2-loop HTLpt. In one loop HTLpt the thermal dilepton rate has been calculated from a 1-loop graph, with one hard bare quark propagator with asymptotic mass, and one HTL quark propagator which is taken to be asymptotically hard ($M_\infty=\sqrt{2}m_{\rm{th}}^q$). The imaginary part of one loop HTLpt diagram corresponds to Compton ($qg\rightarrow q\gamma^*$) and annihilation ($q{\bar q}\rightarrow g\gamma^*$) processes. Imaginary part of the 2-loop HTLpt leads to two processes: bremsstrahlung and off-shell annihilation. The dilepton rate was calculated from this two processes including multiple scattering  and interference terms  all of which are contained in the LPM resummation. Because of a strong collinear singularity in this two processes there is collinear enhancement, powers of $T^2/M_\infty^2\sim 1/g^2\sim 1/\alpha_s$ are generated so that they contribute at the leading order in the strong coupling. Higher order diagrams with a ladder topology involve the same enhancement mechanism and therefore they also contribute at leading order. Then the complete leading order rate  in strong coupling  is  sum of Born, Compton, annihilation, bremsstrahlung and off-shell annihilation. We have also compared the HTLpt result with the nonperturbative calculations: LQCD and nonperturbative model with Gribov action.

 In subsec~\ref{photon_chap} we have discussed the real photon production rate. We first discuss the hard photon rate from 2-loop BPT (with hard propagators and bare vertices) and imaginary part of which lead  to Compton ($qg\rightarrow q\gamma$) and annihilation ($q{\bar q}\rightarrow g\gamma$) processes. The photon rate for these processes was found to be logarithmically infrared divergent when soft momentum transfer cut-off goes to zero. In some calculation the soft cut-off was replaced  by thermal quark mass $m_{\rm{th}}^q$. This means that even the production of energetic photons is sensitive to many-body effects of the QGP, since the exchange of soft quarks plays an important role in the production
  mechanism. A systematic treatment of many-body effects was provided by the HTL resummation technique developed by Braaten and Pisarski which was independently used 
  by Kapusta et al and Baier et al to regulate the soft photon production from QGP using a cut-off between soft and hard momentum transfer. When both the hard and soft 
  contributions are added the cut-off gets cancelled. We note further note that  the contributions beyond the leading logarithm become important and can even dominate the 
  leading order ${\cal O}(\alpha \alpha_s)$ coming from 1-loop HTLpt. Such contributions come from higher order diagrams, i.e., 2-loop diagrams with soft exchanged gluons 
  in the HTLpt and imaginary part of which leads to  processes like bremsstrahlung and inelastic pair annihilation. 
  Usually one expects that the imaginary part of 2-loop diagrams should be of the order $e^2g^4$ due to additional 3-point QCD vertices and also multiplied by a factor 
  $T^2/M_\infty^2$. However, due to a strong collinear singularity in the processes appearing from 2-loop HTLpt there is collinear enhancement, powers of  
  $T^2/M_\infty^2\sim 1/g^2\sim 1/\alpha_s$ are generated so that they contribute at the leading order ${\cal O}(\alpha\alpha_s)$. We also added a discussion on photon production from higher than 2-loop. Through power counting it is shown that the 3-loop diagram is proportional to the 2-loop diagram times a 
  factor $g^2T/m_s$, where $m_s$ is the infrared cut-off for the additional exchanged gluon. For longitudinal gluons the $m_s$ is provided by the static electric screening  of order $gT$. On the hand, for a transverse gluon this cut-off is provided by the non-perturbative magnetic screening mass of the order $m_s = g^2T$. Hence, the 3-loop 
  contribution becomes the same order as the 2-loop. This means that the infinite number of higher-order diagrams contribute to the same order in $\alpha\alpha_s$ as the  2-loop HTL diagram. This argument is basically in the same line that has been utilised by Linde  to indicate that the BPT in QCD breaks down. We also note that the static magnetic fields are not screened at leading order of HTLpt since 1-loop HTL  transverse gluon self-energy  in all gauges vanishes in the infrared limit $\om\rightarrow 0$.
 
 In section~\ref{meson}, we analyse the structure of temporal correlation functions and their spectral functions within the context of HTLpt at zero spatial momentum. The spectral functions, derived from the analysis of temporal correlators, are closely related to quark-antiquark annihilation processes in the quark-gluon plasma. In the vector channel, this is related to the dilepton production at high temperatures, which has been studied in the HTL-approximation and discussed in details in subsec~\ref{bpy_soft}. The temperature dependence of the pseudoscalar correlator is related to the chiral condensate. Thermal effects on this, as well as the pseudoscalar masses and dispersion relations, impact the appearance or suppression of a disoriented chiral condensate, potentially leading to observable effects in relativistic heavy-ion collisions. Furthermore, we discussed the relationship between the quark number susceptibility (QNS) and the temporal Euclidean correlation function. The leading-order QNS in HTLpt serves as a response to conserved density fluctuation, exhibiting a similar trend to available lattice data with improved lattice actions in the literature, although there are deviations to some extent. The same HTL QNS is used to compute the temporal part of the Euclidian correlation in the vector current, showing good agreement with improved lattice gauge theory within the quenched approximation on lattices up to size $128^3 \times 48$ for a quark mass $\sim 0.1T$. Notably, the quantitative difference between the recent quenched approximation data  and the full QCD data with improved (asqtad) lattice action for QNS is within $5\%$ in the temperature range $T_c\leq T\leq  3T_c$. Leaving aside the difference in ingredients in various lattice calculations, one can expect that the HTLpt and lattice calculations are in close proximity for quantities associated with the conserved density fluctuation. 

In section~\ref{chapter:thermodynamics}, we have discussed various thermodynamic quantities in 1-loop, 2-loop and 3-loop order in HTLpt relevant for QGP at high temperature and density. At extremely high temperatures, a significant correction occurs while going from LO to NLO pressure in HTLpt. It's been observed that due to the logarithmic evolution of the coupling,  a very high temperatures is necessary for the LO and NLO predictions to overlap, attributable to overcounting issues at LO. This overcounting leads to an order-$g^2$ perturbative coefficient twice as large as expected, a problem rectified at NLO, albeit resulting in a reasonably large correction (approximately $5\%)$ at the indicated temperatures. The NLO result from two-loop HTLpt result deviates from the pQCD result through order $\alpha_s^3\ln\alpha_s$ at low temperatures, indicating a modest improvement over pQCD in terms of convergence and sensitivity to the renormalisation scale. Both LO and NLO HTLpt results exhibit less sensitivity to the choice of renormalisation scale compared to weak coupling results at successive orders of approximation. The LO HTLpt prediction for QNS appears to agree reasonably well with available Wuppertal-Budapest LQCD data, obtained using tree-level improved Symanzik action and stout smeared staggered fermionic action with light quark masses $\sim 0.035m_s$, where $m_s$ is the strange quark mass near its physical value. However, the NLO QNS surpasses the LO one at higher temperatures and exceeds the free gas value at lower temperatures. Fourth-order QNS results in HTLpt are found to be considerably below the pQCD result which is accurate to $\alpha_s^3\ln\alpha_s$, and the LQCD results. It's worth noting that although the 2-loop calculation improves upon the LO results by rectifying overcounting, it does so by pushing the problem to higher orders in $g$. This asymmetry in loop and coupling expansion in HTLpt results in contributions from higher orders in coupling at a given loop order, leading to overcounting at ${\cal O}(g^4)$ and ${\cal O}(g^5)$). A NNLO HTLpt (3-loop) calculation addresses this issue for various thermodynamic quantities through ${\cal O}(g^5)$, ensuring that the HTLpt results, when expanded in a strict power series in $g$, reproduce the perturbative result order-by-order through ${\cal O}(g^5)$. Extending the NLO expansion to NNLO provides new estimates for a comprehensive set of thermodynamic quantities and various order QNS with remarkable accuracy, agreeing with Lattice QCD data down to a temperature of 200 MeV, and exhibiting complete analyticity with no fit parameters and gauge independence.

In section~\ref{damp} we have discussed the  dampting rate of  heavy fermions (muon and quark) and massless  partons for both QED and QCD plasma. 
We started this section in discussing the damping rate for hard heavy 
lepton (muon) with  $M, E, p\ge T$ in QED plasma. In BPT the muon  self-energy in lowest order comes from 1-loop and imaginary part of which leads to process  that describes
the emission or absorption of photon from a bare muon. However,  such process is forbidden due to energy-momentum conservation. The hard contribution to damping rate
of muon comes from the two-loop muon self-energy in BPT. Using naive power counting of the 2-loop diagrams,  one finds find that the muon damping rate  is order of 
$e^4 (\sim \alpha^2)$ and quadratically divergent due to the exchange of massless photon. One can improve the BPT result by using HTLpt where the photon propagator 
in one-loop self-energy  of muon is replaced by an effective HTL photon propagator but one does not need the effective muon propagator and vertices as the muon is hard ($M\gg T$). 
In BPT the damping rate for hard  muon was proportional to $e^4$ but in HTLpt  it becomes $e^2$. This reduction is due to strong infrared singularity in the damping rate.  The quadratically infrared divergence in BPT  becomes logarithmically infrared divergent in HTLpt. One can obtain damping rate for heavy quarks in QCD by replacing the QED coupling $e^2$ by $C_Fg^2$. Also, one needs an infrared ultra-soft cut-off  for the integration over exchanged momentum $q$ as $q\sim m_{\rm{mag}}\sim g^2T$ and an upper limit or ultra-violet cut-off for $q$-integration as  $q\sim eT$ in QCD. The heavy fermion formalism can be extended 
to light fermion (electron and quarks) with momentum $p\geq T$ and setting velocity $v=1$ with thermal energy $E=p\geq 3T$ in Eq.~\eqref{dm25}. 
The gluon damping rate can also be obtained from the light quark result by replacing $C_F$ by $C_A$ for adjoint representation. It can also be obtained by explicitly
calculating the  gluon self-energy with one effective gluon propagator. The hard photon damping rate can be obtained from hard photon production rate discussed in 
subsec~\ref{photon_chap} as they are related.

We also discussed the damping rate for fermion and gauge boson with soft momenta, $p~\sim gT$ and  these particles are collective modes known as long wavelength 
plasmon and plasmino mode. The soft damping of massless quark is to be gauge independent and in leading order in $g$ it is  obtained at rest as
$\gamma_{\rm q}(0) = a (g^2T/12\pi)$,
where the coefficient $a=5.63$ and $5.71$ for two and three flavour, respectively. We note that the damping mechanism is happening due to elastic ($2\rightarrow 2$) 
and inelastic ($2\rightarrow 3$) processes which appear by appropriately cutting the quark self-energy diagram with effective propagator and vertices.
Historically, way back in middle of 1980's, there was a huge controversy in computing the soft gluon damping rate in QCD perturbation theory at finite temperature. 
Much focus was made on plasmon mode.  It was found that the leading order gluon damping rate is gauge dependent both in magnitude and sign as discussed 
in subsec~\ref{bpt}. This led to a doubt on the applicability of QCD perturbation theory at finite temperature. This inconsistency also led Braaten and Pisarski 
to develop HTL resummation and point out that those calculations were incomplete --  there are higher-loop diagrams which contribute to the same order 
in the coupling constant $g$ as the one-loop diagram. To obtain the soft gluon damping rate one needs soft-gluon self-energy where the gluon propagators and vertices 
are HTL resumed ones. The gluon damping rate at rest was found as
$\gamma_{\rm g}(0) =a (g^2T/8\pi)$ where $a=6.63568$.  The hard damping rates are logarithmically infrared divergent but soft damping rates are infrared finite 
even though there is no magnetic screening of static magnetic field. The reason is that the damping rates at zero momentum are caused by scatterings, 
the exchange of transverse gluons do not cause any divergence because the magnetic processes are suppressed for zero external momentum.

In section~\ref{el} we have discussed the energy loss of heavy and massless particles in a medium. We note that the energy loss is related to the interaction rate 
vis-a-vis damping rate of particle in a medium, which are associated to the scattering of a high energy particle with medium medium particles. We will first calculate the 
collisional energy loss of a muon ($M,E,p > T$) in QED plasma. The collisional energy loss of a particle was first 
calculated by Bjorken. We know that the Bjorken formula for collisional energy loss for heavy quark  has uncertainty due to ambiguities in the choice of $q_{\min}$ and $q_{\max}$ 
of the  integration over the exchange momentum $q$. This ambiguity has been removed in which the effect of screening is taken into account by considering a 
separation scale $q_c$ between hard ($q \sim T$) momenta transfer and soft ($q\sim eT$) momenta transfer, akin to the photon production rate. This should be  
chosen so that $eT\ll q_c \ll T$  in the $e\rightarrow 0$. The hard contribution ($q>q_c$) is calculated from the tree level processes\footnote{The Compton process 
$\mu\gamma\rightarrow \mu\gamma$ would not contribute to leading order in $T/M$.}($\mu e^\pm\rightarrow \mu e^\pm$) which are obtained from the imaginary part of two loop muon 
self-energy in BPT. On the other hand, the soft contribution ($q_c < q$) from one-loop muon self-energy diagram 
where the bare photon propagator is replaced by the effective photon propagator in HTL approximation. When hard and soft contributions are added the separation scale $q_c$
gets cancelled. The energy loss for a heavy lepton with mass $M \gg T$ has been computed  in three regimes of velocity or energy:  for subthermal velocities $v \ll \sqrt{3T/M}$; in the intermediate region
where $v \gg \sqrt{3T/M}$, and $E < 0.54 M^2 /T$  and in the ultrarelativistic region $E > 0.54 M^2 /T$. 
We have also discussed the collisional energy loss of a relativistic heavy quark traversing through a QGP to leading order in $g$ and $T/M$ for two kinematical  
regimes: i) $E\ll M^2/T$ and ii) $E\gg M^2/T$. We note that the leading order in strong coupling $g$, the energy loss of 
heavy quark originates from elastic scattering processes: $Qq({\bar q}) \rightarrow Qq({\bar q})$ and $Qg \rightarrow Qg$ and the calculation will be almost parallel to that heavy lepton 
(muon) means that QED result can be generalised to QCD with simple group theoretical factors. We also note that  the heavy fermion energy loss by computing the 
electric field induced by a classical source consisting of a point charge moving at fixed velocity is equal to the quantum field theory description. The agreement holds only after 
using the small energy of the exchanged particles, which is equivalent to a high-temperature approximation. We note that the collisional energy loss does not take into account the 
fluctuations of the induced chromoelectric field. We have obtained the collisional energy gain of a heavy quark by taking into account the chromo-electromagnetic field fluctuations 
which reduces the net collisional energy loss.

We have also discussed the collisional energy loss of massless partons.  After Bjorken formula, the semiclassical approach was used for light quark energy loss following the same procedure as obtained for heavy quark but the  $q$-integration, over exchanged momenta, was found to be proportional to $\ln(q_{max}/q_{min})$ where the electric screening can be used as infrared cut-off, $q_{\min}=m^g_D$ and whereas it shows 
logarithmic utraviolet divergence as $q_{\max}=p$ for light quarks with $E=p$ and $v=1$. However, we note that the result is incomplete since the high temperature approximation has been used for the dielectric functions which is valid only for energies and momenta of the exchanged gluon $\om,\, q \ll T$.  It also  suffered from 
an ambiguity associated with the choice of upper limit $q_{\max}$. Further, the soft contribution to the energy loss of a light quark or gluon can be calculated similarly 
as the one of a heavy quark for $ E\gg M^2/T$ following one loop calculation of quark self-energy with soft gluon propagator. The only difference is that one introduces 
another separation scale $T\ll {\tilde q}\ll E$ apart from the separation scale $q_c$ between hard ($q\sim T$) momenta  and soft ($q\sim gT$) momenta transfer.  The hard
contribution comes from the tree level diagrams for $qq\rightarrow qq$ and $qg\rightarrow qg$ scatterings. The total energy loss of massless quark is obtained by adding hard and soft contributions where the separation scale $q_c$ cancels out but the scale $\tilde q$ remains. We also note that
the energy loss of a high energy gluon can be obtained in multiplying the energy loss of high energy massless quark  by $9/4$. 

The radiative energy loss occurs due to deceleration of a charged particle in the medium, emitting a gluon is what known as bremsstrahlung process. Several approaches to radiative energy loss have been developed, each based on the modelling of the medium. One such approach, developed by Arnold, Moore and Yaffe, is known as AMY scheme. In this scheme, the medium is assumed to be composed of quark gluon quasi-particles, with dispersion relations and interactions described by the HTL effective theory. Consequently, all quasi-particles in the medium possess thermal masses on the order of $\sim gT$, and their scattering is predominantly governed by soft interactions. The hard jet is assumed to have a virtuality scale comparable to the Debye mass or thermal mass, but with energy $E\gg T$. Following this, one identifies and resums the collinear enhanced contributions emanating from the scattering and induced radiation off the hard parton. The entire calculation is carried out at the temperature scale $T$, assumed to be large ($T\rightarrow \infty$) so that the effective coupling $g (T )$ is small.  After the calculation of the single gluon emission kernel, the AMY scheme employs rate equations to account for multiple emissions.
Unlike other approaches, the AMY formalism naturally includes feedback from the medium, distinguishing it from others in the approaches. A consistent method is utilised to derive an integral equation, akin in form to a linearised kinetic equation, whose solution determines the rate of bremsstrahlung.
The rates for processes such as a quark decaying into a quark and gluon, or a gluon decaying into a quark-antiquark pair, are employed to formulate a Fokker-Planck equation describing the evolution of the distribution of hard partons with time. Two sets of equations describe the changes in the distribution of the sum of quarks with antiquarks, and gluons. Subsequently, the radiative energy loss of hard partons and the transport coefficient $\hat q$ were obtained in ${\cal O}(\alpha_s^2)$.

We now note that the QCD thermodynamics  and other quantities have been calculated in ${\cal O}(g^5\sim \alpha_s^{5/2})$ within 3-loop HTLpt 
and both collisional and radiative energy loss  are up to ${\cal O}(\alpha_s^2)$ in HTLpt. But other quantities like the  dilepton  and photon production rate within 2-loop HTLpt,
mesonic spectral and correlation function,  and damping rate within 1-loop HTLpt are computed in leading order in $\alpha_s$. We suggest that the calculation of quark pressure 
can be pushed to ${\cal O}(g^6)$ within 4-loop order but not for  gluon  due to Linde problem. The other calculations like 
the dilepton rate, mesonic spectral and correlation and damping rate can also be pushed to higher order. But for photon production rate cannot be pushed beyond 2-loop order
as it reqires to resum infinite number of diagrams.  The damping rates for light partons have been calculated in rest 
($p=0$) and the calculations can be extended to finite $p$.

	\section*{Acknowledgements}
	\vspace{-0.2cm}
	It is a great pleasure to thank  Raktim Abir (RA), Aritra Bandyopadhyay (AB), Aritra Das (AD), Bithika Karmakar (BK), Chowdhury Aminul Islam (CAI), Purnendu Chakraborty (PC) and Ritesh Ghosh (RG) for collaboration, various discussions and numerous help during the course of this review. We are thankful to AB, BK, CAI and RG for critically reading the manuscript. MGM would like to acknowledge and thank Markus Thoma who introduced to him the HTL approximation when he was an Alexander von Humboldt Fellow in $1999$ in Institute f\"ur Theoretische Physik, Universit\"at  Giessen, Germany. MGM also had a long collaboration with him with more than 20 papers. The authors would also thank collaborators Jen Andersen, Michael Strickland and Nan Su for works on QCD thermodynamics and other various observables in LO, NLO and NLO in HTLpt. NH would like to acknowledge the discussion with Sumit and Sreemoyee Sarkar. NH is supported in part by the SERB- MATRICS under Grant No. MTR/2021/000939. MGM would also like to acknowledge the support received from the Department of Atomic Energy, Government of India for the project TPAES in Theory division of Saha Institute of Nuclear Physics.
	

	\appendix
	\renewcommand*{\thesection}{\Alph{section}}
	
	\vspace{-0.2cm}
	\section{Appendix}\label{appendix}
	\vspace{-0.2cm}
\subsection*{Braaten-Pisarski-Yuan (BPY) Prescription}
\label{bpy_disc}
Lets consider a complex function $f(z)$ having branch cut
\be
f(z) =\frac{1}{2\pi i} \oint \frac{f(\xi)\ d\xi}{\xi-z} \, . \label{bpy_0}
\ee
Considering $\xi=x+i\epsilon$, one can write
\bea
f(z) =\frac{1}{2\pi i} \int_{-\infty}^{+\infty} \frac{f(x+i\epsilon) -f(x-i\epsilon)}{x-z} \,  dx 
= \frac{1}{2\pi i} \int_{-\infty}^{+\infty} \frac{\textrm{Disc} f(x+i\epsilon)}{x-z} dx \, . \label{bpy_01}
\eea
where the discontinuity is related to the imaginary part of a complex function as
\be
\textrm{Disc} f(x+i\epsilon) = f(x+i\epsilon)-f(x-i\epsilon) = 2i \, \textrm{ Im} f(x+i\epsilon) \, . \label{bpy_02}
\ee
Combining~\eqref{bpy_01}  and~\eqref{bpy_02}, one can write
\be
f(z) = \frac{1}{\pi}  \int_{-\infty}^{+\infty} \frac{\textrm{\cal Im} \,  f(x+i\epsilon)}{x-z} dx 
=  \int_{-\infty}^{+\infty} \frac{ \rho (x)}{x-z} dx \label{bpy_03}
\ee
where the spectral density $\rho$ is defined as
\be
\rho(x) = \frac{1}{\pi} \textrm{Im} \,  f(x+i\epsilon) \, . \label{bpy_04}
\ee
The spectral density $\rho_1 (\omega_1)$ is related to the any complex function $F_1(k_0)$ as given in \eqref{bpy_03}
\be
F_1(k_0) =\int_{-\infty}^{+\infty} \frac{\rho_1(\omega_1) d\omega_1}{\omega_1-k_0-i\epsilon_1} \, . \label{bpy1}
\ee
We note that $K\equiv (k_0,\bm{\vec k})$ is the fermionic momentum with $k_0 =(2m+1) i\pi T$.  Now, lets have,
\bea
\int_{0}^{1/T} d\tau_1 \ e^{x\tau_1}  &=& \frac{ e^{x/T}-1} {x}  \, \, 
\Rightarrow \,\, \frac{1}{x} = \frac{1}{e^{x/T}-1} \int_{0}^{1/T} d\tau_1 \  e^{x\tau_1}  \, , \label{bpy2}
\eea
where $T$ is the temperature.
Now, considering $x=(\omega_1-k_0-i\epsilon_1)$ one can write~\eqref{bpy2} as
\be
\frac{1}{\omega_1-k_0-i\epsilon_1} =  \frac{1}{e^{(\omega_1-k_0)/T}-1} \int_{0}^{1/T} d\tau_1 \  e^{(\omega_1-k_0-i\epsilon_1)\tau_1}  \, . \label{bpy3}
\ee
Combining \eqref{bpy3} with \eqref{bpy1}, one gets
\be
F_1(k_0) = \int_{-\infty}^{+\infty} \frac{\rho_1(\omega_1) d\omega_1}{e^{(\omega_1-k_0)/T}-1} \int_{0}^{1/T} 
d\tau_1 \  e^{(\omega_1-k_0-i\epsilon_1)\tau_1}  \, . \label{bpy4}
\ee 
Now, using $e^{-k_0/T} = e^{-(2m+1)i\pi} = -1$, one can write as
\bea
F_1(k_0) = - \int_{-\infty}^{+\infty} \frac{\rho_1(\omega_1) d\omega_1}{e^{\omega_1/T}+1} \int_{0}^{1/T} d\tau_1 \  e^{(\omega_1-k_0-i\epsilon_1)\tau_1} 
= -  \int_{-\infty}^{+\infty} n_F(\omega_1) \ \rho_1(\omega_1) \, d\omega_1 \int_{0}^{1/T} d\tau_1 \  e^{(\omega_1-k_0-i\epsilon_1)\tau_1}  \, . \label{bpy4a}
\eea
Similarly, one can write another complex function $F_2(q_0)$ as
\bea
F_2(q_0) &=& -  \int_{-\infty}^{+\infty} n_F(\omega_2) \ \rho_2(\omega_2) \, d\omega_2 \int_{0}^{1/T} d\tau_2 \  e^{(\omega_2-q_0-i\epsilon_2)\tau_2}  \, , \label{bpy4b}
\eea
where $q_0=(p_0-k_0)$ and $P$ is the bosonic momentum with $p_0=2mi\pi T$.

We would like to compute the imaginary part of the product of two complex functions $T\sum_{k_0} F_1(k_0)F_2(q_0)$ :
\vspace*{-0.in}
\bea
\textrm{Im} \, \, \, T\sum_{k_0} F_1(k_0)F_2(q_0)  &=& \textrm{Im} \, \int_{-\infty}^{+\infty} d\omega_1 \int_{-\infty}^{+\infty} d\omega_2\, 
n_F(\omega_1) n_F(\omega_2)  \rho_1(\omega_1)  \rho_2(\omega_2) \nn
&&\times  \int_{0}^{1/T} d\tau_1 \int_{0}^{1/T} d\tau_2 \, \, e^{(\omega_1-i\epsilon_1)\tau_1}\,\, e^{(\omega_2-p_0-i\epsilon_2)\tau_2} \, 
\left[T\sum_{k_0} e^{-k_0(\tau_1-\tau_2)}\right] . \label{bpy5}
\eea
As the bracketed term in Eq.~\eqref{bpy5} is $\delta(\tau_2-\tau_1)$ we perform $\tau_2$-integration using $\delta$-function, then performing the $\tau_1$ integration and using $e^{-p_0/t} =e^{-2\pi i mT}=1$, one can write
\be
\textrm{Im} \, \, \, \! T\sum_{k_0} F_1(k_0)F_2(q_0) =\! \int_{-\infty}^{+\infty}\! d\omega_1 \int_{-\infty}^{+\infty}\! d\omega_2\, 
n_F(\omega_1) n_F(\omega_2)  \rho_1(\omega_1)  \rho_2(\omega_2) 
 \left (e^{(\omega_1+\omega_2)/T} -1\right )   \textrm{Im} \left (\frac{1}{ \omega_1+\omega_2 -p_0 -i\epsilon'}\right) \!.\ \ \   \label{bpy6}
\ee
where $\epsilon'=\epsilon_1+\epsilon_2 $. Now using  
\be
p_0 =\omega+i\epsilon'' \, ; \, \,\,
\frac{1}{ \omega_1+\omega_2 -\om -i\epsilon'' -i\epsilon'} = \frac{1}{ \omega_1+\omega_2 -\omega -i\epsilon}\, \, \,{\mbox{and}}\,\,\,
\textrm{Im} \left ( \frac{1}{\omega_1+\omega_2-\omega-i\epsilon}\right ) = -\pi \delta \left (\omega_1+\omega_2-\omega \right ) \, , \label{bpy10}
\ee
one gets
\be
\textrm{Im} \, \,  T\sum_{k_0} F_1(k_0)F_2(q_0) 
= \pi  \left (1-e^{\beta \omega} \right ) \int_{-\infty}^{+\infty} d\omega_1 \int_{-\infty}^{+\infty} d\omega_2\, \, \,
n_F(\omega_1) n_F(\omega_2) 
 \rho_1(\omega_1)  \rho_2(\omega_2) \, \delta \left(\omega_1+\omega_2-\omega \right ) \, , \label{bpy11}
\ee
where $\beta=1/T$. 
We finally obtain following \eqref{bpy_02} and \eqref{bpy11},  the discontinuity from the imaginary part of a product of two complex functions~\cite{Braaten:1990wp} as
\be
\textrm{Disc} \, \, T\sum_{k_0} F_1(k_0)F_2(q_0) 
=2 \pi i  \left (1-e^{\beta \omega} \right ) \int_{-\infty}^{+\infty} d\omega_1 \int_{-\infty}^{+\infty} d\omega_2
n_F(\omega_1) n_F(\omega_2) 
 \rho_1(\omega_1)  \rho_2(\omega_2) \, \delta \left(\omega_1+\omega_2-\omega \right ) \, . \label{bpy12}
\ee
	%
	\setlength{\bibsep}{0pt plus 0.32ex}
	\bibliography{HTL_rev}{}

\begin{thebibliography}{100}
\expandafter\ifx\csname url\endcsname\relax
  \def\url#1{\texttt{#1}}\fi
\expandafter\ifx\csname urlprefix\endcsname\relax\def\urlprefix{URL }\fi
\expandafter\ifx\csname href\endcsname\relax
  \def\href#1#2{#2} \def\path#1{#1}\fi

\bibitem{Matsubara:1955ws}
T.~Matsubara, {A New approach to quantum statistical mechanics}, Prog. Theor.
  Phys. 14 (1955) 351--378.
\newblock \href {https://doi.org/10.1143/PTP.14.351}
  {\path{doi:10.1143/PTP.14.351}}.

\bibitem{Schwinger:1960qe}
J.~S. Schwinger, {Brownian motion of a quantum oscillator}, J. Math. Phys. 2
  (1961) 407--432.
\newblock \href {https://doi.org/10.1063/1.1703727}
  {\path{doi:10.1063/1.1703727}}.

\bibitem{Keldysh:1964ud}
L.~V. Keldysh, {Diagram technique for nonequilibrium processes}, Zh. Eksp.
  Teor. Fiz. 47 (1964) 1515--1527.

\bibitem{Ghiglieri:2020dpq}
J.~Ghiglieri, A.~Kurkela, M.~Strickland, A.~Vuorinen, {Perturbative Thermal
  QCD: Formalism and Applications}, Phys. Rept. 880 (2020) 1--73.
\newblock \href {http://arxiv.org/abs/2002.10188} {\path{arXiv:2002.10188}},
  \href {https://doi.org/10.1016/j.physrep.2020.07.004}
  {\path{doi:10.1016/j.physrep.2020.07.004}}.

\bibitem{Mustafa:2022got}
M.~G. Mustafa, {An introduction to thermal field theory and some of its
  application}, Eur. Phys. J. ST 232~(9) (2023) 1369--1457.
\newblock \href {http://arxiv.org/abs/2207.00534} {\path{arXiv:2207.00534}},
  \href {https://doi.org/10.1140/epjs/s11734-023-00868-8}
  {\path{doi:10.1140/epjs/s11734-023-00868-8}}.

\bibitem{Gross:1980br}
D.~J. Gross, R.~D. Pisarski, L.~G. Yaffe, {QCD and Instantons at Finite
  Temperature}, Rev. Mod. Phys. 53 (1981) 43.
\newblock \href {https://doi.org/10.1103/RevModPhys. 53.43}
  {\path{doi:10.1103/RevModPhys. 53.43}}.

\bibitem{Appelquist:1981vg}
T.~Appelquist, R.~D. Pisarski, {High-Temperature Yang-Mills Theories and
  Three-Dimensional Quantum Chromodynamics}, Phys. Rev. D 23 (1981) 2305.
\newblock \href {https://doi.org/10.1103/PhysRevD.23.2305}
  {\path{doi:10.1103/PhysRevD.23.2305}}.

\bibitem{Nadkarni:1982kb}
S.~Nadkarni, {Dimensional Reduction in Hot QCD}, Phys. Rev. D 27 (1983) 917.
\newblock \href {https://doi.org/10.1103/PhysRevD.27.917}
  {\path{doi:10.1103/PhysRevD.27.917}}.

\bibitem{Nadkarni:1988fh}
S.~Nadkarni, {Dimensional Reduction in Finite Temperature Quantum
  Chromodynamics. 2.}, Phys. Rev. D 38 (1988) 3287.
\newblock \href {https://doi.org/10.1103/PhysRevD.38.3287}
  {\path{doi:10.1103/PhysRevD.38.3287}}.

\bibitem{Nadkarni:1988pb}
S.~Nadkarni, {Large Scale Structure of the Deconfined Phase}, Phys. Rev. Lett.
  60 (1988) 491--494.
\newblock \href {https://doi.org/10.1103/PhysRevLett.60.491}
  {\path{doi:10.1103/PhysRevLett.60.491}}.

\bibitem{Braaten:1989mz}
E.~Braaten, R.~D. Pisarski, {Soft Amplitudes in Hot Gauge Theories: A General
  Analysis}, Nucl. Phys. B337 (1990) 569.
\newblock \href {https://doi.org/10.1016/0550-3213(90)90508-B}
  {\path{doi:10.1016/0550-3213(90)90508-B}}.

\bibitem{Braaten:1991gm}
E.~Braaten, R.~D. Pisarski, {Simple effective Lagrangian for hard thermal
  loops}, Phys. Rev. D45 (1992) 1827--1830.
\newblock \href {https://doi.org/10.1103/PhysRevD.45.R1827}
  {\path{doi:10.1103/PhysRevD.45.R1827}}.

\bibitem{Braaten:1990az}
E.~Braaten, R.~D. Pisarski, {Deducing Hard Thermal Loops From Ward Identities},
  Nucl. Phys. B339 (1990) 310--324.
\newblock \href {https://doi.org/10.1016/0550-3213(90)90351-D}
  {\path{doi:10.1016/0550-3213(90)90351-D}}.

\bibitem{Taylor:1990ia}
J.~Taylor, S.~Wong, {The Effective Action of Hard Thermal Loops in {QCD}},
  Nucl. Phys. B346 (1990) 115--128.
\newblock \href {https://doi.org/10.1016/0550-3213(90)90240-E}
  {\path{doi:10.1016/0550-3213(90)90240-E}}.

\bibitem{Frenkel:1991ts}
J.~Frenkel, J.~C. Taylor, {Hard thermal QCD, forward scattering and effective
  actions}, Nucl. Phys. B 374 (1992) 156--168.
\newblock \href {https://doi.org/10.1016/0550-3213(92)90480-Y}
  {\path{doi:10.1016/0550-3213(92)90480-Y}}.

\bibitem{Barton:1989fk}
G.~Barton, {On the Finite Temperature Quantum Electrodynamics of Free Electrons
  and Photons}, Annals Phys. 200 (1990) 271.
\newblock \href {https://doi.org/10.1016/0003-4916(90)90276-T}
  {\path{doi:10.1016/0003-4916(90)90276-T}}.

\bibitem{Linde:1978px}
A.~D. Linde, {Phase Transitions in Gauge Theories and Cosmology}, Rept. Prog.
  Phys. 42 (1979) 389.
\newblock \href {https://doi.org/10.1088/0034-4885/42/3/001}
  {\path{doi:10.1088/0034-4885/42/3/001}}.

\bibitem{Linde:1980ts}
A.~D. Linde, {Infrared Problem in Thermodynamics of the Yang-Mills Gas}, Phys.
  Lett. B96 (1980) 289.
\newblock \href {https://doi.org/10.1016/0370-2693(80)90769-8}
  {\path{doi:10.1016/0370-2693(80)90769-8}}.

\bibitem{Karsch:1997gj}
F.~Karsch, A.~Patkos, P.~Petreczky, {Screened perturbation theory}, Phys. Lett.
  B401 (1997) 69--73.
\newblock \href {http://arxiv.org/abs/hep-ph/9702376}
  {\path{arXiv:hep-ph/9702376}}, \href
  {https://doi.org/10.1016/S0370-2693(97)00392-4}
  {\path{doi:10.1016/S0370-2693(97)00392-4}}.

\bibitem{Chiku:1998kd}
S.~Chiku, T.~Hatsuda, {Optimized perturbation theory at finite temperature},
  Phys. Rev. D 58 (1998) 076001.
\newblock \href {http://arxiv.org/abs/hep-ph/9803226}
  {\path{arXiv:hep-ph/9803226}}, \href
  {https://doi.org/10.1103/PhysRevD.58.076001}
  {\path{doi:10.1103/PhysRevD.58.076001}}.

\bibitem{Andersen:2000yj}
J.~O. Andersen, E.~Braaten, M.~Strickland, {Screened perturbation theory to
  three loops}, Phys. Rev. D63 (2001) 105008.
\newblock \href {http://arxiv.org/abs/hep-ph/0007159}
  {\path{arXiv:hep-ph/0007159}}, \href
  {https://doi.org/10.1103/PhysRevD.63.105008}
  {\path{doi:10.1103/PhysRevD.63.105008}}.

\bibitem{Andersen:2001ez}
J.~Andersen, M.~Strickland, {Mass expansions of screened perturbation theory},
  Phys.Rev. D64 (2001) 105012.
\newblock \href {http://arxiv.org/abs/hep-ph/0105214}
  {\path{arXiv:hep-ph/0105214}}, \href
  {https://doi.org/10.1103/PhysRevD.64.105012}
  {\path{doi:10.1103/PhysRevD.64.105012}}.

\bibitem{Andersen:2008bz}
J.~O. Andersen, L.~Kyllingstad, {Four-loop Screened Perturbation Theory}, Phys.
  Rev. D78 (2008) 076008.
\newblock \href {http://arxiv.org/abs/0805.4478} {\path{arXiv:0805.4478}},
  \href {https://doi.org/10.1103/PhysRevD.78.076008}
  {\path{doi:10.1103/PhysRevD.78.076008}}.

\bibitem{Blaizot:1999ip}
J.~Blaizot, E.~Iancu, A.~Rebhan, {The Entropy of the QCD plasma}, Phys. Rev.
  Lett. 83 (1999) 2906--2909.
\newblock \href {http://arxiv.org/abs/hep-ph/9906340}
  {\path{arXiv:hep-ph/9906340}}, \href
  {https://doi.org/10.1103/PhysRevLett.83.2906}
  {\path{doi:10.1103/PhysRevLett.83.2906}}.

\bibitem{Blaizot:1999ap}
J.~Blaizot, E.~Iancu, A.~Rebhan, {Selfconsistent hard thermal loop
  thermodynamics for the quark gluon plasma}, Phys. Lett. B470 (1999) 181--188.
\newblock \href {http://arxiv.org/abs/hep-ph/9910309}
  {\path{arXiv:hep-ph/9910309}}, \href
  {https://doi.org/10.1016/S0370-2693(99)01306-4}
  {\path{doi:10.1016/S0370-2693(99)01306-4}}.

\bibitem{Blaizot:2000fc}
J.~Blaizot, E.~Iancu, A.~Rebhan, {Approximately selfconsistent resummations for
  the thermodynamics of the quark gluon plasma. 1. Entropy and density}, Phys.
  Rev. D63 (2001) 065003.
\newblock \href {http://arxiv.org/abs/hep-ph/0005003}
  {\path{arXiv:hep-ph/0005003}}, \href
  {https://doi.org/10.1103/PhysRevD.63.065003}
  {\path{doi:10.1103/PhysRevD.63.065003}}.

\bibitem{Andersen:1999fw}
J.~O. Andersen, E.~Braaten, M.~Strickland, {Hard thermal loop resummation of
  the free energy of a hot gluon plasma}, Phys. Rev. Lett. 83 (1999)
  2139--2142.
\newblock \href {http://arxiv.org/abs/hep-ph/9902327}
  {\path{arXiv:hep-ph/9902327}}, \href
  {https://doi.org/10.1103/PhysRevLett.83.2139}
  {\path{doi:10.1103/PhysRevLett.83.2139}}.

\bibitem{Andersen:1999sf}
J.~O. Andersen, E.~Braaten, M.~Strickland, {Hard thermal loop resummation of
  the thermodynamics of a hot gluon plasma}, Phys. Rev. D61 (2000) 014017.
\newblock \href {http://arxiv.org/abs/hep-ph/9905337}
  {\path{arXiv:hep-ph/9905337}}, \href
  {https://doi.org/10.1103/PhysRevD.61.014017}
  {\path{doi:10.1103/PhysRevD.61.014017}}.

\bibitem{Andersen:1999va}
J.~O. Andersen, E.~Braaten, M.~Strickland, {Hard thermal loop resummation of
  the free energy of a hot quark - gluon plasma}, Phys. Rev. D61 (2000) 074016.
\newblock \href {http://arxiv.org/abs/hep-ph/9908323}
  {\path{arXiv:hep-ph/9908323}}, \href
  {https://doi.org/10.1103/PhysRevD.61.074016}
  {\path{doi:10.1103/PhysRevD.61.074016}}.

\bibitem{Andersen:2002ey}
J.~O. Andersen, E.~Braaten, E.~Petitgirard, M.~Strickland, {HTL perturbation
  theory to two loops}, Phys. Rev. D66 (2002) 085016.
\newblock \href {http://arxiv.org/abs/hep-ph/0205085}
  {\path{arXiv:hep-ph/0205085}}, \href
  {https://doi.org/10.1103/PhysRevD.66.085016}
  {\path{doi:10.1103/PhysRevD.66.085016}}.

\bibitem{Andersen:2003zk}
J.~O. Andersen, E.~Petitgirard, M.~Strickland, {Two loop HTL thermodynamics
  with quarks}, Phys. Rev. D70 (2004) 045001.
\newblock \href {http://arxiv.org/abs/hep-ph/0302069}
  {\path{arXiv:hep-ph/0302069}}, \href
  {https://doi.org/10.1103/PhysRevD.70.045001}
  {\path{doi:10.1103/PhysRevD.70.045001}}.

\bibitem{Andersen:2009tw}
J.~O. Andersen, M.~Strickland, N.~Su, {Three-loop HTL Free Energy for QED},
  Phys. Rev. D80 (2009) 085015.
\newblock \href {http://arxiv.org/abs/0906.2936} {\path{arXiv:0906.2936}},
  \href {https://doi.org/10.1103/PhysRevD.80.085015}
  {\path{doi:10.1103/PhysRevD.80.085015}}.

\bibitem{Andersen:2009tc}
J.~O. Andersen, M.~Strickland, N.~Su, {Gluon Thermodynamics at Intermediate
  Coupling}, Phys. Rev. Lett. 104 (2010) 122003.
\newblock \href {http://arxiv.org/abs/0911.0676} {\path{arXiv:0911.0676}},
  \href {https://doi.org/10.1103/PhysRevLett.104.122003}
  {\path{doi:10.1103/PhysRevLett.104.122003}}.

\bibitem{Andersen:2010ct}
J.~O. Andersen, M.~Strickland, N.~Su, {Three-loop HTL gluon thermodynamics at
  intermediate coupling}, JHEP 1008 (2010) 113.
\newblock \href {http://arxiv.org/abs/1005.1603} {\path{arXiv:1005.1603}},
  \href {https://doi.org/10.1007/JHEP08(2010)113}
  {\path{doi:10.1007/JHEP08(2010)113}}.

\bibitem{Andersen:2010wu}
J.~O. Andersen, L.~E. Leganger, M.~Strickland, N.~Su, {NNLO hard-thermal-loop
  thermodynamics for QCD}, Phys. Lett. B696 (2011) 468--472.
\newblock \href {http://arxiv.org/abs/1009.4644} {\path{arXiv:1009.4644}},
  \href {https://doi.org/10.1016/j.physletb.2010.12.070}
  {\path{doi:10.1016/j.physletb.2010.12.070}}.

\bibitem{Andersen:2011sf}
J.~O. Andersen, L.~E. Leganger, M.~Strickland, N.~Su, {Three-loop HTL QCD
  thermodynamics}, JHEP 1108 (2011) 053.
\newblock \href {http://arxiv.org/abs/1103.2528} {\path{arXiv:1103.2528}},
  \href {https://doi.org/10.1007/JHEP08(2011)053}
  {\path{doi:10.1007/JHEP08(2011)053}}.

\bibitem{Andersen:2011ug}
J.~O. Andersen, L.~E. Leganger, M.~Strickland, N.~Su, {The QCD trace anomaly},
  Phys. Rev. D84 (2011) 087703.
\newblock \href {http://arxiv.org/abs/1106.0514} {\path{arXiv:1106.0514}},
  \href {https://doi.org/10.1103/PhysRevD.84.087703}
  {\path{doi:10.1103/PhysRevD.84.087703}}.

\bibitem{Haque:2012my}
N.~Haque, M.~G. Mustafa, M.~Strickland, {Two-loop HTL pressure at finite
  temperature and chemical potential}, Phys. Rev. D87 (2013) 105007.
\newblock \href {http://arxiv.org/abs/1212.1797} {\path{arXiv:1212.1797}},
  \href {https://doi.org/10.1103/PhysRevD.87.105007}
  {\path{doi:10.1103/PhysRevD.87.105007}}.

\bibitem{Mogliacci:2013mca}
S.~Mogliacci, J.~O. Andersen, M.~Strickland, N.~Su, A.~Vuorinen, {Equation of
  State of hot and dense QCD: Resummed perturbation theory confronts lattice
  data}, JHEP 1312 (2013) 055.
\newblock \href {http://arxiv.org/abs/1307.8098} {\path{arXiv:1307.8098}},
  \href {https://doi.org/10.1007/JHEP12(2013)055}
  {\path{doi:10.1007/JHEP12(2013)055}}.

\bibitem{Haque:2013qta}
N.~Haque, M.~G. Mustafa, M.~Strickland, {Quark Number Susceptibilities from
  Two-Loop Hard Thermal Loop Perturbation Theory}, JHEP 1307 (2013) 184.
\newblock \href {http://arxiv.org/abs/1302.3228} {\path{arXiv:1302.3228}},
  \href {https://doi.org/10.1007/JHEP07(2013)184}
  {\path{doi:10.1007/JHEP07(2013)184}}.

\bibitem{Haque:2013sja}
N.~Haque, J.~O. Andersen, M.~G. Mustafa, M.~Strickland, N.~Su, {Three-loop
  HTLpt Pressure and Susceptibilities at Finite Temperature and Density}, Phys.
  Rev. D89 (2014) 061701.
\newblock \href {http://arxiv.org/abs/1309.3968} {\path{arXiv:1309.3968}},
  \href {https://doi.org/10.1103/PhysRevD.89.061701}
  {\path{doi:10.1103/PhysRevD.89.061701}}.

\bibitem{Haque:2014rua}
N.~Haque, A.~Bandyopadhyay, J.~O. Andersen, M.~G. Mustafa, M.~Strickland,
  et~al., {Three-loop HTLpt thermodynamics at finite temperature and chemical
  potential}, JHEP 1405 (2014) 027.
\newblock \href {http://arxiv.org/abs/1402.6907} {\path{arXiv:1402.6907}},
  \href {https://doi.org/10.1007/JHEP05(2014)027}
  {\path{doi:10.1007/JHEP05(2014)027}}.

\bibitem{Andersen:2015eoa}
J.~O. Andersen, N.~Haque, M.~G. Mustafa, M.~Strickland, {Three-loop
  hard-thermal-loop perturbation theory thermodynamics at finite temperature
  and finite baryonic and isospin chemical potential}, Phys. Rev. D 93~(5)
  (2016) 054045.
\newblock \href {http://arxiv.org/abs/1511.04660} {\path{arXiv:1511.04660}},
  \href {https://doi.org/10.1103/PhysRevD.93.054045}
  {\path{doi:10.1103/PhysRevD.93.054045}}.

\bibitem{Moller:1960cva}
C.~M\o{}ller, P.~T. Matthews, J.~S. Schwinger, N.~Fukuda, J.~J. Sakurai (Eds.),
  {Proceedings, 3rd Brandeis University Summer Institute in Theoretical
  Physics}: {Waltham, MA, USA, 1960}, Brandeis U., Waltham, MA, 1960.

\bibitem{Das:1997gg}
A.~K. Das, {Finite Temperature Field Theory}, World Scientific, New York, 1997.

\bibitem{Kapusta:2006pm}
J.~I. Kapusta, C.~Gale, {Finite-temperature field theory: Principles and
  applications}, Cambridge Monographs on Mathematical Physics, Cambridge
  University Press, 2011.
\newblock \href {https://doi.org/10.1017/CBO9780511535130}
  {\path{doi:10.1017/CBO9780511535130}}.

\bibitem{Pisarski:1987wc}
R.~D. Pisarski, {Computing Finite Temperature Loops with Ease}, Nucl. Phys. B
  309 (1988) 476--492.
\newblock \href {https://doi.org/10.1016/0550-3213(88)90454-3}
  {\path{doi:10.1016/0550-3213(88)90454-3}}.

\bibitem{Bellac:2011kqa}
M.~L. Bellac, {Thermal Field Theory}, Cambridge Monographs on Mathematical
  Physics, Cambridge University Press, 2011.
\newblock \href {https://doi.org/10.1017/CBO9780511721700}
  {\path{doi:10.1017/CBO9780511721700}}.

\bibitem{Mallik:2016anp}
S.~Mallik, S.~Sarkar, {Hadrons at Finite Temperature}, Cambridge University
  Press, Cambridge, 2016.
\newblock \href {https://doi.org/10.1017/9781316535585}
  {\path{doi:10.1017/9781316535585}}.

\bibitem{Weldon:1982bn}
H.~A. Weldon, {Effective Fermion Masses of Order gT in High Temperature Gauge
  Theories with Exact Chiral Invariance}, Phys. Rev. D 26 (1982) 2789.
\newblock \href {https://doi.org/10.1103/PhysRevD.26.2789}
  {\path{doi:10.1103/PhysRevD.26.2789}}.

\bibitem{Weldon:1982aq}
H.~A. Weldon, {Covariant Calculations at Finite Temperature: The Relativistic
  Plasma}, Phys. Rev. D26 (1982) 1394.
\newblock \href {https://doi.org/10.1103/PhysRevD.26.1394}
  {\path{doi:10.1103/PhysRevD.26.1394}}.

\bibitem{lopez}
J.~A. Lopez, J.~C. Parikh, P.~J. Siemens, Texas A @ M preprint, 1985.

\bibitem{Kajantie:1982xx}
K.~Kajantie, J.~I. Kapusta, {Behavior of Gluons at High Temperature}, Annals
  Phys. 160 (1985) 477.
\newblock \href {https://doi.org/10.1016/0003-4916(85)90153-8}
  {\path{doi:10.1016/0003-4916(85)90153-8}}.

\bibitem{Heinz:1986kh}
U.~W. Heinz, K.~Kajantie, T.~Toimela, {Damping of Plasma Oscillations in Hot
  Gluon Matter}, Phys. Lett. B 183 (1987) 96--100.
\newblock \href {https://doi.org/10.1016/0370-2693(87)91424-9}
  {\path{doi:10.1016/0370-2693(87)91424-9}}.

\bibitem{Hansson:1987um}
T.~H. Hansson, I.~Zahed, {Electric and Magnetic Properties of Hot Gluons},
  Phys. Rev. Lett. 58 (1987) 2397.
\newblock \href {https://doi.org/10.1103/PhysRevLett.58.2397}
  {\path{doi:10.1103/PhysRevLett.58.2397}}.

\bibitem{Kobes:1987bi}
R.~Kobes, G.~Kunstatter, {Stability of Plasma Oscillations in Hot Gluonic
  Matter}, Phys. Rev. Lett. 61 (1988) 392.
\newblock \href {https://doi.org/10.1103/PhysRevLett.61.392}
  {\path{doi:10.1103/PhysRevLett.61.392}}.

\bibitem{Braaten:1995cm}
E.~Braaten, A.~Nieto, {Effective field theory approach to high temperature
  thermodynamics}, Phys. Rev. D51 (1995) 6990--7006.
\newblock \href {http://arxiv.org/abs/hep-ph/9501375}
  {\path{arXiv:hep-ph/9501375}}, \href
  {https://doi.org/10.1103/PhysRevD.51.6990}
  {\path{doi:10.1103/PhysRevD.51.6990}}.

\bibitem{Braaten:1995jr}
E.~Braaten, A.~Nieto, {Free energy of QCD at high temperature}, Phys. Rev. D53
  (1996) 3421--3437.
\newblock \href {http://arxiv.org/abs/hep-ph/9510408}
  {\path{arXiv:hep-ph/9510408}}, \href
  {https://doi.org/10.1103/PhysRevD.53.3421}
  {\path{doi:10.1103/PhysRevD.53.3421}}.

\bibitem{Kajantie:2002wa}
K.~Kajantie, M.~Laine, K.~Rummukainen, Y.~Schroder, {The Pressure of hot QCD up
  to} $g^6$ ln(1/$g$), Phys. Rev. D67 (2003) 105008.
\newblock \href {http://arxiv.org/abs/hep-ph/0211321}
  {\path{arXiv:hep-ph/0211321}}, \href
  {https://doi.org/10.1103/PhysRevD.67.105008}
  {\path{doi:10.1103/PhysRevD.67.105008}}.

\bibitem{Espinosa:2003af}
O.~Espinosa, E.~Stockmeyer, {An Operator approach for Matsubara sums}, Phys.
  Rev. D 69 (2004) 065004.
\newblock \href {http://arxiv.org/abs/hep-ph/0305001}
  {\path{arXiv:hep-ph/0305001}}, \href
  {https://doi.org/10.1103/PhysRevD.69.065004}
  {\path{doi:10.1103/PhysRevD.69.065004}}.

\bibitem{Espinosa:2005gq}
O.~Espinosa, {The Thermal operator representation for Matsubara sums}, Phys.
  Rev. D 71 (2005) 065009.
\newblock \href {http://arxiv.org/abs/hep-ph/0501273}
  {\path{arXiv:hep-ph/0501273}}, \href
  {https://doi.org/10.1103/PhysRevD.71.065009}
  {\path{doi:10.1103/PhysRevD.71.065009}}.

\bibitem{silin}
V.~Silin, {On the electronmagnetic properties of a relativistic plasma}, Sov.
  Phys. JETP 11 (1960) 1136.

\bibitem{Fradkin}
E.~S. Fradkin, {Proc. Lebedev Inst.}, Vol.~29, 1965, p.~6.

\bibitem{Klimov:1981ka}
V.~Klimov, {Spectrum of Elementary Fermi Excitations in Quark Gluon Plasma. (In
  Russian)}, Sov. J. Nucl. Phys. 33 (1981) 934--935.

\bibitem{Weldon:1989ys}
H.~A. Weldon, {Dynamical Holes in the Quark - Gluon Plasma}, Phys. Rev. D 40
  (1989) 2410.
\newblock \href {https://doi.org/10.1103/PhysRevD.40.2410}
  {\path{doi:10.1103/PhysRevD.40.2410}}.

\bibitem{Kalashnikov:1979kq}
O.~K. Kalashnikov, V.~V. Klimov, {Infrared Behavior of the Polarization
  Operator in Scalar Electrodynamics at Finite Temperature}, Phys. Lett. B 95
  (1980) 423--425.
\newblock \href {https://doi.org/10.1016/0370-2693(80)90182-3}
  {\path{doi:10.1016/0370-2693(80)90182-3}}.

\bibitem{Klimov:1982bv}
V.~Klimov, {Collective Excitations in a Hot Quark Gluon Plasma}, Sov. Phys.
  JETP 55 (1982) 199--204.

\bibitem{Petitgirard:1991mf}
E.~Petitgirard, {Massive fermion dispersion relation at finite temperature}, Z.
  Phys. C 54 (1992) 673--678.
\newblock \href {https://doi.org/10.1007/BF01559497}
  {\path{doi:10.1007/BF01559497}}.

\bibitem{Baym:1992eu}
G.~Baym, J.-P. Blaizot, B.~Svetitsky, {Emergence of new quasiparticles in
  quantum electrodynamics at finite temperature}, Phys. Rev. D 46 (1992)
  4043--4051.
\newblock \href {https://doi.org/10.1103/PhysRevD.46.4043}
  {\path{doi:10.1103/PhysRevD.46.4043}}.

\bibitem{Kobes:1990xf}
R.~Kobes, G.~Kunstatter, A.~Rebhan, {QCD plasma parameters and the gauge
  dependent gluon propagator}, Phys. Rev. Lett. 64 (1990) 2992--2995.
\newblock \href {https://doi.org/10.1103/PhysRevLett.64.2992}
  {\path{doi:10.1103/PhysRevLett.64.2992}}.

\bibitem{Blaizot:1995kg}
J.-P. Blaizot, E.~Iancu, R.~R. Parwani, {On the screening of static
  electromagnetic fields in hot QED plasmas}, Phys. Rev. D 52 (1995)
  2543--2562.
\newblock \href {http://arxiv.org/abs/hep-ph/9504408}
  {\path{arXiv:hep-ph/9504408}}, \href
  {https://doi.org/10.1103/PhysRevD.52.2543}
  {\path{doi:10.1103/PhysRevD.52.2543}}.

\bibitem{Luttinger:1960ua}
J.~M. Luttinger, J.~C. Ward, {Ground state energy of a many fermion system.
  2.}, Phys. Rev. 118 (1960) 1417--1427.
\newblock \href {https://doi.org/10.1103/PhysRev.118.1417}
  {\path{doi:10.1103/PhysRev.118.1417}}.

\bibitem{deDominicis:1964zz}
C.~de~Dominicis, P.~C. Martin, {Stationary Entropy Principle and
  Renormalization in Normal and Superfluid Systems. I. Algebraic Formulation},
  J. Math. Phys. 5 (1964) 14--30.
\newblock \href {https://doi.org/10.1063/1.1704062}
  {\path{doi:10.1063/1.1704062}}.

\bibitem{Baym:1962sx}
G.~Baym, {Selfconsistent approximation in many body systems}, Phys. Rev. 127
  (1962) 1391--1401.
\newblock \href {https://doi.org/10.1103/PhysRev.127.1391}
  {\path{doi:10.1103/PhysRev.127.1391}}.

\bibitem{Vanderheyden:1998ph}
B.~Vanderheyden, G.~Baym, {Selfconsistent approximations in relativistic
  plasmas: Quasiparticle analysis of the thermodynamic properties}, J. Statist.
  Phys. 93 (1998) 843.
\newblock \href {http://arxiv.org/abs/hep-ph/9803300}
  {\path{arXiv:hep-ph/9803300}}, \href
  {https://doi.org/10.1023/B:JOSS.0000033166.37520.ae}
  {\path{doi:10.1023/B:JOSS.0000033166.37520.ae}}.

\bibitem{Pisarski:1990ds}
R.~D. Pisarski, {Resummation and the gluon damping rate in hot QCD}, Nucl.
  Phys. A525 (1991) 175--188.
\newblock \href {https://doi.org/10.1016/0375-9474(91)90325-Z}
  {\path{doi:10.1016/0375-9474(91)90325-Z}}.

\bibitem{Carignano:2019ofj}
S.~Carignano, M.~E. Carrington, J.~Soto, {The HTL Lagrangian at NLO: the photon
  case}, Phys. Lett. B 801 (2020) 135193.
\newblock \href {http://arxiv.org/abs/1909.10545} {\path{arXiv:1909.10545}},
  \href {https://doi.org/10.1016/j.physletb.2019.135193}
  {\path{doi:10.1016/j.physletb.2019.135193}}.

\bibitem{Haque:2018eph}
N.~Haque, {Quark mass dependent collective excitations and quark number
  susceptibilities within the hard thermal loop approximation}, Phys. Rev. D
  98~(1) (2018) 014013.
\newblock \href {http://arxiv.org/abs/1804.04996} {\path{arXiv:1804.04996}},
  \href {https://doi.org/10.1103/PhysRevD.98.014013}
  {\path{doi:10.1103/PhysRevD.98.014013}}.

\bibitem{Sumit:2022bor}
Sumit, N.~Haque, B.~K. Patra, {NLO quark self-energy and dispersion relation
  using the hard thermal loop resummation}, JHEP 05 (2023) 171.
\newblock \href {http://arxiv.org/abs/2201.07173} {\path{arXiv:2201.07173}},
  \href {https://doi.org/10.1007/JHEP05(2023)171}
  {\path{doi:10.1007/JHEP05(2023)171}}.

\bibitem{Haque:2010rb}
N.~Haque, M.~G. Mustafa, {A Modified Hard Thermal Loop Perturbation Theory}
  (2010).
\newblock \href {http://arxiv.org/abs/1007.2076} {\path{arXiv:1007.2076}}.

\bibitem{Haque:2011iz}
N.~Haque, M.~G. Mustafa, M.~H. Thoma, {Conserved Density Fluctuation and
  Temporal Correlation Function in HTL Perturbation Theory}, Phys. Rev. D84
  (2011) 054009.
\newblock \href {http://arxiv.org/abs/1103.3394} {\path{arXiv:1103.3394}},
  \href {https://doi.org/10.1103/PhysRevD.84.054009}
  {\path{doi:10.1103/PhysRevD.84.054009}}.

\bibitem{Chakraborty:2001kx}
P.~Chakraborty, M.~G. Mustafa, M.~H. Thoma, {Quark number susceptibility in
  hard thermal loop approximation}, Eur. Phys. J. C23 (2002) 591--596.
\newblock \href {http://arxiv.org/abs/hep-ph/0111022}
  {\path{arXiv:hep-ph/0111022}}, \href {https://doi.org/10.1007/s100520200899}
  {\path{doi:10.1007/s100520200899}}.

\bibitem{Chakraborty:2002yt}
P.~Chakraborty, M.~G. Mustafa, M.~H. Thoma, {Chiral susceptibility in hard
  thermal loop approximation}, Phys. Rev. D67 (2003) 114004.
\newblock \href {http://arxiv.org/abs/hep-ph/0210159}
  {\path{arXiv:hep-ph/0210159}}, \href
  {https://doi.org/10.1103/PhysRevD.67.114004}
  {\path{doi:10.1103/PhysRevD.67.114004}}.

\bibitem{Chakraborty:2003uw}
P.~Chakraborty, M.~G. Mustafa, M.~H. Thoma, {Quark number susceptibility,
  thermodynamic sum rule, and hard thermal loop approximation}, Phys. Rev. D68
  (2003) 085012.
\newblock \href {http://arxiv.org/abs/hep-ph/0303009}
  {\path{arXiv:hep-ph/0303009}}, \href
  {https://doi.org/10.1103/PhysRevD.68.085012}
  {\path{doi:10.1103/PhysRevD.68.085012}}.

\bibitem{McLerran:1984ay}
L.~D. McLerran, T.~Toimela, {Photon and Dilepton Emission from the Quark -
  Gluon Plasma: Some General Considerations}, Phys. Rev. D 31 (1985) 545.
\newblock \href {https://doi.org/10.1103/PhysRevD.31.545}
  {\path{doi:10.1103/PhysRevD.31.545}}.

\bibitem{Baier:1988xv}
R.~Baier, B.~Pire, D.~Schiff, {Dilepton production at finite temperature:
  Perturbative treatment at order $\alpha_s$}, Phys. Rev. D 38 (1988) 2814.
\newblock \href {https://doi.org/10.1103/PhysRevD.38.2814}
  {\path{doi:10.1103/PhysRevD.38.2814}}.

\bibitem{Braaten:1990wp}
E.~Braaten, R.~D. Pisarski, T.-C. Yuan, {Production of Soft Dileptons in the
  Quark - Gluon Plasma}, Phys. Rev. Lett. 64 (1990) 2242.
\newblock \href {https://doi.org/10.1103/PhysRevLett.64.2242}
  {\path{doi:10.1103/PhysRevLett.64.2242}}.

\bibitem{Greiner:2010zg}
C.~Greiner, N.~Haque, M.~G. Mustafa, M.~H. Thoma, {Low Mass Dilepton Rate from
  the Deconfined Phase}, Phys. Rev. C83 (2011) 014908.
\newblock \href {http://arxiv.org/abs/1010.2169} {\path{arXiv:1010.2169}},
  \href {https://doi.org/10.1103/PhysRevC.83.014908}
  {\path{doi:10.1103/PhysRevC.83.014908}}.

\bibitem{Ghisoiu:2014mha}
I.~Ghisoiu, M.~Laine, {Interpolation of hard and soft dilepton rates}, JHEP 10
  (2014) 083.
\newblock \href {http://arxiv.org/abs/1407.7955} {\path{arXiv:1407.7955}},
  \href {https://doi.org/10.1007/JHEP10(2014)083}
  {\path{doi:10.1007/JHEP10(2014)083}}.

\bibitem{Ghiglieri:2014kma}
J.~Ghiglieri, G.~D. Moore, {Low Mass Thermal Dilepton Production at NLO in a
  Weakly Coupled Quark-Gluon Plasma}, JHEP 12 (2014) 029.
\newblock \href {http://arxiv.org/abs/1410.4203} {\path{arXiv:1410.4203}},
  \href {https://doi.org/10.1007/JHEP12(2014)029}
  {\path{doi:10.1007/JHEP12(2014)029}}.

\bibitem{Ghiglieri:2015nba}
J.~Ghiglieri, {The thermal dilepton rate at NLO at small and large invariant
  mass}, Nucl. Part. Phys. Proc. 276-278 (2016) 305--308.
\newblock \href {http://arxiv.org/abs/1510.00525} {\path{arXiv:1510.00525}},
  \href {https://doi.org/10.1016/j.nuclphysbps.2016.05.070}
  {\path{doi:10.1016/j.nuclphysbps.2016.05.070}}.

\bibitem{Aurenche:1998nw}
P.~Aurenche, F.~Gelis, R.~Kobes, H.~Zaraket, {Bremsstrahlung and photon
  production in thermal QCD}, Phys. Rev. D58 (1998) 085003.
\newblock \href {http://arxiv.org/abs/hep-ph/9804224}
  {\path{arXiv:hep-ph/9804224}}, \href
  {https://doi.org/10.1103/PhysRevD.58.085003}
  {\path{doi:10.1103/PhysRevD.58.085003}}.

\bibitem{Aurenche:1999ec}
P.~Aurenche, F.~Gelis, R.~Kobes, H.~Zaraket, {Two loop Compton and annihilation
  processes in thermal QCD}, Phys. Rev. D60 (1999) 076002.
\newblock \href {http://arxiv.org/abs/hep-ph/9903307}
  {\path{arXiv:hep-ph/9903307}}, \href
  {https://doi.org/10.1103/PhysRevD.60.076002}
  {\path{doi:10.1103/PhysRevD.60.076002}}.

\bibitem{Karsch:2000gi}
F.~Karsch, M.~Mustafa, M.~Thoma, {Finite temperature meson correlation
  functions in HTL approximation}, Phys. Lett. B497 (2001) 249--258.
\newblock \href {http://arxiv.org/abs/hep-ph/0007093}
  {\path{arXiv:hep-ph/0007093}}, \href
  {https://doi.org/10.1016/S0370-2693(00)01322-8}
  {\path{doi:10.1016/S0370-2693(00)01322-8}}.

\bibitem{Thoma:1997dk}
M.~H. Thoma, C.~T. Traxler, {Production of energetic dileptons with small
  invariant masses from the quark - gluon plasma}, Phys. Rev. D56 (1997)
  198--202.
\newblock \href {http://arxiv.org/abs/hep-ph/9701354}
  {\path{arXiv:hep-ph/9701354}}, \href
  {https://doi.org/10.1103/PhysRevD.56.198}
  {\path{doi:10.1103/PhysRevD.56.198}}.

\bibitem{Aurenche:2002pc}
P.~Aurenche, F.~Gelis, H.~Zaraket, {Enhanced thermal production of hard
  dileptons by $3 \rightarrow 2$ processes}, JHEP 0207 (2002) 063.
\newblock \href {http://arxiv.org/abs/hep-ph/0204145}
  {\path{arXiv:hep-ph/0204145}}, \href
  {https://doi.org/10.1088/1126-6708/2002/07/063}
  {\path{doi:10.1088/1126-6708/2002/07/063}}.

\bibitem{Aurenche:2002wq}
P.~Aurenche, F.~Gelis, G.~Moore, H.~Zaraket, {Landau-Pomeranchuk-Migdal
  resummation for dilepton production}, JHEP 0212 (2002) 006.
\newblock \href {http://arxiv.org/abs/hep-ph/0211036}
  {\path{arXiv:hep-ph/0211036}}, \href
  {https://doi.org/10.1088/1126-6708/2002/12/006}
  {\path{doi:10.1088/1126-6708/2002/12/006}}.

\bibitem{Carrington:2007gt}
M.~Carrington, A.~Gynther, P.~Aurenche, {Energetic di-leptons from the Quark
  Gluon Plasma}, Phys. Rev. D77 (2008) 045035.
\newblock \href {http://arxiv.org/abs/0711.3943} {\path{arXiv:0711.3943}},
  \href {https://doi.org/10.1103/PhysRevD.77.045035}
  {\path{doi:10.1103/PhysRevD.77.045035}}.

\bibitem{Kapusta:1991qp}
J.~I. Kapusta, P.~Lichard, D.~Seibert, {High-energy photons from quark - gluon
  plasma versus hot hadronic gas}, Phys. Rev. D44 (1991) 2774--2788.
\newblock \href {https://doi.org/10.1103/PhysRevD.47.4171,
  10.1103/PhysRevD.44.2774} {\path{doi:10.1103/PhysRevD.47.4171,
  10.1103/PhysRevD.44.2774}}.

\bibitem{Baier:1991em}
R.~Baier, H.~Nakkagawa, A.~Niegawa, K.~Redlich, {Production rate of hard
  thermal photons and screening of quark mass singularity}, Z. Phys. C 53
  (1992) 433--438.
\newblock \href {https://doi.org/10.1007/BF01625902}
  {\path{doi:10.1007/BF01625902}}.

\bibitem{Arnold:2001ba}
P.~B. Arnold, G.~D. Moore, L.~G. Yaffe, {Photon emission from ultrarelativistic
  plasmas}, JHEP 11 (2001) 057.
\newblock \href {http://arxiv.org/abs/hep-ph/0109064}
  {\path{arXiv:hep-ph/0109064}}, \href
  {https://doi.org/10.1088/1126-6708/2001/11/057}
  {\path{doi:10.1088/1126-6708/2001/11/057}}.

\bibitem{Arnold:2001ms}
P.~B. Arnold, G.~D. Moore, L.~G. Yaffe, {Photon emission from quark gluon
  plasma: Complete leading order results}, JHEP 0112 (2001) 009.
\newblock \href {http://arxiv.org/abs/hep-ph/0111107}
  {\path{arXiv:hep-ph/0111107}}, \href
  {https://doi.org/10.1088/1126-6708/2001/12/009}
  {\path{doi:10.1088/1126-6708/2001/12/009}}.

\bibitem{Peitzmann:2001mz}
T.~Peitzmann, M.~H. Thoma, {Direct photons from relativistic heavy ion
  collisions}, Phys. Rept. 364 (2002) 175--246.
\newblock \href {http://arxiv.org/abs/hep-ph/0111114}
  {\path{arXiv:hep-ph/0111114}}, \href
  {https://doi.org/10.1016/S0370-1573(02)00012-1}
  {\path{doi:10.1016/S0370-1573(02)00012-1}}.

\bibitem{Mustafa:2004hf}
M.~G. Mustafa, M.~H. Thoma, P.~Chakraborty, {Screening of a moving parton in
  the quark gluon plasma}, Phys. Rev. C71 (2005) 017901.
\newblock \href {http://arxiv.org/abs/hep-ph/0403279}
  {\path{arXiv:hep-ph/0403279}}, \href
  {https://doi.org/10.1103/PhysRevC.71.017901}
  {\path{doi:10.1103/PhysRevC.71.017901}}.

\bibitem{Mustafa:2005je}
M.~G. Mustafa, P.~Chakraborty, M.~H. Thoma, {Dynamical screening in a quark
  gluon plasma}, J. Phys. Conf. Ser. 50 (2006) 438--441.
\newblock \href {http://arxiv.org/abs/hep-ph/0504174}
  {\path{arXiv:hep-ph/0504174}}, \href
  {https://doi.org/10.1088/1742-6596/50/1/066}
  {\path{doi:10.1088/1742-6596/50/1/066}}.

\bibitem{Chakraborty:2006md}
P.~Chakraborty, M.~G. Mustafa, M.~H. Thoma, {Wakes in the quark-gluon plasma},
  Phys. Rev. D74 (2006) 094002.
\newblock \href {http://arxiv.org/abs/hep-ph/0606316}
  {\path{arXiv:hep-ph/0606316}}, \href
  {https://doi.org/10.1103/PhysRevD.74.094002}
  {\path{doi:10.1103/PhysRevD.74.094002}}.

\bibitem{Chakraborty:2007ug}
P.~Chakraborty, M.~G. Mustafa, R.~Ray, M.~H. Thoma, {Wakes in a Collisional
  Quark-Gluon Plasma}, J. Phys. G34 (2007) 2141--2152.
\newblock \href {http://arxiv.org/abs/0705.1447} {\path{arXiv:0705.1447}},
  \href {https://doi.org/10.1088/0954-3899/34/10/004}
  {\path{doi:10.1088/0954-3899/34/10/004}}.

\bibitem{Laine:2006ns}
M.~Laine, O.~Philipsen, P.~Romatschke, M.~Tassler, {Real-time static potential
  in hot QCD}, JHEP 0703 (2007) 054.
\newblock \href {http://arxiv.org/abs/hep-ph/0611300}
  {\path{arXiv:hep-ph/0611300}}, \href
  {https://doi.org/10.1088/1126-6708/2007/03/054}
  {\path{doi:10.1088/1126-6708/2007/03/054}}.

\bibitem{Dumitru:2007hy}
A.~Dumitru, Y.~Guo, M.~Strickland, {The Heavy-quark potential in an anisotropic
  (viscous) plasma}, Phys. Lett. B662 (2008) 37--42.
\newblock \href {http://arxiv.org/abs/0711.4722} {\path{arXiv:0711.4722}},
  \href {https://doi.org/10.1016/j.physletb.2008.02.048}
  {\path{doi:10.1016/j.physletb.2008.02.048}}.

\bibitem{Dumitru:2009ni}
A.~Dumitru, Y.~Guo, A.~Mocsy, M.~Strickland, {Quarkonium states in an
  anisotropic QCD plasma}, Phys. Rev. D79 (2009) 054019.
\newblock \href {http://arxiv.org/abs/0901.1998} {\path{arXiv:0901.1998}},
  \href {https://doi.org/10.1103/PhysRevD.79.054019}
  {\path{doi:10.1103/PhysRevD.79.054019}}.

\bibitem{Thakur:2013nia}
L.~Thakur, U.~Kakade, B.~K. Patra, {Dissociation of Quarkonium in a Complex
  Potential}, Phys. Rev. D 89~(9) (2014) 094020.
\newblock \href {http://arxiv.org/abs/1401.0172} {\path{arXiv:1401.0172}},
  \href {https://doi.org/10.1103/PhysRevD.89.094020}
  {\path{doi:10.1103/PhysRevD.89.094020}}.

\bibitem{Pisarski:1993rf}
R.~Pisarski, {Damping rates for moving particles in hot QCD}, Phys. Rev. D47
  (1993) 5589--5600.
\newblock \href {https://doi.org/10.1103/PhysRevD.47.5589}
  {\path{doi:10.1103/PhysRevD.47.5589}}.

\bibitem{Braaten:1992gd}
E.~Braaten, R.~D. Pisarski, {Calculation of the quark damping rate in hot QCD},
  Phys. Rev. D 46 (1992) 1829--1834.
\newblock \href {https://doi.org/10.1103/PhysRevD.46.1829}
  {\path{doi:10.1103/PhysRevD.46.1829}}.

\bibitem{Peigne:1993ky}
S.~Peigne, E.~Pilon, D.~Schiff, {The Heavy fermion damping rate puzzle}, Z.
  Phys. C60 (1993) 455--460.
\newblock \href {http://arxiv.org/abs/hep-ph/9306219}
  {\path{arXiv:hep-ph/9306219}}, \href {https://doi.org/10.1007/BF01560043}
  {\path{doi:10.1007/BF01560043}}.

\bibitem{Thoma:1995ju}
M.~H. Thoma, {Applications of high temperature field theory to heavy ion
  collisions}, Quark Gluon Plasma 2, ed. R.C. Hwa (World Scientific, Singapore,
  1995)  51.

\bibitem{Thoma:2000dc}
M.~H. Thoma, {New developments and applications of thermal field theory} (10
  2000).
\newblock \href {http://arxiv.org/abs/hep-ph/0010164}
  {\path{arXiv:hep-ph/0010164}}.

\bibitem{Thoma:1990fm}
M.~H. Thoma, M.~Gyulassy, {Quark Damping and Energy Loss in the High
  Temperature {QCD}}, Nucl. Phys. B351 (1991) 491--506.
\newblock \href {https://doi.org/10.1016/S0550-3213(05)80031-8}
  {\path{doi:10.1016/S0550-3213(05)80031-8}}.

\bibitem{Thoma:1993vs}
M.~H. Thoma, {Parton interaction rates in the quark - gluon plasma}, Phys. Rev.
  D 49 (1994) 451--459.
\newblock \href {http://arxiv.org/abs/hep-ph/9308257}
  {\path{arXiv:hep-ph/9308257}}, \href
  {https://doi.org/10.1103/PhysRevD.49.451}
  {\path{doi:10.1103/PhysRevD.49.451}}.

\bibitem{Thoma:1994fd}
M.~H. Thoma, {Damping rate of a hard photon in a relativistic plasma}, Phys.
  Rev. D 51 (1995) 862--865.
\newblock \href {http://arxiv.org/abs/hep-ph/9405309}
  {\path{arXiv:hep-ph/9405309}}, \href
  {https://doi.org/10.1103/PhysRevD.51.862}
  {\path{doi:10.1103/PhysRevD.51.862}}.

\bibitem{Abada:2011cc}
A.~Abada, N.~Daira-Aifa, {Photon Damping in One-Loop HTL Perturbation Theory},
  JHEP 1204 (2012) 071.
\newblock \href {http://arxiv.org/abs/1112.6065} {\path{arXiv:1112.6065}},
  \href {https://doi.org/10.1007/JHEP04(2012)071}
  {\path{doi:10.1007/JHEP04(2012)071}}.

\bibitem{Braaten:1989kk}
E.~Braaten, R.~D. Pisarski, {Resummation and Gauge Invariance of the Gluon
  Damping Rate in Hot QCD}, Phys. Rev. Lett. 64 (1990) 1338.
\newblock \href {https://doi.org/10.1103/PhysRevLett.64.1338}
  {\path{doi:10.1103/PhysRevLett.64.1338}}.

\bibitem{Braaten:1990it}
E.~Braaten, R.~D. Pisarski, {Calculation of the gluon damping rate in hot QCD},
  Phys. Rev. D42 (1990) 2156--2160.
\newblock \href {https://doi.org/10.1103/PhysRevD.42.2156}
  {\path{doi:10.1103/PhysRevD.42.2156}}.

\bibitem{Braaten:1991jj}
E.~Braaten, M.~H. Thoma, {Energy loss of a heavy fermion in a hot plasma},
  Phys. Rev. D44 (1991) 1298--1310.
\newblock \href {https://doi.org/10.1103/PhysRevD.44.1298}
  {\path{doi:10.1103/PhysRevD.44.1298}}.

\bibitem{Braaten:1991we}
E.~Braaten, M.~H. Thoma, {Energy loss of a heavy quark in the quark - gluon
  plasma}, Phys. Rev. D44 (1991) R 2625--2630.
\newblock \href {https://doi.org/10.1103/PhysRevD.44.R2625}
  {\path{doi:10.1103/PhysRevD.44.R2625}}.

\bibitem{Thoma:1991jum}
M.~H. Thoma, {Collisional energy loss of high-energy jets in the quark gluon
  plasma}, Phys. Lett. B 273 (1991) 128--132.
\newblock \href {https://doi.org/10.1016/0370-2693(91)90565-8}
  {\path{doi:10.1016/0370-2693(91)90565-8}}.

\bibitem{Arnold:2002ja}
P.~B. Arnold, G.~D. Moore, L.~G. Yaffe, {Photon and gluon emission in
  relativistic plasmas}, JHEP 06 (2002) 030.
\newblock \href {http://arxiv.org/abs/hep-ph/0204343}
  {\path{arXiv:hep-ph/0204343}}, \href
  {https://doi.org/10.1088/1126-6708/2002/06/030}
  {\path{doi:10.1088/1126-6708/2002/06/030}}.

\bibitem{Jeon:2003gi}
S.~Jeon, G.~D. Moore, {Energy loss of leading partons in a thermal QCD medium},
  Phys. Rev. C 71 (2005) 034901.
\newblock \href {http://arxiv.org/abs/hep-ph/0309332}
  {\path{arXiv:hep-ph/0309332}}, \href
  {https://doi.org/10.1103/PhysRevC.71.034901}
  {\path{doi:10.1103/PhysRevC.71.034901}}.

\bibitem{Mustafa:2004dr}
M.~G. Mustafa, {Energy loss of charm quarks in the quark-gluon plasma:
  Collisional versus radiative}, Phys. Rev. C72 (2005) 014905.
\newblock \href {http://arxiv.org/abs/hep-ph/0412402}
  {\path{arXiv:hep-ph/0412402}}, \href
  {https://doi.org/10.1103/PhysRevC.72.014905}
  {\path{doi:10.1103/PhysRevC.72.014905}}.

\bibitem{Mustafa:2003vh}
M.~G. Mustafa, M.~H. Thoma, {Quenching of hadron spectra due to the collisional
  energy loss of partons in the quark gluon plasma}, Acta Phys. Hung. A22
  (2005) 93--102.
\newblock \href {http://arxiv.org/abs/hep-ph/0311168}
  {\path{arXiv:hep-ph/0311168}}, \href
  {https://doi.org/10.1556/APH.22.2005.1-2.10}
  {\path{doi:10.1556/APH.22.2005.1-2.10}}.

\bibitem{Chakraborty:2006db}
P.~Chakraborty, M.~G. Mustafa, M.~H. Thoma, {Energy gain of heavy quarks by
  fluctuations in the QGP}, Phys. Rev. C75 (2007) 064908.
\newblock \href {http://arxiv.org/abs/hep-ph/0611355}
  {\path{arXiv:hep-ph/0611355}}, \href
  {https://doi.org/10.1103/PhysRevC.75.064908}
  {\path{doi:10.1103/PhysRevC.75.064908}}.

\bibitem{Qin:2007rn}
G.-Y. Qin, J.~Ruppert, C.~Gale, S.~Jeon, G.~D. Moore, M.~G. Mustafa, {Radiative
  and collisional jet energy loss in the quark-gluon plasma at RHIC}, Phys.
  Rev. Lett. 100 (2008) 072301.
\newblock \href {http://arxiv.org/abs/0710.0605} {\path{arXiv:0710.0605}},
  \href {https://doi.org/10.1103/PhysRevLett.100.072301}
  {\path{doi:10.1103/PhysRevLett.100.072301}}.

\bibitem{Ghiglieri:2015ala}
J.~Ghiglieri, G.~D. Moore, D.~Teaney, {Jet-Medium Interactions at NLO in a
  Weakly-Coupled Quark-Gluon Plasma}, JHEP 03 (2016) 095.
\newblock \href {http://arxiv.org/abs/1509.07773} {\path{arXiv:1509.07773}},
  \href {https://doi.org/10.1007/JHEP03(2016)095}
  {\path{doi:10.1007/JHEP03(2016)095}}.

\bibitem{Guo:2024mgh}
Y.~Guo, L.~Qiu, R.~Zhao, M.~Strickland, {Energy loss of a heavy fermion in a
  collisional QED plasma} (3 2024).
\newblock \href {http://arxiv.org/abs/2403.06739} {\path{arXiv:2403.06739}}.

\bibitem{Mrowczynski:2000ed}
S.~Mrowczynski, M.~H. Thoma, {Hard loop approach to anisotropic systems}, Phys.
  Rev. D62 (2000) 036011.
\newblock \href {http://arxiv.org/abs/hep-ph/0001164}
  {\path{arXiv:hep-ph/0001164}}, \href
  {https://doi.org/10.1103/PhysRevD.62.036011}
  {\path{doi:10.1103/PhysRevD.62.036011}}.

\bibitem{Romatschke:2003ms}
P.~Romatschke, M.~Strickland, {Collective modes of an anisotropic quark gluon
  plasma}, Phys. Rev. D68 (2003) 036004.
\newblock \href {http://arxiv.org/abs/hep-ph/0304092}
  {\path{arXiv:hep-ph/0304092}}, \href
  {https://doi.org/10.1103/PhysRevD.68.036004}
  {\path{doi:10.1103/PhysRevD.68.036004}}.

\bibitem{Romatschke:2004jh}
P.~Romatschke, M.~Strickland, {Collective modes of an anisotropic quark-gluon
  plasma II}, Phys. Rev. D70 (2004) 116006.
\newblock \href {http://arxiv.org/abs/hep-ph/0406188}
  {\path{arXiv:hep-ph/0406188}}, \href
  {https://doi.org/10.1103/PhysRevD.70.116006}
  {\path{doi:10.1103/PhysRevD.70.116006}}.

\bibitem{Rebhan:2004ur}
A.~Rebhan, P.~Romatschke, M.~Strickland, {Hard-loop dynamics of non-Abelian
  plasma instabilities}, Phys. Rev. Lett. 94 (2005) 102303.
\newblock \href {http://arxiv.org/abs/hep-ph/0412016}
  {\path{arXiv:hep-ph/0412016}}, \href
  {https://doi.org/10.1103/PhysRevLett.94.102303}
  {\path{doi:10.1103/PhysRevLett.94.102303}}.

\bibitem{Schenke:2006fz}
B.~Schenke, M.~Strickland, {Fermionic Collective Modes of an Anisotropic
  Quark-Gluon Plasma}, Phys. Rev. D74 (2006) 065004.
\newblock \href {http://arxiv.org/abs/hep-ph/0606160}
  {\path{arXiv:hep-ph/0606160}}, \href
  {https://doi.org/10.1103/PhysRevD.74.065004}
  {\path{doi:10.1103/PhysRevD.74.065004}}.

\bibitem{Rebhan:2008uj}
A.~Rebhan, M.~Strickland, M.~Attems, {Instabilities of an anisotropically
  expanding non-Abelian plasma: 1D+3V discretized hard-loop simulations}, Phys.
  Rev. D78 (2008) 045023.
\newblock \href {http://arxiv.org/abs/0802.1714} {\path{arXiv:0802.1714}},
  \href {https://doi.org/10.1103/PhysRevD.78.045023}
  {\path{doi:10.1103/PhysRevD.78.045023}}.

\bibitem{Attems:2012js}
M.~Attems, A.~Rebhan, M.~Strickland, {Instabilities of an anisotropically
  expanding non-Abelian plasma: 3D+3V discretized hard-loop simulations}, Phys.
  Rev. D87 (2013) 025010.
\newblock \href {http://arxiv.org/abs/1207.5795} {\path{arXiv:1207.5795}},
  \href {https://doi.org/10.1103/PhysRevD.87.025010}
  {\path{doi:10.1103/PhysRevD.87.025010}}.

\bibitem{Litim:1999ns}
D.~F. Litim, C.~Manuel, {Mean field dynamics in nonAbelian plasmas from
  classical transport theory}, Phys. Rev. Lett. 82 (1999) 4981--4984.
\newblock \href {http://arxiv.org/abs/hep-ph/9902430}
  {\path{arXiv:hep-ph/9902430}}, \href
  {https://doi.org/10.1103/PhysRevLett.82.4981}
  {\path{doi:10.1103/PhysRevLett.82.4981}}.

\bibitem{Litim:2001db}
D.~F. Litim, C.~Manuel, {Semiclassical transport theory for nonAbelian
  plasmas}, Phys. Rept. 364 (2002) 451--539.
\newblock \href {http://arxiv.org/abs/hep-ph/0110104}
  {\path{arXiv:hep-ph/0110104}}, \href
  {https://doi.org/10.1016/S0370-1573(02)00015-7}
  {\path{doi:10.1016/S0370-1573(02)00015-7}}.

\bibitem{Litim:1999id}
D.~F. Litim, C.~Manuel, {Effective transport equations for nonAbelian plasmas},
  Nucl. Phys. B 562 (1999) 237--274.
\newblock \href {http://arxiv.org/abs/hep-ph/9906210}
  {\path{arXiv:hep-ph/9906210}}, \href
  {https://doi.org/10.1016/S0550-3213(99)00531-3}
  {\path{doi:10.1016/S0550-3213(99)00531-3}}.

\bibitem{Litim:2001je}
D.~F. Litim, C.~Manuel, {Transport theory for a two flavor color
  superconductor}, Phys. Rev. Lett. 87 (2001) 052002.
\newblock \href {http://arxiv.org/abs/hep-ph/0103092}
  {\path{arXiv:hep-ph/0103092}}, \href
  {https://doi.org/10.1103/PhysRevLett.87.052002}
  {\path{doi:10.1103/PhysRevLett.87.052002}}.

\bibitem{Drewes:2017zyw}
M.~Drewes, B.~Garbrecht, P.~Hernandez, M.~Kekic, J.~Lopez-Pavon, J.~Racker,
  N.~Rius, J.~Salvado, D.~Teresi, {ARS Leptogenesis}, Int. J. Mod. Phys. A
  33~(05n06) (2018) 1842002.
\newblock \href {http://arxiv.org/abs/1711.02862} {\path{arXiv:1711.02862}},
  \href {https://doi.org/10.1142/S0217751X18420022}
  {\path{doi:10.1142/S0217751X18420022}}.

\bibitem{Biondini:2017rpb}
S.~Biondini, et~al., {Status of rates and rate equations for thermal
  leptogenesis}, Int. J. Mod. Phys. A 33~(05n06) (2018) 1842004.
\newblock \href {http://arxiv.org/abs/1711.02864} {\path{arXiv:1711.02864}},
  \href {https://doi.org/10.1142/S0217751X18420046}
  {\path{doi:10.1142/S0217751X18420046}}.

\bibitem{Besak:2012qm}
D.~Besak, D.~Bodeker, {Thermal production of ultrarelativistic right-handed
  neutrinos: Complete leading-order results}, JCAP 03 (2012) 029.
\newblock \href {http://arxiv.org/abs/1202.1288} {\path{arXiv:1202.1288}},
  \href {https://doi.org/10.1088/1475-7516/2012/03/029}
  {\path{doi:10.1088/1475-7516/2012/03/029}}.

\bibitem{Anisimov:2010gy}
A.~Anisimov, D.~Besak, D.~Bodeker, {Thermal production of relativistic Majorana
  neutrinos: Strong enhancement by multiple soft scattering}, JCAP 03 (2011)
  042.
\newblock \href {http://arxiv.org/abs/1012.3784} {\path{arXiv:1012.3784}},
  \href {https://doi.org/10.1088/1475-7516/2011/03/042}
  {\path{doi:10.1088/1475-7516/2011/03/042}}.

\bibitem{Ghiglieri:2016xye}
J.~Ghiglieri, M.~Laine, {Neutrino dynamics below the electroweak crossover},
  JCAP 07 (2016) 015.
\newblock \href {http://arxiv.org/abs/1605.07720} {\path{arXiv:1605.07720}},
  \href {https://doi.org/10.1088/1475-7516/2016/07/015}
  {\path{doi:10.1088/1475-7516/2016/07/015}}.

\bibitem{Ghiglieri:2018wbs}
J.~Ghiglieri, M.~Laine, {Precision study of GeV-scale resonant leptogenesis},
  JHEP 02 (2019) 014.
\newblock \href {http://arxiv.org/abs/1811.01971} {\path{arXiv:1811.01971}},
  \href {https://doi.org/10.1007/JHEP02(2019)014}
  {\path{doi:10.1007/JHEP02(2019)014}}.

\bibitem{Ghiglieri:2017gjz}
J.~Ghiglieri, M.~Laine, {GeV-scale hot sterile neutrino oscillations: a
  derivation of evolution equations}, JHEP 05 (2017) 132.
\newblock \href {http://arxiv.org/abs/1703.06087} {\path{arXiv:1703.06087}},
  \href {https://doi.org/10.1007/JHEP05(2017)132}
  {\path{doi:10.1007/JHEP05(2017)132}}.

\bibitem{Graf:2010tv}
P.~Graf, F.~D. Steffen, {Thermal axion production in the primordial quark-gluon
  plasma}, Phys. Rev. D 83 (2011) 075011.
\newblock \href {http://arxiv.org/abs/1008.4528} {\path{arXiv:1008.4528}},
  \href {https://doi.org/10.1103/PhysRevD.83.075011}
  {\path{doi:10.1103/PhysRevD.83.075011}}.

\bibitem{Kiessig:2011ga}
C.~Kiessig, M.~Plumacher, {Hard-Thermal-Loop Corrections in Leptogenesis II:
  Solving the Boltzmann Equations}, JCAP 1209 (2012) 012.
\newblock \href {http://arxiv.org/abs/1111.1235} {\path{arXiv:1111.1235}},
  \href {https://doi.org/10.1088/1475-7516/2012/09/012}
  {\path{doi:10.1088/1475-7516/2012/09/012}}.

\bibitem{Kiessig:2011fw}
C.~Kiessig, M.~Plumacher, {Hard-Thermal-Loop Corrections in Leptogenesis I:
  CP-Asymmetries}, JCAP 1207 (2012) 014.
\newblock \href {http://arxiv.org/abs/1111.1231} {\path{arXiv:1111.1231}},
  \href {https://doi.org/10.1088/1475-7516/2012/07/014}
  {\path{doi:10.1088/1475-7516/2012/07/014}}.

\bibitem{Du:2020odw}
Q.~Du, M.~Strickland, U.~Tantary, B.-W. Zhang, {Two-loop HTL-resummed
  thermodynamics for $ \mathcal{N} $ = 4 supersymmetric Yang-Mills theory},
  JHEP 09 (2020) 038.
\newblock \href {http://arxiv.org/abs/2006.02617} {\path{arXiv:2006.02617}},
  \href {https://doi.org/10.1007/JHEP09(2020)038}
  {\path{doi:10.1007/JHEP09(2020)038}}.

\bibitem{Du:2021jai}
Q.~Du, M.~Strickland, U.~Tantary, {$ \mathcal{N} $ = 4 supersymmetric
  Yang-Mills thermodynamics to order \ensuremath{\lambda}$^{2}$}, JHEP 08
  (2021) 064, [Erratum: JHEP 02, 053 (2022)].
\newblock \href {http://arxiv.org/abs/2105.02101} {\path{arXiv:2105.02101}},
  \href {https://doi.org/10.1007/JHEP08(2021)064}
  {\path{doi:10.1007/JHEP08(2021)064}}.

\bibitem{Andersen:2021bgw}
J.~O. Andersen, Q.~Du, M.~Strickland, U.~Tantary, {N=4 supersymmetric
  Yang-Mills thermodynamics from effective field theory}, Phys. Rev. D 105~(1)
  (2022) 015006.
\newblock \href {http://arxiv.org/abs/2111.12160} {\path{arXiv:2111.12160}},
  \href {https://doi.org/10.1103/PhysRevD.105.015006}
  {\path{doi:10.1103/PhysRevD.105.015006}}.

\bibitem{Jackson:2019yao}
G.~Jackson, M.~Laine, {Testing thermal photon and dilepton rates}, JHEP 11
  (2019) 144.
\newblock \href {http://arxiv.org/abs/1910.09567} {\path{arXiv:1910.09567}},
  \href {https://doi.org/10.1007/JHEP11(2019)144}
  {\path{doi:10.1007/JHEP11(2019)144}}.

\bibitem{Ghiglieri:2016tvj}
J.~Ghiglieri, O.~Kaczmarek, M.~Laine, F.~Meyer, {Lattice constraints on the
  thermal photon rate}, Phys. Rev. D 94~(1) (2016) 016005.
\newblock \href {http://arxiv.org/abs/1604.07544} {\path{arXiv:1604.07544}},
  \href {https://doi.org/10.1103/PhysRevD.94.016005}
  {\path{doi:10.1103/PhysRevD.94.016005}}.

\bibitem{Ruuskanen:1992hh}
P.~V. Ruuskanen, {Photons and lepton pairs: The Deep probes of quark - gluon
  plasma}, in: {NATO Advanced Study Institute on Particle Production in Highly
  Excited Matter}, 1992, pp. 593--613.

\bibitem{Gutbrod1993ParticlePI}
P.~Ruuskanen,
  \href{https://api.semanticscholar.org/CorpusID:118262819}{Particle production
  in highly excited matter, (ed.: Hans h. gutbrod and johann rafelski}, (Plenum
  Press, New York, 1993).
\newline\urlprefix\url{https://api.semanticscholar.org/CorpusID:118262819}

\bibitem{Feinberg:1976ua}
E.~L. Feinberg, {Direct Production of Photons and Dileptons in Thermodynamical
  Models of Multiple Hadron Production}, Nuovo Cim. A 34 (1976) 391.

\bibitem{Weldon:1990iw}
H.~A. Weldon, {Reformulation of finite temperature dilepton production}, Phys.
  Rev. D 42 (1990) 2384--2387.
\newblock \href {https://doi.org/10.1103/PhysRevD.42.2384}
  {\path{doi:10.1103/PhysRevD.42.2384}}.

\bibitem{Gale:1990pn}
C.~Gale, J.~I. Kapusta, {Vector dominance model at finite temperature}, Nucl.
  Phys. B357 (1991) 65--89.
\newblock \href {https://doi.org/10.1016/0550-3213(91)90459-B}
  {\path{doi:10.1016/0550-3213(91)90459-B}}.

\bibitem{Kobes:1985kc}
R.~L. Kobes, G.~W. Semenoff, {Discontinuities of Green Functions in Field
  Theory at Finite Temperature and Density}, Nucl. Phys. B 260 (1985) 714--746.
\newblock \href {https://doi.org/10.1016/0550-3213(85)90056-2}
  {\path{doi:10.1016/0550-3213(85)90056-2}}.

\bibitem{Kobes:1986za}
R.~L. Kobes, G.~W. Semenoff, {Discontinuities of Green Functions in Field
  Theory at Finite Temperature and Density. 2}, Nucl. Phys. B 272 (1986)
  329--364.
\newblock \href {https://doi.org/10.1016/0550-3213(86)90006-4}
  {\path{doi:10.1016/0550-3213(86)90006-4}}.

\bibitem{Ruuskanen:1991au}
P.~V. Ruuskanen, {Electromagnetic probes of quark - gluon plasma in
  relativistic heavy ion collisions}, Nucl. Phys. A 544 (1992) 169--182.
\newblock \href {https://doi.org/10.1016/0375-9474(92)90572-2}
  {\path{doi:10.1016/0375-9474(92)90572-2}}.

\bibitem{Cleymans:1986na}
J.~Cleymans, J.~Fingberg, K.~Redlich, {Transverse Momentum Distribution of
  Dileptons in Different Scenarios for the {QCD} Phase Transition}, Phys. Rev.
  D35 (1987) 2153.
\newblock \href {https://doi.org/10.1103/PhysRevD.35.2153}
  {\path{doi:10.1103/PhysRevD.35.2153}}.

\bibitem{Cleymans:1993jm}
J.~Cleymans, I.~Dadic, J.~Joubert, {Lepton pair production from a quark - gluon
  plasma to first order in alpha-s: numerical evaluation}, Phys. Rev. D49
  (1994) 230--237.
\newblock \href {https://doi.org/10.1103/PhysRevD.49.230}
  {\path{doi:10.1103/PhysRevD.49.230}}.

\bibitem{Cleymans:1992gb}
J.~Cleymans, I.~Dadic, {Lepton pair production from a quark - gluon plasma to
  first order in alpha-s}, Phys. Rev. D47 (1993) 160--172.
\newblock \href {https://doi.org/10.1103/PhysRevD.47.160}
  {\path{doi:10.1103/PhysRevD.47.160}}.

\bibitem{Alam:1996fd}
J.~Alam, B.~Sinha, S.~Raha, {Electromagnetic probes of quark gluon plasma},
  Phys. Rept. 273 (1996) 243--362.
\newblock \href {https://doi.org/10.1016/0370-1573(95)00084-4}
  {\path{doi:10.1016/0370-1573(95)00084-4}}.

\bibitem{hove}
L.~V. Hove, {The Occurrence of Singularities in the Elastic Frequency
  Distribution of a Crystal}, Phys. Rev. 89  1189.

\bibitem{Kapusta:1984rz}
J.~I. Kapusta, {ELECTRON POSITRON PAIR PRODUCTION AS A PROBE OF CHIRAL SYMMETRY
  IN A HOT QCD PLASMA}, Phys. Lett. B 136 (1984) 201--203.
\newblock \href {https://doi.org/10.1016/0370-2693(84)91181-X}
  {\path{doi:10.1016/0370-2693(84)91181-X}}.

\bibitem{Pisarski:1988vd}
R.~D. Pisarski, {Scattering Amplitudes in Hot Gauge Theories}, Phys. Rev. Lett.
  63 (1989) 1129.
\newblock \href {https://doi.org/10.1103/PhysRevLett.63.1129}
  {\path{doi:10.1103/PhysRevLett.63.1129}}.

\bibitem{Mustafa:1999dt}
M.~G. Mustafa, A.~Schafer, M.~H. Thoma, {Nonperturbative dilepton production
  from a quark gluon plasma}, Phys. Rev. C61 (2000) 024902.
\newblock \href {http://arxiv.org/abs/hep-ph/9908461}
  {\path{arXiv:hep-ph/9908461}}, \href
  {https://doi.org/10.1103/PhysRevC.61.024902}
  {\path{doi:10.1103/PhysRevC.61.024902}}.

\bibitem{Mustafa:2002pb}
M.~G. Mustafa, M.~H. Thoma, {Can Van Hove singularities be observed in
  relativistic heavy ion collisions?}, Pramana 60 (2003) 711--724.
\newblock \href {http://arxiv.org/abs/hep-ph/0201060}
  {\path{arXiv:hep-ph/0201060}}, \href {https://doi.org/10.1007/BF02705170}
  {\path{doi:10.1007/BF02705170}}.

\bibitem{Mustafa:1999cp}
M.~Mustafa, A.~Schafer, M.~Thoma, {Gluon condensate, quark propagation, and
  dilepton production in the quark gluon plasma}, Nucl. Phys. A661 (1999)
  653--656.
\newblock \href {http://arxiv.org/abs/hep-ph/9906391}
  {\path{arXiv:hep-ph/9906391}}, \href
  {https://doi.org/10.1016/S0375-9474(99)85110-0}
  {\path{doi:10.1016/S0375-9474(99)85110-0}}.

\bibitem{Peshier:1999dt}
A.~Peshier, M.~H. Thoma, {Quark dispersion relation and dilepton production in
  the quark gluon plasma}, Phys. Rev. Lett. 84 (2000) 841--844.
\newblock \href {http://arxiv.org/abs/hep-ph/9907268}
  {\path{arXiv:hep-ph/9907268}}, \href
  {https://doi.org/10.1103/PhysRevLett.84.841}
  {\path{doi:10.1103/PhysRevLett.84.841}}.

\bibitem{Bandyopadhyay:2015wua}
A.~Bandyopadhyay, N.~Haque, M.~G. Mustafa, M.~Strickland, {Dilepton rate and
  quark number susceptibility with the Gribov action}, Phys. Rev. D 93~(6)
  (2016) 065004.
\newblock \href {http://arxiv.org/abs/1508.06249} {\path{arXiv:1508.06249}},
  \href {https://doi.org/10.1103/PhysRevD.93.065004}
  {\path{doi:10.1103/PhysRevD.93.065004}}.

\bibitem{Landau:1953ivy}
L.~D. Landau, I.~Pomeranchuk, {The Limits of Applicability of the Theory of
  Bremsstrahlung by Electrons and of the Creation of Pairs at Large Energies},
  Dokl. Akad. Nauk SSSR 92 (1953).
\newblock \href {https://doi.org/10.1016/b978-0-08-010586-4.50080-8}
  {\path{doi:10.1016/b978-0-08-010586-4.50080-8}}.

\bibitem{Landau:1965ksp}
L.~D. Landau, I.~Pomeranchuk, {Electron-Cascade Processes at Ultra-High
  Energies}, Dokl. Akad. Nauk SSSR 92 (1965).
\newblock \href {https://doi.org/10.1016/b978-0-08-010586-4.50081-x}
  {\path{doi:10.1016/b978-0-08-010586-4.50081-x}}.

\bibitem{Migdal:1956tc}
A.~B. Migdal, {Bremsstrahlung and pair production in condensed media at
  high-energies}, Phys. Rev. 103 (1956) 1811--1820.
\newblock \href {https://doi.org/10.1103/PhysRev.103.1811}
  {\path{doi:10.1103/PhysRev.103.1811}}.

\bibitem{Altherr:1992th}
T.~Altherr, P.~Ruuskanen, {Low mass dileptons at high momenta in
  ultrarelativistic heavy ion collisions}, Nucl. Phys. B380 (1992) 377--390.
\newblock \href {https://doi.org/10.1016/0550-3213(92)90249-B}
  {\path{doi:10.1016/0550-3213(92)90249-B}}.

\bibitem{Laine:2013vma}
M.~Laine, {NLO thermal dilepton rate at non-zero momentum}, JHEP 11 (2013) 120.
\newblock \href {http://arxiv.org/abs/1310.0164} {\path{arXiv:1310.0164}},
  \href {https://doi.org/10.1007/JHEP11(2013)120}
  {\path{doi:10.1007/JHEP11(2013)120}}.

\bibitem{Peshier:1994zf}
A.~Peshier, B.~Kampfer, O.~Pavlenko, G.~Soff, {An Effective model of the quark
  - gluon plasma with thermal parton masses}, Phys. Lett. B337 (1994) 235--239.
\newblock \href {https://doi.org/10.1016/0370-2693(94)90969-5}
  {\path{doi:10.1016/0370-2693(94)90969-5}}.

\bibitem{Levai:1997yx}
P.~Levai, U.~W. Heinz, {Massive gluons and quarks and the equation of state
  obtained from SU(3) lattice QCD}, Phys. Rev. C57 (1998) 1879--1890.
\newblock \href {http://arxiv.org/abs/hep-ph/9710463}
  {\path{arXiv:hep-ph/9710463}}, \href
  {https://doi.org/10.1103/PhysRevC.57.1879}
  {\path{doi:10.1103/PhysRevC.57.1879}}.

\bibitem{Gribov:1978npb}
V.~N. Gribov, {Quantization of non-Abelian gauge theories}, Nucl. Phys. B 1
  (1978). 139 (1978) 1.
\newblock \href {https://doi.org/10.1103/PhysRevD.93.065004}
  {\path{doi:10.1103/PhysRevD.93.065004}}.

\bibitem{Zwanziger:1989mf}
D.~Zwanziger, {Local and Renormalizable Action From the Gribov Horizon}, Nucl.
  Phys. B 323 (1989) 513--544.
\newblock \href {https://doi.org/10.1016/0550-3213(89)90122-3}
  {\path{doi:10.1016/0550-3213(89)90122-3}}.

\bibitem{Vandersickel:2012tz}
N.~Vandersickel, D.~Zwanziger, {The Gribov problem and QCD dynamics}, Phys.
  Rept. 520 (2012) 175--251.
\newblock \href {http://arxiv.org/abs/1202.1491} {\path{arXiv:1202.1491}},
  \href {https://doi.org/10.1016/j.physrep.2012.07.003}
  {\path{doi:10.1016/j.physrep.2012.07.003}}.

\bibitem{Kitazawa:2009uw}
M.~Kitazawa, F.~Karsch, {Spectral Properties of Quarks at Finite Temperature in
  Lattice QCD}, Nucl. Phys. A 830 (2009) 223C--226C.
\newblock \href {http://arxiv.org/abs/0908.3079} {\path{arXiv:0908.3079}},
  \href {https://doi.org/10.1016/j.nuclphysa.2009.09.024}
  {\path{doi:10.1016/j.nuclphysa.2009.09.024}}.

\bibitem{Kaczmarek:2012mb}
O.~Kaczmarek, F.~Karsch, M.~Kitazawa, W.~Soldner, {Thermal mass and dispersion
  relations of quarks in the deconfined phase of quenched QCD}, Phys. Rev. D 86
  (2012) 036006.
\newblock \href {http://arxiv.org/abs/1206.1991} {\path{arXiv:1206.1991}},
  \href {https://doi.org/10.1103/PhysRevD.86.036006}
  {\path{doi:10.1103/PhysRevD.86.036006}}.

\bibitem{Ding:2010ga}
H.-T. Ding, A.~Francis, O.~Kaczmarek, F.~Karsch, E.~Laermann, et~al., {Thermal
  dilepton rate and electrical conductivity: An analysis of vector current
  correlation functions in quenched lattice QCD}, Phys. Rev. D83 (2011) 034504.
\newblock \href {http://arxiv.org/abs/1012.4963} {\path{arXiv:1012.4963}},
  \href {https://doi.org/10.1103/PhysRevD.83.034504}
  {\path{doi:10.1103/PhysRevD.83.034504}}.

\bibitem{Karsch:2001uw}
F.~Karsch, E.~Laermann, P.~Petreczky, S.~Stickan, I.~Wetzorke, {A Lattice
  calculation of thermal dilepton rates}, Phys. Lett. B530 (2002) 147--152.
\newblock \href {http://arxiv.org/abs/hep-lat/0110208}
  {\path{arXiv:hep-lat/0110208}}, \href
  {https://doi.org/10.1016/S0370-2693(02)01326-6}
  {\path{doi:10.1016/S0370-2693(02)01326-6}}.

\bibitem{Asakawa:2000tr}
M.~Asakawa, T.~Hatsuda, Y.~Nakahara, {Maximum entropy analysis of the spectral
  functions in lattice QCD}, Prog. Part. Nucl. Phys. 46 (2001) 459--508.
\newblock \href {http://arxiv.org/abs/hep-lat/0011040}
  {\path{arXiv:hep-lat/0011040}}, \href
  {https://doi.org/10.1016/S0146-6410(01)00150-8}
  {\path{doi:10.1016/S0146-6410(01)00150-8}}.

\bibitem{Nakahara:1999vy}
Y.~Nakahara, M.~Asakawa, T.~Hatsuda, {Hadronic spectral functions in lattice
  QCD}, Phys. Rev. D 60 (1999) 091503.
\newblock \href {http://arxiv.org/abs/hep-lat/9905034}
  {\path{arXiv:hep-lat/9905034}}, \href
  {https://doi.org/10.1103/PhysRevD.60.091503}
  {\path{doi:10.1103/PhysRevD.60.091503}}.

\bibitem{Kim:2015poa}
T.~Kim, M.~Asakawa, M.~Kitazawa, {Dilepton production spectrum above Tc with a
  lattice quark propagator}, Phys. Rev. D 92~(11) (2015) 114014.
\newblock \href {http://arxiv.org/abs/1505.07195} {\path{arXiv:1505.07195}},
  \href {https://doi.org/10.1103/PhysRevD.92.114014}
  {\path{doi:10.1103/PhysRevD.92.114014}}.

\bibitem{Wright:1992tf}
E.~L. Wright, et~al., {Interpretation of the Cosmic Microwave Background
  radiation anisotropy detected by the COBE differential microwave radiometer},
  Astrophys. J. Lett. 396 (1992) L13--L18.
\newblock \href {https://doi.org/10.1086/186506} {\path{doi:10.1086/186506}}.

\bibitem{Shuryak:1978ij}
E.~V. Shuryak, {Quark-Gluon Plasma and Hadronic Production of Leptons, Photons
  and Psions}, Phys. Lett. B 78 (1978) 150.
\newblock \href {https://doi.org/10.1016/0370-2693(78)90370-2}
  {\path{doi:10.1016/0370-2693(78)90370-2}}.

\bibitem{Kajantie:1981wg}
K.~Kajantie, H.~I. Miettinen, {Temperature Measurement of Quark-Gluon Plasma
  Formed in High-Energy Nucleus-Nucleus Collisions}, Z. Phys. C 9 (1981) 341.
\newblock \href {https://doi.org/10.1007/BF01548770}
  {\path{doi:10.1007/BF01548770}}.

\bibitem{Halzen:1981kz}
F.~Halzen, H.~C. Liu, {Experimental Signatures of Phase Transition to Quark
  Matter in High-energy Collisions of Nuclei}, Phys. Rev. D 25 (1982) 1842.
\newblock \href {https://doi.org/10.1103/PhysRevD.25.1842}
  {\path{doi:10.1103/PhysRevD.25.1842}}.

\bibitem{Kajantie:1982nj}
K.~Kajantie, P.~V. Ruuskanen, {Shielding of Quark Mass Singularities in Photon
  Emission From Hot Quark - Gluon Plasma}, Phys. Lett. B 121 (1983) 352--354.
\newblock \href {https://doi.org/10.1016/0370-2693(83)91385-0}
  {\path{doi:10.1016/0370-2693(83)91385-0}}.

\bibitem{Hwa:1985xg}
R.~C. Hwa, K.~Kajantie, {Diagnosing Quark Matter by Measuring the Total Entropy
  and the Photon Or Dilepton Emission Rates}, Phys. Rev. D 32 (1985) 1109.
\newblock \href {https://doi.org/10.1103/PhysRevD.32.1109}
  {\path{doi:10.1103/PhysRevD.32.1109}}.

\bibitem{Staadt:1985uc}
G.~Staadt, W.~Greiner, J.~Rafelski, {Photons From Strange Quark Annihilation in
  Quark - Gluon Plasma}, Phys. Rev. D 33 (1986) 66.
\newblock \href {https://doi.org/10.1103/PhysRevD.33.66}
  {\path{doi:10.1103/PhysRevD.33.66}}.

\bibitem{Neubert:1989hu}
M.~Neubert, {Photon Production in Ultrarelativistic Heavy Ion Collisions at
  200-{GeV}/u}, Z. Phys. C 42 (1989) 231--242.
\newblock \href {https://doi.org/10.1007/BF01555862}
  {\path{doi:10.1007/BF01555862}}.

\bibitem{Weldon:1983jn}
H.~A. Weldon, {Simple Rules for Discontinuities in Finite Temperature Field
  Theory}, Phys. Rev. D 28 (1983) 2007.
\newblock \href {https://doi.org/10.1103/PhysRevD.28.2007}
  {\path{doi:10.1103/PhysRevD.28.2007}}.

\bibitem{Gelis:1997zv}
F.~Gelis, {Cutting rules in the real time formalisms at finite temperature},
  Nucl. Phys. B 508 (1997) 483--505.
\newblock \href {http://arxiv.org/abs/hep-ph/9701410}
  {\path{arXiv:hep-ph/9701410}}, \href
  {https://doi.org/10.1016/S0550-3213(97)00511-7}
  {\path{doi:10.1016/S0550-3213(97)00511-7}}.

\bibitem{Gale:1987ki}
C.~Gale, J.~I. Kapusta, {Dilepton radiation from high temperature nuclear
  matter}, Phys. Rev. C 35 (1987) 2107--2116.
\newblock \href {https://doi.org/10.1103/PhysRevC.35.2107}
  {\path{doi:10.1103/PhysRevC.35.2107}}.

\bibitem{Braaten:1991dd}
E.~Braaten, T.~C. Yuan, {Calculation of screening in a hot plasma}, Phys. Rev.
  Lett. 66 (1991) 2183--2186.
\newblock \href {https://doi.org/10.1103/PhysRevLett.66.2183}
  {\path{doi:10.1103/PhysRevLett.66.2183}}.

\bibitem{Leontovich:1961ne}
M.~Leontovich, {Generalization of the Kramers-Kronig formulas to media with
  spatial dispersion}, Sov. Phys. JETP 13 (1961) 634--637.

\bibitem{Thoma:2000ne}
M.~H. Thoma, {Leontovich relations in thermal field theory}, Eur. Phys. J. C 16
  (2000) 513--518.
\newblock \href {http://arxiv.org/abs/hep-ph/0004146}
  {\path{arXiv:hep-ph/0004146}}, \href {https://doi.org/10.1007/s100520000440}
  {\path{doi:10.1007/s100520000440}}.

\bibitem{Shuryak:1992bt}
E.~V. Shuryak, L.~Xiong, {Dilepton and photon production in the 'hot glue'
  scenario}, Phys. Rev. Lett. 70 (1993) 2241--2244.
\newblock \href {http://arxiv.org/abs/hep-ph/9301218}
  {\path{arXiv:hep-ph/9301218}}, \href
  {https://doi.org/10.1103/PhysRevLett.70.2241}
  {\path{doi:10.1103/PhysRevLett.70.2241}}.

\bibitem{Kampfer:1994rr}
B.~Kampfer, O.~P. Pavlenko, {Photon production in an expanding and chemically
  equilibrating gluon enriched plasma}, Z. Phys. C 62 (1994) 491--497.
\newblock \href {https://doi.org/10.1007/BF01555909}
  {\path{doi:10.1007/BF01555909}}.

\bibitem{Strickland:1994rf}
M.~Strickland, {Thermal photons and dileptons from nonequilibrium quark - gluon
  plasma}, Phys. Lett. B 331 (1994) 245--250.
\newblock \href {https://doi.org/10.1016/0370-2693(94)91045-6}
  {\path{doi:10.1016/0370-2693(94)91045-6}}.

\bibitem{Traxler:1995kx}
C.~T. Traxler, M.~H. Thoma, {Photon emission from a parton gas at chemical
  nonequilibrium}, Phys. Rev. C 53 (1996) 1348--1352.
\newblock \href {http://arxiv.org/abs/hep-ph/9507444}
  {\path{arXiv:hep-ph/9507444}}, \href
  {https://doi.org/10.1103/PhysRevC.53.1348}
  {\path{doi:10.1103/PhysRevC.53.1348}}.

\bibitem{Srivastava:1996qd}
D.~K. Srivastava, M.~G. Mustafa, B.~Muller, {Expanding quark - gluon plasmas:
  Transverse flow, chemical equilibration and electromagnetic radiation},
  Phys.Rev. C56 (1997) 1064--1074.
\newblock \href {http://arxiv.org/abs/nucl-th/9611041}
  {\path{arXiv:nucl-th/9611041}}, \href
  {https://doi.org/10.1103/PhysRevC.56.1064}
  {\path{doi:10.1103/PhysRevC.56.1064}}.

\bibitem{Pal:1998jr}
D.~Pal, M.~G. Mustafa, {Soft electromagnetic radiations from equilibrating
  quark gluon plasma}, Phys. Rev. C60 (1999) 034905.
\newblock \href {http://arxiv.org/abs/nucl-th/9808049}
  {\path{arXiv:nucl-th/9808049}}, \href
  {https://doi.org/10.1103/PhysRevC.60.034905}
  {\path{doi:10.1103/PhysRevC.60.034905}}.

\bibitem{Baier:1997xc}
R.~Baier, M.~Dirks, K.~Redlich, D.~Schiff, {Thermal photon production rate from
  nonequilibrium quantum field theory}, Phys. Rev. D 56 (1997) 2548--2554.
\newblock \href {http://arxiv.org/abs/hep-ph/9704262}
  {\path{arXiv:hep-ph/9704262}}, \href
  {https://doi.org/10.1103/PhysRevD.56.2548}
  {\path{doi:10.1103/PhysRevD.56.2548}}.

\bibitem{Dutta:2001ii}
D.~Dutta, S.~S.~V. Suryanarayana, A.~K. Mohanty, K.~Kumar, R.~K. Choudhury,
  {Hard photon production from unsaturated quark gluon plasma at two loop
  level}, Nucl. Phys. A 710 (2002) 415--438.
\newblock \href {http://arxiv.org/abs/hep-ph/0104134}
  {\path{arXiv:hep-ph/0104134}}, \href
  {https://doi.org/10.1016/S0375-9474(02)01166-1}
  {\path{doi:10.1016/S0375-9474(02)01166-1}}.

\bibitem{Steffen:2001pv}
F.~D. Steffen, M.~H. Thoma, {Hard thermal photon production in relativistic
  heavy ion collisions}, Phys. Lett. B 510 (2001) 98--106, [Erratum:
  Phys.Lett.B 660, 604--606 (2008)].
\newblock \href {http://arxiv.org/abs/hep-ph/0103044}
  {\path{arXiv:hep-ph/0103044}}, \href
  {https://doi.org/10.1016/S0370-2693(01)00525-1}
  {\path{doi:10.1016/S0370-2693(01)00525-1}}.

\bibitem{Srivastava:1999ekv}
D.~K. Srivastava, {Photon production in relativistic heavy ion collisions using
  rates with two loop calculations from quark matter}, Eur. Phys. J. C 10
  (1999) 487--490, [Erratum: Eur.Phys.J.C 20, 399--400 (2001)].
\newblock \href {http://arxiv.org/abs/nucl-th/0103023}
  {\path{arXiv:nucl-th/0103023}}, \href {https://doi.org/10.1007/s100520050772}
  {\path{doi:10.1007/s100520050772}}.

\bibitem{Mustafa:2000sg}
M.~G. Mustafa, M.~H. Thoma, {Bremsstrahlung from an equilibrating quark - gluon
  plasma}, Phys. Rev. C 62 (2000) 014902.
\newblock \href {http://arxiv.org/abs/hep-ph/0001230}
  {\path{arXiv:hep-ph/0001230}}, \href
  {https://doi.org/10.1103/PhysRevC.62.014902}
  {\path{doi:10.1103/PhysRevC.62.014902}}.

\bibitem{Mustafa:2001pf}
M.~G. Mustafa, M.~H. Thoma, {Erratum: Bremsstrahlung from an equilibrating
  quark gluon plasma}, Phys. Rev. C 63 (2001) 069902.
\newblock \href {http://arxiv.org/abs/hep-ph/0103293}
  {\path{arXiv:hep-ph/0103293}}, \href
  {https://doi.org/10.1103/PhysRevC.63.069902}
  {\path{doi:10.1103/PhysRevC.63.069902}}.

\bibitem{Aurenche:1999tq}
P.~Aurenche, F.~Gelis, H.~Zaraket, {KLN theorem, magnetic mass, and thermal
  photon production}, Phys. Rev. D61 (2000) 116001.
\newblock \href {http://arxiv.org/abs/hep-ph/9911367}
  {\path{arXiv:hep-ph/9911367}}, \href
  {https://doi.org/10.1103/PhysRevD.61.116001}
  {\path{doi:10.1103/PhysRevD.61.116001}}.

\bibitem{Mustafa:2008nk}
M.~G. Mustafa, B.~Kampfer, {Gamma flashes from relativistic electron-positron
  plasma droplets}, Phys. Rev. A 79 (2009) 020103.
\newblock \href {http://arxiv.org/abs/0809.2460} {\path{arXiv:0809.2460}},
  \href {https://doi.org/10.1103/PhysRevA.79.020103}
  {\path{doi:10.1103/PhysRevA.79.020103}}.

\bibitem{Yaresko:2010xe}
R.~Yaresko, M.~G. Mustafa, B.~Kampfer, {Relativistic Expansion of
  Electron-Positron-Photon Plasma Droplets and Photon Emission}, Phys. Plasmas
  17 (2010) 103302.
\newblock \href {http://arxiv.org/abs/1008.3495} {\path{arXiv:1008.3495}},
  \href {https://doi.org/10.1063/1.3499410} {\path{doi:10.1063/1.3499410}}.

\bibitem{ParticleDataGroup:2012pjm}
J.~Beringer, et~al., {Review of Particle Physics (RPP)}, Phys. Rev. D 86 (2012)
  010001.
\newblock \href {https://doi.org/10.1103/PhysRevD.86.010001}
  {\path{doi:10.1103/PhysRevD.86.010001}}.

\bibitem{Bazavov:2012ka}
A.~Bazavov, N.~Brambilla, X.~Garcia~i Tormo, P.~Petreczky, J.~Soto, et~al.,
  {Determination of $\alpha_s$ from the QCD static energy}, Phys. Rev. D86
  (2012) 114031.
\newblock \href {http://arxiv.org/abs/1205.6155} {\path{arXiv:1205.6155}},
  \href {https://doi.org/10.1103/PhysRevD.86.114031}
  {\path{doi:10.1103/PhysRevD.86.114031}}.

\bibitem{Kinoshita:1962ur}
T.~Kinoshita, {Mass singularities of Feynman amplitudes}, J. Math. Phys. 3
  (1962) 650--677.
\newblock \href {https://doi.org/10.1063/1.1724268}
  {\path{doi:10.1063/1.1724268}}.

\bibitem{lee1964degenerate}
T.-D. Lee, M.~Nauenberg, Degenerate systems and mass singularities, Physical
  Review 133~(6B) (1964) B1549.

\bibitem{Detar:1987hib}
C.~E. Detar, J.~B. Kogut, {Measuring the Hadronic Spectrum of the Quark
  Plasma}, Phys. Rev. D 36 (1987) 2828.
\newblock \href {https://doi.org/10.1103/PhysRevD.36.2828}
  {\path{doi:10.1103/PhysRevD.36.2828}}.

\bibitem{Hashimoto:1992np}
T.~Hashimoto, A.~Nakamura, I.~Stamatescu, {Temperature dependent structure in
  the mesonic channels of QCD}, Nucl. Phys. B400 (1993) 267--308.
\newblock \href {https://doi.org/10.1016/0550-3213(93)90407-G}
  {\path{doi:10.1016/0550-3213(93)90407-G}}.

\bibitem{Boyd:1994np}
G.~Boyd, S.~Gupta, F.~Karsch, E.~Laermann, {Spatial and temporal hadron
  correlators below and above the chiral phase transition}, Z. Phys. C64 (1994)
  331--338.
\newblock \href {http://arxiv.org/abs/hep-lat/9405006}
  {\path{arXiv:hep-lat/9405006}}, \href {https://doi.org/10.1007/BF01557406}
  {\path{doi:10.1007/BF01557406}}.

\bibitem{Shuryak:1993kg}
E.~V. Shuryak, {Correlation functions in the QCD vacuum}, Rev. Mod. Phys. 65
  (1993) 1--46.
\newblock \href {https://doi.org/10.1103/RevModPhys.65.1}
  {\path{doi:10.1103/RevModPhys.65.1}}.

\bibitem{Boyd:1995cw}
G.~Boyd, S.~Gupta, F.~Karsch, E.~Laermann, B.~Petersson, K.~Redlich, {Hadron
  properties just before deconfinement}, Phys. Lett. B 349 (1995) 170--176.
\newblock \href {http://arxiv.org/abs/hep-lat/9501029}
  {\path{arXiv:hep-lat/9501029}}, \href
  {https://doi.org/10.1016/0370-2693(95)00220-F}
  {\path{doi:10.1016/0370-2693(95)00220-F}}.

\bibitem{QCD-TARO:2001jaq}
P.~de~Forcrand, M.~Garcia~Perez, T.~Hashimoto, S.~Hioki, H.~Matsufuru,
  O.~Miyamura, A.~Nakamura, I.~O. Stamatescu, T.~Takaishi, T.~Umeda, {Mesons
  above the deconfining transition} (1 2001).
\newblock \href {http://arxiv.org/abs/hep-lat/9901017}
  {\path{arXiv:hep-lat/9901017}}.

\bibitem{Brown:1991kk}
G.~E. Brown, M.~Rho, {Scaling effective Lagrangians in a dense medium}, Phys.
  Rev. Lett. 66 (1991) 2720--2723.
\newblock \href {https://doi.org/10.1103/PhysRevLett.66.2720}
  {\path{doi:10.1103/PhysRevLett.66.2720}}.

\bibitem{Friman:1997tc}
B.~Friman, H.~J. Pirner, {P wave polarization of the rho meson and the dilepton
  spectrum in dense matter}, Nucl. Phys. A 617 (1997) 496--509.
\newblock \href {http://arxiv.org/abs/nucl-th/9701016}
  {\path{arXiv:nucl-th/9701016}}, \href
  {https://doi.org/10.1016/S0375-9474(97)00050-X}
  {\path{doi:10.1016/S0375-9474(97)00050-X}}.

\bibitem{Rapp:1997fs}
R.~Rapp, G.~Chanfray, J.~Wambach, {Rho meson propagation and dilepton
  enhancement in hot hadronic matter}, Nucl. Phys. A 617 (1997) 472--495.
\newblock \href {http://arxiv.org/abs/hep-ph/9702210}
  {\path{arXiv:hep-ph/9702210}}, \href
  {https://doi.org/10.1016/S0375-9474(97)00137-1}
  {\path{doi:10.1016/S0375-9474(97)00137-1}}.

\bibitem{Thoma:1999nm}
M.~H. Thoma, S.~Leupold, U.~Mosel, {Dilepton production from rho mesons in a
  quark gluon plasma}, Eur. Phys. J. A 7 (2000) 219--223.
\newblock \href {http://arxiv.org/abs/nucl-th/9905016}
  {\path{arXiv:nucl-th/9905016}}, \href {https://doi.org/10.1007/s100500050384}
  {\path{doi:10.1007/s100500050384}}.

\bibitem{Islam:2014sea}
C.~A. Islam, S.~Majumder, N.~Haque, M.~G. Mustafa, {Vector meson spectral
  function and dilepton production rate in a hot and dense medium within an
  effective QCD approach}, JHEP 02 (2015) 011.
\newblock \href {http://arxiv.org/abs/1411.6407} {\path{arXiv:1411.6407}},
  \href {https://doi.org/10.1007/JHEP02(2015)011}
  {\path{doi:10.1007/JHEP02(2015)011}}.

\bibitem{Jarrell:1996rrw}
M.~Jarrell, J.~E. Gubernatis, {Bayesian inference and the analytic continuation
  of imaginary-time quantum Monte Carlo data}, Phys. Rept. 269 (1996) 133--195.
\newblock \href {https://doi.org/10.1016/0370-1573(95)00074-7}
  {\path{doi:10.1016/0370-1573(95)00074-7}}.

\bibitem{Thoma:1997bi}
M.~H. Thoma, {QCD perturbation theory at finite temperature / density and its
  application}, Nucl. Phys. A 638 (1998) 317C--328C.
\newblock \href {http://arxiv.org/abs/hep-ph/9801266}
  {\path{arXiv:hep-ph/9801266}}, \href
  {https://doi.org/10.1016/S0375-9474(98)00372-8}
  {\path{doi:10.1016/S0375-9474(98)00372-8}}.

\bibitem{Schaffner-Bielich:1998mra}
J.~Schaffner-Bielich, J.~Randrup, {DCC dynamics with the SU(3) linear sigma
  model}, Phys. Rev. C 59 (1999) 3329--3342.
\newblock \href {http://arxiv.org/abs/nucl-th/9812032}
  {\path{arXiv:nucl-th/9812032}}, \href
  {https://doi.org/10.1103/PhysRevC.59.3329}
  {\path{doi:10.1103/PhysRevC.59.3329}}.

\bibitem{Hatsuda:1994pi}
T.~Hatsuda, T.~Kunihiro, {QCD phenomenology based on a chiral effective
  Lagrangian}, Phys. Rept. 247 (1994) 221--367.
\newblock \href {http://arxiv.org/abs/hep-ph/9401310}
  {\path{arXiv:hep-ph/9401310}}, \href
  {https://doi.org/10.1016/0370-1573(94)90022-1}
  {\path{doi:10.1016/0370-1573(94)90022-1}}.

\bibitem{Kunihiro:1991qu}
T.~Kunihiro, {Quark number susceptibility and fluctuations in the vector
  channel at high temperatures}, Phys. Lett. B271 (1991) 395--402.
\newblock \href {https://doi.org/10.1016/0370-2693(91)90107-2}
  {\path{doi:10.1016/0370-2693(91)90107-2}}.

\bibitem{Florkowski:1993bq}
W.~Florkowski, B.~L. Friman, {Spatial dependence of the finite temperature
  meson correlation function}, Z. Phys. A 347 (1994) 271--276.
\newblock \href {https://doi.org/10.1007/BF01289794}
  {\path{doi:10.1007/BF01289794}}.

\bibitem{Thoma:1994yw}
M.~H. Thoma, {Damping of a Yukawa fermion at finite temperature}, Z. Phys. C 66
  (1995) 491--494.
\newblock \href {http://arxiv.org/abs/hep-ph/9406242}
  {\path{arXiv:hep-ph/9406242}}, \href {https://doi.org/10.1007/BF01556376}
  {\path{doi:10.1007/BF01556376}}.

\bibitem{Kraemmer:1994az}
U.~Kraemmer, A.~Rebhan, H.~Schulz, {Resummations in hot scalar
  electrodynamics}, Annals Phys. 238 (1995) 286--331.
\newblock \href {http://arxiv.org/abs/hep-ph/9403301}
  {\path{arXiv:hep-ph/9403301}}, \href {https://doi.org/10.1006/aphy.1995.1023}
  {\path{doi:10.1006/aphy.1995.1023}}.

\bibitem{Blaizot:1993be}
J.~P. Blaizot, E.~Iancu, {Soft collective excitations in hot gauge theories},
  Nucl. Phys. B 417 (1994) 608--673.
\newblock \href {http://arxiv.org/abs/hep-ph/9306294}
  {\path{arXiv:hep-ph/9306294}}, \href
  {https://doi.org/10.1016/0550-3213(94)90486-3}
  {\path{doi:10.1016/0550-3213(94)90486-3}}.

\bibitem{Baier:1993zb}
R.~Baier, S.~Peigne, D.~Schiff, {Soft photon production rate in resummed
  perturbation theory of high temperature QCD}, Z. Phys. C62 (1994) 337--342.
\newblock \href {http://arxiv.org/abs/hep-ph/9311329}
  {\path{arXiv:hep-ph/9311329}}, \href {https://doi.org/10.1007/BF01560248}
  {\path{doi:10.1007/BF01560248}}.

\bibitem{Alberico:2004we}
W.~M. Alberico, A.~Beraudo, A.~Molinari, {Meson correlation functions in a QCD
  plasma}, Nucl. Phys. A 750 (2005) 359--390.
\newblock \href {http://arxiv.org/abs/hep-ph/0411346}
  {\path{arXiv:hep-ph/0411346}}, \href
  {https://doi.org/10.1016/j.nuclphysa.2004.12.070}
  {\path{doi:10.1016/j.nuclphysa.2004.12.070}}.

\bibitem{Alberico:2006wc}
W.~M. Alberico, A.~Beraudo, P.~Czerski, A.~Molinari, {Finite momentum meson
  correlation functions in a QCD plasma}, Nucl. Phys. A 775 (2006) 188--211.
\newblock \href {http://arxiv.org/abs/hep-ph/0605060}
  {\path{arXiv:hep-ph/0605060}}, \href
  {https://doi.org/10.1016/j.nuclphysa.2006.06.006}
  {\path{doi:10.1016/j.nuclphysa.2006.06.006}}.

\bibitem{Czerski:2008zz}
P.~Czerski, {HTL meson correlation functions at finite momentum and chemical
  potential}, Nucl. Phys. A 807 (2008) 11--27.
\newblock \href {https://doi.org/10.1016/j.nuclphysa.2008.03.012}
  {\path{doi:10.1016/j.nuclphysa.2008.03.012}}.

\bibitem{Blaizot:2001vr}
J.~Blaizot, E.~Iancu, A.~Rebhan, {Quark number susceptibilities from HTL
  resummed thermodynamics}, Phys. Lett. B523 (2001) 143--150.
\newblock \href {http://arxiv.org/abs/hep-ph/0110369}
  {\path{arXiv:hep-ph/0110369}}, \href
  {https://doi.org/10.1016/S0370-2693(01)01316-8}
  {\path{doi:10.1016/S0370-2693(01)01316-8}}.

\bibitem{Blaizot:2002xz}
J.~Blaizot, E.~Iancu, A.~Rebhan, {Comparing different hard thermal loop
  approaches to quark number susceptibilities}, Eur. Phys. J. C27 (2003)
  433--438.
\newblock \href {http://arxiv.org/abs/hep-ph/0206280}
  {\path{arXiv:hep-ph/0206280}}, \href
  {https://doi.org/10.1140/epjc/s2002-01103-5}
  {\path{doi:10.1140/epjc/s2002-01103-5}}.

\bibitem{Toimela:1984xy}
T.~Toimela, {Perturbative {QED} and {QCD} at Finite Temperatures and
  Densities}, Int. J. Theor. Phys. 24 (1985) 901, [Erratum: Int.J.Theor.Phys.
  26, 1021 (1987)].
\newblock \href {https://doi.org/10.1007/BF00671334}
  {\path{doi:10.1007/BF00671334}}.

\bibitem{Allton:2005gk}
C.~Allton, M.~Doring, S.~Ejiri, S.~Hands, O.~Kaczmarek, et~al., {Thermodynamics
  of two flavor QCD to sixth order in quark chemical potential}, Phys. Rev. D71
  (2005) 054508.
\newblock \href {http://arxiv.org/abs/hep-lat/0501030}
  {\path{arXiv:hep-lat/0501030}}, \href
  {https://doi.org/10.1103/PhysRevD.71.054508}
  {\path{doi:10.1103/PhysRevD.71.054508}}.

\bibitem{Petreczky:2009cr}
P.~Petreczky, P.~Hegde, A.~Velytsky, {Quark number fluctuations at high
  temperatures}, PoS LAT2009 (2009) 159.
\newblock \href {http://arxiv.org/abs/0911.0196} {\path{arXiv:0911.0196}}.

\bibitem{Bazavov:2009zn}
A.~Bazavov, et~al., {Equation of state and QCD transition at finite
  temperature}, Phys. Rev. D 80 (2009) 014504.
\newblock \href {http://arxiv.org/abs/0903.4379} {\path{arXiv:0903.4379}},
  \href {https://doi.org/10.1103/PhysRevD.80.014504}
  {\path{doi:10.1103/PhysRevD.80.014504}}.

\bibitem{Bernard:2004je}
C.~Bernard, et~al., {QCD thermodynamics with three flavors of improved
  staggered quarks}, Phys. Rev. D71 (2005) 034504.
\newblock \href {http://arxiv.org/abs/hep-lat/0405029}
  {\path{arXiv:hep-lat/0405029}}, \href
  {https://doi.org/10.1103/PhysRevD.71.034504}
  {\path{doi:10.1103/PhysRevD.71.034504}}.

\bibitem{Gavai:2001fr}
R.~V. Gavai, S.~Gupta, {Quark number susceptibilities, strangeness and
  dynamical confinement}, Phys. Rev. D64 (2001) 074506.
\newblock \href {http://arxiv.org/abs/hep-lat/0103013}
  {\path{arXiv:hep-lat/0103013}}, \href
  {https://doi.org/10.1103/PhysRevD.64.074506}
  {\path{doi:10.1103/PhysRevD.64.074506}}.

\bibitem{Gavai:2001ie}
R.~V. Gavai, S.~Gupta, P.~Majumdar, {Susceptibilities and screening masses in
  two flavor QCD}, Phys. Rev. D65 (2002) 054506.
\newblock \href {http://arxiv.org/abs/hep-lat/0110032}
  {\path{arXiv:hep-lat/0110032}}, \href
  {https://doi.org/10.1103/PhysRevD.65.054506}
  {\path{doi:10.1103/PhysRevD.65.054506}}.

\bibitem{Cheng:2009zi}
M.~Cheng, S.~Ejiri, P.~Hegde, F.~Karsch, O.~Kaczmarek, et~al., {Equation of
  State for physical quark masses}, Phys. Rev. D81 (2010) 054504.
\newblock \href {http://arxiv.org/abs/0911.2215} {\path{arXiv:0911.2215}},
  \href {https://doi.org/10.1103/PhysRevD.81.054504}
  {\path{doi:10.1103/PhysRevD.81.054504}}.

\bibitem{Petreczky:2009at}
P.~Petreczky, {Lattice QCD at finite temperature : Present status}, Nucl. Phys.
  A830 (2009) 11C--18C.
\newblock \href {http://arxiv.org/abs/0908.1917} {\path{arXiv:0908.1917}},
  \href {https://doi.org/10.1016/j.nuclphysa.2009.10.086}
  {\path{doi:10.1016/j.nuclphysa.2009.10.086}}.

\bibitem{Peshier:1999ww}
A.~Peshier, B.~Kampfer, G.~Soff, {The Equation of state of deconfined matter at
  finite chemical potential in a quasiparticle description}, Phys. Rev. C61
  (2000) 045203.
\newblock \href {http://arxiv.org/abs/hep-ph/9911474}
  {\path{arXiv:hep-ph/9911474}}, \href
  {https://doi.org/10.1103/PhysRevC.61.045203}
  {\path{doi:10.1103/PhysRevC.61.045203}}.

\bibitem{Peshier:2002ww}
A.~Peshier, B.~Kampfer, G.~Soff, {From QCD lattice calculations to the equation
  of state of quark matter}, Phys. Rev. D66 (2002) 094003.
\newblock \href {http://arxiv.org/abs/hep-ph/0206229}
  {\path{arXiv:hep-ph/0206229}}, \href
  {https://doi.org/10.1103/PhysRevD.66.094003}
  {\path{doi:10.1103/PhysRevD.66.094003}}.

\bibitem{Pisarski:2000eq}
R.~D. Pisarski, {Quark gluon plasma as a condensate of SU(3) Wilson lines},
  Phys. Rev. D62 (2000) 111501.
\newblock \href {http://arxiv.org/abs/hep-ph/0006205}
  {\path{arXiv:hep-ph/0006205}}, \href
  {https://doi.org/10.1103/PhysRevD.62.111501}
  {\path{doi:10.1103/PhysRevD.62.111501}}.

\bibitem{Ratti:2005jh}
C.~Ratti, M.~A. Thaler, W.~Weise, {Phases of QCD: Lattice thermodynamics and a
  field theoretical model}, Phys. Rev. D73 (2006) 014019.
\newblock \href {http://arxiv.org/abs/hep-ph/0506234}
  {\path{arXiv:hep-ph/0506234}}, \href
  {https://doi.org/10.1103/PhysRevD.73.014019}
  {\path{doi:10.1103/PhysRevD.73.014019}}.

\bibitem{Ghosh:2006qh}
S.~K. Ghosh, T.~K. Mukherjee, M.~G. Mustafa, R.~Ray, {Susceptibilities and
  speed of sound from PNJL model}, Phys. Rev. D73 (2006) 114007.
\newblock \href {http://arxiv.org/abs/hep-ph/0603050}
  {\path{arXiv:hep-ph/0603050}}, \href
  {https://doi.org/10.1103/PhysRevD.73.114007}
  {\path{doi:10.1103/PhysRevD.73.114007}}.

\bibitem{Ghosh:2007wy}
S.~K. Ghosh, T.~K. Mukherjee, M.~G. Mustafa, R.~Ray, {PNJL model with a Van der
  Monde term}, Phys. Rev. D77 (2008) 094024.
\newblock \href {http://arxiv.org/abs/0710.2790} {\path{arXiv:0710.2790}},
  \href {https://doi.org/10.1103/PhysRevD.77.094024}
  {\path{doi:10.1103/PhysRevD.77.094024}}.

\bibitem{Mukherjee:2006hq}
S.~Mukherjee, M.~G. Mustafa, R.~Ray, {Thermodynamics of the PNJL model with
  nonzero baryon and isospin chemical potentials}, Phys. Rev. D75 (2007)
  094015.
\newblock \href {http://arxiv.org/abs/hep-ph/0609249}
  {\path{arXiv:hep-ph/0609249}}, \href
  {https://doi.org/10.1103/PhysRevD.75.094015}
  {\path{doi:10.1103/PhysRevD.75.094015}}.

\bibitem{Roessner:2006xn}
S.~Roessner, C.~Ratti, W.~Weise, {Polyakov loop, diquarks and the two-flavour
  phase diagram}, Phys. Rev. D75 (2007) 034007.
\newblock \href {http://arxiv.org/abs/hep-ph/0609281}
  {\path{arXiv:hep-ph/0609281}}, \href
  {https://doi.org/10.1103/PhysRevD.75.034007}
  {\path{doi:10.1103/PhysRevD.75.034007}}.

\bibitem{Sasaki:2006ww}
C.~Sasaki, B.~Friman, K.~Redlich, {Susceptibilities and the Phase Structure of
  a Chiral Model with Polyakov Loops}, Phys. Rev. D75 (2007) 074013.
\newblock \href {http://arxiv.org/abs/hep-ph/0611147}
  {\path{arXiv:hep-ph/0611147}}, \href
  {https://doi.org/10.1103/PhysRevD.75.074013}
  {\path{doi:10.1103/PhysRevD.75.074013}}.

\bibitem{DeGrand:2006zz}
T.~DeGrand, C.~DeTar, Lattice Methods for Quantum Chromodynamics, World
  Scientific, 2006.

\bibitem{Gavai:2003mf}
R.~V. Gavai, S.~Gupta, {Pressure and nonlinear susceptibilities in QCD at
  finite chemical potentials}, Phys. Rev. D68 (2003) 034506.
\newblock \href {http://arxiv.org/abs/hep-lat/0303013}
  {\path{arXiv:hep-lat/0303013}}, \href
  {https://doi.org/10.1103/PhysRevD.68.034506}
  {\path{doi:10.1103/PhysRevD.68.034506}}.

\bibitem{Borsanyi:2012cr}
S.~Borsanyi, G.~Endrodi, Z.~Fodor, S.~Katz, S.~Krieg, et~al., {QCD equation of
  state at nonzero chemical potential: continuum results with physical quark
  masses at order $\mu^2$}, JHEP 1208 (2012) 053.
\newblock \href {http://arxiv.org/abs/1204.6710} {\path{arXiv:1204.6710}},
  \href {https://doi.org/10.1007/JHEP08(2012)053}
  {\path{doi:10.1007/JHEP08(2012)053}}.

\bibitem{Shuryak:1977ut}
E.~V. Shuryak, {Theory of Hadronic Plasma}, Sov. Phys. JETP 47 (1978) 212--219.

\bibitem{Chin:1978gj}
S.~Chin, {Transition to Hot Quark Matter in Relativistic Heavy Ion Collision},
  Phys. Lett. B78 (1978) 552--555.
\newblock \href {https://doi.org/10.1016/0370-2693(78)90637-8}
  {\path{doi:10.1016/0370-2693(78)90637-8}}.

\bibitem{Kapusta:1979fh}
J.~I. Kapusta, {Quantum Chromodynamics at High Temperature}, Nucl. Phys. B148
  (1979) 461--498.
\newblock \href {https://doi.org/10.1016/0550-3213(79)90146-9}
  {\path{doi:10.1016/0550-3213(79)90146-9}}.

\bibitem{Toimela:1982hv}
T.~Toimela, {The Next Term in the Thermodynamic Potential of {QCD}}, Phys.
  Lett. B124 (1983) 407.
\newblock \href {https://doi.org/10.1016/0370-2693(83)91484-3}
  {\path{doi:10.1016/0370-2693(83)91484-3}}.

\bibitem{Arnold:1994ps}
P.~B. Arnold, C.-X. Zhai, {The Three loop free energy for pure gauge QCD},
  Phys. Rev. D50 (1994) 7603--7623.
\newblock \href {http://arxiv.org/abs/hep-ph/9408276}
  {\path{arXiv:hep-ph/9408276}}, \href
  {https://doi.org/10.1103/PhysRevD.50.7603}
  {\path{doi:10.1103/PhysRevD.50.7603}}.

\bibitem{Arnold:1994eb}
P.~B. Arnold, C.-x. Zhai, {The Three loop free energy for high temperature QED
  and QCD with fermions}, Phys. Rev. D51 (1995) 1906--1918.
\newblock \href {http://arxiv.org/abs/hep-ph/9410360}
  {\path{arXiv:hep-ph/9410360}}, \href
  {https://doi.org/10.1103/PhysRevD.51.1906}
  {\path{doi:10.1103/PhysRevD.51.1906}}.

\bibitem{Zhai:1995ac}
C.-x. Zhai, B.~M. Kastening, {The Free energy of hot gauge theories with
  fermions through $g^5$}, Phys. Rev. D52 (1995) 7232--7246.
\newblock \href {http://arxiv.org/abs/hep-ph/9507380}
  {\path{arXiv:hep-ph/9507380}}, \href
  {https://doi.org/10.1103/PhysRevD.52.7232}
  {\path{doi:10.1103/PhysRevD.52.7232}}.

\bibitem{Vuorinen:2003fs}
A.~Vuorinen, {The Pressure of QCD at finite temperatures and chemical
  potentials}, Phys. Rev. D68 (2003) 054017.
\newblock \href {http://arxiv.org/abs/hep-ph/0305183}
  {\path{arXiv:hep-ph/0305183}}, \href
  {https://doi.org/10.1103/PhysRevD.68.054017}
  {\path{doi:10.1103/PhysRevD.68.054017}}.

\bibitem{Ipp:2006ij}
A.~Ipp, K.~Kajantie, A.~Rebhan, A.~Vuorinen, {The Pressure of deconfined QCD
  for all temperatures and quark chemical potentials}, Phys. Rev. D74 (2006)
  045016.
\newblock \href {http://arxiv.org/abs/hep-ph/0604060}
  {\path{arXiv:hep-ph/0604060}}, \href
  {https://doi.org/10.1103/PhysRevD.74.045016}
  {\path{doi:10.1103/PhysRevD.74.045016}}.

\bibitem{Vuorinen:2002ue}
A.~Vuorinen, {Quark number susceptibilities of hot QCD up to} $g^6 \ln g$,
  Phys. Rev. D67 (2003) 074032.
\newblock \href {http://arxiv.org/abs/hep-ph/0212283}
  {\path{arXiv:hep-ph/0212283}}, \href
  {https://doi.org/10.1103/PhysRevD.67.074032}
  {\path{doi:10.1103/PhysRevD.67.074032}}.

\bibitem{Rebhan:2003wn}
A.~Rebhan, P.~Romatschke, {HTL quasiparticle models of deconfined QCD at finite
  chemical potential}, Phys. Rev. D68 (2003) 025022.
\newblock \href {http://arxiv.org/abs/hep-ph/0304294}
  {\path{arXiv:hep-ph/0304294}}, \href
  {https://doi.org/10.1103/PhysRevD.68.025022}
  {\path{doi:10.1103/PhysRevD.68.025022}}.

\bibitem{Cassing:2007nb}
W.~Cassing, {Dynamical quasiparticles properties and effective interactions in
  the sQGP}, Nucl. Phys. A795 (2007) 70--97.
\newblock \href {http://arxiv.org/abs/0707.3033} {\path{arXiv:0707.3033}},
  \href {https://doi.org/10.1016/j.nuclphysa.2007.08.010}
  {\path{doi:10.1016/j.nuclphysa.2007.08.010}}.

\bibitem{Gardim:2009mt}
F.~Gardim, F.~Steffens, {Thermodynamics of Quasi-Particles at Finite Chemical
  Potential}, Nucl.Phys. A825 (2009) 222--244.
\newblock \href {http://arxiv.org/abs/0905.0667} {\path{arXiv:0905.0667}},
  \href {https://doi.org/10.1016/j.nuclphysa.2009.05.001}
  {\path{doi:10.1016/j.nuclphysa.2009.05.001}}.

\bibitem{Kurkela:2009gj}
A.~Kurkela, P.~Romatschke, A.~Vuorinen, {Cold Quark Matter}, Phys. Rev. D81
  (2010) 105021.
\newblock \href {http://arxiv.org/abs/0912.1856} {\path{arXiv:0912.1856}},
  \href {https://doi.org/10.1103/PhysRevD.81.105021}
  {\path{doi:10.1103/PhysRevD.81.105021}}.

\bibitem{Andersen:2012wr}
J.~O. Andersen, S.~Mogliacci, N.~Su, A.~Vuorinen, {Quark number
  susceptibilities from resummed perturbation theory}, Phys. Rev. D87 (2013)
  074003.
\newblock \href {http://arxiv.org/abs/1210.0912} {\path{arXiv:1210.0912}},
  \href {https://doi.org/10.1103/PhysRevD.87.074003}
  {\path{doi:10.1103/PhysRevD.87.074003}}.

\bibitem{Petreczky:2012rq}
P.~Petreczky, {Lattice QCD at non-zero temperature}, J.Phys. G39 (2012) 093002.
\newblock \href {http://arxiv.org/abs/1203.5320} {\path{arXiv:1203.5320}},
  \href {https://doi.org/10.1088/0954-3899/39/9/093002}
  {\path{doi:10.1088/0954-3899/39/9/093002}}.

\bibitem{Borsanyi:2012rr}
S.~Borsanyi, {Thermodynamics of the QCD transition from lattice}, Nucl. Phys.
  A904-905 (2013) 270c--277c.
\newblock \href {http://arxiv.org/abs/1210.6901} {\path{arXiv:1210.6901}},
  \href {https://doi.org/10.1016/j.nuclphysa.2013.01.072}
  {\path{doi:10.1016/j.nuclphysa.2013.01.072}}.

\bibitem{Bhattacharyya:2010jd}
A.~Bhattacharyya, P.~Deb, A.~Lahiri, R.~Ray, {Susceptibilities with multi-quark
  interactions in PNJL model}, Phys. Rev. D82 (2010) 114028.
\newblock \href {http://arxiv.org/abs/1008.0768} {\path{arXiv:1008.0768}},
  \href {https://doi.org/10.1103/PhysRevD.82.114028}
  {\path{doi:10.1103/PhysRevD.82.114028}}.

\bibitem{Bhattacharyya:2010ef}
A.~Bhattacharyya, P.~Deb, A.~Lahiri, R.~Ray, {Correlation between conserved
  charges in PNJL Model with multi-quark interactions}, Phys. Rev. D83 (2011)
  014011.
\newblock \href {http://arxiv.org/abs/1010.2394} {\path{arXiv:1010.2394}},
  \href {https://doi.org/10.1103/PhysRevD.83.014011}
  {\path{doi:10.1103/PhysRevD.83.014011}}.

\bibitem{Cheng:2008zh}
M.~Cheng, P.~Hendge, C.~Jung, F.~Karsch, O.~Kaczmarek, et~al., {Baryon Number,
  Strangeness and Electric Charge Fluctuations in QCD at High Temperature},
  Phys. Rev. D79 (2009) 074505.
\newblock \href {http://arxiv.org/abs/0811.1006} {\path{arXiv:0811.1006}},
  \href {https://doi.org/10.1103/PhysRevD.79.074505}
  {\path{doi:10.1103/PhysRevD.79.074505}}.

\bibitem{Borsanyi:2010cj}
S.~Borsanyi, G.~Endrodi, Z.~Fodor, A.~Jakovac, S.~D. Katz, et~al., {The QCD
  equation of state with dynamical quarks}, JHEP 1011 (2010) 077.
\newblock \href {http://arxiv.org/abs/1007.2580} {\path{arXiv:1007.2580}},
  \href {https://doi.org/10.1007/JHEP11(2010)077}
  {\path{doi:10.1007/JHEP11(2010)077}}.

\bibitem{Borsanyi:2012uq}
S.~Borsanyi, S.~Durr, Z.~Fodor, C.~Hoelbling, S.~D. Katz, et~al., {QCD
  thermodynamics with continuum extrapolated Wilson fermions I}, JHEP 1208
  (2012) 126.
\newblock \href {http://arxiv.org/abs/1205.0440} {\path{arXiv:1205.0440}},
  \href {https://doi.org/10.1007/JHEP08(2012)126}
  {\path{doi:10.1007/JHEP08(2012)126}}.

\bibitem{Huovinen:2009yb}
P.~Huovinen, P.~Petreczky, {QCD Equation of State and Hadron Resonance Gas},
  Nucl. Phys. A837 (2010) 26--53.
\newblock \href {http://arxiv.org/abs/0912.2541} {\path{arXiv:0912.2541}},
  \href {https://doi.org/10.1016/j.nuclphysa.2010.02.015}
  {\path{doi:10.1016/j.nuclphysa.2010.02.015}}.

\bibitem{KorthalsAltes:1999xb}
C.~Korthals-Altes, A.~Kovner, M.~A. Stephanov, {Spatial 't Hooft loop, hot QCD
  and Z(N) domain walls}, Phys. Lett. B469 (1999) 205--212.
\newblock \href {http://arxiv.org/abs/hep-ph/9909516}
  {\path{arXiv:hep-ph/9909516}}, \href
  {https://doi.org/10.1016/S0370-2693(99)01242-3}
  {\path{doi:10.1016/S0370-2693(99)01242-3}}.

\bibitem{KorthalsAltes:2000gs}
C.~Korthals-Altes, A.~Kovner, {Magnetic Z(N) symmetry in hot QCD and the
  spatial Wilson loop}, Phys. Rev. D62 (2000) 096008.
\newblock \href {http://arxiv.org/abs/hep-ph/0004052}
  {\path{arXiv:hep-ph/0004052}}, \href
  {https://doi.org/10.1103/PhysRevD.62.096008}
  {\path{doi:10.1103/PhysRevD.62.096008}}.

\bibitem{Zwanziger:2004np}
D.~Zwanziger, {Equation of state of gluon plasma from fundamental modular
  region}, Phys. Rev. Lett. 94 (2005) 182301.
\newblock \href {http://arxiv.org/abs/hep-ph/0407103}
  {\path{arXiv:hep-ph/0407103}}, \href
  {https://doi.org/10.1103/PhysRevLett.94.182301}
  {\path{doi:10.1103/PhysRevLett.94.182301}}.

\bibitem{Vuorinen:2006nz}
A.~Vuorinen, L.~G. Yaffe, {Z(3)-symmetric effective theory for SU(3) Yang-Mills
  theory at high temperature}, Phys. Rev. D74 (2006) 025011.
\newblock \href {http://arxiv.org/abs/hep-ph/0604100}
  {\path{arXiv:hep-ph/0604100}}, \href
  {https://doi.org/10.1103/PhysRevD.74.025011}
  {\path{doi:10.1103/PhysRevD.74.025011}}.

\bibitem{deForcrand:2008aw}
P.~de~Forcrand, A.~Kurkela, A.~Vuorinen, {Center-Symmetric Effective Theory for
  High-Temperature SU(2) Yang-Mills Theory}, Phys. Rev. D77 (2008) 125014.
\newblock \href {http://arxiv.org/abs/0801.1566} {\path{arXiv:0801.1566}},
  \href {https://doi.org/10.1103/PhysRevD.77.125014}
  {\path{doi:10.1103/PhysRevD.77.125014}}.

\bibitem{Fukushima:2013xsa}
K.~Fukushima, N.~Su, {Stabilizing perturbative Yang-Mills thermodynamics with
  Gribov quantization}, Phys. Rev. D88 (2013) 076008.
\newblock \href {http://arxiv.org/abs/1304.8004} {\path{arXiv:1304.8004}},
  \href {https://doi.org/10.1103/PhysRevD.88.076008}
  {\path{doi:10.1103/PhysRevD.88.076008}}.

\bibitem{Borsanyi:2011sw}
S.~Borsanyi, Z.~Fodor, S.~D. Katz, S.~Krieg, C.~Ratti, et~al., {Fluctuations of
  conserved charges at finite temperature from lattice QCD}, JHEP 1201 (2012)
  138.
\newblock \href {http://arxiv.org/abs/1112.4416} {\path{arXiv:1112.4416}},
  \href {https://doi.org/10.1007/JHEP01(2012)138}
  {\path{doi:10.1007/JHEP01(2012)138}}.

\bibitem{Bazavov:2013dta}
A.~Bazavov, H.~T. Ding, P.~Hegde, O.~Kaczmarek, F.~Karsch, et~al., {Strangeness
  at high temperatures: from hadrons to quarks}, Phys. Rev. Lett. 111, 082301
  (2013) 082301.
\newblock \href {http://arxiv.org/abs/1304.7220} {\path{arXiv:1304.7220}},
  \href {https://doi.org/10.1103/PhysRevLett.111.082301}
  {\path{doi:10.1103/PhysRevLett.111.082301}}.

\bibitem{Bazavov:2013uja}
A.~Bazavov, H.-T. Ding, P.~Hegde, F.~Karsch, C.~Miao, et~al., {Quark number
  susceptibilities at high temperatures}, Phys. Rev. D88~(9) (2013) 094021.
\newblock \href {http://arxiv.org/abs/1309.2317} {\path{arXiv:1309.2317}},
  \href {https://doi.org/10.1103/PhysRevD.88.094021}
  {\path{doi:10.1103/PhysRevD.88.094021}}.

\bibitem{Datta:2014zqa}
S.~Datta, R.~Gavai, S.~Gupta, {QCD at finite chemical potential with $N_t =
  8$}, PoS LATTICE2013 (2014) 202.

\bibitem{Borsanyi:2013hza}
S.~Borsanyi, Z.~Fodor, S.~Katz, S.~Krieg, C.~Ratti, et~al., {Freeze-out
  parameters: lattice meets experiment}, Phys. Rev. Lett. 111 (2013) 062005.
\newblock \href {http://arxiv.org/abs/1305.5161} {\path{arXiv:1305.5161}},
  \href {https://doi.org/10.1103/PhysRevLett.111.062005}
  {\path{doi:10.1103/PhysRevLett.111.062005}}.

\bibitem{Borsanyi:2018grb}
S.~Borsanyi, Z.~Fodor, J.~N. Guenther, S.~K. Katz, K.~K. Szabo, A.~Pasztor,
  I.~Portillo, C.~Ratti, {Higher order fluctuations and correlations of
  conserved charges from lattice QCD}, JHEP 10 (2018) 205.
\newblock \href {http://arxiv.org/abs/1805.04445} {\path{arXiv:1805.04445}},
  \href {https://doi.org/10.1007/JHEP10(2018)205}
  {\path{doi:10.1007/JHEP10(2018)205}}.

\bibitem{Borsanyi:2024anr}
S.~Bors\'anyi, Z.~Fodor, J.~N. Guenther, S.~D. Katz, P.~Parotto, A.~P\'asztor,
  D.~Peszny\'ak, K.~K. Szab\'o, C.~H. Wong, {High order fluctuations of
  conserved charges in the continuum limit}, EPJ Web Conf. 296 (2024) 14006.
\newblock \href {https://doi.org/10.1051/epjconf/202429614006}
  {\path{doi:10.1051/epjconf/202429614006}}.

\bibitem{Bazavov:2012vg}
A.~Bazavov, H.~Ding, P.~Hegde, O.~Kaczmarek, F.~Karsch, et~al., {Freeze-out
  Conditions in Heavy Ion Collisions from QCD Thermodynamics}, Phys. Rev. Lett.
  109 (2012) 192302.
\newblock \href {http://arxiv.org/abs/1208.1220} {\path{arXiv:1208.1220}},
  \href {https://doi.org/10.1103/PhysRevLett.109.192302}
  {\path{doi:10.1103/PhysRevLett.109.192302}}.

\bibitem{Haque:2020eyj}
N.~Haque, M.~Strickland, {Next-to-next-to leading-order hard-thermal-loop
  perturbation-theory predictions for the curvature of the QCD phase transition
  line}, Phys. Rev. C 103~(3) (2021) 031901.
\newblock \href {http://arxiv.org/abs/2011.06938} {\path{arXiv:2011.06938}},
  \href {https://doi.org/10.1103/PhysRevC.103.L031901}
  {\path{doi:10.1103/PhysRevC.103.L031901}}.

\bibitem{Cea:2015cya}
P.~Cea, L.~Cosmai, A.~Papa, {Critical line of 2+1 flavor QCD: Toward the
  continuum limit}, Phys. Rev. D 93~(1) (2016) 014507.
\newblock \href {http://arxiv.org/abs/1508.07599} {\path{arXiv:1508.07599}},
  \href {https://doi.org/10.1103/PhysRevD.93.014507}
  {\path{doi:10.1103/PhysRevD.93.014507}}.

\bibitem{Bonati:2015bha}
C.~Bonati, M.~D'Elia, M.~Mariti, M.~Mesiti, F.~Negro, F.~Sanfilippo, {Curvature
  of the chiral pseudocritical line in QCD: Continuum extrapolated results},
  Phys. Rev. D 92~(5) (2015) 054503.
\newblock \href {http://arxiv.org/abs/1507.03571} {\path{arXiv:1507.03571}},
  \href {https://doi.org/10.1103/PhysRevD.92.054503}
  {\path{doi:10.1103/PhysRevD.92.054503}}.

\bibitem{Bonati:2018nut}
C.~Bonati, M.~D'Elia, F.~Negro, F.~Sanfilippo, K.~Zambello, {Curvature of the
  pseudocritical line in QCD: Taylor expansion matches analytic continuation},
  Phys. Rev. D 98~(5) (2018) 054510.
\newblock \href {http://arxiv.org/abs/1805.02960} {\path{arXiv:1805.02960}},
  \href {https://doi.org/10.1103/PhysRevD.98.054510}
  {\path{doi:10.1103/PhysRevD.98.054510}}.

\bibitem{Borsanyi:2020fev}
S.~Borsanyi, Z.~Fodor, J.~N. Guenther, R.~Kara, S.~D. Katz, P.~Parotto,
  A.~Pasztor, C.~Ratti, K.~K. Szabo, {QCD Crossover at Finite Chemical
  Potential from Lattice Simulations}, Phys. Rev. Lett. 125~(5) (2020) 052001.
\newblock \href {http://arxiv.org/abs/2002.02821} {\path{arXiv:2002.02821}},
  \href {https://doi.org/10.1103/PhysRevLett.125.052001}
  {\path{doi:10.1103/PhysRevLett.125.052001}}.

\bibitem{HotQCD:2018pds}
A.~Bazavov, et~al., {Chiral crossover in QCD at zero and non-zero chemical
  potentials}, Phys. Lett. B 795 (2019) 15--21.
\newblock \href {http://arxiv.org/abs/1812.08235} {\path{arXiv:1812.08235}},
  \href {https://doi.org/10.1016/j.physletb.2019.05.013}
  {\path{doi:10.1016/j.physletb.2019.05.013}}.

\bibitem{Bonati:2014rfa}
C.~Bonati, M.~D'Elia, M.~Mariti, M.~Mesiti, F.~Negro, F.~Sanfilippo, {Curvature
  of the chiral pseudocritical line in QCD}, Phys. Rev. D 90~(11) (2014)
  114025.
\newblock \href {http://arxiv.org/abs/1410.5758} {\path{arXiv:1410.5758}},
  \href {https://doi.org/10.1103/PhysRevD.90.114025}
  {\path{doi:10.1103/PhysRevD.90.114025}}.

\bibitem{Lebedev:1989ev}
V.~V. Lebedev, A.~V. Smilga, {Spectrum of Quark - Gluon Plasma}, Annals Phys.
  202 (1990) 229--270.
\newblock \href {https://doi.org/10.1016/0003-4916(90)90225-D}
  {\path{doi:10.1016/0003-4916(90)90225-D}}.

\bibitem{Lebedev:1990un}
V.~V. Lebedev, A.~V. Smilga, {On anomalous damping in quark - gluon plasma},
  Phys. Lett. B 253 (1991) 231--236.
\newblock \href {https://doi.org/10.1016/0370-2693(91)91389-D}
  {\path{doi:10.1016/0370-2693(91)91389-D}}.

\bibitem{Lebedev:1990kt}
V.~V. Lebedev, A.~V. Smilga, {Anomalous damping in plasma}, Physica A 181
  (1992) 187--220.

\bibitem{Braaten:1990ee}
E.~Braaten, {Diagnosis and treatment of the plasmon problem of hot QCD}, Nucl.
  Phys. B Proc. Suppl. 23 (1991) 351--361.
\newblock \href {https://doi.org/10.1016/0920-5632(91)90703-H}
  {\path{doi:10.1016/0920-5632(91)90703-H}}.

\bibitem{Burgess:1991wc}
C.~P. Burgess, A.~L. Marini, {The Damping of energetic gluons and quarks in
  high temperature QCD}, Phys. Rev. D 45 (1992) 17--20.
\newblock \href {http://arxiv.org/abs/hep-th/9109051}
  {\path{arXiv:hep-th/9109051}}, \href
  {https://doi.org/10.1103/PhysRevD.45.R17}
  {\path{doi:10.1103/PhysRevD.45.R17}}.

\bibitem{Baier:1991dy}
R.~Baier, G.~Kunstatter, D.~Schiff, {High temperature fermion propagator:
  Resummation and gauge dependence of the damping rate}, Phys. Rev. D 45 (1992)
  R4381--R4384.
\newblock \href {https://doi.org/10.1103/PhysRevD.45.R4381}
  {\path{doi:10.1103/PhysRevD.45.R4381}}.

\bibitem{Nakkagawa:1992ew}
H.~Nakkagawa, A.~Niegawa, B.~Pire, {Resolution of the gauge dependence problem
  of the fermion damping rate in hot gauge theories}, Phys. Lett. B 294 (1992)
  396--402.
\newblock \href {https://doi.org/10.1016/0370-2693(92)91540-P}
  {\path{doi:10.1016/0370-2693(92)91540-P}}.

\bibitem{Rebhan:1992ca}
A.~Rebhan, {Comment on `Damping of energetic gluons and quarks in high
  temperature QCD'}, Phys. Rev. D 46 (1992) 482--483.
\newblock \href {http://arxiv.org/abs/hep-ph/9203211}
  {\path{arXiv:hep-ph/9203211}}, \href
  {https://doi.org/10.1103/PhysRevD.46.482}
  {\path{doi:10.1103/PhysRevD.46.482}}.

\bibitem{Kobes:1992ys}
R.~Kobes, G.~Kunstatter, K.~Mak, {Fermion damping in hot gauge theories}, Phys.
  Rev. D 45 (1992) 4632--4639.
\newblock \href {https://doi.org/10.1103/PhysRevD.45.4632}
  {\path{doi:10.1103/PhysRevD.45.4632}}.

\bibitem{Altherr:1992ti}
T.~Altherr, E.~Petitgirard, T.~del Rio~Gaztelurrutia, {Damping rate of a moving
  fermion}, Phys. Rev. D 47 (1993) 703--710.
\newblock \href {https://doi.org/10.1103/PhysRevD.47.703}
  {\path{doi:10.1103/PhysRevD.47.703}}.

\bibitem{Baier:1992bv}
R.~Baier, H.~Nakkagawa, A.~Niegawa, {An Issue in evaluating the damping rate of
  an energetic fermion in a hot plasma}, Can. J. Phys. 71 (1993) 205--207.
\newblock \href {https://doi.org/10.1139/p93-032} {\path{doi:10.1139/p93-032}}.

\bibitem{Gyulassy:1990ye}
M.~Gyulassy, M.~Plumer, {Jet Quenching in Dense Matter}, Phys. Lett. B 243
  (1990) 432--438.
\newblock \href {https://doi.org/10.1016/0370-2693(90)91409-5}
  {\path{doi:10.1016/0370-2693(90)91409-5}}.

\bibitem{Bjorken:1982tu}
J.~D. Bjorken, {Energy Loss of Energetic Partons in Quark - Gluon Plasma:
  Possible Extinction of High p(t) Jets in Hadron - Hadron Collisions} (8
  1982).

\bibitem{Ichimaru73}
S.~Ichimaru, {Basic Principles of Plasma Physics}, Benjamin, Reading, New York,
  1973.
\newblock \href {https://doi.org/10.1201/9780429502118}
  {\path{doi:10.1201/9780429502118}}.

\bibitem{Mrowczynski:1991da}
S.~Mrowczynski, {Energy loss of a high-energy parton in the quark - gluon
  plasma}, Phys. Lett. B 269 (1991) 383--388.
\newblock \href {https://doi.org/10.1016/0370-2693(91)90188-V}
  {\path{doi:10.1016/0370-2693(91)90188-V}}.

\bibitem{Koike:1992xs}
Y.~Koike, T.~Matsui, {Passage of high-energy partons through a quark - gluon
  plasma}, Phys. Rev. D 45 (1992) 3237--3251.
\newblock \href {https://doi.org/10.1103/PhysRevD.45.3237}
  {\path{doi:10.1103/PhysRevD.45.3237}}.

\bibitem{Svetitsky:1987gq}
B.~Svetitsky, {Diffusion of charmed quarks in the quark-gluon plasma}, Phys.
  Rev. D37 (1988) 2484--2491.
\newblock \href {https://doi.org/10.1103/PhysRevD.37.2484}
  {\path{doi:10.1103/PhysRevD.37.2484}}.

\bibitem{GolamMustafa:1997id}
M.~Golam~Mustafa, D.~Pal, D.~Kumar~Srivastava, {Propagation of charm quarks in
  equilibrating quark - gluon plasma}, Phys. Rev. C 57 (1998) 889--898,
  [Erratum: Phys.Rev.C 57, 3499--3499 (1998)].
\newblock \href {http://arxiv.org/abs/nucl-th/9706001}
  {\path{arXiv:nucl-th/9706001}}, \href
  {https://doi.org/10.1103/PhysRevC.57.3499}
  {\path{doi:10.1103/PhysRevC.57.3499}}.

\bibitem{Kalman61}
G.~Kalman, A.~Ron, {Interaction of a test particle with a plasma: Part II.
  Energy loss of the test particle}, Ann. Phys. (NY) 16 (1961) 118.
\newblock \href {https://doi.org/10.1016/0003-4916(61)90183-X}
  {\path{doi:10.1016/0003-4916(61)90183-X}}.

\bibitem{Adil:2006ei}
A.~Adil, M.~Gyulassy, W.~A. Horowitz, S.~Wicks, {Collisional Energy Loss of Non
  Asymptotic Jets in a QGP}, Phys. Rev. C 75 (2007) 044906.
\newblock \href {http://arxiv.org/abs/nucl-th/0606010}
  {\path{arXiv:nucl-th/0606010}}, \href
  {https://doi.org/10.1103/PhysRevC.75.044906}
  {\path{doi:10.1103/PhysRevC.75.044906}}.

\bibitem{Wang:2001cs}
E.~Wang, X.-N. Wang, {Parton energy loss with detailed balance}, Phys. Rev.
  Lett. 87 (2001) 142301.
\newblock \href {http://arxiv.org/abs/nucl-th/0106043}
  {\path{arXiv:nucl-th/0106043}}, \href
  {https://doi.org/10.1103/PhysRevLett.87.142301}
  {\path{doi:10.1103/PhysRevLett.87.142301}}.

\bibitem{Gunion:1981qs}
J.~F. Gunion, G.~Bertsch, {HADRONIZATION BY COLOR BREMSSTRAHLUNG}, Phys. Rev. D
  25 (1982) 746.
\newblock \href {https://doi.org/10.1103/PhysRevD.25.746}
  {\path{doi:10.1103/PhysRevD.25.746}}.

\bibitem{Srivastava:1996rz}
D.~K. Srivastava, M.~G. Mustafa, B.~Muller, {Chemical equilibration of an
  expanding quark - gluon plasma}, Phys. Lett. B 396 (1997) 45--49.
\newblock \href {http://arxiv.org/abs/hep-ph/9608424}
  {\path{arXiv:hep-ph/9608424}}, \href
  {https://doi.org/10.1016/S0370-2693(97)00090-7}
  {\path{doi:10.1016/S0370-2693(97)00090-7}}.

\bibitem{Das:2010hs}
S.~K. Das, J.-e. Alam, {Soft gluon multiplicity distribution revisited}, Phys.
  Rev. D 82 (2010) 051502.
\newblock \href {http://arxiv.org/abs/1007.4405} {\path{arXiv:1007.4405}},
  \href {https://doi.org/10.1103/PhysRevD.82.051502}
  {\path{doi:10.1103/PhysRevD.82.051502}}.

\bibitem{Abir:2010kc}
R.~Abir, C.~Greiner, M.~Martinez, M.~G. Mustafa, {Generalisation of
  Gunion-Bertsch Formula for Soft Gluon Emission}, Phys. Rev. D 83 (2011)
  011501.
\newblock \href {http://arxiv.org/abs/1011.4638} {\path{arXiv:1011.4638}},
  \href {https://doi.org/10.1103/PhysRevD.83.011501}
  {\path{doi:10.1103/PhysRevD.83.011501}}.

\bibitem{Dokshitzer:1991fd}
Y.~L. Dokshitzer, V.~A. Khoze, S.~I. Troian, {On specific QCD properties of
  heavy quark fragmentation ('dead cone')}, J. Phys. G 17 (1991) 1602--1604.
\newblock \href {https://doi.org/10.1088/0954-3899/17/10/023}
  {\path{doi:10.1088/0954-3899/17/10/023}}.

\bibitem{Dokshitzer:2001zm}
Y.~L. Dokshitzer, D.~E. Kharzeev, {Heavy quark colorimetry of QCD matter},
  Phys. Lett. B 519 (2001) 199--206.
\newblock \href {http://arxiv.org/abs/hep-ph/0106202}
  {\path{arXiv:hep-ph/0106202}}, \href
  {https://doi.org/10.1016/S0370-2693(01)01130-3}
  {\path{doi:10.1016/S0370-2693(01)01130-3}}.

\bibitem{Abir:2011jb}
R.~Abir, C.~Greiner, M.~Martinez, M.~G. Mustafa, J.~Uphoff, {Soft gluon
  emission off a heavy quark revisited}, Phys. Rev. D 85 (2012) 054012.
\newblock \href {http://arxiv.org/abs/1109.5539} {\path{arXiv:1109.5539}},
  \href {https://doi.org/10.1103/PhysRevD.85.054012}
  {\path{doi:10.1103/PhysRevD.85.054012}}.

\bibitem{Mustafa:1997pm}
M.~G. Mustafa, D.~Pal, D.~K. Srivastava, M.~Thoma, {Radiative energy loss of
  heavy quarks in a quark gluon plasma}, Phys. Lett. B 428 (1998) 234--240.
\newblock \href {http://arxiv.org/abs/nucl-th/9711059}
  {\path{arXiv:nucl-th/9711059}}, \href
  {https://doi.org/10.1016/S0370-2693(98)00429-8}
  {\path{doi:10.1016/S0370-2693(98)00429-8}}.

\bibitem{Abir:2012pu}
R.~Abir, U.~Jamil, M.~G. Mustafa, D.~K. Srivastava, {Heavy quark energy loss
  and D-mesons in RHIC and LHC energies}, Phys. Lett. B 715 (2012) 183--189.
\newblock \href {http://arxiv.org/abs/1203.5221} {\path{arXiv:1203.5221}},
  \href {https://doi.org/10.1016/j.physletb.2012.07.044}
  {\path{doi:10.1016/j.physletb.2012.07.044}}.

\bibitem{Armesto:2003jh}
N.~Armesto, C.~A. Salgado, U.~A. Wiedemann, {Medium induced gluon radiation off
  massive quarks fills the dead cone}, Phys. Rev. D 69 (2004) 114003.
\newblock \href {http://arxiv.org/abs/hep-ph/0312106}
  {\path{arXiv:hep-ph/0312106}}, \href
  {https://doi.org/10.1103/PhysRevD.69.114003}
  {\path{doi:10.1103/PhysRevD.69.114003}}.

\bibitem{Wiedemann:2000ez}
U.~A. Wiedemann, {Transverse dynamics of hard partons in nuclear media and the
  QCD dipole}, Nucl. Phys. B 582 (2000) 409--450.
\newblock \href {http://arxiv.org/abs/hep-ph/0003021}
  {\path{arXiv:hep-ph/0003021}}, \href
  {https://doi.org/10.1016/S0550-3213(00)00286-8}
  {\path{doi:10.1016/S0550-3213(00)00286-8}}.

\bibitem{Wiedemann:2000za}
U.~A. Wiedemann, {Gluon radiation off hard quarks in a nuclear environment:
  Opacity expansion}, Nucl. Phys. B 588 (2000) 303--344.
\newblock \href {http://arxiv.org/abs/hep-ph/0005129}
  {\path{arXiv:hep-ph/0005129}}, \href
  {https://doi.org/10.1016/S0550-3213(00)00457-0}
  {\path{doi:10.1016/S0550-3213(00)00457-0}}.

\bibitem{Wiedemann:2000tf}
U.~A. Wiedemann, {Jet quenching versus jet enhancement: A Quantitative study of
  the BDMPS-Z gluon radiation spectrum}, Nucl. Phys. A 690 (2001) 731--751.
\newblock \href {http://arxiv.org/abs/hep-ph/0008241}
  {\path{arXiv:hep-ph/0008241}}, \href
  {https://doi.org/10.1016/S0375-9474(01)00362-1}
  {\path{doi:10.1016/S0375-9474(01)00362-1}}.

\bibitem{Gyulassy:1999zd}
M.~Gyulassy, P.~Levai, I.~Vitev, {Jet quenching in thin quark gluon plasmas. 1.
  Formalism}, Nucl. Phys. B 571 (2000) 197--233.
\newblock \href {http://arxiv.org/abs/hep-ph/9907461}
  {\path{arXiv:hep-ph/9907461}}, \href
  {https://doi.org/10.1016/S0550-3213(99)00713-0}
  {\path{doi:10.1016/S0550-3213(99)00713-0}}.

\bibitem{Gyulassy:2000fs}
M.~Gyulassy, P.~Levai, I.~Vitev, {NonAbelian energy loss at finite opacity},
  Phys. Rev. Lett. 85 (2000) 5535--5538.
\newblock \href {http://arxiv.org/abs/nucl-th/0005032}
  {\path{arXiv:nucl-th/0005032}}, \href
  {https://doi.org/10.1103/PhysRevLett.85.5535}
  {\path{doi:10.1103/PhysRevLett.85.5535}}.

\bibitem{Gyulassy:2000er}
M.~Gyulassy, P.~Levai, I.~Vitev, {Reaction operator approach to nonAbelian
  energy loss}, Nucl. Phys. B 594 (2001) 371--419.
\newblock \href {http://arxiv.org/abs/nucl-th/0006010}
  {\path{arXiv:nucl-th/0006010}}, \href
  {https://doi.org/10.1016/S0550-3213(00)00652-0}
  {\path{doi:10.1016/S0550-3213(00)00652-0}}.

\bibitem{Gyulassy:2001nm}
M.~Gyulassy, P.~Levai, I.~Vitev, {Jet tomography of Au+Au reactions including
  multigluon fluctuations}, Phys. Lett. B 538 (2002) 282--288.
\newblock \href {http://arxiv.org/abs/nucl-th/0112071}
  {\path{arXiv:nucl-th/0112071}}, \href
  {https://doi.org/10.1016/S0370-2693(02)01990-1}
  {\path{doi:10.1016/S0370-2693(02)01990-1}}.

\bibitem{Djordjevic:2003zk}
M.~Djordjevic, M.~Gyulassy, {Heavy quark radiative energy loss in QCD matter},
  Nucl. Phys. A 733 (2004) 265--298.
\newblock \href {http://arxiv.org/abs/nucl-th/0310076}
  {\path{arXiv:nucl-th/0310076}}, \href
  {https://doi.org/10.1016/j.nuclphysa.2003.12.020}
  {\path{doi:10.1016/j.nuclphysa.2003.12.020}}.

\bibitem{Djordjevic:2004nq}
M.~Djordjevic, M.~Gyulassy, S.~Wicks, {The Charm and beauty of RHIC and LHC},
  Phys. Rev. Lett. 94 (2005) 112301.
\newblock \href {http://arxiv.org/abs/hep-ph/0410372}
  {\path{arXiv:hep-ph/0410372}}, \href
  {https://doi.org/10.1103/PhysRevLett.94.112301}
  {\path{doi:10.1103/PhysRevLett.94.112301}}.

\bibitem{Majumder:2010qh}
A.~Majumder, M.~Van~Leeuwen, {The Theory and Phenomenology of Perturbative QCD
  Based Jet Quenching}, Prog. Part. Nucl. Phys. 66 (2011) 41--92.
\newblock \href {http://arxiv.org/abs/1002.2206} {\path{arXiv:1002.2206}},
  \href {https://doi.org/10.1016/j.ppnp.2010.09.001}
  {\path{doi:10.1016/j.ppnp.2010.09.001}}.

\end{thebibliography}
	

	\end{document}